\documentclass[preprint,12pt,number,sort&compress]{elsarticle}
\pdfoutput=1
\usepackage{graphicx}
\usepackage{latexsym,amssymb,amsmath,bm,bbm,times}
\usepackage[caption=false]{subfig}
\usepackage{xcolor}
\usepackage{mathrsfs}
\usepackage{hyperref}
\usepackage[hang]{footmisc}
\usepackage{gensymb}
\usepackage{siunitx}
\journal{Annual Review of Heat Transfer}
\begin{document}
\begin{frontmatter}
\title{\textbf{The dependent scattering effect on radiative properties of micro/nanoscale discrete disordered media}}
\author{B. X. ~Wang}
\author{C. Y. ~Zhao\corref{cor1}}
\ead{Changying.zhao@sjtu.edu.cn}
%\thanks{}
\address{Institute of Engineering Thermophysics, School of Mechanical Engineering, Shanghai Jiao Tong University, Shanghai, 200240, P. R. China}
\cortext[cor1]{Corresponding author}
		
\begin{abstract}
The dependent scattering effect (DSE), which arises from the wave nature of electromagnetic radiation, is a critical mechanism affecting the radiative properties of micro/nanoscale discrete disordered media (DDM). In the last a few decades, the approximate nature of radiative transfer equation (RTE) leads to a plethora of investigations of the DSE in various DDM, ranging from fluidized beds, photonic glass, colloidal suspensions and snow packs, etc. In this article, we give a general overview on the theoretical, numerical and experimental methods and progresses in the study of the DSE. We first present a summary of the multiple scattering theory of electromagnetic waves, including the analytic wave theory and Foldy-Lax equations, as well as its relationship with the RTE. Then we describe in detail the physical mechanisms that are critical to DSE and relevant theoretical considerations as well as numerical modeling methods. Experimental approaches to probe the radiative properties and relevant progresses in the experimental investigations of the DSE are also discussed. In addition, we give a brief review on the studies on the DSE and other relevant interference phenomena in mesoscopic physics and atomic physics, especially the coherent backscattering cone, Anderson localization, as well as the statistics and correlations in disordered media. We expect this review can provide profound and interdisciplinary insights to the understanding and manipulation of the DSE in disordered media for thermal engineering applications.
\end{abstract}

\end{frontmatter}

\tableofcontents
%
%\documentclass[article,jeh]{beg_32}             %  final version
%% Use option "equation" for numbering equation as section
%
%%\count0=115
%\usepackage[hang]{footmisc}
%\usepackage{gensymb}
%\usepackage{siunitx}
%\setlength{\footnotemargin}{0in}
%\frenchspacing
%\fancypagestyle{plain}{%
%  \fancyhf{}
%  \fancyhead[R]{\small {\it \jname}, x(x):\thepage--\pageref{LastPage} (\myyear\today)}
%  \fancyfoot[R]{\small\bf\thepage }
%  \fancyfoot[L]{\fottitle}
%  }
%%\fancypage{\fbox}{}
%\renewcommand{\dmyy}{20}
%\renewcommand{\myyear}{2020}
%\renewcommand{\today}{}

\newpage
\section{Introduction}
It is known that when the size of microscopic inhomogeneities in a heterogeneous medium is comparable to the wavelength of electromagnetic waves, significant wave interference phenomena can occur \cite{bornandwolf}. As a consequence, this kind of heterogeneous media can usually interact with electromagnetic waves in a much more complicated and stronger manner than conventional bulk and homogeneous materials that typically possess inhomogeneities on the scale much smaller than the wavelength, and thus they have a great potential in controlling the propagation of electromagnetic waves. In particular, micro/nanoscale disordered media, which have inhomogeneities with characteristic sizes ranging from a few tens of nanometers to several hundred microns, can significantly affect the propagation of thermal radiation, whose wavelength usually lies in the range of $100\mathrm{nm}$-$100\mathrm{\mu m}$ for objects in heat transfer applications.
Therefore, micro/nanoscale disordered media, such as porous dielectric media with micropores and voids \cite{mindianeyCMS2017}, particulate media containing micro- and nanoparticles \cite{garciaADMA2007}, colloidal suspensions of nanoparticles \cite{rojasochoaPRL2004}, many kinds of coatings \cite{kulkarniActMat2003}, foams \cite{luAM1998}, fibers \cite{jiangACSAMI2018} and soot aggregates \cite{sunIJHMT2011}, have been widely applied in controlling thermal radiation transfer. For example, porous silicon carbide (SiC) material can be utilized as high-efficiency solar absorbers \cite{tanSE2014}, porous zirconia ($\mathrm{ZrO_2}$) coatings are enormously used to provide thermal protection for the metallic components of gas turbines \cite{yangIJHMT2013}, and polymer films containing randomly distributed silica ($\mathrm{SiO_2}$) nanoparticles show an excellent performance for radiative cooling \cite{zhaiScience2017}, etc. 

Since thermal radiation transfer in those micro/nanoscale disordered media is strongly affected by the microscopic structures, in order to tailor their radiative properties and functionalities, a full understanding of underlying physical mechanisms of electromagnetic wave transport as well as the relationship between micro/nanostructures and radiative properties is of critical importance. 
However, in such media, radiation is scattered and absorbed in a very complicated way, which brings difficulties to theoretical and experimental investigations. Conventionally, the propagation of radiation is described by the radiative transfer equation (RTE) in the mesoscopic scale. The radiative properties entering the RTE, including the scattering coefficient $\kappa_s$, absorption coefficient $\kappa_a$ and phase function $P(\mathbf{\Omega}',\mathbf{\Omega})$ (where $\mathbf{\Omega}'$ and $\mathbf{\Omega}$ denote incident and scattered directions, respectively), depend on the microstructures as well as the permittivity and permeability of the composing materials. 
In particular, for disordered media consisting of discrete scatterers, i.e., discrete disordered media (DDM), the radiative properties are usually theoretically predicted under the independent scattering approximation (ISA), i.e., in which each discrete inclusion is assumed to scatter electromagnetic waves independently as if no other inclusions exist, i.e., without any inter-scatterer interference effects \cite{lagendijk1996resonant,VanRossum1998,tsang2004scattering,sheng2006introduction,mishchenko2010polarimetric}.
ISA is valid only when the scatterers are far-apart from each other (i.e., the far-field assumption) and no interparticle correlations exist (i.e., independent scatterers) \cite{VanRossum1998,mishchenko2006multiple,tsang2004scattering,sheng2006introduction,mishchenko2010polarimetric}. When the two conditions are violated, the scattered waves from different scatterers interfere substantially and consequently, ISA fails  \cite{garciaPRA2008,wangIJHMT2015,Naraghi2015,wangIJHMT2018}. In this circumstance, we call the radiation scattering process from a scatterer is ``dependent" of the presence of other scatterers. This fact leads many researchers to the considerations on the dependent scattering effect (DSE) in order to correctly predict the radiative properties of DDM. 
%\cite{tien1987thermal,kumar1990dependent,leeJTHT1992,ivezicIJHMT1996,durantJOSAA2007,garciaPRA2008,nguyenOE2013,wangIJHMT2015,Naraghi2015,maJQSRT2017,wangPRA2018,wangIJHMT2018,wangJAP2018}

%In the 1930s, Ryde and Cooper ``have laid the basis for and developed a quantitative theory of approximation of the interference, and they find good agreement between calculations and experiment up to small spacing between particles." (We did not found there are such theories in these two papers).
It was generally well-known to paint and paper coating technologists for a long time that high-concentration packing of pigment particles in a white paint layer can lead to a decrease in its opacity (or "hiding power"), e.g., Refs.\cite{tinsleyJOCCA1949,stieg1959effect}, due to the interference of light scattered by neighboring particles. In his noted book on light scattering \cite{hulst1957}, Hendrik C. van Hulst mentioned that the mutual distance between the particles of three times the particle radius may be a sufficient condition for independent scattering. In the 1960s, several experimental works were carried out for optically scattering turbid media composed of dielectric particles, like $\mathrm{TiO_2}$ and polystyrene (PS) particles, and found at certain particle concentrations the dependent scattering effect became prominent, but no clear criterion that could quantitatively determine the departure from ISA was obtained \cite{churchillDFC1960,hardingJOSA1960,blevinJOSA1961a,rozenberg1962optical}. For example, Churchill \textit{et al.} \cite{churchillDFC1960} found a critical value of $\delta/d\sim1.7$, above which no interference effect was observed, where $d$ is the particle diameter and $\delta$ is the the center-to-center distance. Harding \textit{et al.} \cite{hardingJOSA1960} intended to achieve optimal scattering properties at minimum cost for paint films, and they revealed critical concentrations above which the ability of a particle to scatter light may be precipitously reduced because of optical ``overlap" with its neighbors. Blevin \textit{et al.} \cite{blevinJOSA1961a} experimentally showed that a high concentrations the reflectance falls due to interparticle interferences. These works were published in optics and chemistry journals. On the other hand, the general theory of multiple scattering of classical waves, including electromagnetic and acoustic waves, was already established even earlier by physicists \cite{foldyPR1945,laxRMP1951,laxPR1952,twerskyJASA1952}, which can provide rigorous treatments for the dependent scattering effect. However, it was not noticed and applied by these works due to the lack of easy-to-use practical formulas. 

To the best of our knowledge, in the thermal radiation heat transfer community, the DSE was firstly investigated by Hoyt C. Hottel\footnote{He was a professor of chemical engineering who mainly studied radiative heat transfer and combustion.} and co-workers in the 1970s \cite{hottelJHT1970, hottelAIAAJ1971}. They experimentally measured the bidirectional reflectance and transmittance spectra of monodisperse PS nanosphere suspensions in water confined between parallel glass slides at different optical thicknesses, where sphere diameter was in the range of $0.102-0.53\mathrm{\mu m}$ and volume fraction was varied from $1.3\times10^{-6}$ to $0.295$. By comparing experimental data with ISA predictions, empirical criteria for the dependent scattering regime were also obtained, where the critical parameter was the clearance-wavelength ratio ($c/\lambda$), where the clearance $c=\delta-d$ for spherical particles and $\lambda$ is the wavelength of incident light \cite{hottelAIAAJ1971}. In particular, an experimental correlation between the effective scattering efficiency $Q_\mathrm{s}$ and $c/\lambda$ was established accordingly. 

Later, in the 1980s and 1990s, Chang-Lin Tien and co-workers conducted comprehensive studies on the dependent scattering effect in particulate media \cite{brewsterJHT1982,yamadaJHT1986,cartignyJHT1986,drolenJTHT1987,tienARHT1987,tienJHT1988,kumar1990dependent}. Notably, Tien and Drolen published the very first article of \textit{Annual Review of Heat Transfer}, which reviewed the independent and dependent scattering of thermal radiation transfer in these media \cite{tienARHT1987}. Later many authors carried out investigations into this effect, Refs. \cite{singhIJHMT1992,wangIJHMT2015,maJQSRT2017}, to name a few. In fact, in the last several decades, many efforts have been made  in the study of this mechanism, not only in the field of radiative heat transfer, but also in the communities of optics, photonics, biomedical engineering, astrophysics, paint industry (visibility), meteorology (atmospheric sciences), remote sensing and mesoscopic physics, since the DSE can happen not only for thermal radiation, but also for any types of electromagnetic waves, including visible light, terahertz waves and microwaves, only if the packing density is high enough and the distance between adjacent scatterers is comparable with or smaller than the wavelength. Moreover, in the past years, significant progresses are made in micro- and nano-fabrication, nano-optics and photonics, lasers, modulators and detectors, as well as computational capabilities of modern computers, which further reshape the way of studying radiative transfer and the dependent scattering effect.

Since the DSE is a very generic effect, references on this topic are disseminated in the literature covering a wide range of disciplines. In this regard, the goal of the present article is three-fold. Firstly of all, this article intends to give an overview of the physical mechanisms of the DSE and related interference phenomena. Secondly, this article aims to systematically summarize the theoretical and numerical modeling methods as well as experimental approaches to investigate the DSE. Thirdly, this article is also devoted to introducing and discussing important advances about the DSE in the last several decades. As a result, this article is organized as follows. In Section \ref{concepts}, we give an introduction of basic concepts and definitions related to dependent scattering in DDM (or discrete random media). In Section \ref{theory}, the multiple scattering theory of electromagnetic waves is briefly summarized, which provides a fundamental tool to understand and deal with the complicated radiative transfer phenomena in discrete disordered media. In this section, we do not present the technical details of these theoretical treatments and the derivation procedures because a variety of literature can be referred to. More importantly, our main aim in this review is to depict a general picture of the dependent scattering effect in the context of radiative transfer in disordered media by emphasizing the physical significance of these theoretical methods rather than technical details. In Section \ref{theoryandnumerical}, we first describe some basic mechanisms involved in the DSE, including the far-field DSE, near-field DSE, recurrent scattering, structural correlations and the effect of absorbing host media, and summarize relevant theoretical models that deal with them. Then numerical methods to model the DSE are summarized, including the supercell method, the representative volume element method and the direct numerical simulation method. In Section \ref{expsec}, we review the experimental investigations of the DSE, including the coherent transmittance measurement and the measurements of total, angle-resolved and time-resolved transmittance and reflectance. In Section \ref{mesointerference}, the dependent scattering effects and other related interference phenomena in mesoscopic physics and atomic physics are introduced, including the coherent backscattering cone and Anderson localization, the statistics and correlations in disordered media, the breakdown of mean-field optics and the structural correlations in cold atoms, in order to establish a bridge among different communities that all investigate disordered media. We expect this review can provide profound and interdisciplinary insights to the understanding and manipulation of the DSE in thermal engineering applications.

\section{Basic concepts and definitions}\label{concepts}
Let us start from some basic concepts and definitions. In this section, we first discuss the concept of micro/nanoscale discrete disordered media (DDM), in which several restrictions will be made. Then, in the framework of the radiative transfer equation, we give the definition of the mesoscopic radiative properties and show they are important to the understanding of thermal radiation transport in DDM. On this basis, we proceed to a discussion on the concepts of independent, dependent and multiple scattering, which sometimes are not clearly distinguished in many works.
\subsection{Micro/nanoscale discrete disordered media}\label{DDM_definition}
%The definition of DRM (or using discrete disordered media?) should be clarified in a comprehensive way. It is necessary to refer to the books of Tsang, and papers and monographs by Mishchenko, and recent papers by Mishchenko. The figures are, actually, not necessary, and examples are more compelling than these non-informative figures.
Consider the concept of micro/nanoscale discrete disordered media. By saying ``disordered", we generally refer to the media that are inhomogeneous (or equivalently, heterogeneous) and these inhomogeneities distributed in a disordered way \cite{teixeiraIEEETAP2008}. ``Micro/nanoscale" describes the characteristic size of the inhomogeneities, and "discrete" means that the inhomogeneities are distributed discretely, i.e., not interconnected. An additional implication of "disorder" is the physical process in which the propagation of radiation is forced to deviate from a straight trajectory due to the existence of disordered inhomogeneities in the medium. In a microscopic view, when electromagnetic radiation is illuminated upon an obstacle, which can be an atom, a molecule, a solid or liquid particle, it drives the electric charges in the obstacle into oscillatory motion that can then radiate secondary electromagnetic waves in all directions. This secondary radiation is then called the radiation ``scattered" by the obstacle \cite{bohrenandhuffman}. Therefore the scattering of thermal radiation plays a dominant role in the response of such disordered media. For instance, this is the case for media composed of weakly absorbing dielectrics. Nevertheless, this does not necessarily imply that the total absorptance of the medium is low. An intuitive example is that a thick-enough ($300\mathrm{\mu m}$) porous coating composed of very low-absorbing 8\%wt-yttria stabilized zirconia (8YSZ) material (refractive index is around 2.1) can still achieve a total absorptance of 20\% percent or higher, and in Ref. \cite{wangIJHMT2018} it was shown that a medium consisting of highly-scattering low-absorbing metallic nanoparticles can exhibit substantial total absorption. Since in these media, the multiple scattering of radiation substantially enhances the path length of photons and leads to significant absorption although the intrinsic absorption of the material is low. This prominent feature of DDM actually finds its applications in solar steam generation \cite{hoganNL2014} and solar cells \cite{Vynck2012,galvezJPCC2012}.

\begin{figure}[htbp]
	\centering
	\subfloat{
		\label{photonicglass}
		\includegraphics[width=0.385\linewidth]{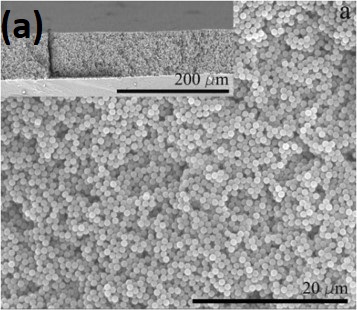}
	}
	\subfloat{
		\label{porouszirconia}
		\includegraphics[width=0.45\linewidth]{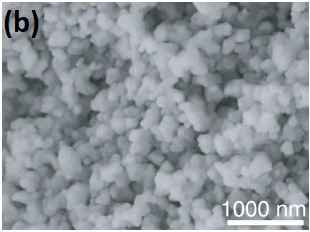}
	}
	\caption{Microstructures of typical discrete disordered media. (a) The SEM image of a coating consisting of 780nm-diameter PS spheres prepared from charged colloidal suspensions. Reprinted with permission from Ref. \cite{garciaADMA2007}. Copyright 2017, Wiley-VCH Verlag GmbH \& Co. KGaA, Weinheim.  (b) The SEM image of a nanoporous zirconia coating. Reprinted with permission from Ref. \cite{svenssonPRL2011}.  Copyright 2011 by the American Physical Society.}\label{complexscatteringmediaexamples}
	
\end{figure}

In disordered media, the microscopic structures distribute randomly without any long-range order, although in which short-range order is possible \cite{liuJOSAB2018,Liew2011,wiersma2013disordered}. They are usually formed in typical bottom-up fabrication methods, for example, spraying, self-assembly, spin coating and so on. Fig.\ref{complexscatteringmediaexamples} shows two examples of DDM. Fig.\ref{photonicglass} is the scanning electron microscopy (SEM) image of a so-called photonic glass consisting of 780nm-diameter polystyrene spheres, which is prepared from the self-assembly process of charged colloidal suspensions. Fig.\ref{porouszirconia} demonstrates a nanoporous coating composed of zirconia parciles that is used to increase light absorption in gas by exploiting the path-length enhancement brought by multiple scattering, where the size of pores is around 115 nm. 

It is important to study radiation transfer in DDM from both fundamental and applied aspects. The fabrication cost of DDM is relatively low compared to ordered micro/nanostructures, and moreover they are usually tolerant to fabrication errors. They are thus more feasible to be widely applied in engineering situations. They also exhibit unique light and radiation transport properties brought by the interplay between electromagnetic interferences and disorder, including weak and strong (Anderson) localization \cite{wiersma1997localization,wiersma2013disordered,Segev2013}, position-dependent diffusion coefficient \cite{yamilovPRL2014}, sub-diffusive \cite{sebbahPRB1993} and super-diffusive \cite{barthelemyNature2008,bertolottiPRL2010} transport behaviors, multimode excitation and broadband operation capabilities \cite{Cao1999,changSR2018}, and noniridescence \cite{xiaoSciAdv2017} (i.e., optical and radiative properties are almost isotropic at all observation directions). Furthermore, by sophisticatedly designing the single scatterer properties and the disordered patterns in these media, many exotic phenomena that only exist in ordered DDM previously can be also realized, for instance, the photonic bandgap \cite{Edagawa2008}. In practice, DDM with engineered micro/nanostructures also exhibit promising applications such as thermal radiation harvesting \cite{Vynck2012,galvezJPCC2012,Liew2016ACSPh,liuJOSAB2018}, conversion \cite{fendEnergy2004,steinfeldSE2005,hoganNL2014,mindianeyCMS2017}, and management \cite{howellPECS1996,siegelMSEA1998,shiLSA2018}. More generally, the study of interference phenomena in DDM have received growing attention and given rise to a rapidly developing field called ``disordered photonics'' \cite{wiersma2013disordered,Rotter2017}, along with applications like  quantum optics \cite{garciaAnnPhys2017}, structural coloration \cite{xiaoSciAdv2017}, and random lasers \cite{Cao1999}, etc. 

Throughout the article, we interchangeably use the terms ``disordered media" and ``random media". This is because for light propagation in random media containing temporally moving scatterers (e.g., nanoparticle suspensions in solutions, flows of blood cells, a cloud of water droplets, air bubbles in water, cold atomic gases, etc.), the scattering processes are usually much faster than the random motions of scatterers, making them seem to reside in fixed positions. In other words, we deal with random media in (quasi-)quenched disorder. In fact, the dynamic responses probed by the dynamic light scattering techniques for various random media indicate the random motion of particles in them occurs in a time scale even much longer than the entire radiative transfer process \cite{pinePRL1988,boasJOSAA1997}\footnote{A simple estimation can be done as follows. Consider a nonabsorbing, optically thick slab with thickness $L=10l_s$, where $l_s$ is the scattering mean free path for this isotropically scattering medium. The average time scale $\tau$ for a pulse to transmit can be estimated under the diffusion equation as $\tau\sim L^2/D$, where $D\sim c_0l_s/3$ is the diffusion coefficient and $c_0$ is the velocity of light in vacuum. We have $\tau\sim l_s/\mathrm{(1\mu m)} ~\mathrm{ns}$. For typical highly scattering media, $l_s$ is on the order of tens to hundreds of microns, leading to $\tau$ on the scale of tens to hundreds of nanoseconds. On the other hand, dynamic light scattering experiments reveal that particle diffusion time under Brownian motion is typically on the scale of several milliseconds \cite{pinePRL1988,goldburgAJP1999}. Therefore the particle movement is indeed very slow compared to the radiative transfer process. Details on the diffusion equation can be found in Section \ref{rteandde} and Section \ref{time_resolved}.}.

\subsection{Radiative properties}\label{RTE_intro}
%Since the characteristic size of the internal structures of micro/nanoscale DDM is usually comparable with or even smaller than the wavelength of thermal radiation, significant electromagnetic wave interferences can occur. In this situation, micro/nanoscale DDM can interact with thermal radiation much more strongly and complicated than conventional bulk materials. As a consequence, it is crucial to establish the relationship between their micro/nanostructures and radiative properties based on the first-principle Maxwell equations with considerations on the electromagnetic wave nature of thermal radiation, rather than the traditional ray tracing treating optically large structures.
In order to design and utilize DDM in practical thermal radiation applications, it is crucial to first understand the physical mechanisms that affect their radiative properties and then provide feasible as well as accurate theoretical modeling methods accordingly. In general, radiative properties for a specific sample of DDM can include the reflectance $R$, absorptance $A$, transmittance $T$, emittance $E$, scattering coefficient $\kappa_s$, absorption coefficient $\kappa_a$, extinction coefficient $\kappa_e=\kappa_s+\kappa_a$ and scattering phase function $P(\mathbf{\Omega}',\mathbf{\Omega})$ \cite{modest2013radiative,howell2015thermal}. Alternatively, for highly scattering, weakly absorbing DDM, two length scales are also important, namely, the photon scattering mean free path $l_s=1/\kappa_s$ and the photon transport mean free path $l_\mathrm{tr}=1/[\kappa_s(1-g)]$, where $g=(1/4\pi)\int_0^\pi P(\mathbf{\Omega}',\mathbf{\Omega})\cos\theta\sin\theta d\theta$ is the asymmetry factor with $\theta$ denoting the angle between $\mathbf{\Omega}'$ and $\mathbf{\Omega}$ \cite{VanRossum1998,lagendijk1996resonant,sheng2006introduction,vanTiggelenRMP2000}. 

%\subsection{Macroscopic and mesoscopic radiative properties}
As a matter of fact, when it comes to the theoretical study of the radiative properties, there is an essential distinction between disordered and ordered media. Generally, theories for ordered structures are more well-established because Bloch theorem for periodic photonic systems can be conveniently applied and studying a single unit cell in combination with the periodic boundary condition is enough to capture the properties of the whole system in most cases, which largely reduces the difficulties of modeling \cite{joannopoulos2011photonic}. In this situation, only macroscopic radiative properties describing the whole radiative response of a sample are of main concern, namely, the first four parameters ($R$, $A$, $T$ and $E$)\footnote{The microscopic electric and magnetic field distributions are also definitely important when studying the underlying physics.}.

In the meanwhile, for disordered media, it is much more demanding to theoretically predict their radiative properties, because the presence of disorder in microscopic structures without any long-range order can induce intricate electromagnetic wave interference behaviors, which not only take place locally between adjacent scatterers, but also occur in the long range that can involve a substantial proportion of the \textit{entire} system \cite{sheng2006introduction,akkermans2007mesoscopic,Naraghi2016,mishchenkoOSAC2019} (as can be seen through the mesoscopic interference phenomena later on). Therefore, the last four parameters ($\kappa_s$, $\kappa_a$, $\kappa_e$ and $P(\mathbf{\Omega}',\mathbf{\Omega})$), are exploited to relate the microscopic scale with the macroscopic scale. In other words, they are used to characterize the propagation of radiation in disordered media in the mesoscopic scale. This is a scale comparable with the scattering/transport mean free path $l\sim 1/\kappa_s$ of photons, which should be much smaller than the macroscopic sample size but much larger than the wavelength \cite{VanRossum1998}\footnote{If the mean free path is comparable or even smaller than the wavelength, we say the system might enter the strong localization regime, as indicated by the Ioffe-Regel criterion \cite{sheng2006introduction}, which will be discussed in latter sections.}.

\textit{The mesoscopic radiative properties are important.} Actually, the mesoscopic radiative properties are defined in the framework of radiative transfer equation (RTE). The RTE for unpolarized radiation (i.e., scalar RTE) under local thermodynamic equilibrium in a statistically homogeneous disordered medium (i.e., spatially constant radiative properties, for simplicity) can be expressed as \cite{tsang2000scattering1,modest2013radiative}:
\begin{equation}\label{RTE_eq}
\frac{dI}{ds}=\kappa_aI_{b}-\kappa_eI+\frac{\kappa_s}{4\pi}\int_{4\pi}I(\mathbf{\Omega}')P(\mathbf{\Omega}',\mathbf{\Omega})d\mathbf{\Omega}',
\end{equation}
where  $I$ is the spectral radiative intensity (here we omit the subscript $\lambda$ wavelength-dependency for convenience),  $I_{b}$ is the spectral radiative intensity of blackbody with a temperature at the local position, $s$ is the transport path length.

When the polarization states of radiation are taken into account, the RTE can be written in its vectorial form as:
\begin{equation}\label{vec_rte}
\frac{d\mathbf{I}}{ds}=\bm{\kappa}_a\mathbf{I}_{b}-\bm{\kappa}_e\mathbf{I}+\frac{1}{4\pi}\int_{4\pi}\mathbf{I}(\mathbf{\Omega}')\mathbf{Z}(\mathbf{\Omega}',\mathbf{\Omega})d\mathbf{\Omega}'
\end{equation}
where $\mathbf{I}=[I~Q~U~V]^\mathrm{T}$ represents the Stokes vector that can describe the polarization properties of radiation. Here $I$ is the radiative intensity the same as that in the scalar RTE,  $Q$ and $U$ describe the degree of linear polarization and its orientation, while $V$ characterizes the degree of circular polarization. $\mathbf{I}_{b}$ is the vector form of blackbody radiation as $[I_b~0~0~0]^\mathrm{T}$ because conventional thermal emission is unpolarized. $\bm{\kappa}_a$, $\bm{\kappa}_e$ and $\mathbf{Z}(\mathbf{\Omega}',\mathbf{\Omega})$ are the absorption, extinction and (unnormalized) phase matrices, respectively \cite{mishchenkoAO2002,mishchenko2006multiple,zhaoJQSRT2015}. For a detailed description of the Stokes vector and its connection to the monochromatic transverse electromagnetic waves as well as these matrices, one can refer to Ref.\cite{mishchenko2006multiple} and references therein. 

In this article, we focus on scalar radiative transfer processes and scalar radiative properties, on which most studies of the DSE are conducted. One can refer to Refs. \cite{tishkovetsJQSRT2006,tishkovetsJQSRT2011,tsang2004scattering,doicuJQSRT2019c} for a few discussions on the role of DSE in vectorial (polarized) radiative transfer in densely packed DDM.  And in the following, the term concerning thermally emitted radiation $\kappa_aI_{b}$ is omitted, since throughout the article we only treat room temperature DDM with thermal emission far smaller than the incident radiation energy (namely, the cold medium assumption \cite{modest2013radiative}).  

In RTE, radiation is treated as classical particles or energy bundles without phase, i.e., the wave nature in the transport process is usually omitted. More specifically, it is derived phenomenologically in its initial stage from energy conservation arguments, see Refs. \cite{chandrasekhar1950radiative,modest2013radiative}. However, in the microscopic sense, it is shown RTE is actually an approximate form of Bethe-Salpeter equation for electromagnetic waves  \cite{ishimaru1978book,tsang1985theory,lagendijk1996resonant,VanRossum1998,tsang2004scattering,sheng2006introduction,mishchenko2006multiple}. The latter is an exact equation accounting for all interference phenomena for the transport of electromagnetic field correlation function $\langle\mathbf{E}\mathbf{E}^*\rangle$ in random media, which is originally taken from quantum field theory and exactly equivalent to Maxwell's equations in this context. RTE only takes the ladder diagrams in Bethe-Salpeter equation into account, neglecting all kinds of \textit{microscopic and mesoscopic} interference effects occurring outside the scatterers, and thus can break down in some circumstances \cite{mishchenko2006multiple,mishchenkoPhysrep2016,VanRossum1998,vanTiggelenRMP2000}. In this article, we are mainly dedicated to these interference effects that influence the mesoscopic radiative properties of DDM, especially the dependent scattering effect. And we also would like to emphasize that, throughout this article, RTE and radiative transfer (RT) are different concepts, as mentioned by many other authors \cite{mishchenko2006multiple,tsangJQSRT2019}. The RTE is only a phenomenological equation that only considers the transport of energy bundles without wave effects, while all the physical processes of wave propagation in disordered media governed by Maxwell's equations are implied when we speak of RT.

\subsection{Independent, dependent and multiple scattering}\label{indanddep}

Micro/nanostructure-based theoretical methods to obtain radiative properties of DDM mainly include the Monte Carlo ray-tracing method based geometric optics approximation (MC-RT-GOA) \cite{tancrezIJHMT2004}, and Mie theory combined with independent scattering approximation \cite{dombrovskyIPT2007,bohrenandhuffman}. The former is only applicable for DDM whose characteristic size of microstructures is much larger than the radiation wavelength, for instance, some metallic and ceramic foams \cite{tancrezIJHMT2004} with $\sim100\mathrm{\mu m}$-sized microstructures, where geometric optics approximation is justified. This method is already well-developed by many researchers and can achieve a satisfactory accuracy provided the microstructures are given \cite{tancrezIJHMT2004,randrianalisoaIJHMT2014}. The latter method is more suitable for DDM containing wavelength- and subwavelength-scale discrete micro/nanostructures, which can be treated as spheres or cylinders. In this method, Mie theory is able to account for all electromagnetic interference effects inside an individual scatterer, and ISA assumes that different scatterers in the DDM scatter electromagnetic waves independently, without any consideration of inter-scatterer interference effects. In this approximation, the scattering coefficient, for instance, is calculated through
\begin{equation}
\kappa_s=\sum_{i=1}^{N}n_iC_{\mathrm{s},i},
\end{equation}
where $n_i$ and $C_{\mathrm{s},i}$ are the number density and scattering cross section of the $i$-th type scatterer, respectively, if there are $N$ kinds of scatterers in the entire medium. The scattering cross section can be calculated from Mie theory or other methods treating single scattering problems (which will be discussed below), which can take all electromagnetic wave interference phenomena involving an invididual scatterer into account. Similarly, the absorption coefficient and scattering phase function can be also calculated.

ISA works well for DDM containing dilutely distributed scatterers. For example, for very dilute soot aggregates (volume fraction is around 1\%), radiative properties calculated from Mie theory combined with ISA can result in a good agreement with the experimental measurements \cite{sunIJHMT2011}. Generally, it is widely demonstrated that the ISA is rigorously valid when the following two conditions are simultaneously satisfied \cite{wangJAP2018,wangIJHMT2018}: (1) the scatterers are far-apart from each other (i.e., the far-field assumption, which means the normalized distance $k\delta\gg1$, where $k=2\pi/\lambda$ is the wave number and $\delta$ is the average center-to-center distance between scatterers)\footnote{A question naturally arises, that is, how far is far enough? At least two criteria should be fulfilled. The first is obvious, which requires the inter-scatterer clearance should be much larger than the wavelength $k(\delta-2a)\gg1$. The second criterion depends on the scattering strength of the scatterer. For monodisperse, randomly distributed scatterers, it can be estimated from the scattering mean free path that $kl_s=k/(n_0C_s)\gg1$. Using $n_0\approx1/\delta^3$, we have $\delta\gg\sqrt[3]{C_s/k}$. } and (2) no positional correlations exist  (i.e., scatterers are independently distributed) \cite{mishchenko2006multiple,tsang2004scattering,mishchenkoOE2007,tishkovetsJQSRT2011,mishchenkoPhysrep2016}. In this circumstance, we can assume that at the microscopic scale, the scatterers scatter electromagnetic waves independently without the need to take inter-particle interferences into account  \cite{lagendijk1996resonant,VanRossum1998,tsang2004scattering,sheng2006introduction,akkermans2007mesoscopic}, i.e., the ISA is justified. 

%Without loss of generality, for micro/nanoscale DDM composed of discrete scatterers, such as materials made of randomly distributed/packed zinc oxide ($\mathrm{ZnO}$), silica ($\mathrm{SiO_2}$) and titanium dioxide ($\mathrm{TiO_2}$) nanoparticles \cite{wiersma1997localization,storzerPRL2006,zhaiScience2017,baoSEMSC2017,linACSPhoton2017,xiaoSciAdv2017}, 

%Tien and Drolen \cite{tien1987thermal} said "Dependent scattering occurs when the scattering from a single particle is effected by the presence of its neighbors. This pertains to many engineering heat transfer applications including fluidized and packed beds, microsphere insulations, packed-sphere heat regenerators, reactor fuel pellets, deposited soot layers, conglomerated soot particles, paint layers and other colloidal suspensions." They used size parameter and the clearance-to-wavelength ratio as key parameters to construct a "phase diagram" for independent and dependent scattering regimes.

%In other words, RTE is rigorously valid to describe mesoscopic radiation transport only when ISA is applicable to calculate mesoscopic radiative properties.
%\subsection{The dependent scattering effect (DSE)}\label{DSE_intro}

%However, the large coherence time, elastic scattering and time-reversible characteristics of light guarantee the interference phenomena even in the macroscopic scale. 

However, when one or both of the above far-field and independent-scatterer conditions is violated, the scattered waves from different scatterers can interfere with each other substantially, leading to the failure of ISA \cite{garciaPRA2008,Naraghi2015}. In practice, ISA breaks down for a variety of micro/nanoscale DDM, in which the volume fraction of scatterers (voids, particles or generally, permittivity fluctuations) exceeds 5\% \cite{mishchenkoOL2013,tsangJQSRT2019}. In these media, the distance between adjacent scatterers is usually comparable to or even smaller than the wavelength of radiation, and leads to remarkable interference effects of scattered electromagnetic waves from different scatterers. In addition, the inter-scatterer positional correlations are also important (i.e., the positions of scatterers are not independent of each other any more) \cite{fradenPRL1990,rojasochoaPRL2004}. If we consider a complex scattering medium consisting of randomly packed hard spheres, structural correlations arise because the existence of one hard sphere would create an exclusion volume into which other particles are not allowed to penetrate, which leads to definite phase differences among scattered waves preserving over ensemble average, and modifies the radiative properties. 

\begin{figure}[htbp]
	\flushleft
	\includegraphics[width=1\linewidth]{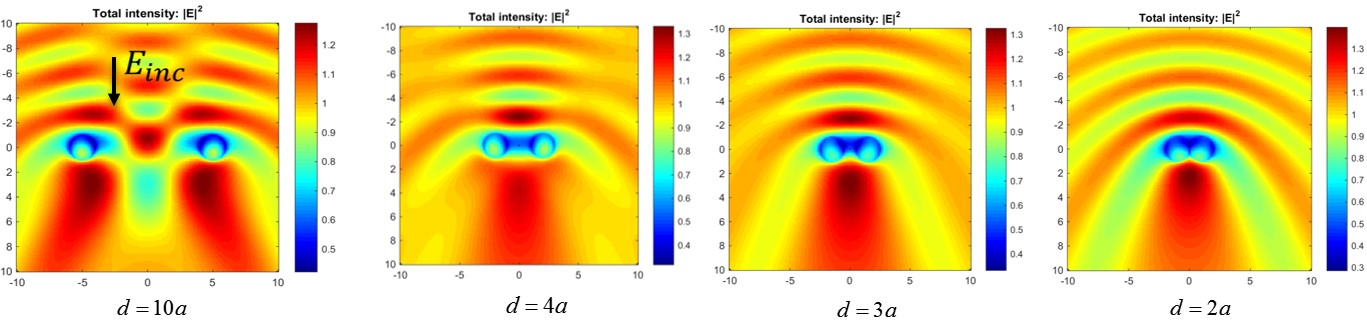}\label{twosphereefield}
	\caption{A schematic of the DSE involving two particles, where the variation of electric intensity distribution with the distance between the two particles. Here the refractive index of the particle is chosen to be $n=1.59$ (polystyrene) and the size parameter is $x=ka=2\pi a/\lambda=1$, where $a$ is the radius and $\lambda$ is the wavelength. }\label{dse_schematic}
	
\end{figure}

If ISA breaks down, we say the dependent scattering regime is entered. Thus the dependent scattering effect (DSE) can be defined as a generalization for those \textit{microscopic} interference effects that are not possible to explain under ISA \cite{yamadaJHT1986,aernoutsOE2014,vanTiggelenRMP2000}. Figure \ref{dse_schematic} presents an intuitive schematic of the DSE involving two particles when the far-field approximation is gradually violated by decreasing the inter-particle distance, where the electric field intensity distribution is calculated in the two-sphere system using an exact electromagnetic solving method\footnote{Here we use the multiple sphere $T$-matrix method, which will be introduced in Section \ref{multipoleFLE}.}. It is found that with the decrease of the particle distance, the scattered electromagnetic fields of them are strongly coupled, and the total scattering cross section varies with the distance significantly (not shown here). Note the DSE generally can involve a group of scatterers (the number of scatterers is much larger than two) distributed in a range covering several wavelengths \cite{vanTiggelenRMP2000}.

%It should be noted here it is a definition that is usually used in the thermal science and engineering \cite{yamadaJHT1986,tienARHT1987} and remote sensing communities \cite{tsang2004scattering}, and its meaning is broader than that in mesoscopic physics, for example, van Tiggelen \textit{et al.} \cite{Vantiggelen1990JPCM} only regarded the multiple scattering trajectories that visit the same particle more than once and result in a closed loop, or called ``recurrent scattering''\cite{Aubry2014PRL}, as the dependent scattering mechanism.

%Although, conventionally, the term DSE is not used to describe the interference phenomena in these media, we can notice that the underlying mechanism is still originating from the multiple scattering and interference of electromagnetic waves. Therefore, our research on DSE can also provide a fundamental understanding on these media.
Although in theory the RTE cannot be applied when it enters the dependent scattering regime since it is derived under the ISA \cite{VanRossum1998}, it is still possible for us to retain the form of RTE by correcting the mesoscopic radiative properties considering DSE. This is key assumption of the present article and most reviewed references herein, which is valid in a perturbative fashion \cite{tsang2004scattering}. In this assumption, it is postulated that the DSE gives rise to some correction to the radiative properties perturbatively by adding some small (or moderate) terms to the diagrammatic expansion. This \textit{a priori} postulation is viable in thermal engineering applications using moderate-refractive index materials, usually away from electromagnetic resonances. To the best of our knowledge and experiences, this is the only feasible treatment for the large scale DDM, where the mesoscopic radiative properties are significantly affected by the DSE.

In addition, the long coherence time, elastic scattering and time-reversible characteristics of light in conventional disordered media guarantee the interference phenomena in both microscopic and mesoscopic scales, and even in the macroscopic scale. In Section \ref{mesointerference}, we further briefly introduce several remarkable mechanisms that are also originated from wave interferences, which although different from the DSE, are indeed very relevant. These mechanisms usually take place in the scale of several scattering/transport mean free paths, which thus do not directly impact the \textit{mesoscopic radiative properties}, although they have an influence on the \textit{macroscopic radiative properties}. Hence these mechanisms can be called ``mesoscopic interference phenomena", and correspondingly, the DSE can be considered as the \textit{microscopic interference phenomenon} since it plays a role at the scale of several wavelengths. This classification will be further discussed in Fig.\ref{RTE_regime_vanTiggelen} of Section \ref{rteandde}, where for wave interferences taking place at the length scale of $L\sim \lambda$, we say they belong to the dependent scattering mechanism, while at the length scale of $L\gg \lambda$ with intermediate scattering strength (i.e., $l/\lambda$ is substantially larger than unity but away from the weak scattering limit, i.e., $l$ is on the scale of several $\lambda$), mesoscopic interferences like the weak localization occur. In that sense, we can expect that the effect of mesoscopic interferences is not as significant as the DSE in weakly scattering media.

The concept of dependent scattering is usually confused with the multiple scattering effect, due to the fact that the concept of ``multiple scattering" is not used consistently in the literature. Formally speaking, the DSE is a phenomenon resulting from the latter one, because multiple scattering of electromagnetic waves can lead to the DSE. However, in the context of many papers, ``multiple scattering" indicates a process that happens at the scale of several scattering mean free paths and specifically describes the form of ballistic transport of energy and intensity, namely, a pure geometric optics effect, for example, in Ref. \cite{drolenJTHT1987}. In that sense, it can take place with or without the DSE, as stated by Kalkman \textit{et al.} \cite{kalkmanOE2010}: ``In addition to \textit{multiple scattering}, the scattering coefficient is also influenced by \textit{coherent light scattering effects}, i.e. due to close packing of particles the coherent addition of light can lead to a reduction in the scattering rate. This effect is called \textit{dependent scattering}; a dependence of the scattering strength on the separation between the particles." This is the most prevailing case. Another situation is that many references did not actually distinguish these two concepts (or even did not notice the difference). When the ``multiple scattering" effect is mentioned, it might generally indicate both the dependent scattering effect due to wave interferences and the pure geometric optics, like in Ref. \cite{berryJOSA1962}. This case is frequently encountered in many early works. As already stated by Auger and Stout \cite{augerJCTR2012}, the authors of many pioneering works (e.g., Ref. \cite{tinsleyJOCCA1949}) actually never mentioned either dependent or multiple scattering phenomena lead to the reduction of observed overall scattering strength. In this article, we carefully avoid using vague terms. More specifically, we imply the geometric optics effect by using ``multiple scattering of (light or electromagnetic or radiation) intensity" and include all effects by saying ``multiple scattering of (electromagnetic) waves" or ``multiple wave scattering", in order to make a clear distinction between the intensity and wave aspects. If we do not make an explicit discrimination, we are referring to the multiple scattering of intensity, in order to emphasize the role of the DSE.

In a nutshell, it is critical to take DSE into account and develop an effective theoretical model of mesoscopic radiative properties directly based on the first-principle Maxwell's equations and the micro/nanostructures, in order to accurately predict, understand and manipulate the radiative transfer process in micro/nanoscale DDM. This is also crucial to nowadays applications of micro/nanoscale DDM in thermal science and engineering, optics and photonics, atmospheric sciences as well as biomedical engineering. And since our main concern in this article is the mesoscopic radiative properties, hereafter by saying "radiative properties", we specify the mesoscopic ones for brevity.

\section{Multiple scattering theory of electromagnetic waves and radiative transfer equation}\label{theory}
%In this section, we do not aim to give detailed formulas of multiple scattering theory but provide some crucial equations in this theory. Details can be found from Mischenko's papers and books, as well as mine.
When thermal radiation propagates in DDM, it undergoes scattering in a very complicated way. To better understand the dependent scattering effect originating from the electromagnetic interferences in the radiative transfer process, one should resort to the fundamental theories and methods for the treatment of electromagnetic scattering by a single particle as well as particle groups. In this section, we first summarize the single and multiple scattering theories of electromagnetic waves in DDM, which are directly based on the first-principle Maxwell's equations. On this basis, we proceed to an introduction on the relationship between Maxwell's equations and the RTE, and then discuss briefly the applicability of the RTE. It is worth noting that recently Doicu and Mishchenko \cite{doicuJQSRT2018,doicuJQSRT2019,doicuJQSRT2019a,doicuJQSRT2019b,doicuJQSRT2019c,doicuJQSRT2019d} published a series of reviews summarizing the multiple scattering theory of electromagnetic waves in random media as well as the connection between this theory and the RTE, which contain many technical details that can be referred to. 

Before we dive into the remainder of this article, several crucial postulations should be made. (1) We assume the radiation must have sufficient spatial and temporal coherence to permit coherent effects \cite{capetaPRA2011,augerJCTR2012}. (2) As mentioned in the Section \ref{DDM_definition}, we only consider the static multiple scattering problem in this article, by assuming the scattering processes are much faster than the random movement of particles in the disordered media. To be more precisely, the dynamic positional fluctuations of the random media during the photon scattering process should be very small compared to the wavelength, i.e., $kv_p\ll1/\tau$, where $v_p$ is the velocity of particle motion, $\tau$ is the time scale of an individual photon scattering event, otherwise considerable inelastic effects like Doppler frequency shifts and decoherence will occur \cite{labeyriePRL2003}. For nonresonant scattering, we have $\tau\sim l_s/c_0$ with $c_0$ denoting the velocity of light, leading to a condition as $v_p/c_0\ll1/(kl_s)$. Since for most disordered media $kl_s$ is substantially larger than 1, this condition gives a more stringent criterion than the conventional one, $v_p\ll c_0$ \cite{stephenPRL1987,stephenPRB1988}. Moreover, for resonant multiple scattering, owing to the time delay brought by resonances, $\tau$ becomes much larger \cite{lagendijk1996resonant}, resulting in a much smaller upper limit for $v_p$. Therefore, under this assumption, at a given moment the multiple wave scattering can be described by assuming that the scatterers are all fixed and solving the corresponding (quasi)instantaneous problem in the frequency domain without any considerations of the inelastic effects \cite{mishchenkoPhysrep2016}.  Apparently, for certain densely packed rigid DDM where all scatterers are stationary, there is no need to use this assumption. (3) On this basis of assumption (2), we further apply the ergodicity hypothesis for the random motions of scatterers and therefore the time-averaged signals over a sufficiently long period of time can be replaced by the ensemble-averaged ones over all system states, such as positions, sizes and orientations of the scatterers, with appropriate probability functions characterizing all the system states (See \ref{en_avg_def} for an introduction of ensemble average). This assumption is important for conventional (i.e., not ultrafast) detection techniques which usually take a long period of time, thus permitting to study the dynamic problem in a static manner \cite{mishchenko2006multiple}. Similarly, for those fixed DDM, there is no need to use this assumption. But it is always postulated in such stationary media that the ensemble average over all system states can be achieved by characterizing a large amount of different samples from the same fabrication process or different zones in a single sample \cite{etemadPRL1986,fengPRL1988}. This is important to eliminate strong statistical fluctuations. (4) We assume there are not any quantum \cite{kaiserJMO2009,smolkaPRL2009,ottPRL2010} or nonlinear effects \cite{alberucciOL2018,angelaniPRL2006,contiPRA2007}, for both the radiation sources and the DDM. (5) Here we only work in three dimensions (3D) for generality and brevity, and the same problem in one- (1D) and two-dimensions (2D) can be largely simplified after symmetry considerations. See Refs. \cite{sheng2006introduction,leeJAP1990,leeJQSRT1992,leeJTHT1994,leeJQSRT2019,tsang2000scattering1,tsang2004scattering2} for more details.

%$\omega^{-1}\ll\tau_{s}\ll\tau_\mathrm{mov}\ll\tau_\mathrm{meas}$

%(4) The harmonic oscillations of the electromagnetic field are much faster than the scattering process, namely, $1/\omega\ll l/c$. Otherwise the time-averaged Poynting vector cannot be well-defined.

\subsection{Maxwell's equations and single scattering}\label{single_scatter}
From classical and semi-classical viewpoints, light and thermal radiation are treated as electromagnetic waves. The electromagnetic wave is represented by the electric field $\mathbf{E}$ and magnetic induction $\mathbf{B}$, which are both space- and time-dependent. The responses of matter over electromagnetic waves, described as electric displacement $\mathbf{D}$, electric current density $\mathbf{j}$, electric charge density $\rho$, magnetic vector $\mathbf{H}$, are related with the field vectors under the framework of Maxwell's equations.
In this article, we are mainly concerned with conventional nonmagnetic materials with relative magnetic permeability $\mu=1$ and non-unity scalar permittivity $\varepsilon(\omega)$ (i.e., isotropic materials) with possible frequency dispersions. These materials include most conventional materials which we encountered in the study of thermal radiation transfer, including dielectric materials like zirconia, titania and polystyrene, and metallic materials like silver and gold. 

The scattering of electromagnetic waves by single homogeneous or multi-layered spherical particles with arbitrary electric and magnetic properties is one of the earliest solved problems in electromagnetic scattering, which was first done by Gustav Mie over 100 years ago \cite{mie1908}, and Ludvig Lorenz and others independently developed the theory of plane wave scattering by a dielectric sphere (For a historical review, see Ref.\cite{wriedt2012mie}.). Along with the rapid development of nanofabrication and nanophotonics in the last a few years, the anomalous scattering properties of single dielectric particles are theoretically and experimentally studied by many authors very extensively \cite{tribelskyPRL2006,kivsharScience2016}, giving rise to the booming of nanoscale light scattering study based on Mie theory. The basic idea behind Mie theory is to rigorously solve the boundary value problem of Maxwell's equations in spherical coordinates. In this condition, the solution of Maxwell's equations can be formally expanded into a linear combination of vector spherical harmonics (VSHs) or vector spherical wave functions (VSWFs) \cite{bohrenandhuffman,tsang2000scattering1}. The VSWFs can be found in any monographs or seminal papers on electromagnetic scattering, like Refs. \cite{mackowskiJOSAA1996,mackowskiJQSRT2013,tsang2000scattering1,bohrenandhuffman,hulst1957}, which are also listed in \ref{vswf_appendix}. 
The extinction $C_{\text{e}}$ and scattering $C_{\text{s}}$ cross sections of a single homogeneous sphere with a complex refractive index of $\tilde{m}$ and radius $a$ placed in vacuum illuminated by a plane wave is formally given by \cite{Gomez-MedinaPRA2012,bohrenandhuffman,tsang2000scattering1}
\begin{equation}\label{cext_eq}
C_{\text{e}}=\frac{2\pi}{k^2}\sum_{n=1}^{\infty}(2n+1)\mathrm{Re}(a_n+b_n),
\end{equation}
\begin{equation}
C_{\text{s}}=\frac{2\pi}{k^2}\sum_{n=1}^{\infty}(2n+1)(|a_n|^2+|b_n|^2),
\end{equation}
where 
\begin{equation}\label{Mie_an}
a_n=\frac{\tilde{m}^2j_n(\tilde{m}x)[xj_n(x)]'-j_n(x)[\tilde{m}xj_n(\tilde{m}x)]'}{\tilde{m}^2j_n(\tilde{m}x)[xh_n(x)]'-h_n(x)[\tilde{m}xj_n(\tilde{m}x)]'},
\end{equation}
\begin{equation}\label{Mie_bn}
b_n=\frac{j_n(\tilde{m}x)[xj_n(x)]'-j_n(x)[\tilde{m}xj_n(\tilde{m}x)]'}{j_n(\tilde{m}x)[xh_n(x)]'-h_n(x)[\tilde{m}xj_n(\tilde{m}x)]'},
\end{equation}
and $k=2\pi/\lambda$ is the wavenumber of plane wave with wavelength $\lambda$ in vacuum and $x=ka$ is the corresponding size parameter which describes the relative size of the sphere over the wavelength. $j_n(z)$ and $h_n(z)$ are spherical Bessel functions and Hankel functions of the first kind of order $n$, with respect to the argument $z$ \cite{bohrenandhuffman}.  

The differential scattering cross section as a function of the polar angle for an unpolarized, plane-wave illumination is given by
\begin{equation}\label{pf_isa}
\frac{dC_{\text{s}}}{d\theta_{\text{s}}}=\frac{\pi}{k^2}(|S_1(\theta_{\text{s}})|^2+|S_2(\theta_{\text{s}})|^2),
\end{equation}
where 
\begin{equation}\label{s1_isa}
S_1(\theta_{\text{s}})=\sum_{n=1}^{\infty}\frac{2n+1}{n(n+1)}[a_n\pi_n(\cos{\theta_{\text{s}}})+b_n\tau_n(\cos{\theta_{\text{s}}})]
\end{equation}
and
\begin{equation}\label{s2_isa}
S_2(\theta_{\text{s}})=\sum_{n=1}^{\infty}\frac{2n+1}{n(n+1)}[a_n\tau_n(\cos{\theta_{\text{s}}})+b_n\pi_n(\cos{\theta_{\text{s}}})]
\end{equation}
are elements of amplitude scattering matrix and $\theta_s$ is the polar scattering angle with respect to the incident wavevector, in which $\pi_n$ and $\tau_n$ are special functions that can be found in standard textbooks \cite{bohrenandhuffman,tsang2000scattering1}, also listed in \ref{vswf_appendix} for the readers' convenience. The normalized differential scattering cross section is also called scattering phase function used in ISA and RTE. Therefore, the scattering asymmetry factor for the single scattering phase function, defined as the mean cosine of scattering angle, $\langle\cos\theta_{\text{s}}\rangle$, is calculated through \cite{bohrenandhuffman}

\begin{equation}
g=\frac{1}{C_{\text{s}}}\int_{0}^{\pi}\frac{dC_{\text{s}}}{d\theta_{\text{s}}}\cos\theta_{\text{s}}\sin\theta_{\text{s}} d\theta_{\text{s}}.
\end{equation}
Mie theory for multilayered spherical particles and infinitely long cylinders (the 2D version) can be similarly derived \cite{bohrenAO1985}. 
Moreover, since Mie theory belongs to the family of separation of variables methods (SVM) to solve the electromagnetic properties of regular geometries, more generally, analytical solutions for arbitrary spheroids with various aspect ratios can be also derived based on spheroidal wave functions using SVM \cite{mishchenko1999light}.  

For a single irregular particle, numerical methods are more suitable to calculate its scattering and absorption properties. Here we briefly describe some widely-used numerical methods for irregular particles. The \textit{T}-matrix method is essentially a generalization of Mie theory to calculate the scattering and absorption properties of a single non-spherical particle based on the extended boundary condition method (EBCM) and VSWF expansion technique. It is originally proposed by Waterman \cite{watermanPIEEE1965} and further developed by the light scattering community \cite{petersonPRD1974,mishchenko1999light,tsang2000scattering1}. The basic idea of this approach is to expand the incident and scattered waves into VSWFs and relate the expansion coefficients using the $T$-matrix. More precisely, consider a particle centered at the origin, and the incident and scattered electric field can be expanded into VSWFs as
\begin{equation}\label{inc_expansion_tmat}
\mathbf{E}_{\text{inc}}(\mathbf{r})=\sum_{mnp} a_{mnp}^\mathrm{inc}\mathbf{N}^{(1)}_{mnp}(\mathbf{r}),
\end{equation}
\begin{equation}\label{sca_expansion_tmat}
\mathbf{E}_{\text{s}}(\mathbf{r})=\sum_{mnp} a_{mnp}^\mathrm{s}\mathbf{N}^{(3)}_{mnp}(\mathbf{r}),
\end{equation}
where $a_{mnp}^\mathrm{s}$ and $a^\mathrm{inc}_{m'n'p'}$ are the expansion coefficients. $\mathbf{N}^{(1)}_{mnp}(\mathbf{r})$ and $\mathbf{N}^{(3)}_{mnp}(\mathbf{r})$ are the type-1 and type-3 VSWFs, respectively \cite{mackowskiJOSAA1996,mackowskiJQSRT2013,tsang2000scattering1,bohrenandhuffman,hulst1957,mishchenko1999light}, whose expressions can be found in \ref{vswf_appendix}. $n$ and $m$ are integers denoting the order and degree of VSWFs with $n\geq 1$ and $|m|\leq n$. The subscript $p$ can only be $1$ or $2$, which denotes magnetic (TM) or electric (TE) modes respectively. 

For a given incident field, the expansion coefficients $a_{mnp}^\mathrm{inc}$ can be solved directly by using the orthogonality of VSWFs. For example, given a plane wave propagating in the $z$ direction, namely, $\mathbf{E}_{\text{inc}}(\mathbf{r})=\mathbf{E}_{\mathrm{inc},0}\exp{(i\mathbf{k}\cdot\mathbf{r})}=\mathbf{E}_{\mathrm{inc},0}\exp{(ikz)}$, the expansion coefficients are obtained as \cite{tsang2000scattering1,mishchenko1999light}
\begin{equation}
a_{mn1}^\mathrm{inc}=-i^n\sqrt{\frac{4\pi (2n+1)(n+m)!}{n(n+1)(n-m)!}}\mathbf{E}_{\mathrm{inc},0}\cdot\mathbf{B}_{-m,n}(0,0)
\end{equation}
and
\begin{equation}
a_{mn2}^\mathrm{inc}=-i^n\sqrt{\frac{4\pi (2n+1)(n+m)!}{n(n+1)(n-m)!}}\mathbf{E}_{\mathrm{inc},0}\cdot\mathbf{C}_{-m,n}(0,0)
\end{equation}
for $m=\pm1$ (for other $m$-s the coefficients are zero in this circumstance), where $\mathbf{B}_{mn}(\theta,\phi)$ and $\mathbf{C}_{mn}(\theta,\phi)$ are VSHs defined in spherical coordinates, with detailed expressions presented in \ref{vswf_appendix}. According to the linearity of Maxwell's equations, the relation between the expansion coefficients of scattered and incident fields should be linear, and a corresponding transition matrix (i.e., $T$-matrix) can be defined \cite{mishchenko1999light}:
\begin{equation}\label{tmat_def}
a_{mnp}^\mathrm{s}=\sum_{m'n'p'}T_{mnpm'n'p'}a^\mathrm{inc}_{m'n'p'}.
\end{equation}
As a consequence, if the $T$-matrix of an object is known, the expansion coefficients of the scattered field $a_{mnp}^\mathrm{s}$ can be solved from Eq.(\ref{tmat_def}), and subsequently the single scattering properties, including the scattering/extinction cross sections, phase function and so on, can be instantly calculated (by using the far-field asymptotic forms of VSWFs. See \ref{vswf_appendix}.). Therefore, the central task of the $T$-matrix method is to solve the $T$-matrix of arbitrarily shaped particles. 

In fact, for a homogeneous, isotropic spherical particle, the $T$-matrix is equivalent to the Mie coefficient obeying the following relation: 
\begin{equation}
T_{mnpm'n'p'}=\begin{cases}
b_n\delta_{mm'}\delta_{nn'}\delta_{pp'} &{p=1}\\
a_n\delta_{mm'}\delta_{nn'}\delta_{pp'} & {p=2}
\end{cases}
\end{equation}
where $\delta$ is the Kronecker delta. However, for arbitrary nonspherical particles, above relation is not valid. In this condition, the EBCM is employed to calculate the $T$-matrix. EBCM indicates the imaginary spherical boundaries circumscribing (radius $R_{>}$) and inscribing ($R_{<}$) the nonspherical particle, in order to use VSWF expansion at those boundaries. By relating the incident field with the internal field in the particle, whose expansion coefficients are $a_{mnp}^\mathrm{int}$, to points inside the inscribing sphere, we have 
\begin{equation}
a_{mnp}^\mathrm{inc}=\sum_{m'n'p'}Q_{mnpm'n'p'}a^\mathrm{int}_{m'n'p'}.
\end{equation}
Similarly, by relating the scattered field with the internal field in the particle to points outside the circumscribing sphere, we can obtain 
\begin{equation}
a_{mnp}^\mathrm{s}=-\sum_{m'n'p'}Q'_{mnpm'n'p'}a^\mathrm{int}_{m'n'p'}.
\end{equation}
Here the elements $Q_{mnpm'n'p'}$ and $Q'_{mnpm'n'p'}$ can be numerically evaluated by simple surface integrals over $S$, only involving the particle size, shape and refractive index. For more details one can refer to Tsang \textit{et al.} \cite{tsang2000scattering1} and Mishchenko \textit{et al.} \cite{mishchenko1999light}. Therefore, the $T$-matrix can be solved straightforwardly through a matrix inversion and multiplication procedure according to its original definition in Eq. (\ref{tmat_def}). 

%\begin{figure}[htbp]
%	\centering
%	\includegraphics[width=0.4\linewidth]{tmatrix_schematic.jpg}
%	
%	\caption{A schematic of EBCM in the $T$-matrix method. $S$ is the boundary of a nonspherical particle, where its smallest circumscribing sphere has a radius of $R_{>}$ and a concentric inscribing sphere with a radius of $R_{<}$ is also shown. Reprinted from Ref. \cite{mishchenko1999light}.}\label{tmatrix_schematic}
%\end{figure}

The discrete dipole approximation (DDA) \cite{yurkinJQSRT2007} is also a very general numerical electromagnetic method to solve the light scattering problem by arbitrary particles. The principle of this numerical method is to discretize an individual scatterer into a periodic grid (usually in the form of a cubic lattice) of fictitious dipoles, calculate the electromagnetic field for this set of dipoles and then sum up the electromagnetic fields generated by all dipoles to obtain the scattering properties of this scatterer. This technique was firstly proposed by Purcell and Pennypacker \cite{purcellAPJ1973} and further developed by Draine and coworkers \cite{draineAPJ1988,draineAPJ1993,draineJOSA1994}. In general, when considering the formalism of this method, it is somewhat equivalent to the coupled-dipole model (CDM), which will be discussed later in Section \ref{cdm_intro} for the study of multiple scattering of electromagnetic waves, while there are still some nontrivial differences between DDA and CDM, which have been revisited recently in Ref. \cite{markelJQSRT2019}. And for more theoretical and numerical considerations regarding the practical implementation for realistic scatterers to improve the accuracy and computation speed of DDA, e.g., different types of dipole unit cell, renormalized polarizability models, the Fourier transform (FFT) technique and the fast multipole method (FMM), see Refs. \cite{mishchenko1999light,yurkinJQSRT2007,yurkinJQSRT2011}.

After introducing the analytical and numerical methods for treating electromagnetic scattering of a single particle, in the following subsection, we will briefly summarize the rigorous theories for the treatment of multiple scattering of electromagnetic waves. These theories, including the analytic wave theory and the Foldy-Lax equations, are originally developed for both classical (e.g., electromagnetic and acoustic waves) and quantum waves (e.g., electrons) \cite{laxRMP1951,mahan2013many}. Therefore they are very general in tackling with multiple scattering problems. The equivalence between the analytic wave theory and the Foldy-Lax equations will also be discussed. 

\subsection{Analytic wave theory}\label{analytic_wave_theory}
%Perhaps we should follows from the essay by Leung Tsang \cite{tsangJQSRT2019}. 
% What is "analytic wave theory" for and where does this term come from? Original papers and historical development should be briefly introduced.
The analytic wave theory stemmed from the work by Frisch \cite{frisch1968wave}, which presented a Feynman diagrammatic representation and Bethe-Salpeter equation technique, borrowed from quantum field theory (QFT), for the treatment of multiple scattering of waves. This method was then used and further developed by Ishimaru \cite{ishimaruPIEEE1977,ishimaru1978book}, Barabanenkov \cite{barabanenkov1971status,barabanenkovJEWA1995}, Tsang and Kong \cite{tsangJAP1980,tsang1985theory,tsang2004scattering}, Lagendijk \cite{lagendijk1996resonant}, Nieuwenhuizen \cite{VanRossum1998}, Sheng \cite{sheng2006introduction}, Mishchenko \cite{mishchenko2006multiple} and their coworkers, to name a few. In this subsection, we attempt to give a brief introduction to this theory, and this theory is applied to derive several analytical models of the DSE in Section \ref{models}.

%Later, we will demonstrate how the framework of multiple scattering theory leads to the quasicrystalline approximation (QCA) \cite{laxRMP1951,laxPR1952,lagendijk1996resonant,VanRossum1998, tsang2004scattering,sheng2006introduction}, which we will use as the main analytical tool to develop a theoretical model of radiative properties properties used throughout this thesis. 

\subsubsection{Dyson equation}\label{dyson_eq_sec}
Let us start from the the general case of an infinite nonmagnetic three-dimensional (3D) medium, where the spatial distribution of permittivity $\varepsilon(\mathbf{r})$ is inhomogeneous and can be generally described as $\varepsilon(\mathbf{r})=1+\delta \varepsilon(\mathbf{r})$, where $\delta \varepsilon(\mathbf{r})$ is the fluctuational part of the permittivity due to random morphology of the inhomogeneous medium. Electromagnetic wave propagation in such media is described by the vectorial Helmholtz equation \cite{lagendijk1996resonant,tsang2004scattering,wangIJHMT2018}: 
\begin{equation}
\nabla \times \nabla \times \mathbf{E}(\mathbf{r})-k^{2}\varepsilon(\mathbf{r}) \mathbf{E}(\mathbf{r})=0.
\end{equation}

Let $k^2=\omega^2/c_0^2$  be the wavenumber in the background medium and $V(\mathbf{r})=k^2\delta \varepsilon(\mathbf{r})=\omega^2\delta \varepsilon(\mathbf{r})/c_0^2$ be the disordered ``potential" inducing electromagnetic scattering, where $c_0$ is the speed of light in the background medium. Then we have an alternative form of vectorial Helmholtz equation convenient for electromagnetic scattering problems in random media:
\begin{equation}
\nabla \times \nabla \times \mathbf{E}(\mathbf{r})-k^{2} \mathbf{E}(\mathbf{r})=V(\mathbf{r})\mathbf{E}(\mathbf{r}).
\end{equation}

To solve the equation, we can introduce the dyadic Green's function for this random medium which satisfies
\begin{equation}
\nabla \times \nabla \times \mathbf{G}(\mathbf{r},\mathbf{r}')-k^{2} \mathbf{G}(\mathbf{r},\mathbf{r}')=V(\mathbf{r})\mathbf{G}(\mathbf{r},\mathbf{r}')+\mathbf{I}\delta(\mathbf{r},\mathbf{r}').
\end{equation}

In the meanwhile, the Green's function in the homogeneous background medium is\footnote{In vacuum, it is the free-space Green's function, whose expression in the real domain is given in Eq.(\ref{free_space_green_function}).}
\begin{equation}
\nabla \times \nabla \times \mathbf{G}_0(\mathbf{r},\mathbf{r}')-k^{2} \mathbf{G}_0(\mathbf{r},\mathbf{r}')=\mathbf{I}\delta(\mathbf{r},\mathbf{r}'),
\end{equation}
where $\mathbf{I}$ is the identity matrix. Taking the Fourier transform with respect to $\mathbf{r}$ and $\mathbf{r}'$ to the reciprocal space  in terms of the momentum vetors $\mathbf{p}$ and $\mathbf{p}'$ and letting $V(\mathbf{r},\mathbf{r}')=k^2\delta \epsilon(\mathbf{r})\delta(\mathbf{r}-\mathbf{r}')$ using the Dirac delta function, we can write down the solution for dyadic Green's function in the disordered  media as
\begin{equation}\label{lp_eq1}
\mathbf{G}(\mathbf{p},\mathbf{p}')=\mathbf{G}_0(\mathbf{p},\mathbf{p}')+\mathbf{G}_0(\mathbf{p},\mathbf{p}_2)\mathbf{V}(\mathbf{p}_2,\mathbf{p}_1)\mathbf{G}(\mathbf{p}_1,\mathbf{p}'),
\end{equation} 
where the dummy variables $\mathbf{p}_1$ and $\mathbf{p}_2$ will be integrated out and we don't write this integral explicitly here as well as below. This equation is known as the Lippmann-Schwinger equation \cite{VanRossum1998,mishchenko2006multiple,sheng2006introduction}. By introducing the $T$-operator $\mathbf{T}$, Eq.(\ref{lp_eq1}) is transformed into the following form
\begin{equation}\label{lp_eq2}
\mathbf{G}(\mathbf{p},\mathbf{p}')=\mathbf{G}_0(\mathbf{p},\mathbf{p}')+\mathbf{G}_0(\mathbf{p},\mathbf{p}_2)\mathbf{T}(\mathbf{p}_2,\mathbf{p}_1)\mathbf{G}_0(\mathbf{p}_1,\mathbf{p}').
\end{equation} 
Based on Eqs.(\ref{lp_eq1}) and (\ref{lp_eq2}), it can be easily shown that the $T$-operator is given by
\begin{equation}
\mathbf{T}(\mathbf{p},\mathbf{p}')=\mathbf{V}(\mathbf{p},\mathbf{p}')+\mathbf{V}(\mathbf{p},\mathbf{p}_2)\mathbf{G}_0(\mathbf{p}_2,\mathbf{p}_1)\mathbf{T}(\mathbf{p}_1,\mathbf{p}').
\end{equation}
%\begin{equation}
%\mathbf{T}(\mathbf{p},\mathbf{p}')=[\mathbf{I}-\mathbf{V}(\mathbf{p},\mathbf{p}')\mathbf{G}_0(\mathbf{p},\mathbf{p}')]^{-1}\mathbf{V}(\mathbf{p},\mathbf{p}').
%\end{equation}
If the medium only contains one discrete scatterer, $\mathbf{T}(\mathbf{p},\mathbf{p}')$ is then known as the $T$-operator for the single scatterer. Obviously for a random medium composed of many scatterers, Eq.(\ref{lp_eq1}) still applies. However, if each scatterer can be described by its own $T$-operator, it is more convenient to transform Eq.(\ref{lp_eq1}) into the form only involving the $T$-operators of the individual particles, rather than the ``scattering potential'' $\mathbf{V}$. This is most suitable for a random medium consisting of well-defined, discrete scatterers. Since the $T$-operator of the $j$-th scatterer is analogously given by \cite{tsang2004scattering,sheng2006introduction}
\begin{equation}
\mathbf{T}_j(\mathbf{p},\mathbf{p}')=\mathbf{V}_j(\mathbf{p},\mathbf{p}')+\mathbf{V}_j(\mathbf{p},\mathbf{p}_2)\mathbf{G}_0(\mathbf{p}_2,\mathbf{p}_1)\mathbf{T}(\mathbf{p}_1,\mathbf{p}'),
\end{equation}
where $\mathbf{V}_j(\mathbf{p},\mathbf{p}')$ is the scattering potential of the $j$-th scatterer, which constitutes the scattering potential of the system simply as $\mathbf{V}(\mathbf{p},\mathbf{p}')=\sum_{j=1}^N\mathbf{V}_j(\mathbf{p},\mathbf{p}')$ \cite{tsang2004scattering}. After some manipulations, the $T$-operator of the full system is then given by
\begin{equation}\label{t_op_expansion}
\begin{split}
\mathbf{T}(\mathbf{p},\mathbf{p}')&=\sum_{i=1}^N\mathbf{T}_j(\mathbf{p},\mathbf{p}')+\sum_{i=1}^{N}\sum_{j\neq i}^{N}\mathbf{T}_i(\mathbf{p},\mathbf{p}_1)\mathbf{G}_0(\mathbf{p}_1,\mathbf{p}_2)\mathbf{T}_j(\mathbf{p}_2,\mathbf{p}')\\&+\sum_{i=1}^{N}\sum_{j\neq i}^{N}\sum_{l\neq j}^{N}\mathbf{T}_i(\mathbf{p},\mathbf{p}_1)\mathbf{G}_0(\mathbf{p}_1,\mathbf{p}_2)\mathbf{T}_j(\mathbf{p}_2,\mathbf{p}_3)\mathbf{G}_0(\mathbf{p}_3,\mathbf{p}_4)\mathbf{T}_j(\mathbf{p}_4,\mathbf{p}')...,
\end{split}
\end{equation} 
where  $\mathbf{p}_3$ and $\mathbf{p}_4$ are also dummy variables to be integrated out. This formula can be understood as the sum of all multiple wave scattering paths at different orders, in which the $T$-operators of scatterers visited by these paths are connected by Green's functions, or the propagators \cite{tsang2004scattering}.

In this fashion, the Lippmann-Schwinger equation for a medium consisting of $N$ discrete scatterers is rewritten as
\begin{equation}\label{lp_eq3}
\mathbf{G}(\mathbf{p},\mathbf{p}')=\mathbf{G}_0(\mathbf{p},\mathbf{p}')+\mathbf{G}_0(\mathbf{p},\mathbf{p}_2)\sum_{j=1}^N\mathbf{T}_j(\mathbf{p}_2,\mathbf{p}_1)\mathbf{G}_j(\mathbf{p}_1,\mathbf{p}'),
\end{equation} 
where the Green's function with respect to each scatterer $\mathbf{G}_j(\mathbf{p},\mathbf{p}')$ is given by
\begin{equation}
\mathbf{G}_j(\mathbf{p},\mathbf{p}')=\mathbf{G}_0(\mathbf{p},\mathbf{p}')+\mathbf{G}_0(\mathbf{p},\mathbf{p}_2)\sum_{i=1,i\neq j}^N\mathbf{T}_i(\mathbf{p}_2,\mathbf{p}_1)\mathbf{G}_i(\mathbf{p}_1,\mathbf{p}').
\end{equation}
This equation is also known as Foldy-Lax equations for multiple scattering of classical waves \cite{foldyPR1945,laxRMP1951,mishchenko2006multiple}, which will be discussed in the Section \ref{foldy_lax_eqs}.

Then, to obtain a statistically meaningful description of the random medium, it is necessary to take ensemble average of the full system to eliminate the impact of a specific configuration.  Taking ensemble average of Eq.(\ref{lp_eq3}), we obtain
\begin{equation}\label{lp_eq4}
\langle\mathbf{G}(\mathbf{p},\mathbf{p}')\rangle=\mathbf{G}_0(\mathbf{p},\mathbf{p}')+\mathbf{G}_0(\mathbf{p},\mathbf{p}')\langle\mathbf{T}(\mathbf{p},\mathbf{p}')\rangle\mathbf{G}_0(\mathbf{p},\mathbf{p}'),
\end{equation} 
where $\langle\mathbf{G}(\mathbf{p},\mathbf{p}')\rangle$ denotes ensemble averaged amplitude Green's function, and $\langle\mathbf{T}(\mathbf{p},\mathbf{p}')\rangle$ is the ensemble averaged \textit{T}-operator of the full system by invoking Eq.(\ref{t_op_expansion}), 
\begin{equation}\label{lp_eq5}
\begin{split}
\langle\mathbf{T}(\mathbf{p},\mathbf{p}')\rangle=\langle\sum_{i=1}^N\mathbf{T}_j(\mathbf{p},\mathbf{p}')\rangle+\langle\sum_{i=1}^{N}\sum_{j\neq i}^{N}\mathbf{T}_j(\mathbf{p},\mathbf{p}_1)\mathbf{G}_0(\mathbf{p}_1,\mathbf{p}_2)\mathbf{T}_j(\mathbf{p}_2,\mathbf{p}')\rangle+....
\end{split}
\end{equation} 
After some manipulations of identifying and retaining only irreducible terms in the ensemble-averaged $T$-operator, we then obtain the well-known Dyson equation for the coherent, or mean component of the (electric) field as \cite{lagendijk1996resonant,VanRossum1998,tsang2004scattering}
\begin{equation}
\langle\mathbf{G}(\mathbf{p},\mathbf{p}')\rangle=\mathbf{G}_0(\mathbf{p},\mathbf{p}')+\mathbf{G}_0(\mathbf{p},\mathbf{p}')\bm{\Sigma}(\mathbf{p},\mathbf{p}')\langle\mathbf{G}(\mathbf{p},\mathbf{p}')\rangle,
\end{equation} 
where $\bm{\Sigma}(\mathbf{p},\mathbf{p}')$ is the so-called self-energy (or mass operator) containing all irreducible multiple scattering expansion terms in $T$-operator $\langle\mathbf{T}(\mathbf{p},\mathbf{p}')\rangle$. If we express the multiple wave scattering processes involving many particles into Feynman diagrams according to Eq.(\ref{lp_eq5}), where particles (represented by $\mathbf{T}_j$) are connected by the propagator ($\mathbf{G}_0$) and their relationships (including positional correlations between different particles, turning back to the same particle, etc.),  then the irreducible terms stand for those multiple wave scattering diagrams that cannot be divided without breaking the innate particle connections, including the same particle or particle correlations. For more details on irreducible and reducible diagrams, see Refs. \cite{VanRossum1998,tsang2004scattering,tsangJQSRT2019}. 
%We will show the Feynman diagrammatic technique to obtain the self-energy later in Section \ref{diagrammatic_tech}.

For a statistically homogeneous medium having translational symmetry after ensemble average, $\bm{\Sigma}(\mathbf{p},\mathbf{p}')=\bm{\Sigma}(\mathbf{p})\delta(\mathbf{p}-\mathbf{p}')$ and $\langle\mathbf{G}(\mathbf{p},\mathbf{p}')\rangle=\langle\mathbf{G}(\mathbf{p})\rangle\delta(\mathbf{p}-\mathbf{p}')$ \cite{sheng2006introduction}. In the momentum representation the free-space dyadic Green's function is $ \mathbf{G}_{0}(\mathbf{p})=-{1}/(k^{2}\mathbf{I}-p^{2}(\mathbf{I}-\mathbf{\hat{p}}\mathbf{\hat{p}}))$, and thus the averaged amplitude Green's function is
\begin{equation}
\langle\mathbf{G}(\mathbf{p})\rangle=-\frac{1}{k^{2}\mathbf{I}-p^{2}(\mathbf{I}-\mathbf{\hat{p}}\mathbf{\hat{p}})-\bm{\Sigma}(\mathbf{p})}, 
\end{equation}
where $\mathbf{\hat{p}}=\mathbf{p}/p$ is the unit vector in the momentum space. Through this equation, self-energy $\bm{\Sigma}(\mathbf{p}) $ provides a renormalization for the electromagnetic wave propagation in random media, and determines the effective (renormalized) permittivity as \cite{lagendijk1996resonant}
\begin{equation}\label{epsilon_eff}
\bm{\varepsilon}_\mathrm{eff}(\mathbf{p})=\mathbf{I}-\frac{\bm{\Sigma}(\mathbf{p})}{k^2}.
\end{equation} 
For a statistically isotropic random medium, the obtained momentum-dependent effective permittivity tensor is decomposed into a transverse part and a longitudinal part as $
\bm{\varepsilon}(\mathbf{p})=\varepsilon^{\bot}(\mathbf{p})(\mathbf{I}-\mathbf{\hat{p}}\mathbf{\hat{p}})+\varepsilon^{\parallel}(\mathbf{p})\mathbf{\hat{p}}\mathbf{\hat{p}}$, where $\varepsilon^{\bot}(\mathbf{p})=1-\Sigma^{\bot}(\mathbf{p})/{k^{2}}$ and $\varepsilon^{\parallel}(\mathbf{p})=1-\Sigma^{\parallel}(\mathbf{p})/{k^{2}}$ determine the effective permittivities of transverse and longitudinal modes in momentum space \cite{lagendijk1996resonant}. Therefore, by determining the poles of the amplitude Green's function we can obtain the dispersion relation which corresponds to collective excitations of the disordered medium  (i.e., like ``the band structure" in periodic systems). An equivalent and frequently used method to find the collective excitations is to resort to the spectral function \cite{sheng2006introduction,lagendijk1996resonant,Page1996}:
\begin{equation}
\begin{split}
\mathbf{S(\omega,\mathbf{p})}=&-\mathrm{Im}\langle\mathbf{G}(\omega,\mathbf{p})\rangle=\frac{\mathrm{Im}\varepsilon^{\parallel}(\omega,\mathbf{p})/k^2}{[\mathrm{Re}\varepsilon^{\parallel}(\omega,\mathbf{p})]^2+[\mathrm{Im}\varepsilon^{\parallel}(\omega,\mathbf{p})]^2}\mathbf{\hat{p}}\mathbf{\hat{p}}\\
&+\frac{\mathrm{Im}\varepsilon^{\bot}(\omega,\mathbf{p})/k^2}{[\mathrm{Re}\varepsilon^{\bot}(\omega,\mathbf{p})-p^2/k^2]^2+[\mathrm{Im}\varepsilon^{\bot}(\omega,\mathbf{p})]^2}(\mathbf{I}-\mathbf{\hat{p}}\mathbf{\hat{p}}).
\end{split}
\end{equation}
Since longitudinal modes are usually not propagating, here we mainly consider the transverse modes. By finding the (local) maxima of the transverse compoment of the spectral function in real momentum space, the transverse (propagating) mode's wavenumber can be calculated as $K^{\bot }=\sqrt{\varepsilon^{\bot}(\omega,p_\mathrm{max})}k$, where $p_\mathrm{max}$ is the momentum value that makes the spectral function maximal. Then the effective refractive index of the disordered medium is directly obtained from the real part of effective wavenumber as
\begin{equation}
n_\mathrm{eff}=\frac{\mathrm{Re}K^\bot}{k},
\end{equation}
and the extinction coefficient of this disordered medium is given by the imaginary part of the effective wavenumber:
\begin{equation}
\kappa_\mathrm{e}=2\mathrm{Im}K^\bot.
\end{equation}

A scrutiny of the transverse component of the spectral function tells us that there is a difference between the real part of $K^{\bot}$ and $p_\mathrm{max}$. More precisely, it reaches maximum when $\mathrm{Re}\varepsilon^{\bot}(\omega,\mathbf{p})-p^2/k^2$ approaches zero, that is $(\mathrm{Re}K^\bot)^2-(\mathrm{Im}K^\bot)^2=p_\mathrm{max}^2$. For nonabsorbing, purely scattering media, the scattering coefficient $\kappa_s$ is equal to the extinction coefficient $\kappa_e$ and this momentum mismatch profoundly stands for a propagating momentum shell broadening due to scattering scaling as $\kappa_s$. This broadening can be rather large for strongly scattering media \cite{lagendijk1996resonant}, while for weakly scattering media, we can apply the on-shell approximation, namely, $\mathrm{Re}K^{\bot}\approx p_\mathrm{max}$, which is employed in a recent work of ours \cite{wangPRA2018}.

Note after above treatments, we have implicitly assumed that the propagation of waves in disordered media is dominated by only one (transverse) mode, and thus a distinct wavenumber (propagation constant) can be well defined, so do the related radiative properties. However, it is not proved rigorously. Generally, the wave propagation behavior in disordered media is determined by the contributions of several modes with different wavenumbers and the spatial dispersion (nonlocality) is sometimes important, especially when it comes to the reflection and transmission at boundary of a disordered medium with strong fluctuations at the scale of wavelength \cite{barreraPRB2007,gowerRSPA2019}. Nevertheless, for conventional disordered media, we can safely adopt this single-mode treatment.

According to the above analysis on the Dyson equation, the first task to determine the radiative properties of DDM is to derive the self-energy. As a first-order perturbative approximation, ISA gives a self-energy that is simply the ensemble average of the sum of the $T$-operators of all scatterers per unit volume. If we only consider an ensemble of point dipole scatterers as the ideal case, the self-energy is given by \cite{VanRossum1998,devriesRMP1998}
\begin{equation}\label{self_energy_dipole_ISA}
\Sigma_\mathrm{ISA}=n_0t_0, 
\end{equation}
where $t_0=-k^2\alpha$ is the $T$-operator of a single electric dipole scatterer in the momentum representation, $\alpha$ is the dipole polarizability, and $n_0$ is the number density of particles. Note for the infinitely small scatterers, the self-energy is momentum-independent.

\subsubsection{Bethe-Salpeter equation}
It is noted that Dyson equation and the self-energy only provide a characterization for coherent electromagnetic field propagation in random media, i.e., the first moment of electromagnetic field $\langle\mathbf{G}(\mathbf{p})\rangle$, while a more relevant quantity to our concern is the radiation intensity in disordered media which directly determines the phase function of each scattering process in terms of energy transport. This is exactly governed by the Bethe-Salpeter equation \cite{lagendijk1996resonant,VanRossum1998,tsang2004scattering}, which describes the second moment of the electromagnetic field $\langle \mathbf{G}\mathbf{G}^*\rangle$ in random media. In the operator notation, Bethe-Salpeter equation is written as
\begin{equation}
\langle\mathbf{G}\mathbf{G}^*\rangle=\langle\mathbf{G}\rangle\langle\mathbf{G}^*\rangle+\langle\mathbf{G}\rangle\langle\mathbf{G^*}\rangle\bm{\Gamma}\langle\mathbf{G}\mathbf{G}^*\rangle,
\end{equation} 
where $\bm{\Gamma}$ is the irreducible vertex representing the renormalized scattering center for the incoherent part of radiation intensity due to random fluctuations of the disordered media. It can be understood as the differential scattering coefficient as well as (non-normalized) scattering phase function relevant in the radiative transfer equation if the momentum shell broadening of the transport processes can be neglected, which means the radiation intensity is concentrated on the momentum shell $p=K$. This is the case for random media containing dilute scatterers or weak scatterers. 
Moreover, since in most circumstances one only needs to consider transverse electromagnetic waves, and the transverse component of the irreducible vertex can be written in the momentum representation as \cite{Cherroret2016,sheng2006introduction,lagendijk1996resonant}
\begin{equation}\label{trans_comp}
\Gamma^{\bot}(K\hat{\mathbf{p}},K\hat{\mathbf{p}}')=(\mathbf{I}-\hat{\mathbf{p}}\hat{\mathbf{p}})\bm{\Gamma}(K\hat{\mathbf{p}},K\hat{\mathbf{p}}')(\mathbf{I}-\hat{\mathbf{p}}'\hat{\mathbf{p}}')
\end{equation}
under the \textit{on-shell} approximation. For isotropic media, the scattering properties (scattering coefficient and phase function) do not rely on the incident direction but only the solid angle $\Omega_s$ between the incident and scattering directions, which can be described by the polar scattering angle $\theta_s=\arccos(\hat{\mathbf{p}}\cdot\hat{\mathbf{p}}')$ and the azimuth angle $\varphi_s$. The choice of the latter depends on the definition of local frame of spherical coordinates with respect to the incident direction $\hat{\mathbf{p}}'$. Therefore, the differential scattering coefficient with respect to the incident direction $\hat{\mathbf{p}}'$ can be obtained as \cite{barabanenkov1975multiple}
\begin{equation}
\begin{split}
\frac{d\kappa_s}{d\Omega_s}=\frac{1}{(2\pi)^2}\Gamma^{\bot}(K\hat{\mathbf{p}},K\hat{\mathbf{p}}').
\end{split}
\end{equation}
From this equation, the scattering coefficient can be obtained by integrating over $\Omega_s$ (or equivalently, $\theta_s$ and $\phi_s$). 

The irreducible vertex and the self-energy are not independent of each other. For a nonabsorbing medium, the scattering coefficient should be equal to the extinction coefficient as required by the energy conservation, which leads to the following equation:
\begin{equation}\label{ward_identity}
\frac{1}{(2\pi)^2}\int_{0}^{2\pi}\Gamma^{\bot}(K\hat{\mathbf{p}},K\hat{\mathbf{p}}')d\hat{\mathbf{p}}=-\frac{\mathrm{Im}\Sigma}{n_\mathrm{eff}k}.
\end{equation}
This equation is also known as the Ward-Takahashi identity \cite{Cherroret2016,sheng2006introduction}, originally established in QFT \cite{kugoPLB1992}. Note there is a general form for this identity that does not require the on-shell approximation, which will not be shown here for simplicity \cite{sheng2006introduction}. 

For an ensemble of point dipole scatterers, the irreducible vertex has a simple expression under the ISA:
\begin{equation}\label{vertex_dipole_ISA}
\bm{\Gamma}(K\hat{\mathbf{p}},K\hat{\mathbf{p}}')=n_0|t_0|^2\mathbf{I}\otimes\mathbf{I}.
\end{equation}
where $\otimes$ means the tensor product. By taking the transverse component of the irreducible vertex and integrating over the azimuth angle $\varphi_s$ with incident direction $\hat{\mathbf{p}}'$ fixed, the differential scattering cross coefficient with respect to the polar angle is then
\begin{equation}
\begin{split}
\frac{d\kappa_s}{d\theta_s}=\frac{1+\cos^2\theta_s}{4\pi}n_0|t_0|^2.
\end{split}
\end{equation}
By combining the above equation and Eq.(\ref{self_energy_dipole_ISA}), one can directly examine that for nonabsorbing scatterers, the Ward-Takahashi identity Eq.(\ref{ward_identity}) is apparently fulfilled, which, in this case, equivalent to the Optical Theorem in the scattering theory \cite{ishimaru1978book,devriesRMP1998,lagendijk1996resonant,bornandwolf}.

Last but not least, we note that Dyson and Bethe-Salpeter equations, and the self-energy and irreducible intensity vertex, are directly derived from the vectorial Helmholtz equation. This means that it is a first-principle method for electromagnetic waves and can treat, generally, any disordered medium. Therefore, for continuous heterogeneous media where no discrete scatterers exist, the self-energy and irreducible intensity vertex can still be calculated, for instance, the well-known bilocal approximation for the self-energy of Gaussian random media \cite{tsang2004scattering}. In contrast, this kind of random media cannot be conveniently treated by Foldy-Lax equations. And it should be borne in mind that the irreducible vertex is a much more complex quantity, compared to the differential scattering coefficient, because it describes all interference phenomena in multiple wave scattering process and cannot always be simply interpreted as the differential scattering coefficient especially when the on-shell approximation does not apply.

\subsection{The Foldy-Lax equations (FLEs)}\label{foldy_lax_eqs}
As mentioned in the previous section, the FLEs are also general equations describing multiple scattering of both classical and quantum waves in disordered media containing discrete scatterers. This set of equations obtained its name from the early workers Foldy \cite{foldyPR1945} and Lax \cite{laxRMP1951}. Generally speaking, FLEs are the derivation of the analytic wave theory \cite{tsang2004scattering}. However, due to the explicit formalism and easy numerical implementation, FLEs become a more widely used method than the analytic wave theory, especially in engineering.
In this section, we briefly introduce the basics of FLEs along with the multipole expansion method and coupled-dipole model, which are important for practical use. The readers can refer to the extensive review regarding the FLEs for the treatment of DDM with formal definitions and general theoretical derivations by Mishchenko \textit{et al.} \cite{mishchenkoPhysrep2016}.

\begin{figure}[htbp]
	\centering
	\includegraphics[width=0.6\linewidth]{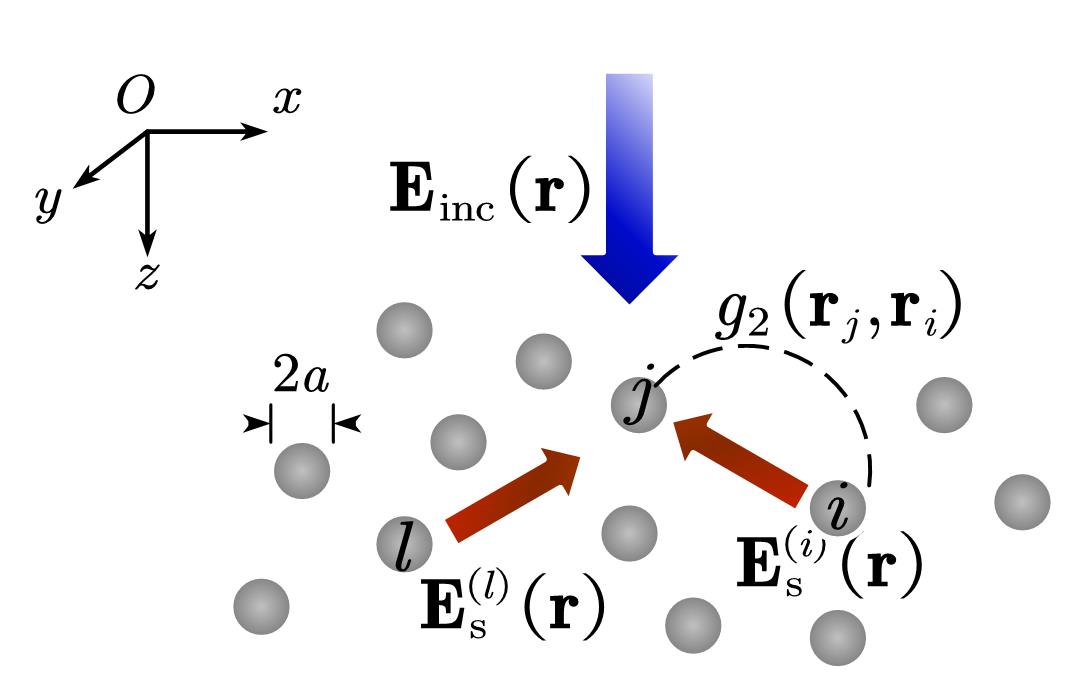}
	\caption{A schematic of Foldy-Lax equations for multiple scattering of electromagnetic waves in randomly distributed spherical particles in three dimensions. The particles are numbered as $i$, $j$, $l$, etc. The dashed line denotes $g_2(\mathbf{r}_j,\mathbf{r}_i)$, the pair distribution function between the two particles. The blue thick arrow indicates the propagation direction of the incident wave, while the red thin arrows stand for the propagation directions of the partial scattered waves from particle $i$ to $j$ and from $l$ to $j$.}
	\label{system_config}
\end{figure}

The well-known FLEs depicting the multiple scattering process of electromagnetic waves among $N$ discrete scatterers read \cite{laxRMP1951,varadanJOSAA1985,mackowskiJOSAA1996,tsang2004scattering}
\begin{equation}\label{fl_eq}
\mathbf{E}_{\text{exc}}^{(j)}(\mathbf{r})=\mathbf{E}_{\text{inc}}(\mathbf{r})+\sum_{i=1\atop i\neq j}^{N}\mathbf{E}_{\text{s}}^{(i)}(\mathbf{r}),
\end{equation}
where $\mathbf{E}_{\text{inc}}(\mathbf{r})$ is the incident electric field, $\mathbf{E}_{\text{exc}}^{(j)}(\mathbf{r})$ is the electric component of the so-called exciting field impinging on the vicinity of particle $j$, and $\mathbf{E}_{\text{s}}^{(i)}(\mathbf{r})$ is  electric component of partial scattered waves from particle $i$. We also show FLEs schematically in Fig.\ref{system_config}, where the scatterers are assumed to be spheres. Obviously, the scatterers can have arbitrary geometries. The physical significance of FLEs is straightforward. This series of equations describe that the exciting field impinging on the vicinity of particle $j$ is the sum of the incident field $\mathbf{E}_{\text{inc}}(\mathbf{r})$ and all partial scattered field from all other particles.

\subsubsection{The multipole (VSWF) expansion of FLEs}\label{multipoleFLE}
To solve FLEs for a disordered medium consisting of spheres as shown in Fig.\ref{system_config}, it is convenient to expand the electric fields in VSWFs to utilize the spherical boundary condition of individual particles, following the way usually done for a single spherical particle in Mie theory. The expansion coefficients then naturally correspond to multipoles supported by the particles \cite{lambPRB1980,tsangJAP1982,mackowskiJOSAA1996,tsang2004scattering,xuJQSRT2001}. As a matter of fact, in the spirit of extended boundary condition method, this expansion is still applicable for non-spherical particles if their $T$-matrix elements are given. Here we only consider spherical particles as a simple introduction. Using this technique, the exciting field $\mathbf{E}_{\text{exc}}^{(j)}(\mathbf{r})$ is expressed as
\begin{equation}\label{exc_expansion}
\mathbf{E}_{\text{exc}}^{(j)}(\mathbf{r})=\sum_{mnp} c_{mnp}^{(j)}\mathbf{N}^{(1)}_{mnp}(\mathbf{r}-\mathbf{r}_j).
\end{equation}
Based on the expansion coefficients of the exciting field, the scattered field from particle $i$ propagating to arbitrary position $\mathbf{r}$ can be obtained through its $T$-matrix elements $T_{mnp}$ as \cite{mackowskiJOSAA1996,tsang2004scattering}
\begin{equation}\label{sca_expansion}
\mathbf{E}_{\text{s}}^{(i)}(\mathbf{r})=\sum_{mnp} c_{mnp}^{(i)}T_{mnp}\mathbf{N}^{(3)}_{mnp}(\mathbf{r}-\mathbf{r}_i).
\end{equation}
Inserting Eqs.(\ref{exc_expansion}) and (\ref{sca_expansion}) into Eq.(\ref{fl_eq}), we obtain
\begin{equation}\label{fl_eq2}
\begin{split}
\sum_{mnp} c_{mnp}^{(j)}\mathbf{N}^{(1)}_{mnp}(\mathbf{r}-\mathbf{r}_j)=\mathbf{E}_{\text{inc}}(\mathbf{r})+\sum_{i=1\atop i\neq j}^{N}\sum_{mnp} c_{mnp}^{(i)}T_{mnp}\mathbf{N}^{(3)}_{mnp}(\mathbf{r}-\mathbf{r}_i).
\end{split}
\end{equation}

To solve this equation, we need to translate the VSWFs centered at $\mathbf{r}_i$ to their counterparts centered at $\mathbf{r}_j$. Using translation addition theorem for VSWFs (see Ref. \cite{wangPRA2018}), Eq.(\ref{fl_eq2}) becomes
\begin{equation}\label{fl_eq3}
\begin{split}
\sum_{mnp} c_{mnp}^{(j)}\mathbf{N}^{(1)}_{mnp}(\mathbf{r}-\mathbf{r}_j)=\mathbf{E}_{\text{inc}}(\mathbf{r})+\sum_{i=1\atop i\neq j}^{N}\sum_{mnp\mu\nu q} c_{\mu \nu q}^{(i)}T_{\mu\nu q}A_{mnp\mu\nu q}^{(3)}(\mathbf{r}_j-\mathbf{r}_i)\mathbf{N}^{(1)}_{m np}(\mathbf{r}-\mathbf{r}_j), 
\end{split}
\end{equation}
where $A_{mnp\mu\nu q}^{(3)}(\mathbf{r}_j-\mathbf{r}_i)$ can translate the outgoing VSWFs centered at $\mathbf{r}_i$ to regular VSWFs centered at $\mathbf{r}_j$, whose explicit expressions are listed in \ref{vswf_appendix}. We further expand the incident waves into regular VSWFs centered at $\mathbf{r}_j$ with expansion coefficients $a_{mnp}^{(j)}$, use the orthogonal relation of VSWFs with different orders and degrees, and obtain the following equation:
\begin{equation}\label{fl_eq4}
c_{mnp}^{(j)}=a_{mnp}^{(j)}+\sum_{i=1\atop i\neq j}^{N}\sum_{\mu\nu q}c_{\mu\nu q}^{(i)}T_{\mu\nu q}A_{mnp\mu\nu q}^{(3)}(\mathbf{r}_j-\mathbf{r}_i).
\end{equation}
The matrix form of this equation can be directly exploited in numerical calculation provided the particle positions are known. 

Actually, Eq.(\ref{fl_eq4}) is the governing equation of the well-established Fortran code multiple-sphere \textit{T}-matrix method (MSTM) developed by Mackowski and Mishchenko \cite{mackowskiJOSAA1996,mackowskiJQSRT2013} to exactly solve the multiple scattering of electromagnetic waves for a group of spherical particles. This code is further extended to optically active media \cite{mackowskiJQSRT2011, mackowskiJQSRT2013}. MSTM is widely used by many authors in the fields of thermal radiation transfer \cite{maJQSRT2017}, astrophysics \cite{pitmanJQSRT2017} and nanophotonics \cite{Naraghi2015}, together with a similar Fortran code developed by Xu, the generalized multiparticle Mie-solution (GMM) \cite{xuJQSRT2001}. Other similar algorithms also using the multipole expansion include Stout \textit{et al}'s recursive transfer matrix method \cite{stoutJMO2002} and Chew \textit{et al}'s recursive T-matrix method \cite{wangIEEETAP1993}, etc. Notably, exploiting the CUDA-acceleration feature of graphic processing units (GPUs) for parallel computing, Egel \textit{et al.} \cite{egelJQSRT2017} recently built a Matlab toolbox using the same governing equation, which can be at least twice faster than MSTM code, according to their testing cases. 

For nonspherical particles, the generalization of this multipole expansion method of FLEs is then the multiple particle \textit{T}-matrix method \cite{varadanPRD1979}, which uses the full-form \textit{T}-matrix of nonspherical particles. However, it should be noted that a severe limitation of the multiple particle \textit{T}-matrix method is its incapability to deal with elongated particles placed in the near-field of each other or of an interface \cite{bertrand2019global}. This is because the VSWF decomposition of the electromagnetic field is formally valid only in a uniform background beyond the smallest sphere that circumscribes the entire particle. Recent efforts to solve this limitation include Refs. \cite{doicuJQSRT2010,forestiereJQSRT2011,egelOE2016,egelJQSRT2017b,theobaldPRA2017,bertrand2019global}, to name a few. As a variant, the fast multi-particle scattering (FMPS) algorithm \cite{gimbutasJCP2013,laiOE2014,blankrotIEEEJMMCT2019} has been developed in recent years to accelerate the computation speed of the multi-particle scattering problem by using the combination of the integral equation technique to discretize each well-separated nonspherical particle \cite{martinEABE2003} (or closely placed particle clusters in order to avoid the overlapping of the enclosing sphere due to the same reason above), the Debye scalar potential representation \footnote{Instead of using the VSWF representation in Eqs.(\ref{exc_expansion})-(\ref{fl_eq4}), this scalar potential representation can reduce the complexity by avoiding the heavy use of vector algebra and vector translation functions \cite{gumerovJCP2007}.} and the fast multipole method (FMM) \cite{greengardJCP1987,enghetaIEEETAP1992,chengJCP2006,markkanenJQSRT2017}. This algorithm permits to compute the electromagnetic field  for ensembles containing several thousand or more particles on a single CPU \cite{gimbutasJCP2013}.

\subsubsection{The coupled-dipole model (CDM)}\label{cdm_intro}
Since Eq. (\ref{fl_eq4}) is a general equation which includes all multipolar excitations in the spheres, a much simpler method which only considers the electric dipolar excitations is also frequently used. This is the well-known coupled-dipole model. As mentioned in Section \ref{single_scatter}, the governing equation of this method and the DDA method is the same, despite the fact that they are used for different purposes. This coupled-dipole model, although very simplified, still preserves the essence of multiple scattering physics, and therefore usually acts as a prototype model for studying many complicated mechanisms, for instance, the recurrent scattering mechanism \cite{Cherroret2016}. Moreover, it is also very suitable for a group of small particles in which only dipolar modes are excited, not necessarily being spheres. Along with the rapid development of nanofabriction, plasmonics and nanophotonics, the coupled-dipole model is now already verified experimentally for a variety of nanostructures, including one-dimensional chains of metallic (typically gold and silver which support surface plasmon polaritons) nanoparticles \cite{maierPRB2002}, ordered \cite{gunnarssonJPCB2005,barnesPRL2008} and disordered \cite{barnesOL2009} two-dimensional arrays of metallic nanoparticles, and three-dimensional nanoparticle aggregates \cite{taylorJPCC2012}, etc. The nanoparticles can be nanorods \cite{barnesOL2009}, nanospheres \cite{maierPRB2002} and many other geometries \cite{kravetsPRB2014}, only if they show an electric-dipole electromagnetic response. Moreover, for ultracold two-level atoms, in the limit of low excitation intensity of light, where atoms are not saturated, and assuming no Zeeman degeneracy of hyperfine sublevels or other internal quantum effects, for example, $\mathrm{Sr}$ atoms with zero electronic angular momentum \cite{haveyPhysrep2017}, the response of a single two-level atom over light can be also treated as a linear point dipole. In this circumstance,  the coupled-dipole model is also suitable to describe the light-matter interaction in ultracold atomic clouds \cite{guerinPRL2016}, which will be discussed in Section \ref{light_atom_interaction}.

Since we are working in a single frequency/wavelength (the frequency domain), we will abbreviate the frequency dependency of all quantities appearing below. In vacuum or any homogeneous, isotropic host medium, the coupled-dipole model has the following form \cite{wangIJHMT2018}: 
\begin{equation}\label{coupled-dipole_eq}
\mathbf{d}_j=\alpha\left[\mathbf{E}_\mathrm{inc}(\mathbf{r}_j)+k^2\sum_{i=1,i\neq j}^{N}\mathbf{G}_0(\mathbf{r}_j,\mathbf{r}_i)\mathbf{d}_i\right],
\end{equation}
where $\alpha$ is the polarizability of the particle.  $\mathbf{E}_\mathrm{inc}(\mathbf{r}_j)$ is the incident field impinging on the $j$-th particle. For instance, for a plane wave illumination along the $z$-axis, we have $\mathbf{E}_\mathrm{inc}(\mathbf{r}_j)=\mathbf{E}_0\exp{(i\mathbf{k}\cdot\mathbf{r}_j)}$ with $\mathbf{k}=k\hat{\mathbf{z}}$. $\mathbf{d}_i$ is the excited dipole moment of $i$-th particle. $\mathbf{G}_{0}(\omega,\mathbf{r}_j,\mathbf{r}_i)$ is the free-space dyadic Green's function and describes the propagation of scattered field of $j$-th dipole to $i$-th dipole as \cite{Cherroret2016} 
\begin{equation}\label{free_space_green_function}
\begin{split}
\mathbf{G}_{0}(\mathbf{r}_j,\mathbf{r}_i)=\frac{\exp{(ikr)}}{4\pi r}\left(\frac{i}{kr}-\frac{1}{k^2r^2}+1\right)\mathbf{I}+\frac{\exp{(ikr)}}{4\pi r}\left(-\frac{3i}{kr}+\frac{3}{k^2r^2}-1\right)\mathbf{\hat{r}}\mathbf{\hat{r}}-\frac{\delta({\mathbf{r}})}{3k^2},
\end{split}
\end{equation}
where the Dirac delta function $\delta(\mathbf{r})$ is responsible for the so-called local field in the scatterers \cite{Cherroret2016}. $\mathbf{I}$ is identity matrix and $\hat{\mathbf{r}}$ is the unit vector of $\mathbf{r}=\mathbf{r}_j-\mathbf{r}_i$. After the EM responses of all scatterers (namely, all dipole moments $\mathbf{d}_i$ with respect to a specific incident field) based on above multiple wave scattering equations are calculated, the total scattered field of the random cluster of particles at an arbitrary position $\mathbf{r}\neq\mathbf{r}_j$, where $\mathbf{r}_j$ denotes the position of scatterers, is computed as
\begin{equation}
\mathbf{E}_s(\mathbf{r})=k^2\sum_{i=1}^{N}\mathbf{G}_0(\mathbf{r},\mathbf{r}_j)\mathbf{d}_j,
\end{equation}
where the Green's function $\mathbf{G}_0(\mathbf{r},\mathbf{r}_j)$ then describes the propagation of the scattered field of $j$-th dipole to a given position $\mathbf{r}$, similarly. 
%The knowledge of scattered field enables us to obtain the mescoscopic and macroscopic transport parameters and radiative properties. More precisely, 
And the total extinction cross section of the group of scatterers can also be calculated:
\begin{equation}
C_\mathrm{e}=k\sum_{j=1}^N\mathrm{Im}(\mathbf{d}_j\cdot\mathbf{E}_{\mathrm{exc},j}^*),
\end{equation}
where $\mathbf{E}_{\mathrm{exc},j}$ is the exciting field imping on $j$-th particle and given by $\mathbf{E}_{\mathrm{exc},j}=\mathbf{d}_j/\alpha$. The total absorption cross section is calculated as
\begin{equation}\label{cabs_cdm}
C_\mathrm{a}=k\sum_{j=1}^N\mathrm{Im}(\mathbf{d}_j\cdot\mathbf{E}_{\mathrm{exc},j}^*-\frac{k^3}{6\pi}|\mathbf{d}_j|^2).
\end{equation}
Therefore, the total scattering cross section of the system of dipoles is directly given by $C_\mathrm{s}=C_\mathrm{e}-C_\mathrm{a}$.

\subsection{Radiative transfer and diffusion equations}\label{rteandde}
Above equations are all related to the propagation of electromagnetic waves in the framework of Maxwell's equations, while the radiative properties are defined in the framework of the RTE that treats radiation as energy bundles (like classical particles). In this subsection, we attempt to briefly discuss how to establish the relationship between Maxwell's equations and the equation of radiative transfer, as well as the hydrodynamic limit, the diffusion equation.

\subsubsection{Radiative transfer equation}
As mentioned before, RTE is derived phenomenologically in its initial stage from energy conservation considerations \cite{chandrasekhar1950radiative}. In the 1970s-1980s, there were continuous efforts to establish the relationship between the RTE and Maxwell's equations, and it was finally shown that this connection can be made by means of the analytical wave theory \cite{ishimaru1978book,tsang1985theory,collettJOSAA1977,fanteJOSAA1981}. To put it simply, the RTE can be derived by retaining the so-called ladder diagrams in the Bethe-Salpeter equation in which the field correlation function $\langle\mathbf{E}(\mathbf{r})\mathbf{E}^*(\mathbf{r}')\rangle$ is expressed into the specific intensity. Specifically, the ladder diagrams are constructed by connecting the multiple wave scattering trajectories (according Eq.(\ref{lp_eq5})) to their complex-conjugated counterparts with exactly the same ordering of scatterers, and in each of these trajectories, all scatterers are assumed to be visited only once. This procedure results in diagrams looking like a series of ``ladders", which then give them the name. As a result, this ladder approximation only describes the transport of radiation intensity, and at the intensity (or more formally, irreducible vertex) level, it is actually equivalent to the ISA \cite{VanRossum1998}. All kinds of \textit{microscopic and mesoscopic} wave interference effects occurring outside the scatterers are neglected in the RTE \cite{lagendijk1996resonant,VanRossum1998,vanTiggelenRMP2000,tsang2004scattering,mishchenko2006multiple,sheng2006introduction,mishchenkoPhysrep2016}. It is also noted that according to the derivation procedure, the RTE describes the transport at length and time scales much larger than the wavelength and the period of light, and assumes weak scattering, i.e., scattering/transport mean free path is much larger than the wavelength \cite{lagendijk1996resonant,VanRossum1998,tsang2004scattering,sheng2006introduction,ishimaru1978book}. 
Details of the derivation of the RTE using the diagrammatic technique in the analytic wave theory can be found in many monographs and papers, e.g., Refs.\cite{cazeJOSAA2015,tsang2004scattering,VanRossum1998,tsang2004scattering,vynckPRA2016} and thus are not shown here. For a historical review about the relationship between the analytic wave theory and the RTE, see the Van de Hulst essay paper by Tsang \cite{tsangJQSRT2019}. Remarkably, Doicu and Mishchenko presented a series of reviews to describe how to derive the RTE from Maxwell's equations using different methods and discuss relevant interference effects \cite{doicuJQSRT2018,doicuJQSRT2019,doicuJQSRT2019a,doicuJQSRT2019b,doicuJQSRT2019c,doicuJQSRT2019d}, including the far-field Foldy-Lax equations \cite{doicuJQSRT2018} and the analytic wave theory (or directly dubbed "Dyson and Bethe–Salpeter equations") \cite{doicuJQSRT2019}.

\begin{figure}[htbp]
	\centering
	%	\subfloat{
	%		\label{photonicglass}
	%		\includegraphics[width=0.4\linewidth]{photonicglass}
	%	}
	\subfloat{
		\includegraphics[width=0.6\linewidth]{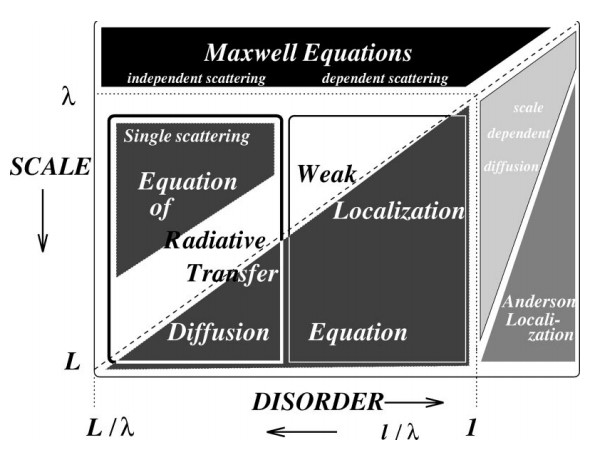}
	}
	\caption{Schematic diagram of different regimes in the theoretical treatment of radiative transport. Reprinted from Ref. \cite{vanTiggelenRMP2000}. Copyright 2000 by the American Physical Society.}\label{RTE_regime_vanTiggelen}
	
\end{figure}

To show the connection between the RTE and Maxwell's equations more intuitively, in Fig. \ref{RTE_regime_vanTiggelen}, a schematic diagram of different regimes in the theoretical treatment of radiative transfer, proposed by van Tiggelen \textit{et al.} \cite{vanTiggelenRMP2000}, is presented. In this figure, different regimes of radiative transfer are distinguished by three parameters, i.e., the wavelength of radiation $\lambda$, the degree of disorder characterized by the mean free path $l$ divided by the wavelength, and the length scale of radiative transfer quantified by the sample size $L$ divided by the wavelength.  It is clearly indicated by this figure that the RTE is most suitable for the situations in which $l\gg\lambda$ (weak scattering or equivalently, weak disorder) and $L\gg\lambda$. When the disorder (or equivalently, scattering strength) is increased with intermediate scattering strength (i.e., $l/\lambda$ is sufficiently larger than unity but away from the weak scattering limit, i.e., $l$ is at the scale of several $\lambda$s), mesoscopic interferences like the weak localization occur at the length scale of $L\gg \lambda$. Weak localization, also known as the coherent backscattering, can be expressed as a series of most-crossed diagrams \cite{vandermarkPRB1988,akkermansPRL1986,akkermans1988theoretical} (or so-called cyclical terms \cite{mishchenkoMNRAS1992,mishchenko2006multiple}) in the diagrammatic representation of the irreducible intensity vertex in the Bethe-Salpeter equation, which cannot be considered in the RTE (see Section \ref{mesointerference}). Moreover, when the disorder continues to increase to result in $l\lesssim\lambda$, strong localization (or known as Anderson localization) can occur. And at small length scales and intermediate disorder (or scattering strength), the DSE is important, manifested as the wave interferences that take place at the length scale $L\sim \lambda$,

As mentioned in Section \ref{concepts}, in a rigorous sense, the RTE cannot be applied in the dependent scattering regime since it is derived under the ISA, which is also implied by Fig.\ref{RTE_regime_vanTiggelen}. However, it is quite useful in practice to retain the form of RTE by correcting the mesoscopic radiative properties by considering DSE. This is key assumption of the present article and most reviewed references herein \cite{tsang2004scattering}. It is thus postulated that the DSE introduces some additional perturbative terms to the diagrammatic expansion of the RTE and results in a modification to the mesoscopic radiative properties. In this manner, the numerical techniques to solve the RTE in large-scale disordered media can be exploited, which is very valuable in practice. For instance, the dense media radiative transfer theory (DMRT), developed by Tsang and coworkers \cite{tsangJQSRT2019,tsang1985theory,tsang2004scattering} in the spirit of this postulation, has been widely applied in the field of microwave remote sensing for the prediction of radiative transfer in dense media like terrestrial snow with densely packed ice grains, whose validity has been confirmed by extensive experimental measurements \cite{mandtWRM1992,tsangRS2000,liangIEEETGRS2008,picardGMD2013}. As a consequence, this postulation is also viable in many other practical applications like thermal engineering where moderate-refractive index materials are used, usually away from electromagnetic resonances\footnote{When there are electromagnetic resonances like Mie or some internal resonances in the densely packed scatterers, it is not quite clear whether this postulation can effectively work. By now there are very few works on resonant multiple scattering in dense DDM in the regime described by the RTE ($L\sim l$). Most works about resonant multiple wave scattering are either on dilute DDM, e.g., Refs. \cite{vanalbabaPRL1991, Cherroret2016,lagendijk1996resonant} or in the diffusive regime, e.g., Refs. \cite{storzerPRL2006,aubryPRA2017}, to name a few. In particular, in highly scattering DDM in the diffusive transport regime, it is instructive to note that the diffusion coefficient should be significantly modified by introducing a position dependence to account for \textit{mesoscopic} wave interferences like weak and strong localization mechanisms and the theory agrees well with experiments and numerical simulations \cite{tiggelenPRL2000,tianPRL2010,yamilovPRL2014,huNaturephys2008,zhangPRB2009,haberko2018transition}. This may provide some implications on developing and justifying similar schemes in the RTE regime for microscopic interferences under resonant multiple wave scattering.}. To the best of our knowledge and experiences, this is the only feasible approach to treat large-scale DDM when the mesoscopic radiative properties are significantly affected by the DSE. 

Although the RTE is an approximate equation that describes radiative transfer in disordered media, closed-form analytical solution is usually not available in most cases \cite{howell2015thermal}. In fact, it is far from trivial to solve this equation with high accuracy and fast computation speed simultaneously for large-scale disordered media under realistic conditions (e.g., spatially gradient radiative properties, complicated scattering phase functions, irregular boundary conditions, etc.), which is still an active area under intensive investigations by researchers from a variety of fields including thermal sciences \cite{modest2013radiative,howell2015thermal}, atmospheric sciences \cite{thomas2002radiative,cloughJQSRT2005}, astrophysics \cite{peraiah2002introduction,chandrasekhar1950radiative}, remote sensing \cite{tsang1985theory,myneniIEEEGRS1997}, computer graphics \cite{jaraboACMTG2018,zhangACMTG2019}, applied mathematics \cite{kimSIAMJSC2002,edstromSIAMR2005}, to name a few. Typical numerical methods include the discrete ordinates method (DOM, or the $S_N$ method) \cite{fivelandJTHT1988,edstromSIAMR2005,lehardyJCP2017}, the method of spherical harmonics (the $P_N$ method) \cite{modestJQSRT2008,hermelineJCP2016}, the Monte Carlo method \cite{howellJHT1998,whitneyBASI2011,wangJQSRT2017}, the Chebyshev spectral method \cite{kimJCP1999,kimSIAMJSC2002} and the lattice Boltzmann method \cite{maPRE2011,minkJQSRT2020}, etc. The inverse RTE problem, that aims to reconstruct the radiative properties of disordered media from macroscopically measured signals (reflectance, transmittance, etc.) and usually ill-posed or ill-conditioned (e.g., the crosstalk between scattering and absorption coefficients), is even more challenging, which requires more elaborate computational methods \cite{mccormickNSE1992,maJQSRT2016,smirnovSIAMJSC2019}, some \textit{a priori} information about the media, and as will be mentioned in Section \ref{expsec}, more experimental data, e.g.,  by modulating the illumination \cite{renSIAMJSC2006}.

\subsubsection{Diffusion equation}
A well-known limit of the RTE is the celebrated diffusion equation of photons, which follows from the diffusive transport behavior of other classical particles in Brownian motion, like macroscopic heat transport by phonons \cite{ishimaru1978book,VanRossum1998,sheng2006introduction}. This equation applies for highly scattering media with sufficiently low absorption, and the thickness of the sample should be much larger than the scattering mean free path, namely, $L\gg l_s$. As a result, the photons are scattered so many times before exiting the sample that they ``forget" their initial transport direction, and therefore the long-time (long path-length) transport behavior can be described as isotropic, even if the scattering phase function itself is also anisotropic \cite{VanRossum1998}. The resulting diffusive scale of photon migration mean free path is called transport mean free path $l_\mathrm{tr}=l_{s}/(1-g)$. In this circumstance, we say the radiative transfer enters the ``diffusive regime", which is important and exhibits distinctions from the regime of radiative transfer equation, as will be shown below.

The diffusion equation takes the following form (a derivation from RTE to diffusion equation is given, for instance, in Ref.\cite{continiAO1997}):
\begin{equation}\label{diffusion_eq}
\Big(-\nabla \cdot  D\nabla+v_E\kappa_a\Big)\phi(\mathbf{r})=Q(\mathbf{r}),
\end{equation}
where $D$ is called the diffusion coefficient, $\phi(\mathbf{r})=\frac{1}{4\pi}\int_{4\pi} I(\mathbf{r},\mathbf{\Omega})d\mathbf{\Omega}$ is the average diffuse intensity, $Q(\mathbf{r})$ is the source function and $v_E$ is the energy transport velocity. For nonabsorbing disordered media, the diffusion coefficient is given by $D=v_El_\mathrm{tr}/3$. In the meantime, when the absorption coefficient is nonzero, $D$ should somehow depend on it, while the exact formula is still under debate \cite{continiAO1997}. A heuristic form is $D=v_E/[3(\kappa_\mathrm{tr}+\kappa_a)]$, where $\kappa_\mathrm{tr}=1/l_\mathrm{tr}$ is the transport or reduced scattering coefficient. Pierrat \textit{et al.} \cite{pierratJOSAA2006} proposed a method to determine the exact diffusion coefficient based on finding the diffusion eigenmode of the RTE. Nevertheless, for very weakly absorbing media, the difference is somewhat not substantial. 

The diffusion equation has achieved a success in describing radiation transport in highly scattering media, such as $\mathrm{TiO_2}$ colloidal films \cite{storzerPRL2006}, porous gallium phosphide (GaP) networks \cite{schuurmansScience1999}, and is the fundamental basis for understanding radiative transfer in such disordered media. For example, Eldridge \textit{et al.} \cite{eldridgeJACS2008,eldridgeJACS2009} examined this relation with experimental data of scattering coefficient and transmittance of TBC samples and a good agreement was found. 

A useful property in the diffusion equation in nonabsorbing media is that the transmission of a thick slab geometry of thickness $L$ obeys the well-known Ohm’s law as
\begin{equation}\label{ohm_eq}
T=\frac{l_\mathrm{tr}+z_e}{L+2z_e},
\end{equation}
where $z_e$ is the extrapolation length given by
\begin{equation}\label{extrap_length}
z_e=\frac{2}{3}\bar{R}l_\mathrm{tr}.
\end{equation}
Here $\bar{R}$ is related to average reflection coefficient at the boundary surface, whose expression can be found in literature, for instance, Ref.\cite{continiAO1997}. For very thick samples, we have $L\gg l_\mathrm{tr}$, and therefore $T\sim l_\mathrm{tr}/L$ is obtained, which provides a convenient method to measure the transport mean free path, as will be demonstrated in Section \ref{expsec}.

Besides, it is interesting to note in Fig. \ref{RTE_regime_vanTiggelen} that the diffusion equation can be  even applied in the regime where the RTE breaks down. This is because the diffusion equation is also the hydrodynamic limit of the Bethe-Salpeter equation \cite{vanTiggelenRMP2000,sheng2006introduction}, and a derivation can be found in Refs.\cite{barabanenkovPLA1991,lagendijk1996resonant,VanRossum1998,sheng2006introduction,akkermans2007mesoscopic,cazeJOSAA2015,Cherroret2016}, to name a few. In this sense, in the diffusive regime ($L\gg l$), the diffusion equation is not subject to the limitations of the RTE, such as the weak scattering condition. In particular, the diffusion equation is able to describe weak localization effect through a renormalized diffusion coefficient \cite{sheng2006introduction}. Moreover, as the scattering strength or disorder continues to increase, approaching the threshold of localization transition (i.e., $l\sim\lambda$), the diffusion equation can be further extended by introducing a position-dependent diffusion coefficient to consider the position-dependent return probability of waves via the looped multiple-scattering paths \cite{tiggelenPRL2000,tianPRL2010,yamilovPRL2014}.

To summarize, in this section, we review the basic theories for describing single and multiple scattering of electromagnetic waves in DDM based on the first-principle Maxwell's equations. For single scattering, we introduce the Mie theory, the $T$-matrix method and the DDA. For multiple scattering of electromagnetic waves, we describe the analytical wave theory (Dyson and Bethe-Salpeter equations) and the Foldy-Lax equations. We also show that the diagrammatic technique can be applied to obtain the self-energy and irreducible intensity vertex in a perturbative way, and these two quantities can be used to give radiative properties. The simple coupled-dipole model is also introduced as a popular and easily accessible tool to investigate the underlying physics in a group consisting of a large number of scatterers. On this basis, we proceed to a discussion on the relationship between Maxwell's equations and the RTE as well as the diffusion equation along with their applicabilities, in order to demonstrate the role of wave interferences in radiative transfer and understand the nature of the DSE.

\section{Theoretical and numerical treatments on the DSE}\label{theoryandnumerical}
According to the discussion in Section \ref{indanddep}, it is necessary to explore the underlying mechanisms in the DSE, and develop corresponding analytical and semi-analytical methods to predict the effect of dependent scattering mechanism on the mesoscopic radiative properties. In this section, we first describe some basic mechanisms involved in the DSE, including the far-field DSE, near-field DSE, recurrent scattering, structural correlations and the effect of absorbing host media, and summarize relevant theoretical models that deal with them, most of which have closed-form analytical formulas. Then numerical methods to model the DSE are summarized, including the supercell method, the representative volume element method and the direct numerical simulation method. 

\subsection{Theoretical mechanisms of the DSE}\label{mechanisms}
\begin{figure}[htbp]
	\flushleft
	\subfloat{
		\label{farfieldinterference}
		\includegraphics[width=0.48\linewidth]{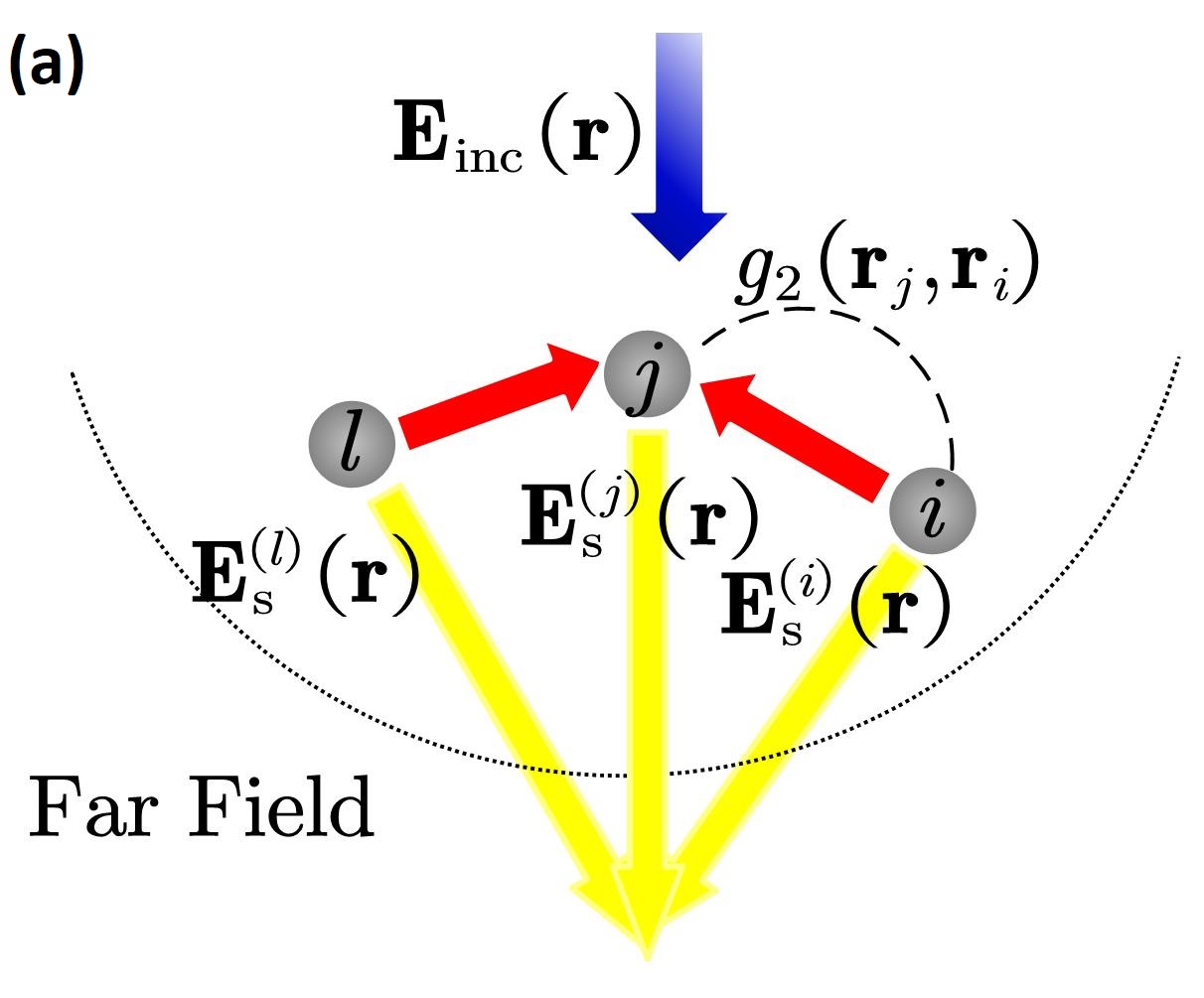}
	}
	\subfloat{
		\label{nearfieldcoupling}
		\includegraphics[width=0.48\linewidth]{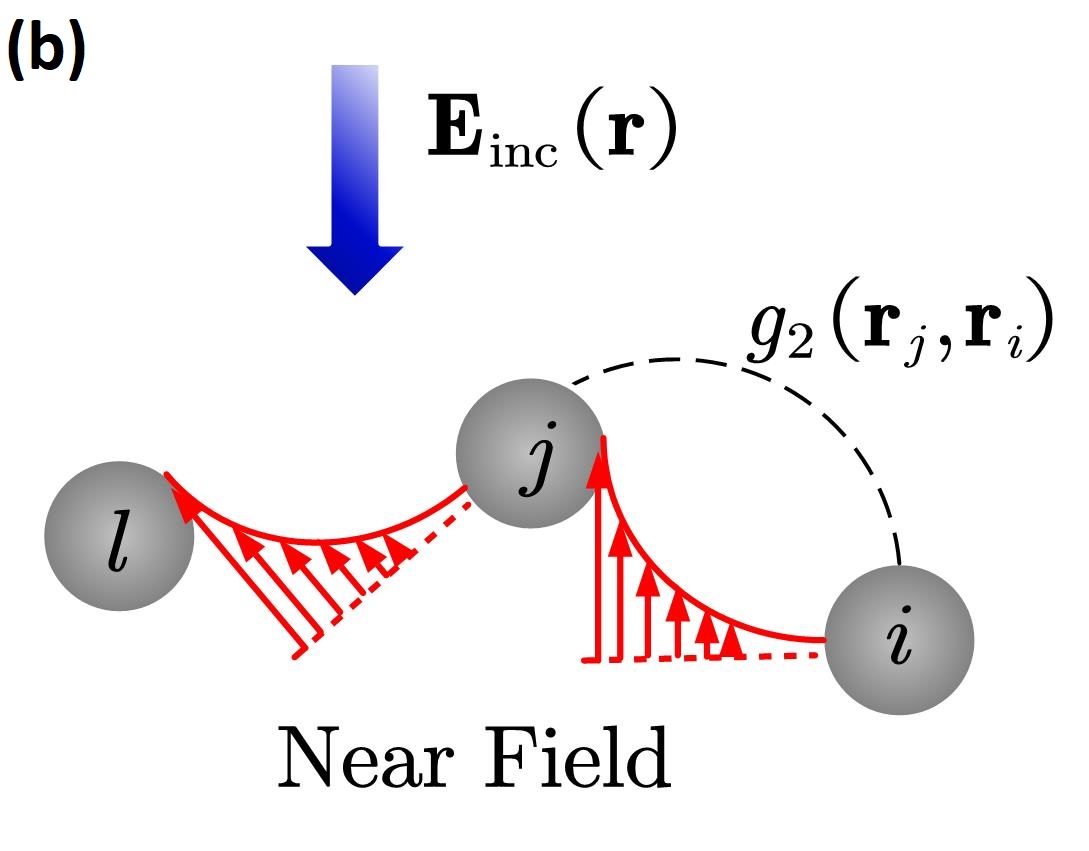}
	}
	\caption{Schematic of the \textbf{(a)} far-field and \textbf{(b)} near-field DSEs in a discrete disordered medium. Here for simplicity, only three scatterers are presented, which are numbered as $i$, $j$, $l$. The dashed line denotes $g_2(\mathbf{r}_j,\mathbf{r}_i)$, the pair distribution function between the two scatterers. In \textbf{(a)}, the dotted line indicates a schematic boundary of the far-field region outside of the entire medium, and the scatterers are also distributed in the far field region of each other. The blue thick arrow indicates the incident wave $\mathbf{E}_\mathrm{inc}$, while the red thick arrows stand for the partially scattered, propagating waves from particle $i$ to $j$ and from $l$ to $j$. The yellow thick arrows represent scattered waves that propagate to the far field out of the medium $\mathbf{E}^{i}_\mathrm{s}(\mathbf{r})$, $\mathbf{E}^{j}_\mathrm{s}(\mathbf{r})$ and $\mathbf{E}^{l}_\mathrm{s}(\mathbf{r})$. In \textbf{(b)}, the scatterers are located in the near fields of each other, and the multiple wave scattering is dominated by the near-field tunneling of evanescent waves, as shown by the red thin arrows combined with a curve representing the exponential decay of wave amplitudes.
}\label{ffnf_dse}
\end{figure}

The DSE, which can occur in the near field as well as the far field among adjacent scatterers \cite{akkermans2007mesoscopic,sheng2006introduction,VanRossum1998}, will significantly affect the radiative properties in different length scales. Accordingly, the DSE can be roughly classified into two categories, i.e., the far-field and near-field DSEs. Figure \ref{ffnf_dse} schematically shows their differences. In the following, we will briefly review theoretical considerations on these two categories of dependent scattering mechanisms. And in particular, we also introduce the recurrent scattering mechanism, the role of structural correlations and the absorption of the background medium on the DSE.

\subsubsection{The far-field DSE}\label{farfieldinterference_sec}
The far-field DSE, depicted in Fig.\ref{farfieldinterference}, mainly considers the interference of scattered electromagnetic waves from different scatterers in the far field, while different scatterers can also interact with each other through far-field scattered waves. The main contribution to the far-field DSE, in conventional cases, stems from the inter-particle correlations, since these correlations lead to constructive or destructive interferences among the far-field scattered waves. Therefore, theoretical models on this mechanism usually relies on some description of the distribution of scatterers. More specifically, the structure factor, which is the Fourier transform of the two-particle correlation function, usually provides a first-order consideration of the far-field DSE, leading to the well-known interference approximation (ITA). In this method, the single scattering property of an individual scatterer is assumed to be not affected, namely, no interparticle interactions through scattered waves (denoted by red thick arrows in Fig.\ref{farfieldinterference}). As a result, the ITA is not able to fully capture the far-field DSE, especially when the packing density is becomes large, structural correlations are strong or the scatterers are highly scattering. 

A conspicuous mechanism that is not accounted for in the first-order correction of the far-field DSE is the deformation of local electromagnetic field impinging on each scatterer, due to the scattered waves from adjacent scatterers. It means that the local incident field with respect to each scatterer is no longer the same as the externally incident field as the ITA and ISA assume. This mechanism becomes appreciable at moderate and high packing densities for strongly scattering particles. To tackle with this mechanism, many authors have introduced a homogenized environment with some effective refractive index surrounding each scatterer to modify the ITA model \cite{rojasochoaPRL2004,reuferAPL2007,xiaoSciAdv2017}. However, this type of methods is not capable of explicitly demonstrating how the local incident field is altered by other scatterers. Moreover, they cannot consider some circumstances in which, for a plane wave illumination, the local field can be deformed into neither plane-wave-like nor spherical-wave-like, resulting in the invalidity of the assumption of the existence of an effective refractive index for the surrounding background. The latter mechanism is seldom discussed, and will substantially affect the mesoscopic radiative properties, which was elucidated in Ref. \cite{wangPRA2018} based on a dependent scattering model derived from the quasicrystalline approximation (QCA) by our group. But the QCA approach cannot tackle with the resonances well. The details of theoretical models mentioned here will be further discussed in Section \ref{models}.

\subsubsection{The near-field DSE}
As the concentration of scattering particles in disordered media rises, they are inclined to step into the near fields of each other \cite{Liew2011,Naraghi2015} (i.e., the clearance $c$ between scatterers is comparable or even smaller than the wavelength, approximately, $kc\lesssim1$). In this circumstance, near-field interaction (NFI) among scatterers, which is negligible when the scatterers are in the far fields of each other, can also contribute to radiation energy tunneling and thus transport properties. Specifically, the NFI indicates the electromagnetic coupling of different scatterers through their scattered near fields, which are mainly composed of electromagnetic wave components with large wave numbers ($k>k_0$, where $k_0$ is the wave number of free-space radiation). Moreover, the NFI is purely vectorial, namely, containing both transverse and longitudinal components \cite{lagendijk1996resonant}, while in the far-field DSE, the far-field scattered waves are always transverse spherical waves. This can be intuitively understood from the dyadic Green's function in Eq.(\ref{free_space_green_function}), in which the transverse component contains a term slowly decaying with the distance $r$ as $1/r$ (i.e., a spherical wave) and the longitudinal component contains two terms decaying as $1/r^2$ and $1/r^3$\footnote{We have explicitly presented the expressions of transverse and longitudinal components in the dyadic Green's function in Eqs.(\ref{g_bot}) and (\ref{g_parallel}) for the convenience of calculation.}. As a consequence, the NFI leads to a more intricate picture of dependent scattering than the far-field interaction.  Figure \ref{nearfieldcoupling} schematically shows the possible tunneling of evanescent waves between nearby scatterers, where the red thin arrows combined with a curve representing the exponential decay of the amplitudes of evanescent waves.

The near-field DSE in densely packed DDM is still difficult to fully capture by now. Recently, there has been growing interest in addressing this mechanism thanks to the development of computational and experimental capabilities. It was numerically and experimentally shown by Naraghi \textit{et al.} that the NFI can enhance total transmission of disordered media by adding channels of transport \cite{Naraghi2015}. It is also demonstrated that the longitudinal component of the NFI is a hindering factor for Anderson localization in three dimensions \cite{Skipetrov2014,escalanteADP2017,schirmacherPRL2018}, while interestingly Silies \textit{et al.} revealed that near-field coupling  assists the formation of localized modes \cite{Silies2016}. Pierrat \textit{et al.} reported that NFI of a dipolar emitter with more than one particle creates optical modes confined in a small volume around it and give rise to strong fluctuations in local density of states (LDOS) \cite{Pierrat2010}.  
Notably, Tishkovets and coworkers \cite{tishkovetsJQSRT2006,petrovaIcarus2007,tishkovetsJQSRT2008,tishkovetsJQSRT2011,tishkovetsLSR2013} carried out a series of theoretical analysis on the near-field DSE in clusters of densely packed scatterers with sizes comparable or smaller than the wavelength, where the near-field mutual shielding effect, the inhomogeneous field and the negative values of the degree of linear polarization in the backscattering direction were discussed. Nevertheless, a clear elucidation of the near-field DSE and its influence on the mesoscopic radiative properties is still rare, although several phenomenological models are developed to qualitatively address the effect of NFI, which will be presented in Section \ref{nfmodels}. Note the methods based on the coherent-potential approximation (CPA) are believed to account for the near-field interactions to some degree in densely packed DDM, although in an implicit way through an effective refractive index.

In addition, since NFI is much stronger than far-field interaction, it is promising to utilize NFI to achieve extreme light-matter interaction. Therefore, the near-field DSE still needs to be systematically and quantitatively investigated. Recently, Shen and Dogariu \cite{shenOptica2019} investigated the phase and effective interaction volume of a nanoparticle, which provided some important insights for the near-field DSE.

\begin{figure}
	\centering
	\includegraphics[width=0.8\linewidth]{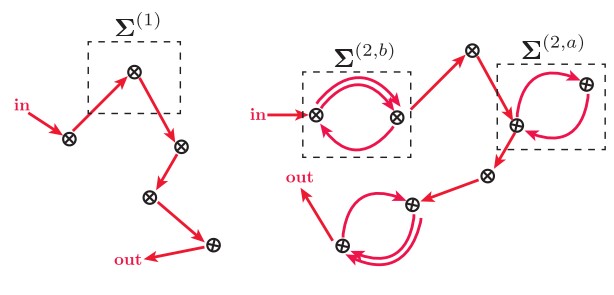}
	\caption{Schematic of the recurrent scattering mechanism in a random medium containing discrete scatterers. The $\otimes$ indicates a scatterer and the lines stand for the propagation of waves. Left: a multiple scattering path visiting independent scatterers, in which the propagating wave is never scattered more than once by the same scatterer. The contribution to the self-energy is $\bm{\Sigma}^{(1)}$, namely, the independent scattering approximation. Right: a multiple scattering path involving the repeated scattering between pairs of scatterers. There are two types of contributions to the self-energy, $\bm{\Sigma}^{(2,a)}$ and $\bm{\Sigma}^{(2,b)}$. The former describes binary processes in which the radiation incident on one scatterer eventually returns to the same one, and the latter describes all processes in which the radiation incident on one scatterer emerges from the second. Reprinted with permission from Ref. \cite{Cherroret2016}.  Copyright 2016 by the American Physical Society.  }\label{recurrent_scattering_schematic}
\end{figure}

\subsubsection{Recurrent scattering mechanism}\label{rsintro}
Another mechanism in the DSE worth discussing is the recurrent scattering mechanism. When the scattering strength increases, the probability of a multiply scattered wave propagating back to a scatterer which it formerly visited also grows, leading to a closed-loop-like scattering trajectory. For very strongly scattering media, for example, cold atomic clouds near the atomic bare resonance and metallic nanoparticles near the localized surface plasmonic resonances, the influence of recurrent scattering is significant\cite{Cherroret2016,wangIJHMT2018}. However, the analytical calculation of the recurrent scattering mechanism is still very troublesome and can only be done for very simplified cases, for example, recurrent scattering between two point scatterers \cite{Cherroret2016,Vantiggelen1990JPCM,vantiggelenPRB1994}. Figure \ref{recurrent_scattering_schematic} shows a typical recurrent scattering scheme between a pair of nearby scatterers \cite{Cherroret2016}, which can result in analytical expressions for the self-energy.  In particular, there are two types of contributions to the self-energy, $\bm{\Sigma}^{(2,a)}$ and $\bm{\Sigma}^{(2,b)}$. The former describes two-scatterer processes in which the radiation incident on one scatterer eventually returns to the same one, and the latter describes all processes in which the radiation incident on one scatterer emerges from the second. In the diagrammatic representation of self-energy, this mechanism amounts to a group of self-connected diagrams. The calculation for this self-energy is presented in Section \ref{rsmodel}, which was conducted by Cherroret \textit{et al.} \cite{Cherroret2016}. It has been shown that for resonantly scattering particles, the impact of this two-scatterer recurrent scattering process becomes significant even in a dilute disordered medium ($4\pi n_0/k^3\ll1$) \cite{Cherroret2016,Vantiggelen1990JPCM,vantiggelenPRB1994}.

%Recently, it was found that NFI between closely packed random scatterers brings rich phenomena like inducing photon tunneling \cite{Naraghi2015} and “phase transition” \cite{Naraghi2016}, leading to unusual structural colors \cite{Liew2011} and affecting quantum emission \cite{Sapienza2011}.

%Above deviation is unambiguously originated from the near-field DSE. This effect can be phenomenologically divided into two mechanisms: 
%(1) The local electromagnetic field impinging on each scatterer is deformed, due to the scattered waves from adjacent scatterers, meaning that this field is no longer the same as the externally incident field as ISA assumed. In some circumstances, for a plane wave illumination, the local field can be deformed to be non plane wave, which will be discussed in Chapter \ref{chap3}. (2) The near-field interaction (NFI) between adjacent scatterers can induce the tunneling of evanescent waves, leading to a modification of radiative transport properties. These mechanisms are, to the best of our knowledge, not systematically and quantitatively studied before. 
%
%In a short summary, when the ISA is not applicable in DDM containing densely distributed scatterers or with substantial structural correlations, the DSE is important to the mesoscopic radiative properties. So far, a thorough theoretical understanding of the DSE is still lacking, especially the modification of local incident field and the near-field DSE.
 
\subsubsection{The role of structural correlations}\label{structural_correlation_sec}
As mentioned in previouse sections, one important factor that leads to the dependent scattering mechanism is known as the structural correlations, which describe the possible reminiscence of order (usually short- or medium-ranged) existing in the spatial variation of the dielectric constant in disordered media \cite{tsang2004scattering2}. They can give rise to definite phase differences among the scattered waves \cite{laxRMP1951,laxPR1952,fradenPRL1990,tsang2004scattering2,rojasochoaPRL2004,Froufe-PerezPNAS2017,liuJOSAB2018}, which can well preserve over the ensemble average procedure. Hence constructive or destructive interferences among the scattered waves occur and thus affect the transport properties of light remarkably. This is also called ``partial coherence'' by Lax \cite{laxRMP1951,laxPR1952}. The most well-known type of structural correlations is the hard-sphere positional correlation in disordered media consisting of purely hard spheres without any additional inter-particle interactions \cite{wertheimPRL1963,fradenPRL1990}. This is due to the fact that hard spheres cannot deform or penetrate into each other, and the structural correlations emerge when the concentration of spheres is substantial (usually under a volume fraction of $f_v>5\%$ \cite{mishchenkoOL2013}). Moreover, when the correlation length of particle positions is comparable with or smaller than the wavelength, the structural correlations play a very important role in determining the microscopic interferences and thus radiative properties \cite{fradenPRL1990,mishchenkoJQSRT1994,rojasochoaPRL2004}.

Generally speaking, the structural correlations are not only affected by the packing density (or volume fraction), but also by the interaction potential between particles. In fact, for some types of specially-designed structural correlations, even if the concentration of particles is not very high, the strong positional correlation will give rise to very significant interference phenomena, for instance, in the so-called short-range ordered hyperuniform media \cite{leseurOptica2016,Froufe-Perez2016,Froufe-PerezPNAS2017}. Several typical kinds of interaction potential among particles, for example, the surface adhesive potential \cite{Baxter1968,Frenkel2002} and the interparticle Coulombic electrostatic potential \cite{rojasochoaPRL2004,bresselJSQRT2013}, can be realized experimentally. By controlling the interaction potential and thus the structural correlations, a flexible manipulation of the radiative properties of random media can be achieved \cite{wangJAP2018}. 

\subsubsection{The effect of an absorbing host medium}
In most theoretical considerations of the multiple scattering of waves, the background medium is assumed to be nonabsorbing. This assumption holds for many applications of visible light in atmospheric sciences and optics. On the other hand, in many other applications, e.g., infrared spectroscopy, the host medium may become absorptive. As a matter of fact, dependent scattering in an absorbing host medium is still an unsolved problem both theoretically and experimentally. As light scattering by a single spherical particle embedded in an absorbing background medium was not formally solved theoretically until the 2000s \cite{sudiartaJOSAA2001,fuAO2001,yangAO2002,videenAO2003,yinJOSAA2006} and is still under intensive theoretical and experimental investigation \cite{aernoutsOE2014b,mishchenkoOL2017,mishchenkoJQSRT2018,mishchenkoOSAC2019b}, a full theory for the multiple scattering of electromagnetic waves in an absorbing host medium was still not well-established. Such a theory is nowadays in need for many applications, for example, the use of highly-scattering nanoparticles to enhance the light absorption of thin-film silicon solar cells \cite{leeOE2010,nagelOE2010}, as well as microsphere enhanced subdiffraction optical imaging \cite{chenAPR2019}. Recently, progresses on this problem were made by Mishchenko and coworkers \cite{mishchenkoOE2008,mishchenkoJQSRT2008}, who established the Foldy-Lax equations of multiple scatterers, solved the coherent electromagnetic field and derived the RTE in a weakly absorbing host media. Their derivation was based on the far-field approximation, namely, with no considerations of the DSE. Durant and coworkers \cite{durantJOSAA2007a,durantJOSAA2007} were the first to investigate the DSE in an absorbing matrix. They developed an analytical formula using the diagrammatic expansion method (as will be shown to be an extension of the Keller's approach) to predict the extinction coefficient, which considered the DSE and was verified by full-wave numerical simulations.

\subsection{Theoretical models of radiative properties considering the DSE}\label{models}
In this subsection, we present some notable theoretical models that are used to predict the radiative properties of DDM with considerations of the DSE, some of which are already mentioned in the previous subsection. We will give a brief discussion on their physical significance, derivation procedure, advantages and limitations. Most of the models can result in closed-form formulas for the extinction coefficient $\kappa_e$, while for scattering and absorption coefficients as well as the scattering phase function, many models did not have analytical expressions. This is because the latter quantities require additional derivations on the incoherent intensity (e.g., the irreducible vertex in the analytic wave theory), while deriving the extinction coefficient only needs the knowledge of the coherent field (e.g., the self-energy in the analytic wave theory), which is much easier to deal with. 

Here we focus on the theoretical models developed for homogeneous spherical scatterers due to its high availability, which  can be easily extended to the cases of other shapes, like multilayered spheres and cylinders in the 2D case. However, for scatterers of more complicated shapes like spheroids, cubes and pyramids, no general closed-form theoretical models are available due to many difficulties, one of which is to obtain the solution of pair distribution function describing structural correlations. Moreover, we only consider a single species of particles, and the particles are disorderedly distributed in vacuum without loss of generality. In addition, for most models, we mainly give analytical expressions for electric dipoles with a concise derivation process to present the underlying physics, while for high-order multipolar excitations, we basically provide the formulas derived in the literature.

%We also discuss the Foldy's approximation here.
\subsubsection{Interference approximation (ITA)}\label{ITA_model}
As discussed in Section \ref{farfieldinterference_sec} and Section \ref{structural_correlation_sec}, since the scatterers are randomly distributed in the medium and have finite sizes, it is pivot to take the inter-particle correlations into account in the analysis of the dependent scattering mechanism \cite{fradenPRL1990,rojasochoaPRL2004}, especially when the volume concentration is substantial. The interference approximation \cite{dickJOSAA1999}, also known as the collective scattering approximation that takes the ``collective scattering" due to structural correlations into account \cite{Naraghi2015}, is regarded as the first-order correction to the DSE. This method only considers the correlation between a pair of particles, which is described by the pair distribution function (PDF), $g_2(\mathbf{r}_1,\mathbf{r}_2)$ (already schematically shown in Figs.\ref{system_config} and \ref{farfieldinterference}). Specifically, it is the conditional probability density function of finding a particle centered at the position $\mathbf{r}_1$ when a fixed particle is seated at $\mathbf{r}_2$.  When assuming the random medium is statistically homogeneous and isotropic, the PDF only depends on the distance between the pair of particles, i.e., $g_2(\mathbf{r}_1,\mathbf{r}_2)=g_2(|\mathbf{r}_1-\mathbf{r}_2|)$ \cite{wertheimPRL1963,tsang2004scattering2}. There are already several approximate analytical solutions of the PDF for some specific random systems, e.g., Refs.\cite{wertheimPRL1963,Baxter1968}. A well-known approximation for hard spheres is the Percus-Yevick (P-Y) approximation \cite{wertheimPRL1963}. On the other hand, the PDF $g_2(r)$ between pairs of particles can also be obtained experimentally by analyzing the statistical correlations in the microscopic structures \cite{xiaoeSciAdv2019}. On this basis, the structure factor can be calculated through a Fourier transform process for the pair correlation function $h_2(r)=g_2(r)-1$ as
\begin{equation}\label{structure_factor}
S(\mathbf{q})=1+n_0\int d\mathbf{r}h_2(r)\exp{(-i\mathbf{q}\cdot\mathbf{r})}.
\end{equation}

In the ITA, this structure factor is then used as a correction to the ISA to derive the mesoscopic radiative properties. In this method, the single scattering property of an individual scatterer is assumed to be not affected. Then in a statistically homogeneous and isotropic disordered medium, the differential scattering coefficient in this method can be given by
\begin{equation}
\frac{d\kappa_\text{s,ITA}}{d\theta_\text{s}}=n_0S(\mathbf{q})\frac{dC_{\text{s}}}{d\theta_{\text{s}}},
\end{equation}
where $\frac{dC_{\text{s}}}{d\theta_{\text{s}}}$ is the single particle differential scattering cross section. Here  $\theta_\text{s}$ is the polar scattering angle, and the dependency on azimuth angle is integrated out (namely, an azimuthal symmetry is assumed here), and $\mathbf{q}$ is chosen to be the difference between scattered wavevector and incident wavevector, i.e., $q=2k\sin\theta_s$. For spherical and homogeneous Mie scatterers (Eqs.(\ref{pf_isa}-\ref{s2_isa})), we have
\begin{equation}
\frac{d\kappa_\text{s,ITA}}{d\theta_\text{s}}=\frac{n_0\pi}{k^2}S(2k\sin\theta_s)(|S_1(\theta_{\text{s}})|^2+|S_2(\theta_{\text{s}})|^2),
\end{equation}
where the scattering amplitudes $S_1(\theta_{\text{s}})$ and $S_2(\theta_{\text{s}})$ are given by Eqs.(\ref{s1_isa}-\ref{s2_isa}). The total scattering coefficient is obtained by directly integrating the differential scattering coefficient over $\theta_\text{s}$ as $\kappa_\mathrm{s,ITA}=\int_{0}^{\pi}\frac{d\kappa_\mathrm{s,ITA}}{d\theta_\text{s}}\sin\theta_\text{s}d\theta_\text{s}$, and the scattering phase function and asymmetry factor can also be accordingly computed. In the long wavelength limit $q\rightarrow 0$, the structure factor is given by \cite{wertheimPRL1963,tsang2004scattering2}
\begin{equation}
S(q=0)=\frac{(1-f_v)^4}{(1+2f_v)^2},
\end{equation}
which results in the well-known Twersky's formula for the scattering coefficient as \cite{twerskyJASA1978}
\begin{equation}\label{twersky_model}
\kappa_\mathrm{s,Twersky}=n_0C_\mathrm{s}\frac{(1-f_v)^4}{(1+2f_v)^2}.
\end{equation}

As discussed in Section \ref{farfieldinterference_sec}, the ITA model accounts for the first-order far-field interference effect, which is the most widely used method \cite{fradenPRL1990,mishchenkoJQSRT1994,rojasochoaPRL2004,yamadaJHT1986,holthoffPRE1997,conleyPRL2014,liuJOSAB2018}. For example, Tien and coworkers \cite{cartignyJHT1986,yamadaJHT1986,kumar1990dependent} implemented this approach to consider the DSE on radiative properties of packed beds containing spherical particles. Garcia\textit{ et al.} \cite{garciaPRA2008} also employed it and showed the theoretical prediction for total transmission of the photonic glass can achieve a much better agreement with the experimental result than ISA  \cite{garciaPRA2008} (Results are shown later in Fig.\ref{garciaPRA2008a}.). However, it was demonstrated in this experiment, a quantitative agreement is still not available, especially in the long wavelength range, where DSE is expected to be more prominent since the distance between particles becomes comparable or smaller than the wavelength, resulting in more particles involved in the multiple wave scattering process. Similarly, Conley \textit{et al.} \cite{conleyPRL2014} calculated the decay rate of diffuse radiation transport in a 2D dielectric medium (refractive index is 3.5) containing randomly distributed circular air holes whose volume fraction is 20\%, and they found the prediction of this model deviated from the numerically exact result significantly, where the discrepancy could reach an order of magnitude in some cases that exhibit strong positional correlations among the scatterers. 

% As a summary, this method cannot apply to the situations with strong structural correlations and many-particle scattering becomes prominent.

%%We also discuss here the effect of the structural correlations, by taking the role of interparticle forces into account. We also discuss the dynamic light scattering (DLS) and diffusing wave spectroscopy (DWS).
\subsubsection{Local field correction (Maxwell-Garnett Approximation)}\label{local_field_cor}
The local field correction, also known as the Lorenz-Lorentz relation (LLR) takes the following form in 3D random media for the self-energy:
\begin{equation}\label{LLR}
\Sigma_\mathrm{LLR}=\frac{n_0t_0}{1+n_0t_0/(3k^2)}=-\frac{n_0\alpha k^2}{1-n_0\alpha/3}.
\end{equation}
This model is originally derived from the mean-field assumption in atomic and molecular optics (see Refs.\cite{bornandwolf,hulst1957}) using a local field concept for an ideal cubic array of electric dipoles and therefore was usually regarded as not capable of considering the dependent scattering effect in disordered media. On the contrary, it was shown by Lagendijk and coworkers using the diagrammatic expansion that this formula indeed takes positional correlations among randomly distributed point dipolar scatterers ($a\rightarrow0$) into account (namely, non-overlapping condition for different scatterers) into infinite scattering orders \cite{Lagendijk1997PRL}\footnote{Note in Ref.\cite{Lagendijk1997PRL}, the definition of ``dependent scattering" is equivalent to recurrent scattering, in which the same scatterer is visited more than once, different from ours. Our definition of dependent scattering is a more general one.}. This can be understood by noting that Eq.(\ref{LLR}) is alternatively rewritten as 
\begin{equation}
\Sigma_\mathrm{LLR}=n_0t_0\left[1+\left(-\frac{1}{3}\right)\frac{n_0t_0}{k^2}+\left(-\frac{1}{3}\right)^2\left(\frac{n_0t_0}{k^2}\right)^2+\left(-\frac{1}{3}\right)^3\left(\frac{n_0t_0}{k^2}\right)^3+...\right],
\end{equation} 
which is actually a summation over all scattering orders with the prefactors indicating two-particle, three-particle, four-particle correlations and so on.

According to Eq.(\ref{epsilon_eff}) the effective permittivity is then given by $\varepsilon_\mathrm{eff}=1-\Sigma/k^2$, and equivalently, we have
\begin{equation}\label{epslion_eff_LLR}
\frac{\varepsilon_\mathrm{eff}-1}{\varepsilon_\mathrm{eff}+2}=\frac{1}{3}n_0\alpha,
\end{equation}
which is indeed the Clausius-Mossotti relation. If the polarizability $\alpha$ is calculated from the electrostatic approximation for a particle much smaller than the wavelength as $\alpha_\mathrm{ES}=4\pi a^3(\varepsilon_p-1)/(\varepsilon_p+2)$\footnote{Note for nonabsorbing particles, this formula violates the optical theorem and thus cannot be used when the scattering is significant \cite{malletPRB2005,devriesRMP1998}.}, the well-known Maxwell-Garnett approximation (MGA) is obtained:
\begin{equation}
\frac{\varepsilon_\mathrm{eff}-1}{\varepsilon_\mathrm{eff}+2}=f_v\frac{\varepsilon_\mathrm{p}-1}{\varepsilon_\mathrm{p}+2}.
\end{equation}
Above equation is only valid for very small scatterers with only electric dipole excitation with negligible scattering cross sections. When the size of the particle increases, the scattering becomes significant that can introduce additional imaginary part into the effective permittivity. In this situation, the polarizability is expressed in the first-order electric Mie coefficient $a_1$ as \cite{bohrenandhuffman}:
\begin{equation}\label{alphamie}
\alpha_\mathrm{ED}=\frac{6\pi i}{k^3}a_1=\frac{6\pi i}{k^3}\frac{m^2j_1(mx)[xj_1(x)]'-j_1(x)[mxj_1(mx)]'}{m^2j_1(mx)[xh_1(x)]'-h_1(x)[mxj_1(mx)]'},
\end{equation}
where $m=\sqrt{\varepsilon_p}$ is the complex refractive index of the particle. In combination with Eq.(\ref{epslion_eff_LLR}), the model is called the extended Maxwell-Garnett theory (EMGT). Analogously, by considering the magnetic dipole excitation described by the Mie coefficient $b_1$, this model can further be extended to result in an effective permeability \cite{grimesPRB1991}:
\begin{equation}
\frac{\mu_\mathrm{eff}-1}{\mu_\mathrm{eff}+2}=\frac{2\pi n_0 i}{k^3}b_1.
\end{equation}
Although this model was originally derived for cubic lattice \cite{grimesPRB1991} and not formally derived for disordered media, it can be deduced from the full-wave equations describing an ensemble of coupled magnetic dipoles \cite{chaumetJQSRT2009} in analogy of the procedure of Ref. \cite{Lagendijk1997PRL} for magnetic field instead. Ruppin \cite{ruppinOC2000} presented an extensive summary and evaluation for these EMGTs. These formulas have recently received a lot of attention due to the invention of metamaterials and metasurfaces, for example, negative-refractive-index metamaterials from dielectric particles supporting both electric and magnetic dipoles \cite{wheelerPRB2009}. 

Therefore, the effective refractive index is calculated as $n_\mathrm{eff}=\mathrm{Re}\sqrt{\varepsilon_\mathrm{eff}\mu_\mathrm{eff}}$ and the extinction coefficient is given by $\kappa_e=2k\mathrm{Im}\sqrt{\varepsilon_\mathrm{eff}\mu_\mathrm{eff}}$. Since this model is derived for an ensemble of point scatterers with infinitesimal exclusion volumes, it is not capable of considering the structural correlations for finite-size particles, which may lead to a broadening and shift for the resonances in the spectra \cite{wangPhotonAsia2019}. This effect is taken into account by the QCA (See Section \ref{qca_formula}), which is actually equivalent to the local field correction in the point scatterer limit ($a\rightarrow0$).

%%Maxwell Garnett model and Bruggeman model
\subsubsection{Keller's approach}
Developed by Keller and coworkers \cite{karalJMP1964,keller1964stochastic,kellerJMP1966}, this approach is a perturbative formula for the effective propagation constant up to the second order of number density $n_0$. In this approach, the self-energy for an ensemble of electric dipoles in the reciprocal space is given by
%\begin{equation}
%\begin{split}
%\mathbf{\Sigma}_\mathrm{Keller}(\omega,\mathbf{p})=&n_0t+n_0^2t^2\int{d^3\mathbf{r}\mathbf{G}_0(\mathbf{r})h_2(\mathbf{r})\exp{(i\mathbf{p}\cdot\mathbf{r})}}=n_0t-\frac{n_0^2t^2}{3k_0^2}\mathbf{I}\\
%&+4\pi n_0^2t^2\int{r^2drh_2(r)\left[G_0^{\bot}(r)j_0(pr)\mathbf{I}+\left(G_0^{\parallel}(r)-G_0^{\bot}(r)\right)
%	\left(\begin{matrix}
%	&\frac{j_1(pr)}{pr} &  & \\
%	& &\frac{j_1(pr)}{pr} &  \\
%	&  & &\frac{j_0(pr)pr-j_1(pr)}{pr}
%	\end{matrix}\right)
%	\right]}
%\end{split}
%\end{equation}

\begin{equation}
\begin{split}
\mathbf{\Sigma}_\mathrm{Keller}(\omega,\mathbf{p})=n_0t_0\mathbf{I}+n_0^2t^2\int_{-\infty}^\infty{d^3\mathbf{r}\mathbf{G}_0(\mathbf{r})h_2(\mathbf{r})\exp{(i\mathbf{p}\cdot\mathbf{r})}}.
\end{split}
\end{equation}
If we only consider the transverse waves and set $p=K$, namely, the effective propagation constant, we can obtain the self-energy under the on-shell approximation as
\begin{equation}
\Sigma_\mathrm{Keller}^\bot(K)=n_0t-\frac{n_0^2t^2}{3k^2}+4\pi n_0^2t^2\int r^2drh_2(r)\left[G_0^{\bot}(r)j_0(Kr)+\left(G_0^{\parallel}(r)-G_0^{\bot}(r)\right)\frac{j_1(Kr)}{Kr}\right],
\end{equation}
where the second term arises from the singular part of the Green's function, indicating the local contact between particles and the third term is the leading-order contribution of the two-particle correlations. $G_0^{\bot}(r)$ and $G_0^{\parallel}(r)$ are transverse and longitudinal components of the Green's function, which are given by
\begin{equation}\label{g_bot}
G_0^{\bot}(r)=-(\frac{i}{kr}-\frac{1}{(kr)^2}+1)\frac{\exp{(ikr)}}{4\pi r}
\end{equation} 
and
\begin{equation}\label{g_parallel}
G_0^{\parallel}(r)=2(\frac{i}{kr}-\frac{1}{(kr)^2})\frac{\exp{(ikr)}}{4\pi r}.
\end{equation}

In this situation it is noted that the self-energy and thus effective permittivity depend on the momentum (wavevector), which is known as the spatial dispersion or nonlocality \cite{durantJOSAA2007a,hespelJOSAA2001,barreraPRB2007}\footnote{Although the nonlocality is considered, this method still assumes that the wave propagation behavior is mainly determined by a single mode. See the discussion in Section \ref{dyson_eq_sec}.}. The effective propagation constant (wavenumber) $K$ can therefore be solved self-consistently by using the relation 
\begin{equation}
K^2=k^2-\Sigma_\mathrm{Keller}^\bot(K).
\end{equation}
For scalar waves, the formula becomes simpler without the singular part of Green's function \cite{durantJOSAA2007a}. More generally by taking multipoles into account, the propagation constant is given by \cite{hespelJOSAA2001,durantJOSAA2007a,ishimaruJOSA1982}
\begin{equation}
K^{2}=k^{2}+i \frac{4 \pi n_0\mathrm{S}_{k}(0)}{k} +\left(i  \frac{4 \pi n_0\mathrm{S}_{k}(0)}{k}\right)^{2} \frac{1}{K} \int_{0}^{\infty} e^{i k r} \sin (K r) g_{2}(r) \mathrm{d} r,
\end{equation}
where $S_k(0)$ is the forward scattering amplitude with respect to the background medium with a wavenumber $k$. Hespel \textit{et al.} \cite{hespelJOSAA2001} reported experimental measurements of the extinction coefficient in a suspension of PS spheres. It was found that the Keller model is in good agreement with the data provided that nonlocal effects are properly taken into account. Moreover, the local version using $p=k$ of this model can lead to substantial deviations from experimental results especially for large particles (size parameter$\sim1$) and for high volume densities. They also examined the simple criterion establishing the regime of independent scattering previously introduced by Hottel \textit{et al.} \cite{hottelAIAAJ1971}, which was shown to be not consistent with their experimental data. Notably, Derode \textit{et al.} \cite{derodePRE2006} tested this formula using acoustic waves in random media composed of metallic rods in water, where the scatterer densities were (6\% and 14\%), and the agreement was quite good. Chanal \textit{et al.} \cite{chanalJOSAA2006} numerically examined Keller's formula for 2D random media with a wide range of particle sizes ($a/\lambda=1/40, 1/20, 1/10, 1/5$) and volume fractions (up to 50\%). The refractive index of the particles is 2.25 with varying imaginary parts ranging from 0 to 0.2. They found that this formula can achieve a good agreement with numerical results up to a volume fraction of 30\%. Durant \textit{et al.} further extended this model for absorbing host media \cite{durantJOSAA2007a} and verified it numerically \cite{durantJOSAA2007}.

%If we further assume the long-wavelength limit so that the self-energy does not dependent on the momentum, we have the popular form as (by letting $p\rightarrow0$)
%\begin{equation}
%\Sigma_\mathrm{Keller}^\bot(p=0)=n_0t-\frac{n_0^2t^2}{3k^2}+ n_0^2t^2\int_0^\infty rdrh_2(r)\exp{(ikr)}
%\end{equation}
%This formula is also known as the bilocal approximation in random media described by permittivity fluctuations. 

%Barrera et al. \cite{barreraPRB2007} detailed the discussion of the nonlocal permittivity of random media.

\subsubsection{Quasicrystalline approximation (QCA)}\label{qca_formula}
Above formulas only involve the treatment of two-particle correlations. In much denser random media, high-order position correlations involving three or more particles simultaneously become important. However, there are no closed-form formulas for the correlation functions and it is also difficult to analytically calculate third- and high-order diagrams in the analytic wave theory. Therefore, approximations on the correlations are necessary, among which the quasicrystalline approximation is the mostly used \cite{laxPR1952,tsangJAP1982,wangPRA2018}. In this method, three- and higher-order correlations are treated as a hierarchy of pair distribution functions, e.g., $g_3(\mathbf{r}_1,\mathbf{r}_2,\mathbf{r}_3)=g_2(\mathbf{r}_1,\mathbf{r}_2)g_2(\mathbf{r}_2,\mathbf{r}_3)$, $g_4(\mathbf{r}_1,\mathbf{r}_2,\mathbf{r}_3,\mathbf{r}_4)=g_2(\mathbf{r}_1,\mathbf{r}_2)g_2(\mathbf{r}_2,\mathbf{r}_3)g_2(\mathbf{r}_3,\mathbf{r}_4)$ and so forth, where $g_3$ and $g_4$ indicate three-particle and four-particle distribution functions, respectively \cite{laxPR1952,bringi1982coherent,tsangJAP1982,tishkovetsJQSRT2011}. This approximating method permits to solve the propagation problem of coherent electromagnetic field (mean field) in random media in closed-form formulas. In this one respect, QCA is actually a perturbative approach only containing multiple wave scattering diagrams with cascading two-particle statistics, although it still takes infinite scattering orders into account \cite{maAO1988,varadanJOSAA1985}. 

QCA was initially proposed by Lax \cite{laxPR1952} for both quantum and classical waves, and examined by exact numerical simulations as well as experiments to be satisfactorily accurate for the DSE in moderately dense random media \cite{westJOSAA1994,Nashashibi1999}. It is also widely used in the prediction of optical and radiative properties of disordered materials for applications in remote sensing \cite{liangIEEETGRS2008} as well as thermal radiation transfer \cite{prasherJAP2007,wangIJHMT2015}. More generally, its validity for ultrasonic waves propagation in acoustical random media is also frequently verified numerically and experimentally \cite{meulenJASA2001}.

In the low-frequency limit for electric-dipole particles, the self-energy under QCA is expressed in a self-consisted way in the reciprocal space as
\begin{equation}
\bm{\Sigma}\left(\mathbf{p} \right)=n_0t_0 \mathbf{I}+n_0t_0\int_{-\infty}^\infty\mathbf{G}_0\left( \mathbf{p} \right)H_2\left( \mathbf{p}\right)\bm{\Sigma}\left(\mathbf{p} \right),
\end{equation}
where $H_2(\mathbf{q})$ is defined as the Fourier transform of pair correlation function $h_2(r)$ as
\begin{equation}\label{Hq_eq}
H_2(\mathbf{q})=\int_{-\infty}^{\infty}d^3\mathbf{r}h_2(r)\exp{(-i\mathbf{q}\cdot\mathbf{r})}.
\end{equation}
Note it differs with the structure factor in Eq.(\ref{structure_factor}) by unity. Letting $\bm{\Sigma}\left(\mathbf{p} \right)=\Sigma\mathbf{I}$ be $\mathbf{p}$ independent, which is applicable for small particles meaning the scattering properties are not spatially dispersive (i.e., locality is assumed), we have
\begin{equation}\label{qca1}
\Sigma\mathbf{I}=n_0t_0\mathbf{I}-\frac{n_0t_0\Sigma}{3k^2}\mathbf{I}+n_0t_0\Sigma\int_{-\infty}^\infty{d\mathbf{r}\mathrm{PV}\mathbf{G}_0\left( \mathbf{r} \right)h_2(r)},
\end{equation}
where we have separated the singular part of Green's function and defined the principal value (PV) of Green's function as $\mathrm{PV}\mathbf{G}_0\left( \mathbf{r} \right)$. The integral is also transformed from reciprocal domain to space domain. Then the self-energy is solved as 
\begin{equation}
\Sigma=\frac{n_0t_0}{1+n_0t_0/(3k^2)+2n_0t_0\int_0^{\infty}{drr\exp(ikr)\left[g_2(r)-1\right]}/3}.
\end{equation}
This equation in the point scatterer limit ($a\rightarrow0$) is equivalent to the LLR formula in Section \ref{local_field_cor}. Therefore according to Eq.(\ref{epsilon_eff}) the effective propagation constant $K$ is given by 
\begin{equation}
K^2=k^2-\frac{1}{1/(n_0t_0)+1/(3k^2)+2\int_0^\infty{drr\exp(ikr)\left[g_2(r)-1\right]}/3}.
\end{equation}

After the effective propagation constant for the coherent wave is calculated, the differential scattering coefficient for the incoherent wave can be derived. It is determined by the irreducible intensity vertex $\bm{\Gamma}$. Again here we only consider two-particle statistics. 
%Note the derivation can be simplified using $\Sigma$ in the second line of Fig.X is due to our locality assumption for the scattering process, which reduces the correlation function into only two particle positions. 
The irreducible intensity vertex is solved as
\begin{equation}\label{gamma_qca}
\Gamma(\mathbf{p},\mathbf{p}')=\left[n_0|C|^2+n_0^2|C|^2H_2(\mathbf{p}-\mathbf{p}')\right]\mathbf{I}\otimes\mathbf{I},
\end{equation}
where $C=\Sigma/n_0$. Afterwards, we take the on-shell approximation, which implies the photons transport with a fixed momentum value $p=K$ and those excitations with other momentum values are negligible. It gives
\begin{equation}\label{gamma_qca2}
\Gamma(K\hat{\mathbf{p}},K\hat{\mathbf{p}}')=\left[n_0|C|^2+n_0^2|C|^2H_2(K\hat{\mathbf{p}}-K\hat{\mathbf{p}}')\right]\mathbf{I}\otimes\mathbf{I}.
\end{equation}
Since the pair correlation function $H_2(K\hat{\mathbf{p}}-K\hat{\mathbf{p}}')$ only depends on the difference between $\hat{\mathbf{p}}$ and $\hat{\mathbf{p}}'$, the present medium is isotropic. And for unpolarized radiation transport, the azimuth symmetry is preserved. By taking the transverse component of the irreducible intensity vertex, the differential scattering coefficient can be obtained by integrating over the azimuth angle $\varphi_s$ with incident direction $\hat{\mathbf{p}}'$ fixed as \cite{barabanenkov1975multiple}
%\begin{equation}\label{pf_qca1}
%\begin{split}
%\frac{d\kappa_s}{d\theta_s}=\frac{1}{(4\pi)^2}\int_{0}^{2\pi}\Gamma^{\bot}(K\hat{\mathbf{p}},K\hat{\mathbf{p}}')d\varphi_s.
%\end{split}
%\end{equation}
%This leads to the following result of differential scattering coefficient that is not dependent on the specific incident direction $\hat{\mathbf{p}}'$ but the relative polar scattering angle $\theta_s$ between $\hat{\mathbf{p}}$ and $\hat{\mathbf{p}}'$ as
\begin{equation}\label{pf_qca2}
\begin{split}
\frac{d\kappa_s}{d\theta_s}=\frac{n_0|C|^2(1+\cos^2\theta_s)}{4\pi}\left\{1+n_0H_2[2K\sin(\theta_s/2)]\right\}.
\end{split}
\end{equation}
Therefore the scattering coefficient can be calculated accordingly.

For spherical scatterers supporting high-order multipoles, the equations to calculate the effective propagation constant under QCA can also be developed \cite{tsang2004scattering}. Here we only present the main formulas as follows.
\begin{equation}
{K} - k =  - \frac{{i\pi {n_0}}}{{{k^2}}}\sum\limits_{n = 1}^{{N_{\max }}} {(2n + 1)(T_n^{(M)}X_n^{(M)} + T_n^{(N)}X_n^{(N)})}, 
\end{equation}
where $T_n^{(M)}$ and $T_n^{(N)}$ are $T$-matrix elements and for spheres, $T_n^{(M)}=-b_n$ and $T_n^{(N)}=-a_n$. $X_n^{(M)}$, and $N_\mathrm{max}$ is the maximum expansion order for the multipolar modes. Here $X_n^{(N)}$ can be understood as the ensemble-averaged excitation amplitudes for the multipolar modes, which can be calculated as 
\begin{equation}
\begin{array}{l}
X_\upsilon ^{(M)} =  - 2\pi {n_0}\sum\limits_{n = 1}^{{N_{\max }}} {\sum\limits_{p = |n - \upsilon |}^{|n + \upsilon |} {(2n + 1)} } [{L_p}(k,{K}|d) + {M_p}(k,{K}|d)]\\
{\rm{            }} \times [T_n^{(M)}X_n^{(M)}a(1,n| - 1,\upsilon |p)A(n,\upsilon ,p){\rm{            }} + T_n^{(N)}X_n^{(N)}a(1,n| - 1,\upsilon |p,p - 1)B(n,\upsilon ,p)],
\end{array}
\end{equation}

\begin{equation}
\begin{array}{l}
X_\upsilon ^{(N)} =  - 2\pi {n_0}\sum\limits_{n = 1}^{{N_{\max }}} {\sum\limits_{p = |n - \upsilon |}^{|n + \upsilon |} {(2n + 1)} } [{L_p}(k,{K_\mathrm{eff}}|d) + {M_p}(k,{K}|d)]\\
{\rm{            }} \times [T_n^{(M)}X_n^{(M)}a(1,n| - 1,\upsilon |p,p - 1)B(n,\upsilon ,p){\rm{            }} + T_n^{(N)}X_n^{(N)}a(1,n| - 1,\upsilon |p)A(n,\upsilon ,p)],
\end{array}
\end{equation}

where ${L_p}(k,{K}|D)$  and ${M_p}(k,{K}|d)$  are given as:
\begin{equation}
{M_p}(k,{K_\mathrm{eff}}|d) = \int_d^\infty  {{r^2}} [g_2(r) - 1]{h_p}(kr){j_p}({K_\mathrm{eff}}r)dr
\end{equation}
and
\begin{equation}
{L_p}(k,{K}|d) = -\frac{{{d^2}}}{{K^2 - {k^2}}} \times [k{h_p}^\prime (kd){j_p}({K}d) - {K}{h_p}(kd){j_p}^\prime ({K}d)]
\end{equation}
These formulas are derived by Tsang and Kong \cite{tsangJAP1980,tsang2004scattering,tsangRS2000}, and by applying the distorted Born approximation (DBA) \cite{maAO1988}, namely, considering the first-order scattering for a thin layer, the scattering phase function and scattering coefficient can be obtained \cite{tsang2004scattering}.  This is called the dense media radiative transfer theory (DMRT) \cite{tsang2004scattering,tsangJQSRT2019}, which indeed follows from the original theory for electron transport in disordered materials (more specifically, liquid metals) \cite{gyorffyPRB1970}. Recently, we have re-derived the QCA formulas for a random system containing dual-dipolar particles in which only electric and magnetic dipoles are excited \cite{wangPRA2018}. Specifically, in terms of the intensity
transport, we have obtained a Bethe-Salpeter-type equation for this
system. By applying the far-field and on-shell approximations as
well as Fourier transform techniques, we have finally obtained
the scattering phase function and scattering coefficient, without resorting to the DBA because full multiple scattering series of radiation intensity is accounted for. Our treatment is based on more explicit arguments and can be easily extended to multipolar excitations.
%Recent experimental tests for QCA. M\'arquez-Islas and Garc\'ia-Valenzuela \cite{marquez-islasAO2018}
\subsubsection{Recurrent scattering models}\label{rsmodel}
%\begin{figure}[htbp]
%	\centering
%	\subfloat{
%		\label{vantiggelenPRB1994}
%		\includegraphics[width=0.8\linewidth]{vantiggelenPRB1994}
%	}
%	\subfloat{
%		\label{leseurOptica2016}
%		\includegraphics[width=0.8\linewidth]{leseurOptica2016}
%	}
%	\caption{Second order diagrammatic expansions. (a) All second order diagrams including recurrent scattering and structural correlations. Ref.\cite{vantiggelenPRB1994}.(b) Only those second order diagrams that involves structural correlations. Reprinted from Ref.\cite{leseurOptica2016}.}\label{twoordermodels}
%\end{figure}

The only recurrent scattering formula is derived by van Tiggelen and coworkers \cite{Vantiggelen1990JPCM,vantiggelenPRB1994,Cherroret2016} for each pair of (uncorrelated) scatterers, as schematically depicted in Fig.\ref{recurrent_scattering_schematic}. Recently, this formula was extended to consider the correlations between particles \cite{kwongPRA2019,wang2018role}. For very small dipole scatterers, by assuming that $\mathbf{\Sigma}$ is $\mathbf{p}$-independent, we have the formula in the following form
\begin{equation}
\begin{split}
\mathbf{\Sigma}_\mathrm{rec}=&n_0t\mathbf{I}+n_0^2t_0^2\int{d^3\mathbf{r}\mathbf{G}_0(\mathbf{r})h_2(\mathbf{r})}+n_0^2t_0^3\int{d^3\mathbf{r}\frac{\mathbf{G}_0^2(\mathbf{r})[1+h_2(\mathbf{r})]}{\mathbf{I}-t^2\mathbf{G}_0^2(\mathbf{r})}}\\&+n_0^2t_0^4\int{d^3\mathbf{r}\frac{\mathbf{G}_0^3(\mathbf{r})[1+h_2(\mathbf{r})]}{\mathbf{I}-t^2\mathbf{G}_0^2(\mathbf{r})}}.
\end{split}
\end{equation}
The transverse component of the self-energy is then given as
\begin{equation}
\begin{split}
\Sigma_\mathrm{rec}^\bot=&n_0t_0-\frac{n_0^2t_0^2}{3k^2}+\frac{2n_0^2t_0^2}{3}\int rdrh_2(r)\exp{(ikr)}\\&+4\pi n_0^2t_0^3\int{r^2dr\left[\frac{2}{3}\frac{G_0^{\bot 2}(r)}{1-t_0^2G_0^{\bot 2}(r)}+\frac{1}{3}\frac{G_0^{\parallel 2}(r)}{1-t_0^2G_0^{\parallel 2}(r)}\right]\left[1+h_2(r)\right]}\\&+
4\pi n_0^2t_0^4\int{r^2dr\left[\frac{2}{3}\frac{G_0^{\bot 3}(r)}{1-t_0^2G_0^{\bot 2}(r)}+\frac{1}{3}\frac{G_0^{\parallel 3}(r)}{1-t_0^2G_0^{\parallel 2}(r)}\right]\left[1+h_2(r)\right]}.
\end{split}
\end{equation}
Therefore, the effective propagation constant and the effective permittivity can be obtained by using the Dyson equation. Moreover, van Tiggelen and coworkers \cite{Vantiggelen1990JPCM,vantiggelenPRB1994,Cherroret2016} also derived the corresponding irreducible intensity vertex for uncorrelated particles, while for correlated particles, by now no similar formulas are obtained.

This recurrent scattering model also belongs to the perturbative approach in the second order of the particle number density $n_0$ under the framework of the analytical wave theory, which is valid in very dilute random media, because in denser media, recurrent scattering between three or more particles might become prominent. More precisely, it was shown in Refs. \cite{Vantiggelen1990JPCM,vantiggelenPRB1994,Cherroret2016} the criterion for the diluteness is $4\pi n_0/k^3\ll1$. Recently, Kwong \textit{et al.} \cite{kwongPRA2019} used numerical calculations to examine the validity range of this model.

\subsubsection{Coherent potential approximation (CPA) and its modifications}
The concept of coherent potential starts from very simple assumptions, which was firstly developed for disordered electronic systems \cite{laxRMP1951,sovenPR1967,rothPRL1972}. Consider that in a renormalized (effective) disordered medium with an effective Green's function $\mathrm{G}_e$ (or called "modified propagator" \cite{laxRMP1951}), the Lippman-Schwinger equation (Eq.(\ref{lp_eq2})) is rewritten in the operator form as
\begin{equation}\label{cpa_eq1}
\mathbf{G}=\mathbf{G}_e+\mathbf{G}_e\overline{\mathbf{T}}\mathbf{G}_e
\end{equation} 
where $\langle\overline{\mathbf{T}}\rangle$ is the \textit{T}-operator of the full system in the effective medium. Taking ensemble average of Eq.(\ref{cpa_eq1}), we obtain
\begin{equation}\label{cpa_eq2}
\langle\mathbf{G}\rangle=\mathbf{G}_e+\mathbf{G}_e\langle\overline{\mathbf{T}}\rangle\mathbf{G}_e.
\end{equation} 
According to the definition of the effective Green's function, we have $\langle\mathbf{G}\rangle=\mathbf{G}_e$, and thus the ensemble averaged \textit{T}-operator in the effective medium should be zero, i.e., 
\begin{equation}\label{cpa_eq3}
\begin{split}
\langle\overline{\mathbf{T}}\rangle=0.
\end{split}
\end{equation} 
In this sense, in the effective medium, \textit{T}-operator vanishes, leading to a zero scattering condition. By expanding the many-particle \textit{T}-operator into the multiple wave scattering series of individual particles' \textit{T}-operators, we have
\begin{equation}\label{cpa_eq4}
\begin{split}
\langle\overline{\mathbf{T}}\rangle=\langle\sum_{i=1}^N\overline{\mathbf{T}}_i\rangle+\langle\sum_{i=1}^{N}\sum_{j\neq i}^{N}\overline{\mathbf{T}}_i\mathbf{G}_e\overline{\mathbf{T}}_j\rangle+\langle\sum_{i=1}^{N}\sum_{j\neq i}^{N}\sum_{k\neq j}^{N}\overline{\mathbf{T}}_i\mathbf{G}_e\overline{\mathbf{T}}_j\mathbf{G}_e\overline{\mathbf{T}}_k\rangle...=0,
\end{split}
\end{equation} 
where $\overline{\mathbf{T}}_j$ is the $T$-operator of $j$-th scatterer in the effective medium. And in this circumstance, the self-energy in the effective medium $\overline{\bm{\Sigma}}$ is also zero according to the Dyson equation. 

At first sight, it seems that above equation involving ensemble averaged $T$-operators is still difficult to solve and no difference is found compared with the original multiple wave scattering series for a disordered medium except for a modified Green's function. Indeed, the attractive point of the CPA lies in the zero scattering condition, because of which it is reasonable to assume that in the effective medium, all particles are weakly scattering and can be regarded as independent scatterers. As a consequence, it would be a good approximation by letting 
\begin{equation}\label{cpa_eq5}
\begin{split}
\langle\overline{\mathbf{T}}\rangle\approx\langle\sum_{i=1}^N\overline{\mathbf{T}}_i\rangle=0
\end{split}
\end{equation} 
and then the effective propagation constant can be solved self-consistently. This is the basic formula of the CPA for calculation of radiative properties \cite{sheng2006introduction}. For uncorrelated disordered media, this formula is rather accurate up to the third order of the $T$-operator expansion \cite{sheng2006introduction}. The inaccuracies arise from its inability to capture microstructural correlations and recurrent scattering effects. In this circumstance, high-order techniques used in the previous models like QCA and recurrent scattering expansions can also be employed to further improve the accuracy. For example, Tsang and Kong \cite{tsangJAP1980} extended the QCA model by using the CPA approach, and derived the QCA-CP model for an ensemble of spherical particles.

Recently, this approach has received substantial attention in the design and modeling of dielectric ordered and disordered metamaterials. For instance, Slovick and coworkers developed a generalized effective medium formula \cite{slovickPRB2014} based on the CPA (which they called the ``zero-scattering condition") and proposed a design for a negative-index metamaterial using electric and magnetic dipolar excitations \cite{slovickPRB2017}. This model was further employed in the design of negative index metamaterials with high-order multipoles (electric quadrupole) \cite{huangJQSRT2018}, which was numerically validated by the plane-wave expansion (PWE) method as well as FDTD simulations.

\begin{figure}[htbp]
	\centering
	\includegraphics[width=0.6\linewidth]{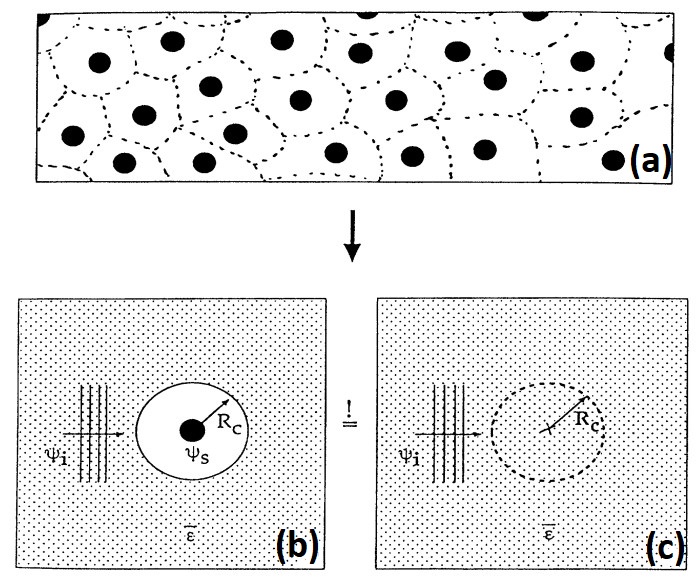}
	\caption{The energy-density CPA approach. (a) In a random medium composed of dielectric spheres, the basic scattering unit may be regarded as a coated sphere, as represented by the dashed lines. To calculate the effective dielectric constant $\overline{\varepsilon}$, a coated sphere of radius $R_c=R/f^{1/3}$ is embedded in a uniform medium. The self-consistent condition for the determination of $\overline{\varepsilon}$ is that the energy of (b) a coated sphere  is equal to the energy of (c) a sphere with radius $R_c$ and dielectric constant $\overline{\varepsilon}$. Reprinted with permission from Ref.\cite{buschPRL1995}. Copyright 1995 by the American Physical Society. }
	\label{energy_density_CPA}
\end{figure}

Another important improvement for the basic CPA formula is the energy-based CPA (ECPA) proposed by Soukoulis and coworkers \cite{soukoulisPRB1994,buschPRL1995,buschPRB1996}. In order to account for the short-range correlations in disordered media consisting of densely packed spheres (volume fraction up to 0.6), Soukoulis \textit{et al.} \cite{soukoulisPRB1994} first modified the CPA approach by considering a coated sphere as the basic scattering unit. However, for low volume fraction this approach undesirably gives a phase velocity higher than the velocity of light near Mie resonances. Furthermore, Busch and Soukoulis \cite{buschPRL1995,buschPRB1996} improved this coated CPA model by using the heuristic idea that in a random medium the energy density should be uniform when averaged over the correlation length of the microstructure, as schematically shown in Fig.\ref{energy_density_CPA} with the coated sphere represented by the dashed lines. To calculate the effective dielectric constant $\overline{\varepsilon}$, the coated sphere of radius $R_c=a/f^{1/3}$ is embedded in a uniform medium. The self-consistent condition for the determination of $\overline{\varepsilon}$ is that the energy of a coated sphere is equal to the energy of a sphere with radius $R_c$ and dielectric constant $\overline{\varepsilon}$, schematically illustrated in Fig.\ref{energy_density_CPA}(b-c), i.e., 
\begin{equation}\label{ed_cpa}
\int^{R_c}_0d^3\mathbf{r}\rho_E^{(1)}(\mathbf{r})=\int^{R_c}_0d^3\mathbf{r}\rho_E^{(2)}(\mathbf{r}),
\end{equation}
whereas the energy density $\rho_E(\mathbf{r})$ is given by 
\begin{equation}\label{ed_cpa2}
\rho_E(\mathbf{r})=\frac{1}{2}[\varepsilon(\mathbf{r})|\mathbf{E}(\mathbf{r})|^2+\mu|\mathbf{H}(\mathbf{r})|^2].
\end{equation}
Since the energy densities (and surely the electromagnetic fields) implicitly depend on the effective permittivity, from Eqs.(\ref{ed_cpa}-\ref{ed_cpa2}), the effective permittivity can be determined self-consistently. After this procedure, the self-energy $\Sigma_\mathrm{ECPA}$ of the random medium can be calculated with respect to the effective permittivity under the ISA. Thus the scattering coefficient is given by \cite{buschPRL1995}
\begin{equation}
\kappa_\mathrm{s,ECPA}=\frac{\sqrt{2}\mathrm{Im}\Sigma_\mathrm{ECPA}}{\left[(k_e^2-\mathrm{Re}\Sigma_\mathrm{ECPA})^2+\sqrt{(k_e^2-\mathrm{Re}\Sigma_\mathrm{ECPA})^2+(\mathrm{Im}\Sigma_\mathrm{ECPA})^2}\right]^{1/2}},
\end{equation}
where $k_e=k_0n_\mathrm{ECPA}$ is the wave number in the effective medium and $n_\mathrm{ECPA}=\sqrt{\overline{\varepsilon}}$ is the effective refractive index.

%{\color{red}``There the solution of the corresponding self-consistent equation can disappear or jump abruptly or even get multiple solutions. This new scheme can easily follow the unique solution of the self-consistent equation. We feel that the integration over all angles in Eq.(\ref{ed_cpa}) is responsible for the well behaved solution."}

The ECPA approach is most suitable for disordered media composed of extremely dense-packed monodisperse particles exhibiting Mie resonances, in which other dependent scattering models like QCA and recurrent scattering models do not apply or exhibit divergences in the calculation, and the solution of other CPA methods can disappear or jump abruptly or have multiple solutions \cite{buschPRL1995}. As a notable example for the use of this approach, Maret and coworkers \cite{schertelPRMat2019,aubryPRA2017} recently carried out a series of numerical and experimental works on photonic glasses containing densely assembled ($f_v\sim0.5$) monodisperse spherical $\mathrm{TiO_2}$ or PS nanoparticles near strong Mie resonances, and showed the good predication capability of the ECPA approach for the light transport properties without any fitting parameters\footnote{Note in these works, different from the original ECPA paper \cite{buschPRL1995}, the ITA is used to calculate the transport mean free path with respect to the effective medium with $n_\mathrm{ECPA}$. Maret and coworkers claimed that this model (ECPA combined with ITA) ``takes into account resonant Mie scattering, short-range positional correlations, optical near-field coupling of randomly packed, spherical scatterers."}. It was also demonstrated that this approach can quantitatively predict the reflectance spectra of these photonic glass samples with different thicknesses as structural color materials \cite{schertelADOM2019}. 

%{\color{red}In their calculation, they found that ``short-range structural correlations accounted for by structure factor $S(\theta)$ mainly influence the height and position of the first resonance, while Mie scattering in an effective medium represented by the form factor $F(\theta)$ is related to all resonances. $n_\mathrm{eff}$ influences both, the height as well as the position of each resonance. Changing $n_\mathrm{ECPA}$ to the often used Garnett effective index $n_\mathrm{MG}$ smears out strongly resonant behavior."\cite{schertelADOM2019}}

\subsubsection{Phenomenological models for near-field DSE}\label{nfmodels}
%Early works that proposed the use of an effective medium instead of the background medium to account for the effect of adjacent particles as a phenomenological consideration for the near-field DSE include Refs. \cite{gateJOSA1973}.

%As the concentration of scattering particles in disordered media rises, they are inclined to step into the near fields of each other \cite{Liew2011,Naraghi2015} (i.e., the distance $r$ between scatterers is comparable or even smaller than the wavelength, approximately, $kr\lesssim1$). In this circumstance, near-field interaction (NFI) among scatterers, which is negligible when the scatterers are in the far field of each other, can also contribute to radiation energy tunneling and thus transport properties, making the far-field approximation invalid. 

\begin{figure}[htbp]
	\centering
	\subfloat{
		\label{liewOE2011a}
		\includegraphics[width=0.46\linewidth]{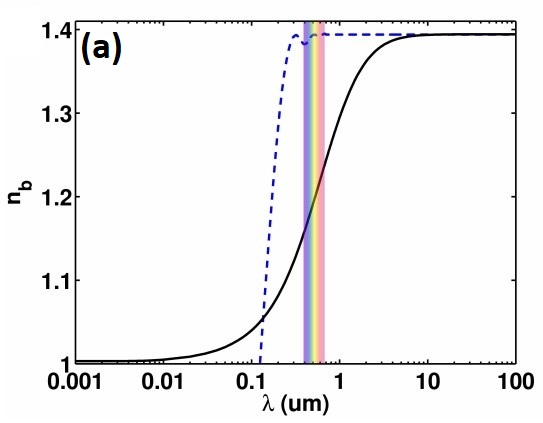}
	}
	\subfloat{
		\label{liewOE2011b}
		\includegraphics[width=0.45\linewidth]{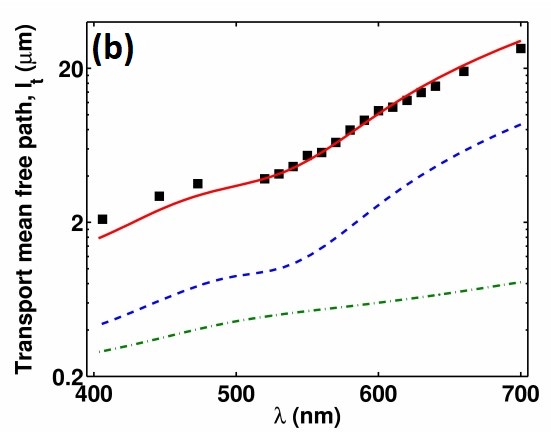}
	}
	\caption{A phenomenological near-field DSE model. (a) Near-field effects on form factors can be included in an effective background refractive index nb, whose value is calculated from Eq. (\ref{liew_nf_model}). It approaches the refractive index of air at short wavelength, and that of a homogenized medium at long-wavelength. The wavelength range of the experimental measurement is highlighted with color. For comparison,	the value of $n_b$ obtained from Ref. \cite{Peng2007} is plotted with blue dashed line. (b) Measured (black square) and estimated (lines) transport mean free path $l_\mathrm{tr}$ vs. wavelength $\lambda$. Green dash-dots curve represents $l_\mathrm{tr}$ estimated without short-range order and near-field effects, blue dashed line is with short-range order but no near-field effects, and red solid curve is with both. The experiment was done by measuring the coherent backscattering cone. Reprinted with permission from Ref. \cite{Liew2011}. Copyright 2011 Optical Society of America.}\label{liewOE2011}
	
\end{figure}

Previous models are mostly based on the far-field approximations of electromagnetic scattering, or include the near-field interaction implicitly, e.g., the CPA-based methods indeed account for the near-field interactions to some extent by using an effective refractive index. It is not easy to explicitly determine the role played by the near-field interaction among scatterers. To do this, several authors developed phenomenological models that explicitly take the near-field interactions into consideration, for instance, Liew \textit{et al.} \cite{Liew2011} also proposed a similar model to predict the near-field DSE in densely packed polystyrene spheres (volume fraction 64\%) by using a near-field-dependent effective refractive index of the background. The value of background refractive index $n_b$ for a particle is obtained by averaging the actual refractive index surrounding the particle with a weighting factor from an exponentially-decaying evanescent field as
\begin{equation}\label{liew_nf_model}
n_b=\frac{\int_0^\infty n(r)\exp{(-\beta r/\lambda)}r^2dr}{\int_0^\infty \exp{(-\beta r/\lambda)}r^2dr},
\end{equation}
where $r$ is the distance from the particle's surface, $\lambda$ is the wavelength of light in vacuum, $n(r)$ is the ensemble-averaged refractive index distribution based on the packing geometry, and $\beta$ is a fitting parameter that is expected to depend on the refractive indices of the particles and the local packing geometry. Their experimental results lead to a fitting parameter $\beta=14.1\pm3.2$ with a relative standard error around 9\%, indicating the applicability of this phenomenological model, as shown in Fig.\ref{liewOE2011b} with the corresponding background refractive index presented in Fig.\ref{liewOE2011a}. It was demonstrated that near-field effects together with the short-range order reduce the scattering strength by one order of magnitude in random close-packed structures. Peng and Dinsmore \cite{Peng2007} proposed a similar model by modifying the ITA, which used an effective background medium with a refractive index $n_\mathrm{eff}$, to consider the effect of neighboring particles. To obtain $n_\mathrm{eff}$, they proposed a near-field coupling distance $r_c$ of neighboring particles, which was estimated as $r_c=(n_s/n_b)\lambda/2$, where $n_s$ and $n_b$ are the refractive indices of particle and matrix materials. Then the effective index can be calculated from the volume-averaged refractive index within a spherical region of radius $r_c$ surrounding a typical scatterer. This model agreed well with their experimental data of high-concentration films of randomly packed ZnS-PS core-shell microspheres. The calculated effective index by this model for closely packed PS spheres is also presented in Fig.\ref{liewOE2011a} by Ref. \cite{Liew2011} for comparison.

Recently, Naraghi and Dogariu \cite{Naraghi2015} proposed a phenomenological model that added an evanescent-wave-scattering correction to predict the near-field DSE. Their model of transport mean free path reads
\begin{equation}\label{nfscatteringmodel}
l _ { \mathrm { CS } + \mathrm { NF } } ^ { * } = \frac { 1 } { n _ { 0 } C_s \left( 1 - g \right) } + \overline { \left( \frac { P _ { \mathrm { NF } } } { n _ { 0 } C_{ \mathrm { NF } } \left( 1 - g _ { \mathrm { NF } } \right) } \right) },
\end{equation}
where the first term in the RHS is exactly the transport mean free path using the ITA (which was called collective scattering (CS) by the authors). In the second term, $C_\mathrm{NF}$ and  $g_\mathrm{NF}$ are the scattering cross section and asymmetry factor of evanescent waves impinging on a spherical particle \cite{bekshaevOE2013}, and $P_\mathrm{NF}=n_0\lambda^3\exp(-\kappa_\mathrm{NF} d)$ is the probability function for evanescent wave transfer, where $\kappa_\mathrm{NF}$ is the characteristic attenuation coefficient of the evanescent waves. Because the decay rate of the evanescent waves depends on the incident angle, an average process $\overline{(...)}$ is taken over the angular domain defined by the refractive indices of the
particle and its surrounding medium. Although this model qualitatively captures the physical significance of near-field DSE especially in the high-concentration range, its prediction on transport mean free path showed a substantial deviation from experimental data in the intermediate and large volume fraction range \cite{Naraghi2015} (comparison is shown in Fig.\ref{naraghiPRL2015ltr} in Section \ref{expsec}). 

\subsection{Numerical modeling of the DSE}

Nowadays, with the rapid increase of computation resources, it is already possible to employ numerical electromagnetic methods to directly compute the macroscopic radiative properties based on the microscopic structures of DDM, which naturally consider the electromagnetic interferences including the DSE. For example, Auger\textit{ et al.} \cite{augerJCTR2009} investigated the effect of dependent scattering on the radiative properties of white paints, namely, coatings made of densely packed $\mathrm{TiO_2}$ nanoparticles, by using the MSTM code, which exactly solves the electromagnetic field propagation in a group of homogeneous or multilayered spherical particles. Lallich \textit{et al.} \cite{lallichJHT2009} used the discrete dipole approximation (DDA) to simulate electromagnetic wave scattering in silica aerogels. Liu \textit{et al.} \cite{qiuJQSRT2015} implemented the finite-difference time-domain (FDTD) method to directly calculate the macroscopic radiative properties of a microporous aluminum foam. Roughly speaking, there are three types of methods to numerically model electromagnetic wave propagation in DDM and obtain their macroscopic and mesoscopic radiative properties, which are discussed as follows.

\subsubsection{The supercell method}
\begin{figure}[htbp]
	\flushleft
	\subfloat{
		\label{dyachenkoACSPhoton2014simulation}
		\includegraphics[width=0.46\linewidth]{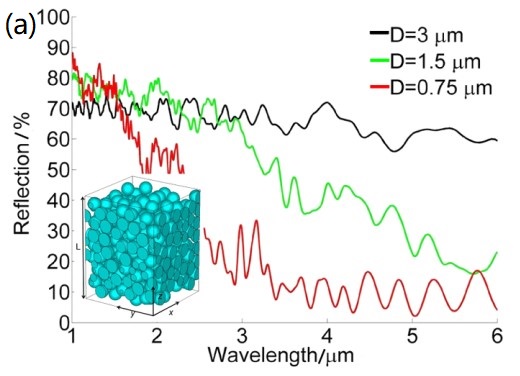}
	}
	\subfloat{
		\label{dyachenkoACSPhoton2014exp}
		\includegraphics[width=0.45\linewidth]{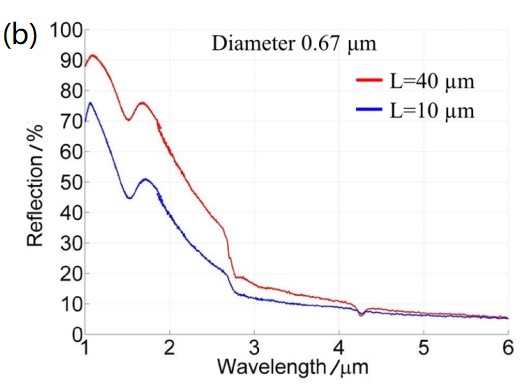}
	}
	\caption{The RVE method using the periodic boundary condition. (a) Calculated hemispherical diffuse reflectance from disordered photonic structures with sphere diameter $d = 0.75, 1.5$ and $3 \mathrm{\mu m}$ and thicknesses $L = 18 \mathrm{\mu m}$. The inset shows the schematic of the disordered photonic structure. (b) Experimental data of hemispherical diffuse reflectance. Reprinted with permission from Ref. \cite{dyachenkoACSPhoton2014}. Copyright 2014 American Chemical Society.}\label{supercellmethod}
	
\end{figure}

The first type can be dubbed "the supercell method". In this method, a supercell containing a large amount of scatterers is chosen in combination with the periodic boundary condition to mimic the whole disordered medium \cite{qiuJQSRT2015,liuJOSAB2018}. For example, Dyachenko \textit{et al.} \cite{dyachenkoACSPhoton2014} computed the reflectance spectra of photonic glass slabs composed of ZrO$_2$ microspheres using the FDTD method. They chose a unit cell expanding $18 \mathrm{\mu m}\times18 \mathrm{\mu m}$ in the lateral directions ($xOy$ plane in their paper) implemented with the periodic boundary condition, while the thickness in the propagation direction ($z$-axis in their paper) was varied with the perfect matching layer (PML) boundary condition. Random structures of the microspheres were obtained using event-driven molecular dynamics simulations of a monodisperse hard sphere system implemented as a freely available software package DynamO. Fig. \ref{dyachenkoACSPhoton2014simulation} shows their simulation results for different microsphere diameters with a fixed thickness of $L = 18 \mathrm{\mu m}$. Experimental data was also shown for comparison in Fig.\ref{dyachenkoACSPhoton2014exp}. It can be seen that this method can qualitatively reproduce experimental data, while significant oscillations emerge in the simulated spectra due to the inevitable interferences brought by the periodic boundary condition. More importantly, this method is not capable of obtaining radiative properties unless an inverse algorithm is implemented to the reflectance/transmittance spectra \cite{chenJQSRT2018}.

\subsubsection{The representative volume element method}
The second method can be called the representative volume element (RVE) method, which is properly selected in order to obtain the ``effective" mesoscopic radiative properties by taking a RVE-scale average \cite{yuOE2014,augerJCTR2012,augerJCTR2015}. Tsang \textit{et al.} \cite{tsangOL1992,zurkJOSAA1995} described this method in detail in order to simulate the extinction coefficient of dense media with randomly distributed dielectric spheres that occupy up to 25\% by volume and size parameter $ka = 0.2$. They firstly generated a random cluster of $N$ (about 2000-4000) particles by means of Monte Carlo method (specifically, using the Metropolis algorithm) and the multiple wave scattering Foldy-Lax equations for this distribution of particle positions were solved iteratively. This procedure were repeated for $N_r$ random realizations of particle positions, which can range from hundreds to thousands, in order to perform ensemble average for the field quantities. Then the electromagnetic field of each realization of cluster can be divided into two parts, namely, a coherent field and a fluctuating incoherent field
\begin{equation}
{{\bf{E}}}_l({{\bf{r}}_s}) = \left\langle {{{\bf{E}}}({{\bf{r}}_s})} \right\rangle  + \delta {{\bf{E}}}_l({{\bf{r}}_s}),
\end{equation}
where ${\bf{r}}_s$ denotes an arbitrary observation point, $l=1,2,3,...,N_r$ stands for a specific configuration. The coherent field was obtained through an ensemble average procedure performed over all realizations, i.e., $\left\langle {{{\bf{E}}}({{\bf{r}}_s})} \right\rangle=\sum_{l=1}^{N_r}{\bf{E}}_l({\bf{r}}_s)/N_r$, and $ \delta {{\bf{E}}}_l({{\bf{r}}_s})$ indicates the incoherent field for the $l$-th configuration. Following from this process, the total ensemble averaged intensity was also divided into coherent and incoherent components as
\begin{equation}
\left\langle {{I}} \right\rangle  = {\left| {\left\langle {{{\bf{E}}}({{\bf{r}}_s})} \right\rangle } \right|^2} + \left\langle {{{\left| {\delta {{\bf{E}}_l}({{\bf{r}}_s})} \right|}^2}} \right\rangle.
\end{equation}
Unlike the incoherent field, the incoherent intensity does not vanish after ensemble average. The incoherent intensity is vital for understanding energy transport in random media. In fact, when the incident intensity propagates and undergoes scattering in the random media, the scattered intensity is diffused into different directions, which then becomes a part of incoherent intensity, while the unscattered intensity which still propagates ballistically constitutes the coherent intensity. As a consequence, the incoherent intensity is indeed a manifestation of scattering strength for the disordered media. Therefore, Tsang \textit{et al.} \cite{tsangOL1992} integrated the incoherent intensity in the far field over all solid angles to obtain the scattering coefficient as
\begin{equation}
\kappa_s  = \frac{1}{V_0}\int_{\bf{\Omega }} {\left\langle {{{\left| {\delta {{\bf{E}}_l}({{\bf{r}}_s})} \right|}^2}} \right\rangle d{\bf{\Omega }}},
\end{equation}
where ${\bf{\Omega }}$ is the solid angle with respect to  ${\bf{r}}_s$, and $V_0$ is the occupied volume of the entire cluster.
\begin{figure}[htbp]
	\flushleft
	\subfloat{
		\label{tsangOL1992}
		\includegraphics[width=0.47\linewidth]{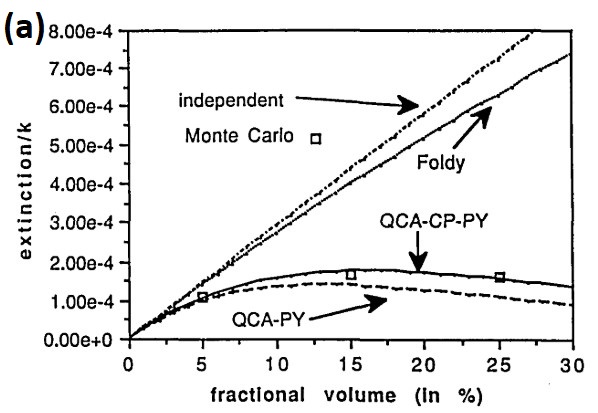}
	}
	\subfloat{
		\label{augerJCTR2009}
		\includegraphics[width=0.45\linewidth]{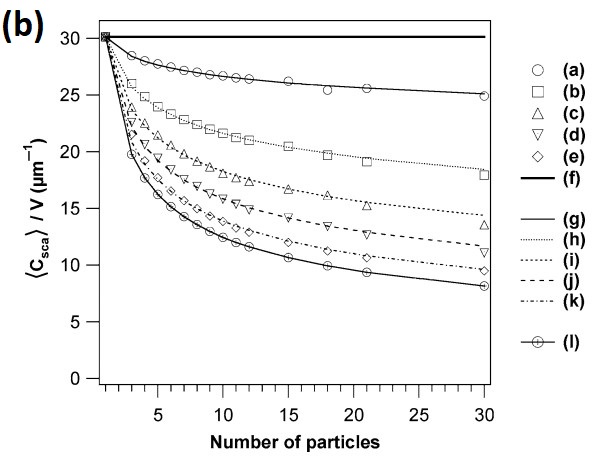}
	}
	\caption{The supercell method. (a) Extinction rate normalized to the free-space wave number as a function of the fractional volume of scatterers. The plots show calculations based on independent scattering, Foldy's formula, QCA using P-Y approximation (QCA-PY), QCA under coherent potential under P-Y approximation (QCA-CP-PY), and Monte Carlo simulations. Other parameters are $\varepsilon=3.2$ and $ka=0.2$. Reprinted with permission from Ref.\cite{tsangOL1992}. Copyright 1992 Optical Society of America. (b) Ensemble averaged scattering cross section per unit volume as function of the number of particles in the cell. (a-e) Volume fraction of particles $f_v$ = 0.01, 0.05, 0.10, 0.15, 0.20 respectively. (f) Independent scattering approximation, (g-k) fitted curves using a power low function (l) interpolation $f_v$ = 0.25. Reprinted by permission from Springer Nature Customer Service Centre GmbH: Springer Nature, \textit{Journal of Coatings Technology and Research}, Ref. \cite{augerJCTR2009}, Copyright (2009). }\label{RVEmethod}	
\end{figure} 
Their results were shown in Fig.\ref{tsangOL1992}, where the predictions of several theoretical models were also presented for comparison.

It should be noted that this method of determining scattering coefficient in the dependent scattering regime requires that the size of the cluster $r_c$ is much smaller than the scattering mean free path $l_s$ in order to avoid the occurrence of multiple scattering of intensity inside the cluster. Moreover, the size of the cluster should also be much larger than the wavelength $\lambda$, which guarantees that microscopic interference effects can be adequately described. Since a cluster should contain an enough number of scatterers, $r_c\gg a$ should be satisfied, where $a$ is the size of the scatterer. As a result, a stringent condition for this method is that
\begin{equation}
a,\lambda\ll r_c\ll l_s.
\end{equation}
Therefore, this method is most applicable for the case of small particles and weak scattering strength.

Another implementation of RVE method seems to need further rationalization, which is also used by many researchers as an \textit{ad-hoc} approach. Here we briefly describe this method on the bais of the work done by Auger \textit{et al.} \cite{augerJCTR2009}, which investigated the scattering efficiency of white paint films as a function of the volume fraction and spatial state of dispersion of rutile titanium dioxide pigments. To model the radiative properties, they proposed a unit cell composed of tens of particles and calculated the total scattering cross section using the full-wave recursive \textit{T}-matrix method. By increasing the number of particles in the unit cell, they showed that the ensemble averaged scattering cross section per unit volume $\langle C_\mathrm{s}\rangle/V_0$ (or scattering coefficient) reaches a pseudo asymptotic plateau, as shown in Fig.\ref{augerJCTR2009}. 

Even these methods require significant computing resources, especially for scatterers small compared to the wavelength, not to mention they neglect the dependent scattering effects between the neighboring supercells and RVEs, which are also very important to the mesoscopic radiative properties, and therefore are physically unreasonable and even questionable \cite{mishchenkoOSAC2019}. 

%{\color{red}Mishchenko \cite{mishchenkoOSAC2019} also commented: Previous studies's treatment is:``then postulating the FOSA for a small volume element and deriving (essentially postulating) the phenomenological radiative transfer equation by considering
%'incoherent multiple scattering' by small volume elements serving as building blocks of the particulate medium; and finally by speculating how the single-scattering properties of the individual particles and of the small volume elements can change as a consequence of hypothetical 'packing density' effects. Again, this questionable approach is based on the lack of recognition that from the fundamental perspective of electromagnetics, the entire particulate medium is a unified scattering object and must be treated as such from the outset. Our first-principles analysis appears to imply that the FOSA and RTT may be the only
%notable manifestations of the independent scattering regime, all other cases of electromagnetic scattering by particulate media belonging to the category of dependent scattering. If so, the
%terms 'independent scattering' and 'dependent scattering' have limited heuristic value, and their use can probably be avoided altogether by referring directly to the FOSA and the RTT as
%opposed to any other scattering scenario. }

\subsubsection{Direct numerical simulation method}
The third type of method is then the direct numerical simulation method. This method actually acts as virtual experiments \cite{mishchenkoPhysrep2016,mishchenkoOSAC2019,zurkJOSAA1995}. Note the possibility to conduct this method relies on the semi-analytical form of the multipole expansion of the FLEs, especially the MSTM and related codes for multiple spherical particle groups. For other DDM with more complicated micro/nanostructures, it is still difficult to do so. On the basis of previous works on direct numerical simulations, we can summarize the implementation procedure of this method to derive radiative properties, which is described as follows \cite{wangJQSRT2018}.

\textit{Step 1.} Firstly, a random distribution of spherical particles should be generated according to their interacting potential. For the hard-sphere system, the well-known Metropolis algorithm can be utilized \cite{tsang2004scattering2}, while for the sticky-sphere system, we can use the Kranendonk-Frenk algorithm \cite{Frenkel2002,tsang2004scattering2}. To model a realistic random medium (like a coating), a large slab geometry containing several thousands of spheres is needed, and its lateral size (perpendicular to the propagation direction of radiation) should be substantially larger than the thickness (along the propagation direction) in order to avoid side effects induced by the boundaries \cite{mackowskiJQSRT2013,Naraghi2015}.

\textit{Step 2.} A direct numerical algorithm (e.g., the semi-analytical MSTM code) is then implemented to calculate the electromagnetic field distribution at a large plane which is put in the downstream direction of the slab to collect all transmitted radiation \cite{Naraghi2015}. The position of this plane should be carefully chosen. If the distance of the plane with the downstream boundary of the slab is too large, then a very large plane is necessary to collect all the transmitted radiation, which needs substantially more computing resources. On the contrary, when the distance is too small, the contribution of evanescent waves is then included, which may be substantial and lead to unphysical values of total transmittance. By integrating the Poynting vector over the plane, we are able to obtain the total transmitted energy and thus the total transmittance $\mathcal{T}$. 

\textit{Step 3.} Another transmittance, i.e., the transmittance of coherent waves $\mathcal{T}_c$, is also calculated by evaluating the electromagnetic field distribution on a small plane. This small plane is placed in the extreme far-field (e.g., 10 times the wavelength) in the downstream direction of the slab. The size of this plane should also be chosen carefully \cite{aubryPRA2017}. If it is too large, the plane will collect the incoherent waves while if it is too small, the plane will miss a part of the coherent intensity \cite{aubryPRA2017}.  
%(\textit{How to avoid the coherent forward scattering peak since it does not affect the radiative transfer process but always occur in the forward direction?} \cite{rouabahJOSAA2014,tsang2004scattering2})

\textit{Step 4.} Since we need the ensemble averaged value of transmittance to minimize the random fluctuations, above Steps 1-3 should be iterated for many times for different random configurations. Typically, several hundreds of configurations are recommended \cite{mackowskiJQSRT2013}. In this step, we can get the ensemble averaged total transmittance $T=\langle\mathcal{T}\rangle$ and coherent transmittance $T_c=\langle\mathcal{T}_c\rangle$. 

\textit{Step 5.} By varying the thickness $L$ of the slab, we can obtain $T(L)$ and $T_c(L)$ as functions of $L$. It is known that for a thick enough random medium without absorption, it can enter the diffusive transport regime, if the strong (Anderson) localization does not occur \cite{sheng2006introduction,akkermans2007mesoscopic}. In this regime, the transmittance scales with the slab thickness following the Ohm's law as $T(L)\propto l_\mathrm{tr}/L$, where $l_\mathrm{tr}=1/[\kappa_s(1-g)]$ is the transport mean free path \cite{sapienzaPRL2007}. Therefore, we can numerically obtain $l_\mathrm{tr}$ from this relationship \cite{wiersma1997localization,Naraghi2015}. On the other hand, according to Beer's law, we have $T_c(L)=\exp{(-\kappa_sL)}$. Thus the scattering coefficient $\kappa_s$ is acquired. Consequently, we can numerically determine the scattering coefficient and asymmetry factor. 

\begin{figure}[htbp]
	\flushleft
	\subfloat{
		\label{mackowskiJQSRT2013a}
		\includegraphics[width=0.46\linewidth]{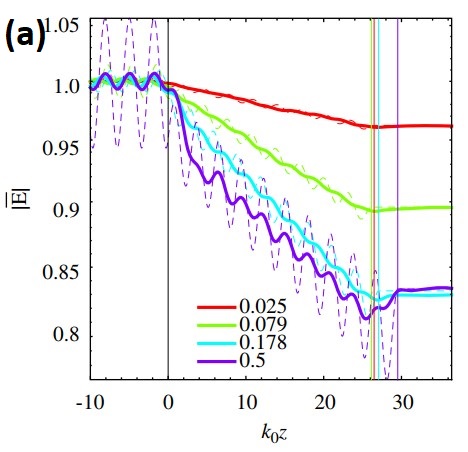}
	}
	\subfloat{
		\label{mackowskiJQSRT2013b}
		\includegraphics[width=0.45\linewidth]{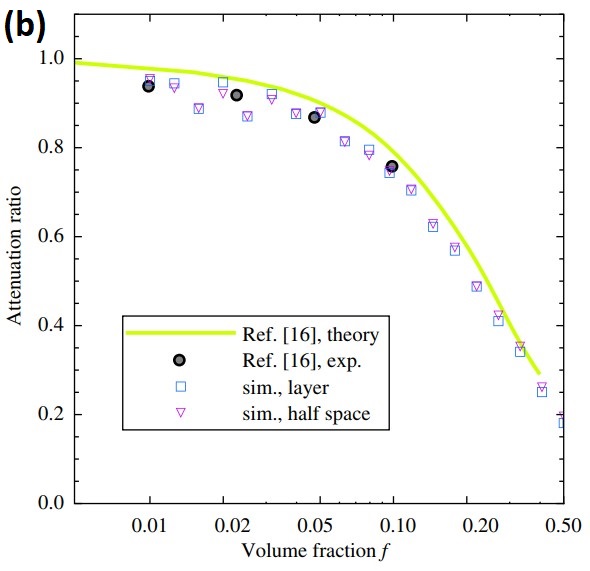}
	}
	\caption{Radiative properties extracted from direct numerical simulations. \textbf{(a)} Spatial profile of $\mathbf{E}_\mathrm{coh}(z)$ vs. $k_0z$ for different volume fractions with $k_0a=2.645$ and $\tilde{m}=1.194$, where $k_0$ is the wavenumber in free space. \textbf{(b)} Attenuation ratio $\gamma$ vs. volume fraction $f$, for simulation results extracted through the layer model and the half-space model, along with Varadan \textit{et al}'s \cite{varadanRS1983} (denoted by ``Ref. [16]" in the figure) experimental results and theoretical predictions based QCA. Reprinted from Ref. \cite{mackowskiJQSRT2013}, Copyright (2013), with permission from Elsevier. }\label{DNS_mackowskiJQSRT2013}	
\end{figure}

Another method to determine the effective refractive index is to fit the simulated spatial profile of coherent field with the analytical solution of the electric field profile of a plane wave impinging on a slab with a homogeneous complex refractive index as fitting parameter. An example is given in Fig.\ref{DNS_mackowskiJQSRT2013}.  In Fig.\ref{mackowskiJQSRT2013a}, spatial profile of the coherent field $\mathbf{E}_\mathrm{coh}(z)$ is shown against the propagation direction $k_0z$ for different volume fractions. The extracted attenuation ratio (i.e., 2 times the imaginary part of the effective refractive index) is given in Fig.\ref{mackowskiJQSRT2013b}, compared with theoretical predictions based on QCA as well as experiments. Good agreement is observed. Similar treatment is also employed by Pattelli \textit{et al.} recently \cite{pattelliOptica2018}. 

%Similar works Chanal et al \cite{chanalJOSAA2006} numerically simulated 2D random media with a wide range of particle sizes and refractive indices and determined the effective propagation constant. The particle positions were randomly generated using the Metropolis algorithm. They also compared the numerical result with Foldy-Twersky model and Keller's formula.

\begin{figure}[htbp]
	\flushleft
	\subfloat{
		\label{naraghiPRL2015}
		\includegraphics[width=0.4\linewidth]{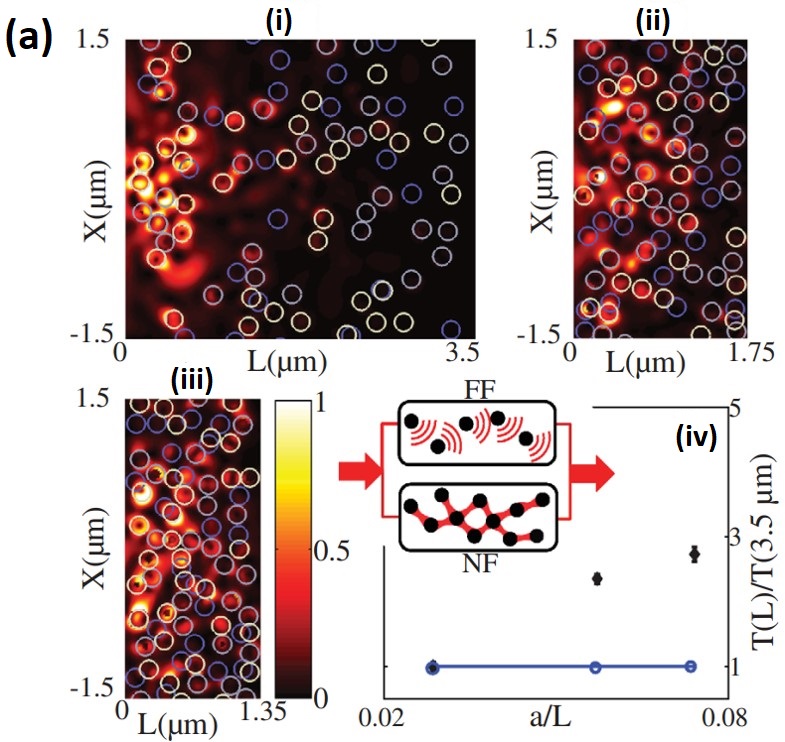}
	}
	\subfloat{
		\label{aubryPRA2017}
		\includegraphics[width=0.47\linewidth]{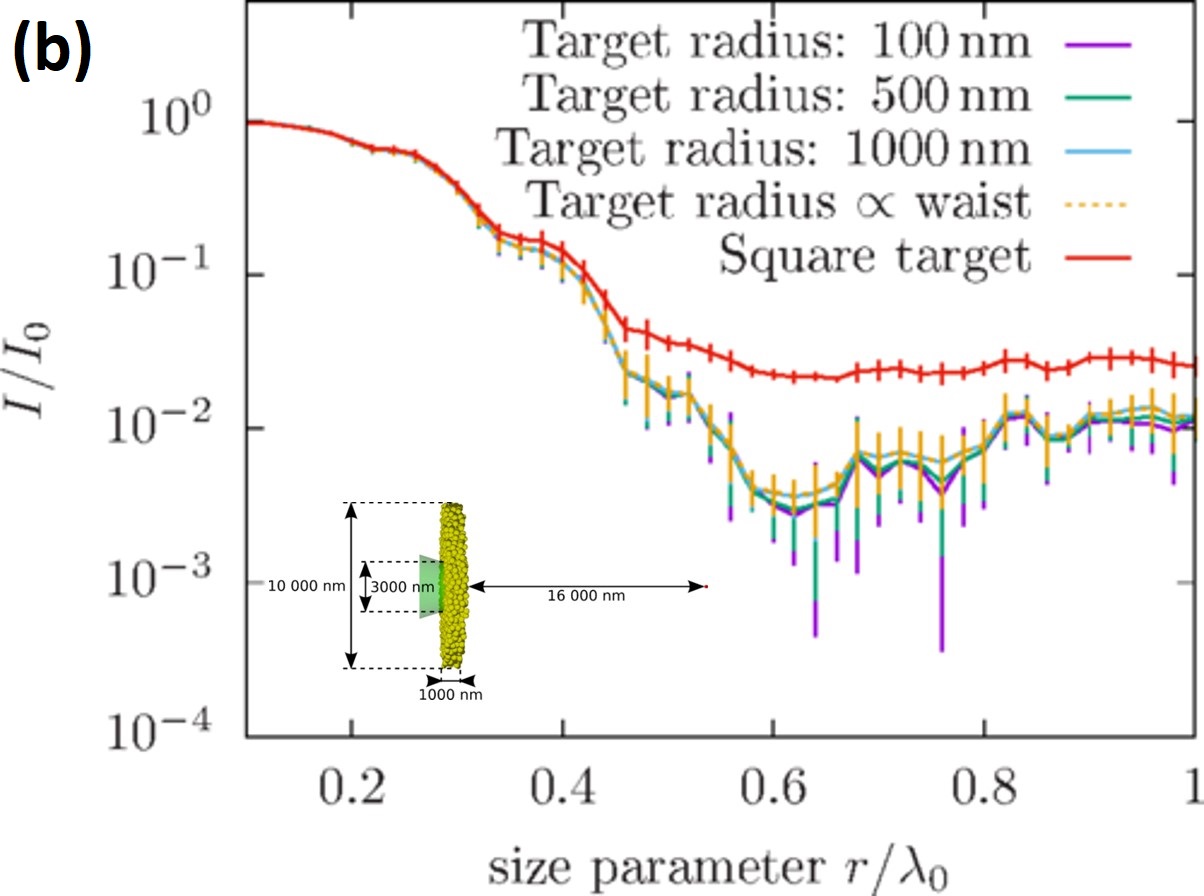}
	}
	\caption{Direct numerical simulations for near-field and resonant effects. \textbf{(a)} Direct numerical simulation of large slabs containing randomly dispersed $a=100~nm$ radius TiO$_2$ nanospheres. (i)–(iii) Intensity distributions in the cross-sectional areas of 3D slabs with reducing lengths as indicated. The media. Rings colored in gray denote particles located in the considered cross section, while the white and blue ones indicate particles situated at 100 nm above and, respectively, below that plane. (iv) Total transmission as a function of inverse thickness. The blue and black symbols designate the ISA and the results of MSTM calculations, respectively. The inset illustrates the appearance of additional transmission channels due to near-field coupling. Reprinted with permission from Ref. \cite{Naraghi2015} Copyright (2015) by the American Physical Society. \textbf{(b)} Direct simulation of cylindrical slabs with a diameter	of 10000 nm , a thickness of $L=$ 1000 nm containing about 2270 particles with a radius $r$ = 160 nm (filling fraction of	50\%). Average over five different slabs of the MSTM calculated transmission as a function of the size parameter: influence of the target size on the transmission values. The target is either circular (violet, green, blue, and yellow lower curves) or square (red upper curve, square area $10 640^2 \mathrm{nm^2}$). Inset: The geometry for MSTM simulation. Reprinted with permission from Ref. \cite{aubryPRA2017} Copyright (2017) by the American Physical Society.}\label{DNS}	
\end{figure}

Besides extracting the radiative properties, it is instructive to apply the direct numerical simulation method to investigate the applicability of the RTE \cite{mishchenkoPhysrep2016}. Recent works include \cite{greenAO2001,rouxJOSAA2001,schaferOL2008,voitOL2009,maJQSRT2017}. Direct numerical simulation method also provides a platform to investigate the near-field DSE in unprecedented details. Naraghi \textit{et al.} \cite{nalitovPRL2015} presented a direct numerical simulation of densely packed TiO$_2$ nanospheres based on MSTM to investigate the near-field coupling in the disordered media, as shown in Fig.\ref{naraghiPRL2015}. It can be seen in the simulated electric intensity distributions that by increasing the volume fraction, extra energy transport channels open in the media, that lead to an increase of transmission. In addition, this method can more accurately investigate the resonant multiple wave scattering phenomena, like the fine structure of resonances shown in Fig.\ref{aubryPRA2017}.

%Furthermore, direct numerical simulation method actually allows us to investigate into more details of the validity of theoretical models. In terms of the effective field amplitudes in the theory, we can simply fix the position of a specific particle at $\mathbf{r}_j$, and calculate its ensemble-averaged exciting field amplitudes $\langle c_{mp}^{(j)}\rangle_j$ for many random configurations of other particles. Using Eqs.(\ref{fl_eq4}-\ref{aprox1}), we can numerically obtain $C_{mp}$. Note here the effective propagation constant can be determined by evaluating the propagation of coherent wave through the slab by means of the method used by Mackowski \textit{et al.} \cite{mackowskiJQSRT2013}. Note the the effective propagation constant should fulfill the relationship of $\mathrm{Im}K=\kappa_s/2$, which can act as a criterion of numerical validity. In addition, by fixing the positions of two particles, we can further examine the validity of Eq.(\ref{aprox2}), i.e., the key assumption of QCA. 

Currently, the direct numerical simulation method can only deal with spherical particle groups or well-separated nonspherical particles \footnote{For nonspherical particles, the main limitation of the multiple particle \textit{T}-matrix method is its incapability to deal with elongated particles placed in the near-field of each other, which requires further investigation \cite{bertrand2019global}, as mentioned above. And there are also fast solvers for multiple wave scattering in 2D for arbitrary geometries, and the same limitation exists \cite{laiOE2014,blankrotIEEEJMMCT2019}. However, the performance of all available numerical algorithms degrades for very densely packing, high-refractive-index particles.}, still needs tremendous computing resources and is very time-consuming, for example, it takes about 30 hours to simulate the electromagnetic propagation in a disordered medium containing 12000 spherical $\mathrm{TiO_2}$ nanoparticles on a supercomputer cluster with 260 to 340 processors \cite{Naraghi2016}, and \textit{hundreds to thousands} of such simulations are necessary to perform an ensemble average process over many random configurations of particles in order to obtain a \textit{single} data point of transmittance for the disordered medium \cite{mackowskiJQSRT2013}. Hence it is by now still impractical to carry out direct numerical simulations to determine the mesoscopic radiative properties of DDM. 

To summarize, in this section, we describe some basic mechanisms involved in the DSE, including the far-field DSE, near-field DSE, recurrent scattering, structural correlations and the effect of absorbing host media. Concrete theoretical models that deal with these mechanisms are then introduced with detailed analytical formulas presented together with brief derivation procedures. Furthermore, numerical methods to model the DSE are summarized, including the supercell method, the representative volume element method and the direct numerical simulation method. For the mentioned theoretical models, by now the CPA-based models have been shown to be most suitable for the treatment of densely packed DDM \cite{schertelPRMat2019,aubryPRA2017} although they require a self-consistently solved effective refractive index. Among the perturbative expansion-based models, QCA is usually believed to be the most accurate and can be safely applied in moderately dense DDM, although the second-order perturbative Keller's formula sometimes can achieve better predictions due to the deviations brought by high-order diagrams in QCA. ITA, due to the simplicity is the most widely used one. Since all these models cannot explicitly account for the effects of recurrent scattering and near-field interactions, the two-particle recurrent scattering model and phenomenological near-field models are somewhat valuable in identifying the role of near-field DSE. In terms of the reviewed numerical methods, the supercell method is simple but fails to directly provide radiative properties. The RVE method can provide a direct determination of radiative properties but require stringent conditions for the volume element, which also needs further formal rationalization. Despite that the direct numerical simulation method is accurate, by now the mainstream computation capabilities can only deal with relatively small samples consisting of spherical scatterers (around tens of thousands of spheres). Therefore, more rapid, general and robust numerical methods are expected in the future.

We also note that the effect of dependent scattering on the absorption of DDM is rarely studied, especially lacking suitable theoretical models \cite{kumar1990dependent,ma1990enhanced,prasherJAP2007,weiAO2012}. This situation is a consequence of the difficulty in deriving analytical formulas for the scattering coefficient, as mentioned in the beginning of this subsection. In fact, when the particle density is small (typically volume fraction $f_v<0.05$) and absorption coefficient is much larger than the scattering coefficient, i.e., $\kappa_a\gg\kappa_s$, this ignorance gives rise to no substantial discrepancies because multiple wave scattering is weak. However, when particle density continues to increase or $\kappa_s$ is comparable with or much larger than $\kappa_a$, a careful consideration of DSE on total absorption is necessary because the interparticle interference of scattered waves may lead to a redistribution in particle absorption. This issue is becoming important as the recent growing interests in nanofluids, as well as other nanoparticle-based solar absorbers, which usually utilize plasmonic resonances of metallic particles to enhance solar absorption \cite{taylorJAP2013,saidICHMT2013,xuanRSCA2014,hoganNL2014,liuNanoscale2017,gaoJAP2017} and find their applications in concentrating solar power, like direct-steam generation, photocatalysis, solar thermochemistry and solar desalination applications \cite{psaltisNature2006,ericksonNPhoton2011}. In some researches on the structural coloration based on disordered photonic structures, predictions based on ISA interpreted the experimental results poorly, partly due to the extremely-high absorption predicted by ISA for densely packed nanostructures. This may smear out the reflectance peak indeed observed by the experiments \cite{xiaoSciAdv2017}. Wei \textit{et al.} \cite{weiAO2012} recently investigated the DSE on the absorption coefficient for very dense nanofluids ($f_v$ up to 0.74) containing very small metallic particles (radius $a=15\mathrm{nm}$) by using several different dependent-scattering models as well as developing a modified QCA model. However, their model, as well as the model proposed by Prasher \textit{et al.} \cite{prasherJAP2007} can only treat very small and strongly absorbing particles. Therefore it is necessary to explore in depth the role of dependent scattering mechanism on light absorption in disordered media consisting of highly scattering scatterers.  Recently progresses have been made to develop more general theoretical models using the diagrammatic expansion method to consider the DSE on the absorption in DDM \cite{wangIJHMT2018,sheremet2020absorption}, which all indicated that the absorption coefficient and total absorptance can be flexibly tailored by means of structural correlations \cite{liuJOSAB2018,bigourdanOE2019}.

\section{Experimental investigations of the DSE}\label{expsec}
As already mentioned in the Introduction section, the experimental investigation of the DSE started early in the 1960s and many efforts have been made to delineate the boundary between independent and dependent scattering \cite{hottelJHT1970,hottelAIAAJ1971,tienARHT1987,ishimaruJOSA1982}. Later, experimental studies concentrated on the measurement of radiative transport properties in order to understand the complicated microscopic and mesoscopic scattering phenomena. Nowadays, it becomes more and more attractive to utilize the DSE to manipulate the multiple scattering of waves and radiative transfer behaviors to achieve desirable functionalities like radiation insulation \cite{wangIJHMT2015}, structural coloration \cite{xiaoSciAdv2017}, display and imaging \cite{sunBOE2014}, radiative cooling \cite{baoSEMSC2017,zhaiScience2017} and Anderson localization (See Section \ref{mesointerference}).

In this section, we attempt to review experimental methods and representative works on the DSE. We have to emphasize that the vast majority of available experimental works investigate the DSE in an indirect manner, that is, by comparing experimental measured radiative properties with the predictions of ISA. Moreover, the radiative properties, including scattering and absorption coefficients $\kappa_a$ and $\kappa_s$, scattering phase function $P(\mathbf{\Omega}',\mathbf{\Omega})$ and asymmetry factor $g$, should also be deduced from experimental measurements of (angular-resolved or hemispherical) reflectance and transmittance in the time, frequency or continuous wave domain using appropriate inverse radiative transfer models \cite{eldridgeJACS2008,eldridgeJACS2009}. Since the identification of radiative properties is restricted by the available experimental data and very sensitive to measurement uncertainties and noises, and especially dependent on the choice of phase functions \cite{baillisJOSAA2004}, the analysis of the DSE in these experimental works remains case by case. Moreover, since many different samples are needed to reduce the random fluctuations brought by disorder, the sample preparation procedure is also very time-consuming. As a result, nowadays, it is still difficult to directly measure the effect of dependent scattering in an exact manner. However, with careful sample preparation, appropriate experimental techniques and elaborated radiative transfer and DSE models, it is still possible for us to establish an in-depth understanding of the role of dependent scattering.

On the other hand, there are also a great deal of experimental studies on DSE that make comparisons directly between the measured macroscopic quantities (e.g., hemispherical or angular resolved reflectance and transmittance) and theoretically predicted ones (through a DSE model of radiative properties combined with a forward RTE calculation), without any inverse procedures to obtain the mesoscopic radiative properties, for instance, Refs. \cite{yamadaJHT1986,wangIJHMT2015}. This scheme avoids the possible ill-posed retrieving problem in regard to the mesoscopic radiative properties, while it is not very straightforward for the understanding of how the DSE influences the radiative transfer and multiple scattering processes.

In this section, we classify experimental investigations of the DSE according to the adopted experimental methods, which measure different macroscopic quantities, including the coherent transmittance, hemispherical reflectance and transmittance, angular-resolved reflectance and transmittance, and time-resolved responses. In order to make the discussion self-consistent and introductory, for each method, we first introduce the measurement principle and corresponding inverse radiative transfer models, followed by a discussion on the results of representative experimental works that examine the DSE using this method. A description of typical experimental setups is also provided.

\subsection{Measurement of coherent transmittance}

By measuring the coherent transmittance (or called the collimated or unscattered/ballistic transmittance), the extinction coefficient of the disordered medium can be obtained directly using the Beer's law as
\begin{equation}
\kappa_e=-\frac{1}{L}\ln{\left(\frac{I_c}{I_0}\right)},
\end{equation}
where $I_c$ is the coherent/collimated intensity that transmits through the medium without any scattering (i.e., the ballistic component of the intensity), $I_0$ is the transmitted intensity in the absence of the disordered medium and $L$ is the thickness of the disordered medium.  

%Similar experiments have been carried out for a variety of DDM like Intralipid-20\% phantoms (20\% i.v. fat emulsion, typically used for biomedical experiments to simulate tissues) in water \cite{zaccantiAO2003}, 

\begin{figure}[htbp]
	\centering
	\subfloat{
		\label{ishimaruJOSA1982}
		\includegraphics[width=0.42\linewidth]{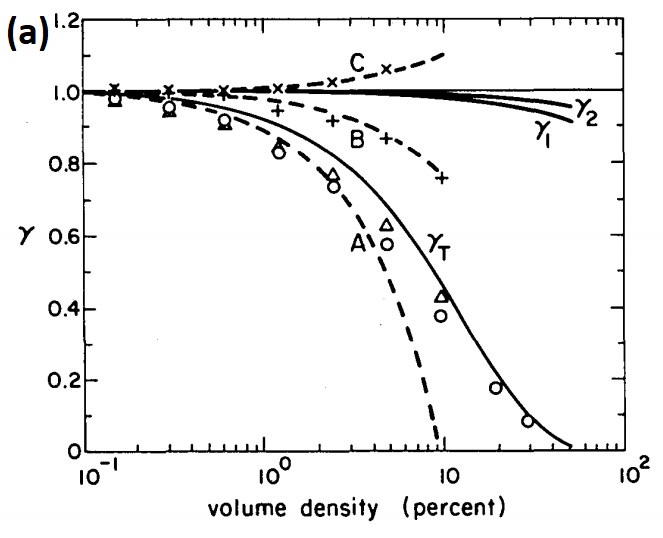}
	}
	\subfloat{
		\label{zaccantiAO2003}
		\includegraphics[width=0.5\linewidth]{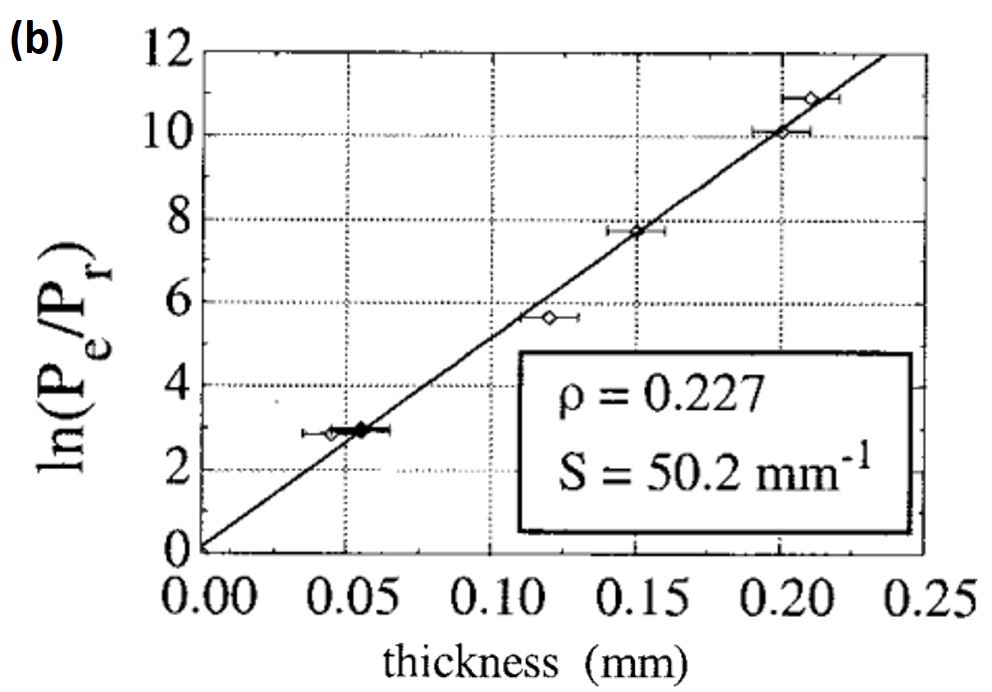}
	}
\\
	\subfloat{
	\label{ishimaruJOSA1982config}
	\includegraphics[width=0.5\linewidth]{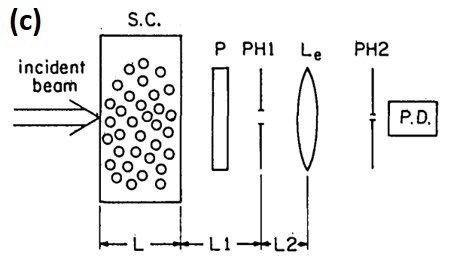}
}
	\subfloat{
	\label{zaccantiAO2003config}
	\includegraphics[width=0.45\linewidth]{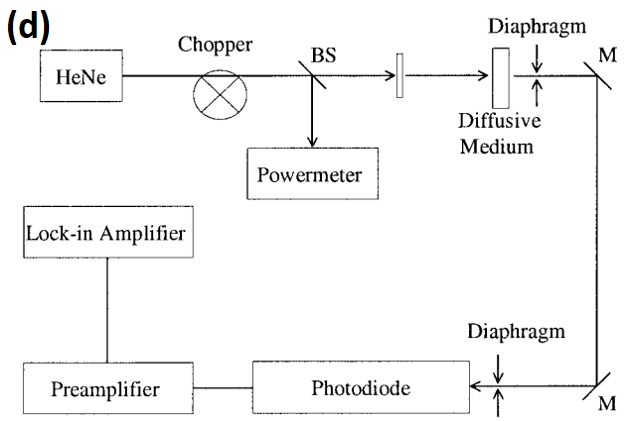}
}
\caption{Extinction coefficient from measured coherent transmittance. \textbf{(a)} Experimental results of the ratio $\gamma=\kappa/\kappa_0$ for , where $\kappa_0$ is the extinction coefficient under ISA. Twersky's formula is $\gamma_T=\kappa_T/\kappa_0$. $\gamma_1$ is the calculated ratio $\kappa_e/\kappa_0$ for 0.091, 0.109, and 0.481 $\mathrm{\mu m}$ and $\gamma_2$ is the same ratio for 1.101 $\mathrm{\mu m}$: $\circ$, particle size 0.091 $\mathrm{\mu m}$; $ka = 0.529$; $\triangle$, 0.109 $\mathrm{\mu m}$ (0.681); $+$, 0.481 $\mathrm{\mu m}$ (3.518); and X, 1.101 $\mathrm{\mu m}$ (7.280). Dashed curves are $\kappa_h/\kappa_0$ calculated from Keller's approach with the hole correction approximation for pair distribution function: A, 0.109 $\mathrm{\mu m}$; B, 0.481 $\mathrm{\mu m}$; and C, 1.101 $\mathrm{\mu m}$. \textbf{(c)} Experimental setup for \textbf{(a)}. S.C. indicates sample cell, P is a polarizer, PH1 stands for pinhole 1 with a diameter of 3 mm and PH2 for pinhole 2 with a diameter of 25 $\mathrm{\mu m}$, $\mathrm{L_e}$ is a $10\times$ microscope objective lens, P.D. is a photodiode. Reprinted with permission from Ref. \cite{ishimaruJOSA1982}. Copyright 1982 Optical Society of America. \textbf{(b)} An example of measuring the extinction coefficient through the thickness dependence of coherent transmittance. $P_r$ is the measured attenuated power, $P_e$ is the incident power, $\rho$ is the volume fraction and $S$ is the fitted extinction coefficient. \textbf{(d)} Experimental setup for \textbf{(b)}. BS: beam splitter. M: Mirror. Reprinted with permission from Ref. \cite{zaccantiAO2003}. Copyright 2003 Optical Society of America.}\label{experiment1}	
\end{figure} 

As mentioned in the Introduction section, based on this method, Hottel and co-workers \cite{hottelJHT1970, hottelAIAAJ1971} experimentally measured the extinction coefficient of monodisperse PS nanosphere suspensions in water confined between parallel glass slides at different optical thicknesses. Ishimaru and Kuga \cite{ishimaruJOSA1982} carried out a comprehensive study using a similar scheme. The investigated size parameter of the PS spheres can vary from 0.529 to 82.793 and the volume fraction can be controlled from 0.1\% to 40\%. They found that for suspensions composed of spheres with the size parameter $x<1$, the ratio between the measured extinction coefficient and the ISA prediction decreases as the volume fraction, while for the cases with $x>1$, the trend is reversed. The results compared with different theoretical models are shown in Fig.\ref{ishimaruJOSA1982}. It is interesting to note that the Keller's approach can give fairly good agreement with the experimental data, despite the use of the relatively rough hole correction (HC) approximation for the pair distribution function. 

The experimental setup is given in Fig.\ref{ishimaruJOSA1982config}. In this setup, the incident beam, which was a He-Ne laser (wavelength $\lambda = 0.6328 \mathrm{\mu m}$), passed through the sample cell containing monodisperse PS suspensions in water. The transmitted light was collected by a detector, which was composed of an input aperture, a 10$\times$ microscope objective lens, a $25~\mathrm{\mu m}$ pinhole, and a photo diode with a field of view (FOV) of $0.085\degree$. And a polarizer was put between the sample and the detector to pass the same polarization with the incident beam. Moreover, since the random motion of the spheres in the suspensions could induce fluctuations of the detected intensity signal, the time constant of the low-pass filter was set to about 0.84 sec, which led to an ensemble average carried over approximately 1000 measurements.

It should be noted that this method becomes impractical for optically thick materials because in this circumstance a substantial amount of multiply scattered light can reach the collimated direction while the ballistic component is small, leading to a considerable error in the measured coherent transmittance and thus extinction coefficient. Therefore, a careful analysis on the thickness dependence of coherent transmittance should be employed. An example taken from Ref.\cite{zaccantiAO2003} is presented in Fig.\ref{zaccantiAO2003}, which shows the measured transmitted intensity in logarithmic scale, $\ln{(I_0/I_c)}$ (represented by $\ln{(P_e/P_r)}$ in the figure). It is seen that this quantity grows linearly with the sample thickness, which indicates the error due to multiply scattered intensity in the measured coherent transmittance is small. The experimental setup in Ref.\cite{zaccantiAO2003} is illustrated in Fig.\ref{zaccantiAO2003config}, which exhibited some improvements over the setup in Fig.\ref{ishimaruJOSA1982}. In this experimental configuration, a chopped 10-mW He–Ne laser was used to illuminate the medium, which was finally detected by a photodiode and measured with a lock-in amplifier. A beam splitter was put after the chopper to monitor the light source continuously. To reduce the contribution of multiply scattered intensity as much as possible, two 3-mm-diameter diaphragms D1 and D2 were used, approximately equal to the diameter of the laser beam. As a result, only photons scattered within an angle $\alpha=1.7~\mathrm{mrad}$ can be received, which was only slightly larger than the divergence of the incident laser beam. It was shown by Zaccanti et al \cite{zaccantiAO2003} that for moderate values of optical thickness $\tau_e$, the relative error for $\kappa_e$ due to the multiply scattered light was expected to be around $1\times 10^{-5}$. Similar measurements have been carried out by many researchers to study the effect of dependent scattering, e.g., Refs \cite{mandtWRM1992,westJOSAA1994,hespelJOSAA2001}, to name a few.

Apart from the above method of direct measurement of the coherent transmittance under continuous-wave illumination, a time-domain method that only detects the early-time transmitted photons, i.e., ballistic photons, under pulsed laser illumination, has also been developed to obtain the extinction coefficient \cite{wangOL1995}. This is achieved with the aid of an ultrafast time gate to separate the unscattered photons and scatterer photons, a temporal version the spatial pinhole. Ultrafast optical Kerr gate (OKG) is the most popular time gate, which has a time window on the order of hundreds of femtoseconds to several picoseconds depending on the pump pulse duration and the relaxation time of the optical Kerr medium. This method is more accurate \cite{tongOpteng2011}, and capable of dealing with DDM samples with very high optical thicknesses (even several hundred) \cite{dabzacOE2012}. It also lays the foundation of the well-known time-gated ballistic imaging technique in turbid media \cite{wangScience1991,xuOE2015}. Other time-domain methods will be discussed in Section \ref{time_resolved}.

\subsection{Measurement of total reflectance and transmittance}\label{total_ref_trans}

\begin{figure}[htbp]
	\centering
	\subfloat{
		\label{aernoutsOE2014}
		\includegraphics[width=0.5\linewidth]{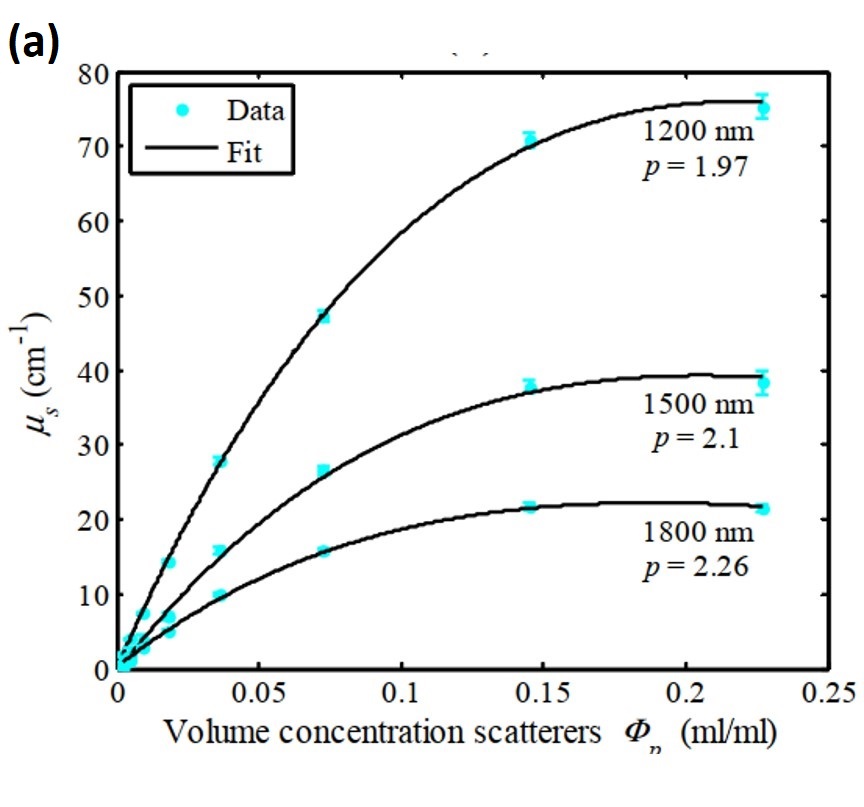}
	}
	\subfloat{
	\label{aernoutsOE2013config}
	\includegraphics[width=0.46\linewidth]{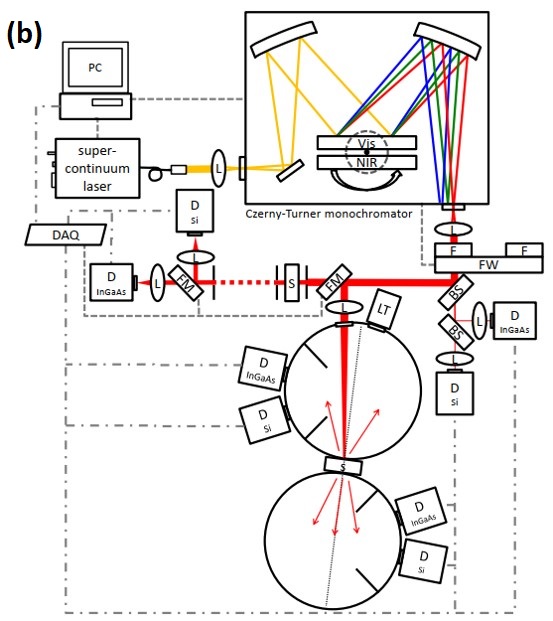}
		}

	\caption{Radiative properties measured from total reflectance and transmittance. \textbf{(a)} Effect of dependent scattering with increasing volume concentration of scattering	particles (here denoted by $\phi_p$ in the original figure) on the bulk scattering coefficient at three wavelengths. The average (cyan dots) and standard deviation (cyan error bars) of the $\mathrm{\mu s}$ measurements are plotted at the considered volume concentrations of the scattering particles. A modified Twersky equation (Eq.(\ref{fitting_DSE})) with variable packing dimension $p$ is fitted to the data for the different wavelengths (solid lines). Reprinted with permission from Ref. \cite{aernoutsOE2014}. Copyright 2014 Optical Society of America. \textbf{(b)} The double integrating sphere (DIS) setup for \textbf{(a)}. BS: Polka-dot beam splitter. D: detector. DAQ: data acquisition card. F: long-pass filter. FM: motorized flip mirror. FW: motorized filter wheel. L: convex lens. LT: light trap. S: sample. Reprinted with permission from Ref. \cite{aernoutsOE2013}. Copyright 2013 Optical Society of America.}\label{experiment2}	
\end{figure} 

The measurement of coherent transmittance can only give the extinction coefficient. To further investigate the role of the DSE in radiative properties, more experimental information is necessary. A natural idea is to measure the total (or equivalently, hemispherical) reflectance and transmittance via integrating spheres. For instance, Aernouts \textit{et al.} \cite{aernoutsOE2014} measured the total reflectance and transmittance of $\mathrm{Intralipid}^{\textregistered}$ phantoms in the 600-1850 nm wavelength range with different volume fractions at two sample thicknesses, and identified the scattering and absorption coefficients based on the inverse adding-doubling (IAD) algorithm to inversely solve the RTE \cite{prahl2011iad}. They found that in the volume fraction range less than 0.25, the scattering coefficient slowly increases with the volume fraction, much slower than the linear increase predicted by ISA, as shown in Fig.\ref{aernoutsOE2014}. They also showed that DSE is more prominent in the long wavelength range. By modifying the Twersky's model (Eq.(\ref{twersky_model})), they proposed an empirical model with a wavelength-dependent fitting parameter $p(\lambda)$ (which is exactly 3 in the Twersky's model) for the dependent scattering coefficient as
\begin{equation}\label{fitting_DSE}
\kappa _ { s } \left( \lambda , f _ { v } \right) = \frac { \kappa _ { s } ^ \mathrm{ind} ( \lambda ) } { 0.227 } f _ { v } \left\{ \frac { \left( 1 - f _ { v } \right) ^ { p ( \lambda ) + 1 } } { \left[ 1 + f _ { v } ( p ( \lambda ) - 1 ) \right] ^ { p ( \lambda ) - 1 } } \right\},
\end{equation}
where $f _ { v } $ is the volume fraction of scatterers and $\kappa _ { s } ^ \mathrm{ind} ( \lambda ) $ is the scattering coefficient calculated under ISA. From Fig.\ref{aernoutsOE2014}, it can be observed that the agreement is good at different wavelengths and volume concentrations. Moreover, they demonstrated that the asymmetry factor decreases with the volume fraction. 

The experimental setup used in above work is based on the double integrating sphere (DIS) configuration, which allows the simultaneous measurement on total reflectance and transmittance. The schematic is presented in Fig.\ref{aernoutsOE2013config}, which consists of an additional measurement path for the coherent (unscattered) transmittance. By using a supercontinuum laser with 4 W total output power and spectral broadening over the range 450-2400 nm, this experimental setup can measure the total transmittance and reflectance in the 500-2250 nm wavelength range. The high output power offers a possibility to measure very low reflectance and transmittance for thick or highly absorbing samples. More specifically, the white laser is focused by a lens into a high-precision Czerny-Turner monochromator. The beam diffracted from the gratings in the monochromator is then focused into a long-pass filter, which can be controlled by a motorized filter wheel, in order to block higher-order diffracted light. Then the light beam is split by a beam splitter in a sample path and a reference path, and the latter is split again by another beam splitter to two detectors for monitoring laser stability, including an InGaAs detector for wavelengths above 1050 nm and a Si detector for wavelengths below 1050 nm. On the other hand, the light beam of the sample path is sent to a motorized flip mirror to reflect the light towards the DIS measurement path, or pass the light to the unscattered transmittance measurement path. In the DIS measurement path, the light beam is focused by a lens on the center of the sample, which is positioned between two integrating spheres. Each integrating sphere is equipped with two detectors similar to those used in the reference path. The sample is illuminated under an angle of $9\degree$, which makes it possible to measure both diffuse and total reflectance by including or excluding the specular reflected light, respectively. This setup was validated by the authors of Ref.\cite{aernoutsOE2013} on a set of 57 liquid optical phantoms, which were designed to cover a wide range of absorption and scattering properties.

A great deal of works have been carried out for different materials at different spectral ranges to obtain radiative properties by measuring total reflectance and transmittance. Auger \textit{et al.} \cite{augerJCTR2015} measured reflectance spectra of white paints composed of densely packed hollow polymer spheres at different sample thicknesses and extracted the corresponding scattering coefficient. They also revealed that the scattering coefficient does not scale linearly with the volume fraction of scatterers as ISA predicted, but obeys a second-order polynomial relation. Fricke and coworkers \cite{kuhnRSI1993,gobelWRM1995} investigated the DSE in TiO$_2$ powders by performing hemispherical reflectance and transmittance measurements in the midinfrared and identifying the scattering coefficient based on a 3-flux approximation for the solution of the RTE. Lallich \textit{et al.} \cite{lallichJHT2009} experimentally identified the scattering coefficient and albedo (the ratio between scattering and extinction coefficients) of silica aerogels and revealed that the DSE indeed plays a role in its radiative properties. Our group has also conducted similar works for thermal barrier coatings (TBCs) consisting of porous yttria-stabilized zirconia with various microstructures \cite{wangIJHMT2015,yangIJHMT2013,yangJHT2015}. However, a problem in the identification of radiative properties is that the obtained result strongly relies on the concrete treatment of scattering phase functions \cite{baillisJOSAA2004}, and at times there is a crosstalk between the retrieved scattering and absorption coefficients \cite{eldridgeJACS2008,mazzamutoNJP2016}.

\begin{figure}[htbp]
	\subfloat{
		\label{wiersmaNature1999}
		\includegraphics[width=0.5\linewidth]{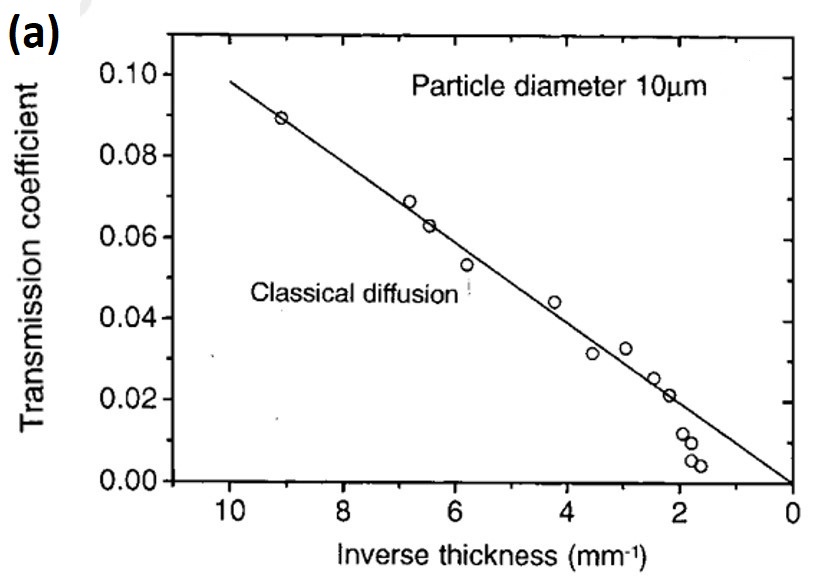}
	}
	\subfloat{
		\label{garciaPRA2008config}
		\includegraphics[width=0.46\linewidth]{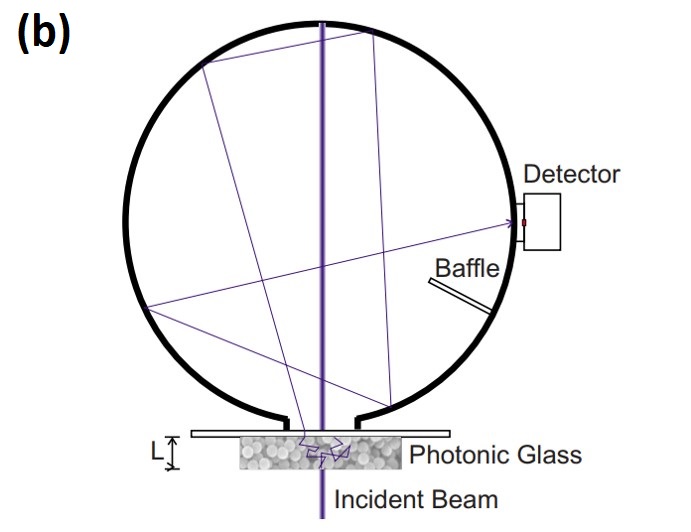}
	}
	
	\subfloat{
		\label{garciaPRA2008a}
		\includegraphics[width=0.46\linewidth]{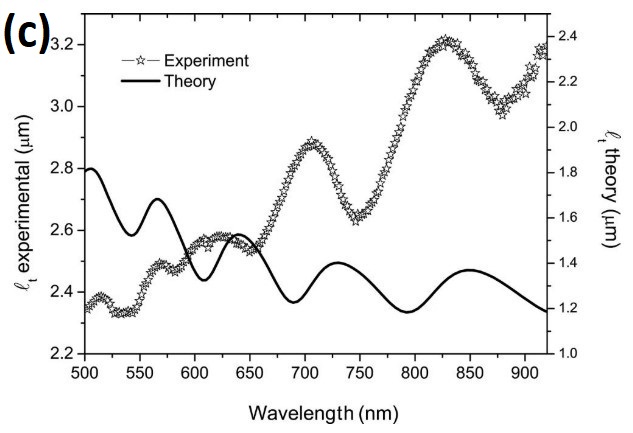}
	}
	\subfloat{
		\label{garciaPRA2008b}
		\includegraphics[width=0.45\linewidth]{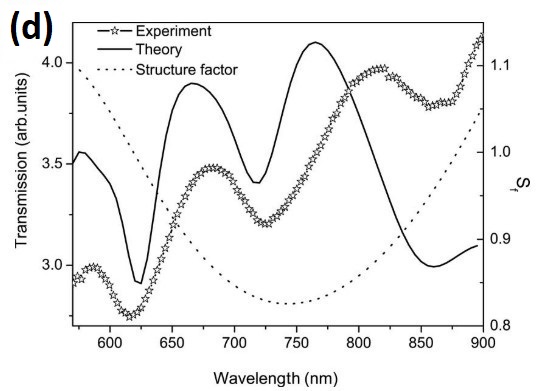}
	}
	\caption{Transport mean free path measured from total transmittance in the diffusive regime. \textbf{(a)} The thickness dependence of total transmittance in the diffusive regime. Reprinted by permission from Springer Nature Customer Service Centre GmbH: Springer Nature, \textit{Nature}, Ref. \cite{wiersma1997localization}, Copyright (1997). \textbf{(b)} Typical experimental setup for the measurement of total transmittance in the diffusive regime. \textbf{(c)} Comparison of experimental data and the prediction of ISA for the transport mean free path $l_\mathrm{tr}$. \textbf{(d)} Comparison of experimental data and the theoretical prediction considering the first-order far-field interference (using structure factor) for total transmission, where the structure factor is also shown. \textbf{(b-d)} are all reprinted with permission from Ref. \cite{garciaPRA2008} Copyright (2008) by the	American Physical Society.}\label{garciaPRA2008}
	
\end{figure}

Sometimes it is a good idea to directly measure the transport mean free path $l_\mathrm{tr}$ (or equivalently the transport scattering coefficient $\kappa_\mathrm{tr}=\kappa_s(1-g)=1/l_\mathrm{tr}$), since this quantity does not require a knowledge of the phase function. Moreover, it is noted the diffusion equation can still be used even in the regime when the RTE substantially breaks down. In the diffusive regime ($L\gg l_\mathrm{tr}$) with negligible absorption, according to the Ohm's law (Eq.(\ref{ohm_eq})), we have
\begin{equation}
T^{-1}=\frac{L}{l_\mathrm{tr}+z_e}+\frac{2z_e}{l_\mathrm{tr}+z_e}.
\end{equation}
Therefore, by measuring the thickness dependence of the inverse of total transmittance, the transport mean free path can be obtained through the slope of the linear relationship. Another way is that by approximating the Ohm's law directly as $T\approx l_\mathrm{tr}/L$ because $L$ is much greater than $l_\mathrm{tr}$ and $z_e$. An example is given in Fig.\ref{wiersmaNature1999} where the solid line denotes the relationship $T=l_\mathrm{tr}/L$, and a good agreement with the experimental data is observed. Note the deviation occurs at large thickness (small inverse thickness) because of the non-negligible contribution of absorption in very thick samples. A simple experimental setup for the total transmittance is presented in Fig.\ref{garciaPRA2008config}. In this configuration, an incident beam is directly focused to the sample, whose back is in contact with a standard integrating sphere in order to collect all transmitted light. 

Based on this method, Garcia \textit{et al.} \cite{garciaPRA2008} measured the transport mean free path $l_\mathrm{tr}$ of a photonic glass (PG) composed of densely packed PS spheres with a diameter of $d$ = 550 nm and volume fraction of 55\%, which exhibit strong Mie resonances. A comparison with the result of ISA (together with Mie theory) shown in Fig. \ref{garciaPRA2008a} indicates that ISA severely breaks down in this situation, where the discrepancy is not only quantitatively large but also leads to a different trend of wavelength-dependence. They further compared experimental results with the prediction of the ITA using a structure factor under the P-Y approximation for hard spheres, which were found to achieve a better agreement as shown in Fig.\ref{garciaPRA2008b}. However, the experimental data demonstrated that ITA is still not capable of accurately predicting the radiative properties under significant DSE, especially in the long-wavelength range ($\sim$ 800 nm - 900 nm).

The advantage of measuring the thickness dependence of transmittance and thus directly retrieving the transport mean free path is that it might even work when RTE itself breaks down because the diffusion equation is also the hydrodynamic limit of the Bethe-Salpeter equation \cite{vanTiggelenRMP2000} (See Section \ref{rteandde}). Moreover, it does not require a detailed choice of the scattering phase function, unlike the aforementioned inverse methods based on RTE. Another remarkable point is that by measuring the thickness dependence of transmittance, it is possible to identify the onset of Anderson localization. More specifically, during the transition from the diffusive transport regime to the Anderson localization regime, $T\propto1/L^2$ based on the prediction of the scaling theory localization \cite{andersonPMB1985,wiersma1997localization} and in the Anderson localized regime, $T\propto\exp{(-L/l_\mathrm{loc})}$, where $l_\mathrm{loc}$ is the localization length which describes the spatial extent of localized modes (see Section \ref{mesophys}). Note the exponential decays should be carefully distinguished from absorption and inelastic scattering \cite{wiersma1997localization,scheffoldNature1999,wiersmaNature1999reply,sperlingNJP2016,haberko2018transition} (see more discussions in Section \ref{anderson_loc_sec}) .

\subsection{Measurement of angle-resolved reflectance and transmittance}\label{angleresolvedmeasurement}

\begin{figure}[htbp]
	\centering
	\subfloat{
		\label{milandriJQSRT2002}
		\includegraphics[width=0.5\linewidth]{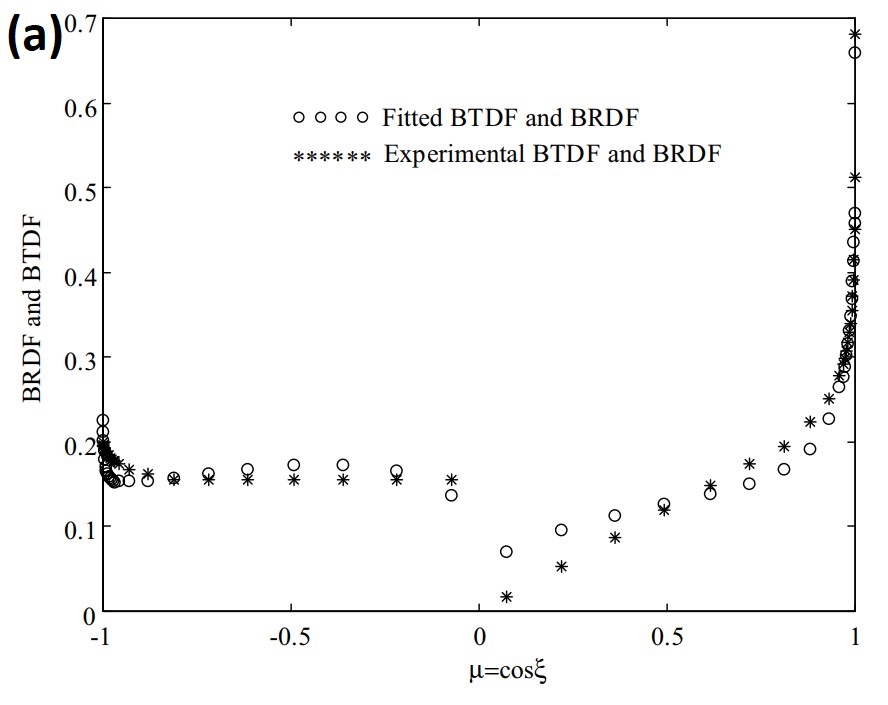}
	}
	\subfloat{
		\label{mishchenkoOL2013}
		\includegraphics[width=0.42\linewidth]{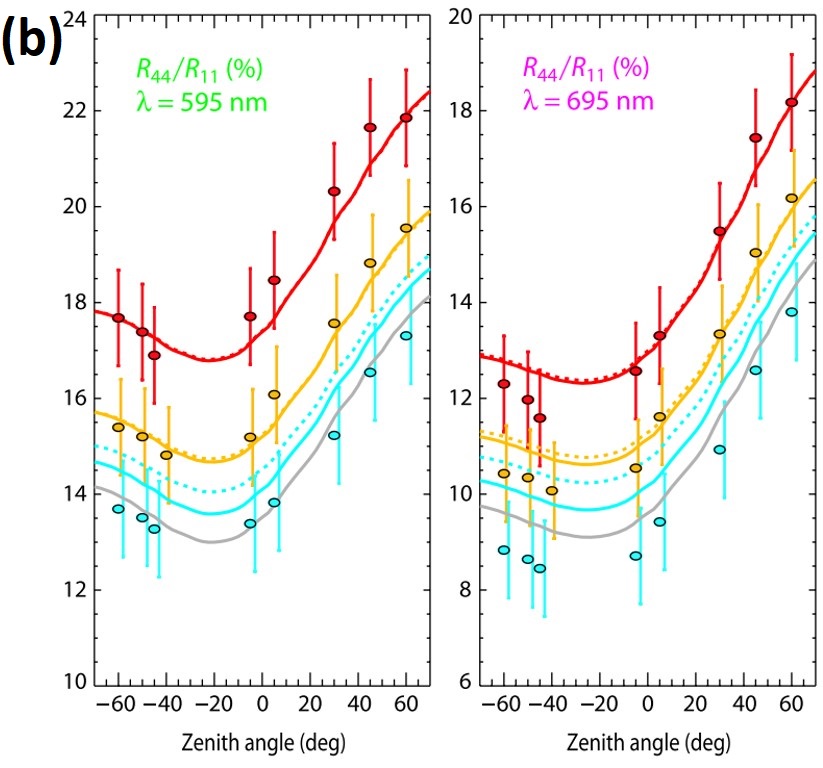}
	}\\
	\subfloat{
	\label{baillisJOSAA2004config}
	\includegraphics[width=0.8\linewidth]{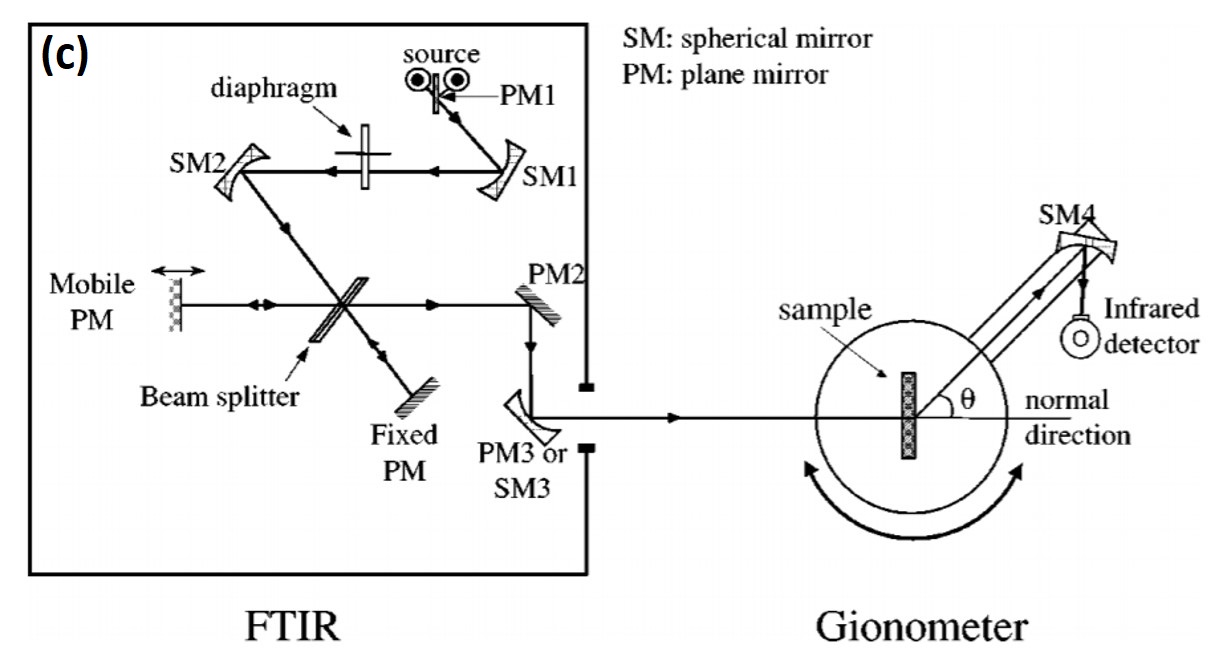}
}
\caption{Measurement of angular-resolved reflectance and transmittance. \textbf{(a)} Comparison between the experimental data and best-fitted results (retried radiative properties not shown here) for the angle-resolved reflectance and transmittance, denoted by the bidirectional transmittance and reflectance distribution functions (BRTF and BRDF). Reprinted from Ref. \cite{milandriJQSRT2002}, Copyright (2002), with permission from Elsevier. \textbf{(b)} Comparison of experimentally measured reflection Mueller matrix element ratios (symbols, "Laboratory data") with the predictions of RTE using the radiative properties given by ISA (dashed lines, "RTE") and ITA (solid lines, "RTE+SSF"). Different colors indicate different volume fractions of particles including $f_v=2\%$ (red color), 5\% (yellow color) and 10\% (blue color). Reprinted with permission from Ref. \cite{mishchenkoOL2013}. Copyright 2013 Optical Society of America. (c) Schematic of the experimental apparatus used to measure the angle-resolved spectral transmittance and reflectance. Reprinted with permission from Ref. \cite{baillisJOSAA2004}. Copyright 2004 Optical Society of America.}\label{experiment3}	
\end{figure}

The measurement of angle-resolved reflectance and transmittance is also a popular method to determine the radiative properties, which can obtain more optical information from a single sample, although it requires more sensitive detectors or stronger radiation sources than the total reflectance and transmittance measurement. There are a variety of different designs for the experimental setup, and for a review, see Ref. \cite{germer2014angleresolved}. Based on the angle-resolved reflectance and transmittance measurement for a single sample and appropriate inverse RTE models, the radiative properties can be retrieved. A typical example is given in Fig.\ref{milandriJQSRT2002} taken from Ref. \cite{milandriJQSRT2002}, in which the discrete ordinates method to solve the RTE combined with a proper assumption of the form of the scattering phase function can result in a good fitting to the experimental data of angle-resolved transmittance of porous silica fibers (the wavelength of radiation is $4\mathrm{\mu m}$).

Figure \ref{baillisJOSAA2004config} shows a typical experimental setup for measuring the spectral angle-resolved transmittance and reflectance, used in Refs.\cite{baillisJOSAA2004,randrianalisoaJTHT2006} to retrieve the infrared radiative properties of porous fused quartz containing randomly distributed bubbles. In this setup, a radiation source is generated from a Fourier transform infrared (FTIR) spectrometer operating in the spectral range from 1.5 to 25 $\mathrm{\mu m}$. Two spherical mirrors (SM1 and SM2) and a diaphragm in the FTIR are combined to achieve a reduced beam diameter and thus good resolution. The mobile plane mirror (PM) and fixed PM as well as a beam splitter are standard interferometer configuration in the FTIR components. The output beam is then focused to the sample. The detection system, which consists of a spherical mirror to collect the transmitted radiation and focus it to the liquid-nitrogen-cooled mercury cadmium telluride (MCT) detector, is mounted on a rotating arm of the goniometer. This configuration enabled the angle-resolved measurement. In this setup, the incident radiation is modulated and the detection is synchronized using a lock-in amplifier (not shown in the figure) in order to rule out the effects of thermal radiation from the environment.

By adding appropriate polarizers, the angle-resolved reflectance and transmittance measurement can be further extended to the measurement of the Mueller matrix of a disordered medium, which is a $4\times4$ matrix $\mathbf{M}(\theta,\theta_0)$ that specifies the transformation of the Stokes parameters of the incident beam into those of the scattered light, with $\theta_0$ and $\theta$ denoting the incident and scattering polar angles (here we simply assume a symmetry in the azimuthal angles) \cite{mishchenko2011polarimetric}. The first term $M_{11}(\theta,\theta_0)$ is then the normalized angle-resolved scattered intensity that is closely related to the angle-resolved reflectance and transmittance, and other matrix elements describe how the polarization of incident beam is transformed by the disordered media. For example, Mishchenko \textit{et al.} \cite{mishchenkoOL2013} presented a well-controlled laboratory experiment on the reflection Mueller matrix (backscattering component of the full Muller matrix) $\mathbf{R}(\theta)$ of PS suspensions in water with different volume fractions ranging from 2\% to 10\%, where the incident angle is fixed as $\theta_0=20^{\circ}$.  They indicated that the ISA combined with the vector RTE can be applied safely to packing densities up to $\sim$2\%, while at higher densities (especially larger than 5\%), the effect of dependent scattering should be taken into account with caution. They further demonstrated that by means of the ITA (namely, based on the so-called static structure factor, SSF) as a simple consideration of the DSE, a better agreement with the experimental data can be achieved. Typical results for the reflection Mueller matrix elements $R_{11}(\theta)$ and $R_{44}(\theta)$ are given in Fig.\ref{mishchenkoOL2013}. Recently, Riviere and coworkers \cite{riviereJQSRT2013,ceolato2018spectropolarimetric} have developed a spectro-polarimetric scatterometer that can realize the fast measurement of the spectral and angle-resolved Mueller matrix. It mainly consists of a supercontinuum laser, wide-band polarizers, a charged-coupled device (CCD) camera/InGaAs-sensor based spectrophotometer and a goniometric platform, and can cover the spectral range of 480 to 2500 nm with an angle resolution of $1\degree$. Details can be found in Ref. \cite{ceolato2018spectropolarimetric}.

\begin{figure}[htbp]
	\centering
	\subfloat{
		\label{fiebig}
		\includegraphics[width=0.46\linewidth]{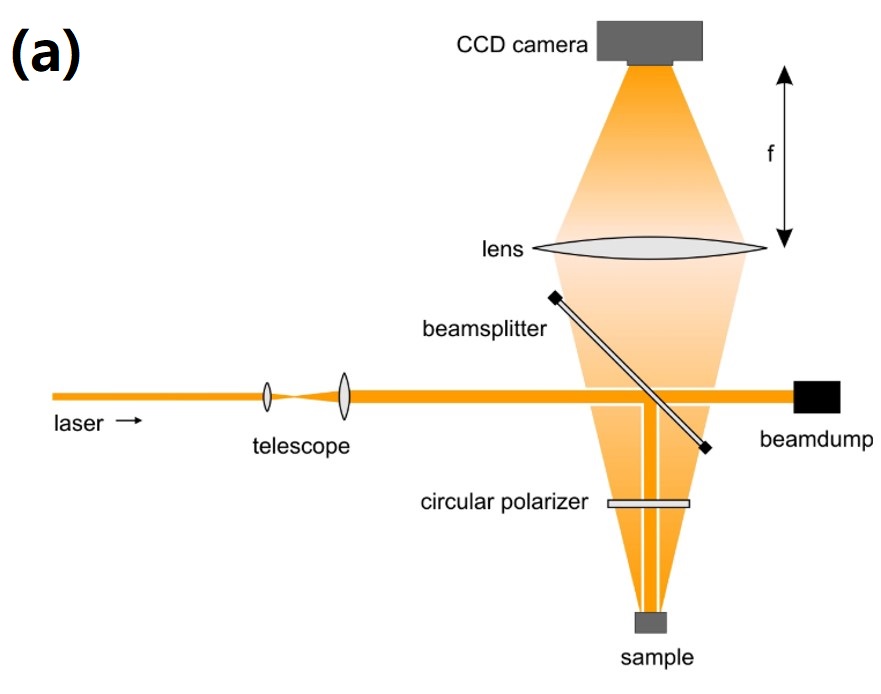}
	}
	\subfloat{
		\label{naraghiPRL2015ltr}
		\includegraphics[width=0.46\linewidth]{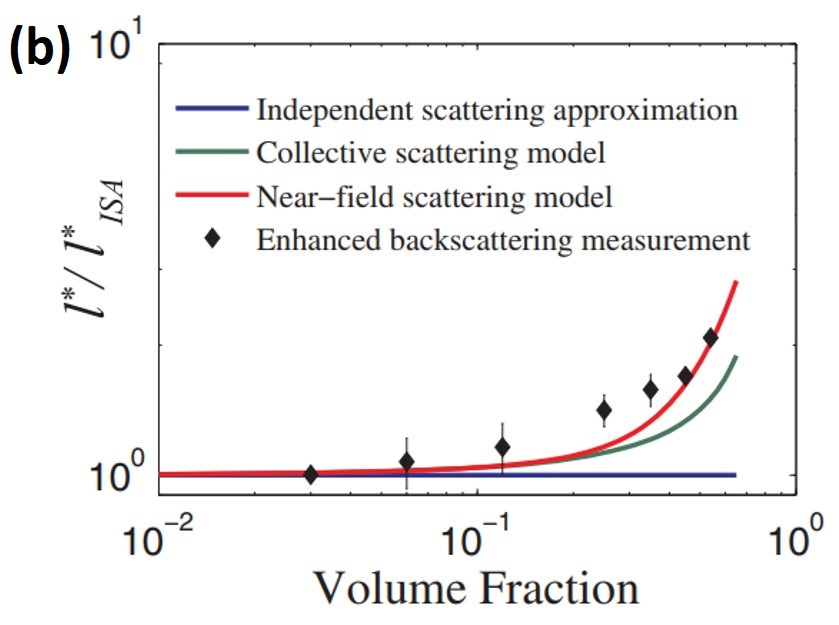}
	}
	\caption{Radiative properties measured from the coherent backscattering cone. \textbf{(a)} Typical experimental setup to measure the coherent backscattering cone. Reprinted with permission from Ref. \cite{fiebig2010coherent}. Copyright (2010) The Author(s). \textbf{(b)} Transport mean free path extracted from the experimentally measured coherent backscattering cone, compared with predictions of different transport models. The inset shows the fitting performance of Eq.(\ref{cb_lineshape}) in the diffusive regime. Reprinted with permission from Ref. \cite{Naraghi2015} Copyright (2015) by the American Physical Society. }\label{experiment4}	
\end{figure} 

%(Mention the measurement of CB cone requires an exact determination of the exact backscattering direction, and previous methods do not apply.) 

In the diffusive regime, measuring angle-resolved reflectance constitutes a useful scheme to determine the transport mean free path. This method utilizes the lineshape (angular profile) of the coherent backscattering (CB) cone (the physics of which is briefly introduced in Section \ref{CB}), which is analytically formulated as \cite{akkermansPRL1986,Aegerter2008,Naraghi2015}
%\begin{equation}
%\alpha ( \theta ) = \frac { 3 } { 8 \pi } \left[ 1 + \frac { 2 z _ { 0 } } { l_\mathrm{tr} } + \frac { 1 } { \left( 1 + q _ { \perp } l_\mathrm{tr} \right) ^ { 2 } } \left( 1 + \frac { 1 - \exp \left( - 2 q _ { \perp } z _ { 0 } \right) } { q _ { \perp } l_\mathrm{tr} } \right) \right]
%\end{equation}

\begin{equation}\label{cb_lineshape}
I (q) = \frac { 3/7 } { (1+ql_\mathrm{tr})^2 } \left[ 1 + \frac { 1 - \exp \left( - 4/3ql_\mathrm{tr} \right) } {ql_\mathrm{tr} }\right],
\end{equation}
where $q=2\pi\sin\theta_b/\lambda_0$ with $\theta_b$ the backscattering polar angle. To correctly identify the CB cone, the determination of the scattered intensity in the exact backscattering direction ($\theta_b=0\degree$ or equivalently, $\theta_s=180\degree$) is important, which brings difficulties for the experimental setup presented in Fig.\ref{baillisJOSAA2004config} \cite{baillisJOSAA2004}. To this end, an additional beam splitter is usually necessary to separate the incident and exactly backscattered beams. Figure.\ref{fiebig} shows a typical small-angle setup to measure scattered intensity profile at and near the exact backscattering direction, taken from Ref. \cite{fiebig2010coherent}. In this small angle setup a high-resolution, thermoelectrically cooled (to reduce electronic noise) monochromatic CCD camera was placed on the backside of the sample. Since the direct backscattering direction is blocked by the camera, a beam splitter is adopted to direct the laser beam onto the sample, while transmitted beam from the beam splitter is absorbed completely by a beamdump, in order to reduce the backreflection that could significantly disturb the image on the camera. A circular polarizer is implemented to block the single-scattering intensity \cite{aubryPRA2017}. Another setup is the one using 256 photosensitive diodes attached to an arc to realize an angle-resolved reflectance measurement, which can work in a wide angle range (almost covering the entire range of backscattering angles except for the very tip of the CB cone at $\theta_b=0$). Although the angle-resolution of this setup is lower than that of the small-angle setup near the cone, it is sufficient for highly scattering materials like $\mathrm{TiO}_2$ powders with a broad CB cone. More details of this approach can be found in Refs. \cite{grossRSI2007,fiebig2010coherent}.

Based on the above method, by fitting the experimentally measured backscattering cone with Eq.(\ref{cb_lineshape}), the transport mean free path can be obtained with a good agreement. In Fig.\ref{naraghiPRL2015ltr}, the transport mean free path of a dense media containing TiO$_2$ nanoparticles measured from the lineshape of CB cone as a function of particle volume fraction is shown, done by Naraghi and Dogariu \cite{Naraghi2015}. It was found that the ISA and the ITA (``collective scattering model") fail to predict the transport mean free path while the near-field scattering model (Eq.(\ref{nfscatteringmodel})) can give rise to a much better agreement.

\subsection{Measurement of time-resolved reflectance and transmittance}\label{time_resolved}

%{\color{red}The backscattering method does not probe beyond a few mean-free-path lengths into the medium, nor does it provide a quantitative means for assessing the role of absorptive processes.
%Our experimental method permits probing of random medium on length scales from a few to a few million mean-free-path lengths. It provides a quantitative means for separation of elastic scattering and absorptive processes. It has the ability to distinguish between the systems which exhibit simple diffusive behavior for photons and those which do not. A method to detect departure from simple diffusive photon transport is important. (quote from Ref.\cite{watsonPRL1987})

%The first sign of a deviation from classical diffusion is expected to occur for long photon paths inside a sample. As already mentioned those paths correspond to very small angles in the coherent backscattering cone making it very hard to unambiguously identify any effect on the shape of the cone by those paths. To be able to measure a path length distribution on our multiple scattering samples we use a single photon counting method \cite{watsonPRL1987,drakePRL1989}. (quote from Storzer PhD thesis)
%}

Above experimental methods all deal with the static radiative properties of the disordered media. On the other hand, it is also important to employ time-resolved measurements to probe the radiative transport dynamics occurring inside the disordered media, thanks to the rapid developments of ultrafast optical technologies \cite{weiner2011ultrafast}. Because of the strong multiple scattering of waves, a short pulse propagation in disordered media is strongly delayed, and the dynamics crucially reflects the microscopic structures and optical properties \cite{watsonPRL1987,vreekerPLA1988,drakePRL1989,genackEPL1990,yooPRL1990,kopRSI1995,kopPRL1997,chabanovPRL2003,johnsonPRE2003,bestemyanovQE2004,cheikhOL2006,aegerterEPL2006,aegerterJOSAA2007,contiNaturephys2008,boryckiOptica2016,sperlingNJP2016,cobusPRL2016,lyonsNaturephoton2019}. This method is, to some extent, most information-rich and helpful to decouple different radiative properties by fully identify the time-dependent transport process.

By now many time-resolved measurements have been carried out in the diffusive regime due to the relative simplicity and high efficiency in interpreting experimental data \cite{johnsonPRE2003,duran-ledezmaAO2018,lyonsNaturephoton2019}  (see Eq.(\ref{time_trans})) and the possibility to search anomalous transport regimes of fundamental interest like Anderson localization by identifying the deviation from the diffusion equation  \cite{watsonPRL1987,drakePRL1989,chabanovPRL2003,aegerterEPL2006,aegerterJOSAA2007,storzerPRL2006,contiNaturephys2008,sperlingNJP2016,cobusPRL2016}. There are also a great deal of works in the RTE regime (or called ``subdiffusive regime")  \cite{guoJQSRT2002,wanJQSRT2004,calbaJOSAA2008,guoHTR2013}. Here we mainly discuss the time-resolved measurement in the diffusive regime for thick enough samples $L\gg l_s$. In this regime, the dynamic diffusion coefficient $D_\mathrm{dyn}$ can be obtained from the dynamics which is described by the time-domain diffusion equation as \cite{continiAO1997}
\begin{equation}\label{time_diffusion_eq}
\Big(\frac{\partial}{\partial t}-\nabla \cdot  D_\mathrm{dyn}\nabla+v_E\kappa_a\Big)\phi(\mathbf{r},t)=Q(\mathbf{r},t),
\end{equation}
where $\phi(\mathbf{r},t)=\frac{1}{4\pi}\int_{4\pi} I(\mathbf{r},\hat{\mathbf{s}},t)d\Omega$ is the average diffuse intensity, $Q(\mathbf{r},t)$ is the source function and $v_E$ is the energy transport velocity, as mentioned above. Note this equation is most suitable for long-time dynamics when the diffusive transport regime is established in time domain. It is shown by Greffet and coworkers \cite{elaloufiJOSAA2003,pierratJOSAA2006} that the dynamic diffusion coefficient is independent of the absorption coefficient as
\begin{equation}
D_\mathrm{dyn}=\frac{1}{3}v_\mathrm{E}l_\mathrm{tr},
\end{equation}
different from the static diffusion coefficient that should incorporate the absorption coefficient. By using the extrapolation boundary condition, the solution of this equation can be obtained and the resultant time-resolved total transmittance $T(t)$ for a slab geometry with thickness $L$ is given by \cite{pattersonAO1989,continiAO1997,garciaPRA2008}
\begin{equation}\label{time_trans}
T(t)=\frac{\exp \left(-\kappa_{a} v_E t\right)}{2(4 \pi D_\mathrm{dyn})^{1 / 2} t^{3 / 2}} \sum_{m=-\infty}^{+\infty}\left[z_{1, m} \exp \left(-\frac{z_{1, m}^{2}}{4  D_\mathrm{dyn} t}\right)-z_{2, m} \exp \left(-\frac{z_{2, m}^{2}}{4  D_\mathrm{dyn} t}\right)\right],
\end{equation}
where $z_{1,m}=L(1-2m)-4mz_e-z_0$ and $z_{2,m}=L(1-2m)-(4m-2)z_e+z_0$ for an integer $m$, $z_e$ is the extrapolation length given in Eq.(\ref{extrap_length}) and $z_0$ is the location of the imaginary sources into the medium, approximated by $z_0=l_\mathrm{tr}$. In fact, at long times, the first term in the summation dominates, such that an exponential decay with time is expected in the diffusive regime. On the other hand, in anomalous transport regimes like the onset of Anderson localization, the diffusion coefficient, in a renormalized spirit, is scale dependent and therefore temporally varying as $D_\mathrm{dyn}(t)$, leading to a non-exponential decay even at long times \cite{cheungPRL2004}.

\begin{figure}[htbp]
	\centering
	\subfloat{
		\label{storzerPRL2006}
		\includegraphics[width=0.48\linewidth]{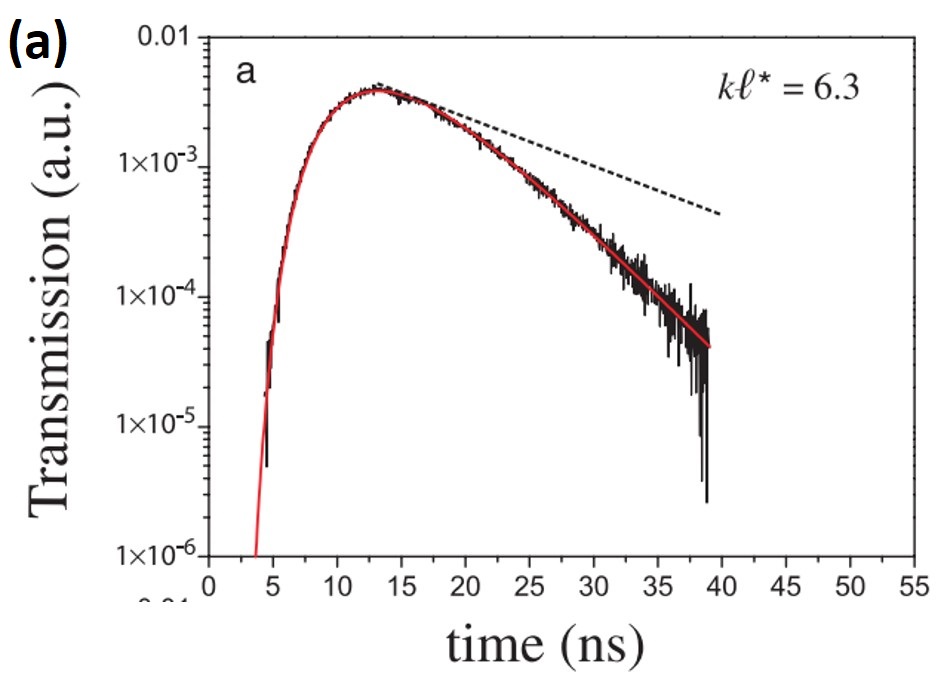}
	}
	\subfloat{
		\label{storzerPRE2006}
		\includegraphics[width=0.44\linewidth]{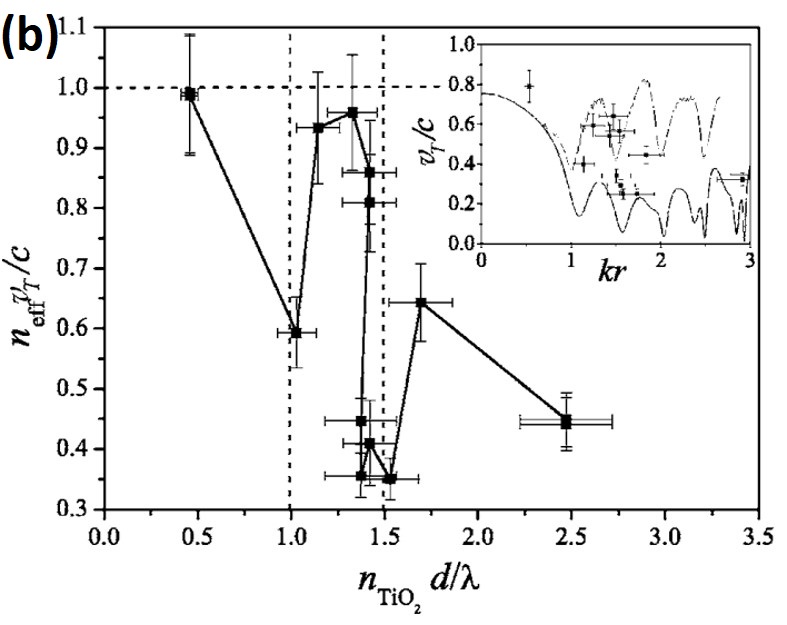}
	}\\
	\subfloat{
	\label{storzer2006config}
	\includegraphics[width=0.5\linewidth]{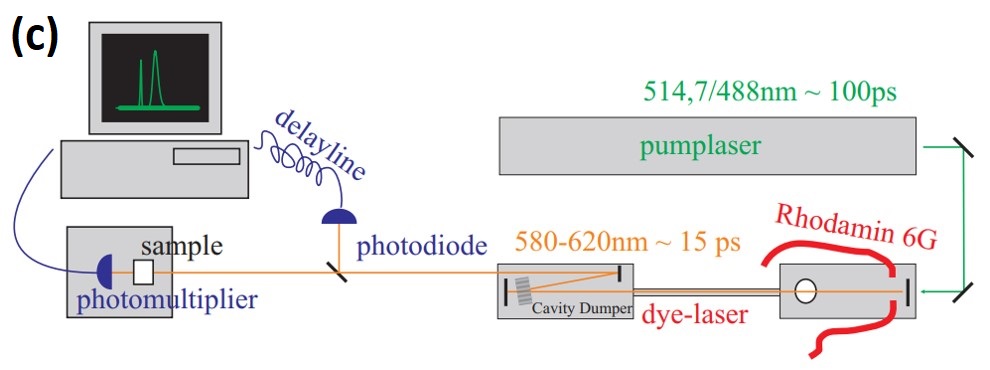}
}
	\subfloat{
	\label{sperling2016config}
	\includegraphics[width=0.46\linewidth]{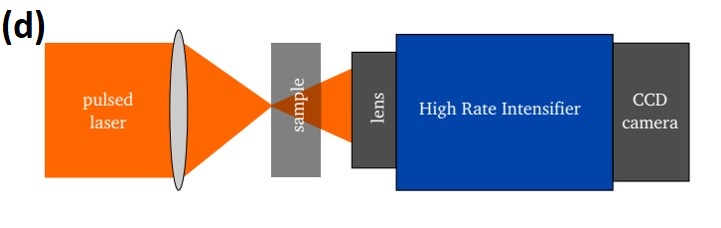}
}
	\caption{Time-resolved measurement. \textbf{(a)} Time-resolved transmittance (or time-of-flight distribution) for a $kl_\mathrm{tr}=6.3$ sample composed of densely packed TiO$_2$ powders. Experimental data (black solid line) can be well-fitted by the time-resolved solution of the diffusion equation (red solid line). Reprinted with permission from Ref. \cite{storzerPRL2006}. Copyright (2006) by the	American Physical Society. \textbf{(b)} Transport velocity extracted from a combined measurement of time-resolved transmittance and CB cone, compared with predictions of the ISA (inset) and the ECPA model (solid line). Reprinted with permission from Ref. \cite{storzerPRE2006}. Copyright (2006) by the	American Physical Society. \textbf{(c)} Experimental setup for the time-solved transmittance in \textbf{(a-b)}. Reprinted with permission from Ref. \cite{storzer2006anderson}. Copyright (2006) The Author(s). \textbf{(d)} Experimental setup for measuring the time-resolved spatial transmission profile. Reprinted from Ref. \cite{sperling2015experimental}. }\label{experiment5}	
\end{figure} 

In general, by fitting Eq.(\ref{time_trans}) to the experimentally measured time-resolved transmittance, it is possible to retrieve the dynamic diffusion coefficient and absorption coefficient simultaneously. For instance, St\"orzer \textit{et al.} \cite{storzerPRL2006} measured the diffusion coefficient and absorption coefficient of densely packed TiO$_2$ powders, and the fitting between the theoretical formula and experiment data was quite good, as shown in Fig.\ref{storzerPRL2006}. Furthermore, a combination of time-resolved transmittance measurement and the CB cone measurement enables us to obtain the transport velocity $v_E$ in the disordered media \cite{storzerPRE2006,sapienzaPRL2007}. The transport velocity is an important indicator for resonant multiple scattering of waves, which usually takes place in high-refractive indexmay strongly reduce the transport velocity. The measured result of St\"orzer \textit{et al.}'s \cite{storzerPRE2006} is given in Fig.\ref{storzerPRE2006}, compared with the predictions of the ISA (inset) and the ECPA model (solid line). It is found that the ECPA approach can lead to a better agreement with the experimental data due to the consideration of the DSE (For details of the calculation of transport velocity in disordered media, see Refs.\cite{vanalbabaPRL1991,buschPRB1996,lagendijk1996resonant}). Similarly, Sapienza \textit{et al.} \cite{sapienzaPRL2007} prepared highly monodisperse ($\sim$2\% diameter dispersion) dielectric spheres supporting Mie resonances packed at high volume fractions and carried out time-resolved and static transmittance measurements, which further confirmed the resonant multiple scattering of waves can induce a significant reduction of energy transport velocity.   

Here we proceed to a brief discussion on typical experimental setups for the time-resolved transmittance of DDM in the diffusive regime. Figure \ref{storzer2006config} shows the experimental setup used by St\"orzer \textit{et al.} \cite{storzer2006anderson}, which is based on the measurement of time-of-flight distribution of transmitted single photons \cite{watsonPRL1987,vreekerPLA1988,drakePRL1989}. It consists of a picosecond laser system and a single photon detection system. In this setup, the picosecond pulses are generated by a Rhodamin 6G dye laser pumped by an Ar$^+$ laser with a mode locker crystal. The laser system can produce pulses with a width of about 20 ps with a repetition rate of up to 10MHz in the wavelength range of 580 to 620 nm. Behind the laser system, a beam splitter divides the incident pulse into the sample path and the reference path. A photomultiplier, which is capable of detecting single photons, is put behind the sample. To ensure that only one photon is detected during each measurement cycle to avoid the saturation effect \cite{torricelliPRL2005}, the sample should be thick enough and the counting rate of the photomultiplier should not be too high (which also should be large enough, typically lower than 1/10 of the repetition rate of the pulsed laser, to guarantee the signal to noise ratio). Using the signal of a counting event, the computer starts a time measurement, which is subsequently stopped by the reference signal of the photodiode. A delay line is put after the photodiode in the reference path such that each single photon measurement is started and stopped by the same pulse. Moreover, since the incident pulse is not a perfect delta pulse, one also needs to measure a reference time-of-flight signal without the sample. By deconvoluting the measurement result for the sample with respect to the reference pulse, the final time-resolved transmittance can be acquired. Recently this experimental setup has been further improved by Sperling \textit{et al.} \cite{sperlingNaturephoton2013,sperling2015experimental,sperlingNJP2016}, using a femtosecond laser system with a pulse width about 250 fs and a repetition rate of 75MHz, which has increased power and can achieve better time resolution. And it can be tuned in wavelength range of 700-1000 nm, with a frequency-doubled optical parametric oscillator (OPO) to further cover the wavelength range of $\sim$550-650 nm \cite{sperlingNaturephoton2013,sperling2015experimental,sperlingNJP2016}. Since the time-of-flight distribution is equivalent to the path length distribution of photon paths in the sample, this experimental approach also provides a route to study the long multiple scattering trajectories that are hardly accessible in static measurements \cite{aegerterEPL2006}, which are particularly important to the lineshape of the CB cone near the cusp. 

%{\color{red}According to the description of Aegerter et al \cite{aegerterEPL2006}:
%The time-resolved transmission is measured using a single photon counting method \cite{watsonPRL1987,drakePRL1989}. Here a pulsed dye-laser with a pulse width of $\sim$20 ps and a beam waist below 1 mm is used to illuminate the sample. Then, for each pulse the flight time of a single photon passing through the sample is measured. A histogram of many such pulses leads to a time-of-flight
%distribution, which is the also the path length distribution of photon paths inside the sample. Due to the presence of after-pulses and the finite width of the pulse, we have to correct the time-of-flight measurements by a deconvolution of the data in Fourier space with a background measurement performed in the absence of a sample. After such a deconvolution, the data can
%be compared directly to the analytic theory of diffusion of a delta pulse through a slab of length $L$. At long times, the first term in the sum dominates, such that an exponential decay of the transmission with time is expected. For a localizing sample,
%where $D$ is scale dependent, this is no longer true and the time dependence in $D(t)$ leads to a non-exponential decay. 
%}
Another approach of the time-resolved measurement is to directly record the time-dependent spatial profile of transmission, i.e., the positions of transmitted photons at certain time points, whose advantage over the time-of-flight distribution measurement is the independence with respect to absorption. A typical experimental setup by Sperling \textit{et al.} is given in Fig.\ref{sperling2016config} \cite{sperling2015experimental}. The pulse laser is focused by a lens onto the sample surface, and the transmitted beam is magnified by another lens and imaged onto the time-gated intensified CCD camera (ICCD), which consists of a high rate image intensifier, a monochromatic CCD camera and an ultrafast gate \cite{sperling2015experimental}. Similar experimental setups are also used in Refs.\cite{sapienzaPRL2007,garciaPRA2008}, in which the time-resolved profile of the transmitted photons is recorded by a streak camera to achieve a high temporal resolution, while the pulses are provided by a Ti:Sapphire laser (2 ps pulse duration), tunable within 700–920 nm. Since in these two setups the single-photon timing method is the time-correlated single-photon counting (TCSPC) technique \cite{selbJBO2006,continiOE2006}, the dynamic range (the ratio between maximum and minimum detectable signal intensities) of the time-gated ICCD and streak cameras suffers from the noise and saturation effects brought by the large numbers of early-photons \cite{pifferiJBO2016,alayedSensors2017}. They are also expensive, complex and bulky. The same problem is also encountered by the above time-of-light distribution measurement. In recent years, it is shown that solid-state single-photon avalanche diodes (SPADs) can be used as an alternative for building fast time-gated detectors. SPADs exhibit ultrafast transition time ($\sim$200 ps) of turning the gate on and off and thus can record late photons without being saturated by early photons, which permit to detect long-lived photons with improved signal quality \cite{alayedSensors2017,tosiOE2011}. A typical experimental measurement of time-resolved transmission profile using SPAD camera has been presented by Lyons \textit{et al} for detecting an object in a scattering medium \cite{lyonsNaturephoton2019}.

%Recently, novel techniques, like the combination of wide-field spatial imaging and ultrafast imaging (sub-picosecond) based on an all-optical gating technique because traditional electronic or pump-probe gating techniques are limited by low-time resolutions or a long acquisition time \cite{pattelliLight2016}. 
%{\color{red}They said: ``This experimental setup is based on the cross-correlation gating technique. Two synchronous, collinear probe and gate pulses at different wavelengths impinge on a 2-mm-thick $\beta$-barium borate (BBO) crystal. When the two beams overlap both spatially and temporally, a sum-frequency signal is generated with intensity proportional to the cross-correlation of the original pulses as a function of the delay-line position. This technique has been used to fully characterize ultrafast pulse propagation, generally disregarding the spatial distribution of the converted signal. To study the dynamics of light emerging from a complex specimen, we image its exit surface on the BBO crystal. There, the original image at the probe wavelength is upconverted to a different (signal) wavelength by sum-frequency generation with a gate pulse. Nonlinear upconversion is a polarization- and wavevector-sensitive process, and both of these aspects must be taken into account when designing an imaging apparatus."}
%
%This technique further provide time-resolved transmission profile with higher spatial-temporal resolutions and improved acquisition times. 

Remarkably, the path-length distribution of photons in DDM can be measured through the low-coherence interferometric methods without the need of ultrafast laser and detectors, which is somewhat similar to the time-of-flight distribution of photons. Therefore, this class of method also allows to obtain the time-resolved photon dynamics in DDM \cite{popescuOL1999,popescuPRE2000,tualleOC2001,weissPRE2013}. Recently, this method is further put forward to extract the time-dependent Green's function and thus achieve spatio-temporal imaging using only low-coherence white light illumination, from which the local optical/radiative properties of DDM can be determined in a high spatial resolution that enables to identify microscale heterogeneities \cite{badonPRL2015,badonOptica2016}. 

%For this point, Kim et al \cite{kim2013fourier} said:``\textit{In essence, OCT can provide the path length distribution of light scattered by a diffusive tissue, with a time resolution given by the coherence length divided by the speed of light, which can easily reach in tens of femtoseconds. Using OCT for time-resolved diffusion measurements is superior to using modelocked Ti:Saph lasers in the point of view of not only time resolution, but also sensitivity and dynamic range (due to the interferometric, heterodyne detection).}"

%The combination of the time-resolved response measurement and the measurement of the CB cone \cite{yooAO1989,yooPRA1989,cobusPRL2016} is useful for the further identification of the contribution of long-multiple-scattering trajectories.
%
%Time-resolved interferometric method \cite{kopPRL1997,kopRSI1995}
%
%Phase conjugation of a light field arising from enhanced backscattering in a multiple scattering medium \cite{reilPRL2005}
%
%Time-resolved measurement of optical phase-space distributions as a new probe for investigating the propagation of light in disordered media \cite{waxPRL2000}.
%
%Time-resolved anisotropic multiple scattering by Wiersma et al \cite{wiersmaPRL1999}

\begin{figure}[htbp]
	\flushleft
	\subfloat{
		\label{sankarana}
		\includegraphics[width=0.46\linewidth]{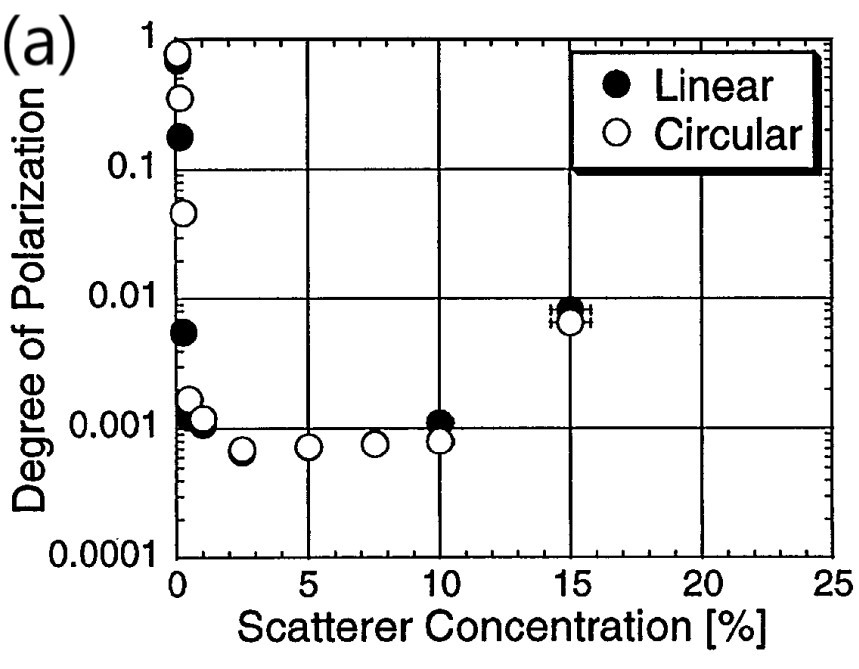}
	}
	\subfloat{
		\label{sankaranb}
		\includegraphics[width=0.46\linewidth]{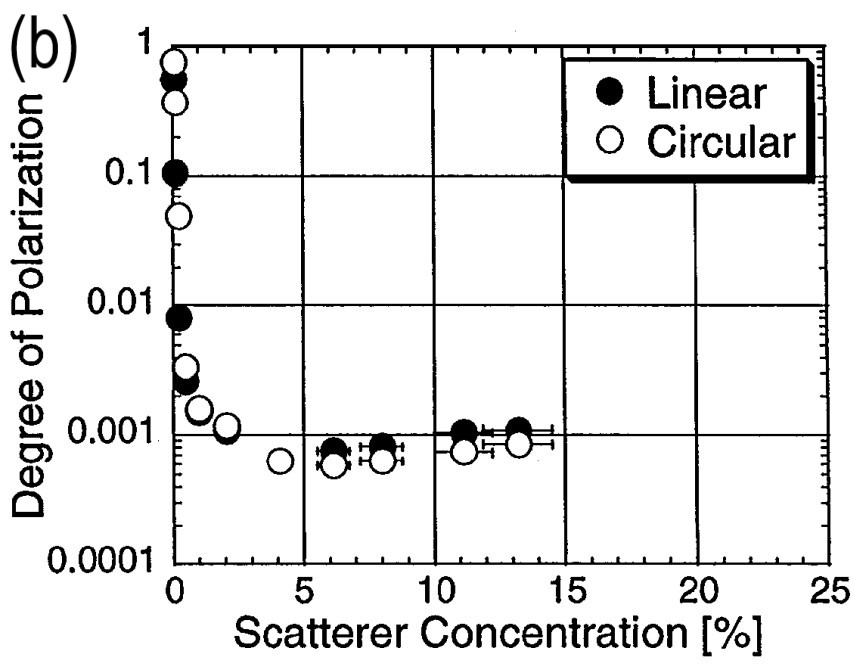}
	}
	\caption{Variation of degrees of linear and circular polarization in PS sphere suspensions with sphere diameters of \textbf{(a)} 0.48 $\mathrm{\mu m}$ and \textbf{(b)} 0.99 $\mathrm{\mu m}$ with the scatterer concentration. Reprinted from Ref. \cite{sankaranOL2000}. Copyright 2000 Optical Society of America.}\label{sankaran}
	
\end{figure}

In summary, we have reviewed various experimental methods to measure/retrieve the radiative properties of DDM and described representative experimental works on the DSE. These experimental works imply that the DSE is indeed a critical factor that affects the radiative properties of DDM containing a substantial volume fraction of scatterers, and should be quantitatively taken into account carefully in practical applications. However, almost all experimental works about DSE are conducted in an indirect manner, which are based on the measurement of macroscopic reflectance and transmittance spectra and a subsequent comparison between experimental data and ISA predictions. In other words, there are no experimental works that can directly and quantitatively determine the DSE without the comparison with ISA predictions, although the transition from independent scattering to dependent scattering is not sharp but continuous. As a consequence, it is necessary to develop a specific method based on experimental observables to predict the DSE straightforwardly. A possible idea is to further exploit the polarization properties of radiation, as implied by a seminal paper by Sankaran \textit{et al.} \cite{sankaranOL2000}, where it was shown that the variation of the degrees of linear and circular polarization of radiation with the scatterer (PS sphere) density undergo a prominent transition at the specific scatterer volume concentration when the DSE becomes to play a role, shown in Fig.\ref{sankaran}.

It should be noted that a number of experimental approaches which are also frequently used to characterize the optical/radiative properties of disordered media but not extensively employed in the study of the DSE, are not introduced. These methods include the measurement of spatially resolved intensity \cite{groenhuisAO1983a,groenhuisAO1983b,farrellMP1992,durianJOSAA1999,zaccantiAO2003}, the optical coherence tomography (OCT) \cite{kim2013fourier,nguyenOE2013,almasianJBO2015}, the photon density wave spectroscopy (PDWS) \cite{fishkinPRE1996,sunRSI2002} (also known as the temporal frequency-domain photon-migration spectroscopy, FDPM \cite{osullivanJBO2012}) and the spatial frequency domain imaging (SFDI) method \cite{cucciaOL2005,cucciaJBO2009,osullivanJBO2012,kanickBOE2014,giouxJBO2019}, etc. Notably, Nguyen \textit{et al.} \cite{nguyenOE2013} applied the transmission OCT to measure the extinction coefficient and the backscattering OCT to identify the scattering coefficient of silica particle suspensions, where they found the DSE plays a role and used the ITA method to model the scattering coefficient. Moreover, since we only deal with static multiple wave scattering problem as mentioned in Section \ref{theory}, we do not review the rapid advances in dynamic light scattering techniques \cite{berne2000dynamic}, including the photon correlation spectroscopy (PCS) \cite{gulariJCP1979,puseyJCP1984}, spatially-resolved or time-resolved diffusing wave spectroscopy (DWS) \cite{pinePRL1988,maretZPB1987,goldburgAJP1999,viasnoffRSI2002,morinAO2002,cheikhOL2006} and the laser Doppler methods (LDM) \cite{yehAPL1964,kienleAO1996}, etc., which can also provide a diagnosis of the radiative/optical properties with high sensitivity \cite{savoScience2017}. It is worth mentioning that the analytical frameworks of these techniques usually neglect the dependent scattering effect in dense DDM, although recently a substantial attention has been paid to this aspect \cite{bresselJSQRT2013,hassAO2013,kimPNAS2019}, mainly using the ITA model (See Section \ref{ITA_model} for the ITA model).

%the power spectrum of low-frequency fluctuations in the transmitted intensity of light \cite{rimbergPRB1988}
%dynamic intensity-intensity correlations \cite{freundPRL1988}

%{\color{red}
%
%The ergodic hypothesis was introduced by James Clerk Maxwell (1831–79) and Ludwig Boltzmann (1844–1906) as a basic underlying principle of statistical mechanics. The details of the ergodic theory, its relation to the famous Poincaré recurrence theorem (Poincaré, 1890), and its applications to statistical mechanics and kinetic theory are described by Khinchin (1949), Uhlenbeck and Ford (1963), and Farquhar (1964). Interesting discussions of the ergodic hypothesis and specific examples of nonergodic scattering media can be found in Pusey and van Megen (1989), Joosten et al. (1990), Xue et al. (1992), Nisato et al. (2000), and Scheffold et al. (2001). (quote from Mishchenko \cite{mishchenko2006multiple}: )
%}

%{\color{red}Note there are also experimental works on exotic mesoscopic interference phenomena in some highly scattering DDM such as Anderson localization \cite{wiersma1997localization, storzerPRL2006}, position-dependent diffusion coefficient \cite{yamilovPRL2014}, sub-diffusive \cite{sebbahPRB1993} and super-diffusive \cite{barthelemyNature2008,bertolottiPRL2010} transport behaviors, recurrent scattering \cite{Aubry2014PRL} and statistics and correlations of intensity \cite{Chabanov2000}, which will not be discussed in detail here.
%}

\section{Dependent scattering and other related interference phenomena in mesoscopic and atomic physics}\label{mesointerference}

The study of radiative properties of disordered media is closely related to the field of mesoscopic physics, which  also investigates the physics of propagation, scattering and interference of quantum and classical waves in disordered materials. In this section, we will introduce some important phenomena that all arise from wave interferences in disordered materials, including the coherent backscattering cone, the Anderson localization of light and the statistics as well as correlations induced by disorder. In fact, due to their close relation to the DSE, we have unavoidably mentioned these phenomena in previous sections. The study of these phenomena can provide a different viewpoint that focuses on the universal behavior of disordered media instead of the microscopic details of interparticle electromagnetic interactions, and offer new methods to obtain more information about the radiative properties, as we have already seen in Section \ref{expsec}. We expect the brief discussion on this discipline can be helpful for the study of the DSE in DDM.

On the other hand, in atomic physics, understanding light propagation and scattering in disordered cold atomic clouds is crucial with important applications in quantum information science. Since cold atoms are extremely scattering near resonance, researchers tried to study the resonant multiple and dependent scattering phenomena of radiation in them, due to the advantages of cold atomic systems over conventional micro/nanoscale scattering media including well-controlled systems and widely tunable parameters \cite{baudouin2014cold,haveyPhysrep2017}. Remarkably, for dilute cold atomic clouds, radiative transfer equation is also widely applied to the description of multiple scattering of photons \cite{baudouin2014cold,sokolovJOSAB2019}. Moreover, the prominent collective effects like subradiance and superradiance due to the multiple wave scattering and interferences are also of fundamental importance \cite{guerin2017light}. In this section, we attempt to give a basic description of light-atom interactions and introduce two remarkable interference phenomena, including the breakdown of the mean-field optics and the significant role of structural correlation in atomic gases, in which multiple wave scattering plays a crucial role. Due to the strong light-matter interaction and high-accuracy experiments, these phenomena are much easier to observe and control in cold atomic gases than in conventional DDM. We expect this section can establish a bridge among different communities and inspire further studies on dependent scattering.

\subsection{Mesoscopic physics}\label{mesophys}
Initially, mesoscopic physics mainly investigated the quantum transport phenomena involving electrons in disordered electronic materials with the consideration of wave nature of electrons in mesoscopic-scale samples, which are smaller than the coherence length of electrons, leading to the emergence of quantum interference effects \cite{akkermans2007mesoscopic}. Later on, due to the easiness of experimental implementation and observation in optics, light transport in disordered dielectric materials offers a good platform for mimicking quantum transport behaviors of electrons, for instance, the direct observation of Anderson localization \cite{lagendijkPT2009} and the straightforward measurement of intensity correlations and statistics \cite{Dogariu2015,berkovitsPhysRep1994}, etc. 

\subsubsection{Coherent backscattering cone}\label{CB}
\begin{figure}
	\centering
	\subfloat{
		\label{mishchenkoJQSRT2006}
		\includegraphics[width=0.42\linewidth]{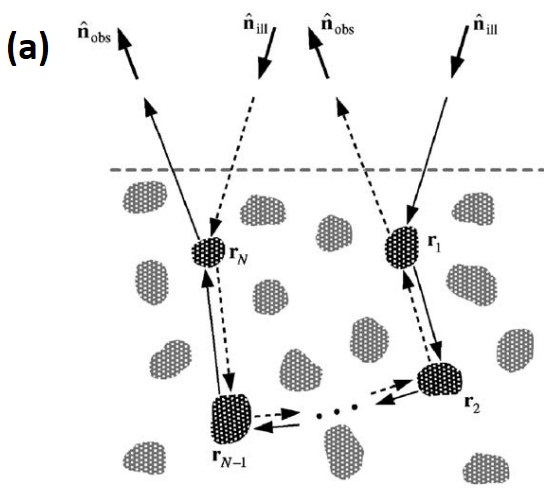}
	}
	\hspace{0.01in}
	\subfloat{
		\label{CBE_data}
		\includegraphics[width=0.5\linewidth]{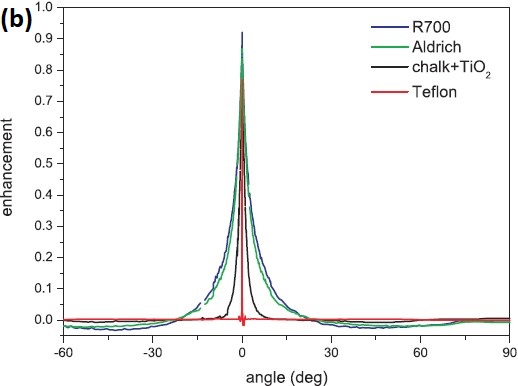}
	}
	\caption{\textbf{(a)} The schematic of the mechanism of the coherent backscattering cone in a random medium containing discrete scatterers. The solid lines denote a random multiple scattering trajectory while dashed lines stand for its time-reversed counterparts. Reprinted from Ref. \cite{mishchenkoJQSRT2006} Copyright (2016), with permission from Elsevier. \textbf{(b)} Experimental data of coherent backscattering enhancement for different disordered media. Reprinted with permission from Ref. \cite{fiebigEPL2008}. Copyright (2008) EPL Association.}\label{CBE}
\end{figure}

Even in the limit of extremely dilute media, the ladder approximation that leads to the RTE is not rigorously exact, because the well-known coherent backscattering cone, or called the weak localization phenomenon \cite{mishchenko2006multiple,wolfPRL1985,Naraghi2015}, can emerge. This is due to the constructive interference between each multiple scattering trajectory and its time-reversed counterpart in the exact backscattering direction, as shown in Fig.\ref{mishchenkoJQSRT2006}. Note the interference between each multiple scattering trajectory and its time-reversal counterpart is always constructive when time-reversal symmetry is conserved \cite{VanRossum1998,mishchenko2006multiple,sheng2006introduction}. 

%Nevertheless, the effect of this phenomenon can be conveniently calculated as a correction on the basis of RTE \cite{VanRossum1998,mishchenko2006multiple,sheng2006introduction}.
%The latter condition is called ladder approximation because it results in the Feynmann diagrammatic representation of field correlation function resembling a series of ladders \cite{VanRossum1998,mishchenko2006multiple,sheng2006introduction}.

When the scattering strength becomes stronger (i.e., mean free path $l$ becomes smaller), the angular width of the coherent backscattering cone broadens as $(kl)^{-1}$, shown in Fig.\ref{CBE_data}, and thus affects the overall reflectance and transmittance (due to energy conservation) more significantly. As a consequence, for these media, the interference between each multiple scattering trajectory and its time-reversed counterpart should be taken into account, resulting a series of most-crossed diagrams \cite{vandermarkPRB1988,akkermansPRL1986,akkermans1988theoretical} (or so-called cyclical terms \cite{mishchenkoMNRAS1992,mishchenko2006multiple}) in the diagrammatic representation of the irreducible intensity vertex in the Bethe-Salpeter equation. In practice, the conventional RTE is firstly used to solve the intensity distribution, and an extra calculation is made to obtain the contribution of the crossed diagrams based on the results of the first step (that is, by using the property of time-reversal invariance) \cite{wolfPRL1985,akkermans1988theoretical,mishchenko2006multiple}. Moreover, besides using the Ohm's law, another method to obtain the transport mean free path is to fit the analytical formula of coherent backscattering cone (Eq.(\ref{cb_lineshape}))\cite{Naraghi2015}, as already discussed in Section \ref{angleresolvedmeasurement}. The peak enhancement factor, ideally, is 2. However, it strongly depends on the thickness and the absorption of the sample \cite{akkermansPRL1986,etemadPRL1987,sheng2006introduction}, and other factors that affect the coherence of multiple wave scattering. 

CB is also frequently discussed in the field of astrophysics. The opposition effect is believed to be a manifestation of it \cite{mishchenkoMNRAS1992,mishchenko2006multiple,mishchenkoPhysrep2016}, like the one exhibited by the rings of Saturn, which are composed of particles that are covered by small, submicron-sized $\mathrm{H_2O}$ grains.

%{\color{red}Sheng commented \cite{sheng2006introduction}: It is interesting to note that the coherent backscattering effect was “discovered” independently in two different fields: as the basis of weak localization of electrons in dirty metals in solid-state physics (Langer and Neal 1966) and in optics as a reflection enhancement effect for electromagnetic wave propagation in a turbulent atmosphere (deWolf 1971). The word “discovered” was put in quotation marks to mean within the realm of recent literature. In any case, the parallel discoveries do reflect the common wave character of the coherent backscattering effect, regardless of whether the wave is quantum or classical in nature. This similarity, however, does not carry over to the ease of actual demonstration. In the quantum case, the difficulty in observing an electronic wave function means that there are so far only indirect demonstrations of the effect. But in the case of the electromagnetic wave there have been several experiments involving laser light reflection from disordered samples in which the coherent backscattering effect was directly observed and quantitatively measured \cite{wolfPRL1985,Albada1985} (Tsang and Ishimaru 1984; van Albada and Lagendijk 1985; Wolf and Maret 1985).}

\subsubsection{Anderson localization of light}\label{anderson_loc_sec}
\begin{figure}
	\centering
	\subfloat{
		\label{andersonloc1}
		\includegraphics[width=0.24\linewidth]{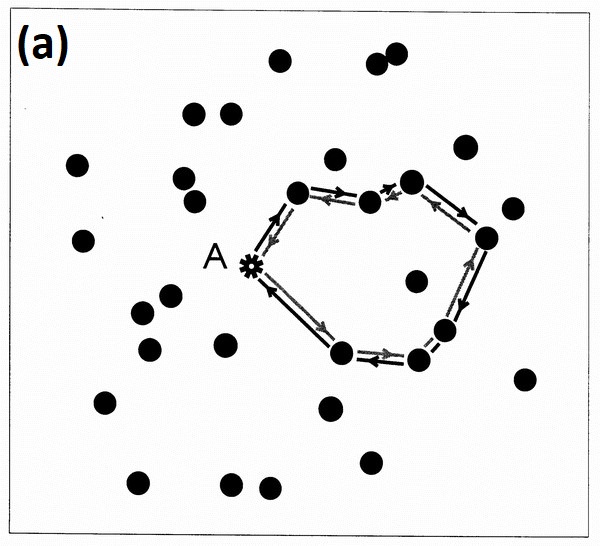}
	}
	\hspace{0.01in}
	\subfloat{
		\label{andersonloc2}
		\includegraphics[width=0.7\linewidth]{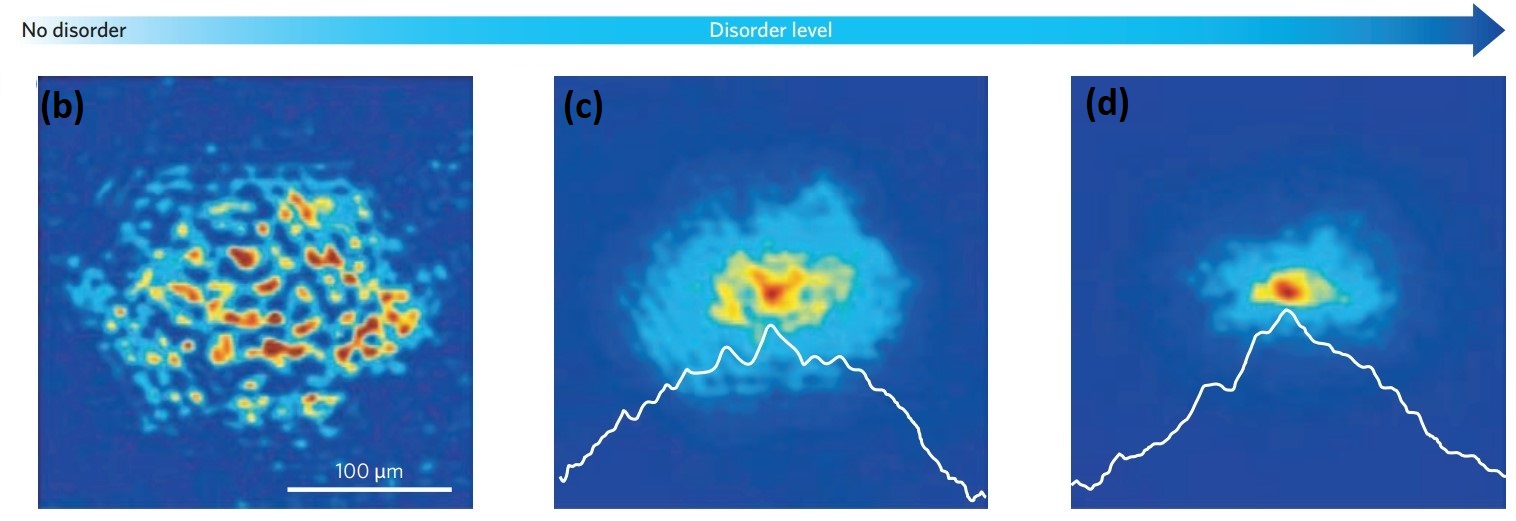}
	}
\caption{\textbf{(a)} The schematic diagram of the mechanism of Anderson localization in a disordered medium. Here the light source is denoted by a star symbol at position $A$ and the spheres denote the scattering elements. A multiple scattering path that returns to the light source forms a loop. Its time-reversed counterpart propagates in exactly the opposite direction along this loop. The two paths thus will acquire exactly the same phase, and interfering constructively in \textit{A}.  Reprinted by permission from Springer Nature Customer Service Centre GmbH: Springer Nature, \textit{Nature}, Ref. \cite{wiersma1997localization}, Copyright 1997. \textbf{(b-d)} Experimental observation of the transition from \textbf{(b)} ballistic transport to \textbf{(c)} diffusion, and then to \textbf{(d)} Anderson localization with the increase of disorder in a 2D photonic lattice, in which the white thin lines denote the ensemble-averaged intensity profile in space. Reprinted by permission from Springer Nature Customer Service Centre GmbH: Springer Nature, \textit{Nature Photonics}, Ref. \cite{Segev2013}, Copyright 2013, and \textit{Nature}, Ref. \cite{Schwartz2007}, Copyright 2007.}\label{andersonloc}
\end{figure}

When the scattering strength continues to increase, making the Ioffe-Regel condition satisfied, i.e., $kl\leq 1$ \cite{lagendijk1996resonant}, Anderson localization can possibly occur. In this regime, the wave packets are exponentially localized and a halt of light diffusion is induced. Anderson localization can be formed by the constructive interference between a closed multiple scattering loop with its time-reversed counterpart given a strong scattering strength, as shown in Fig.\ref{andersonloc1}. By now, Anderson localization of light in one and two dimensions has been theoretically and experimentally confirmed \cite{zhangPRB1994,sheinfuxScience2017,shiNL2018,caselliAPL2017}. For example, in Figs. \ref{andersonloc}(b-d) an experimental observation of the transition from ballistic transport to diffusion in two dimensions is presented, where the ensemble-averaged intensity follows a Gaussian distribution in space, and then Anderson localization, where the ensemble-averaged intensity is exponentially localized, with the increase of disorder in a 2D photonic lattice \cite{Segev2013,Schwartz2007}. Note the difference between the Anderson localization scheme and the recurrent scattering mechanism shown in Fig.\ref{recurrent_scattering_schematic}, because the latter takes place in the microscopic scale while the former occurs in the mesoscopic scale\footnote{However, in the Anderson localization regime, it is hard to tell the phenomenon is mesoscopic or microscopic because the wavelength and mean free path are comparable with each other. For convenience, we regard it as a mesoscopic phenomenon.}.  

On the other hand, Anderson localization of light in 3D is still under intensive theoretical and experimental pursuit but has not been unambiguously observed \cite{akkermans2007mesoscopic,wiersma1997localization,sperlingNJP2016,storzerPRL2006,aegerterEPL2006,aegerterJOSAA2007,skipetrovNJP2016}. As mentioned in Section \ref{total_ref_trans}, by measuring the thickness dependence of transmittance, it is possible to identify the onset of Anderson localization. During the transition from the diffusive transport regime to the Anderson localization regime, the total transmittance would follow $T\propto1/L^2$ based on the prediction of the single-parameter scaling theory of localization \cite{andersonPMB1985,wiersma1997localization}. And in the Anderson localized regime, the relation becomes $T\propto\exp{(-L/l_\mathrm{loc})}$, where $l_\mathrm{loc}$ is the localization length which describes the spatial extent of localized modes. The first experimental claim on 3D Anderson localization of light in GaAs powders \cite{wiersma1997localization} used this method, which, however, was refuted and shown to be the result of weak absorption \cite{scheffoldNature1999,wiersmaNature1999reply}. Later, to further search for 3D Anderson localization, time-resolved measurement was conducted in TiO$_2$ powders due to the advantage to separate scattering and absorption effects of this method \cite{storzerPRL2006,sperlingNaturephoton2013}, as discussed in Section \ref{time_resolved}. By quantifying the deviation of time-resolved transmission from the diffusion equation (more specifically, non-exponential decay at long times), these works also claimed the observation of 3D Anderson localization. However, this claim was later put in question \cite{scheffoldNaturephoton2013} and finally refuted by the authors themselves \cite{sperlingNJP2016}, by demonstrating that the non-diffusive behavior was a consequence of fluorescent emission due to the impurities in the sample at high input laser powers. Therefore, by now there are no unambiguous evidence of 3D Anderson localization of light. 

Remarkably, recently it has been theoretically shown that 3D Anderson localization transition may be found in certain disordered structures with strong short-range correlations (i.e., hyperuniform networks) near the photonic band edges \cite{haberko2018transition}, following from the early theoretical proposal of John \cite{johnPRL1987} but without any defect modes. Based on similar considerations, it has been numerically shown 3D Anderson localization can occur in quasiperiodic structures (i.e., icosahedral quasicrystal) without any additional disorder \cite{jeonNaturephys2017} whereas they cannot be regarded as disordered materials due to the existence of long-range order. 
It would be interesting to examine these theoretical proposals experimentally, which can provide profound implications for wave physics.

\subsubsection{Statistics and correlations in disordered media}
Due to the existence of disorder, the optical responses of disordered media vary from sample to sample, from position to position, and therefore it is more practical to investigate the statistics and correlations in their responses and obtain some general and global properties of them. These mesoscopic phenomena, including the statistical distribution of intensity speckles, spatial and spectral correlations of intensity fluctuations, are comprehensively discussed in the mesoscopic physics community \cite{sheng2006introduction,akkermans2007mesoscopic}. Here we introduce several well-known statistic phenomena of light scattering in disordered media. For more details on this topic, see Ref.\cite{Dogariu2015,berkovitsPhysRep1994}. 

\begin{figure}[htbp]
	\flushleft
	\includegraphics[width=1\linewidth]{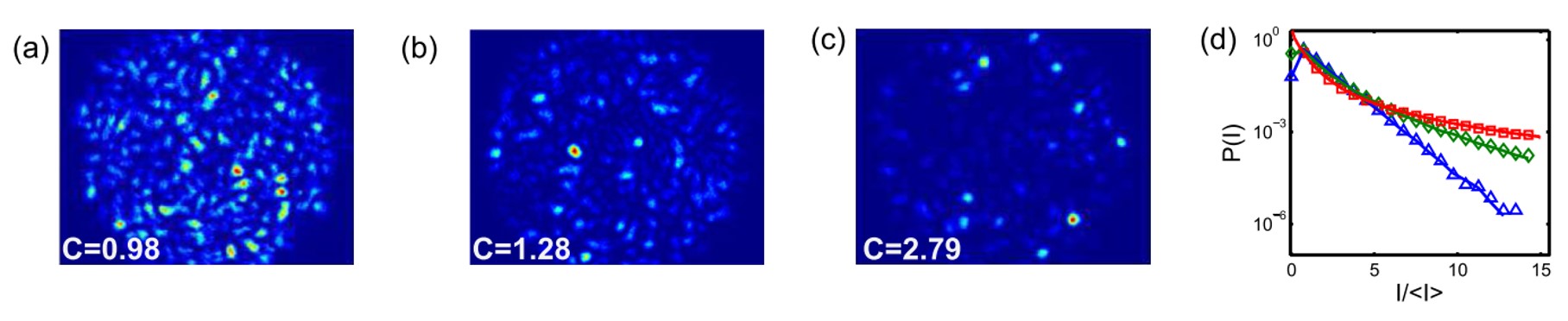}
	\caption{Speckle patterns and intensity distribution in disordered media. \textbf{(a-c)} Images of the speckle patterns with different statistics by controlling the contrast parameter $C=\sqrt{\langle I^2 \rangle/\langle I \rangle^2-1}$ using a spatial light modulator (SLM) to modulate the incident wavefront, and \textbf{(d)} the corresponding intensity distribution function. \textbf{(a)} A speckle pattern with a standard Rayleigh statistics, which has a contrast of $C=0.98$ and a negative exponential intensity distribution [\textbf{(d)}, blue triangles]. \textbf{(b)} A speckle pattern with super-Rayleigh statistics, which has a contrast of $C=1.28$, and an intensity distribution that decays slower than the negative exponential [\textbf{(d)}, green diamonds]. \textbf{(c)} A speckle pattern with super-Rayleigh statistics with a higher contrast of $C=2.79$ [\textbf{(d)}, red squares]. Reprinted from Ref. \cite{brombergPRL2014} Copyright (2014) by the American Physical Society.} \label{brombergPRL2014}
	
\end{figure}

The first is the amplitude and intensity distribution in the speckle pattern, which forms due to the complex interference of randomly scattered waves traveling different path lengths in disordered media. The amplitude follows a Rayleigh distribution, and the intensity is described by a negative exponential probability distribution as
\begin{equation}
P ( I ) = \frac { 1 } { \langle I \rangle _ { \mathrm { c } } } \exp \left( - \frac { I } { \langle I \rangle _ { \mathrm { c } } } \right).
\end{equation}
This is generally known as the Rayleigh statistics in random media, where $\langle I \rangle _ { \mathrm { c } }$ is configuration averaged intensity and $P(I)$ is the probability distribution function of intensity. The condition for this statistics is that the speckle pattern should arise from the interferences of a large number of scattered waves with independently varying amplitudes and phases, and the phases should uniformly distributed in the range of $0-2\pi$ \cite{akkermans2007mesoscopic,brombergPRL2014}. Therefore, if these conditions are not fulfilled, for instance, when the scattering events are strongly correlated or the scattering strength is too weak to cover the phase range, or the scattering strength is too strong to exhibit Anderson localization, the speckles can exhibit non-Rayleigh statistics. An experimental example of the breakdown of Rayleigh statistics and the emergence of super-Rayleigh statistics is given in Fig.\ref{brombergPRL2014}. This is achieved by generating strongly correlated phases in the wavefront of incident light using a spatial light modulator (SLM), which can result in high-contrast speckle patterns that exhibit non-Rayleigh statistics.

%Speckle contrast can also be used to extract the diffusion coefficient (if transport mean free path is known using static transmission measurement) \cite{curryOL2011}.

%The transmission statistics of different modes
 
%Another well-known statistical phenomenon is the universal %conductance fluctuations.

\begin{figure}
	\centering
	\subfloat{
		\label{emilianiPRL2003}
		\includegraphics[width=0.56\linewidth]{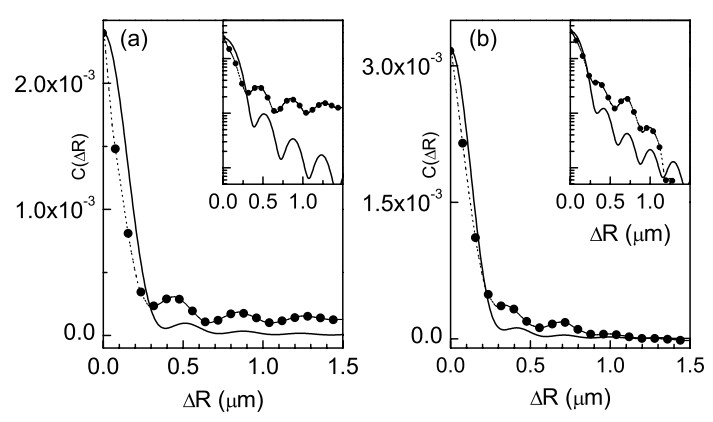}
	}
	\subfloat{
		\label{carminatiPRA2010}
		\includegraphics[width=0.4\linewidth]{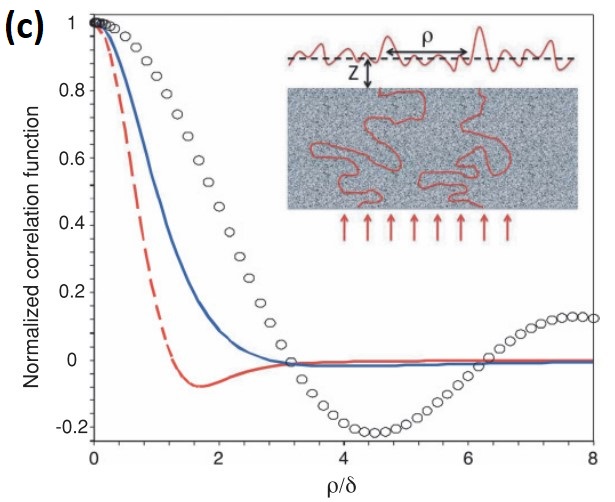}
	}
	\caption{Spatial correlation function of intensity. \textbf{(a-b)} SNOM measurement of averaged radial profile of correlation function (dots), \textbf{(a)} probe wavelength of 780 nm, and \textbf{(b)} probe wavelength of 632 nm. The corresponding theoretical fits using Eq.(\ref{spatialcorrelationfunction}) are shown in solid lines. Reprinted from Ref. \cite{emilianiPRL2003} Copyright (2003) by the American Physical Society. \textbf{(c)} Normalized field spatial correlation function in a plane at a distance $z$ vs $\rho/\delta$. $\delta$ is a reference length scale. Black markers: Far-field regime. Blue solid line: near-field intermediate regime. Red dashed line: extreme near-field regime. Reprinted from Ref. \cite{carminatiPRA2010} Copyright (2010) by the	American Physical Society. }\label{correlations}
\end{figure}

Besides the statistic distribution of intensities, there are also universal correlations for intensities at different positions and frequencies. One frequently used intensity correlation function is defined for the intensity fluctuations in the speckle patterns as $C=\langle\delta I\delta I'\rangle$, where $\delta I=I-\langle I\rangle$ is the fluctuation of intensity short-range intensity and $\delta I'$ stands for the fluctuation of intensity at a different spatial or frequency position \cite{shapiroPRL1986,stephenPRL1987,fengPRL1988,genackPRL1990,emilianiPRL2003}. In this article, we mainly introduce spatial correlations. The correlation function $C$ contains three components, which describe short range ($C_1$), long range ($C_2$), and infinite range ($C_3$) correlations, respectively \cite{emilianiPRL2003}. In particular, short-range spatial correlations describe the averaged size of a speckle spot, while multiple scattering can induce long-range spatial correlations. In the limit of weak scattering, the main contribution to the spatial correlation is the short-range one, which is given by the square of field correlation function as $C_1(\mathbf{R})=|E(\mathbf{r})E^*(\mathbf{r}+\delta\mathbf{r})|^2$. Shapiro theoretically showed that \cite{shapiroPRL1986} 
\begin{equation}\label{spatialcorrelationfunction}
C_1\propto \left(\frac{\sin{(k\delta r)}}{kr}\right)^2\exp(-\frac{\delta r}{l}).
\end{equation}
Using a scanning near-field optical microscope (SNOM), Emiliani \textit{et al.} \cite{emilianiPRL2003} directly measured the 2D short-range intensity correlation function of a disordered dielectric structure of microporous silica glass with randomly oriented and interconnected pores around 200 nm, which is shown in Fig.\ref{emilianiPRL2003}. They found that Eq.(\ref{spatialcorrelationfunction}) can indeed capture the short-range intensity correlation behavior. However, Carminati \cite{carminatiPRA2010} theoretically demonstrated that in the deep near field, spatial correlation length of the field correlation function heavily depends on the local microscopic environment and thus does not show any universal behavior as described by Eq.(\ref{spatialcorrelationfunction}), shown in Fig.\ref{carminatiPRA2010}. The dynamic fluctuations of speckle patterns stemming from the movement of scattering particles in random media enable a novel imaging method for soft materials especially for biological materials called laser speckle contrast imaging (LSCI) \cite{boasJBO2010}, which is widely applied in the real-time imaging of blood flows in retina, skin, brain and many other kinds of tissues and organs \cite{biersJBO2013,heemanJBO2019}. Similar techniques also include the diffuse correlation spectroscopy (DCS) \cite{boryckiOptica2016,chengOL2018}, which can also provide a route to measure radiative/optical properties.

%Cheng et al \cite{chengOL2018} said: ``Diffuse correlation spectroscopy (DCS) offers a non-invasive way to directly and continuously measure blood flow at the bedside [1]. \textbf{In DCS, a tissue of interest is illuminated by coherent near-infrared light which causes a speckle interference pattern to form by multiplying scattered light though the tissue. Dynamic scattering due to moving red blood cells causes the speckle pattern to fluctuate rapidly, and is quantified by the temporal intensity autocorrelation curve $g_2(\tau)$ of a single speckle \cite{boasPRL1995,boasJOSAA1997}.} The DCS-measured blood flow has been used as a biomarker of patient well-being and as an indicator of hemodynamic and metabolic functional responses [1,5–7]. The traditional DCS technique uses continuous wave (CW) light with a coherence length longer than the spread of the path length distribution of the photon trajectories. Recently, we have demonstrated the feasibility of performing DCS measurements in the time domain by using a train of laser pulses as the light source." 

%Frequency correlations in reflection from random media \cite{knotheJOSAA2015}; speckle autocorrelation spectroscopy \cite{watsonPRB1990}; Spectral speckle correlations \cite{genackEPL1990,vanalbabaPRL1991,mikhailovskayaEPJAP2019}.
%
%(spectral-temporal) measurement of the scattering medium transfer function at a single speckle grain via scanning the frequency of a continuous-wave laser \cite{websterOL2004,gerkeJOSAA2005} 

In recent years, the rapid development of photonics further accelerates the advance of mesoscopic physics and gives rise to a brand new field called "disordered photonics" \cite{wiersma2013disordered}. Fascinating achievements were made thanks to the availability of high-speed spatial light modulators (SLMs), including the imaging, focusing and multiplexing through disordered media. More specifically, it has received great attention recently that the statistics and correlations of the speckle patterns due to wave interference can be exploited to realize efficient spatial/temporal imaging and focusing through the scattering disordered media \cite{katzNaturephoton2014,salhovOL2018,sternOL2019,mccabeNaturecomms2011,aulbachPRL2011}, show promising application to biological and medical imaging and laser therapy (like, deliver light to a specific position in human body), as well as in-situ optical/infrared diagnosis of turbid or semitransparent coatings used in industry thermal barrier coatings. Moreover, engineering disorder itself to manipulate light transport is also a possible way, for instance, by controlling fabrication process \cite{riboliNaturemat2014} or using spatially modulated pumping  (all-optical spatial light modulator) \cite{bruckOptica2016}. It is of great interest to find its applications in thermal radiation control. Detailed introduction is, however, out of the scope of this article. For more details, see Refs. \cite{wiersma2013disordered,Rotter2017}.

%\subsection{DSE in acoustics}

\subsection{Light-atom interactions} \label{light_atom_interaction}

As seen from previous sections, it is still not fully understood how structural correlations affect the radiative properties of disordered media, even for an ideally simple system composed of randomly distributed spherical scatterers \cite{Naraghi2015,Naraghi2016,wangPRA2018,pattelliOptica2018,ramezanpourJQSRT2019}. In particular, the interplay between structural correlations and near-field as well as far-field electromagnetic interactions among scatterers is still not very clear \cite{Naraghi2015,Naraghi2016,escalanteADP2017}, especially near single scatterer internal resonances (like Mie resonances for dielectric nanoparticles) at high packing densities  \cite{lagendijk1996resonant,sapienzaPRL2007,garciaPRA2008,aubryPRA2017,tallonPRL2017}.

On the other hand, the last three decades have witnessed the rapid development of laser cooling and trapping of atoms to realize ultracold atomic gases with extremely low temperatures on the order of a few nanokelvin \cite{cohentannoudji2011} in atomic physics, which stimulate a wide range of exciting applications like high-precision atomic clocks, quantum information processing, quantum computing and quantum simulation of condensed matter systems and so on \cite{cohentannoudji2011}. Regarding the very high resonant scattering cross section  of a single two-level cold atom at the dipole transition (on the order of $\sim\lambda_0^2$, where $\lambda_0$ is the wavelength of transition, thus surprisingly larger than the size of a single atom), cold atomic systems are capable of achieving strong light-matter interactions \cite{cohentannoudji2011}. 

In particular, disordered cold atomic gases offer new opportunities for theoretical and experimental study of multiple wave scattering, and they are advantageous over conventional discrete random media due to the precisely controllable and highly tunable system parameters in an unprecedented way \cite{baudouin2014cold,haveyPhysrep2017}. 
%For instance, the ultranarrow radiative linewidth ($\gamma\ll\omega_0$, where $\gamma$ is the linewidth and $\omega_0$ is the resonance frequency) and strong optical resonances can provide more detailed observations of the multiple scattering phenomena.
Classical multiple wave scattering phenomena, for instance, coherent backscattering \cite{labeyrieJOB2000}, slow diffusion \cite{labeyriePRL2003}, Anderson localization \cite{Skipetrov2015,moreira2019localization} and random lasing \cite{baudouinNaturephys2013} were investigated, along with some extraordinary phenomena like L\'evy flights of photons \cite{pereiraPRL2004,mercadierPRA2013}, thermal decoherence \cite{labeyriePRL2006} and non-Lorentzian transmission spectra \cite{zhuPRA2016,cormanPRA2017}, etc. Collective and cooperative effects that are difficult to observe in conventional condensed matter systems also emerge in cold atomic systems, such as the collective polaritonic modes \cite{Schilder2016}, superradiant and subradiant collective modes \cite{guerin2017light}, collective Lamb shift \cite{keaveneyPRL2012,meirPRL2014} and so on. Moreover, the easy incorporation of nonlinearity can provide a platform for the study of multiple scattering of interacting photons, where more intriguing many-body physics can take place \cite{changNaturephoton2014}, not to mention that the quantum nature of cold atoms (for instance quantum statistical correlations in Fermi-Dirac gases \cite{ruostekoskiPRL1999,ruostekoskiPRA2000}) might also affect light propagation, which very much enriches the underlying physics.

In this subsection, we present a short introduction of the DSE in cold atomic systems, mainly including the breakdown of mean-field optics in dense cold atomic clouds and the role of structural correlations.

\subsubsection{Breakdown of the mean-field optics}
\begin{figure}
	\centering
	\subfloat{
		\label{javanainenPRL2014}
		\includegraphics[width=0.61\linewidth]{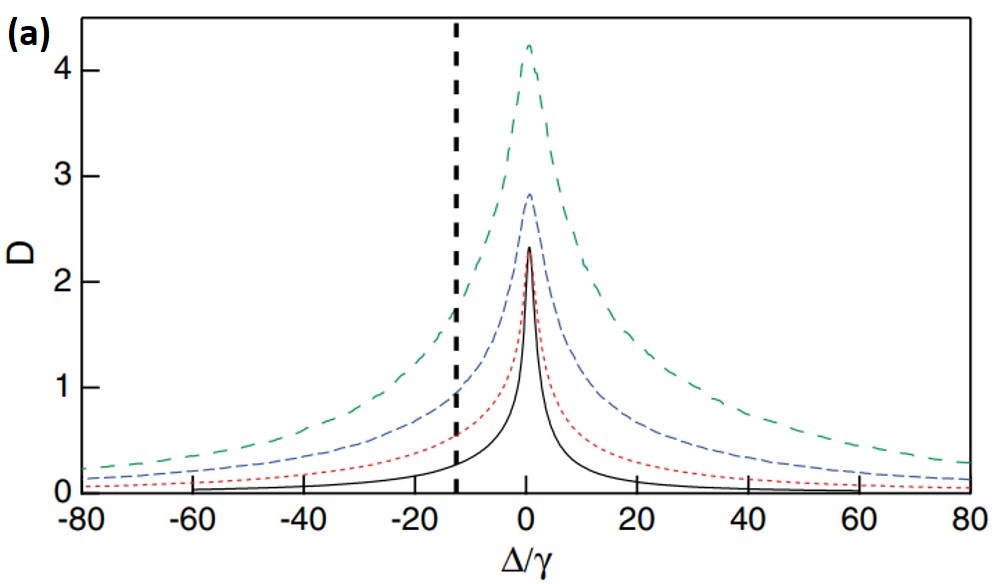}
	}
	\subfloat{
		\label{schilderPRA2017}
		\includegraphics[width=0.35\linewidth]{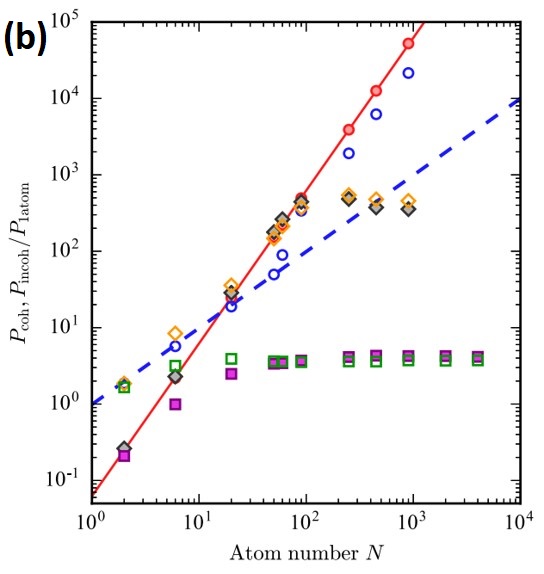}
	}
	\caption{Breakdown of mean-field optics in cold atomic clouds. \textbf{(a)} Optical depth $D=-\ln{T}$ versus detuning $\Delta=\omega-\omega_0$ in a homogeneously broadened sample for normalized thicknesses $kh = $0.25, 0.5, 1.0, and 2.0, from bottom to top; the corresponding atom numbers are $N =$ 128, 256, 512, and 1024. The dashed vertical line shows where the center of the line would be if the naive Lorentz-Lorenz shift applied. $\gamma$ is the linewidth of the single atom resonance. Reprinted from Ref. \cite{javanainenPRL2014} Copyright (2014) by the	American Physical Society. \textbf{(b)} Coherent (plain symbols) and incoherent (open symbols) scattered powers calculated for a linearly polarized incident plane wave propagating along the long axis of the cloud for various	frequency detunings $\Delta$, as a function of the number of atoms $N$. All powers are normalized to the power scattered by a single atom at the same detuning. The plain (dashed) line connects the values of the coherent (incoherent) power for $\Delta=-10^4\gamma$. Circles, $\Delta=-500\gamma$; diamonds, $\Delta=-5\gamma$; squares, $\Delta=0$. Reprinted from Ref. \cite{schilderPRA2017} Copyright (2017) by the	American Physical Society. }\label{meanfield}
\end{figure}

Traditionally, for inhomogeneously broadened atomic gases, e.g., thermal gases with inhomogeneous Doppler broadening, the mean-field description of light-atom interaction (in the low intensity limit without any nonlinear optical phenomena) is valid. In other words, light propagation in such atomic gases can be described by standard electrodynamics in a medium with an effective permittivity under the local field correction (see Eq. (\ref{LLR})). In this circumstance, the collective resonance peak exhibits a frequency red-shift proportional to the atom number density $n_0$ as
\begin{equation}
\Delta_\mathrm{LL}=-2\pi\frac{n_0}{k^3}\gamma,
\end{equation}
which is called the Lorentz-Lorenz (LL) shift. Here $\gamma$ is the linewidth of the singe atom resonance.  However, for dense and ultracold atomic gases, in which inhomogeneous broadening is negligible, recently it has been numerically \cite{javanainenPRL2014} and experimentally \cite{jenkinsPRL2016} shown that there is no such shift. In Fig.\ref{javanainenPRL2014}, Javanainen et al. \cite{javanainenPRL2014}, based on coupled-dipole simulations for a slab composed of two-level ultracold atoms, demonstrated that at high atom number density $n_0/k^3\sim1$, the atomic clouds do not exhibit any resonance frequency shift. This is because at high density where the distance between adjacent atoms is comparable or even much smaller than the light wavelength, the dipole-dipole interactions become strong and lead to significant light-induced correlations among the atoms \cite{javanainenPRA2017}. This is a situation that mean-field theory (or LL relation) is not able to take into account. Recently, Corman \textit{et al.} \cite{cormanPRA2017} found in some cases a dense ultracold atomic samples can exhibit small blue shifts in resonance frequency both numerically and experimentally, and the line shape even turns into non-Lorentzian, in strikingly contrast to the mean-field optics. They used a theoretical model of effective permittivity proposed by Morice \textit{et al.} \cite{moricePRA1995} to partially include the effect of dipole-dipole interactions. This model, in essence, takes the recurrent scattering mechanism between each pair of atoms into account. The agreement between the theoretical model and coupled-dipole simulations was fairly good up to surface density $n_\mathrm{2D}/k^2\sim0.1$, while theoretical and numerical data deviated from experimental measurement data substantially, especially when it comes to the value of frequency shift. This may be attributed to residual motion of atoms during the experimental probing process, nonlinear effects due to large light intensity and the complex atomic level structure (i.e., not ideal two-level atoms).

The strong resonant scattering in atomic clouds permits to reach a regime where the concept of effective medium from the mean-field optics does not apply. Schilder \textit{et al.} \cite{schilderPRA2017} showed that for dense cold atomic samples with very high density ($n_0/k^3\gg1$), the homogenization by an effective permittivity cannot be reached. This is because the strong resonant dipole-dipole interactions between stationary cold atoms lead to significant correlations and fluctuations of scattered waves, which result in a large incoherent component of intensity $P_\mathrm{incoh}$ that is comparable to or even larger than the coherent component $P_\mathrm{coh}$, as shown by diamond and square symbols in Fig.\ref{schilderPRA2017} for near-resonant atoms. In this regime, it is more appropriate to describe the scattering processes in this medium in terms of collective modes rather than as a sequence of individual scattering events (i.e., radiative transfer) \cite{Schilder2016}.

\subsubsection{Structural correlations in cold atomic gases}
\begin{figure}[htbp]
	\flushleft
	\includegraphics[width=1\linewidth]{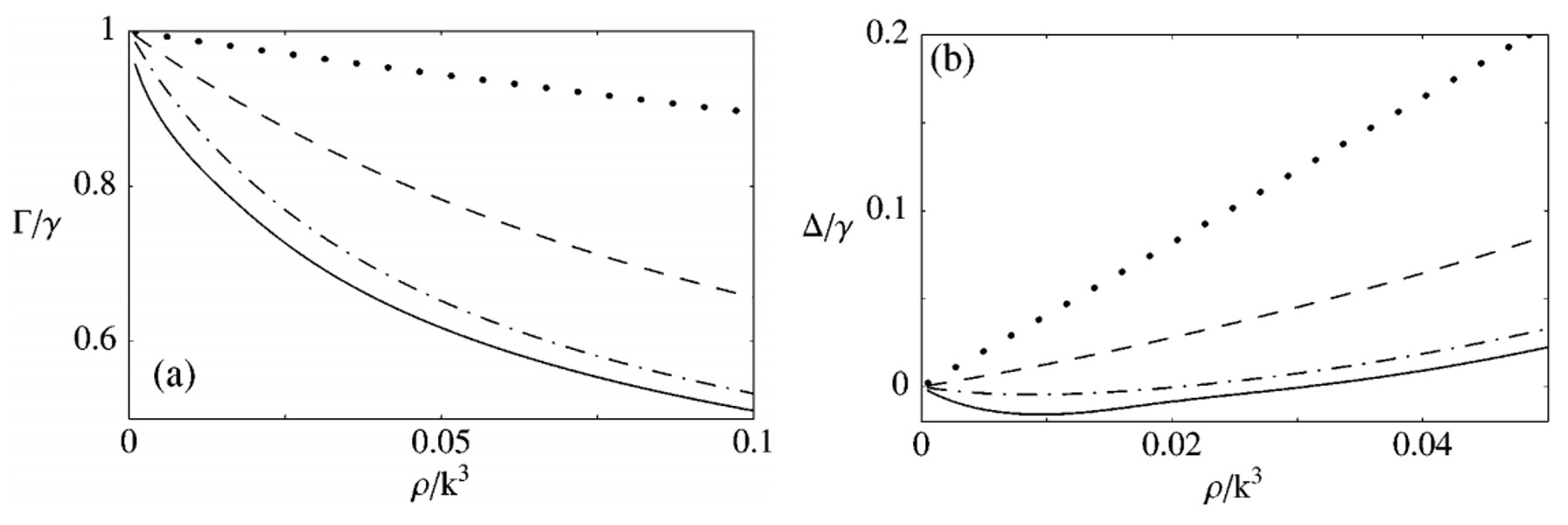}
	\caption{The optical \textbf{(a)} linewidth $\Gamma$ and \textbf{(b)} the line shift $\Delta$ of a Fermi-Dirac gas as a function of the atom number density $\rho/k^3$. The solid line represents $T_\mathrm{gas}=0$, the dash-dotted line $T=150 \mathrm{nK}$ for $\mathrm{^{40}K}$ atoms, the dashed line $T_\mathrm{gas} = 500 \mathrm{nK}$, and the dotted line $T_\mathrm{gas}=2 \mathrm{nK}$. Reprinted from Ref. \cite{ruostekoskiPRL1999} Copyright (1999) by the American Physical Society.}\label{ruostekoskiPRL1999}
	
\end{figure}

Structural correlations indeed exist and can also be engineered in cold atomic gases, which can thus lead to significant dependent scattering effects. It is customary to assume in many theoretical treatments that the positions of atoms are completely independent of each other, especially in very dilute gases, e.g., Refs. \cite{Skipetrov2014,Skipetrov2015,jenneweinPRL2016,Schilder2016,ruostekoskiOE2016,Cherroret2016,javanainenPRA2017,cormanPRA2017}. Strictly speaking this assumption is not true because at very short distances atoms can interact with each other through, like, van der Waals interactions and collisions (like $s$-wave scattering) \cite{cohentannoudji2011}. 
%\footnote{Araújo's PhD thesis \cite{araujothesis2018} said: Also, as the atoms are closer, interactions between them may occur, like van der Waals interactions \cite{grossPhysRep1982}, collisions, and quantum statistics for bosons \cite{moricePRA1995}.} 
More importantly, quantum statistics at low temperature can lead to significant correlations of atom positions in both trapped Bose and Fermi gases \cite{moricePRA1995,parkinsPhysrep1998,ruostekoskiPRL1999,ruostekoskiPRA2000}, where the large de Broglie wavelength of atoms introduces a considerable correlation length. For instance, low-density Fermi gas at zero temperature exhibits a positional correlation length on the order of $k_F^{-1}$, where $k_F\propto(6\pi^2n_0)^{1/3}$ is the Fermi wave number and $n_0$ is atom number density \cite{ruostekoskiPRL1999}. These positional correlations have nontrivial consequences on the optical interactions in atomic gases \cite{moricePRA1995,ruostekoskiPRL1999,ruostekoskiPRA2000}. 
%\footnote{DeMacro and Jin \cite{demacroPRA1998} said: One effect of quantum statistics on light scattering, which has been essential for studying BECs with optical imaging techniques, is simply that the scattered light reflects the spatial profile of the atom cloud, which in a harmonic trap is directly related to the	quantum-statistical occupation of trap energy levels. A second effect of atom quantum statistics on light scattering comes from an enhanced or decreased probability for the scattering of atoms into occupied trap levels. Assuming a two-level atomic system, an atom that scatters a photon receives recoil momentum that projects it onto new harmonic trap levels, which may or may not be already occupied. For bosons an enhanced probability of scattering into occupied final states is predicted to give an extremely broad linewidth \cite{javanainenPRL1994,youPRA1994}. For fermions, the Pauli exclusion principle does not allow scattering into occupied final states, resulting in a blocking effect \cite{javanainenPRA1995,imamogluPRA1994,buschEPL1998} that implies a narrowing of the linewidth.}.
Actually, Ruosteskoski and Javanainen \cite{ruostekoskiPRL1999} theoretically showed that a dramatic narrowing of the linewidth $\Gamma$ can occur in the Fermi gas even at low densities, and the resonance frequency shift $\Delta$ is also considerable, as shown in Fig.\ref{ruostekoskiPRL1999}. They also investigated the effect of temperature $T_\mathrm{gas}$, which can modify the quantum statistical distribution function and thus the pair correlations between atom positions. Recently, the experimental results of Peyrot \textit{et al.} \cite{peyrotPRL2018} also implied that short-range interactions induced correlations as well as light-induced correlations can be a possible source of experimentally measured nonlocality in the optical response in high density atomic ensembles. The negligence of positional correlations may be a possible reason for the deviations between conventional coupled dipole simulations assuming fully disordered atoms and some experimental measurements \cite{pellegrinoPRL2014,jenneweinPRL2016,cormanPRA2017} besides other explanations like residual motion of the atoms during the probing, nonlinear effects and complex atomic level structures, etc. 
%\footnote{They said ``A second possibility could be the influence of the correlations among dipoles induced by the interactions. They are ignored in our treatment of the configuration-averaged field and in all the models developed so far \cite{friedbergPhysRep1973,ruostekoskiOE2016,javanainenPRA2017,javanainenPRL2014,moricePRA1995,ruostekoskiPRA1997}. This assumption is valid for dilute gases, but it could fail at higher densities such as the ones explored here. Going beyond a mean-field approach by including them could lead to a nonlocal response of the gas. The models presented here or in Refs. \cite{friedbergPhysRep1973,ruostekoskiOE2016,javanainenPRA2017,javanainenPRL2014}, which assume a local susceptibility, would then fail—and including the correlations would be a highly nontrivial undertaking." }. 

Another method of creating positional correlations is to directly synthesize the spatial profile of atoms. Recent developments in the manipulation of single atoms make the fabrication of atomic lattices with \textit{arbitrary} distributions feasible. It was already shown that 1D $\mathrm{^{87}}$Rb atomic arrays with desired arrangements in a large scale (more than 50 atoms) can be assembled in an atom by atom way, through a fast, real-time control of an array of tightly focused optical tweezers \cite{endresScience2016}. Two-dimensional $\mathrm{^{87}}$Rb arrays with user-defined geometries can be also fabricated in a similar way \cite{lahayeScience2016}. Other modern quantum simulation techniques, such as nanophotonic atom lattices using dielectric photonic crystals \cite{gonzalezNaturephoton2015} and plasmonic nanoparticle arrays \cite{gullansPRL2012}, are also possible ways. These techniques offer great opportunities for the study of multiple wave scattering of light in controlled structural correlations at the most fundamental level.

As a short summary, in this section, we introduce and discuss the wave interference phenomena in mesoscopic physics, including the coherent backscattering cone and Anderson localization, as well as the statistics and correlations of scattered waves. This field is  closely related to the study of the DSE, and these phenomena are also frequently mentioned in previous sections, because the microscopic and mesoscopic interferences have influences on each other and cooperatively contribute to the radiative transfer processes. And the field of mesoscopic physics provides relevant theoretical and experimental tools for the study of radiative properties, like the measurement of CB cone, the diagrammatic technique and so on. On the other hand, we also introduce two remarkable interference phenomena in light propagation in cold atomic gases, including the breakdown of the mean-field optics and the significant role of structural correlation. The study in this field is fundamentally important to the understanding of dependent scattering effect in disordered media, which can provide physical insights that are difficult to get in conventional DDM composed of dielectric materials, especially the resonant multiple wave scattering behaviors in dense media that are not well understood currently \cite{buschPRL1995,lagendijk1996resonant,aubryPRA2017,Naraghi2015,tallonPRL2017,wangIJHMT2018,pattelliOptica2018}. We expect this section can establish a bridge among different communities and inspire further studies on dependent scattering.

\section{Summary and outlook}
To summarize, in this article, we give a review on the theoretical, numerical and experimental methods and progresses in the study of the DSE in micro/nanoscale DDM. A brief summary of the multiple scattering theory of electromagnetic waves, including the analytic wave theory and Foldy-Lax equations, is presented. Then main physical mechanisms that are critical to DSE and relevant theoretical considerations as well as models are introduced. Numerical modeling methods and experimental approaches are further summarized. We also give a brief review on the studies on the DSE in mesoscopic physics and atomic physics, especially a discussion on other relevant interference phenomena including coherent backscattering, Anderson localization and statistics in disordered media. We expect this review can provide profound and interdisciplinary insights to the understanding and manipulation of the dependent scattering effect.  We hope this review is not only instructive for thermal radiation transfer, but also helpful for biomedical imaging and light atom interactions. 

This article is not an exhaustive overview and it is also not possible for us to make it exhaustive because there are so many works in the last several decades trying to deal with the dependent scattering effect and other relevant interference phenomena in disordered media. We attempt to include the main theoretical, numerical and experimental methods and progresses on this topic, covering almost all relevant topics to the best of our knowledge. We also pay a substantial attention to the advances in mesoscopic and atomic physics, in order establish a bridge among different communities that all tackle with the wave interference phenomena in disordered materials. However, due to the limited pages of this article, we are only able to discuss everything in a very brief manner and the readers need to refer to the original literature to obtain more theoretical, numerical and experimental details. 

Disorder is more the rule than the exception in nature \cite{lopezADOM2018}. Regarding the notorious complexity brought by disorder, it seems that the study of DSE is from a view point of reductionism which attempts to understand how the particles interact with each other in detail, and thus not appropriate. This is in contrast to the research paradigm well-established in condensed matter physics for complex systems \cite{andersonScience1972}, that is to investigate the emergent phenomena at mesoscopic and macroscopic scales as a result of the collective and universal (from wave/quantum interference) behaviors in many-body systems with the presence of disorder \cite{tianPhysicaE2013,akkermans2007mesoscopic,fangPRB2019}, because in many cases in practice, it is hardly possible to correctly characterize and model the microscopic details. However, we would like to emphasize that the study of the DSE is of both fundamental and practical importance, especially in the design the performance of DDM to applications. An intuitive example is that neither the single-parameter scaling theory \cite{abrahamsPRL1979} nor the self-consistent theory of Anderson localization \cite{tiggelenPRL2000} is able to provide direct instructions to the realization of 3D localization transition, while the understanding and tailoring of the detailed near-field interactions among adjacent scatterers is more important to achieve this goal \cite{escalanteADP2017,wang2018role}. This is one important reason why many remarkable progresses are made in recent years to reveal the mechanisms of dependent scattering in densely packed particle systems at the microscopic scale \cite{pattelliOptica2018,aubryPRA2017,Naraghi2015,Skipetrov2014,Skipetrov2015,schertelPRMat2019,wangPRA2018,jacucciADOM2019}, owing to the development of computational and experimental tools.

Another issue worth discussing is the thermal radiation heat transfer in a sufficiently high-temperature dense DDM with substantial temperature gradients throughout the system, namely, when the thermal emission is not negligible \cite{mackowskiJHT2008,mishchenkoJQSRT2017,tsangJQSRT2019}. Several open problems still need further investigation, such as the derivation and applicability of RTE from Maxwell's equations and the fluctuation-dissipation theorem \cite{rytovvolume4}, the definition of emission coefficient $\kappa_\mathrm{em}$, the role of near-field radiative transfer and so on. Although a few of works have been conducted in the settings of many-body thermal radiation transfer \cite{latellaPRB2018,dongPRB2017,zhuPRB2018,benabdallahPRL2013}, the main attention in these works is the overall heat transfer rate, rather than revealing the role of interference effects on radiative properties \cite{chenJQSRT2018b} and the applicability of RTE in this situation.

%\textbf{\textit{Anisotropy and polarization effects.}}

%\subsection{General prospects}
We are very fortunate to witness the explosion of nanoscience and nanotechnologies, which have influenced and are still influencing the world in a revolutionary way. The exotic physics phenomena that emerge in nanoscale systems are extremely attractive and are waiting people to explore with curiosity. In the field of thermal science and engineering, our research is also confronted with the rapid development nanoscience and nanotechnologies, which have already brought us to the newly founded branch of nanoscale thermal science and engineering. In particular, the research of thermal radiation heat transfer is unprecedentedly reshaped, because this discipline is directly affected by the fields of nanophotonics and quantum science and technology, which also deal with, in general, light-matter interaction. 

In particular, the motivation of the present article is very much inspired by the research areas of metamaterials \cite{shalaevNaturephoton2007} and metasurfaces \cite{shalaevScience2013}, disordered photonics and Anderson localization of light \cite{wiersma2013disordered}, photonic crystals \cite{joannopoulos2011photonic}, topological photonics \cite{ozawa2018topological}, ultracold atoms and quantum optics \cite{haveyPhysrep2017,blochNaturephys2005}. The rapid advances in these disciplines, for instance, focusing and molding the flow of light in DDM using the wave-front shaping technique based on spatial light modulators \cite{Vellekoop2010}, assembling ultracold atoms one by one to create photonic structures \cite{endresScience2016}, and especially all-dielectric metamaterials and metasurfaces based on Mie resonances \cite{kivsharScience2016}, will continue to provide us ideas to be implemented in thermal radiation control \cite{wangESEE2019}. As a consequence, we envision that to achieve the tailoring of the transport direction, spatial distribution, polarization states of thermal radiation flow in micro/nanoscale DDM is very promising. In the near future, we also expect it would be possible to design and fabricate disordered metamaterials and metasurfaces that show similar functionalities to current ordered ones \cite{huangJQSRT2018,castrolopezAPLPhoton2017} or even reach the operating regime that is difficult to access using current devices \cite{puttenPRL2011,parkNaturephoton2013,jangNaturephoton2018}, therefore not relying on expensive top-down nanofabrication methods and easily scalable. Moreover, reconfigurable micro/nanoscale devices will be another important direction of research and development, by utilizing phase change materials and materials with temperature-, strain- or electric field-dependent optical properties.

%On the other hand, the discovery of novel nanomaterials, like graphene \cite{bonaccorsoNaturephoton2010}, transition metal dichalcogenides (TMDs) \cite{makNaturephoton2016} and other 2D materials \cite{watanabeNaturephoton2009}, drastically expands the fundamental limits of our scope. The tunability of their electronic and optical properties is also impressive. The unconventional and rich optical phenomena in them can offer an ideal platform for realizing extreme radiative properties and radiative transfer. For instance, it was already theoretically predicted that surface plasmons in graphene nanostructures can greatly enhance the performance of near-field radiative heat transfer \cite{yuNaturecomms2017}. As a matter of fact, there are still many physical mechanisms not fully explored in these 2D materials, and the nanofabrication, compatible integration and high-precision measurement still remain demanding in practice. Nevertheless, we are still looking forward to the applications of these interesting materials in manipulating radiative properties and radiative transfer.

%% The Acknowledgements part is started with the command \acknowledgements;
%% acknowledgements are then done as normal sections before appendix
%% \acknowledgements

\section*{Acknowledgments}
We gratefully acknowledge the financial support from the National Natural Science Foundation of China (Nos. 51636004, 51906144, 51476097 and 51176110), Shanghai Key Fundamental Research Grant (Nos. 18JC1413300 and 16JC1403200), China Postdoctoral Science Foundation (Nos. BX20180187 and 2019M651493), and the Foundation for Innovative Research Groups of the National Natural Science Foundation of China (No. 51521004).

%% appendix sections are then done as normal sections and after Acknowledgements
%\appendix
%\section{Percus-Yevick approximation for hard spheres}
%In this model, $H_2(\mathbf{q})$ is calculated as 
%\begin{equation}
%H_2(\mathbf{q})=\frac{(2\pi)^3C(\mathbf{q})}{1-n_0(2\pi)^3C(\mathbf{q})}
%\end{equation}
%where
%\begin{equation}
%\begin{split}
%C(\mathbf{q})&=C(q)=24f_v[\frac{\alpha+\beta+\delta}{u^2}\cos u-\frac{\alpha+2\beta+4\delta}{u^3}\sin u\\&-2\frac{\beta+6\delta}{u^4}\cos u+\frac{2\beta}{u^4}+\frac{24\delta}{u^5}\sin u+\frac{24\delta}{u^6}(\cos u-1)],
%\end{split}
%\end{equation}
%in which $u=2qa$, $\alpha=(1+2f_v)^2/(1-f_v)^4$, $\beta=-6f_v(1+f_v/2)^2/(1-f_v)^4$, $\delta=f_v(1+2f_v)^2/[2(1-f_v)^2]$ and $f_v$ is the volume fraction of the identical spherical particles. 



%\begin{thebibliography}{-8}


%\small{
%\bibitem{key}

%...

%}

%\end{thebibliography}
\appendix
\section{Definition of ensemble average}\label{en_avg_def}
Generally, an ensemble average of a physical quantity should be carried out with respect to \textit{all} possible states of the system \cite{laxRMP1951,tsang2004scattering2,mishchenko2006multiple} if the ergodicity of the system is assumed \cite{mishchenko2006multiple,mishchenkoPhysrep2016,laxRMP1951}. By saying a random process is ergodic, we mean that time average of each realization is equal to the average across the ensemble of realization. In a present random medium consisting of $N$ identical, homogeneous and isotropic scatterers, the only varying states of particles are their positions, i.e., $\mathbf{r}_j$, where $j=1, 2,...,N$. Therefore, the ensemble average over the whole random medium for a physical quantity $Q(\mathbf{r}_1,\mathbf{r}_2,...,\mathbf{r}_j,...,\mathbf{r}_N)$, which is a function of particle positions, is calculated as \cite{laxRMP1951,tsang2004scattering2}
\begin{equation}
\begin{split}
\langle Q\rangle=\int& Q(\mathbf{r}_1,\mathbf{r}_2,...,\mathbf{r}_j,...,\mathbf{r}_N)d\mathbf{r}_1d\mathbf{r}_2...d\mathbf{r}_j...d\mathbf{r}_N\\&\cdot p(\mathbf{r}_1,\mathbf{r}_2,...,\mathbf{r}_j,...,\mathbf{r}_N)
\end{split}
\end{equation}	
where $p(\mathbf{r}_1,\mathbf{r}_2,...,\mathbf{r}_j,...,\mathbf{r}_N)$ is the joint probability density function of the particle distribution $\mathbf{r}_1,\mathbf{r}_2,...,\mathbf{r}_j,...,\mathbf{r}_N$. If we fix some particle $\mathbf{r}_j$, the ensemble average over other $N-1$ particles is given by
\begin{equation}
\begin{split}
\langle Q\rangle_j=\int& Q(\mathbf{r}_1,\mathbf{r}_2,...,\mathbf{r}_i,...,\mathbf{r}_j,...,\mathbf{r}_N)d\mathbf{r}_1d\mathbf{r}_2...d\mathbf{r}_i...d\mathbf{r}_N\\&\cdot p(\mathbf{r}_1,\mathbf{r}_2,...,\mathbf{r}_i,...,\mathbf{r}_j,...,\mathbf{r}_N)
\end{split}
\end{equation}
where $i\ne j$.	
Similarly, if we fix two particles $\mathbf{r}_i$ and $\mathbf{r}_j$, the ensemble average is expressed as
\begin{equation}
\begin{split}
\langle Q\rangle_{ij}=\int& Q(\mathbf{r}_1,\mathbf{r}_2,...,\mathbf{r}_i,...,\mathbf{r}_j,...,\mathbf{r}_l,...\mathbf{r}_N)d\mathbf{r}_1d\mathbf{r}_2...d\mathbf{r}_l...d\mathbf{r}_N\\&\cdot p(\mathbf{r}_1,\mathbf{r}_2,...,\mathbf{r}_i,...,\mathbf{r}_j,...,\mathbf{r}_l,...\mathbf{r}_N)
\end{split}
\end{equation}
where $i\ne j$, $l\ne j$ and $l\ne i$.
The relation between $\langle Q\rangle_{ij}$ and $\langle Q\rangle_{j}$ can be derived as
\begin{equation}
\langle Q\rangle_j=\int\langle Q\rangle_{ij}p(\mathbf{r}_i|\mathbf{r}_j)d\mathbf{r}_i
\end{equation}
where $p(\mathbf{r}_i|\mathbf{r}_j)$ is the conditional probability density function of $\mathbf{r}_i$ for a fixed $\mathbf{r}_j$. The pair distribution function, is related to $p(\mathbf{r}_i|\mathbf{r}_j)$ as \cite{tsang2004scattering2}
\begin{equation}
p(\mathbf{r}_i|\mathbf{r}_j)=\frac{g_2(\mathbf{r}_i|\mathbf{r}_j)}{V}\frac{N}{N-1}
\end{equation}
where $V$ is the volume occupied by the ensemble of particles. In the thermodynamic limit, $N\rightarrow \infty$, $p(\mathbf{r}_i|\mathbf{r}_j)\approx g_2(\mathbf{r}_i|\mathbf{r}_j)/V$.

\section{VSWFs and translation addition theorem}\label{vswf_appendix}
The regular VSWFs $\mathbf{N}^{(1)}_{mnp}(\mathbf{r})$ for $p=2$ (TE mode) and $p=1$ (TM mode) are defined as \cite{mackowskiJOSAA1996,mackowskiJQSRT2013,tsang2000scattering1,bohrenandhuffman,hulst1957}
\begin{equation}
\mathbf{N}^{(1)}_{mn2}(\mathbf{r})=\sqrt{\frac{(2n+1)(n-m)!}{4\pi n(n+1)(n+m)!}}\nabla\times(\mathbf{r}\psi_{mn}^{(1)}(\mathbf{r})),
\end{equation}
\begin{equation}
\mathbf{N}^{(1)}_{mn1}(\mathbf{r})=\frac{1}{k}\nabla\times\mathbf{N}^{(1)}_{mn2}(\mathbf{r})
\end{equation}
where $k=\omega/c$ is the wave number in free space and $\omega$ is the angular frequency of the electromagnetic wave. $\psi_{mn}^{(1)}(\mathbf{r})$ is regular (type-1) scalar wave function defined as
\begin{equation}
\psi_{mn}^{(1)}(\mathbf{r})=j_n(kr)Y_n^m(\theta,\phi),
\end{equation}
where $j_n(kr)$ is the spherical Bessel function and $Y_n^m(\theta,\phi)$ is spherical harmonics defined as
\begin{equation}
Y_n^m(\theta,\phi)=P_n^m(\cos\theta)\exp(im\phi),
\end{equation}
where we use the convention of quantum mechanics, and $P_n^m(\cos\theta)$ is associated Legendre polynomials.
The outgoing (type-3) VSWFs have can be similarly defined by replacing above spherical Bessel functions with the spherical Hankel functions of the first kind $h_n(kr)$.

It is straightforward to express the VSWFs into the vector spherical harmonics (VSHs) as
\begin{equation}
\mathbf{N}^{(1)}_{mn2}(\mathbf{r})=\sqrt{\frac{(2n+1)(n-m)!}{4\pi n(n+1)(n+m)!}}j_n(kr)\mathbf{C}_{mn}(\theta,\phi),
\end{equation}
\begin{equation}
\mathbf{N}^{(1)}_{mn1}(\mathbf{r})=\sqrt{\frac{(2n+1)(n-m)!}{4\pi n(n+1)(n+m)!}}\left\{\frac{n(n+1)j_n(kr)}{kr}\mathbf{P}_{mn}(\theta,\phi)+\frac{[krj_n(kr)]'}{kr}\mathbf{B}_{mn}(\theta,\phi)\right\},
\end{equation}
where $\mathbf{P}_{mn}(\theta,\phi)$, $\mathbf{B}_{mn}(\theta,\phi)$ and $\mathbf{C}_{mn}(\theta,\phi)$ are VSHs, given by
\begin{equation}
\mathbf{P}_{mn}(\theta,\phi)=\hat{\mathbf{r}}Y_n^m(\theta,\phi),
\end{equation}
\begin{equation}
\mathbf{B}_{mn}(\theta,\phi)=\left[\hat{\bm{\theta}}\frac{d}{d\theta}P_n^m(\cos\theta)+\hat{\bm{\phi}}\frac{im}{\sin\theta}P_n^m(\cos\theta)\right]\exp(im\phi),
\end{equation}
and
\begin{equation}
\mathbf{C}_{mn}(\theta,\phi)=\left[\hat{\bm{\theta}}\frac{im}{\sin\theta}P_n^m(\cos\theta)-\hat{\bm{\phi}}\frac{d}{d\theta}P_n^m(\cos\theta)\right]\exp(im\phi).
\end{equation}

The translation addition theorem of VSWFs, which transforms the VSWFs centered in $\mathbf{r}_i$ into those centered in $\mathbf{r}_j$, is given by
\begin{equation}
\mathbf{N}^{(3)}_{\mu\nu q}(\mathbf{r}-\mathbf{r}_i)=\sum_{\mu\nu q}A_{mnp\mu\nu q}^{(3)}(\mathbf{r}_i-\mathbf{r}_j)\mathbf{N}^{(1)}_{mnp}(\mathbf{r}-\mathbf{r}_j),
\end{equation}
which is valid for $|\mathbf{r}_i-\mathbf{r}_j|>|\mathbf{r}-\mathbf{r}_j|$, and therefore should be used in the vicinity of $\mathbf{r}_j$. The coefficient $A_{\mu qmp}^{(3)}$ is generally given by \cite{tsang2004scattering2}
\begin{equation}\label{translation_coef3}
\begin{split}
A_{mn1\mu\nu 1}^{(3)}(\mathbf{r})=A_{mn2\mu\nu 2}^{(3)}(\mathbf{r})=\frac{\gamma_{\mu\nu}}{\gamma_{mn}}(-1)^{m}\sum_{l}a(\mu,\nu|-m,n|l)a(\nu,n,l)h_l(kr)Y_{l}^{\mu-m}(\theta,\phi),
\end{split}
\end{equation}
\begin{equation}\label{translation_coef4}
\begin{split}
A_{mn1\mu\nu 2}^{(3)}(\mathbf{r})=A_{mn2\mu \nu 1}^{(3)}(\mathbf{r})=\frac{\gamma_{\mu\nu}}{\gamma_{mn}}(-1)^{m+1}\sum_{l}a(\mu,\nu|-m,n|l,l-1)b(\nu,n,l)h_l(kr)Y_{l}^{\mu-m}(\theta,\phi),
\end{split}
\end{equation}
where $\gamma_{mn}$ is defined as
\begin{equation}
\gamma_{mn}=\sqrt{\frac{(2n+1)(n-m)!}{4\pi n(n+1)(n+m)!}}.
\end{equation}
The coefficients $a(\mu,\nu|-m,n|l)$ and $a(\mu,\nu|-m,n|l,l-1)$ are given by
\begin{equation}
\begin{split}
a(\mu,\nu|-m,n|l)=&(-1)^{\mu-m}\left( 2l+1 \right) \left( \begin{matrix}
\nu&	 n&		l\\
\mu&    -m&		\mu-m\\
\end{matrix} \right) \\&\times\left( \begin{matrix}
\nu&	n&		l\\
0&		0&		0\\
\end{matrix} \right)\Big[\frac{(\nu+\mu)!(n-m)!(l-\mu+m)!}{(\nu-\mu)!(n+m)!(l+\mu-m)!}\Big]^{1/2},
\end{split}
\end{equation}
\begin{equation}
\begin{split}
a(\mu,\nu|-m,n|l,l-1)=&(-1)^{\mu-m}\left( 2l+1 \right) \left( \begin{matrix}
\nu&	 n&		l\\
\mu&    -m&		\mu-m\\
\end{matrix} \right) \\&\times\left( \begin{matrix}
\nu&	n&		l-1\\
0&		0&		0\\
\end{matrix} \right)\Big[\frac{(\nu+\mu)!(n-m)!(l-\mu+m)!}{(\nu-\mu)!(n+m)!(l+\mu-m)!}\Big]^{1/2},
\end{split}
\end{equation}
in which the variables in the form $\left( \begin{matrix}
j_1&	j_2&		j_3\\
m_1&		m_2&	m_3\\
\end{matrix} \right)$ are Wigner-3$j$ symbols. They can be found in Ref. \cite{abramowitz1964handbook,tsang2004scattering2} and not shown in detail here. Other coefficients $a(\nu,n,l)$ and $b(\nu,n,l)$ are given as \cite{tsang2004scattering2}
\begin{equation}
\begin{split}
a(\nu,n,l)=&\frac{i^{n+l-\nu}}{2n(n+1)}\Big[2n(n+1)(2n+1)+(n+1)(\nu+n-l)
(\nu+l-n+1)\\&-n(\nu+n+l+2)(n+l-\nu+1)\Big],
\end{split}
\end{equation}
\begin{equation}
\begin{split}
b(\nu,n,l)=-\frac{(2n+1)i^{n+l-\nu}}{2n(n+1)}\Big[(\nu+n+l+1)(n+l-\nu)
(\nu+l-n)(\nu+n-l+1)\Big]^{1/2}.
\end{split}
\end{equation}

We further give the expressions of the far-field approximation for outgoing VSWFs, which are usually used to calculate far-field scattered field and thus intensity-related quantities. For outgoing (type-3) VSWFs $\mathbf{N}^{(3)}_{mnp}(\mathbf{r}-\mathbf{r}_j)$ centered at $\mathbf{r}_j$, their far-field forms (when $r\gg r_j$) are given by \cite{mackowskiJOSAA1996,tsang2000scattering1,bohrenandhuffman}
\begin{equation}
\begin{split}
\mathbf{N}^{(3)}_{mn2}(\mathbf{r}-\mathbf{r}_j)\approx i^{-n}\sqrt{\frac{(2n+1)(n-m)!}{4\pi n(n+1)(n+m)!}}\frac{\exp (kr)}{kr}\exp (-\mathbf{k}_s\cdot\mathbf{r}_j)\mathbf{C}_{mn}(\theta,\phi),
\end{split}
\end{equation}
\begin{equation}
\begin{split}
\mathbf{N}^{(3)}_{mn1}(\mathbf{r}-\mathbf{r}_j)\approx i^{-n}\sqrt{\frac{(2n+1)(n-m)!}{4\pi n(n+1)(n+m)!}}\frac{\exp(kr)}{kr}\exp (-\mathbf{k}_s\cdot\mathbf{r}_j)\mathbf{B}_{mn}(\theta,\phi),
\end{split}
\end{equation}
where $\mathbf{B}_{mn}(\theta,\phi)$ and $\mathbf{C}_{mn}(\theta,\phi)$ are vector spherical harmonics. For homogeneous spheres, only $m=\pm1$ are needed. In this circumstance, we have
\begin{equation}
\mathbf{B}_{1n}(\theta,\phi)=-[\hat{\bm{\theta}}\tau_n(\cos\theta)+\hat{\bm{\phi}}\pi_n(\cos\theta)]\exp(i\phi),
\end{equation}
\begin{equation}
\mathbf{B}_{-1n}(\theta,\phi)=\frac{1}{n(n+1)}[\hat{\bm{\theta}}\tau_n(\cos\theta)-\hat{\bm{\phi}}\pi_n(\cos\theta)]\exp(-i\phi),
\end{equation}
\begin{equation}
\mathbf{C}_{1n}(\theta,\phi)=-[\hat{\bm{\theta}}i\pi_n(\cos\theta)-\hat{\bm{\phi}}\tau_n(\cos\theta)]\exp(i\phi),
\end{equation}
\begin{equation}
\mathbf{C}_{-1n}(\theta,\phi)=-\frac{1}{n(n+1)}[\hat{\bm{\theta}}i\pi_n(\cos\theta)+\hat{\bm{\phi}}\tau_n(\cos\theta)]\exp(-i\phi),
\end{equation}
where $\tau_n$ and $\pi_n$ are functions frequently used in standard light scattering monographs, defined as \cite{bohrenandhuffman}
\begin{equation}
\tau_n(\cos\theta)=-\frac{dP_n^1(\cos\theta)}{d\theta},
\end{equation}
\begin{equation}
\pi_n(\cos\theta)=-\frac{P_n^1(\cos\theta)}{\sin\theta}.
\end{equation}


\bibliographystyle{apsrev4-1}
\bibliography{References}

\begin{thebibliography}{530}%
\makeatletter
\providecommand \@ifxundefined [1]{%
 \@ifx{#1\undefined}
}%
\providecommand \@ifnum [1]{%
 \ifnum #1\expandafter \@firstoftwo
 \else \expandafter \@secondoftwo
 \fi
}%
\providecommand \@ifx [1]{%
 \ifx #1\expandafter \@firstoftwo
 \else \expandafter \@secondoftwo
 \fi
}%
\providecommand \natexlab [1]{#1}%
\providecommand \enquote  [1]{``#1''}%
\providecommand \bibnamefont  [1]{#1}%
\providecommand \bibfnamefont [1]{#1}%
\providecommand \citenamefont [1]{#1}%
\providecommand \href@noop [0]{\@secondoftwo}%
\providecommand \href [0]{\begingroup \@sanitize@url \@href}%
\providecommand \@href[1]{\@@startlink{#1}\@@href}%
\providecommand \@@href[1]{\endgroup#1\@@endlink}%
\providecommand \@sanitize@url [0]{\catcode `\\12\catcode `\$12\catcode
  `\&12\catcode `\#12\catcode `\^12\catcode `\_12\catcode `\%12\relax}%
\providecommand \@@startlink[1]{}%
\providecommand \@@endlink[0]{}%
\providecommand \url  [0]{\begingroup\@sanitize@url \@url }%
\providecommand \@url [1]{\endgroup\@href {#1}{\urlprefix }}%
\providecommand \urlprefix  [0]{URL }%
\providecommand \Eprint [0]{\href }%
\providecommand \doibase [0]{http://dx.doi.org/}%
\providecommand \selectlanguage [0]{\@gobble}%
\providecommand \bibinfo  [0]{\@secondoftwo}%
\providecommand \bibfield  [0]{\@secondoftwo}%
\providecommand \translation [1]{[#1]}%
\providecommand \BibitemOpen [0]{}%
\providecommand \bibitemStop [0]{}%
\providecommand \bibitemNoStop [0]{.\EOS\space}%
\providecommand \EOS [0]{\spacefactor3000\relax}%
\providecommand \BibitemShut  [1]{\csname bibitem#1\endcsname}%
\let\auto@bib@innerbib\@empty
%</preamble>
\bibitem [{\citenamefont {Born}\ and\ \citenamefont
  {Wolf}(2013)}]{bornandwolf}%
  \BibitemOpen
  \bibfield  {author} {\bibinfo {author} {\bibfnamefont {M.}~\bibnamefont
  {Born}}\ and\ \bibinfo {author} {\bibfnamefont {E.}~\bibnamefont {Wolf}},\
  }\href@noop {} {\emph {\bibinfo {title} {Principles of optics:
  electromagnetic theory of propagation, interference and diffraction of
  light}}}\ (\bibinfo  {publisher} {Elsevier},\ \bibinfo {year}
  {2013})\BibitemShut {NoStop}%
\bibitem [{\citenamefont {Min-Dianey}\ \emph {et~al.}(2017)\citenamefont
  {Min-Dianey}, \citenamefont {Zhang}, \citenamefont {M'Bouana}, \citenamefont
  {Su},\ and\ \citenamefont {Xia}}]{mindianeyCMS2017}%
  \BibitemOpen
  \bibfield  {author} {\bibinfo {author} {\bibfnamefont {K.~A.~A.}\
  \bibnamefont {Min-Dianey}}, \bibinfo {author} {\bibfnamefont {H.~C.}\
  \bibnamefont {Zhang}}, \bibinfo {author} {\bibfnamefont {N.~L.~P.}\
  \bibnamefont {M'Bouana}}, \bibinfo {author} {\bibfnamefont {C.~S.}\
  \bibnamefont {Su}}, \ and\ \bibinfo {author} {\bibfnamefont {X.~L.}\
  \bibnamefont {Xia}},\ }\href {\doibase
  https://doi.org/10.1016/j.commatsci.2017.05.016} {\bibfield  {journal}
  {\bibinfo  {journal} {Computational Materials Science}\ }\textbf {\bibinfo
  {volume} {136}},\ \bibinfo {pages} {306 } (\bibinfo {year}
  {2017})}\BibitemShut {NoStop}%
\bibitem [{\citenamefont {Garc\'{\i}a}\ \emph {et~al.}(2007)\citenamefont
  {Garc\'{\i}a}, \citenamefont {Sapienza}, \citenamefont {Blanco},\ and\
  \citenamefont {Lopez}}]{garciaADMA2007}%
  \BibitemOpen
  \bibfield  {author} {\bibinfo {author} {\bibfnamefont {P.}~\bibnamefont
  {Garc\'{\i}a}}, \bibinfo {author} {\bibfnamefont {R.}~\bibnamefont
  {Sapienza}}, \bibinfo {author} {\bibfnamefont {A.}~\bibnamefont {Blanco}}, \
  and\ \bibinfo {author} {\bibfnamefont {C.}~\bibnamefont {Lopez}},\ }\href
  {\doibase 10.1002/adma.200602426} {\bibfield  {journal} {\bibinfo  {journal}
  {Advanced Materials}\ }\textbf {\bibinfo {volume} {19}},\ \bibinfo {pages}
  {2597} (\bibinfo {year} {2007})}\BibitemShut {NoStop}%
\bibitem [{\citenamefont {Rojas-Ochoa}\ \emph {et~al.}(2004)\citenamefont
  {Rojas-Ochoa}, \citenamefont {Mendez-Alcaraz}, \citenamefont {S\'aenz},
  \citenamefont {Schurtenberger},\ and\ \citenamefont
  {Scheffold}}]{rojasochoaPRL2004}%
  \BibitemOpen
  \bibfield  {author} {\bibinfo {author} {\bibfnamefont {L.~F.}\ \bibnamefont
  {Rojas-Ochoa}}, \bibinfo {author} {\bibfnamefont {J.~M.}\ \bibnamefont
  {Mendez-Alcaraz}}, \bibinfo {author} {\bibfnamefont {J.~J.}\ \bibnamefont
  {S\'aenz}}, \bibinfo {author} {\bibfnamefont {P.}~\bibnamefont
  {Schurtenberger}}, \ and\ \bibinfo {author} {\bibfnamefont {F.}~\bibnamefont
  {Scheffold}},\ }\href {\doibase 10.1103/PhysRevLett.93.073903} {\bibfield
  {journal} {\bibinfo  {journal} {Phys. Rev. Lett.}\ }\textbf {\bibinfo
  {volume} {93}},\ \bibinfo {pages} {073903} (\bibinfo {year}
  {2004})}\BibitemShut {NoStop}%
\bibitem [{\citenamefont {Kulkarni}\ \emph {et~al.}(2003)\citenamefont
  {Kulkarni}, \citenamefont {Wang}, \citenamefont {Nakamura}, \citenamefont
  {Sampath}, \citenamefont {Goland}, \citenamefont {Herman}, \citenamefont
  {Allen}, \citenamefont {Ilavsky}, \citenamefont {Long}, \citenamefont
  {Frahm},\ and\ \citenamefont {Steinbrech}}]{kulkarniActMat2003}%
  \BibitemOpen
  \bibfield  {author} {\bibinfo {author} {\bibfnamefont {A.}~\bibnamefont
  {Kulkarni}}, \bibinfo {author} {\bibfnamefont {Z.}~\bibnamefont {Wang}},
  \bibinfo {author} {\bibfnamefont {T.}~\bibnamefont {Nakamura}}, \bibinfo
  {author} {\bibfnamefont {S.}~\bibnamefont {Sampath}}, \bibinfo {author}
  {\bibfnamefont {A.}~\bibnamefont {Goland}}, \bibinfo {author} {\bibfnamefont
  {H.}~\bibnamefont {Herman}}, \bibinfo {author} {\bibfnamefont
  {J.}~\bibnamefont {Allen}}, \bibinfo {author} {\bibfnamefont
  {J.}~\bibnamefont {Ilavsky}}, \bibinfo {author} {\bibfnamefont
  {G.}~\bibnamefont {Long}}, \bibinfo {author} {\bibfnamefont {J.}~\bibnamefont
  {Frahm}}, \ and\ \bibinfo {author} {\bibfnamefont {R.}~\bibnamefont
  {Steinbrech}},\ }\href {\doibase
  http://dx.doi.org/10.1016/S1359-6454(03)00030-2} {\bibfield  {journal}
  {\bibinfo  {journal} {Acta Materialia}\ }\textbf {\bibinfo {volume} {51}},\
  \bibinfo {pages} {2457 } (\bibinfo {year} {2003})}\BibitemShut {NoStop}%
\bibitem [{\citenamefont {Lu}\ \emph {et~al.}(1998)\citenamefont {Lu},
  \citenamefont {Stone},\ and\ \citenamefont {Ashby}}]{luAM1998}%
  \BibitemOpen
  \bibfield  {author} {\bibinfo {author} {\bibfnamefont {T.}~\bibnamefont
  {Lu}}, \bibinfo {author} {\bibfnamefont {H.}~\bibnamefont {Stone}}, \ and\
  \bibinfo {author} {\bibfnamefont {M.}~\bibnamefont {Ashby}},\ }\href
  {\doibase https://doi.org/10.1016/S1359-6454(98)00031-7} {\bibfield
  {journal} {\bibinfo  {journal} {Acta Materialia}\ }\textbf {\bibinfo {volume}
  {46}},\ \bibinfo {pages} {3619 } (\bibinfo {year} {1998})}\BibitemShut
  {NoStop}%
\bibitem [{\citenamefont {Jiang}\ \emph {et~al.}(2018)\citenamefont {Jiang},
  \citenamefont {Liu}, \citenamefont {Li}, \citenamefont {Kuang}, \citenamefont
  {Xu}, \citenamefont {Chen}, \citenamefont {Huang}, \citenamefont {Jia},
  \citenamefont {Zhao}, \citenamefont {Hitz}, \citenamefont {Zhou},
  \citenamefont {Yang}, \citenamefont {Cui},\ and\ \citenamefont
  {Hu}}]{jiangACSAMI2018}%
  \BibitemOpen
  \bibfield  {author} {\bibinfo {author} {\bibfnamefont {F.}~\bibnamefont
  {Jiang}}, \bibinfo {author} {\bibfnamefont {H.}~\bibnamefont {Liu}}, \bibinfo
  {author} {\bibfnamefont {Y.}~\bibnamefont {Li}}, \bibinfo {author}
  {\bibfnamefont {Y.}~\bibnamefont {Kuang}}, \bibinfo {author} {\bibfnamefont
  {X.}~\bibnamefont {Xu}}, \bibinfo {author} {\bibfnamefont {C.}~\bibnamefont
  {Chen}}, \bibinfo {author} {\bibfnamefont {H.}~\bibnamefont {Huang}},
  \bibinfo {author} {\bibfnamefont {C.}~\bibnamefont {Jia}}, \bibinfo {author}
  {\bibfnamefont {X.}~\bibnamefont {Zhao}}, \bibinfo {author} {\bibfnamefont
  {E.}~\bibnamefont {Hitz}}, \bibinfo {author} {\bibfnamefont {Y.}~\bibnamefont
  {Zhou}}, \bibinfo {author} {\bibfnamefont {R.}~\bibnamefont {Yang}}, \bibinfo
  {author} {\bibfnamefont {L.}~\bibnamefont {Cui}}, \ and\ \bibinfo {author}
  {\bibfnamefont {L.}~\bibnamefont {Hu}},\ }\href {\doibase
  10.1021/acsami.7b15125} {\bibfield  {journal} {\bibinfo  {journal} {ACS
  Applied Materials \& Interfaces}\ }\textbf {\bibinfo {volume} {10}},\
  \bibinfo {pages} {1104} (\bibinfo {year} {2018})},\ \bibinfo {note} {pMID:
  29182304},\ \Eprint
  {http://arxiv.org/abs/https://doi.org/10.1021/acsami.7b15125}
  {https://doi.org/10.1021/acsami.7b15125} \BibitemShut {NoStop}%
\bibitem [{\citenamefont {Sun}\ \emph {et~al.}(2011)\citenamefont {Sun},
  \citenamefont {Lou},\ and\ \citenamefont {Zhou}}]{sunIJHMT2011}%
  \BibitemOpen
  \bibfield  {author} {\bibinfo {author} {\bibfnamefont {Y.-P.}\ \bibnamefont
  {Sun}}, \bibinfo {author} {\bibfnamefont {C.}~\bibnamefont {Lou}}, \ and\
  \bibinfo {author} {\bibfnamefont {H.-C.}\ \bibnamefont {Zhou}},\ }\href
  {\doibase https://doi.org/10.1016/j.ijheatmasstransfer.2010.09.049}
  {\bibfield  {journal} {\bibinfo  {journal} {International Journal of Heat and
  Mass Transfer}\ }\textbf {\bibinfo {volume} {54}},\ \bibinfo {pages} {217 }
  (\bibinfo {year} {2011})}\BibitemShut {NoStop}%
\bibitem [{\citenamefont {Wang}\ \emph {et~al.}(2014)\citenamefont {Wang},
  \citenamefont {Tan}, \citenamefont {Ma}, \citenamefont {Shuai}, \citenamefont
  {Tan},\ and\ \citenamefont {Leng}}]{tanSE2014}%
  \BibitemOpen
  \bibfield  {author} {\bibinfo {author} {\bibfnamefont {F.}~\bibnamefont
  {Wang}}, \bibinfo {author} {\bibfnamefont {J.}~\bibnamefont {Tan}}, \bibinfo
  {author} {\bibfnamefont {L.}~\bibnamefont {Ma}}, \bibinfo {author}
  {\bibfnamefont {Y.}~\bibnamefont {Shuai}}, \bibinfo {author} {\bibfnamefont
  {H.}~\bibnamefont {Tan}}, \ and\ \bibinfo {author} {\bibfnamefont
  {Y.}~\bibnamefont {Leng}},\ }\href {\doibase
  https://doi.org/10.1016/j.solener.2014.07.016} {\bibfield  {journal}
  {\bibinfo  {journal} {Solar Energy}\ }\textbf {\bibinfo {volume} {108}},\
  \bibinfo {pages} {348 } (\bibinfo {year} {2014})}\BibitemShut {NoStop}%
\bibitem [{\citenamefont {Yang}\ \emph {et~al.}(2013)\citenamefont {Yang},
  \citenamefont {Zhao},\ and\ \citenamefont {Wang}}]{yangIJHMT2013}%
  \BibitemOpen
  \bibfield  {author} {\bibinfo {author} {\bibfnamefont {G.}~\bibnamefont
  {Yang}}, \bibinfo {author} {\bibfnamefont {C.~Y.}\ \bibnamefont {Zhao}}, \
  and\ \bibinfo {author} {\bibfnamefont {B.~X.}\ \bibnamefont {Wang}},\ }\href
  {\doibase http://dx.doi.org/10.1016/j.ijheatmasstransfer.2013.07.069}
  {\bibfield  {journal} {\bibinfo  {journal} {International Journal of Heat and
  Mass Transfer}\ }\textbf {\bibinfo {volume} {66}},\ \bibinfo {pages} {695 }
  (\bibinfo {year} {2013})}\BibitemShut {NoStop}%
\bibitem [{\citenamefont {Zhai}\ \emph {et~al.}(2017)\citenamefont {Zhai},
  \citenamefont {Ma}, \citenamefont {David}, \citenamefont {Zhao},
  \citenamefont {Lou}, \citenamefont {Tan}, \citenamefont {Yang},\ and\
  \citenamefont {Yin}}]{zhaiScience2017}%
  \BibitemOpen
  \bibfield  {author} {\bibinfo {author} {\bibfnamefont {Y.}~\bibnamefont
  {Zhai}}, \bibinfo {author} {\bibfnamefont {Y.}~\bibnamefont {Ma}}, \bibinfo
  {author} {\bibfnamefont {S.~N.}\ \bibnamefont {David}}, \bibinfo {author}
  {\bibfnamefont {D.}~\bibnamefont {Zhao}}, \bibinfo {author} {\bibfnamefont
  {R.}~\bibnamefont {Lou}}, \bibinfo {author} {\bibfnamefont {G.}~\bibnamefont
  {Tan}}, \bibinfo {author} {\bibfnamefont {R.}~\bibnamefont {Yang}}, \ and\
  \bibinfo {author} {\bibfnamefont {X.}~\bibnamefont {Yin}},\ }\href {\doibase
  10.1126/science.aai7899} {\bibfield  {journal} {\bibinfo  {journal}
  {Science}\ }\textbf {\bibinfo {volume} {355}},\ \bibinfo {pages} {1062}
  (\bibinfo {year} {2017})}\BibitemShut {NoStop}%
\bibitem [{\citenamefont {Lagendijk}\ and\ \citenamefont
  {Van~Tiggelen}(1996)}]{lagendijk1996resonant}%
  \BibitemOpen
  \bibfield  {author} {\bibinfo {author} {\bibfnamefont {A.}~\bibnamefont
  {Lagendijk}}\ and\ \bibinfo {author} {\bibfnamefont {B.~A.}\ \bibnamefont
  {Van~Tiggelen}},\ }\href@noop {} {\bibfield  {journal} {\bibinfo  {journal}
  {Phys. Rep.}\ }\textbf {\bibinfo {volume} {270}},\ \bibinfo {pages} {143}
  (\bibinfo {year} {1996})}\BibitemShut {NoStop}%
\bibitem [{\citenamefont {van Rossum}\ and\ \citenamefont
  {Nieuwenhuizen}(1999)}]{VanRossum1998}%
  \BibitemOpen
  \bibfield  {author} {\bibinfo {author} {\bibfnamefont {M.~C.~W.}\
  \bibnamefont {van Rossum}}\ and\ \bibinfo {author} {\bibfnamefont {T.~M.}\
  \bibnamefont {Nieuwenhuizen}},\ }\href {\doibase 10.1103/RevModPhys.71.313}
  {\bibfield  {journal} {\bibinfo  {journal} {Rev. Mod. Phys.}\ }\textbf
  {\bibinfo {volume} {71}},\ \bibinfo {pages} {313} (\bibinfo {year}
  {1999})}\BibitemShut {NoStop}%
\bibitem [{\citenamefont {Tsang}\ and\ \citenamefont
  {Kong}(2004)}]{tsang2004scattering}%
  \BibitemOpen
  \bibfield  {author} {\bibinfo {author} {\bibfnamefont {L.}~\bibnamefont
  {Tsang}}\ and\ \bibinfo {author} {\bibfnamefont {J.~A.}\ \bibnamefont
  {Kong}},\ }\href@noop {} {\emph {\bibinfo {title} {Scattering of
  Electromagnetic Waves: Advanced Topics}}}\ (\bibinfo  {publisher} {John Wiley
  \& Sons},\ \bibinfo {year} {2004})\BibitemShut {NoStop}%
\bibitem [{\citenamefont {Sheng}(2006)}]{sheng2006introduction}%
  \BibitemOpen
  \bibfield  {author} {\bibinfo {author} {\bibfnamefont {P.}~\bibnamefont
  {Sheng}},\ }\href@noop {} {\emph {\bibinfo {title} {Introduction to Wave
  Scattering, Localization and Mesoscopic Phenomena}}}\ (\bibinfo  {publisher}
  {Springer Science \& Business Media},\ \bibinfo {year} {2006})\BibitemShut
  {NoStop}%
\bibitem [{\citenamefont {Mishchenko}\ \emph {et~al.}(2010)\citenamefont
  {Mishchenko}, \citenamefont {Rosenbush}, \citenamefont {Kiselev},
  \citenamefont {Lupishko}, \citenamefont {Tishkovets}, \citenamefont
  {Kaydash}, \citenamefont {Belskaya}, \citenamefont {Efimov},\ and\
  \citenamefont {Shakhovskoy}}]{mishchenko2010polarimetric}%
  \BibitemOpen
  \bibfield  {author} {\bibinfo {author} {\bibfnamefont {M.~I.}\ \bibnamefont
  {Mishchenko}}, \bibinfo {author} {\bibfnamefont {V.}~\bibnamefont
  {Rosenbush}}, \bibinfo {author} {\bibfnamefont {N.}~\bibnamefont {Kiselev}},
  \bibinfo {author} {\bibfnamefont {D.}~\bibnamefont {Lupishko}}, \bibinfo
  {author} {\bibfnamefont {V.}~\bibnamefont {Tishkovets}}, \bibinfo {author}
  {\bibfnamefont {V.}~\bibnamefont {Kaydash}}, \bibinfo {author} {\bibfnamefont
  {I.}~\bibnamefont {Belskaya}}, \bibinfo {author} {\bibfnamefont {Y.~S.}\
  \bibnamefont {Efimov}}, \ and\ \bibinfo {author} {\bibfnamefont
  {N.}~\bibnamefont {Shakhovskoy}},\ }\href@noop {} {\bibfield  {journal}
  {\bibinfo  {journal} {arXiv preprint arXiv:1010.1171}\ } (\bibinfo {year}
  {2010})}\BibitemShut {NoStop}%
\bibitem [{\citenamefont {Mishchenko}\ \emph {et~al.}(2006)\citenamefont
  {Mishchenko}, \citenamefont {Travis},\ and\ \citenamefont
  {Lacis}}]{mishchenko2006multiple}%
  \BibitemOpen
  \bibfield  {author} {\bibinfo {author} {\bibfnamefont {M.~I.}\ \bibnamefont
  {Mishchenko}}, \bibinfo {author} {\bibfnamefont {L.~D.}\ \bibnamefont
  {Travis}}, \ and\ \bibinfo {author} {\bibfnamefont {A.~A.}\ \bibnamefont
  {Lacis}},\ }\href@noop {} {\emph {\bibinfo {title} {Multiple scattering of
  light by particles: radiative transfer and coherent backscattering}}}\
  (\bibinfo  {publisher} {Cambridge University Press},\ \bibinfo {year}
  {2006})\BibitemShut {NoStop}%
\bibitem [{\citenamefont {Garc\'{\i}a}\ \emph {et~al.}(2008)\citenamefont
  {Garc\'{\i}a}, \citenamefont {Sapienza}, \citenamefont {Bertolotti},
  \citenamefont {Mart\'{\i}n}, \citenamefont {Blanco}, \citenamefont {Altube},
  \citenamefont {Vi\~na}, \citenamefont {Wiersma},\ and\ \citenamefont
  {L\'opez}}]{garciaPRA2008}%
  \BibitemOpen
  \bibfield  {author} {\bibinfo {author} {\bibfnamefont {P.~D.}\ \bibnamefont
  {Garc\'{\i}a}}, \bibinfo {author} {\bibfnamefont {R.}~\bibnamefont
  {Sapienza}}, \bibinfo {author} {\bibfnamefont {J.}~\bibnamefont
  {Bertolotti}}, \bibinfo {author} {\bibfnamefont {M.~D.}\ \bibnamefont
  {Mart\'{\i}n}}, \bibinfo {author} {\bibfnamefont {A.}~\bibnamefont {Blanco}},
  \bibinfo {author} {\bibfnamefont {A.}~\bibnamefont {Altube}}, \bibinfo
  {author} {\bibfnamefont {L.}~\bibnamefont {Vi\~na}}, \bibinfo {author}
  {\bibfnamefont {D.~S.}\ \bibnamefont {Wiersma}}, \ and\ \bibinfo {author}
  {\bibfnamefont {C.}~\bibnamefont {L\'opez}},\ }\href {\doibase
  10.1103/PhysRevA.78.023823} {\bibfield  {journal} {\bibinfo  {journal} {Phys.
  Rev. A}\ }\textbf {\bibinfo {volume} {78}},\ \bibinfo {pages} {023823}
  (\bibinfo {year} {2008})}\BibitemShut {NoStop}%
\bibitem [{\citenamefont {Wang}\ and\ \citenamefont
  {Zhao}(2015)}]{wangIJHMT2015}%
  \BibitemOpen
  \bibfield  {author} {\bibinfo {author} {\bibfnamefont {B.~X.}\ \bibnamefont
  {Wang}}\ and\ \bibinfo {author} {\bibfnamefont {C.~Y.}\ \bibnamefont
  {Zhao}},\ }\href {\doibase 10.1016/j.ijheatmasstransfer.2015.06.017}
  {\bibfield  {journal} {\bibinfo  {journal} {International Journal of Heat and
  Mass Transfer}\ }\textbf {\bibinfo {volume} {89}},\ \bibinfo {pages} {920}
  (\bibinfo {year} {2015})}\BibitemShut {NoStop}%
\bibitem [{\citenamefont {Rezvani~Naraghi}\ \emph {et~al.}(2015)\citenamefont
  {Rezvani~Naraghi}, \citenamefont {Sukhov}, \citenamefont {S\'aenz},\ and\
  \citenamefont {Dogariu}}]{Naraghi2015}%
  \BibitemOpen
  \bibfield  {author} {\bibinfo {author} {\bibfnamefont {R.}~\bibnamefont
  {Rezvani~Naraghi}}, \bibinfo {author} {\bibfnamefont {S.}~\bibnamefont
  {Sukhov}}, \bibinfo {author} {\bibfnamefont {J.~J.}\ \bibnamefont {S\'aenz}},
  \ and\ \bibinfo {author} {\bibfnamefont {A.}~\bibnamefont {Dogariu}},\ }\href
  {\doibase 10.1103/PhysRevLett.115.203903} {\bibfield  {journal} {\bibinfo
  {journal} {Phys. Rev. Lett.}\ }\textbf {\bibinfo {volume} {115}},\ \bibinfo
  {pages} {203903} (\bibinfo {year} {2015})}\BibitemShut {NoStop}%
\bibitem [{\citenamefont {Wang}\ and\ \citenamefont
  {Zhao}(2018{\natexlab{a}})}]{wangIJHMT2018}%
  \BibitemOpen
  \bibfield  {author} {\bibinfo {author} {\bibfnamefont {B.~X.}\ \bibnamefont
  {Wang}}\ and\ \bibinfo {author} {\bibfnamefont {C.~Y.}\ \bibnamefont
  {Zhao}},\ }\href {\doibase 10.1016/j.ijheatmasstransfer.2018.05.004}
  {\bibfield  {journal} {\bibinfo  {journal} {International Journal of Heat and
  Mass Transfer}\ }\textbf {\bibinfo {volume} {125}},\ \bibinfo {pages} {1069 }
  (\bibinfo {year} {2018}{\natexlab{a}})}\BibitemShut {NoStop}%
\bibitem [{\citenamefont {Tinsley}\ \emph {et~al.}(1949)\citenamefont
  {Tinsley}, \citenamefont {Bowman},\ and\ \citenamefont
  {Phil}}]{tinsleyJOCCA1949}%
  \BibitemOpen
  \bibfield  {author} {\bibinfo {author} {\bibfnamefont {S.}~\bibnamefont
  {Tinsley}}, \bibinfo {author} {\bibfnamefont {A.}~\bibnamefont {Bowman}}, \
  and\ \bibinfo {author} {\bibfnamefont {D.}~\bibnamefont {Phil}},\ }\href@noop
  {} {\bibfield  {journal} {\bibinfo  {journal} {J. Oil Colour Chem. Assoc}\
  }\textbf {\bibinfo {volume} {32}},\ \bibinfo {pages} {233} (\bibinfo {year}
  {1949})}\BibitemShut {NoStop}%
\bibitem [{\citenamefont {Stieg~Jr}(1959)}]{stieg1959effect}%
  \BibitemOpen
  \bibfield  {author} {\bibinfo {author} {\bibfnamefont {F.}~\bibnamefont
  {Stieg~Jr}},\ }\href@noop {} {\bibfield  {journal} {\bibinfo  {journal}
  {Official Dig. Fed. Paint Varnish Prod. Clubs}\ }\textbf {\bibinfo {volume}
  {31}},\ \bibinfo {pages} {52} (\bibinfo {year} {1959})}\BibitemShut {NoStop}%
\bibitem [{\citenamefont {Hulst}(1957)}]{hulst1957}%
  \BibitemOpen
  \bibfield  {author} {\bibinfo {author} {\bibfnamefont {H.~C.}\ \bibnamefont
  {Hulst}},\ }\href@noop {} {\emph {\bibinfo {title} {Light scattering by small
  particles}}}\ (\bibinfo  {publisher} {Courier Corporation},\ \bibinfo {year}
  {1957})\BibitemShut {NoStop}%
\bibitem [{\citenamefont {Churchill}\ \emph {et~al.}(1960)\citenamefont
  {Churchill}, \citenamefont {Clark},\ and\ \citenamefont
  {Sliepcevich}}]{churchillDFC1960}%
  \BibitemOpen
  \bibfield  {author} {\bibinfo {author} {\bibfnamefont {S.~W.}\ \bibnamefont
  {Churchill}}, \bibinfo {author} {\bibfnamefont {G.~C.}\ \bibnamefont
  {Clark}}, \ and\ \bibinfo {author} {\bibfnamefont {C.~M.}\ \bibnamefont
  {Sliepcevich}},\ }\href {\doibase 10.1039/DF9603000192} {\bibfield  {journal}
  {\bibinfo  {journal} {Discuss. Faraday Soc.}\ }\textbf {\bibinfo {volume}
  {30}},\ \bibinfo {pages} {192} (\bibinfo {year} {1960})}\BibitemShut
  {NoStop}%
\bibitem [{\citenamefont {Harding}\ \emph {et~al.}(1960)\citenamefont
  {Harding}, \citenamefont {Golding},\ and\ \citenamefont
  {Morgen}}]{hardingJOSA1960}%
  \BibitemOpen
  \bibfield  {author} {\bibinfo {author} {\bibfnamefont {R.~H.}\ \bibnamefont
  {Harding}}, \bibinfo {author} {\bibfnamefont {B.}~\bibnamefont {Golding}}, \
  and\ \bibinfo {author} {\bibfnamefont {R.~A.}\ \bibnamefont {Morgen}},\
  }\href {\doibase 10.1364/JOSA.50.000446} {\bibfield  {journal} {\bibinfo
  {journal} {J. Opt. Soc. Am.}\ }\textbf {\bibinfo {volume} {50}},\ \bibinfo
  {pages} {446} (\bibinfo {year} {1960})}\BibitemShut {NoStop}%
\bibitem [{\citenamefont {Blevin}\ and\ \citenamefont
  {Brown}(1961)}]{blevinJOSA1961a}%
  \BibitemOpen
  \bibfield  {author} {\bibinfo {author} {\bibfnamefont {W.~R.}\ \bibnamefont
  {Blevin}}\ and\ \bibinfo {author} {\bibfnamefont {W.~J.}\ \bibnamefont
  {Brown}},\ }\href {\doibase 10.1364/JOSA.51.000129} {\bibfield  {journal}
  {\bibinfo  {journal} {J. Opt. Soc. Am.}\ }\textbf {\bibinfo {volume} {51}},\
  \bibinfo {pages} {129} (\bibinfo {year} {1961})}\BibitemShut {NoStop}%
\bibitem [{\citenamefont {Rozenberg}(1962)}]{rozenberg1962optical}%
  \BibitemOpen
  \bibfield  {author} {\bibinfo {author} {\bibfnamefont {G.}~\bibnamefont
  {Rozenberg}},\ }in\ \href@noop {} {\emph {\bibinfo {booktitle} {Dokl. Akad.
  Nauk SSSR}}},\ Vol.\ \bibinfo {volume} {145}\ (\bibinfo {year} {1962})\ pp.\
  \bibinfo {pages} {775--777}\BibitemShut {NoStop}%
\bibitem [{\citenamefont {Foldy}(1945)}]{foldyPR1945}%
  \BibitemOpen
  \bibfield  {author} {\bibinfo {author} {\bibfnamefont {L.~L.}\ \bibnamefont
  {Foldy}},\ }\href {\doibase 10.1103/PhysRev.67.107} {\bibfield  {journal}
  {\bibinfo  {journal} {Phys. Rev.}\ }\textbf {\bibinfo {volume} {67}},\
  \bibinfo {pages} {107} (\bibinfo {year} {1945})}\BibitemShut {NoStop}%
\bibitem [{\citenamefont {Lax}(1951)}]{laxRMP1951}%
  \BibitemOpen
  \bibfield  {author} {\bibinfo {author} {\bibfnamefont {M.}~\bibnamefont
  {Lax}},\ }\href {\doibase 10.1103/RevModPhys.23.287} {\bibfield  {journal}
  {\bibinfo  {journal} {Rev. Mod. Phys.}\ }\textbf {\bibinfo {volume} {23}},\
  \bibinfo {pages} {287} (\bibinfo {year} {1951})}\BibitemShut {NoStop}%
\bibitem [{\citenamefont {Lax}(1952)}]{laxPR1952}%
  \BibitemOpen
  \bibfield  {author} {\bibinfo {author} {\bibfnamefont {M.}~\bibnamefont
  {Lax}},\ }\href {\doibase 10.1103/PhysRev.85.621} {\bibfield  {journal}
  {\bibinfo  {journal} {Physical Review}\ }\textbf {\bibinfo {volume} {85}},\
  \bibinfo {pages} {621} (\bibinfo {year} {1952})}\BibitemShut {NoStop}%
\bibitem [{\citenamefont {Twersky}(1952)}]{twerskyJASA1952}%
  \BibitemOpen
  \bibfield  {author} {\bibinfo {author} {\bibfnamefont {V.}~\bibnamefont
  {Twersky}},\ }\href {\doibase 10.1121/1.1906845} {\bibfield  {journal}
  {\bibinfo  {journal} {The Journal of the Acoustical Society of America}\
  }\textbf {\bibinfo {volume} {24}},\ \bibinfo {pages} {42} (\bibinfo {year}
  {1952})},\ \Eprint {http://arxiv.org/abs/https://doi.org/10.1121/1.1906845}
  {https://doi.org/10.1121/1.1906845} \BibitemShut {NoStop}%
\bibitem [{\citenamefont {Hottel}\ \emph {et~al.}(1970)\citenamefont {Hottel},
  \citenamefont {Sarofim}, \citenamefont {Vasalos},\ and\ \citenamefont
  {Dalzell}}]{hottelJHT1970}%
  \BibitemOpen
  \bibfield  {author} {\bibinfo {author} {\bibfnamefont {H.~C.}\ \bibnamefont
  {Hottel}}, \bibinfo {author} {\bibfnamefont {A.~F.}\ \bibnamefont {Sarofim}},
  \bibinfo {author} {\bibfnamefont {I.~A.}\ \bibnamefont {Vasalos}}, \ and\
  \bibinfo {author} {\bibfnamefont {W.~H.}\ \bibnamefont {Dalzell}},\ }\href
  {\doibase 10.1115/1.3449662} {\bibfield  {journal} {\bibinfo  {journal}
  {Journal of Heat Transfer}\ }\textbf {\bibinfo {volume} {92}},\ \bibinfo
  {pages} {285} (\bibinfo {year} {1970})}\BibitemShut {NoStop}%
\bibitem [{\citenamefont {Hottel}\ \emph {et~al.}(1971)\citenamefont {Hottel},
  \citenamefont {Sarofim}, \citenamefont {Dalzeil},\ and\ \citenamefont
  {Vasalos}}]{hottelAIAAJ1971}%
  \BibitemOpen
  \bibfield  {author} {\bibinfo {author} {\bibfnamefont {H.~C.}\ \bibnamefont
  {Hottel}}, \bibinfo {author} {\bibfnamefont {A.~F.}\ \bibnamefont {Sarofim}},
  \bibinfo {author} {\bibfnamefont {W.~H.}\ \bibnamefont {Dalzeil}}, \ and\
  \bibinfo {author} {\bibfnamefont {I.~A.}\ \bibnamefont {Vasalos}},\ }\href
  {\doibase 10.2514/3.49999} {\bibfield  {journal} {\bibinfo  {journal} {AIAA
  Journal}\ }\textbf {\bibinfo {volume} {9}},\ \bibinfo {pages} {1895}
  (\bibinfo {year} {1971})}\BibitemShut {NoStop}%
\bibitem [{\citenamefont {Brewster}\ and\ \citenamefont
  {Tien}(1982)}]{brewsterJHT1982}%
  \BibitemOpen
  \bibfield  {author} {\bibinfo {author} {\bibfnamefont {M.~Q.}\ \bibnamefont
  {Brewster}}\ and\ \bibinfo {author} {\bibfnamefont {C.~L.}\ \bibnamefont
  {Tien}},\ }\href {\doibase 10.1115/1.3245170} {\bibfield  {journal} {\bibinfo
   {journal} {Journal of Heat Transfer}\ }\textbf {\bibinfo {volume} {104}},\
  \bibinfo {pages} {573} (\bibinfo {year} {1982})}\BibitemShut {NoStop}%
\bibitem [{\citenamefont {Yamada}\ \emph {et~al.}(1986)\citenamefont {Yamada},
  \citenamefont {Cartigny},\ and\ \citenamefont {Tien}}]{yamadaJHT1986}%
  \BibitemOpen
  \bibfield  {author} {\bibinfo {author} {\bibfnamefont {Y.}~\bibnamefont
  {Yamada}}, \bibinfo {author} {\bibfnamefont {J.~D.}\ \bibnamefont
  {Cartigny}}, \ and\ \bibinfo {author} {\bibfnamefont {C.~L.}\ \bibnamefont
  {Tien}},\ }\href {\doibase 10.1115/1.3246980} {\bibfield  {journal} {\bibinfo
   {journal} {Journal of Heat Transfer}\ }\textbf {\bibinfo {volume} {108}}
  (\bibinfo {year} {1986}),\ 10.1115/1.3246980}\BibitemShut {NoStop}%
\bibitem [{\citenamefont {Cartigny}\ \emph {et~al.}(1986)\citenamefont
  {Cartigny}, \citenamefont {Yamada},\ and\ \citenamefont
  {Tien}}]{cartignyJHT1986}%
  \BibitemOpen
  \bibfield  {author} {\bibinfo {author} {\bibfnamefont {J.}~\bibnamefont
  {Cartigny}}, \bibinfo {author} {\bibfnamefont {Y.}~\bibnamefont {Yamada}}, \
  and\ \bibinfo {author} {\bibfnamefont {C.}~\bibnamefont {Tien}},\ }\href@noop
  {} {\bibfield  {journal} {\bibinfo  {journal} {Journal of heat transfer}\
  }\textbf {\bibinfo {volume} {108}},\ \bibinfo {pages} {608} (\bibinfo {year}
  {1986})}\BibitemShut {NoStop}%
\bibitem [{\citenamefont {Drolen}\ and\ \citenamefont
  {Tien}(1987)}]{drolenJTHT1987}%
  \BibitemOpen
  \bibfield  {author} {\bibinfo {author} {\bibfnamefont {B.~L.}\ \bibnamefont
  {Drolen}}\ and\ \bibinfo {author} {\bibfnamefont {C.~L.}\ \bibnamefont
  {Tien}},\ }\href {\doibase 10.2514/3.8} {\bibfield  {journal} {\bibinfo
  {journal} {Journal of Thermophysics and Heat Transfer}\ }\textbf {\bibinfo
  {volume} {1}},\ \bibinfo {pages} {63} (\bibinfo {year} {1987})},\ \Eprint
  {http://arxiv.org/abs/https://doi.org/10.2514/3.8}
  {https://doi.org/10.2514/3.8} \BibitemShut {NoStop}%
\bibitem [{\citenamefont {Tien}\ and\ \citenamefont
  {Drolen}(1987)}]{tienARHT1987}%
  \BibitemOpen
  \bibfield  {author} {\bibinfo {author} {\bibfnamefont {C.-L.}\ \bibnamefont
  {Tien}}\ and\ \bibinfo {author} {\bibfnamefont {B.}~\bibnamefont {Drolen}},\
  }\href {\doibase 10.1615/AnnualRevHeatTransfer.v1.30} {\bibfield  {journal}
  {\bibinfo  {journal} {Annual Review of Heat Transfer}\ }\textbf {\bibinfo
  {volume} {1}} (\bibinfo {year} {1987}),\
  10.1615/AnnualRevHeatTransfer.v1.30}\BibitemShut {NoStop}%
\bibitem [{\citenamefont {Tien}(1988)}]{tienJHT1988}%
  \BibitemOpen
  \bibfield  {author} {\bibinfo {author} {\bibfnamefont {C.~L.}\ \bibnamefont
  {Tien}},\ }\href {\doibase 10.1115/1.3250623} {\bibfield  {journal} {\bibinfo
   {journal} {Journal of Heat Transfer}\ }\textbf {\bibinfo {volume} {110}},\
  \bibinfo {pages} {1230} (\bibinfo {year} {1988})}\BibitemShut {NoStop}%
\bibitem [{\citenamefont {Kumar}\ and\ \citenamefont
  {Tien}(1990)}]{kumar1990dependent}%
  \BibitemOpen
  \bibfield  {author} {\bibinfo {author} {\bibfnamefont {S.}~\bibnamefont
  {Kumar}}\ and\ \bibinfo {author} {\bibfnamefont {C.}~\bibnamefont {Tien}},\
  }\href {\doibase 10.1115/1.2910342} {\bibfield  {journal} {\bibinfo
  {journal} {Journal of heat transfer}\ }\textbf {\bibinfo {volume} {112}},\
  \bibinfo {pages} {178} (\bibinfo {year} {1990})}\BibitemShut {NoStop}%
\bibitem [{\citenamefont {Singh}\ and\ \citenamefont
  {Kaviany}(1992)}]{singhIJHMT1992}%
  \BibitemOpen
  \bibfield  {author} {\bibinfo {author} {\bibfnamefont {B.}~\bibnamefont
  {Singh}}\ and\ \bibinfo {author} {\bibfnamefont {M.}~\bibnamefont
  {Kaviany}},\ }\href {\doibase https://doi.org/10.1016/0017-9310(92)90031-M}
  {\bibfield  {journal} {\bibinfo  {journal} {International Journal of Heat and
  Mass Transfer}\ }\textbf {\bibinfo {volume} {35}},\ \bibinfo {pages} {1397 }
  (\bibinfo {year} {1992})}\BibitemShut {NoStop}%
\bibitem [{\citenamefont {Ma}\ \emph {et~al.}(2017)\citenamefont {Ma},
  \citenamefont {Tan}, \citenamefont {Zhao}, \citenamefont {Wang},\ and\
  \citenamefont {Wang}}]{maJQSRT2017}%
  \BibitemOpen
  \bibfield  {author} {\bibinfo {author} {\bibfnamefont {L.}~\bibnamefont
  {Ma}}, \bibinfo {author} {\bibfnamefont {J.}~\bibnamefont {Tan}}, \bibinfo
  {author} {\bibfnamefont {J.}~\bibnamefont {Zhao}}, \bibinfo {author}
  {\bibfnamefont {F.}~\bibnamefont {Wang}}, \ and\ \bibinfo {author}
  {\bibfnamefont {C.}~\bibnamefont {Wang}},\ }\href@noop {} {\bibfield
  {journal} {\bibinfo  {journal} {Journal of Quantitative Spectroscopy and
  Radiative Transfer}\ }\textbf {\bibinfo {volume} {187}},\ \bibinfo {pages}
  {255} (\bibinfo {year} {2017})}\BibitemShut {NoStop}%
\bibitem [{\citenamefont {Teixeira}(2008)}]{teixeiraIEEETAP2008}%
  \BibitemOpen
  \bibfield  {author} {\bibinfo {author} {\bibfnamefont {F.~L.}\ \bibnamefont
  {Teixeira}},\ }\href {\doibase 10.1109/TAP.2008.926767} {\bibfield  {journal}
  {\bibinfo  {journal} {IEEE Transactions on Antennas and Propagation}\
  }\textbf {\bibinfo {volume} {56}},\ \bibinfo {pages} {2150} (\bibinfo {year}
  {2008})}\BibitemShut {NoStop}%
\bibitem [{\citenamefont {Bohren}\ and\ \citenamefont
  {Huffman}(2008)}]{bohrenandhuffman}%
  \BibitemOpen
  \bibfield  {author} {\bibinfo {author} {\bibfnamefont {C.~F.}\ \bibnamefont
  {Bohren}}\ and\ \bibinfo {author} {\bibfnamefont {D.~R.}\ \bibnamefont
  {Huffman}},\ }\href@noop {} {\emph {\bibinfo {title} {Absorption and
  scattering of light by small particles}}}\ (\bibinfo  {publisher} {John Wiley
  \& Sons},\ \bibinfo {year} {2008})\BibitemShut {NoStop}%
\bibitem [{\citenamefont {Hogan}\ \emph {et~al.}(2014)\citenamefont {Hogan},
  \citenamefont {Urban}, \citenamefont {Ayala-Orozco}, \citenamefont
  {Pimpinelli}, \citenamefont {Nordlander},\ and\ \citenamefont
  {Halas}}]{hoganNL2014}%
  \BibitemOpen
  \bibfield  {author} {\bibinfo {author} {\bibfnamefont {N.~J.}\ \bibnamefont
  {Hogan}}, \bibinfo {author} {\bibfnamefont {A.~S.}\ \bibnamefont {Urban}},
  \bibinfo {author} {\bibfnamefont {C.}~\bibnamefont {Ayala-Orozco}}, \bibinfo
  {author} {\bibfnamefont {A.}~\bibnamefont {Pimpinelli}}, \bibinfo {author}
  {\bibfnamefont {P.}~\bibnamefont {Nordlander}}, \ and\ \bibinfo {author}
  {\bibfnamefont {N.~J.}\ \bibnamefont {Halas}},\ }\href {\doibase
  10.1021/nl5016975} {\bibfield  {journal} {\bibinfo  {journal} {Nano Letters}\
  }\textbf {\bibinfo {volume} {14}},\ \bibinfo {pages} {4640} (\bibinfo {year}
  {2014})},\ \bibinfo {note} {pMID: 24960442}\BibitemShut {NoStop}%
\bibitem [{\citenamefont {Vynck}\ \emph {et~al.}(2012)\citenamefont {Vynck},
  \citenamefont {Burresi}, \citenamefont {Riboli},\ and\ \citenamefont
  {Wiersma}}]{Vynck2012}%
  \BibitemOpen
  \bibfield  {author} {\bibinfo {author} {\bibfnamefont {K.}~\bibnamefont
  {Vynck}}, \bibinfo {author} {\bibfnamefont {M.}~\bibnamefont {Burresi}},
  \bibinfo {author} {\bibfnamefont {F.}~\bibnamefont {Riboli}}, \ and\ \bibinfo
  {author} {\bibfnamefont {D.~S.}\ \bibnamefont {Wiersma}},\ }\href {\doibase
  10.1038/nmat3442} {\bibfield  {journal} {\bibinfo  {journal} {Nat. Mater.}\
  }\textbf {\bibinfo {volume} {11}},\ \bibinfo {pages} {1017} (\bibinfo {year}
  {2012})}\BibitemShut {NoStop}%
\bibitem [{\citenamefont {Gálvez}\ \emph {et~al.}(2012)\citenamefont
  {Gálvez}, \citenamefont {Kemppainen}, \citenamefont {Míguez},\ and\
  \citenamefont {Halme}}]{galvezJPCC2012}%
  \BibitemOpen
  \bibfield  {author} {\bibinfo {author} {\bibfnamefont {F.~E.}\ \bibnamefont
  {Gálvez}}, \bibinfo {author} {\bibfnamefont {E.}~\bibnamefont {Kemppainen}},
  \bibinfo {author} {\bibfnamefont {H.}~\bibnamefont {Míguez}}, \ and\
  \bibinfo {author} {\bibfnamefont {J.}~\bibnamefont {Halme}},\ }\href
  {\doibase 10.1021/jp2092708} {\bibfield  {journal} {\bibinfo  {journal} {The
  Journal of Physical Chemistry C}\ }\textbf {\bibinfo {volume} {116}},\
  \bibinfo {pages} {11426} (\bibinfo {year} {2012})},\ \Eprint
  {http://arxiv.org/abs/https://doi.org/10.1021/jp2092708}
  {https://doi.org/10.1021/jp2092708} \BibitemShut {NoStop}%
\bibitem [{\citenamefont {Liu}\ \emph {et~al.}(2018)\citenamefont {Liu},
  \citenamefont {Zhao}, \citenamefont {Wang},\ and\ \citenamefont
  {Fang}}]{liuJOSAB2018}%
  \BibitemOpen
  \bibfield  {author} {\bibinfo {author} {\bibfnamefont {M.~Q.}\ \bibnamefont
  {Liu}}, \bibinfo {author} {\bibfnamefont {C.~Y.}\ \bibnamefont {Zhao}},
  \bibinfo {author} {\bibfnamefont {B.~X.}\ \bibnamefont {Wang}}, \ and\
  \bibinfo {author} {\bibfnamefont {X.}~\bibnamefont {Fang}},\ }\href {\doibase
  10.1364/JOSAB.35.000504} {\bibfield  {journal} {\bibinfo  {journal} {J. Opt.
  Soc. Am. B}\ }\textbf {\bibinfo {volume} {35}},\ \bibinfo {pages} {504}
  (\bibinfo {year} {2018})}\BibitemShut {NoStop}%
\bibitem [{\citenamefont {Liew}\ \emph {et~al.}(2011)\citenamefont {Liew},
  \citenamefont {Forster}, \citenamefont {Noh}, \citenamefont {Schreck},
  \citenamefont {Saranathan}, \citenamefont {Lu}, \citenamefont {Yang},
  \citenamefont {Prum}, \citenamefont {O'Hern}, \citenamefont {Dufresne},\ and\
  \citenamefont {Cao}}]{Liew2011}%
  \BibitemOpen
  \bibfield  {author} {\bibinfo {author} {\bibfnamefont {S.~F.}\ \bibnamefont
  {Liew}}, \bibinfo {author} {\bibfnamefont {J.}~\bibnamefont {Forster}},
  \bibinfo {author} {\bibfnamefont {H.}~\bibnamefont {Noh}}, \bibinfo {author}
  {\bibfnamefont {C.~F.}\ \bibnamefont {Schreck}}, \bibinfo {author}
  {\bibfnamefont {V.}~\bibnamefont {Saranathan}}, \bibinfo {author}
  {\bibfnamefont {X.}~\bibnamefont {Lu}}, \bibinfo {author} {\bibfnamefont
  {L.}~\bibnamefont {Yang}}, \bibinfo {author} {\bibfnamefont {R.~O.}\
  \bibnamefont {Prum}}, \bibinfo {author} {\bibfnamefont {C.~S.}\ \bibnamefont
  {O'Hern}}, \bibinfo {author} {\bibfnamefont {E.~R.}\ \bibnamefont
  {Dufresne}}, \ and\ \bibinfo {author} {\bibfnamefont {H.}~\bibnamefont
  {Cao}},\ }\href {\doibase 10.1364/OE.19.008208} {\bibfield  {journal}
  {\bibinfo  {journal} {Opt. Express}\ }\textbf {\bibinfo {volume} {19}},\
  \bibinfo {pages} {8208} (\bibinfo {year} {2011})}\BibitemShut {NoStop}%
\bibitem [{\citenamefont {Wiersma}(2013)}]{wiersma2013disordered}%
  \BibitemOpen
  \bibfield  {author} {\bibinfo {author} {\bibfnamefont {D.~S.}\ \bibnamefont
  {Wiersma}},\ }\href@noop {} {\bibfield  {journal} {\bibinfo  {journal} {Nat.
  Photon.}\ }\textbf {\bibinfo {volume} {7}},\ \bibinfo {pages} {188} (\bibinfo
  {year} {2013})}\BibitemShut {NoStop}%
\bibitem [{\citenamefont {Svensson}\ \emph {et~al.}(2011)\citenamefont
  {Svensson}, \citenamefont {Adolfsson}, \citenamefont {Lewander},
  \citenamefont {Xu},\ and\ \citenamefont {Svanberg}}]{svenssonPRL2011}%
  \BibitemOpen
  \bibfield  {author} {\bibinfo {author} {\bibfnamefont {T.}~\bibnamefont
  {Svensson}}, \bibinfo {author} {\bibfnamefont {E.}~\bibnamefont {Adolfsson}},
  \bibinfo {author} {\bibfnamefont {M.}~\bibnamefont {Lewander}}, \bibinfo
  {author} {\bibfnamefont {C.~T.}\ \bibnamefont {Xu}}, \ and\ \bibinfo {author}
  {\bibfnamefont {S.}~\bibnamefont {Svanberg}},\ }\href {\doibase
  10.1103/PhysRevLett.107.143901} {\bibfield  {journal} {\bibinfo  {journal}
  {Phys. Rev. Lett.}\ }\textbf {\bibinfo {volume} {107}},\ \bibinfo {pages}
  {143901} (\bibinfo {year} {2011})}\BibitemShut {NoStop}%
\bibitem [{\citenamefont {Wiersma}\ \emph {et~al.}(1997)\citenamefont
  {Wiersma}, \citenamefont {Bartolini}, \citenamefont {Lagendijk},\ and\
  \citenamefont {Righini}}]{wiersma1997localization}%
  \BibitemOpen
  \bibfield  {author} {\bibinfo {author} {\bibfnamefont {D.~S.}\ \bibnamefont
  {Wiersma}}, \bibinfo {author} {\bibfnamefont {P.}~\bibnamefont {Bartolini}},
  \bibinfo {author} {\bibfnamefont {A.}~\bibnamefont {Lagendijk}}, \ and\
  \bibinfo {author} {\bibfnamefont {R.}~\bibnamefont {Righini}},\ }\href@noop
  {} {\bibfield  {journal} {\bibinfo  {journal} {Nature (London)}\ }\textbf
  {\bibinfo {volume} {390}},\ \bibinfo {pages} {671} (\bibinfo {year}
  {1997})}\BibitemShut {NoStop}%
\bibitem [{\citenamefont {Segev}\ \emph {et~al.}(2013)\citenamefont {Segev},
  \citenamefont {Silberberg},\ and\ \citenamefont
  {Christodoulides}}]{Segev2013}%
  \BibitemOpen
  \bibfield  {author} {\bibinfo {author} {\bibfnamefont {M.}~\bibnamefont
  {Segev}}, \bibinfo {author} {\bibfnamefont {Y.}~\bibnamefont {Silberberg}}, \
  and\ \bibinfo {author} {\bibfnamefont {D.~N.}\ \bibnamefont
  {Christodoulides}},\ }\href {\doibase 10.1038/nphoton.2013.30} {\bibfield
  {journal} {\bibinfo  {journal} {Nat. Photon.}\ }\textbf {\bibinfo {volume}
  {7}},\ \bibinfo {pages} {197} (\bibinfo {year} {2013})}\BibitemShut {NoStop}%
\bibitem [{\citenamefont {Yamilov}\ \emph {et~al.}(2014)\citenamefont
  {Yamilov}, \citenamefont {Sarma}, \citenamefont {Redding}, \citenamefont
  {Payne}, \citenamefont {Noh},\ and\ \citenamefont {Cao}}]{yamilovPRL2014}%
  \BibitemOpen
  \bibfield  {author} {\bibinfo {author} {\bibfnamefont {A.~G.}\ \bibnamefont
  {Yamilov}}, \bibinfo {author} {\bibfnamefont {R.}~\bibnamefont {Sarma}},
  \bibinfo {author} {\bibfnamefont {B.}~\bibnamefont {Redding}}, \bibinfo
  {author} {\bibfnamefont {B.}~\bibnamefont {Payne}}, \bibinfo {author}
  {\bibfnamefont {H.}~\bibnamefont {Noh}}, \ and\ \bibinfo {author}
  {\bibfnamefont {H.}~\bibnamefont {Cao}},\ }\href {\doibase
  10.1103/PhysRevLett.112.023904} {\bibfield  {journal} {\bibinfo  {journal}
  {Phys. Rev. Lett.}\ }\textbf {\bibinfo {volume} {112}},\ \bibinfo {pages}
  {023904} (\bibinfo {year} {2014})}\BibitemShut {NoStop}%
\bibitem [{\citenamefont {Sebbah}\ \emph {et~al.}(1993)\citenamefont {Sebbah},
  \citenamefont {Sornette},\ and\ \citenamefont {Vanneste}}]{sebbahPRB1993}%
  \BibitemOpen
  \bibfield  {author} {\bibinfo {author} {\bibfnamefont {P.}~\bibnamefont
  {Sebbah}}, \bibinfo {author} {\bibfnamefont {D.}~\bibnamefont {Sornette}}, \
  and\ \bibinfo {author} {\bibfnamefont {C.}~\bibnamefont {Vanneste}},\ }\href
  {\doibase 10.1103/PhysRevB.48.12506} {\bibfield  {journal} {\bibinfo
  {journal} {Phys. Rev. B}\ }\textbf {\bibinfo {volume} {48}},\ \bibinfo
  {pages} {12506} (\bibinfo {year} {1993})}\BibitemShut {NoStop}%
\bibitem [{\citenamefont {Barthelemy}\ \emph {et~al.}(2008)\citenamefont
  {Barthelemy}, \citenamefont {Bertolotti},\ and\ \citenamefont
  {Wiersma}}]{barthelemyNature2008}%
  \BibitemOpen
  \bibfield  {author} {\bibinfo {author} {\bibfnamefont {P.}~\bibnamefont
  {Barthelemy}}, \bibinfo {author} {\bibfnamefont {J.}~\bibnamefont
  {Bertolotti}}, \ and\ \bibinfo {author} {\bibfnamefont {D.~S.}\ \bibnamefont
  {Wiersma}},\ }\href@noop {} {\bibfield  {journal} {\bibinfo  {journal}
  {Nature}\ }\textbf {\bibinfo {volume} {453}},\ \bibinfo {pages} {495}
  (\bibinfo {year} {2008})}\BibitemShut {NoStop}%
\bibitem [{\citenamefont {Bertolotti}\ \emph {et~al.}(2010)\citenamefont
  {Bertolotti}, \citenamefont {Vynck},\ and\ \citenamefont
  {Wiersma}}]{bertolottiPRL2010}%
  \BibitemOpen
  \bibfield  {author} {\bibinfo {author} {\bibfnamefont {J.}~\bibnamefont
  {Bertolotti}}, \bibinfo {author} {\bibfnamefont {K.}~\bibnamefont {Vynck}}, \
  and\ \bibinfo {author} {\bibfnamefont {D.~S.}\ \bibnamefont {Wiersma}},\
  }\href {\doibase 10.1103/PhysRevLett.105.163902} {\bibfield  {journal}
  {\bibinfo  {journal} {Phys. Rev. Lett.}\ }\textbf {\bibinfo {volume} {105}},\
  \bibinfo {pages} {163902} (\bibinfo {year} {2010})}\BibitemShut {NoStop}%
\bibitem [{\citenamefont {Cao}\ \emph {et~al.}(1999)\citenamefont {Cao},
  \citenamefont {Zhao}, \citenamefont {Ho}, \citenamefont {Seelig},
  \citenamefont {Wang},\ and\ \citenamefont {Chang}}]{Cao1999}%
  \BibitemOpen
  \bibfield  {author} {\bibinfo {author} {\bibfnamefont {H.}~\bibnamefont
  {Cao}}, \bibinfo {author} {\bibfnamefont {Y.}~\bibnamefont {Zhao}}, \bibinfo
  {author} {\bibfnamefont {S.}~\bibnamefont {Ho}}, \bibinfo {author}
  {\bibfnamefont {E.}~\bibnamefont {Seelig}}, \bibinfo {author} {\bibfnamefont
  {Q.}~\bibnamefont {Wang}}, \ and\ \bibinfo {author} {\bibfnamefont
  {R.}~\bibnamefont {Chang}},\ }\href {\doibase 10.1103/PhysRevLett.82.2278}
  {\bibfield  {journal} {\bibinfo  {journal} {Phys. Rev. Lett.}\ }\textbf
  {\bibinfo {volume} {82}},\ \bibinfo {pages} {2278} (\bibinfo {year}
  {1999})}\BibitemShut {NoStop}%
\bibitem [{\citenamefont {Chang}\ \emph {et~al.}(2018)\citenamefont {Chang},
  \citenamefont {Liao}, \citenamefont {Liao}, \citenamefont {Lin},
  \citenamefont {Lin}, \citenamefont {Lin}, \citenamefont {Lin}, \citenamefont
  {Perumal}, \citenamefont {Haider}, \citenamefont {Tai}, \citenamefont {Shen},
  \citenamefont {Chang}, \citenamefont {Huang}, \citenamefont {Lin},\ and\
  \citenamefont {Chen}}]{changSR2018}%
  \BibitemOpen
  \bibfield  {author} {\bibinfo {author} {\bibfnamefont {S.-W.}\ \bibnamefont
  {Chang}}, \bibinfo {author} {\bibfnamefont {W.-C.}\ \bibnamefont {Liao}},
  \bibinfo {author} {\bibfnamefont {Y.-M.}\ \bibnamefont {Liao}}, \bibinfo
  {author} {\bibfnamefont {H.-I.}\ \bibnamefont {Lin}}, \bibinfo {author}
  {\bibfnamefont {H.-Y.}\ \bibnamefont {Lin}}, \bibinfo {author} {\bibfnamefont
  {W.-J.}\ \bibnamefont {Lin}}, \bibinfo {author} {\bibfnamefont {S.-Y.}\
  \bibnamefont {Lin}}, \bibinfo {author} {\bibfnamefont {P.}~\bibnamefont
  {Perumal}}, \bibinfo {author} {\bibfnamefont {G.}~\bibnamefont {Haider}},
  \bibinfo {author} {\bibfnamefont {C.-T.}\ \bibnamefont {Tai}}, \bibinfo
  {author} {\bibfnamefont {K.-C.}\ \bibnamefont {Shen}}, \bibinfo {author}
  {\bibfnamefont {C.-H.}\ \bibnamefont {Chang}}, \bibinfo {author}
  {\bibfnamefont {Y.-F.}\ \bibnamefont {Huang}}, \bibinfo {author}
  {\bibfnamefont {T.-Y.}\ \bibnamefont {Lin}}, \ and\ \bibinfo {author}
  {\bibfnamefont {Y.-F.}\ \bibnamefont {Chen}},\ }\href {\doibase
  10.1038/s41598-018-21228-w} {\bibfield  {journal} {\bibinfo  {journal}
  {Scientific Reports}\ }\textbf {\bibinfo {volume} {8}},\ \bibinfo {pages}
  {2720} (\bibinfo {year} {2018})}\BibitemShut {NoStop}%
\bibitem [{\citenamefont {Xiao}\ \emph {et~al.}(2017)\citenamefont {Xiao},
  \citenamefont {Hu}, \citenamefont {Wang}, \citenamefont {Li}, \citenamefont
  {Tormo}, \citenamefont {Le~Thomas}, \citenamefont {Wang}, \citenamefont
  {Gianneschi}, \citenamefont {Shawkey},\ and\ \citenamefont
  {Dhinojwala}}]{xiaoSciAdv2017}%
  \BibitemOpen
  \bibfield  {author} {\bibinfo {author} {\bibfnamefont {M.}~\bibnamefont
  {Xiao}}, \bibinfo {author} {\bibfnamefont {Z.}~\bibnamefont {Hu}}, \bibinfo
  {author} {\bibfnamefont {Z.}~\bibnamefont {Wang}}, \bibinfo {author}
  {\bibfnamefont {Y.}~\bibnamefont {Li}}, \bibinfo {author} {\bibfnamefont
  {A.~D.}\ \bibnamefont {Tormo}}, \bibinfo {author} {\bibfnamefont
  {N.}~\bibnamefont {Le~Thomas}}, \bibinfo {author} {\bibfnamefont
  {B.}~\bibnamefont {Wang}}, \bibinfo {author} {\bibfnamefont {N.~C.}\
  \bibnamefont {Gianneschi}}, \bibinfo {author} {\bibfnamefont {M.~D.}\
  \bibnamefont {Shawkey}}, \ and\ \bibinfo {author} {\bibfnamefont
  {A.}~\bibnamefont {Dhinojwala}},\ }\href {\doibase 10.1126/sciadv.1701151}
  {\bibfield  {journal} {\bibinfo  {journal} {Science Advances}\ }\textbf
  {\bibinfo {volume} {3}} (\bibinfo {year} {2017}),\
  10.1126/sciadv.1701151}\BibitemShut {NoStop}%
\bibitem [{\citenamefont {Edagawa}\ \emph {et~al.}(2008)\citenamefont
  {Edagawa}, \citenamefont {Kanoko},\ and\ \citenamefont
  {Notomi}}]{Edagawa2008}%
  \BibitemOpen
  \bibfield  {author} {\bibinfo {author} {\bibfnamefont {K.}~\bibnamefont
  {Edagawa}}, \bibinfo {author} {\bibfnamefont {S.}~\bibnamefont {Kanoko}}, \
  and\ \bibinfo {author} {\bibfnamefont {M.}~\bibnamefont {Notomi}},\ }\href
  {\doibase 10.1103/PhysRevLett.100.013901} {\bibfield  {journal} {\bibinfo
  {journal} {Phys. Rev. Lett.}\ }\textbf {\bibinfo {volume} {100}},\ \bibinfo
  {pages} {013901} (\bibinfo {year} {2008})}\BibitemShut {NoStop}%
\bibitem [{\citenamefont {Liew}\ \emph {et~al.}(2016)\citenamefont {Liew},
  \citenamefont {Popoff}, \citenamefont {Sheehan}, \citenamefont {Goetschy},
  \citenamefont {Schmuttenmaer}, \citenamefont {Stone},\ and\ \citenamefont
  {Cao}}]{Liew2016ACSPh}%
  \BibitemOpen
  \bibfield  {author} {\bibinfo {author} {\bibfnamefont {S.~F.}\ \bibnamefont
  {Liew}}, \bibinfo {author} {\bibfnamefont {S.~M.}\ \bibnamefont {Popoff}},
  \bibinfo {author} {\bibfnamefont {S.~W.}\ \bibnamefont {Sheehan}}, \bibinfo
  {author} {\bibfnamefont {A.}~\bibnamefont {Goetschy}}, \bibinfo {author}
  {\bibfnamefont {C.~A.}\ \bibnamefont {Schmuttenmaer}}, \bibinfo {author}
  {\bibfnamefont {A.~D.}\ \bibnamefont {Stone}}, \ and\ \bibinfo {author}
  {\bibfnamefont {H.}~\bibnamefont {Cao}},\ }\href {\doibase
  10.1021/acsphotonics.5b00642} {\bibfield  {journal} {\bibinfo  {journal} {ACS
  Photon.}\ }\textbf {\bibinfo {volume} {3}},\ \bibinfo {pages} {449} (\bibinfo
  {year} {2016})}\BibitemShut {NoStop}%
\bibitem [{\citenamefont {Fend}\ \emph {et~al.}(2004)\citenamefont {Fend},
  \citenamefont {Hoffschmidt}, \citenamefont {Pitz-Paal}, \citenamefont
  {Reutter},\ and\ \citenamefont {Rietbrock}}]{fendEnergy2004}%
  \BibitemOpen
  \bibfield  {author} {\bibinfo {author} {\bibfnamefont {T.}~\bibnamefont
  {Fend}}, \bibinfo {author} {\bibfnamefont {B.}~\bibnamefont {Hoffschmidt}},
  \bibinfo {author} {\bibfnamefont {R.}~\bibnamefont {Pitz-Paal}}, \bibinfo
  {author} {\bibfnamefont {O.}~\bibnamefont {Reutter}}, \ and\ \bibinfo
  {author} {\bibfnamefont {P.}~\bibnamefont {Rietbrock}},\ }\href {\doibase
  https://doi.org/10.1016/S0360-5442(03)00188-9} {\bibfield  {journal}
  {\bibinfo  {journal} {Energy}\ }\textbf {\bibinfo {volume} {29}},\ \bibinfo
  {pages} {823 } (\bibinfo {year} {2004})},\ \bibinfo {note} {solarPACES
  2002}\BibitemShut {NoStop}%
\bibitem [{\citenamefont {Steinfeld}(2005)}]{steinfeldSE2005}%
  \BibitemOpen
  \bibfield  {author} {\bibinfo {author} {\bibfnamefont {A.}~\bibnamefont
  {Steinfeld}},\ }\href {\doibase
  https://doi.org/10.1016/j.solener.2003.12.012} {\bibfield  {journal}
  {\bibinfo  {journal} {Solar Energy}\ }\textbf {\bibinfo {volume} {78}},\
  \bibinfo {pages} {603 } (\bibinfo {year} {2005})},\ \bibinfo {note} {solar
  Hydrogen}\BibitemShut {NoStop}%
\bibitem [{\citenamefont {Howell}\ \emph {et~al.}(1996)\citenamefont {Howell},
  \citenamefont {Hall},\ and\ \citenamefont {Ellzey}}]{howellPECS1996}%
  \BibitemOpen
  \bibfield  {author} {\bibinfo {author} {\bibfnamefont {J.}~\bibnamefont
  {Howell}}, \bibinfo {author} {\bibfnamefont {M.}~\bibnamefont {Hall}}, \ and\
  \bibinfo {author} {\bibfnamefont {J.}~\bibnamefont {Ellzey}},\ }\href
  {\doibase https://doi.org/10.1016/0360-1285(96)00001-9} {\bibfield  {journal}
  {\bibinfo  {journal} {Progress in Energy and Combustion Science}\ }\textbf
  {\bibinfo {volume} {22}},\ \bibinfo {pages} {121 } (\bibinfo {year}
  {1996})}\BibitemShut {NoStop}%
\bibitem [{\citenamefont {Siegel}\ and\ \citenamefont
  {Spuckler}(1998)}]{siegelMSEA1998}%
  \BibitemOpen
  \bibfield  {author} {\bibinfo {author} {\bibfnamefont {R.}~\bibnamefont
  {Siegel}}\ and\ \bibinfo {author} {\bibfnamefont {C.~M.}\ \bibnamefont
  {Spuckler}},\ }\href {\doibase
  http://dx.doi.org/10.1016/S0921-5093(97)00845-9} {\bibfield  {journal}
  {\bibinfo  {journal} {Materials Science and Engineering: A}\ }\textbf
  {\bibinfo {volume} {245}},\ \bibinfo {pages} {150 } (\bibinfo {year}
  {1998})}\BibitemShut {NoStop}%
\bibitem [{\citenamefont {Shi}\ \emph {et~al.}(2018{\natexlab{a}})\citenamefont
  {Shi}, \citenamefont {Tsai}, \citenamefont {Carter}, \citenamefont {Mandal},
  \citenamefont {Overvig}, \citenamefont {Sfeir}, \citenamefont {Lu},
  \citenamefont {Craig}, \citenamefont {Bernard}, \citenamefont {Yang},\ and\
  \citenamefont {Yu}}]{shiLSA2018}%
  \BibitemOpen
  \bibfield  {author} {\bibinfo {author} {\bibfnamefont {N.~N.}\ \bibnamefont
  {Shi}}, \bibinfo {author} {\bibfnamefont {C.-C.}\ \bibnamefont {Tsai}},
  \bibinfo {author} {\bibfnamefont {M.~J.}\ \bibnamefont {Carter}}, \bibinfo
  {author} {\bibfnamefont {J.}~\bibnamefont {Mandal}}, \bibinfo {author}
  {\bibfnamefont {A.~C.}\ \bibnamefont {Overvig}}, \bibinfo {author}
  {\bibfnamefont {M.~Y.}\ \bibnamefont {Sfeir}}, \bibinfo {author}
  {\bibfnamefont {M.}~\bibnamefont {Lu}}, \bibinfo {author} {\bibfnamefont
  {C.~L.}\ \bibnamefont {Craig}}, \bibinfo {author} {\bibfnamefont {G.~D.}\
  \bibnamefont {Bernard}}, \bibinfo {author} {\bibfnamefont {Y.}~\bibnamefont
  {Yang}}, \ and\ \bibinfo {author} {\bibfnamefont {N.}~\bibnamefont {Yu}},\
  }\href {\doibase 10.1038/s41377-018-0033-x} {\bibfield  {journal} {\bibinfo
  {journal} {Light: Science \& Applications}\ }\textbf {\bibinfo {volume}
  {7}},\ \bibinfo {pages} {37} (\bibinfo {year}
  {2018}{\natexlab{a}})}\BibitemShut {NoStop}%
\bibitem [{\citenamefont {Rotter}\ and\ \citenamefont
  {Gigan}(2017)}]{Rotter2017}%
  \BibitemOpen
  \bibfield  {author} {\bibinfo {author} {\bibfnamefont {S.}~\bibnamefont
  {Rotter}}\ and\ \bibinfo {author} {\bibfnamefont {S.}~\bibnamefont {Gigan}},\
  }\href {\doibase 10.1103/RevModPhys.89.015005} {\bibfield  {journal}
  {\bibinfo  {journal} {Rev. Mod. Phys.}\ }\textbf {\bibinfo {volume} {89}},\
  \bibinfo {pages} {015005} (\bibinfo {year} {2017})}\BibitemShut {NoStop}%
\bibitem [{\citenamefont {García}\ and\ \citenamefont
  {Lodahl}(2017)}]{garciaAnnPhys2017}%
  \BibitemOpen
  \bibfield  {author} {\bibinfo {author} {\bibfnamefont {P.~D.}\ \bibnamefont
  {García}}\ and\ \bibinfo {author} {\bibfnamefont {P.}~\bibnamefont
  {Lodahl}},\ }\href {\doibase 10.1002/andp.201600351} {\bibfield  {journal}
  {\bibinfo  {journal} {Annalen der Physik}\ }\textbf {\bibinfo {volume}
  {529}},\ \bibinfo {pages} {1600351} (\bibinfo {year} {2017})}\BibitemShut
  {NoStop}%
\bibitem [{\citenamefont {Pine}\ \emph {et~al.}(1988)\citenamefont {Pine},
  \citenamefont {Weitz}, \citenamefont {Chaikin},\ and\ \citenamefont
  {Herbolzheimer}}]{pinePRL1988}%
  \BibitemOpen
  \bibfield  {author} {\bibinfo {author} {\bibfnamefont {D.~J.}\ \bibnamefont
  {Pine}}, \bibinfo {author} {\bibfnamefont {D.~A.}\ \bibnamefont {Weitz}},
  \bibinfo {author} {\bibfnamefont {P.~M.}\ \bibnamefont {Chaikin}}, \ and\
  \bibinfo {author} {\bibfnamefont {E.}~\bibnamefont {Herbolzheimer}},\ }\href
  {\doibase 10.1103/PhysRevLett.60.1134} {\bibfield  {journal} {\bibinfo
  {journal} {Phys. Rev. Lett.}\ }\textbf {\bibinfo {volume} {60}},\ \bibinfo
  {pages} {1134} (\bibinfo {year} {1988})}\BibitemShut {NoStop}%
\bibitem [{\citenamefont {Boas}\ and\ \citenamefont
  {Yodh}(1997)}]{boasJOSAA1997}%
  \BibitemOpen
  \bibfield  {author} {\bibinfo {author} {\bibfnamefont {D.~A.}\ \bibnamefont
  {Boas}}\ and\ \bibinfo {author} {\bibfnamefont {A.~G.}\ \bibnamefont
  {Yodh}},\ }\href {\doibase 10.1364/JOSAA.14.000192} {\bibfield  {journal}
  {\bibinfo  {journal} {J. Opt. Soc. Am. A}\ }\textbf {\bibinfo {volume}
  {14}},\ \bibinfo {pages} {192} (\bibinfo {year} {1997})}\BibitemShut
  {NoStop}%
\bibitem [{\citenamefont {Goldburg}(1999)}]{goldburgAJP1999}%
  \BibitemOpen
  \bibfield  {author} {\bibinfo {author} {\bibfnamefont {W.~I.}\ \bibnamefont
  {Goldburg}},\ }\href {\doibase 10.1119/1.19101} {\bibfield  {journal}
  {\bibinfo  {journal} {American Journal of Physics}\ }\textbf {\bibinfo
  {volume} {67}},\ \bibinfo {pages} {1152} (\bibinfo {year} {1999})},\ \Eprint
  {http://arxiv.org/abs/https://doi.org/10.1119/1.19101}
  {https://doi.org/10.1119/1.19101} \BibitemShut {NoStop}%
\bibitem [{\citenamefont {Modest}(2013)}]{modest2013radiative}%
  \BibitemOpen
  \bibfield  {author} {\bibinfo {author} {\bibfnamefont {M.~F.}\ \bibnamefont
  {Modest}},\ }\href@noop {} {\emph {\bibinfo {title} {Radiative heat
  transfer}}}\ (\bibinfo  {publisher} {Academic press},\ \bibinfo {year}
  {2013})\BibitemShut {NoStop}%
\bibitem [{\citenamefont {Howell}\ \emph {et~al.}(2015)\citenamefont {Howell},
  \citenamefont {Menguc},\ and\ \citenamefont {Siegel}}]{howell2015thermal}%
  \BibitemOpen
  \bibfield  {author} {\bibinfo {author} {\bibfnamefont {J.~R.}\ \bibnamefont
  {Howell}}, \bibinfo {author} {\bibfnamefont {M.~P.}\ \bibnamefont {Menguc}},
  \ and\ \bibinfo {author} {\bibfnamefont {R.}~\bibnamefont {Siegel}},\
  }\href@noop {} {\emph {\bibinfo {title} {Thermal radiation heat transfer}}}\
  (\bibinfo  {publisher} {CRC press},\ \bibinfo {year} {2015})\BibitemShut
  {NoStop}%
\bibitem [{\citenamefont {van Tiggelen}\ and\ \citenamefont
  {Stark}(2000)}]{vanTiggelenRMP2000}%
  \BibitemOpen
  \bibfield  {author} {\bibinfo {author} {\bibfnamefont {B.}~\bibnamefont {van
  Tiggelen}}\ and\ \bibinfo {author} {\bibfnamefont {H.}~\bibnamefont
  {Stark}},\ }\href {\doibase 10.1103/RevModPhys.72.1017} {\bibfield  {journal}
  {\bibinfo  {journal} {Rev. Mod. Phys.}\ }\textbf {\bibinfo {volume} {72}},\
  \bibinfo {pages} {1017} (\bibinfo {year} {2000})}\BibitemShut {NoStop}%
\bibitem [{\citenamefont {Joannopoulos}\ \emph {et~al.}(2011)\citenamefont
  {Joannopoulos}, \citenamefont {Johnson}, \citenamefont {Winn},\ and\
  \citenamefont {Meade}}]{joannopoulos2011photonic}%
  \BibitemOpen
  \bibfield  {author} {\bibinfo {author} {\bibfnamefont {J.~D.}\ \bibnamefont
  {Joannopoulos}}, \bibinfo {author} {\bibfnamefont {S.~G.}\ \bibnamefont
  {Johnson}}, \bibinfo {author} {\bibfnamefont {J.~N.}\ \bibnamefont {Winn}}, \
  and\ \bibinfo {author} {\bibfnamefont {R.~D.}\ \bibnamefont {Meade}},\
  }\href@noop {} {\emph {\bibinfo {title} {Photonic crystals: molding the flow
  of light}}}\ (\bibinfo  {publisher} {Princeton university press},\ \bibinfo
  {year} {2011})\BibitemShut {NoStop}%
\bibitem [{\citenamefont {Akkermans}\ and\ \citenamefont
  {Montambaux}(2007)}]{akkermans2007mesoscopic}%
  \BibitemOpen
  \bibfield  {author} {\bibinfo {author} {\bibfnamefont {E.}~\bibnamefont
  {Akkermans}}\ and\ \bibinfo {author} {\bibfnamefont {G.}~\bibnamefont
  {Montambaux}},\ }\href@noop {} {\emph {\bibinfo {title} {Mesoscopic Physics
  of Electrons and Photons}}}\ (\bibinfo  {publisher} {Cambridge University
  Press},\ \bibinfo {year} {2007})\BibitemShut {NoStop}%
\bibitem [{\citenamefont {Rezvani~Naraghi}\ and\ \citenamefont
  {Dogariu}(2016)}]{Naraghi2016}%
  \BibitemOpen
  \bibfield  {author} {\bibinfo {author} {\bibfnamefont {R.}~\bibnamefont
  {Rezvani~Naraghi}}\ and\ \bibinfo {author} {\bibfnamefont {A.}~\bibnamefont
  {Dogariu}},\ }\href {\doibase 10.1103/PhysRevLett.117.263901} {\bibfield
  {journal} {\bibinfo  {journal} {Phys. Rev. Lett.}\ }\textbf {\bibinfo
  {volume} {117}},\ \bibinfo {pages} {263901} (\bibinfo {year}
  {2016})}\BibitemShut {NoStop}%
\bibitem [{\citenamefont {Mishchenko}(2018)}]{mishchenkoOSAC2019}%
  \BibitemOpen
  \bibfield  {author} {\bibinfo {author} {\bibfnamefont {M.~I.}\ \bibnamefont
  {Mishchenko}},\ }\href {\doibase 10.1364/OSAC.1.000243} {\bibfield  {journal}
  {\bibinfo  {journal} {OSA Continuum}\ }\textbf {\bibinfo {volume} {1}},\
  \bibinfo {pages} {243} (\bibinfo {year} {2018})}\BibitemShut {NoStop}%
\bibitem [{\citenamefont {Leung~Tsang}\ and\ \citenamefont
  {Ding}(2000)}]{tsang2000scattering1}%
  \BibitemOpen
  \bibfield  {author} {\bibinfo {author} {\bibfnamefont {J.~A.~K.}\
  \bibnamefont {Leung~Tsang}}\ and\ \bibinfo {author} {\bibfnamefont {K.~H.}\
  \bibnamefont {Ding}},\ }\href@noop {} {\emph {\bibinfo {title} {Scattering of
  Electromagnetic Waves, Theories and Applications, vol. 1}}}\ (\bibinfo
  {publisher} {New York: Wiley},\ \bibinfo {year} {2000})\BibitemShut {NoStop}%
\bibitem [{\citenamefont {Mishchenko}(2002)}]{mishchenkoAO2002}%
  \BibitemOpen
  \bibfield  {author} {\bibinfo {author} {\bibfnamefont {M.~I.}\ \bibnamefont
  {Mishchenko}},\ }\href {\doibase 10.1364/AO.41.007114} {\bibfield  {journal}
  {\bibinfo  {journal} {Appl. Opt.}\ }\textbf {\bibinfo {volume} {41}},\
  \bibinfo {pages} {7114} (\bibinfo {year} {2002})}\BibitemShut {NoStop}%
\bibitem [{\citenamefont {Zhao}\ \emph {et~al.}(2015)\citenamefont {Zhao},
  \citenamefont {Tan},\ and\ \citenamefont {Liu}}]{zhaoJQSRT2015}%
  \BibitemOpen
  \bibfield  {author} {\bibinfo {author} {\bibfnamefont {J.}~\bibnamefont
  {Zhao}}, \bibinfo {author} {\bibfnamefont {J.}~\bibnamefont {Tan}}, \ and\
  \bibinfo {author} {\bibfnamefont {L.}~\bibnamefont {Liu}},\ }\href {\doibase
  https://doi.org/10.1016/j.jqsrt.2014.11.005} {\bibfield  {journal} {\bibinfo
  {journal} {Journal of Quantitative Spectroscopy and Radiative Transfer}\
  }\textbf {\bibinfo {volume} {152}},\ \bibinfo {pages} {114 } (\bibinfo {year}
  {2015})}\BibitemShut {NoStop}%
\bibitem [{\citenamefont {Tishkovets}\ and\ \citenamefont
  {Jockers}(2006)}]{tishkovetsJQSRT2006}%
  \BibitemOpen
  \bibfield  {author} {\bibinfo {author} {\bibfnamefont {V.~P.}\ \bibnamefont
  {Tishkovets}}\ and\ \bibinfo {author} {\bibfnamefont {K.}~\bibnamefont
  {Jockers}},\ }\href {\doibase https://doi.org/10.1016/j.jqsrt.2005.10.001}
  {\bibfield  {journal} {\bibinfo  {journal} {Journal of Quantitative
  Spectroscopy and Radiative Transfer}\ }\textbf {\bibinfo {volume} {101}},\
  \bibinfo {pages} {54 } (\bibinfo {year} {2006})}\BibitemShut {NoStop}%
\bibitem [{\citenamefont {Tishkovets}\ \emph {et~al.}(2011)\citenamefont
  {Tishkovets}, \citenamefont {Petrova},\ and\ \citenamefont
  {Mishchenko}}]{tishkovetsJQSRT2011}%
  \BibitemOpen
  \bibfield  {author} {\bibinfo {author} {\bibfnamefont {V.~P.}\ \bibnamefont
  {Tishkovets}}, \bibinfo {author} {\bibfnamefont {E.~V.}\ \bibnamefont
  {Petrova}}, \ and\ \bibinfo {author} {\bibfnamefont {M.~I.}\ \bibnamefont
  {Mishchenko}},\ }\href {\doibase https://doi.org/10.1016/j.jqsrt.2011.04.010}
  {\bibfield  {journal} {\bibinfo  {journal} {Journal of Quantitative
  Spectroscopy and Radiative Transfer}\ }\textbf {\bibinfo {volume} {112}},\
  \bibinfo {pages} {2095 } (\bibinfo {year} {2011})}\BibitemShut {NoStop}%
\bibitem [{\citenamefont {Doicu}\ and\ \citenamefont
  {Mishchenko}(2019{\natexlab{a}})}]{doicuJQSRT2019c}%
  \BibitemOpen
  \bibfield  {author} {\bibinfo {author} {\bibfnamefont {A.}~\bibnamefont
  {Doicu}}\ and\ \bibinfo {author} {\bibfnamefont {M.~I.}\ \bibnamefont
  {Mishchenko}},\ }\href {\doibase https://doi.org/10.1016/j.jqsrt.2019.07.007}
  {\bibfield  {journal} {\bibinfo  {journal} {Journal of Quantitative
  Spectroscopy and Radiative Transfer}\ }\textbf {\bibinfo {volume} {236}},\
  \bibinfo {pages} {106564} (\bibinfo {year} {2019}{\natexlab{a}})}\BibitemShut
  {NoStop}%
\bibitem [{\citenamefont {Chandrasekhar}(1950)}]{chandrasekhar1950radiative}%
  \BibitemOpen
  \bibfield  {author} {\bibinfo {author} {\bibfnamefont {S.}~\bibnamefont
  {Chandrasekhar}},\ }\href@noop {} {\emph {\bibinfo {title} {Radiative
  transfer}}}\ (\bibinfo  {publisher} {Clarendon Press},\ \bibinfo {year}
  {1950})\BibitemShut {NoStop}%
\bibitem [{\citenamefont {Ishimaru}(1978)}]{ishimaru1978book}%
  \BibitemOpen
  \bibfield  {author} {\bibinfo {author} {\bibfnamefont {A.}~\bibnamefont
  {Ishimaru}},\ }\href@noop {} {\emph {\bibinfo {title} {Wave propagation and
  scattering in random media}}},\ Vol.~\bibinfo {volume} {2}\ (\bibinfo
  {publisher} {Academic press New York},\ \bibinfo {year} {1978})\BibitemShut
  {NoStop}%
\bibitem [{\citenamefont {Tsang}\ \emph {et~al.}(1985)\citenamefont {Tsang},
  \citenamefont {Kong},\ and\ \citenamefont {Shin}}]{tsang1985theory}%
  \BibitemOpen
  \bibfield  {author} {\bibinfo {author} {\bibfnamefont {L.}~\bibnamefont
  {Tsang}}, \bibinfo {author} {\bibfnamefont {J.~A.}\ \bibnamefont {Kong}}, \
  and\ \bibinfo {author} {\bibfnamefont {R.~T.}\ \bibnamefont {Shin}},\
  }\href@noop {} {\  (\bibinfo {year} {1985})}\BibitemShut {NoStop}%
\bibitem [{\citenamefont {Mishchenko}\ \emph {et~al.}(2016)\citenamefont
  {Mishchenko}, \citenamefont {Dlugach}, \citenamefont {Yurkin}, \citenamefont
  {Bi}, \citenamefont {Cairns}, \citenamefont {Liu}, \citenamefont {Panetta},
  \citenamefont {Travis}, \citenamefont {Yang},\ and\ \citenamefont
  {Zakharova}}]{mishchenkoPhysrep2016}%
  \BibitemOpen
  \bibfield  {author} {\bibinfo {author} {\bibfnamefont {M.~I.}\ \bibnamefont
  {Mishchenko}}, \bibinfo {author} {\bibfnamefont {J.~M.}\ \bibnamefont
  {Dlugach}}, \bibinfo {author} {\bibfnamefont {M.~A.}\ \bibnamefont {Yurkin}},
  \bibinfo {author} {\bibfnamefont {L.}~\bibnamefont {Bi}}, \bibinfo {author}
  {\bibfnamefont {B.}~\bibnamefont {Cairns}}, \bibinfo {author} {\bibfnamefont
  {L.}~\bibnamefont {Liu}}, \bibinfo {author} {\bibfnamefont {R.~L.}\
  \bibnamefont {Panetta}}, \bibinfo {author} {\bibfnamefont {L.~D.}\
  \bibnamefont {Travis}}, \bibinfo {author} {\bibfnamefont {P.}~\bibnamefont
  {Yang}}, \ and\ \bibinfo {author} {\bibfnamefont {N.~T.}\ \bibnamefont
  {Zakharova}},\ }\href {\doibase 10.1016/j.physrep.2016.04.002} {\bibfield
  {journal} {\bibinfo  {journal} {Physics Reports}\ }\textbf {\bibinfo {volume}
  {632}},\ \bibinfo {pages} {1 } (\bibinfo {year} {2016})}\BibitemShut
  {NoStop}%
\bibitem [{\citenamefont {Tsang}(2019)}]{tsangJQSRT2019}%
  \BibitemOpen
  \bibfield  {author} {\bibinfo {author} {\bibfnamefont {L.}~\bibnamefont
  {Tsang}},\ }\href {\doibase https://doi.org/10.1016/j.jqsrt.2018.10.041}
  {\bibfield  {journal} {\bibinfo  {journal} {Journal of Quantitative
  Spectroscopy and Radiative Transfer}\ }\textbf {\bibinfo {volume} {224}},\
  \bibinfo {pages} {566 } (\bibinfo {year} {2019})}\BibitemShut {NoStop}%
\bibitem [{\citenamefont {Tancrez}\ and\ \citenamefont
  {Taine}(2004)}]{tancrezIJHMT2004}%
  \BibitemOpen
  \bibfield  {author} {\bibinfo {author} {\bibfnamefont {M.}~\bibnamefont
  {Tancrez}}\ and\ \bibinfo {author} {\bibfnamefont {J.}~\bibnamefont
  {Taine}},\ }\href {\doibase https://doi.org/10.1016/S0017-9310(03)00146-7}
  {\bibfield  {journal} {\bibinfo  {journal} {International Journal of Heat and
  Mass Transfer}\ }\textbf {\bibinfo {volume} {47}},\ \bibinfo {pages} {373 }
  (\bibinfo {year} {2004})}\BibitemShut {NoStop}%
\bibitem [{\citenamefont {Dombrovsky}\ \emph {et~al.}(2007)\citenamefont
  {Dombrovsky}, \citenamefont {Tagne}, \citenamefont {Baillis},\ and\
  \citenamefont {Gremillard}}]{dombrovskyIPT2007}%
  \BibitemOpen
  \bibfield  {author} {\bibinfo {author} {\bibfnamefont {L.~A.}\ \bibnamefont
  {Dombrovsky}}, \bibinfo {author} {\bibfnamefont {H.~K.}\ \bibnamefont
  {Tagne}}, \bibinfo {author} {\bibfnamefont {D.}~\bibnamefont {Baillis}}, \
  and\ \bibinfo {author} {\bibfnamefont {L.}~\bibnamefont {Gremillard}},\
  }\href {\doibase http://dx.doi.org/10.1016/j.infrared.2006.11.003} {\bibfield
   {journal} {\bibinfo  {journal} {Infrared Physics \& Technology}\ }\textbf
  {\bibinfo {volume} {51}},\ \bibinfo {pages} {44 } (\bibinfo {year}
  {2007})}\BibitemShut {NoStop}%
\bibitem [{\citenamefont {Randrianalisoa}\ and\ \citenamefont
  {Baillis}(2014)}]{randrianalisoaIJHMT2014}%
  \BibitemOpen
  \bibfield  {author} {\bibinfo {author} {\bibfnamefont {J.}~\bibnamefont
  {Randrianalisoa}}\ and\ \bibinfo {author} {\bibfnamefont {D.}~\bibnamefont
  {Baillis}},\ }\href {\doibase
  https://doi.org/10.1016/j.ijheatmasstransfer.2013.10.071} {\bibfield
  {journal} {\bibinfo  {journal} {International Journal of Heat and Mass
  Transfer}\ }\textbf {\bibinfo {volume} {70}},\ \bibinfo {pages} {264 }
  (\bibinfo {year} {2014})}\BibitemShut {NoStop}%
\bibitem [{\citenamefont {Wang}\ and\ \citenamefont
  {Zhao}(2018{\natexlab{b}})}]{wangJAP2018}%
  \BibitemOpen
  \bibfield  {author} {\bibinfo {author} {\bibfnamefont {B.~X.}\ \bibnamefont
  {Wang}}\ and\ \bibinfo {author} {\bibfnamefont {C.~Y.}\ \bibnamefont
  {Zhao}},\ }\href {\doibase 10.1063/1.5030504} {\bibfield  {journal} {\bibinfo
   {journal} {Journal of Applied Physics}\ }\textbf {\bibinfo {volume} {123}},\
  \bibinfo {pages} {223101} (\bibinfo {year} {2018}{\natexlab{b}})}\BibitemShut
  {NoStop}%
\bibitem [{\citenamefont {Mishchenko}\ \emph {et~al.}(2007)\citenamefont
  {Mishchenko}, \citenamefont {Liu},\ and\ \citenamefont
  {Videen}}]{mishchenkoOE2007}%
  \BibitemOpen
  \bibfield  {author} {\bibinfo {author} {\bibfnamefont {M.~I.}\ \bibnamefont
  {Mishchenko}}, \bibinfo {author} {\bibfnamefont {L.}~\bibnamefont {Liu}}, \
  and\ \bibinfo {author} {\bibfnamefont {G.}~\bibnamefont {Videen}},\ }\href
  {\doibase 10.1364/OE.15.007522} {\bibfield  {journal} {\bibinfo  {journal}
  {Opt. Express}\ }\textbf {\bibinfo {volume} {15}},\ \bibinfo {pages} {7522}
  (\bibinfo {year} {2007})}\BibitemShut {NoStop}%
\bibitem [{\citenamefont {Mishchenko}\ \emph {et~al.}(2013)\citenamefont
  {Mishchenko}, \citenamefont {Goldstein}, \citenamefont {Chowdhary},\ and\
  \citenamefont {Lompado}}]{mishchenkoOL2013}%
  \BibitemOpen
  \bibfield  {author} {\bibinfo {author} {\bibfnamefont {M.~I.}\ \bibnamefont
  {Mishchenko}}, \bibinfo {author} {\bibfnamefont {D.~H.}\ \bibnamefont
  {Goldstein}}, \bibinfo {author} {\bibfnamefont {J.}~\bibnamefont
  {Chowdhary}}, \ and\ \bibinfo {author} {\bibfnamefont {A.}~\bibnamefont
  {Lompado}},\ }\href {\doibase 10.1364/OL.38.003522} {\bibfield  {journal}
  {\bibinfo  {journal} {Opt. Lett.}\ }\textbf {\bibinfo {volume} {38}},\
  \bibinfo {pages} {3522} (\bibinfo {year} {2013})}\BibitemShut {NoStop}%
\bibitem [{\citenamefont {Fraden}\ and\ \citenamefont
  {Maret}(1990)}]{fradenPRL1990}%
  \BibitemOpen
  \bibfield  {author} {\bibinfo {author} {\bibfnamefont {S.}~\bibnamefont
  {Fraden}}\ and\ \bibinfo {author} {\bibfnamefont {G.}~\bibnamefont {Maret}},\
  }\href {\doibase 10.1103/PhysRevLett.65.512} {\bibfield  {journal} {\bibinfo
  {journal} {Phys. Rev. Lett.}\ }\textbf {\bibinfo {volume} {65}},\ \bibinfo
  {pages} {512} (\bibinfo {year} {1990})}\BibitemShut {NoStop}%
\bibitem [{\citenamefont {Aernouts}\ \emph
  {et~al.}(2014{\natexlab{a}})\citenamefont {Aernouts}, \citenamefont {Beers},
  \citenamefont {Watt\'{e}}, \citenamefont {Lammertyn},\ and\ \citenamefont
  {Saeys}}]{aernoutsOE2014}%
  \BibitemOpen
  \bibfield  {author} {\bibinfo {author} {\bibfnamefont {B.}~\bibnamefont
  {Aernouts}}, \bibinfo {author} {\bibfnamefont {R.~V.}\ \bibnamefont {Beers}},
  \bibinfo {author} {\bibfnamefont {R.}~\bibnamefont {Watt\'{e}}}, \bibinfo
  {author} {\bibfnamefont {J.}~\bibnamefont {Lammertyn}}, \ and\ \bibinfo
  {author} {\bibfnamefont {W.}~\bibnamefont {Saeys}},\ }\href {\doibase
  10.1364/OE.22.006086} {\bibfield  {journal} {\bibinfo  {journal} {Opt.
  Express}\ }\textbf {\bibinfo {volume} {22}},\ \bibinfo {pages} {6086}
  (\bibinfo {year} {2014}{\natexlab{a}})}\BibitemShut {NoStop}%
\bibitem [{\citenamefont {Kalkman}\ \emph {et~al.}(2010)\citenamefont
  {Kalkman}, \citenamefont {Bykov}, \citenamefont {Faber},\ and\ \citenamefont
  {van Leeuwen}}]{kalkmanOE2010}%
  \BibitemOpen
  \bibfield  {author} {\bibinfo {author} {\bibfnamefont {J.}~\bibnamefont
  {Kalkman}}, \bibinfo {author} {\bibfnamefont {A.~V.}\ \bibnamefont {Bykov}},
  \bibinfo {author} {\bibfnamefont {D.~J.}\ \bibnamefont {Faber}}, \ and\
  \bibinfo {author} {\bibfnamefont {T.~G.}\ \bibnamefont {van Leeuwen}},\
  }\href {\doibase 10.1364/OE.18.003883} {\bibfield  {journal} {\bibinfo
  {journal} {Opt. Express}\ }\textbf {\bibinfo {volume} {18}},\ \bibinfo
  {pages} {3883} (\bibinfo {year} {2010})}\BibitemShut {NoStop}%
\bibitem [{\citenamefont {Berry}(1962)}]{berryJOSA1962}%
  \BibitemOpen
  \bibfield  {author} {\bibinfo {author} {\bibfnamefont {C.~R.}\ \bibnamefont
  {Berry}},\ }\href {\doibase 10.1364/JOSA.52.000888} {\bibfield  {journal}
  {\bibinfo  {journal} {J. Opt. Soc. Am.}\ }\textbf {\bibinfo {volume} {52}},\
  \bibinfo {pages} {888} (\bibinfo {year} {1962})}\BibitemShut {NoStop}%
\bibitem [{\citenamefont {Auger}\ and\ \citenamefont
  {Stout}(2012)}]{augerJCTR2012}%
  \BibitemOpen
  \bibfield  {author} {\bibinfo {author} {\bibfnamefont {J.-C.}\ \bibnamefont
  {Auger}}\ and\ \bibinfo {author} {\bibfnamefont {B.}~\bibnamefont {Stout}},\
  }\href {\doibase 10.1007/s11998-011-9371-9} {\bibfield  {journal} {\bibinfo
  {journal} {Journal of Coatings Technology and Research}\ }\textbf {\bibinfo
  {volume} {9}},\ \bibinfo {pages} {287} (\bibinfo {year} {2012})}\BibitemShut
  {NoStop}%
\bibitem [{\citenamefont {Doicu}\ and\ \citenamefont
  {Mishchenko}(2018)}]{doicuJQSRT2018}%
  \BibitemOpen
  \bibfield  {author} {\bibinfo {author} {\bibfnamefont {A.}~\bibnamefont
  {Doicu}}\ and\ \bibinfo {author} {\bibfnamefont {M.~I.}\ \bibnamefont
  {Mishchenko}},\ }\href {\doibase https://doi.org/10.1016/j.jqsrt.2018.09.004}
  {\bibfield  {journal} {\bibinfo  {journal} {Journal of Quantitative
  Spectroscopy and Radiative Transfer}\ }\textbf {\bibinfo {volume} {220}},\
  \bibinfo {pages} {123 } (\bibinfo {year} {2018})}\BibitemShut {NoStop}%
\bibitem [{\citenamefont {Doicu}\ and\ \citenamefont
  {Mishchenko}(2019{\natexlab{b}})}]{doicuJQSRT2019}%
  \BibitemOpen
  \bibfield  {author} {\bibinfo {author} {\bibfnamefont {A.}~\bibnamefont
  {Doicu}}\ and\ \bibinfo {author} {\bibfnamefont {M.~I.}\ \bibnamefont
  {Mishchenko}},\ }\href {\doibase https://doi.org/10.1016/j.jqsrt.2018.10.032}
  {\bibfield  {journal} {\bibinfo  {journal} {Journal of Quantitative
  Spectroscopy and Radiative Transfer}\ }\textbf {\bibinfo {volume} {224}},\
  \bibinfo {pages} {25 } (\bibinfo {year} {2019}{\natexlab{b}})}\BibitemShut
  {NoStop}%
\bibitem [{\citenamefont {Doicu}\ and\ \citenamefont
  {Mishchenko}(2019{\natexlab{c}})}]{doicuJQSRT2019a}%
  \BibitemOpen
  \bibfield  {author} {\bibinfo {author} {\bibfnamefont {A.}~\bibnamefont
  {Doicu}}\ and\ \bibinfo {author} {\bibfnamefont {M.~I.}\ \bibnamefont
  {Mishchenko}},\ }\href {\doibase https://doi.org/10.1016/j.jqsrt.2019.03.012}
  {\bibfield  {journal} {\bibinfo  {journal} {Journal of Quantitative
  Spectroscopy and Radiative Transfer}\ }\textbf {\bibinfo {volume} {230}},\
  \bibinfo {pages} {282 } (\bibinfo {year} {2019}{\natexlab{c}})}\BibitemShut
  {NoStop}%
\bibitem [{\citenamefont {Doicu}\ and\ \citenamefont
  {Mishchenko}(2019{\natexlab{d}})}]{doicuJQSRT2019b}%
  \BibitemOpen
  \bibfield  {author} {\bibinfo {author} {\bibfnamefont {A.}~\bibnamefont
  {Doicu}}\ and\ \bibinfo {author} {\bibfnamefont {M.~I.}\ \bibnamefont
  {Mishchenko}},\ }\href {\doibase https://doi.org/10.1016/j.jqsrt.2019.03.011}
  {\bibfield  {journal} {\bibinfo  {journal} {Journal of Quantitative
  Spectroscopy and Radiative Transfer}\ }\textbf {\bibinfo {volume} {230}},\
  \bibinfo {pages} {86 } (\bibinfo {year} {2019}{\natexlab{d}})}\BibitemShut
  {NoStop}%
\bibitem [{\citenamefont {Doicu}\ and\ \citenamefont
  {Mishchenko}(2019{\natexlab{e}})}]{doicuJQSRT2019d}%
  \BibitemOpen
  \bibfield  {author} {\bibinfo {author} {\bibfnamefont {A.}~\bibnamefont
  {Doicu}}\ and\ \bibinfo {author} {\bibfnamefont {M.~I.}\ \bibnamefont
  {Mishchenko}},\ }\href {\doibase https://doi.org/10.1016/j.jqsrt.2019.07.008}
  {\bibfield  {journal} {\bibinfo  {journal} {Journal of Quantitative
  Spectroscopy and Radiative Transfer}\ }\textbf {\bibinfo {volume} {236}},\
  \bibinfo {pages} {106565} (\bibinfo {year} {2019}{\natexlab{e}})}\BibitemShut
  {NoStop}%
\bibitem [{\citenamefont {\ifmmode~\check{C}\else \v{C}\fi{}apeta}\ \emph
  {et~al.}(2011)\citenamefont {\ifmmode~\check{C}\else \v{C}\fi{}apeta},
  \citenamefont {Radi\ifmmode~\acute{c}\else \'{c}\fi{}}, \citenamefont
  {Szameit}, \citenamefont {Segev},\ and\ \citenamefont
  {Buljan}}]{capetaPRA2011}%
  \BibitemOpen
  \bibfield  {author} {\bibinfo {author} {\bibfnamefont {D.}~\bibnamefont
  {\ifmmode~\check{C}\else \v{C}\fi{}apeta}}, \bibinfo {author} {\bibfnamefont
  {J.}~\bibnamefont {Radi\ifmmode~\acute{c}\else \'{c}\fi{}}}, \bibinfo
  {author} {\bibfnamefont {A.}~\bibnamefont {Szameit}}, \bibinfo {author}
  {\bibfnamefont {M.}~\bibnamefont {Segev}}, \ and\ \bibinfo {author}
  {\bibfnamefont {H.}~\bibnamefont {Buljan}},\ }\href {\doibase
  10.1103/PhysRevA.84.011801} {\bibfield  {journal} {\bibinfo  {journal} {Phys.
  Rev. A}\ }\textbf {\bibinfo {volume} {84}},\ \bibinfo {pages} {011801}
  (\bibinfo {year} {2011})}\BibitemShut {NoStop}%
\bibitem [{\citenamefont {Labeyrie}\ \emph {et~al.}(2003)\citenamefont
  {Labeyrie}, \citenamefont {Vaujour}, \citenamefont {M\"uller}, \citenamefont
  {Delande}, \citenamefont {Miniatura}, \citenamefont {Wilkowski},\ and\
  \citenamefont {Kaiser}}]{labeyriePRL2003}%
  \BibitemOpen
  \bibfield  {author} {\bibinfo {author} {\bibfnamefont {G.}~\bibnamefont
  {Labeyrie}}, \bibinfo {author} {\bibfnamefont {E.}~\bibnamefont {Vaujour}},
  \bibinfo {author} {\bibfnamefont {C.~A.}\ \bibnamefont {M\"uller}}, \bibinfo
  {author} {\bibfnamefont {D.}~\bibnamefont {Delande}}, \bibinfo {author}
  {\bibfnamefont {C.}~\bibnamefont {Miniatura}}, \bibinfo {author}
  {\bibfnamefont {D.}~\bibnamefont {Wilkowski}}, \ and\ \bibinfo {author}
  {\bibfnamefont {R.}~\bibnamefont {Kaiser}},\ }\href {\doibase
  10.1103/PhysRevLett.91.223904} {\bibfield  {journal} {\bibinfo  {journal}
  {Phys. Rev. Lett.}\ }\textbf {\bibinfo {volume} {91}},\ \bibinfo {pages}
  {223904} (\bibinfo {year} {2003})}\BibitemShut {NoStop}%
\bibitem [{\citenamefont {Stephen}\ and\ \citenamefont
  {Cwilich}(1987)}]{stephenPRL1987}%
  \BibitemOpen
  \bibfield  {author} {\bibinfo {author} {\bibfnamefont {M.~J.}\ \bibnamefont
  {Stephen}}\ and\ \bibinfo {author} {\bibfnamefont {G.}~\bibnamefont
  {Cwilich}},\ }\href {\doibase 10.1103/PhysRevLett.59.285} {\bibfield
  {journal} {\bibinfo  {journal} {Phys. Rev. Lett.}\ }\textbf {\bibinfo
  {volume} {59}},\ \bibinfo {pages} {285} (\bibinfo {year} {1987})}\BibitemShut
  {NoStop}%
\bibitem [{\citenamefont {Stephen}(1988)}]{stephenPRB1988}%
  \BibitemOpen
  \bibfield  {author} {\bibinfo {author} {\bibfnamefont {M.~J.}\ \bibnamefont
  {Stephen}},\ }\href {\doibase 10.1103/PhysRevB.37.1} {\bibfield  {journal}
  {\bibinfo  {journal} {Phys. Rev. B}\ }\textbf {\bibinfo {volume} {37}},\
  \bibinfo {pages} {1} (\bibinfo {year} {1988})}\BibitemShut {NoStop}%
\bibitem [{\citenamefont {Etemad}\ \emph {et~al.}(1986)\citenamefont {Etemad},
  \citenamefont {Thompson},\ and\ \citenamefont {Andrejco}}]{etemadPRL1986}%
  \BibitemOpen
  \bibfield  {author} {\bibinfo {author} {\bibfnamefont {S.}~\bibnamefont
  {Etemad}}, \bibinfo {author} {\bibfnamefont {R.}~\bibnamefont {Thompson}}, \
  and\ \bibinfo {author} {\bibfnamefont {M.~J.}\ \bibnamefont {Andrejco}},\
  }\href {\doibase 10.1103/PhysRevLett.57.575} {\bibfield  {journal} {\bibinfo
  {journal} {Phys. Rev. Lett.}\ }\textbf {\bibinfo {volume} {57}},\ \bibinfo
  {pages} {575} (\bibinfo {year} {1986})}\BibitemShut {NoStop}%
\bibitem [{\citenamefont {Feng}\ \emph {et~al.}(1988)\citenamefont {Feng},
  \citenamefont {Kane}, \citenamefont {Lee},\ and\ \citenamefont
  {Stone}}]{fengPRL1988}%
  \BibitemOpen
  \bibfield  {author} {\bibinfo {author} {\bibfnamefont {S.}~\bibnamefont
  {Feng}}, \bibinfo {author} {\bibfnamefont {C.}~\bibnamefont {Kane}}, \bibinfo
  {author} {\bibfnamefont {P.~A.}\ \bibnamefont {Lee}}, \ and\ \bibinfo
  {author} {\bibfnamefont {A.~D.}\ \bibnamefont {Stone}},\ }\href {\doibase
  10.1103/PhysRevLett.61.834} {\bibfield  {journal} {\bibinfo  {journal} {Phys.
  Rev. Lett.}\ }\textbf {\bibinfo {volume} {61}},\ \bibinfo {pages} {834}
  (\bibinfo {year} {1988})}\BibitemShut {NoStop}%
\bibitem [{\citenamefont {Kaiser}(2009)}]{kaiserJMO2009}%
  \BibitemOpen
  \bibfield  {author} {\bibinfo {author} {\bibfnamefont {R.}~\bibnamefont
  {Kaiser}},\ }\href {\doibase 10.1080/09500340903082663} {\bibfield  {journal}
  {\bibinfo  {journal} {Journal of Modern Optics}\ }\textbf {\bibinfo {volume}
  {56}},\ \bibinfo {pages} {2082} (\bibinfo {year} {2009})},\ \Eprint
  {http://arxiv.org/abs/https://doi.org/10.1080/09500340903082663}
  {https://doi.org/10.1080/09500340903082663} \BibitemShut {NoStop}%
\bibitem [{\citenamefont {Smolka}\ \emph {et~al.}(2009)\citenamefont {Smolka},
  \citenamefont {Huck}, \citenamefont {Andersen}, \citenamefont {Lagendijk},\
  and\ \citenamefont {Lodahl}}]{smolkaPRL2009}%
  \BibitemOpen
  \bibfield  {author} {\bibinfo {author} {\bibfnamefont {S.}~\bibnamefont
  {Smolka}}, \bibinfo {author} {\bibfnamefont {A.}~\bibnamefont {Huck}},
  \bibinfo {author} {\bibfnamefont {U.~L.}\ \bibnamefont {Andersen}}, \bibinfo
  {author} {\bibfnamefont {A.}~\bibnamefont {Lagendijk}}, \ and\ \bibinfo
  {author} {\bibfnamefont {P.}~\bibnamefont {Lodahl}},\ }\href {\doibase
  10.1103/PhysRevLett.102.193901} {\bibfield  {journal} {\bibinfo  {journal}
  {Phys. Rev. Lett.}\ }\textbf {\bibinfo {volume} {102}},\ \bibinfo {pages}
  {193901} (\bibinfo {year} {2009})}\BibitemShut {NoStop}%
\bibitem [{\citenamefont {Ott}\ \emph {et~al.}(2010)\citenamefont {Ott},
  \citenamefont {Mortensen},\ and\ \citenamefont {Lodahl}}]{ottPRL2010}%
  \BibitemOpen
  \bibfield  {author} {\bibinfo {author} {\bibfnamefont {J.~R.}\ \bibnamefont
  {Ott}}, \bibinfo {author} {\bibfnamefont {N.~A.}\ \bibnamefont {Mortensen}},
  \ and\ \bibinfo {author} {\bibfnamefont {P.}~\bibnamefont {Lodahl}},\ }\href
  {\doibase 10.1103/PhysRevLett.105.090501} {\bibfield  {journal} {\bibinfo
  {journal} {Phys. Rev. Lett.}\ }\textbf {\bibinfo {volume} {105}},\ \bibinfo
  {pages} {090501} (\bibinfo {year} {2010})}\BibitemShut {NoStop}%
\bibitem [{\citenamefont {Alberucci}\ \emph {et~al.}(2018)\citenamefont
  {Alberucci}, \citenamefont {Jisha}, \citenamefont {Bolis}, \citenamefont
  {Beeckman},\ and\ \citenamefont {Nolte}}]{alberucciOL2018}%
  \BibitemOpen
  \bibfield  {author} {\bibinfo {author} {\bibfnamefont {A.}~\bibnamefont
  {Alberucci}}, \bibinfo {author} {\bibfnamefont {C.~P.}\ \bibnamefont
  {Jisha}}, \bibinfo {author} {\bibfnamefont {S.}~\bibnamefont {Bolis}},
  \bibinfo {author} {\bibfnamefont {J.}~\bibnamefont {Beeckman}}, \ and\
  \bibinfo {author} {\bibfnamefont {S.}~\bibnamefont {Nolte}},\ }\href
  {\doibase 10.1364/OL.43.003461} {\bibfield  {journal} {\bibinfo  {journal}
  {Opt. Lett.}\ }\textbf {\bibinfo {volume} {43}},\ \bibinfo {pages} {3461}
  (\bibinfo {year} {2018})}\BibitemShut {NoStop}%
\bibitem [{\citenamefont {Angelani}\ \emph {et~al.}(2006)\citenamefont
  {Angelani}, \citenamefont {Conti}, \citenamefont {Ruocco},\ and\
  \citenamefont {Zamponi}}]{angelaniPRL2006}%
  \BibitemOpen
  \bibfield  {author} {\bibinfo {author} {\bibfnamefont {L.}~\bibnamefont
  {Angelani}}, \bibinfo {author} {\bibfnamefont {C.}~\bibnamefont {Conti}},
  \bibinfo {author} {\bibfnamefont {G.}~\bibnamefont {Ruocco}}, \ and\ \bibinfo
  {author} {\bibfnamefont {F.}~\bibnamefont {Zamponi}},\ }\href {\doibase
  10.1103/PhysRevLett.96.065702} {\bibfield  {journal} {\bibinfo  {journal}
  {Phys. Rev. Lett.}\ }\textbf {\bibinfo {volume} {96}},\ \bibinfo {pages}
  {065702} (\bibinfo {year} {2006})}\BibitemShut {NoStop}%
\bibitem [{\citenamefont {Conti}\ \emph {et~al.}(2007)\citenamefont {Conti},
  \citenamefont {Angelani},\ and\ \citenamefont {Ruocco}}]{contiPRA2007}%
  \BibitemOpen
  \bibfield  {author} {\bibinfo {author} {\bibfnamefont {C.}~\bibnamefont
  {Conti}}, \bibinfo {author} {\bibfnamefont {L.}~\bibnamefont {Angelani}}, \
  and\ \bibinfo {author} {\bibfnamefont {G.}~\bibnamefont {Ruocco}},\ }\href
  {\doibase 10.1103/PhysRevA.75.033812} {\bibfield  {journal} {\bibinfo
  {journal} {Phys. Rev. A}\ }\textbf {\bibinfo {volume} {75}},\ \bibinfo
  {pages} {033812} (\bibinfo {year} {2007})}\BibitemShut {NoStop}%
\bibitem [{\citenamefont {Lee}(1990)}]{leeJAP1990}%
  \BibitemOpen
  \bibfield  {author} {\bibinfo {author} {\bibfnamefont {S.}~\bibnamefont
  {Lee}},\ }\href {\doibase 10.1063/1.347080} {\bibfield  {journal} {\bibinfo
  {journal} {Journal of Applied Physics}\ }\textbf {\bibinfo {volume} {68}},\
  \bibinfo {pages} {4952} (\bibinfo {year} {1990})},\ \Eprint
  {http://arxiv.org/abs/https://doi.org/10.1063/1.347080}
  {https://doi.org/10.1063/1.347080} \BibitemShut {NoStop}%
\bibitem [{\citenamefont {Lee}(1992)}]{leeJQSRT1992}%
  \BibitemOpen
  \bibfield  {author} {\bibinfo {author} {\bibfnamefont {S.-C.}\ \bibnamefont
  {Lee}},\ }\href {\doibase https://doi.org/10.1016/0022-4073(92)90081-E}
  {\bibfield  {journal} {\bibinfo  {journal} {Journal of Quantitative
  Spectroscopy and Radiative Transfer}\ }\textbf {\bibinfo {volume} {48}},\
  \bibinfo {pages} {119 } (\bibinfo {year} {1992})}\BibitemShut {NoStop}%
\bibitem [{\citenamefont {Lee}(1994)}]{leeJTHT1994}%
  \BibitemOpen
  \bibfield  {author} {\bibinfo {author} {\bibfnamefont {S.-C.}\ \bibnamefont
  {Lee}},\ }\href {\doibase 10.2514/3.593} {\bibfield  {journal} {\bibinfo
  {journal} {Journal of Thermophysics and Heat Transfer}\ }\textbf {\bibinfo
  {volume} {8}},\ \bibinfo {pages} {641} (\bibinfo {year} {1994})},\ \Eprint
  {http://arxiv.org/abs/https://doi.org/10.2514/3.593}
  {https://doi.org/10.2514/3.593} \BibitemShut {NoStop}%
\bibitem [{\citenamefont {Lee}(2019)}]{leeJQSRT2019}%
  \BibitemOpen
  \bibfield  {author} {\bibinfo {author} {\bibfnamefont {S.-C.}\ \bibnamefont
  {Lee}},\ }\href {\doibase https://doi.org/10.1016/j.jqsrt.2019.06.032}
  {\bibfield  {journal} {\bibinfo  {journal} {Journal of Quantitative
  Spectroscopy and Radiative Transfer}\ }\textbf {\bibinfo {volume} {235}},\
  \bibinfo {pages} {140 } (\bibinfo {year} {2019})}\BibitemShut {NoStop}%
\bibitem [{\citenamefont {Tsang}\ \emph {et~al.}(2004)\citenamefont {Tsang},
  \citenamefont {Kong}, \citenamefont {Ding},\ and\ \citenamefont
  {Ao}}]{tsang2004scattering2}%
  \BibitemOpen
  \bibfield  {author} {\bibinfo {author} {\bibfnamefont {L.}~\bibnamefont
  {Tsang}}, \bibinfo {author} {\bibfnamefont {J.~A.}\ \bibnamefont {Kong}},
  \bibinfo {author} {\bibfnamefont {K.-H.}\ \bibnamefont {Ding}}, \ and\
  \bibinfo {author} {\bibfnamefont {C.~O.}\ \bibnamefont {Ao}},\ }\href@noop {}
  {\emph {\bibinfo {title} {Scattering of Electromagnetic Waves: Numerical
  Simulations}}},\ Vol.~\bibinfo {volume} {25}\ (\bibinfo  {publisher} {John
  Wiley \& Sons},\ \bibinfo {year} {2004})\BibitemShut {NoStop}%
\bibitem [{\citenamefont {Mie}(1908)}]{mie1908}%
  \BibitemOpen
  \bibfield  {author} {\bibinfo {author} {\bibfnamefont {G.}~\bibnamefont
  {Mie}},\ }\href@noop {} {\bibfield  {journal} {\bibinfo  {journal} {Annalen
  der physik}\ }\textbf {\bibinfo {volume} {330}},\ \bibinfo {pages} {377}
  (\bibinfo {year} {1908})}\BibitemShut {NoStop}%
\bibitem [{\citenamefont {Wriedt}(2012)}]{wriedt2012mie}%
  \BibitemOpen
  \bibfield  {author} {\bibinfo {author} {\bibfnamefont {T.}~\bibnamefont
  {Wriedt}},\ }\enquote {\bibinfo {title} {Mie theory: A review},}\ in\ \href
  {\doibase 10.1007/978-3-642-28738-1_2} {\emph {\bibinfo {booktitle} {The Mie
  Theory: Basics and Applications}}},\ \bibinfo {editor} {edited by\ \bibinfo
  {editor} {\bibfnamefont {W.}~\bibnamefont {Hergert}}\ and\ \bibinfo {editor}
  {\bibfnamefont {T.}~\bibnamefont {Wriedt}}}\ (\bibinfo  {publisher} {Springer
  Berlin Heidelberg},\ \bibinfo {address} {Berlin, Heidelberg},\ \bibinfo
  {year} {2012})\ pp.\ \bibinfo {pages} {53--71}\BibitemShut {NoStop}%
\bibitem [{\citenamefont {Tribelsky}\ and\ \citenamefont
  {Luk'yanchuk}(2006)}]{tribelskyPRL2006}%
  \BibitemOpen
  \bibfield  {author} {\bibinfo {author} {\bibfnamefont {M.~I.}\ \bibnamefont
  {Tribelsky}}\ and\ \bibinfo {author} {\bibfnamefont {B.~S.}\ \bibnamefont
  {Luk'yanchuk}},\ }\href {\doibase 10.1103/PhysRevLett.97.263902} {\bibfield
  {journal} {\bibinfo  {journal} {Phys. Rev. Lett.}\ }\textbf {\bibinfo
  {volume} {97}},\ \bibinfo {pages} {263902} (\bibinfo {year}
  {2006})}\BibitemShut {NoStop}%
\bibitem [{\citenamefont {Kuznetsov}\ \emph {et~al.}(2016)\citenamefont
  {Kuznetsov}, \citenamefont {Miroshnichenko}, \citenamefont {Brongersma},
  \citenamefont {Kivshar},\ and\ \citenamefont
  {Luk’yanchuk}}]{kivsharScience2016}%
  \BibitemOpen
  \bibfield  {author} {\bibinfo {author} {\bibfnamefont {A.~I.}\ \bibnamefont
  {Kuznetsov}}, \bibinfo {author} {\bibfnamefont {A.~E.}\ \bibnamefont
  {Miroshnichenko}}, \bibinfo {author} {\bibfnamefont {M.~L.}\ \bibnamefont
  {Brongersma}}, \bibinfo {author} {\bibfnamefont {Y.~S.}\ \bibnamefont
  {Kivshar}}, \ and\ \bibinfo {author} {\bibfnamefont {B.}~\bibnamefont
  {Luk’yanchuk}},\ }\href@noop {} {\bibfield  {journal} {\bibinfo  {journal}
  {Science}\ }\textbf {\bibinfo {volume} {354}},\ \bibinfo {pages} {aag2472}
  (\bibinfo {year} {2016})}\BibitemShut {NoStop}%
\bibitem [{\citenamefont {Mackowski}\ and\ \citenamefont
  {Mishchenko}(1996)}]{mackowskiJOSAA1996}%
  \BibitemOpen
  \bibfield  {author} {\bibinfo {author} {\bibfnamefont {D.~W.}\ \bibnamefont
  {Mackowski}}\ and\ \bibinfo {author} {\bibfnamefont {M.~I.}\ \bibnamefont
  {Mishchenko}},\ }\href {\doibase 10.1364/JOSAA.13.002266} {\bibfield
  {journal} {\bibinfo  {journal} {J. Opt. Soc. Am. A}\ }\textbf {\bibinfo
  {volume} {13}},\ \bibinfo {pages} {2266} (\bibinfo {year}
  {1996})}\BibitemShut {NoStop}%
\bibitem [{\citenamefont {Mackowski}\ and\ \citenamefont
  {Mishchenko}(2013)}]{mackowskiJQSRT2013}%
  \BibitemOpen
  \bibfield  {author} {\bibinfo {author} {\bibfnamefont {D.}~\bibnamefont
  {Mackowski}}\ and\ \bibinfo {author} {\bibfnamefont {M.}~\bibnamefont
  {Mishchenko}},\ }\href {\doibase
  http://dx.doi.org/10.1016/j.jqsrt.2013.02.008} {\bibfield  {journal}
  {\bibinfo  {journal} {Journal of Quantitative Spectroscopy and Radiative
  Transfer}\ }\textbf {\bibinfo {volume} {123}},\ \bibinfo {pages} {103 }
  (\bibinfo {year} {2013})},\ \bibinfo {note} {peter C. Waterman and his
  scientific legacy}\BibitemShut {NoStop}%
\bibitem [{\citenamefont {G\'omez-Medina}\ \emph {et~al.}(2012)\citenamefont
  {G\'omez-Medina}, \citenamefont {Froufe-P\'erez}, \citenamefont {Y\'epez},
  \citenamefont {Scheffold}, \citenamefont {Nieto-Vesperinas},\ and\
  \citenamefont {S\'aenz}}]{Gomez-MedinaPRA2012}%
  \BibitemOpen
  \bibfield  {author} {\bibinfo {author} {\bibfnamefont {R.}~\bibnamefont
  {G\'omez-Medina}}, \bibinfo {author} {\bibfnamefont {L.~S.}\ \bibnamefont
  {Froufe-P\'erez}}, \bibinfo {author} {\bibfnamefont {M.}~\bibnamefont
  {Y\'epez}}, \bibinfo {author} {\bibfnamefont {F.}~\bibnamefont {Scheffold}},
  \bibinfo {author} {\bibfnamefont {M.}~\bibnamefont {Nieto-Vesperinas}}, \
  and\ \bibinfo {author} {\bibfnamefont {J.~J.}\ \bibnamefont {S\'aenz}},\
  }\href {\doibase 10.1103/PhysRevA.85.035802} {\bibfield  {journal} {\bibinfo
  {journal} {Phys. Rev. A}\ }\textbf {\bibinfo {volume} {85}},\ \bibinfo
  {pages} {035802} (\bibinfo {year} {2012})}\BibitemShut {NoStop}%
\bibitem [{\citenamefont {Bohren}\ and\ \citenamefont
  {Koh}(1985)}]{bohrenAO1985}%
  \BibitemOpen
  \bibfield  {author} {\bibinfo {author} {\bibfnamefont {C.~F.}\ \bibnamefont
  {Bohren}}\ and\ \bibinfo {author} {\bibfnamefont {G.}~\bibnamefont {Koh}},\
  }\href {\doibase 10.1364/AO.24.001023} {\bibfield  {journal} {\bibinfo
  {journal} {Appl. Opt.}\ }\textbf {\bibinfo {volume} {24}},\ \bibinfo {pages}
  {1023} (\bibinfo {year} {1985})}\BibitemShut {NoStop}%
\bibitem [{\citenamefont {Mishchenko}\ \emph {et~al.}(1999)\citenamefont
  {Mishchenko}, \citenamefont {Hovenier},\ and\ \citenamefont
  {Travis}}]{mishchenko1999light}%
  \BibitemOpen
  \bibfield  {author} {\bibinfo {author} {\bibfnamefont {M.~I.}\ \bibnamefont
  {Mishchenko}}, \bibinfo {author} {\bibfnamefont {J.~W.}\ \bibnamefont
  {Hovenier}}, \ and\ \bibinfo {author} {\bibfnamefont {L.~D.}\ \bibnamefont
  {Travis}},\ }\href@noop {} {\emph {\bibinfo {title} {Light scattering by
  nonspherical particles: theory, measurements, and applications}}}\ (\bibinfo
  {publisher} {Academic press},\ \bibinfo {year} {1999})\BibitemShut {NoStop}%
\bibitem [{\citenamefont {Waterman}(1965)}]{watermanPIEEE1965}%
  \BibitemOpen
  \bibfield  {author} {\bibinfo {author} {\bibfnamefont {P.~C.}\ \bibnamefont
  {Waterman}},\ }\href {\doibase 10.1109/PROC.1965.4058} {\bibfield  {journal}
  {\bibinfo  {journal} {Proceedings of the IEEE}\ }\textbf {\bibinfo {volume}
  {53}},\ \bibinfo {pages} {805} (\bibinfo {year} {1965})}\BibitemShut
  {NoStop}%
\bibitem [{\citenamefont {Peterson}\ and\ \citenamefont
  {Str\"om}(1974)}]{petersonPRD1974}%
  \BibitemOpen
  \bibfield  {author} {\bibinfo {author} {\bibfnamefont {B.}~\bibnamefont
  {Peterson}}\ and\ \bibinfo {author} {\bibfnamefont {S.}~\bibnamefont
  {Str\"om}},\ }\href {\doibase 10.1103/PhysRevD.10.2670} {\bibfield  {journal}
  {\bibinfo  {journal} {Phys. Rev. D}\ }\textbf {\bibinfo {volume} {10}},\
  \bibinfo {pages} {2670} (\bibinfo {year} {1974})}\BibitemShut {NoStop}%
\bibitem [{\citenamefont {Yurkin}\ and\ \citenamefont
  {Hoekstra}(2007)}]{yurkinJQSRT2007}%
  \BibitemOpen
  \bibfield  {author} {\bibinfo {author} {\bibfnamefont {M.}~\bibnamefont
  {Yurkin}}\ and\ \bibinfo {author} {\bibfnamefont {A.}~\bibnamefont
  {Hoekstra}},\ }\href {\doibase http://dx.doi.org/10.1016/j.jqsrt.2007.01.034}
  {\bibfield  {journal} {\bibinfo  {journal} {Journal of Quantitative
  Spectroscopy and Radiative Transfer}\ }\textbf {\bibinfo {volume} {106}},\
  \bibinfo {pages} {558 } (\bibinfo {year} {2007})},\ \bibinfo {note} {iX
  Conference on Electromagnetic and Light Scattering by Non-Spherical
  Particles}\BibitemShut {NoStop}%
\bibitem [{\citenamefont {{Purcell}}\ and\ \citenamefont
  {{Pennypacker}}(1973)}]{purcellAPJ1973}%
  \BibitemOpen
  \bibfield  {author} {\bibinfo {author} {\bibfnamefont {E.~M.}\ \bibnamefont
  {{Purcell}}}\ and\ \bibinfo {author} {\bibfnamefont {C.~R.}\ \bibnamefont
  {{Pennypacker}}},\ }\href {\doibase 10.1086/152538} {\bibfield  {journal}
  {\bibinfo  {journal} {The Astrophysical Journal}\ }\textbf {\bibinfo {volume}
  {186}},\ \bibinfo {pages} {705} (\bibinfo {year} {1973})}\BibitemShut
  {NoStop}%
\bibitem [{\citenamefont {{Draine}}(1988)}]{draineAPJ1988}%
  \BibitemOpen
  \bibfield  {author} {\bibinfo {author} {\bibfnamefont {B.~T.}\ \bibnamefont
  {{Draine}}},\ }\href {\doibase 10.1086/166795} {\bibfield  {journal}
  {\bibinfo  {journal} {The Astrophysical Journal}\ }\textbf {\bibinfo {volume}
  {333}},\ \bibinfo {pages} {848} (\bibinfo {year} {1988})}\BibitemShut
  {NoStop}%
\bibitem [{\citenamefont {{Draine}}\ and\ \citenamefont
  {{Goodman}}(1993)}]{draineAPJ1993}%
  \BibitemOpen
  \bibfield  {author} {\bibinfo {author} {\bibfnamefont {B.~T.}\ \bibnamefont
  {{Draine}}}\ and\ \bibinfo {author} {\bibfnamefont {J.}~\bibnamefont
  {{Goodman}}},\ }\href {\doibase 10.1086/172396} {\bibfield  {journal}
  {\bibinfo  {journal} {The Astrophysical Journal}\ }\textbf {\bibinfo {volume}
  {405}},\ \bibinfo {pages} {685} (\bibinfo {year} {1993})}\BibitemShut
  {NoStop}%
\bibitem [{\citenamefont {Draine}\ and\ \citenamefont
  {Flatau}(1994)}]{draineJOSA1994}%
  \BibitemOpen
  \bibfield  {author} {\bibinfo {author} {\bibfnamefont {B.~T.}\ \bibnamefont
  {Draine}}\ and\ \bibinfo {author} {\bibfnamefont {P.~J.}\ \bibnamefont
  {Flatau}},\ }\href {\doibase 10.1364/JOSAA.11.001491} {\bibfield  {journal}
  {\bibinfo  {journal} {J. Opt. Soc. Am. A}\ }\textbf {\bibinfo {volume}
  {11}},\ \bibinfo {pages} {1491} (\bibinfo {year} {1994})}\BibitemShut
  {NoStop}%
\bibitem [{\citenamefont {Markel}(2019)}]{markelJQSRT2019}%
  \BibitemOpen
  \bibfield  {author} {\bibinfo {author} {\bibfnamefont {V.~A.}\ \bibnamefont
  {Markel}},\ }\href {\doibase https://doi.org/10.1016/j.jqsrt.2019.106611}
  {\bibfield  {journal} {\bibinfo  {journal} {Journal of Quantitative
  Spectroscopy and Radiative Transfer}\ }\textbf {\bibinfo {volume} {236}},\
  \bibinfo {pages} {106611} (\bibinfo {year} {2019})}\BibitemShut {NoStop}%
\bibitem [{\citenamefont {Yurkin}\ and\ \citenamefont
  {Hoekstra}(2011)}]{yurkinJQSRT2011}%
  \BibitemOpen
  \bibfield  {author} {\bibinfo {author} {\bibfnamefont {M.~A.}\ \bibnamefont
  {Yurkin}}\ and\ \bibinfo {author} {\bibfnamefont {A.~G.}\ \bibnamefont
  {Hoekstra}},\ }\href {\doibase https://doi.org/10.1016/j.jqsrt.2011.01.031}
  {\bibfield  {journal} {\bibinfo  {journal} {Journal of Quantitative
  Spectroscopy and Radiative Transfer}\ }\textbf {\bibinfo {volume} {112}},\
  \bibinfo {pages} {2234 } (\bibinfo {year} {2011})},\ \bibinfo {note}
  {polarimetric Detection, Characterization, and Remote Sensing}\BibitemShut
  {NoStop}%
\bibitem [{\citenamefont {Mahan}(2013)}]{mahan2013many}%
  \BibitemOpen
  \bibfield  {author} {\bibinfo {author} {\bibfnamefont {G.~D.}\ \bibnamefont
  {Mahan}},\ }\href@noop {} {\emph {\bibinfo {title} {Many-particle physics}}}\
  (\bibinfo  {publisher} {Springer Science \& Business Media},\ \bibinfo {year}
  {2013})\BibitemShut {NoStop}%
\bibitem [{\citenamefont {Frisch}(1968)}]{frisch1968wave}%
  \BibitemOpen
  \bibfield  {author} {\bibinfo {author} {\bibfnamefont {U.}~\bibnamefont
  {Frisch}},\ }\href@noop {} {\bibfield  {journal} {\bibinfo  {journal}
  {Probabilistic methods in applied mathematics}\ ,\ \bibinfo {pages} {75}}
  (\bibinfo {year} {1968})}\BibitemShut {NoStop}%
\bibitem [{\citenamefont {{Ishimaru}}(1977)}]{ishimaruPIEEE1977}%
  \BibitemOpen
  \bibfield  {author} {\bibinfo {author} {\bibfnamefont {A.}~\bibnamefont
  {{Ishimaru}}},\ }\href {\doibase 10.1109/PROC.1977.10612} {\bibfield
  {journal} {\bibinfo  {journal} {Proceedings of the IEEE}\ }\textbf {\bibinfo
  {volume} {65}},\ \bibinfo {pages} {1030} (\bibinfo {year}
  {1977})}\BibitemShut {NoStop}%
\bibitem [{\citenamefont {Barabanenkov}\ \emph {et~al.}(1971)\citenamefont
  {Barabanenkov}, \citenamefont {Kravtsov}, \citenamefont {Rytov},\ and\
  \citenamefont {Tamarski{\u{\i}}}}]{barabanenkov1971status}%
  \BibitemOpen
  \bibfield  {author} {\bibinfo {author} {\bibfnamefont {Y.~N.}\ \bibnamefont
  {Barabanenkov}}, \bibinfo {author} {\bibfnamefont {Y.~A.}\ \bibnamefont
  {Kravtsov}}, \bibinfo {author} {\bibfnamefont {S.~M.}\ \bibnamefont {Rytov}},
  \ and\ \bibinfo {author} {\bibfnamefont {V.~I.}\ \bibnamefont
  {Tamarski{\u{\i}}}},\ }\href {\doibase 10.1070/pu1971v013n05abeh004213}
  {\bibfield  {journal} {\bibinfo  {journal} {Soviet Physics Uspekhi}\ }\textbf
  {\bibinfo {volume} {13}},\ \bibinfo {pages} {551} (\bibinfo {year}
  {1971})}\BibitemShut {NoStop}%
\bibitem [{\citenamefont {Barabanenkov}\ \emph {et~al.}(1995)\citenamefont
  {Barabanenkov}, \citenamefont {Zurk},\ and\ \citenamefont
  {Barabanenkov}}]{barabanenkovJEWA1995}%
  \BibitemOpen
  \bibfield  {author} {\bibinfo {author} {\bibfnamefont {Y.~N.}\ \bibnamefont
  {Barabanenkov}}, \bibinfo {author} {\bibfnamefont {L.}~\bibnamefont {Zurk}},
  \ and\ \bibinfo {author} {\bibfnamefont {M.~Y.}\ \bibnamefont
  {Barabanenkov}},\ }\href {\doibase 10.1163/156939395X00127} {\bibfield
  {journal} {\bibinfo  {journal} {Journal of Electromagnetic Waves and
  Applications}\ }\textbf {\bibinfo {volume} {9}},\ \bibinfo {pages} {1393}
  (\bibinfo {year} {1995})}\BibitemShut {NoStop}%
\bibitem [{\citenamefont {Tsang}\ and\ \citenamefont
  {Kong}(1980)}]{tsangJAP1980}%
  \BibitemOpen
  \bibfield  {author} {\bibinfo {author} {\bibfnamefont {L.}~\bibnamefont
  {Tsang}}\ and\ \bibinfo {author} {\bibfnamefont {J.~A.}\ \bibnamefont
  {Kong}},\ }\href {\doibase 10.1063/1.328200} {\bibfield  {journal} {\bibinfo
  {journal} {Journal of Applied Physics}\ }\textbf {\bibinfo {volume} {51}},\
  \bibinfo {pages} {3465} (\bibinfo {year} {1980})},\ \Eprint
  {http://arxiv.org/abs/https://doi.org/10.1063/1.328200}
  {https://doi.org/10.1063/1.328200} \BibitemShut {NoStop}%
\bibitem [{\citenamefont {Page}\ \emph {et~al.}(1996)\citenamefont {Page},
  \citenamefont {Sheng}, \citenamefont {Schriemer}, \citenamefont {Jones},
  \citenamefont {Jing},\ and\ \citenamefont {Weitz}}]{Page1996}%
  \BibitemOpen
  \bibfield  {author} {\bibinfo {author} {\bibfnamefont {J.~H.}\ \bibnamefont
  {Page}}, \bibinfo {author} {\bibfnamefont {P.}~\bibnamefont {Sheng}},
  \bibinfo {author} {\bibfnamefont {H.~P.}\ \bibnamefont {Schriemer}}, \bibinfo
  {author} {\bibfnamefont {I.}~\bibnamefont {Jones}}, \bibinfo {author}
  {\bibfnamefont {X.}~\bibnamefont {Jing}}, \ and\ \bibinfo {author}
  {\bibfnamefont {D.~A.}\ \bibnamefont {Weitz}},\ }\href {\doibase
  10.1126/science.271.5249.634} {\bibfield  {journal} {\bibinfo  {journal}
  {Science}\ }\textbf {\bibinfo {volume} {271}},\ \bibinfo {pages} {634}
  (\bibinfo {year} {1996})}\BibitemShut {NoStop}%
\bibitem [{\citenamefont {Wang}\ and\ \citenamefont
  {Zhao}(2018{\natexlab{c}})}]{wangPRA2018}%
  \BibitemOpen
  \bibfield  {author} {\bibinfo {author} {\bibfnamefont {B.~X.}\ \bibnamefont
  {Wang}}\ and\ \bibinfo {author} {\bibfnamefont {C.~Y.}\ \bibnamefont
  {Zhao}},\ }\href {\doibase 10.1103/PhysRevA.97.023836} {\bibfield  {journal}
  {\bibinfo  {journal} {Phys. Rev. A}\ }\textbf {\bibinfo {volume} {97}},\
  \bibinfo {pages} {023836} (\bibinfo {year} {2018}{\natexlab{c}})}\BibitemShut
  {NoStop}%
\bibitem [{\citenamefont {Barrera}\ \emph {et~al.}(2007)\citenamefont
  {Barrera}, \citenamefont {Reyes-Coronado},\ and\ \citenamefont
  {Garc\'{\i}a-Valenzuela}}]{barreraPRB2007}%
  \BibitemOpen
  \bibfield  {author} {\bibinfo {author} {\bibfnamefont {R.~G.}\ \bibnamefont
  {Barrera}}, \bibinfo {author} {\bibfnamefont {A.}~\bibnamefont
  {Reyes-Coronado}}, \ and\ \bibinfo {author} {\bibfnamefont {A.}~\bibnamefont
  {Garc\'{\i}a-Valenzuela}},\ }\href {\doibase 10.1103/PhysRevB.75.184202}
  {\bibfield  {journal} {\bibinfo  {journal} {Phys. Rev. B}\ }\textbf {\bibinfo
  {volume} {75}},\ \bibinfo {pages} {184202} (\bibinfo {year}
  {2007})}\BibitemShut {NoStop}%
\bibitem [{\citenamefont {Gower}\ \emph {et~al.}(2019)\citenamefont {Gower},
  \citenamefont {Abrahams},\ and\ \citenamefont {Parnell}}]{gowerRSPA2019}%
  \BibitemOpen
  \bibfield  {author} {\bibinfo {author} {\bibfnamefont {A.~L.}\ \bibnamefont
  {Gower}}, \bibinfo {author} {\bibfnamefont {I.~D.}\ \bibnamefont {Abrahams}},
  \ and\ \bibinfo {author} {\bibfnamefont {W.~J.}\ \bibnamefont {Parnell}},\
  }\href {\doibase 10.1098/rspa.2019.0344} {\bibfield  {journal} {\bibinfo
  {journal} {Proceedings of the Royal Society A: Mathematical, Physical and
  Engineering Sciences}\ }\textbf {\bibinfo {volume} {475}},\ \bibinfo {pages}
  {20190344} (\bibinfo {year} {2019})},\ \Eprint
  {http://arxiv.org/abs/https://royalsocietypublishing.org/doi/pdf/10.1098/rspa.2019.0344}
  {https://royalsocietypublishing.org/doi/pdf/10.1098/rspa.2019.0344}
  \BibitemShut {NoStop}%
\bibitem [{\citenamefont {de~Vries}\ \emph {et~al.}(1998)\citenamefont
  {de~Vries}, \citenamefont {van Coevorden},\ and\ \citenamefont
  {Lagendijk}}]{devriesRMP1998}%
  \BibitemOpen
  \bibfield  {author} {\bibinfo {author} {\bibfnamefont {P.}~\bibnamefont
  {de~Vries}}, \bibinfo {author} {\bibfnamefont {D.~V.}\ \bibnamefont {van
  Coevorden}}, \ and\ \bibinfo {author} {\bibfnamefont {A.}~\bibnamefont
  {Lagendijk}},\ }\href {\doibase 10.1103/RevModPhys.70.447} {\bibfield
  {journal} {\bibinfo  {journal} {Rev. Mod. Phys.}\ }\textbf {\bibinfo {volume}
  {70}},\ \bibinfo {pages} {447} (\bibinfo {year} {1998})}\BibitemShut
  {NoStop}%
\bibitem [{\citenamefont {Cherroret}\ \emph {et~al.}(2016)\citenamefont
  {Cherroret}, \citenamefont {Delande},\ and\ \citenamefont {van
  Tiggelen}}]{Cherroret2016}%
  \BibitemOpen
  \bibfield  {author} {\bibinfo {author} {\bibfnamefont {N.}~\bibnamefont
  {Cherroret}}, \bibinfo {author} {\bibfnamefont {D.}~\bibnamefont {Delande}},
  \ and\ \bibinfo {author} {\bibfnamefont {B.~A.}\ \bibnamefont {van
  Tiggelen}},\ }\href {\doibase 10.1103/PhysRevA.94.012702} {\bibfield
  {journal} {\bibinfo  {journal} {Phys. Rev. A}\ }\textbf {\bibinfo {volume}
  {94}},\ \bibinfo {pages} {012702} (\bibinfo {year} {2016})}\BibitemShut
  {NoStop}%
\bibitem [{\citenamefont {Barabanenkov}(1975)}]{barabanenkov1975multiple}%
  \BibitemOpen
  \bibfield  {author} {\bibinfo {author} {\bibfnamefont {Y.~N.}\ \bibnamefont
  {Barabanenkov}},\ }\href@noop {} {\bibfield  {journal} {\bibinfo  {journal}
  {Soviet Physics Uspekhi}\ }\textbf {\bibinfo {volume} {18}},\ \bibinfo
  {pages} {673} (\bibinfo {year} {1975})}\BibitemShut {NoStop}%
\bibitem [{\citenamefont {Kugo}\ and\ \citenamefont
  {Mitchard}(1992)}]{kugoPLB1992}%
  \BibitemOpen
  \bibfield  {author} {\bibinfo {author} {\bibfnamefont {T.}~\bibnamefont
  {Kugo}}\ and\ \bibinfo {author} {\bibfnamefont {M.~G.}\ \bibnamefont
  {Mitchard}},\ }\href {\doibase https://doi.org/10.1016/0370-2693(92)90496-Q}
  {\bibfield  {journal} {\bibinfo  {journal} {Physics Letters B}\ }\textbf
  {\bibinfo {volume} {282}},\ \bibinfo {pages} {162 } (\bibinfo {year}
  {1992})}\BibitemShut {NoStop}%
\bibitem [{\citenamefont {Varadan}\ \emph {et~al.}(1985)\citenamefont
  {Varadan}, \citenamefont {Ma},\ and\ \citenamefont
  {Varadan}}]{varadanJOSAA1985}%
  \BibitemOpen
  \bibfield  {author} {\bibinfo {author} {\bibfnamefont {V.~V.}\ \bibnamefont
  {Varadan}}, \bibinfo {author} {\bibfnamefont {Y.}~\bibnamefont {Ma}}, \ and\
  \bibinfo {author} {\bibfnamefont {V.~K.}\ \bibnamefont {Varadan}},\ }\href
  {\doibase 10.1364/JOSAA.2.002195} {\bibfield  {journal} {\bibinfo  {journal}
  {J. Opt. Soc. Am. A}\ }\textbf {\bibinfo {volume} {2}},\ \bibinfo {pages}
  {2195} (\bibinfo {year} {1985})}\BibitemShut {NoStop}%
\bibitem [{\citenamefont {Lamb}\ \emph {et~al.}(1980)\citenamefont {Lamb},
  \citenamefont {Wood},\ and\ \citenamefont {Ashcroft}}]{lambPRB1980}%
  \BibitemOpen
  \bibfield  {author} {\bibinfo {author} {\bibfnamefont {W.}~\bibnamefont
  {Lamb}}, \bibinfo {author} {\bibfnamefont {D.~M.}\ \bibnamefont {Wood}}, \
  and\ \bibinfo {author} {\bibfnamefont {N.~W.}\ \bibnamefont {Ashcroft}},\
  }\href {\doibase 10.1103/PhysRevB.21.2248} {\bibfield  {journal} {\bibinfo
  {journal} {Phys. Rev. B}\ }\textbf {\bibinfo {volume} {21}},\ \bibinfo
  {pages} {2248} (\bibinfo {year} {1980})}\BibitemShut {NoStop}%
\bibitem [{\citenamefont {Tsang}\ and\ \citenamefont
  {Kong}(1982)}]{tsangJAP1982}%
  \BibitemOpen
  \bibfield  {author} {\bibinfo {author} {\bibfnamefont {L.}~\bibnamefont
  {Tsang}}\ and\ \bibinfo {author} {\bibfnamefont {J.~A.}\ \bibnamefont
  {Kong}},\ }\href {\doibase 10.1063/1.331611} {\bibfield  {journal} {\bibinfo
  {journal} {Journal of Applied Physics}\ }\textbf {\bibinfo {volume} {53}},\
  \bibinfo {pages} {7162} (\bibinfo {year} {1982})}\BibitemShut {NoStop}%
\bibitem [{\citenamefont {Xu}\ and\ \citenamefont
  {Gustafson}(2001)}]{xuJQSRT2001}%
  \BibitemOpen
  \bibfield  {author} {\bibinfo {author} {\bibfnamefont {Y.~L.}\ \bibnamefont
  {Xu}}\ and\ \bibinfo {author} {\bibfnamefont {B.~A.~S.}\ \bibnamefont
  {Gustafson}},\ }\href {\doibase
  https://doi.org/10.1016/S0022-4073(01)00019-X} {\bibfield  {journal}
  {\bibinfo  {journal} {Journal of Quantitative Spectroscopy and Radiative
  Transfer}\ }\textbf {\bibinfo {volume} {70}},\ \bibinfo {pages} {395 }
  (\bibinfo {year} {2001})}\BibitemShut {NoStop}%
\bibitem [{\citenamefont {Mackowski}\ and\ \citenamefont
  {Mishchenko}(2011)}]{mackowskiJQSRT2011}%
  \BibitemOpen
  \bibfield  {author} {\bibinfo {author} {\bibfnamefont {D.}~\bibnamefont
  {Mackowski}}\ and\ \bibinfo {author} {\bibfnamefont {M.}~\bibnamefont
  {Mishchenko}},\ }\href {\doibase https://doi.org/10.1016/j.jqsrt.2011.02.019}
  {\bibfield  {journal} {\bibinfo  {journal} {Journal of Quantitative
  Spectroscopy and Radiative Transfer}\ }\textbf {\bibinfo {volume} {112}},\
  \bibinfo {pages} {2182 } (\bibinfo {year} {2011})},\ \bibinfo {note}
  {polarimetric Detection, Characterization, and Remote Sensing}\BibitemShut
  {NoStop}%
\bibitem [{\citenamefont {Pitman}\ \emph {et~al.}(2017)\citenamefont {Pitman},
  \citenamefont {Kolokolova}, \citenamefont {Verbiscer}, \citenamefont
  {Mackowski},\ and\ \citenamefont {Joseph}}]{pitmanJQSRT2017}%
  \BibitemOpen
  \bibfield  {author} {\bibinfo {author} {\bibfnamefont {K.~M.}\ \bibnamefont
  {Pitman}}, \bibinfo {author} {\bibfnamefont {L.}~\bibnamefont {Kolokolova}},
  \bibinfo {author} {\bibfnamefont {A.~J.}\ \bibnamefont {Verbiscer}}, \bibinfo
  {author} {\bibfnamefont {D.~W.}\ \bibnamefont {Mackowski}}, \ and\ \bibinfo
  {author} {\bibfnamefont {E.~C.}\ \bibnamefont {Joseph}},\ }\href {\doibase
  https://doi.org/10.1016/j.pss.2017.08.005} {\bibfield  {journal} {\bibinfo
  {journal} {Planetary and Space Science}\ }\textbf {\bibinfo {volume} {149}},\
  \bibinfo {pages} {23 } (\bibinfo {year} {2017})},\ \bibinfo {note} {special
  Issue: Cosmic Dust IX}\BibitemShut {NoStop}%
\bibitem [{\citenamefont {Stout}\ \emph {et~al.}(2002)\citenamefont {Stout},
  \citenamefont {Auger},\ and\ \citenamefont {Lafait}}]{stoutJMO2002}%
  \BibitemOpen
  \bibfield  {author} {\bibinfo {author} {\bibfnamefont {B.}~\bibnamefont
  {Stout}}, \bibinfo {author} {\bibfnamefont {J.-C.}\ \bibnamefont {Auger}}, \
  and\ \bibinfo {author} {\bibfnamefont {J.}~\bibnamefont {Lafait}},\ }\href
  {\doibase 10.1080/09500340210124450} {\bibfield  {journal} {\bibinfo
  {journal} {Journal of Modern Optics}\ }\textbf {\bibinfo {volume} {49}},\
  \bibinfo {pages} {2129} (\bibinfo {year} {2002})},\ \Eprint
  {http://arxiv.org/abs/https://doi.org/10.1080/09500340210124450}
  {https://doi.org/10.1080/09500340210124450} \BibitemShut {NoStop}%
\bibitem [{\citenamefont {{Wang}}\ and\ \citenamefont
  {{Chew}}(1993)}]{wangIEEETAP1993}%
  \BibitemOpen
  \bibfield  {author} {\bibinfo {author} {\bibfnamefont {Y.~.}\ \bibnamefont
  {{Wang}}}\ and\ \bibinfo {author} {\bibfnamefont {W.~C.}\ \bibnamefont
  {{Chew}}},\ }\href {\doibase 10.1109/8.273306} {\bibfield  {journal}
  {\bibinfo  {journal} {IEEE Transactions on Antennas and Propagation}\
  }\textbf {\bibinfo {volume} {41}},\ \bibinfo {pages} {1633} (\bibinfo {year}
  {1993})}\BibitemShut {NoStop}%
\bibitem [{\citenamefont {Egel}\ \emph
  {et~al.}(2017{\natexlab{a}})\citenamefont {Egel}, \citenamefont {Pattelli},
  \citenamefont {Mazzamuto}, \citenamefont {Wiersma},\ and\ \citenamefont
  {Lemmer}}]{egelJQSRT2017}%
  \BibitemOpen
  \bibfield  {author} {\bibinfo {author} {\bibfnamefont {A.}~\bibnamefont
  {Egel}}, \bibinfo {author} {\bibfnamefont {L.}~\bibnamefont {Pattelli}},
  \bibinfo {author} {\bibfnamefont {G.}~\bibnamefont {Mazzamuto}}, \bibinfo
  {author} {\bibfnamefont {D.~S.}\ \bibnamefont {Wiersma}}, \ and\ \bibinfo
  {author} {\bibfnamefont {U.}~\bibnamefont {Lemmer}},\ }\href {\doibase
  https://doi.org/10.1016/j.jqsrt.2017.05.010} {\bibfield  {journal} {\bibinfo
  {journal} {Journal of Quantitative Spectroscopy and Radiative Transfer}\
  }\textbf {\bibinfo {volume} {199}},\ \bibinfo {pages} {103 } (\bibinfo {year}
  {2017}{\natexlab{a}})}\BibitemShut {NoStop}%
\bibitem [{\citenamefont {Varadan}\ \emph {et~al.}(1979)\citenamefont
  {Varadan}, \citenamefont {Bringi},\ and\ \citenamefont
  {Varadan}}]{varadanPRD1979}%
  \BibitemOpen
  \bibfield  {author} {\bibinfo {author} {\bibfnamefont {V.~K.}\ \bibnamefont
  {Varadan}}, \bibinfo {author} {\bibfnamefont {V.~N.}\ \bibnamefont {Bringi}},
  \ and\ \bibinfo {author} {\bibfnamefont {V.~V.}\ \bibnamefont {Varadan}},\
  }\href {\doibase 10.1103/PhysRevD.19.2480} {\bibfield  {journal} {\bibinfo
  {journal} {Phys. Rev. D}\ }\textbf {\bibinfo {volume} {19}},\ \bibinfo
  {pages} {2480} (\bibinfo {year} {1979})}\BibitemShut {NoStop}%
\bibitem [{\citenamefont {Bertrand}\ \emph {et~al.}(2020)\citenamefont
  {Bertrand}, \citenamefont {Devilez}, \citenamefont {Hugonin}, \citenamefont
  {Lalanne},\ and\ \citenamefont {Vynck}}]{bertrand2019global}%
  \BibitemOpen
  \bibfield  {author} {\bibinfo {author} {\bibfnamefont {M.}~\bibnamefont
  {Bertrand}}, \bibinfo {author} {\bibfnamefont {A.}~\bibnamefont {Devilez}},
  \bibinfo {author} {\bibfnamefont {J.-P.}\ \bibnamefont {Hugonin}}, \bibinfo
  {author} {\bibfnamefont {P.}~\bibnamefont {Lalanne}}, \ and\ \bibinfo
  {author} {\bibfnamefont {K.}~\bibnamefont {Vynck}},\ }\href {\doibase
  10.1364/JOSAA.37.000070} {\bibfield  {journal} {\bibinfo  {journal} {J. Opt.
  Soc. Am. A}\ }\textbf {\bibinfo {volume} {37}},\ \bibinfo {pages} {70}
  (\bibinfo {year} {2020})}\BibitemShut {NoStop}%
\bibitem [{\citenamefont {Doicu}\ and\ \citenamefont
  {Wriedt}(2010)}]{doicuJQSRT2010}%
  \BibitemOpen
  \bibfield  {author} {\bibinfo {author} {\bibfnamefont {A.}~\bibnamefont
  {Doicu}}\ and\ \bibinfo {author} {\bibfnamefont {T.}~\bibnamefont {Wriedt}},\
  }\href {\doibase https://doi.org/10.1016/j.jqsrt.2009.10.003} {\bibfield
  {journal} {\bibinfo  {journal} {Journal of Quantitative Spectroscopy and
  Radiative Transfer}\ }\textbf {\bibinfo {volume} {111}},\ \bibinfo {pages}
  {466 } (\bibinfo {year} {2010})}\BibitemShut {NoStop}%
\bibitem [{\citenamefont {Forestiere}\ \emph {et~al.}(2011)\citenamefont
  {Forestiere}, \citenamefont {Iadarola}, \citenamefont {Negro},\ and\
  \citenamefont {Miano}}]{forestiereJQSRT2011}%
  \BibitemOpen
  \bibfield  {author} {\bibinfo {author} {\bibfnamefont {C.}~\bibnamefont
  {Forestiere}}, \bibinfo {author} {\bibfnamefont {G.}~\bibnamefont
  {Iadarola}}, \bibinfo {author} {\bibfnamefont {L.~D.}\ \bibnamefont {Negro}},
  \ and\ \bibinfo {author} {\bibfnamefont {G.}~\bibnamefont {Miano}},\ }\href
  {\doibase https://doi.org/10.1016/j.jqsrt.2011.05.009} {\bibfield  {journal}
  {\bibinfo  {journal} {Journal of Quantitative Spectroscopy and Radiative
  Transfer}\ }\textbf {\bibinfo {volume} {112}},\ \bibinfo {pages} {2384 }
  (\bibinfo {year} {2011})}\BibitemShut {NoStop}%
\bibitem [{\citenamefont {Egel}\ \emph {et~al.}(2016)\citenamefont {Egel},
  \citenamefont {Theobald}, \citenamefont {Donie}, \citenamefont {Lemmer},\
  and\ \citenamefont {Gomard}}]{egelOE2016}%
  \BibitemOpen
  \bibfield  {author} {\bibinfo {author} {\bibfnamefont {A.}~\bibnamefont
  {Egel}}, \bibinfo {author} {\bibfnamefont {D.}~\bibnamefont {Theobald}},
  \bibinfo {author} {\bibfnamefont {Y.}~\bibnamefont {Donie}}, \bibinfo
  {author} {\bibfnamefont {U.}~\bibnamefont {Lemmer}}, \ and\ \bibinfo {author}
  {\bibfnamefont {G.}~\bibnamefont {Gomard}},\ }\href {\doibase
  10.1364/OE.24.025154} {\bibfield  {journal} {\bibinfo  {journal} {Opt.
  Express}\ }\textbf {\bibinfo {volume} {24}},\ \bibinfo {pages} {25154}
  (\bibinfo {year} {2016})}\BibitemShut {NoStop}%
\bibitem [{\citenamefont {Egel}\ \emph
  {et~al.}(2017{\natexlab{b}})\citenamefont {Egel}, \citenamefont {Eremin},
  \citenamefont {Wriedt}, \citenamefont {Theobald}, \citenamefont {Lemmer},\
  and\ \citenamefont {Gomard}}]{egelJQSRT2017b}%
  \BibitemOpen
  \bibfield  {author} {\bibinfo {author} {\bibfnamefont {A.}~\bibnamefont
  {Egel}}, \bibinfo {author} {\bibfnamefont {Y.}~\bibnamefont {Eremin}},
  \bibinfo {author} {\bibfnamefont {T.}~\bibnamefont {Wriedt}}, \bibinfo
  {author} {\bibfnamefont {D.}~\bibnamefont {Theobald}}, \bibinfo {author}
  {\bibfnamefont {U.}~\bibnamefont {Lemmer}}, \ and\ \bibinfo {author}
  {\bibfnamefont {G.}~\bibnamefont {Gomard}},\ }\href {\doibase
  https://doi.org/10.1016/j.jqsrt.2017.08.016} {\bibfield  {journal} {\bibinfo
  {journal} {Journal of Quantitative Spectroscopy and Radiative Transfer}\
  }\textbf {\bibinfo {volume} {202}},\ \bibinfo {pages} {279 } (\bibinfo {year}
  {2017}{\natexlab{b}})}\BibitemShut {NoStop}%
\bibitem [{\citenamefont {Theobald}\ \emph {et~al.}(2017)\citenamefont
  {Theobald}, \citenamefont {Egel}, \citenamefont {Gomard},\ and\ \citenamefont
  {Lemmer}}]{theobaldPRA2017}%
  \BibitemOpen
  \bibfield  {author} {\bibinfo {author} {\bibfnamefont {D.}~\bibnamefont
  {Theobald}}, \bibinfo {author} {\bibfnamefont {A.}~\bibnamefont {Egel}},
  \bibinfo {author} {\bibfnamefont {G.}~\bibnamefont {Gomard}}, \ and\ \bibinfo
  {author} {\bibfnamefont {U.}~\bibnamefont {Lemmer}},\ }\href {\doibase
  10.1103/PhysRevA.96.033822} {\bibfield  {journal} {\bibinfo  {journal} {Phys.
  Rev. A}\ }\textbf {\bibinfo {volume} {96}},\ \bibinfo {pages} {033822}
  (\bibinfo {year} {2017})}\BibitemShut {NoStop}%
\bibitem [{\citenamefont {Gimbutas}\ and\ \citenamefont
  {Greengard}(2013)}]{gimbutasJCP2013}%
  \BibitemOpen
  \bibfield  {author} {\bibinfo {author} {\bibfnamefont {Z.}~\bibnamefont
  {Gimbutas}}\ and\ \bibinfo {author} {\bibfnamefont {L.}~\bibnamefont
  {Greengard}},\ }\href {\doibase https://doi.org/10.1016/j.jcp.2012.01.041}
  {\bibfield  {journal} {\bibinfo  {journal} {Journal of Computational
  Physics}\ }\textbf {\bibinfo {volume} {232}},\ \bibinfo {pages} {22 }
  (\bibinfo {year} {2013})}\BibitemShut {NoStop}%
\bibitem [{\citenamefont {Lai}\ \emph {et~al.}(2014)\citenamefont {Lai},
  \citenamefont {Kobayashi},\ and\ \citenamefont {Greengard}}]{laiOE2014}%
  \BibitemOpen
  \bibfield  {author} {\bibinfo {author} {\bibfnamefont {J.}~\bibnamefont
  {Lai}}, \bibinfo {author} {\bibfnamefont {M.}~\bibnamefont {Kobayashi}}, \
  and\ \bibinfo {author} {\bibfnamefont {L.}~\bibnamefont {Greengard}},\ }\href
  {\doibase 10.1364/OE.22.020481} {\bibfield  {journal} {\bibinfo  {journal}
  {Opt. Express}\ }\textbf {\bibinfo {volume} {22}},\ \bibinfo {pages} {20481}
  (\bibinfo {year} {2014})}\BibitemShut {NoStop}%
\bibitem [{\citenamefont {{Blankrot}}\ and\ \citenamefont
  {{Heitzinger}}(2019)}]{blankrotIEEEJMMCT2019}%
  \BibitemOpen
  \bibfield  {author} {\bibinfo {author} {\bibfnamefont {B.}~\bibnamefont
  {{Blankrot}}}\ and\ \bibinfo {author} {\bibfnamefont {C.}~\bibnamefont
  {{Heitzinger}}},\ }\href {\doibase 10.1109/JMMCT.2019.2950569} {\bibfield
  {journal} {\bibinfo  {journal} {IEEE Journal on Multiscale and Multiphysics
  Computational Techniques}\ }\textbf {\bibinfo {volume} {4}},\ \bibinfo
  {pages} {234} (\bibinfo {year} {2019})}\BibitemShut {NoStop}%
\bibitem [{\citenamefont {Martin}(2003)}]{martinEABE2003}%
  \BibitemOpen
  \bibfield  {author} {\bibinfo {author} {\bibfnamefont {P.}~\bibnamefont
  {Martin}},\ }\href {\doibase https://doi.org/10.1016/S0955-7997(03)00028-6}
  {\bibfield  {journal} {\bibinfo  {journal} {Engineering Analysis with
  Boundary Elements}\ }\textbf {\bibinfo {volume} {27}},\ \bibinfo {pages} {771
  } (\bibinfo {year} {2003})},\ \bibinfo {note} {special issue on
  Acoustics}\BibitemShut {NoStop}%
\bibitem [{\citenamefont {Gumerov}\ and\ \citenamefont
  {Duraiswami}(2007)}]{gumerovJCP2007}%
  \BibitemOpen
  \bibfield  {author} {\bibinfo {author} {\bibfnamefont {N.~A.}\ \bibnamefont
  {Gumerov}}\ and\ \bibinfo {author} {\bibfnamefont {R.}~\bibnamefont
  {Duraiswami}},\ }\href {\doibase https://doi.org/10.1016/j.jcp.2006.11.025}
  {\bibfield  {journal} {\bibinfo  {journal} {Journal of Computational
  Physics}\ }\textbf {\bibinfo {volume} {225}},\ \bibinfo {pages} {206 }
  (\bibinfo {year} {2007})}\BibitemShut {NoStop}%
\bibitem [{\citenamefont {Greengard}\ and\ \citenamefont
  {Rokhlin}(1987)}]{greengardJCP1987}%
  \BibitemOpen
  \bibfield  {author} {\bibinfo {author} {\bibfnamefont {L.}~\bibnamefont
  {Greengard}}\ and\ \bibinfo {author} {\bibfnamefont {V.}~\bibnamefont
  {Rokhlin}},\ }\href {\doibase https://doi.org/10.1016/0021-9991(87)90140-9}
  {\bibfield  {journal} {\bibinfo  {journal} {Journal of Computational
  Physics}\ }\textbf {\bibinfo {volume} {73}},\ \bibinfo {pages} {325 }
  (\bibinfo {year} {1987})}\BibitemShut {NoStop}%
\bibitem [{\citenamefont {{Engheta}}\ \emph {et~al.}(1992)\citenamefont
  {{Engheta}}, \citenamefont {{Murphy}}, \citenamefont {{Rokhlin}},\ and\
  \citenamefont {{Vassiliou}}}]{enghetaIEEETAP1992}%
  \BibitemOpen
  \bibfield  {author} {\bibinfo {author} {\bibfnamefont {N.}~\bibnamefont
  {{Engheta}}}, \bibinfo {author} {\bibfnamefont {W.~D.}\ \bibnamefont
  {{Murphy}}}, \bibinfo {author} {\bibfnamefont {V.}~\bibnamefont {{Rokhlin}}},
  \ and\ \bibinfo {author} {\bibfnamefont {M.~S.}\ \bibnamefont
  {{Vassiliou}}},\ }\href {\doibase 10.1109/8.144597} {\bibfield  {journal}
  {\bibinfo  {journal} {IEEE Transactions on Antennas and Propagation}\
  }\textbf {\bibinfo {volume} {40}},\ \bibinfo {pages} {634} (\bibinfo {year}
  {1992})}\BibitemShut {NoStop}%
\bibitem [{\citenamefont {Cheng}\ \emph {et~al.}(2006)\citenamefont {Cheng},
  \citenamefont {Crutchfield}, \citenamefont {Gimbutas}, \citenamefont
  {Greengard}, \citenamefont {Ethridge}, \citenamefont {Huang}, \citenamefont
  {Rokhlin}, \citenamefont {Yarvin},\ and\ \citenamefont
  {Zhao}}]{chengJCP2006}%
  \BibitemOpen
  \bibfield  {author} {\bibinfo {author} {\bibfnamefont {H.}~\bibnamefont
  {Cheng}}, \bibinfo {author} {\bibfnamefont {W.~Y.}\ \bibnamefont
  {Crutchfield}}, \bibinfo {author} {\bibfnamefont {Z.}~\bibnamefont
  {Gimbutas}}, \bibinfo {author} {\bibfnamefont {L.~F.}\ \bibnamefont
  {Greengard}}, \bibinfo {author} {\bibfnamefont {J.~F.}\ \bibnamefont
  {Ethridge}}, \bibinfo {author} {\bibfnamefont {J.}~\bibnamefont {Huang}},
  \bibinfo {author} {\bibfnamefont {V.}~\bibnamefont {Rokhlin}}, \bibinfo
  {author} {\bibfnamefont {N.}~\bibnamefont {Yarvin}}, \ and\ \bibinfo {author}
  {\bibfnamefont {J.}~\bibnamefont {Zhao}},\ }\href {\doibase
  https://doi.org/10.1016/j.jcp.2005.12.001} {\bibfield  {journal} {\bibinfo
  {journal} {Journal of Computational Physics}\ }\textbf {\bibinfo {volume}
  {216}},\ \bibinfo {pages} {300 } (\bibinfo {year} {2006})}\BibitemShut
  {NoStop}%
\bibitem [{\citenamefont {Markkanen}\ and\ \citenamefont
  {Yuffa}(2017)}]{markkanenJQSRT2017}%
  \BibitemOpen
  \bibfield  {author} {\bibinfo {author} {\bibfnamefont {J.}~\bibnamefont
  {Markkanen}}\ and\ \bibinfo {author} {\bibfnamefont {A.~J.}\ \bibnamefont
  {Yuffa}},\ }\href {\doibase https://doi.org/10.1016/j.jqsrt.2016.11.004}
  {\bibfield  {journal} {\bibinfo  {journal} {Journal of Quantitative
  Spectroscopy and Radiative Transfer}\ }\textbf {\bibinfo {volume} {189}},\
  \bibinfo {pages} {181 } (\bibinfo {year} {2017})}\BibitemShut {NoStop}%
\bibitem [{\citenamefont {Maier}\ \emph {et~al.}(2002)\citenamefont {Maier},
  \citenamefont {Brongersma}, \citenamefont {Kik},\ and\ \citenamefont
  {Atwater}}]{maierPRB2002}%
  \BibitemOpen
  \bibfield  {author} {\bibinfo {author} {\bibfnamefont {S.~A.}\ \bibnamefont
  {Maier}}, \bibinfo {author} {\bibfnamefont {M.~L.}\ \bibnamefont
  {Brongersma}}, \bibinfo {author} {\bibfnamefont {P.~G.}\ \bibnamefont {Kik}},
  \ and\ \bibinfo {author} {\bibfnamefont {H.~A.}\ \bibnamefont {Atwater}},\
  }\href {\doibase 10.1103/PhysRevB.65.193408} {\bibfield  {journal} {\bibinfo
  {journal} {Phys. Rev. B}\ }\textbf {\bibinfo {volume} {65}},\ \bibinfo
  {pages} {193408} (\bibinfo {year} {2002})}\BibitemShut {NoStop}%
\bibitem [{\citenamefont {Gunnarsson}\ \emph {et~al.}(2005)\citenamefont
  {Gunnarsson}, \citenamefont {Rindzevicius}, \citenamefont {Prikulis},
  \citenamefont {Kasemo}, \citenamefont {Käll}, \citenamefont {Zou},\ and\
  \citenamefont {Schatz}}]{gunnarssonJPCB2005}%
  \BibitemOpen
  \bibfield  {author} {\bibinfo {author} {\bibfnamefont {L.}~\bibnamefont
  {Gunnarsson}}, \bibinfo {author} {\bibfnamefont {T.}~\bibnamefont
  {Rindzevicius}}, \bibinfo {author} {\bibfnamefont {J.}~\bibnamefont
  {Prikulis}}, \bibinfo {author} {\bibfnamefont {B.}~\bibnamefont {Kasemo}},
  \bibinfo {author} {\bibfnamefont {M.}~\bibnamefont {Käll}}, \bibinfo
  {author} {\bibfnamefont {S.}~\bibnamefont {Zou}}, \ and\ \bibinfo {author}
  {\bibfnamefont {G.~C.}\ \bibnamefont {Schatz}},\ }\href {\doibase
  10.1021/jp049084e} {\bibfield  {journal} {\bibinfo  {journal} {The Journal of
  Physical Chemistry B}\ }\textbf {\bibinfo {volume} {109}},\ \bibinfo {pages}
  {1079} (\bibinfo {year} {2005})},\ \bibinfo {note} {pMID:
  16851063}\BibitemShut {NoStop}%
\bibitem [{\citenamefont {Augui\'e}\ and\ \citenamefont
  {Barnes}(2008)}]{barnesPRL2008}%
  \BibitemOpen
  \bibfield  {author} {\bibinfo {author} {\bibfnamefont {B.}~\bibnamefont
  {Augui\'e}}\ and\ \bibinfo {author} {\bibfnamefont {W.~L.}\ \bibnamefont
  {Barnes}},\ }\href {\doibase 10.1103/PhysRevLett.101.143902} {\bibfield
  {journal} {\bibinfo  {journal} {Phys. Rev. Lett.}\ }\textbf {\bibinfo
  {volume} {101}},\ \bibinfo {pages} {143902} (\bibinfo {year}
  {2008})}\BibitemShut {NoStop}%
\bibitem [{\citenamefont {Augui\'{e}}\ and\ \citenamefont
  {Barnes}(2009)}]{barnesOL2009}%
  \BibitemOpen
  \bibfield  {author} {\bibinfo {author} {\bibfnamefont {B.}~\bibnamefont
  {Augui\'{e}}}\ and\ \bibinfo {author} {\bibfnamefont {W.~L.}\ \bibnamefont
  {Barnes}},\ }\href {\doibase 10.1364/OL.34.000401} {\bibfield  {journal}
  {\bibinfo  {journal} {Opt. Lett.}\ }\textbf {\bibinfo {volume} {34}},\
  \bibinfo {pages} {401} (\bibinfo {year} {2009})}\BibitemShut {NoStop}%
\bibitem [{\citenamefont {Taylor}\ \emph {et~al.}(2012)\citenamefont {Taylor},
  \citenamefont {Esteban}, \citenamefont {Mahajan}, \citenamefont {Coulston},
  \citenamefont {Scherman}, \citenamefont {Aizpurua},\ and\ \citenamefont
  {Baumberg}}]{taylorJPCC2012}%
  \BibitemOpen
  \bibfield  {author} {\bibinfo {author} {\bibfnamefont {R.~W.}\ \bibnamefont
  {Taylor}}, \bibinfo {author} {\bibfnamefont {R.}~\bibnamefont {Esteban}},
  \bibinfo {author} {\bibfnamefont {S.}~\bibnamefont {Mahajan}}, \bibinfo
  {author} {\bibfnamefont {R.}~\bibnamefont {Coulston}}, \bibinfo {author}
  {\bibfnamefont {O.~A.}\ \bibnamefont {Scherman}}, \bibinfo {author}
  {\bibfnamefont {J.}~\bibnamefont {Aizpurua}}, \ and\ \bibinfo {author}
  {\bibfnamefont {J.~J.}\ \bibnamefont {Baumberg}},\ }\href {\doibase
  10.1021/jp308986c} {\bibfield  {journal} {\bibinfo  {journal} {The Journal of
  Physical Chemistry C}\ }\textbf {\bibinfo {volume} {116}},\ \bibinfo {pages}
  {25044} (\bibinfo {year} {2012})}\BibitemShut {NoStop}%
\bibitem [{\citenamefont {Kravets}\ \emph {et~al.}(2014)\citenamefont
  {Kravets}, \citenamefont {Schedin}, \citenamefont {Pisano}, \citenamefont
  {Thackray}, \citenamefont {Thomas},\ and\ \citenamefont
  {Grigorenko}}]{kravetsPRB2014}%
  \BibitemOpen
  \bibfield  {author} {\bibinfo {author} {\bibfnamefont {V.~G.}\ \bibnamefont
  {Kravets}}, \bibinfo {author} {\bibfnamefont {F.}~\bibnamefont {Schedin}},
  \bibinfo {author} {\bibfnamefont {G.}~\bibnamefont {Pisano}}, \bibinfo
  {author} {\bibfnamefont {B.}~\bibnamefont {Thackray}}, \bibinfo {author}
  {\bibfnamefont {P.~A.}\ \bibnamefont {Thomas}}, \ and\ \bibinfo {author}
  {\bibfnamefont {A.~N.}\ \bibnamefont {Grigorenko}},\ }\href {\doibase
  10.1103/PhysRevB.90.125445} {\bibfield  {journal} {\bibinfo  {journal} {Phys.
  Rev. B}\ }\textbf {\bibinfo {volume} {90}},\ \bibinfo {pages} {125445}
  (\bibinfo {year} {2014})}\BibitemShut {NoStop}%
\bibitem [{\citenamefont {Kupriyanov}\ \emph {et~al.}(2017)\citenamefont
  {Kupriyanov}, \citenamefont {Sokolov},\ and\ \citenamefont
  {Havey}}]{haveyPhysrep2017}%
  \BibitemOpen
  \bibfield  {author} {\bibinfo {author} {\bibfnamefont {D.}~\bibnamefont
  {Kupriyanov}}, \bibinfo {author} {\bibfnamefont {I.}~\bibnamefont {Sokolov}},
  \ and\ \bibinfo {author} {\bibfnamefont {M.}~\bibnamefont {Havey}},\ }\href
  {\doibase https://doi.org/10.1016/j.physrep.2016.12.004} {\bibfield
  {journal} {\bibinfo  {journal} {Physics Reports}\ }\textbf {\bibinfo {volume}
  {671}},\ \bibinfo {pages} {1 } (\bibinfo {year} {2017})},\ \bibinfo {note}
  {mesoscopic coherence in light scattering from cold, optically dense and
  disordered atomic systems}\BibitemShut {NoStop}%
\bibitem [{\citenamefont {Guerin}\ \emph {et~al.}(2016)\citenamefont {Guerin},
  \citenamefont {Ara\'ujo},\ and\ \citenamefont {Kaiser}}]{guerinPRL2016}%
  \BibitemOpen
  \bibfield  {author} {\bibinfo {author} {\bibfnamefont {W.}~\bibnamefont
  {Guerin}}, \bibinfo {author} {\bibfnamefont {M.~O.}\ \bibnamefont
  {Ara\'ujo}}, \ and\ \bibinfo {author} {\bibfnamefont {R.}~\bibnamefont
  {Kaiser}},\ }\href {\doibase 10.1103/PhysRevLett.116.083601} {\bibfield
  {journal} {\bibinfo  {journal} {Phys. Rev. Lett.}\ }\textbf {\bibinfo
  {volume} {116}},\ \bibinfo {pages} {083601} (\bibinfo {year}
  {2016})}\BibitemShut {NoStop}%
\bibitem [{\citenamefont {Collett}\ \emph {et~al.}(1977)\citenamefont
  {Collett}, \citenamefont {Foley},\ and\ \citenamefont
  {Wolf}}]{collettJOSAA1977}%
  \BibitemOpen
  \bibfield  {author} {\bibinfo {author} {\bibfnamefont {E.}~\bibnamefont
  {Collett}}, \bibinfo {author} {\bibfnamefont {J.~T.}\ \bibnamefont {Foley}},
  \ and\ \bibinfo {author} {\bibfnamefont {E.}~\bibnamefont {Wolf}},\ }\href
  {\doibase 10.1364/JOSA.67.000465} {\bibfield  {journal} {\bibinfo  {journal}
  {J. Opt. Soc. Am.}\ }\textbf {\bibinfo {volume} {67}},\ \bibinfo {pages}
  {465} (\bibinfo {year} {1977})}\BibitemShut {NoStop}%
\bibitem [{\citenamefont {Fante}(1981)}]{fanteJOSAA1981}%
  \BibitemOpen
  \bibfield  {author} {\bibinfo {author} {\bibfnamefont {R.~L.}\ \bibnamefont
  {Fante}},\ }\href {\doibase 10.1364/JOSA.71.000460} {\bibfield  {journal}
  {\bibinfo  {journal} {J. Opt. Soc. Am.}\ }\textbf {\bibinfo {volume} {71}},\
  \bibinfo {pages} {460} (\bibinfo {year} {1981})}\BibitemShut {NoStop}%
\bibitem [{\citenamefont {Caz\'{e}}\ and\ \citenamefont
  {Schotland}(2015)}]{cazeJOSAA2015}%
  \BibitemOpen
  \bibfield  {author} {\bibinfo {author} {\bibfnamefont {A.}~\bibnamefont
  {Caz\'{e}}}\ and\ \bibinfo {author} {\bibfnamefont {J.~C.}\ \bibnamefont
  {Schotland}},\ }\href {\doibase 10.1364/JOSAA.32.001475} {\bibfield
  {journal} {\bibinfo  {journal} {J. Opt. Soc. Am. A}\ }\textbf {\bibinfo
  {volume} {32}},\ \bibinfo {pages} {1475} (\bibinfo {year}
  {2015})}\BibitemShut {NoStop}%
\bibitem [{\citenamefont {Vynck}\ \emph {et~al.}(2016)\citenamefont {Vynck},
  \citenamefont {Pierrat},\ and\ \citenamefont {Carminati}}]{vynckPRA2016}%
  \BibitemOpen
  \bibfield  {author} {\bibinfo {author} {\bibfnamefont {K.}~\bibnamefont
  {Vynck}}, \bibinfo {author} {\bibfnamefont {R.}~\bibnamefont {Pierrat}}, \
  and\ \bibinfo {author} {\bibfnamefont {R.}~\bibnamefont {Carminati}},\ }\href
  {\doibase 10.1103/PhysRevA.94.033851} {\bibfield  {journal} {\bibinfo
  {journal} {Phys. Rev. A}\ }\textbf {\bibinfo {volume} {94}},\ \bibinfo
  {pages} {033851} (\bibinfo {year} {2016})}\BibitemShut {NoStop}%
\bibitem [{\citenamefont {van~der Mark}\ \emph {et~al.}(1988)\citenamefont
  {van~der Mark}, \citenamefont {van Albada},\ and\ \citenamefont
  {Lagendijk}}]{vandermarkPRB1988}%
  \BibitemOpen
  \bibfield  {author} {\bibinfo {author} {\bibfnamefont {M.~B.}\ \bibnamefont
  {van~der Mark}}, \bibinfo {author} {\bibfnamefont {M.~P.}\ \bibnamefont {van
  Albada}}, \ and\ \bibinfo {author} {\bibfnamefont {A.}~\bibnamefont
  {Lagendijk}},\ }\href {\doibase 10.1103/PhysRevB.37.3575} {\bibfield
  {journal} {\bibinfo  {journal} {Phys. Rev. B}\ }\textbf {\bibinfo {volume}
  {37}},\ \bibinfo {pages} {3575} (\bibinfo {year} {1988})}\BibitemShut
  {NoStop}%
\bibitem [{\citenamefont {Akkermans}\ \emph {et~al.}(1986)\citenamefont
  {Akkermans}, \citenamefont {Wolf},\ and\ \citenamefont
  {Maynard}}]{akkermansPRL1986}%
  \BibitemOpen
  \bibfield  {author} {\bibinfo {author} {\bibfnamefont {E.}~\bibnamefont
  {Akkermans}}, \bibinfo {author} {\bibfnamefont {P.~E.}\ \bibnamefont {Wolf}},
  \ and\ \bibinfo {author} {\bibfnamefont {R.}~\bibnamefont {Maynard}},\ }\href
  {\doibase 10.1103/PhysRevLett.56.1471} {\bibfield  {journal} {\bibinfo
  {journal} {Phys. Rev. Lett.}\ }\textbf {\bibinfo {volume} {56}},\ \bibinfo
  {pages} {1471} (\bibinfo {year} {1986})}\BibitemShut {NoStop}%
\bibitem [{\citenamefont {Akkermans}\ \emph {et~al.}(1988)\citenamefont
  {Akkermans}, \citenamefont {Wolf}, \citenamefont {Maynard},\ and\
  \citenamefont {Maret}}]{akkermans1988theoretical}%
  \BibitemOpen
  \bibfield  {author} {\bibinfo {author} {\bibfnamefont {E.}~\bibnamefont
  {Akkermans}}, \bibinfo {author} {\bibfnamefont {P.}~\bibnamefont {Wolf}},
  \bibinfo {author} {\bibfnamefont {R.}~\bibnamefont {Maynard}}, \ and\
  \bibinfo {author} {\bibfnamefont {G.}~\bibnamefont {Maret}},\ }\href@noop {}
  {\bibfield  {journal} {\bibinfo  {journal} {J. de Phys. (France)}\ }\textbf
  {\bibinfo {volume} {49}},\ \bibinfo {pages} {77} (\bibinfo {year}
  {1988})}\BibitemShut {NoStop}%
\bibitem [{\citenamefont {Mishchenko}\ and\ \citenamefont
  {Dlugach}(1992)}]{mishchenkoMNRAS1992}%
  \BibitemOpen
  \bibfield  {author} {\bibinfo {author} {\bibfnamefont {M.~I.}\ \bibnamefont
  {Mishchenko}}\ and\ \bibinfo {author} {\bibfnamefont {J.~M.}\ \bibnamefont
  {Dlugach}},\ }\href {\doibase 10.1093/mnras/254.1.15P} {\bibfield  {journal}
  {\bibinfo  {journal} {Monthly Notices of the Royal Astronomical Society}\
  }\textbf {\bibinfo {volume} {254}},\ \bibinfo {pages} {15P} (\bibinfo {year}
  {1992})},\ \Eprint
  {http://arxiv.org/abs/http://oup.prod.sis.lan/mnras/article-pdf/254/1/15P/18523367/mnras254-015P.pdf}
  {http://oup.prod.sis.lan/mnras/article-pdf/254/1/15P/18523367/mnras254-015P.pdf}
  \BibitemShut {NoStop}%
\bibitem [{\citenamefont {Mandt}\ \emph {et~al.}(1992)\citenamefont {Mandt},
  \citenamefont {Kuga}, \citenamefont {Tsang},\ and\ \citenamefont
  {Ishimaru}}]{mandtWRM1992}%
  \BibitemOpen
  \bibfield  {author} {\bibinfo {author} {\bibfnamefont {C.~E.}\ \bibnamefont
  {Mandt}}, \bibinfo {author} {\bibfnamefont {Y.}~\bibnamefont {Kuga}},
  \bibinfo {author} {\bibfnamefont {L.}~\bibnamefont {Tsang}}, \ and\ \bibinfo
  {author} {\bibfnamefont {A.}~\bibnamefont {Ishimaru}},\ }\href {\doibase
  10.1088/0959-7174/2/3/004} {\bibfield  {journal} {\bibinfo  {journal} {Waves
  in Random Media}\ }\textbf {\bibinfo {volume} {2}},\ \bibinfo {pages} {225}
  (\bibinfo {year} {1992})},\ \Eprint
  {http://arxiv.org/abs/https://doi.org/10.1088/0959-7174/2/3/004}
  {https://doi.org/10.1088/0959-7174/2/3/004} \BibitemShut {NoStop}%
\bibitem [{\citenamefont {Tsang}\ and\ \citenamefont
  {Chang}(2000)}]{tsangRS2000}%
  \BibitemOpen
  \bibfield  {author} {\bibinfo {author} {\bibfnamefont {L.}~\bibnamefont
  {Tsang}}\ and\ \bibinfo {author} {\bibfnamefont {T.~C.}\ \bibnamefont
  {Chang}},\ }\href {\doibase 10.1029/1999RS002270} {\bibfield  {journal}
  {\bibinfo  {journal} {Radio Science}\ }\textbf {\bibinfo {volume} {35}},\
  \bibinfo {pages} {731} (\bibinfo {year} {2000})}\BibitemShut {NoStop}%
\bibitem [{\citenamefont {Liang}\ \emph {et~al.}(2008)\citenamefont {Liang},
  \citenamefont {Xu}, \citenamefont {Tsang}, \citenamefont {Andreadis},\ and\
  \citenamefont {Josberger}}]{liangIEEETGRS2008}%
  \BibitemOpen
  \bibfield  {author} {\bibinfo {author} {\bibfnamefont {D.}~\bibnamefont
  {Liang}}, \bibinfo {author} {\bibfnamefont {X.}~\bibnamefont {Xu}}, \bibinfo
  {author} {\bibfnamefont {L.}~\bibnamefont {Tsang}}, \bibinfo {author}
  {\bibfnamefont {K.~M.}\ \bibnamefont {Andreadis}}, \ and\ \bibinfo {author}
  {\bibfnamefont {E.~G.}\ \bibnamefont {Josberger}},\ }\href {\doibase
  10.1109/TGRS.2008.922143} {\bibfield  {journal} {\bibinfo  {journal} {IEEE
  Transactions on Geoscience and Remote Sensing}\ }\textbf {\bibinfo {volume}
  {46}},\ \bibinfo {pages} {3663} (\bibinfo {year} {2008})}\BibitemShut
  {NoStop}%
\bibitem [{\citenamefont {Picard}\ \emph {et~al.}(2013)\citenamefont {Picard},
  \citenamefont {Brucker}, \citenamefont {Roy}, \citenamefont {Dupont},
  \citenamefont {Fily}, \citenamefont {Royer},\ and\ \citenamefont
  {Harlow}}]{picardGMD2013}%
  \BibitemOpen
  \bibfield  {author} {\bibinfo {author} {\bibfnamefont {G.}~\bibnamefont
  {Picard}}, \bibinfo {author} {\bibfnamefont {L.}~\bibnamefont {Brucker}},
  \bibinfo {author} {\bibfnamefont {A.}~\bibnamefont {Roy}}, \bibinfo {author}
  {\bibfnamefont {F.}~\bibnamefont {Dupont}}, \bibinfo {author} {\bibfnamefont
  {M.}~\bibnamefont {Fily}}, \bibinfo {author} {\bibfnamefont {A.}~\bibnamefont
  {Royer}}, \ and\ \bibinfo {author} {\bibfnamefont {C.}~\bibnamefont
  {Harlow}},\ }\href {\doibase 10.5194/gmd-6-1061-2013} {\bibfield  {journal}
  {\bibinfo  {journal} {Geoscientific Model Development}\ }\textbf {\bibinfo
  {volume} {6}},\ \bibinfo {pages} {1061} (\bibinfo {year} {2013})}\BibitemShut
  {NoStop}%
\bibitem [{\citenamefont {van Albada}\ \emph {et~al.}(1991)\citenamefont {van
  Albada}, \citenamefont {van Tiggelen}, \citenamefont {Lagendijk},\ and\
  \citenamefont {Tip}}]{vanalbabaPRL1991}%
  \BibitemOpen
  \bibfield  {author} {\bibinfo {author} {\bibfnamefont {M.~P.}\ \bibnamefont
  {van Albada}}, \bibinfo {author} {\bibfnamefont {B.~A.}\ \bibnamefont {van
  Tiggelen}}, \bibinfo {author} {\bibfnamefont {A.}~\bibnamefont {Lagendijk}},
  \ and\ \bibinfo {author} {\bibfnamefont {A.}~\bibnamefont {Tip}},\ }\href
  {\doibase 10.1103/PhysRevLett.66.3132} {\bibfield  {journal} {\bibinfo
  {journal} {Phys. Rev. Lett.}\ }\textbf {\bibinfo {volume} {66}},\ \bibinfo
  {pages} {3132} (\bibinfo {year} {1991})}\BibitemShut {NoStop}%
\bibitem [{\citenamefont {St\"orzer}\ \emph
  {et~al.}(2006{\natexlab{a}})\citenamefont {St\"orzer}, \citenamefont {Gross},
  \citenamefont {Aegerter},\ and\ \citenamefont {Maret}}]{storzerPRL2006}%
  \BibitemOpen
  \bibfield  {author} {\bibinfo {author} {\bibfnamefont {M.}~\bibnamefont
  {St\"orzer}}, \bibinfo {author} {\bibfnamefont {P.}~\bibnamefont {Gross}},
  \bibinfo {author} {\bibfnamefont {C.~M.}\ \bibnamefont {Aegerter}}, \ and\
  \bibinfo {author} {\bibfnamefont {G.}~\bibnamefont {Maret}},\ }\href
  {\doibase 10.1103/PhysRevLett.96.063904} {\bibfield  {journal} {\bibinfo
  {journal} {Phys. Rev. Lett.}\ }\textbf {\bibinfo {volume} {96}},\ \bibinfo
  {pages} {063904} (\bibinfo {year} {2006}{\natexlab{a}})}\BibitemShut
  {NoStop}%
\bibitem [{\citenamefont {Aubry}\ \emph {et~al.}(2017)\citenamefont {Aubry},
  \citenamefont {Schertel}, \citenamefont {Chen}, \citenamefont {Weyer},
  \citenamefont {Aegerter}, \citenamefont {Polarz}, \citenamefont {C\"olfen},\
  and\ \citenamefont {Maret}}]{aubryPRA2017}%
  \BibitemOpen
  \bibfield  {author} {\bibinfo {author} {\bibfnamefont {G.~J.}\ \bibnamefont
  {Aubry}}, \bibinfo {author} {\bibfnamefont {L.}~\bibnamefont {Schertel}},
  \bibinfo {author} {\bibfnamefont {M.}~\bibnamefont {Chen}}, \bibinfo {author}
  {\bibfnamefont {H.}~\bibnamefont {Weyer}}, \bibinfo {author} {\bibfnamefont
  {C.~M.}\ \bibnamefont {Aegerter}}, \bibinfo {author} {\bibfnamefont
  {S.}~\bibnamefont {Polarz}}, \bibinfo {author} {\bibfnamefont
  {H.}~\bibnamefont {C\"olfen}}, \ and\ \bibinfo {author} {\bibfnamefont
  {G.}~\bibnamefont {Maret}},\ }\href {\doibase 10.1103/PhysRevA.96.043871}
  {\bibfield  {journal} {\bibinfo  {journal} {Phys. Rev. A}\ }\textbf {\bibinfo
  {volume} {96}},\ \bibinfo {pages} {043871} (\bibinfo {year}
  {2017})}\BibitemShut {NoStop}%
\bibitem [{\citenamefont {van Tiggelen}\ \emph {et~al.}(2000)\citenamefont {van
  Tiggelen}, \citenamefont {Lagendijk},\ and\ \citenamefont
  {Wiersma}}]{tiggelenPRL2000}%
  \BibitemOpen
  \bibfield  {author} {\bibinfo {author} {\bibfnamefont {B.~A.}\ \bibnamefont
  {van Tiggelen}}, \bibinfo {author} {\bibfnamefont {A.}~\bibnamefont
  {Lagendijk}}, \ and\ \bibinfo {author} {\bibfnamefont {D.~S.}\ \bibnamefont
  {Wiersma}},\ }\href {\doibase 10.1103/PhysRevLett.84.4333} {\bibfield
  {journal} {\bibinfo  {journal} {Phys. Rev. Lett.}\ }\textbf {\bibinfo
  {volume} {84}},\ \bibinfo {pages} {4333} (\bibinfo {year}
  {2000})}\BibitemShut {NoStop}%
\bibitem [{\citenamefont {Tian}\ \emph {et~al.}(2010)\citenamefont {Tian},
  \citenamefont {Cheung},\ and\ \citenamefont {Zhang}}]{tianPRL2010}%
  \BibitemOpen
  \bibfield  {author} {\bibinfo {author} {\bibfnamefont {C.-S.}\ \bibnamefont
  {Tian}}, \bibinfo {author} {\bibfnamefont {S.-K.}\ \bibnamefont {Cheung}}, \
  and\ \bibinfo {author} {\bibfnamefont {Z.-Q.}\ \bibnamefont {Zhang}},\ }\href
  {\doibase 10.1103/PhysRevLett.105.263905} {\bibfield  {journal} {\bibinfo
  {journal} {Phys. Rev. Lett.}\ }\textbf {\bibinfo {volume} {105}},\ \bibinfo
  {pages} {263905} (\bibinfo {year} {2010})}\BibitemShut {NoStop}%
\bibitem [{\citenamefont {Hu}\ \emph {et~al.}(2008)\citenamefont {Hu},
  \citenamefont {Strybulevych}, \citenamefont {Page}, \citenamefont
  {Skipetrov},\ and\ \citenamefont {van Tiggelen}}]{huNaturephys2008}%
  \BibitemOpen
  \bibfield  {author} {\bibinfo {author} {\bibfnamefont {H.}~\bibnamefont
  {Hu}}, \bibinfo {author} {\bibfnamefont {A.}~\bibnamefont {Strybulevych}},
  \bibinfo {author} {\bibfnamefont {J.~H.}\ \bibnamefont {Page}}, \bibinfo
  {author} {\bibfnamefont {S.~E.}\ \bibnamefont {Skipetrov}}, \ and\ \bibinfo
  {author} {\bibfnamefont {B.~A.}\ \bibnamefont {van Tiggelen}},\ }\href
  {\doibase 10.1038/nphys1101} {\bibfield  {journal} {\bibinfo  {journal}
  {Nature Physics}\ }\textbf {\bibinfo {volume} {4}},\ \bibinfo {pages} {945}
  (\bibinfo {year} {2008})}\BibitemShut {NoStop}%
\bibitem [{\citenamefont {Zhang}\ \emph {et~al.}(2009)\citenamefont {Zhang},
  \citenamefont {Chabanov}, \citenamefont {Cheung}, \citenamefont {Wong},\ and\
  \citenamefont {Genack}}]{zhangPRB2009}%
  \BibitemOpen
  \bibfield  {author} {\bibinfo {author} {\bibfnamefont {Z.~Q.}\ \bibnamefont
  {Zhang}}, \bibinfo {author} {\bibfnamefont {A.~A.}\ \bibnamefont {Chabanov}},
  \bibinfo {author} {\bibfnamefont {S.~K.}\ \bibnamefont {Cheung}}, \bibinfo
  {author} {\bibfnamefont {C.~H.}\ \bibnamefont {Wong}}, \ and\ \bibinfo
  {author} {\bibfnamefont {A.~Z.}\ \bibnamefont {Genack}},\ }\href {\doibase
  10.1103/PhysRevB.79.144203} {\bibfield  {journal} {\bibinfo  {journal} {Phys.
  Rev. B}\ }\textbf {\bibinfo {volume} {79}},\ \bibinfo {pages} {144203}
  (\bibinfo {year} {2009})}\BibitemShut {NoStop}%
\bibitem [{\citenamefont {{Haberko}}\ \emph {et~al.}(2018)\citenamefont
  {{Haberko}}, \citenamefont {{Froufe-P{\'e}rez}},\ and\ \citenamefont
  {{Scheffold}}}]{haberko2018transition}%
  \BibitemOpen
  \bibfield  {author} {\bibinfo {author} {\bibfnamefont {J.}~\bibnamefont
  {{Haberko}}}, \bibinfo {author} {\bibfnamefont {L.~S.}\ \bibnamefont
  {{Froufe-P{\'e}rez}}}, \ and\ \bibinfo {author} {\bibfnamefont
  {F.}~\bibnamefont {{Scheffold}}},\ }\href@noop {} {\bibfield  {journal}
  {\bibinfo  {journal} {arXiv e-prints}\ ,\ \bibinfo {eid} {arXiv:1812.02095}}
  (\bibinfo {year} {2018})},\ \Eprint {http://arxiv.org/abs/1812.02095}
  {arXiv:1812.02095 [physics.optics]} \BibitemShut {NoStop}%
\bibitem [{\citenamefont {Thomas}\ and\ \citenamefont
  {Stamnes}(2002)}]{thomas2002radiative}%
  \BibitemOpen
  \bibfield  {author} {\bibinfo {author} {\bibfnamefont {G.~E.}\ \bibnamefont
  {Thomas}}\ and\ \bibinfo {author} {\bibfnamefont {K.}~\bibnamefont
  {Stamnes}},\ }\href@noop {} {\emph {\bibinfo {title} {Radiative transfer in
  the atmosphere and ocean}}}\ (\bibinfo  {publisher} {Cambridge University
  Press},\ \bibinfo {year} {2002})\BibitemShut {NoStop}%
\bibitem [{\citenamefont {Clough}\ \emph {et~al.}(2005)\citenamefont {Clough},
  \citenamefont {Shephard}, \citenamefont {Mlawer}, \citenamefont {Delamere},
  \citenamefont {Iacono}, \citenamefont {Cady-Pereira}, \citenamefont
  {Boukabara},\ and\ \citenamefont {Brown}}]{cloughJQSRT2005}%
  \BibitemOpen
  \bibfield  {author} {\bibinfo {author} {\bibfnamefont {S.}~\bibnamefont
  {Clough}}, \bibinfo {author} {\bibfnamefont {M.}~\bibnamefont {Shephard}},
  \bibinfo {author} {\bibfnamefont {E.}~\bibnamefont {Mlawer}}, \bibinfo
  {author} {\bibfnamefont {J.}~\bibnamefont {Delamere}}, \bibinfo {author}
  {\bibfnamefont {M.}~\bibnamefont {Iacono}}, \bibinfo {author} {\bibfnamefont
  {K.}~\bibnamefont {Cady-Pereira}}, \bibinfo {author} {\bibfnamefont
  {S.}~\bibnamefont {Boukabara}}, \ and\ \bibinfo {author} {\bibfnamefont
  {P.}~\bibnamefont {Brown}},\ }\href {\doibase
  https://doi.org/10.1016/j.jqsrt.2004.05.058} {\bibfield  {journal} {\bibinfo
  {journal} {Journal of Quantitative Spectroscopy and Radiative Transfer}\
  }\textbf {\bibinfo {volume} {91}},\ \bibinfo {pages} {233 } (\bibinfo {year}
  {2005})}\BibitemShut {NoStop}%
\bibitem [{\citenamefont {Peraiah}(2002)}]{peraiah2002introduction}%
  \BibitemOpen
  \bibfield  {author} {\bibinfo {author} {\bibfnamefont {A.}~\bibnamefont
  {Peraiah}},\ }\href@noop {} {\emph {\bibinfo {title} {An Introduction to
  Radiative Transfer: Methods and applications in astrophysics}}}\ (\bibinfo
  {publisher} {Cambridge University Press},\ \bibinfo {year}
  {2002})\BibitemShut {NoStop}%
\bibitem [{\citenamefont {{Myneni}}\ \emph {et~al.}(1997)\citenamefont
  {{Myneni}}, \citenamefont {{Ramakrishna}}, \citenamefont {{Nemani}},\ and\
  \citenamefont {{Running}}}]{myneniIEEEGRS1997}%
  \BibitemOpen
  \bibfield  {author} {\bibinfo {author} {\bibfnamefont {R.~B.}\ \bibnamefont
  {{Myneni}}}, \bibinfo {author} {\bibfnamefont {R.}~\bibnamefont
  {{Ramakrishna}}}, \bibinfo {author} {\bibfnamefont {R.}~\bibnamefont
  {{Nemani}}}, \ and\ \bibinfo {author} {\bibfnamefont {S.~W.}\ \bibnamefont
  {{Running}}},\ }\href {\doibase 10.1109/36.649788} {\bibfield  {journal}
  {\bibinfo  {journal} {IEEE Transactions on Geoscience and Remote Sensing}\
  }\textbf {\bibinfo {volume} {35}},\ \bibinfo {pages} {1380} (\bibinfo {year}
  {1997})}\BibitemShut {NoStop}%
\bibitem [{\citenamefont {Jarabo}\ \emph {et~al.}(2018)\citenamefont {Jarabo},
  \citenamefont {Aliaga},\ and\ \citenamefont {Gutierrez}}]{jaraboACMTG2018}%
  \BibitemOpen
  \bibfield  {author} {\bibinfo {author} {\bibfnamefont {A.}~\bibnamefont
  {Jarabo}}, \bibinfo {author} {\bibfnamefont {C.}~\bibnamefont {Aliaga}}, \
  and\ \bibinfo {author} {\bibfnamefont {D.}~\bibnamefont {Gutierrez}},\ }\href
  {\doibase 10.1145/3197517.3201282} {\bibfield  {journal} {\bibinfo  {journal}
  {ACM Trans. Graph.}\ }\textbf {\bibinfo {volume} {37}} (\bibinfo {year}
  {2018}),\ 10.1145/3197517.3201282}\BibitemShut {NoStop}%
\bibitem [{\citenamefont {Zhang}\ \emph {et~al.}(2019)\citenamefont {Zhang},
  \citenamefont {Wu}, \citenamefont {Zheng}, \citenamefont {Gkioulekas},
  \citenamefont {Ramamoorthi},\ and\ \citenamefont {Zhao}}]{zhangACMTG2019}%
  \BibitemOpen
  \bibfield  {author} {\bibinfo {author} {\bibfnamefont {C.}~\bibnamefont
  {Zhang}}, \bibinfo {author} {\bibfnamefont {L.}~\bibnamefont {Wu}}, \bibinfo
  {author} {\bibfnamefont {C.}~\bibnamefont {Zheng}}, \bibinfo {author}
  {\bibfnamefont {I.}~\bibnamefont {Gkioulekas}}, \bibinfo {author}
  {\bibfnamefont {R.}~\bibnamefont {Ramamoorthi}}, \ and\ \bibinfo {author}
  {\bibfnamefont {S.}~\bibnamefont {Zhao}},\ }\href {\doibase
  10.1145/3355089.3356522} {\bibfield  {journal} {\bibinfo  {journal} {ACM
  Trans. Graph.}\ }\textbf {\bibinfo {volume} {38}} (\bibinfo {year} {2019}),\
  10.1145/3355089.3356522}\BibitemShut {NoStop}%
\bibitem [{\citenamefont {Kim}\ and\ \citenamefont
  {Moscoso}(2002)}]{kimSIAMJSC2002}%
  \BibitemOpen
  \bibfield  {author} {\bibinfo {author} {\bibfnamefont {A.~D.}\ \bibnamefont
  {Kim}}\ and\ \bibinfo {author} {\bibfnamefont {M.}~\bibnamefont {Moscoso}},\
  }\href {\doibase 10.1137/S1064827500382312} {\bibfield  {journal} {\bibinfo
  {journal} {SIAM Journal on Scientific Computing}\ }\textbf {\bibinfo {volume}
  {23}},\ \bibinfo {pages} {2074} (\bibinfo {year} {2002})},\ \Eprint
  {http://arxiv.org/abs/https://doi.org/10.1137/S1064827500382312}
  {https://doi.org/10.1137/S1064827500382312} \BibitemShut {NoStop}%
\bibitem [{\citenamefont {Edström}(2005)}]{edstromSIAMR2005}%
  \BibitemOpen
  \bibfield  {author} {\bibinfo {author} {\bibfnamefont {P.}~\bibnamefont
  {Edström}},\ }\href {\doibase 10.1137/S0036144503438718} {\bibfield
  {journal} {\bibinfo  {journal} {SIAM Review}\ }\textbf {\bibinfo {volume}
  {47}},\ \bibinfo {pages} {447} (\bibinfo {year} {2005})},\ \Eprint
  {http://arxiv.org/abs/https://doi.org/10.1137/S0036144503438718}
  {https://doi.org/10.1137/S0036144503438718} \BibitemShut {NoStop}%
\bibitem [{\citenamefont {Fiveland}(1988)}]{fivelandJTHT1988}%
  \BibitemOpen
  \bibfield  {author} {\bibinfo {author} {\bibfnamefont {W.~A.}\ \bibnamefont
  {Fiveland}},\ }\href {\doibase 10.2514/3.105} {\bibfield  {journal} {\bibinfo
   {journal} {Journal of Thermophysics and Heat Transfer}\ }\textbf {\bibinfo
  {volume} {2}},\ \bibinfo {pages} {309} (\bibinfo {year} {1988})},\ \Eprint
  {http://arxiv.org/abs/https://doi.org/10.2514/3.105}
  {https://doi.org/10.2514/3.105} \BibitemShut {NoStop}%
\bibitem [{\citenamefont {Hardy}\ \emph {et~al.}(2017)\citenamefont {Hardy},
  \citenamefont {Favennec}, \citenamefont {Rousseau},\ and\ \citenamefont
  {Hecht}}]{lehardyJCP2017}%
  \BibitemOpen
  \bibfield  {author} {\bibinfo {author} {\bibfnamefont {D.~L.}\ \bibnamefont
  {Hardy}}, \bibinfo {author} {\bibfnamefont {Y.}~\bibnamefont {Favennec}},
  \bibinfo {author} {\bibfnamefont {B.}~\bibnamefont {Rousseau}}, \ and\
  \bibinfo {author} {\bibfnamefont {F.}~\bibnamefont {Hecht}},\ }\href
  {\doibase https://doi.org/10.1016/j.jcp.2017.01.019} {\bibfield  {journal}
  {\bibinfo  {journal} {Journal of Computational Physics}\ }\textbf {\bibinfo
  {volume} {334}},\ \bibinfo {pages} {541 } (\bibinfo {year}
  {2017})}\BibitemShut {NoStop}%
\bibitem [{\citenamefont {Modest}\ and\ \citenamefont
  {Yang}(2008)}]{modestJQSRT2008}%
  \BibitemOpen
  \bibfield  {author} {\bibinfo {author} {\bibfnamefont {M.~F.}\ \bibnamefont
  {Modest}}\ and\ \bibinfo {author} {\bibfnamefont {J.}~\bibnamefont {Yang}},\
  }\href {\doibase https://doi.org/10.1016/j.jqsrt.2007.12.018} {\bibfield
  {journal} {\bibinfo  {journal} {Journal of Quantitative Spectroscopy and
  Radiative Transfer}\ }\textbf {\bibinfo {volume} {109}},\ \bibinfo {pages}
  {1641 } (\bibinfo {year} {2008})}\BibitemShut {NoStop}%
\bibitem [{\citenamefont {Hermeline}(2016)}]{hermelineJCP2016}%
  \BibitemOpen
  \bibfield  {author} {\bibinfo {author} {\bibfnamefont {F.}~\bibnamefont
  {Hermeline}},\ }\href {\doibase https://doi.org/10.1016/j.jcp.2016.02.058}
  {\bibfield  {journal} {\bibinfo  {journal} {Journal of Computational
  Physics}\ }\textbf {\bibinfo {volume} {313}},\ \bibinfo {pages} {549 }
  (\bibinfo {year} {2016})}\BibitemShut {NoStop}%
\bibitem [{\citenamefont {Howell}(1998)}]{howellJHT1998}%
  \BibitemOpen
  \bibfield  {author} {\bibinfo {author} {\bibfnamefont {J.}~\bibnamefont
  {Howell}},\ }\href@noop {} {\bibfield  {journal} {\bibinfo  {journal}
  {Journal of heat transfer}\ }\textbf {\bibinfo {volume} {120}},\ \bibinfo
  {pages} {547} (\bibinfo {year} {1998})}\BibitemShut {NoStop}%
\bibitem [{\citenamefont {{Whitney}}(2011)}]{whitneyBASI2011}%
  \BibitemOpen
  \bibfield  {author} {\bibinfo {author} {\bibfnamefont {B.~A.}\ \bibnamefont
  {{Whitney}}},\ }\href@noop {} {\bibfield  {journal} {\bibinfo  {journal}
  {Bulletin of the Astronomical Society of India}\ }\textbf {\bibinfo {volume}
  {39}},\ \bibinfo {pages} {101} (\bibinfo {year} {2011})},\ \Eprint
  {http://arxiv.org/abs/1104.4990} {arXiv:1104.4990 [astro-ph.SR]} \BibitemShut
  {NoStop}%
\bibitem [{\citenamefont {Wang}\ \emph {et~al.}(2017)\citenamefont {Wang},
  \citenamefont {Cui}, \citenamefont {Yang}, \citenamefont {Gao}, \citenamefont
  {Liu},\ and\ \citenamefont {Zhang}}]{wangJQSRT2017}%
  \BibitemOpen
  \bibfield  {author} {\bibinfo {author} {\bibfnamefont {Z.}~\bibnamefont
  {Wang}}, \bibinfo {author} {\bibfnamefont {S.}~\bibnamefont {Cui}}, \bibinfo
  {author} {\bibfnamefont {J.}~\bibnamefont {Yang}}, \bibinfo {author}
  {\bibfnamefont {H.}~\bibnamefont {Gao}}, \bibinfo {author} {\bibfnamefont
  {C.}~\bibnamefont {Liu}}, \ and\ \bibinfo {author} {\bibfnamefont
  {Z.}~\bibnamefont {Zhang}},\ }\href {\doibase
  https://doi.org/10.1016/j.jqsrt.2016.12.002} {\bibfield  {journal} {\bibinfo
  {journal} {Journal of Quantitative Spectroscopy and Radiative Transfer}\
  }\textbf {\bibinfo {volume} {189}},\ \bibinfo {pages} {283 } (\bibinfo {year}
  {2017})}\BibitemShut {NoStop}%
\bibitem [{\citenamefont {Kim}\ and\ \citenamefont
  {Ishimaru}(1999)}]{kimJCP1999}%
  \BibitemOpen
  \bibfield  {author} {\bibinfo {author} {\bibfnamefont {A.~D.}\ \bibnamefont
  {Kim}}\ and\ \bibinfo {author} {\bibfnamefont {A.}~\bibnamefont {Ishimaru}},\
  }\href {\doibase https://doi.org/10.1006/jcph.1999.6247} {\bibfield
  {journal} {\bibinfo  {journal} {Journal of Computational Physics}\ }\textbf
  {\bibinfo {volume} {152}},\ \bibinfo {pages} {264 } (\bibinfo {year}
  {1999})}\BibitemShut {NoStop}%
\bibitem [{\citenamefont {Ma}\ \emph {et~al.}(2011)\citenamefont {Ma},
  \citenamefont {Dong},\ and\ \citenamefont {Tan}}]{maPRE2011}%
  \BibitemOpen
  \bibfield  {author} {\bibinfo {author} {\bibfnamefont {Y.}~\bibnamefont
  {Ma}}, \bibinfo {author} {\bibfnamefont {S.}~\bibnamefont {Dong}}, \ and\
  \bibinfo {author} {\bibfnamefont {H.}~\bibnamefont {Tan}},\ }\href {\doibase
  10.1103/PhysRevE.84.016704} {\bibfield  {journal} {\bibinfo  {journal} {Phys.
  Rev. E}\ }\textbf {\bibinfo {volume} {84}},\ \bibinfo {pages} {016704}
  (\bibinfo {year} {2011})}\BibitemShut {NoStop}%
\bibitem [{\citenamefont {Mink}\ \emph {et~al.}(2020)\citenamefont {Mink},
  \citenamefont {McHardy}, \citenamefont {Bressel}, \citenamefont {Rauh},\ and\
  \citenamefont {Krause}}]{minkJQSRT2020}%
  \BibitemOpen
  \bibfield  {author} {\bibinfo {author} {\bibfnamefont {A.}~\bibnamefont
  {Mink}}, \bibinfo {author} {\bibfnamefont {C.}~\bibnamefont {McHardy}},
  \bibinfo {author} {\bibfnamefont {L.}~\bibnamefont {Bressel}}, \bibinfo
  {author} {\bibfnamefont {C.}~\bibnamefont {Rauh}}, \ and\ \bibinfo {author}
  {\bibfnamefont {M.~J.}\ \bibnamefont {Krause}},\ }\href {\doibase
  https://doi.org/10.1016/j.jqsrt.2019.106810} {\bibfield  {journal} {\bibinfo
  {journal} {Journal of Quantitative Spectroscopy and Radiative Transfer}\
  }\textbf {\bibinfo {volume} {243}},\ \bibinfo {pages} {106810} (\bibinfo
  {year} {2020})}\BibitemShut {NoStop}%
\bibitem [{\citenamefont {McCormick}(1992)}]{mccormickNSE1992}%
  \BibitemOpen
  \bibfield  {author} {\bibinfo {author} {\bibfnamefont {N.~J.}\ \bibnamefont
  {McCormick}},\ }\href {\doibase 10.13182/NSE112-185} {\bibfield  {journal}
  {\bibinfo  {journal} {Nuclear Science and Engineering}\ }\textbf {\bibinfo
  {volume} {112}},\ \bibinfo {pages} {185} (\bibinfo {year} {1992})},\ \Eprint
  {http://arxiv.org/abs/https://doi.org/10.13182/NSE112-185}
  {https://doi.org/10.13182/NSE112-185} \BibitemShut {NoStop}%
\bibitem [{\citenamefont {Ma}\ \emph {et~al.}(2016)\citenamefont {Ma},
  \citenamefont {Zhao}, \citenamefont {Liu}, \citenamefont {Zhang},
  \citenamefont {Li},\ and\ \citenamefont {Jiang}}]{maJQSRT2016}%
  \BibitemOpen
  \bibfield  {author} {\bibinfo {author} {\bibfnamefont {C.}~\bibnamefont
  {Ma}}, \bibinfo {author} {\bibfnamefont {J.}~\bibnamefont {Zhao}}, \bibinfo
  {author} {\bibfnamefont {L.}~\bibnamefont {Liu}}, \bibinfo {author}
  {\bibfnamefont {L.}~\bibnamefont {Zhang}}, \bibinfo {author} {\bibfnamefont
  {X.}~\bibnamefont {Li}}, \ and\ \bibinfo {author} {\bibfnamefont
  {B.}~\bibnamefont {Jiang}},\ }\href {\doibase
  https://doi.org/10.1016/j.jqsrt.2015.08.002} {\bibfield  {journal} {\bibinfo
  {journal} {Journal of Quantitative Spectroscopy and Radiative Transfer}\
  }\textbf {\bibinfo {volume} {172}},\ \bibinfo {pages} {146 } (\bibinfo {year}
  {2016})},\ \bibinfo {note} {eurotherm Conference No. 105: Computational
  Thermal Radiation in Participating Media V}\BibitemShut {NoStop}%
\bibitem [{\citenamefont {Smirnov}\ \emph {et~al.}(2019)\citenamefont
  {Smirnov}, \citenamefont {Klibanov},\ and\ \citenamefont
  {Nguyen}}]{smirnovSIAMJSC2019}%
  \BibitemOpen
  \bibfield  {author} {\bibinfo {author} {\bibfnamefont {A.~V.}\ \bibnamefont
  {Smirnov}}, \bibinfo {author} {\bibfnamefont {M.~V.}\ \bibnamefont
  {Klibanov}}, \ and\ \bibinfo {author} {\bibfnamefont {L.~H.}\ \bibnamefont
  {Nguyen}},\ }\href {\doibase 10.1137/19M1253605} {\bibfield  {journal}
  {\bibinfo  {journal} {SIAM Journal on Scientific Computing}\ }\textbf
  {\bibinfo {volume} {41}},\ \bibinfo {pages} {B929} (\bibinfo {year}
  {2019})},\ \Eprint {http://arxiv.org/abs/https://doi.org/10.1137/19M1253605}
  {https://doi.org/10.1137/19M1253605} \BibitemShut {NoStop}%
\bibitem [{\citenamefont {Ren}\ \emph {et~al.}(2006)\citenamefont {Ren},
  \citenamefont {Bal},\ and\ \citenamefont {Hielscher}}]{renSIAMJSC2006}%
  \BibitemOpen
  \bibfield  {author} {\bibinfo {author} {\bibfnamefont {K.}~\bibnamefont
  {Ren}}, \bibinfo {author} {\bibfnamefont {G.}~\bibnamefont {Bal}}, \ and\
  \bibinfo {author} {\bibfnamefont {A.~H.}\ \bibnamefont {Hielscher}},\ }\href
  {\doibase 10.1137/040619193} {\bibfield  {journal} {\bibinfo  {journal} {SIAM
  Journal on Scientific Computing}\ }\textbf {\bibinfo {volume} {28}},\
  \bibinfo {pages} {1463} (\bibinfo {year} {2006})},\ \Eprint
  {http://arxiv.org/abs/https://doi.org/10.1137/040619193}
  {https://doi.org/10.1137/040619193} \BibitemShut {NoStop}%
\bibitem [{\citenamefont {Contini}\ \emph {et~al.}(1997)\citenamefont
  {Contini}, \citenamefont {Martelli},\ and\ \citenamefont
  {Zaccanti}}]{continiAO1997}%
  \BibitemOpen
  \bibfield  {author} {\bibinfo {author} {\bibfnamefont {D.}~\bibnamefont
  {Contini}}, \bibinfo {author} {\bibfnamefont {F.}~\bibnamefont {Martelli}}, \
  and\ \bibinfo {author} {\bibfnamefont {G.}~\bibnamefont {Zaccanti}},\ }\href
  {\doibase 10.1364/AO.36.004587} {\bibfield  {journal} {\bibinfo  {journal}
  {Appl. Opt.}\ }\textbf {\bibinfo {volume} {36}},\ \bibinfo {pages} {4587}
  (\bibinfo {year} {1997})}\BibitemShut {NoStop}%
\bibitem [{\citenamefont {Pierrat}\ \emph {et~al.}(2006)\citenamefont
  {Pierrat}, \citenamefont {Greffet},\ and\ \citenamefont
  {Carminati}}]{pierratJOSAA2006}%
  \BibitemOpen
  \bibfield  {author} {\bibinfo {author} {\bibfnamefont {R.}~\bibnamefont
  {Pierrat}}, \bibinfo {author} {\bibfnamefont {J.-J.}\ \bibnamefont
  {Greffet}}, \ and\ \bibinfo {author} {\bibfnamefont {R.}~\bibnamefont
  {Carminati}},\ }\href {\doibase 10.1364/JOSAA.23.001106} {\bibfield
  {journal} {\bibinfo  {journal} {J. Opt. Soc. Am. A}\ }\textbf {\bibinfo
  {volume} {23}},\ \bibinfo {pages} {1106} (\bibinfo {year}
  {2006})}\BibitemShut {NoStop}%
\bibitem [{\citenamefont {Schuurmans}\ and\ \citenamefont
  {Vanmaekelbergh}(1999)}]{schuurmansScience1999}%
  \BibitemOpen
  \bibfield  {author} {\bibinfo {author} {\bibfnamefont {F.}~\bibnamefont
  {Schuurmans}}\ and\ \bibinfo {author} {\bibfnamefont {D.}~\bibnamefont
  {Vanmaekelbergh}},\ }\href@noop {} {\bibfield  {journal} {\bibinfo  {journal}
  {Science}\ }\textbf {\bibinfo {volume} {284}},\ \bibinfo {pages} {141}
  (\bibinfo {year} {1999})}\BibitemShut {NoStop}%
\bibitem [{\citenamefont {Eldridge}\ and\ \citenamefont
  {Spuckler}(2008)}]{eldridgeJACS2008}%
  \BibitemOpen
  \bibfield  {author} {\bibinfo {author} {\bibfnamefont {J.~I.}\ \bibnamefont
  {Eldridge}}\ and\ \bibinfo {author} {\bibfnamefont {C.~M.}\ \bibnamefont
  {Spuckler}},\ }\href@noop {} {\bibfield  {journal} {\bibinfo  {journal}
  {Journal of the American Ceramic Society}\ }\textbf {\bibinfo {volume}
  {91}},\ \bibinfo {pages} {1603} (\bibinfo {year} {2008})}\BibitemShut
  {NoStop}%
\bibitem [{\citenamefont {Eldridge}\ \emph {et~al.}(2009)\citenamefont
  {Eldridge}, \citenamefont {Spuckler},\ and\ \citenamefont
  {Markham}}]{eldridgeJACS2009}%
  \BibitemOpen
  \bibfield  {author} {\bibinfo {author} {\bibfnamefont {J.~I.}\ \bibnamefont
  {Eldridge}}, \bibinfo {author} {\bibfnamefont {C.~M.}\ \bibnamefont
  {Spuckler}}, \ and\ \bibinfo {author} {\bibfnamefont {J.~R.}\ \bibnamefont
  {Markham}},\ }\href@noop {} {\bibfield  {journal} {\bibinfo  {journal}
  {Journal of the American Ceramic Society}\ }\textbf {\bibinfo {volume}
  {92}},\ \bibinfo {pages} {2276} (\bibinfo {year} {2009})}\BibitemShut
  {NoStop}%
\bibitem [{\citenamefont {Barabanenkov}\ and\ \citenamefont
  {Ozrin}(1991)}]{barabanenkovPLA1991}%
  \BibitemOpen
  \bibfield  {author} {\bibinfo {author} {\bibfnamefont {Y.}~\bibnamefont
  {Barabanenkov}}\ and\ \bibinfo {author} {\bibfnamefont {V.}~\bibnamefont
  {Ozrin}},\ }\href {\doibase https://doi.org/10.1016/0375-9601(91)90425-8}
  {\bibfield  {journal} {\bibinfo  {journal} {Physics Letters A}\ }\textbf
  {\bibinfo {volume} {154}},\ \bibinfo {pages} {38 } (\bibinfo {year}
  {1991})}\BibitemShut {NoStop}%
\bibitem [{\citenamefont {Reufer}\ \emph {et~al.}(2007)\citenamefont {Reufer},
  \citenamefont {Rojas-Ochoa}, \citenamefont {Eiden}, \citenamefont {Sáenz},\
  and\ \citenamefont {Scheffold}}]{reuferAPL2007}%
  \BibitemOpen
  \bibfield  {author} {\bibinfo {author} {\bibfnamefont {M.}~\bibnamefont
  {Reufer}}, \bibinfo {author} {\bibfnamefont {L.~F.}\ \bibnamefont
  {Rojas-Ochoa}}, \bibinfo {author} {\bibfnamefont {S.}~\bibnamefont {Eiden}},
  \bibinfo {author} {\bibfnamefont {J.~J.}\ \bibnamefont {Sáenz}}, \ and\
  \bibinfo {author} {\bibfnamefont {F.}~\bibnamefont {Scheffold}},\ }\href
  {\doibase 10.1063/1.2800372} {\bibfield  {journal} {\bibinfo  {journal}
  {Applied Physics Letters}\ }\textbf {\bibinfo {volume} {91}},\ \bibinfo
  {pages} {171904} (\bibinfo {year} {2007})},\ \Eprint
  {http://arxiv.org/abs/https://doi.org/10.1063/1.2800372}
  {https://doi.org/10.1063/1.2800372} \BibitemShut {NoStop}%
\bibitem [{\citenamefont {Skipetrov}\ and\ \citenamefont
  {Sokolov}(2014)}]{Skipetrov2014}%
  \BibitemOpen
  \bibfield  {author} {\bibinfo {author} {\bibfnamefont {S.~E.}\ \bibnamefont
  {Skipetrov}}\ and\ \bibinfo {author} {\bibfnamefont {I.~M.}\ \bibnamefont
  {Sokolov}},\ }\href {\doibase 10.1103/PhysRevLett.112.023905} {\bibfield
  {journal} {\bibinfo  {journal} {Phys. Rev. Lett.}\ }\textbf {\bibinfo
  {volume} {112}},\ \bibinfo {pages} {023905} (\bibinfo {year}
  {2014})}\BibitemShut {NoStop}%
\bibitem [{\citenamefont {M.}\ and\ \citenamefont
  {E.}(2017)}]{escalanteADP2017}%
  \BibitemOpen
  \bibfield  {author} {\bibinfo {author} {\bibfnamefont {E.~J.}\ \bibnamefont
  {M.}}\ and\ \bibinfo {author} {\bibfnamefont {S.~S.}\ \bibnamefont {E.}},\
  }\href {\doibase 10.1002/andp.201700039} {\bibfield  {journal} {\bibinfo
  {journal} {Annalen der Physik}\ }\textbf {\bibinfo {volume} {529}},\ \bibinfo
  {pages} {1700039} (\bibinfo {year} {2017})}\BibitemShut {NoStop}%
\bibitem [{\citenamefont {Schirmacher}\ \emph {et~al.}(2018)\citenamefont
  {Schirmacher}, \citenamefont {Abaie}, \citenamefont {Mafi}, \citenamefont
  {Ruocco},\ and\ \citenamefont {Leonetti}}]{schirmacherPRL2018}%
  \BibitemOpen
  \bibfield  {author} {\bibinfo {author} {\bibfnamefont {W.}~\bibnamefont
  {Schirmacher}}, \bibinfo {author} {\bibfnamefont {B.}~\bibnamefont {Abaie}},
  \bibinfo {author} {\bibfnamefont {A.}~\bibnamefont {Mafi}}, \bibinfo {author}
  {\bibfnamefont {G.}~\bibnamefont {Ruocco}}, \ and\ \bibinfo {author}
  {\bibfnamefont {M.}~\bibnamefont {Leonetti}},\ }\href {\doibase
  10.1103/PhysRevLett.120.067401} {\bibfield  {journal} {\bibinfo  {journal}
  {Phys. Rev. Lett.}\ }\textbf {\bibinfo {volume} {120}},\ \bibinfo {pages}
  {067401} (\bibinfo {year} {2018})}\BibitemShut {NoStop}%
\bibitem [{\citenamefont {Silies}\ \emph {et~al.}(2016)\citenamefont {Silies},
  \citenamefont {Mascheck}, \citenamefont {Leipold}, \citenamefont {Kollmann},
  \citenamefont {Schmidt}, \citenamefont {Sartor}, \citenamefont {Yatsui},
  \citenamefont {Kitamura}, \citenamefont {Ohtsu}, \citenamefont {Kalt},
  \citenamefont {Runge},\ and\ \citenamefont {Lienau}}]{Silies2016}%
  \BibitemOpen
  \bibfield  {author} {\bibinfo {author} {\bibfnamefont {M.}~\bibnamefont
  {Silies}}, \bibinfo {author} {\bibfnamefont {M.}~\bibnamefont {Mascheck}},
  \bibinfo {author} {\bibfnamefont {D.}~\bibnamefont {Leipold}}, \bibinfo
  {author} {\bibfnamefont {H.}~\bibnamefont {Kollmann}}, \bibinfo {author}
  {\bibfnamefont {S.}~\bibnamefont {Schmidt}}, \bibinfo {author} {\bibfnamefont
  {J.}~\bibnamefont {Sartor}}, \bibinfo {author} {\bibfnamefont
  {T.}~\bibnamefont {Yatsui}}, \bibinfo {author} {\bibfnamefont
  {K.}~\bibnamefont {Kitamura}}, \bibinfo {author} {\bibfnamefont
  {M.}~\bibnamefont {Ohtsu}}, \bibinfo {author} {\bibfnamefont
  {H.}~\bibnamefont {Kalt}}, \bibinfo {author} {\bibfnamefont {E.}~\bibnamefont
  {Runge}}, \ and\ \bibinfo {author} {\bibfnamefont {C.}~\bibnamefont
  {Lienau}},\ }\href {\doibase 10.1007/s00340-016-6456-2} {\bibfield  {journal}
  {\bibinfo  {journal} {Applied Physics B}\ }\textbf {\bibinfo {volume}
  {122}},\ \bibinfo {pages} {181} (\bibinfo {year} {2016})}\BibitemShut
  {NoStop}%
\bibitem [{\citenamefont {Pierrat}\ and\ \citenamefont
  {Carminati}(2010)}]{Pierrat2010}%
  \BibitemOpen
  \bibfield  {author} {\bibinfo {author} {\bibfnamefont {R.}~\bibnamefont
  {Pierrat}}\ and\ \bibinfo {author} {\bibfnamefont {R.}~\bibnamefont
  {Carminati}},\ }\href {\doibase 10.1103/PhysRevA.81.063802} {\bibfield
  {journal} {\bibinfo  {journal} {Phys. Rev. A}\ }\textbf {\bibinfo {volume}
  {81}},\ \bibinfo {pages} {063802} (\bibinfo {year} {2010})}\BibitemShut
  {NoStop}%
\bibitem [{\citenamefont {Petrova}\ \emph {et~al.}(2007)\citenamefont
  {Petrova}, \citenamefont {Tishkovets},\ and\ \citenamefont
  {Jockers}}]{petrovaIcarus2007}%
  \BibitemOpen
  \bibfield  {author} {\bibinfo {author} {\bibfnamefont {E.~V.}\ \bibnamefont
  {Petrova}}, \bibinfo {author} {\bibfnamefont {V.~P.}\ \bibnamefont
  {Tishkovets}}, \ and\ \bibinfo {author} {\bibfnamefont {K.}~\bibnamefont
  {Jockers}},\ }\href {\doibase https://doi.org/10.1016/j.icarus.2006.11.011}
  {\bibfield  {journal} {\bibinfo  {journal} {Icarus}\ }\textbf {\bibinfo
  {volume} {188}},\ \bibinfo {pages} {233 } (\bibinfo {year}
  {2007})}\BibitemShut {NoStop}%
\bibitem [{\citenamefont {Tishkovets}(2008)}]{tishkovetsJQSRT2008}%
  \BibitemOpen
  \bibfield  {author} {\bibinfo {author} {\bibfnamefont {V.~P.}\ \bibnamefont
  {Tishkovets}},\ }\href {\doibase https://doi.org/10.1016/j.jqsrt.2008.05.008}
  {\bibfield  {journal} {\bibinfo  {journal} {Journal of Quantitative
  Spectroscopy and Radiative Transfer}\ }\textbf {\bibinfo {volume} {109}},\
  \bibinfo {pages} {2665 } (\bibinfo {year} {2008})}\BibitemShut {NoStop}%
\bibitem [{\citenamefont {Tishkovets}\ and\ \citenamefont
  {Petrova}(2013)}]{tishkovetsLSR2013}%
  \BibitemOpen
  \bibfield  {author} {\bibinfo {author} {\bibfnamefont {V.~P.}\ \bibnamefont
  {Tishkovets}}\ and\ \bibinfo {author} {\bibfnamefont {E.~V.}\ \bibnamefont
  {Petrova}},\ }\enquote {\bibinfo {title} {Light scattering by densely packed
  systems of particles: near-field effects},}\ in\ \href {\doibase
  10.1007/978-3-642-21907-8_1} {\emph {\bibinfo {booktitle} {Light Scattering
  Reviews 7: Radiative Transfer and Optical Properties of Atmosphere and
  Underlying Surface}}}\ (\bibinfo  {publisher} {Springer Berlin Heidelberg},\
  \bibinfo {address} {Berlin, Heidelberg},\ \bibinfo {year} {2013})\ pp.\
  \bibinfo {pages} {3--36}\BibitemShut {NoStop}%
\bibitem [{\citenamefont {Shen}\ and\ \citenamefont
  {Dogariu}(2019)}]{shenOptica2019}%
  \BibitemOpen
  \bibfield  {author} {\bibinfo {author} {\bibfnamefont {Z.}~\bibnamefont
  {Shen}}\ and\ \bibinfo {author} {\bibfnamefont {A.}~\bibnamefont {Dogariu}},\
  }\href {\doibase 10.1364/OPTICA.6.000455} {\bibfield  {journal} {\bibinfo
  {journal} {Optica}\ }\textbf {\bibinfo {volume} {6}},\ \bibinfo {pages} {455}
  (\bibinfo {year} {2019})}\BibitemShut {NoStop}%
\bibitem [{\citenamefont {van Tiggelen}\ \emph {et~al.}(1990)\citenamefont {van
  Tiggelen}, \citenamefont {Lagendijk},\ and\ \citenamefont
  {Tip}}]{Vantiggelen1990JPCM}%
  \BibitemOpen
  \bibfield  {author} {\bibinfo {author} {\bibfnamefont {B.~A.}\ \bibnamefont
  {van Tiggelen}}, \bibinfo {author} {\bibfnamefont {A.}~\bibnamefont
  {Lagendijk}}, \ and\ \bibinfo {author} {\bibfnamefont {A.}~\bibnamefont
  {Tip}},\ }\href@noop {} {\bibfield  {journal} {\bibinfo  {journal} {J. Phys.:
  Cond. Mat.}\ }\textbf {\bibinfo {volume} {2}},\ \bibinfo {pages} {7653}
  (\bibinfo {year} {1990})}\BibitemShut {NoStop}%
\bibitem [{\citenamefont {van Tiggelen}\ and\ \citenamefont
  {Lagendijk}(1994)}]{vantiggelenPRB1994}%
  \BibitemOpen
  \bibfield  {author} {\bibinfo {author} {\bibfnamefont {B.~A.}\ \bibnamefont
  {van Tiggelen}}\ and\ \bibinfo {author} {\bibfnamefont {A.}~\bibnamefont
  {Lagendijk}},\ }\href {\doibase 10.1103/PhysRevB.50.16729} {\bibfield
  {journal} {\bibinfo  {journal} {Phys. Rev. B}\ }\textbf {\bibinfo {volume}
  {50}},\ \bibinfo {pages} {16729} (\bibinfo {year} {1994})}\BibitemShut
  {NoStop}%
\bibitem [{\citenamefont {Froufe-P{\'e}rez}\ \emph {et~al.}(2017)\citenamefont
  {Froufe-P{\'e}rez}, \citenamefont {Engel}, \citenamefont {S{\'a}enz},\ and\
  \citenamefont {Scheffold}}]{Froufe-PerezPNAS2017}%
  \BibitemOpen
  \bibfield  {author} {\bibinfo {author} {\bibfnamefont {L.~S.}\ \bibnamefont
  {Froufe-P{\'e}rez}}, \bibinfo {author} {\bibfnamefont {M.}~\bibnamefont
  {Engel}}, \bibinfo {author} {\bibfnamefont {J.~J.}\ \bibnamefont
  {S{\'a}enz}}, \ and\ \bibinfo {author} {\bibfnamefont {F.}~\bibnamefont
  {Scheffold}},\ }\href {\doibase 10.1073/pnas.1705130114} {\bibfield
  {journal} {\bibinfo  {journal} {Proceedings of the National Academy of
  Sciences}\ }\textbf {\bibinfo {volume} {114}},\ \bibinfo {pages} {9570}
  (\bibinfo {year} {2017})}\BibitemShut {NoStop}%
\bibitem [{\citenamefont {Wertheim}(1963)}]{wertheimPRL1963}%
  \BibitemOpen
  \bibfield  {author} {\bibinfo {author} {\bibfnamefont {M.~S.}\ \bibnamefont
  {Wertheim}},\ }\href {\doibase 10.1103/PhysRevLett.10.321} {\bibfield
  {journal} {\bibinfo  {journal} {Phys. Rev. Lett.}\ }\textbf {\bibinfo
  {volume} {10}},\ \bibinfo {pages} {321} (\bibinfo {year} {1963})}\BibitemShut
  {NoStop}%
\bibitem [{\citenamefont {Mishchenko}(1994)}]{mishchenkoJQSRT1994}%
  \BibitemOpen
  \bibfield  {author} {\bibinfo {author} {\bibfnamefont {M.~I.}\ \bibnamefont
  {Mishchenko}},\ }\href {\doibase
  https://doi.org/10.1016/0022-4073(94)90142-2} {\bibfield  {journal} {\bibinfo
   {journal} {Journal of Quantitative Spectroscopy and Radiative Transfer}\
  }\textbf {\bibinfo {volume} {52}},\ \bibinfo {pages} {95 } (\bibinfo {year}
  {1994})}\BibitemShut {NoStop}%
\bibitem [{\citenamefont {Leseur}\ \emph {et~al.}(2016)\citenamefont {Leseur},
  \citenamefont {Pierrat},\ and\ \citenamefont {Carminati}}]{leseurOptica2016}%
  \BibitemOpen
  \bibfield  {author} {\bibinfo {author} {\bibfnamefont {O.}~\bibnamefont
  {Leseur}}, \bibinfo {author} {\bibfnamefont {R.}~\bibnamefont {Pierrat}}, \
  and\ \bibinfo {author} {\bibfnamefont {R.}~\bibnamefont {Carminati}},\ }\href
  {\doibase 10.1364/OPTICA.3.000763} {\bibfield  {journal} {\bibinfo  {journal}
  {Optica}\ }\textbf {\bibinfo {volume} {3}},\ \bibinfo {pages} {763} (\bibinfo
  {year} {2016})}\BibitemShut {NoStop}%
\bibitem [{\citenamefont {Froufe-P\'erez}\ \emph {et~al.}(2016)\citenamefont
  {Froufe-P\'erez}, \citenamefont {Engel}, \citenamefont {Damasceno},
  \citenamefont {Muller}, \citenamefont {Haberko}, \citenamefont {Glotzer},\
  and\ \citenamefont {Scheffold}}]{Froufe-Perez2016}%
  \BibitemOpen
  \bibfield  {author} {\bibinfo {author} {\bibfnamefont {L.~S.}\ \bibnamefont
  {Froufe-P\'erez}}, \bibinfo {author} {\bibfnamefont {M.}~\bibnamefont
  {Engel}}, \bibinfo {author} {\bibfnamefont {P.~F.}\ \bibnamefont
  {Damasceno}}, \bibinfo {author} {\bibfnamefont {N.}~\bibnamefont {Muller}},
  \bibinfo {author} {\bibfnamefont {J.}~\bibnamefont {Haberko}}, \bibinfo
  {author} {\bibfnamefont {S.~C.}\ \bibnamefont {Glotzer}}, \ and\ \bibinfo
  {author} {\bibfnamefont {F.}~\bibnamefont {Scheffold}},\ }\href {\doibase
  10.1103/PhysRevLett.117.053902} {\bibfield  {journal} {\bibinfo  {journal}
  {Phys. Rev. Lett.}\ }\textbf {\bibinfo {volume} {117}},\ \bibinfo {pages}
  {053902} (\bibinfo {year} {2016})}\BibitemShut {NoStop}%
\bibitem [{\citenamefont {Baxter}(1968)}]{Baxter1968}%
  \BibitemOpen
  \bibfield  {author} {\bibinfo {author} {\bibfnamefont {R.~J.}\ \bibnamefont
  {Baxter}},\ }\href {\doibase http://dx.doi.org/10.1063/1.1670482} {\bibfield
  {journal} {\bibinfo  {journal} {J. Chem. Phys.}\ }\textbf {\bibinfo {volume}
  {49}},\ \bibinfo {pages} {2770} (\bibinfo {year} {1968})}\BibitemShut
  {NoStop}%
\bibitem [{\citenamefont {Frenkel}(2002)}]{Frenkel2002}%
  \BibitemOpen
  \bibfield  {author} {\bibinfo {author} {\bibfnamefont {D.}~\bibnamefont
  {Frenkel}},\ }\href {\doibase 10.1126/science.1070865} {\bibfield  {journal}
  {\bibinfo  {journal} {Science}\ }\textbf {\bibinfo {volume} {296}},\ \bibinfo
  {pages} {65} (\bibinfo {year} {2002})}\BibitemShut {NoStop}%
\bibitem [{\citenamefont {Bressel}\ \emph {et~al.}(2013)\citenamefont
  {Bressel}, \citenamefont {Hass},\ and\ \citenamefont
  {Reich}}]{bresselJSQRT2013}%
  \BibitemOpen
  \bibfield  {author} {\bibinfo {author} {\bibfnamefont {L.}~\bibnamefont
  {Bressel}}, \bibinfo {author} {\bibfnamefont {R.}~\bibnamefont {Hass}}, \
  and\ \bibinfo {author} {\bibfnamefont {O.}~\bibnamefont {Reich}},\ }\href
  {\doibase https://doi.org/10.1016/j.jqsrt.2012.11.031} {\bibfield  {journal}
  {\bibinfo  {journal} {Journal of Quantitative Spectroscopy and Radiative
  Transfer}\ }\textbf {\bibinfo {volume} {126}},\ \bibinfo {pages} {122 }
  (\bibinfo {year} {2013})},\ \bibinfo {note} {lasers and interactions with
  particles 2012}\BibitemShut {NoStop}%
\bibitem [{\citenamefont {Sudiarta}\ and\ \citenamefont
  {Chylek}(2001)}]{sudiartaJOSAA2001}%
  \BibitemOpen
  \bibfield  {author} {\bibinfo {author} {\bibfnamefont {I.~W.}\ \bibnamefont
  {Sudiarta}}\ and\ \bibinfo {author} {\bibfnamefont {P.}~\bibnamefont
  {Chylek}},\ }\href {\doibase 10.1364/JOSAA.18.001275} {\bibfield  {journal}
  {\bibinfo  {journal} {J. Opt. Soc. Am. A}\ }\textbf {\bibinfo {volume}
  {18}},\ \bibinfo {pages} {1275} (\bibinfo {year} {2001})}\BibitemShut
  {NoStop}%
\bibitem [{\citenamefont {Fu}\ and\ \citenamefont {Sun}(2001)}]{fuAO2001}%
  \BibitemOpen
  \bibfield  {author} {\bibinfo {author} {\bibfnamefont {Q.}~\bibnamefont
  {Fu}}\ and\ \bibinfo {author} {\bibfnamefont {W.}~\bibnamefont {Sun}},\
  }\href {\doibase 10.1364/AO.40.001354} {\bibfield  {journal} {\bibinfo
  {journal} {Appl. Opt.}\ }\textbf {\bibinfo {volume} {40}},\ \bibinfo {pages}
  {1354} (\bibinfo {year} {2001})}\BibitemShut {NoStop}%
\bibitem [{\citenamefont {Yang}\ \emph {et~al.}(2002)\citenamefont {Yang},
  \citenamefont {Gao}, \citenamefont {Wiscombe}, \citenamefont {Mishchenko},
  \citenamefont {Platnick}, \citenamefont {Huang}, \citenamefont {Baum},
  \citenamefont {Hu}, \citenamefont {Winker}, \citenamefont {Tsay},\ and\
  \citenamefont {Park}}]{yangAO2002}%
  \BibitemOpen
  \bibfield  {author} {\bibinfo {author} {\bibfnamefont {P.}~\bibnamefont
  {Yang}}, \bibinfo {author} {\bibfnamefont {B.-C.}\ \bibnamefont {Gao}},
  \bibinfo {author} {\bibfnamefont {W.~J.}\ \bibnamefont {Wiscombe}}, \bibinfo
  {author} {\bibfnamefont {M.~I.}\ \bibnamefont {Mishchenko}}, \bibinfo
  {author} {\bibfnamefont {S.~E.}\ \bibnamefont {Platnick}}, \bibinfo {author}
  {\bibfnamefont {H.-L.}\ \bibnamefont {Huang}}, \bibinfo {author}
  {\bibfnamefont {B.~A.}\ \bibnamefont {Baum}}, \bibinfo {author}
  {\bibfnamefont {Y.~X.}\ \bibnamefont {Hu}}, \bibinfo {author} {\bibfnamefont
  {D.~M.}\ \bibnamefont {Winker}}, \bibinfo {author} {\bibfnamefont {S.-C.}\
  \bibnamefont {Tsay}}, \ and\ \bibinfo {author} {\bibfnamefont {S.~K.}\
  \bibnamefont {Park}},\ }\href {\doibase 10.1364/AO.41.002740} {\bibfield
  {journal} {\bibinfo  {journal} {Appl. Opt.}\ }\textbf {\bibinfo {volume}
  {41}},\ \bibinfo {pages} {2740} (\bibinfo {year} {2002})}\BibitemShut
  {NoStop}%
\bibitem [{\citenamefont {Videen}\ and\ \citenamefont
  {Sun}(2003)}]{videenAO2003}%
  \BibitemOpen
  \bibfield  {author} {\bibinfo {author} {\bibfnamefont {G.}~\bibnamefont
  {Videen}}\ and\ \bibinfo {author} {\bibfnamefont {W.}~\bibnamefont {Sun}},\
  }\href {\doibase 10.1364/AO.42.006724} {\bibfield  {journal} {\bibinfo
  {journal} {Appl. Opt.}\ }\textbf {\bibinfo {volume} {42}},\ \bibinfo {pages}
  {6724} (\bibinfo {year} {2003})}\BibitemShut {NoStop}%
\bibitem [{\citenamefont {Yin}\ and\ \citenamefont
  {Pilon}(2006)}]{yinJOSAA2006}%
  \BibitemOpen
  \bibfield  {author} {\bibinfo {author} {\bibfnamefont {J.}~\bibnamefont
  {Yin}}\ and\ \bibinfo {author} {\bibfnamefont {L.}~\bibnamefont {Pilon}},\
  }\href {\doibase 10.1364/JOSAA.23.002784} {\bibfield  {journal} {\bibinfo
  {journal} {J. Opt. Soc. Am. A}\ }\textbf {\bibinfo {volume} {23}},\ \bibinfo
  {pages} {2784} (\bibinfo {year} {2006})}\BibitemShut {NoStop}%
\bibitem [{\citenamefont {Aernouts}\ \emph
  {et~al.}(2014{\natexlab{b}})\citenamefont {Aernouts}, \citenamefont
  {Watt\'{e}}, \citenamefont {Beers}, \citenamefont {Delport}, \citenamefont
  {Merchiers}, \citenamefont {Block}, \citenamefont {Lammertyn},\ and\
  \citenamefont {Saeys}}]{aernoutsOE2014b}%
  \BibitemOpen
  \bibfield  {author} {\bibinfo {author} {\bibfnamefont {B.}~\bibnamefont
  {Aernouts}}, \bibinfo {author} {\bibfnamefont {R.}~\bibnamefont {Watt\'{e}}},
  \bibinfo {author} {\bibfnamefont {R.~V.}\ \bibnamefont {Beers}}, \bibinfo
  {author} {\bibfnamefont {F.}~\bibnamefont {Delport}}, \bibinfo {author}
  {\bibfnamefont {M.}~\bibnamefont {Merchiers}}, \bibinfo {author}
  {\bibfnamefont {J.~D.}\ \bibnamefont {Block}}, \bibinfo {author}
  {\bibfnamefont {J.}~\bibnamefont {Lammertyn}}, \ and\ \bibinfo {author}
  {\bibfnamefont {W.}~\bibnamefont {Saeys}},\ }\href {\doibase
  10.1364/OE.22.020223} {\bibfield  {journal} {\bibinfo  {journal} {Opt.
  Express}\ }\textbf {\bibinfo {volume} {22}},\ \bibinfo {pages} {20223}
  (\bibinfo {year} {2014}{\natexlab{b}})}\BibitemShut {NoStop}%
\bibitem [{\citenamefont {Mishchenko}\ \emph {et~al.}(2017)\citenamefont
  {Mishchenko}, \citenamefont {Videen},\ and\ \citenamefont
  {Yang}}]{mishchenkoOL2017}%
  \BibitemOpen
  \bibfield  {author} {\bibinfo {author} {\bibfnamefont {M.~I.}\ \bibnamefont
  {Mishchenko}}, \bibinfo {author} {\bibfnamefont {G.}~\bibnamefont {Videen}},
  \ and\ \bibinfo {author} {\bibfnamefont {P.}~\bibnamefont {Yang}},\ }\href
  {\doibase 10.1364/OL.42.004873} {\bibfield  {journal} {\bibinfo  {journal}
  {Opt. Lett.}\ }\textbf {\bibinfo {volume} {42}},\ \bibinfo {pages} {4873}
  (\bibinfo {year} {2017})}\BibitemShut {NoStop}%
\bibitem [{\citenamefont {Mishchenko}\ and\ \citenamefont
  {Yang}(2018)}]{mishchenkoJQSRT2018}%
  \BibitemOpen
  \bibfield  {author} {\bibinfo {author} {\bibfnamefont {M.~I.}\ \bibnamefont
  {Mishchenko}}\ and\ \bibinfo {author} {\bibfnamefont {P.}~\bibnamefont
  {Yang}},\ }\href {\doibase https://doi.org/10.1016/j.jqsrt.2017.10.014}
  {\bibfield  {journal} {\bibinfo  {journal} {Journal of Quantitative
  Spectroscopy and Radiative Transfer}\ }\textbf {\bibinfo {volume} {205}},\
  \bibinfo {pages} {241 } (\bibinfo {year} {2018})}\BibitemShut {NoStop}%
\bibitem [{\citenamefont {Mishchenko}\ \emph {et~al.}(2019)\citenamefont
  {Mishchenko}, \citenamefont {Yurkin},\ and\ \citenamefont
  {Cairns}}]{mishchenkoOSAC2019b}%
  \BibitemOpen
  \bibfield  {author} {\bibinfo {author} {\bibfnamefont {M.~I.}\ \bibnamefont
  {Mishchenko}}, \bibinfo {author} {\bibfnamefont {M.~A.}\ \bibnamefont
  {Yurkin}}, \ and\ \bibinfo {author} {\bibfnamefont {B.}~\bibnamefont
  {Cairns}},\ }\href {\doibase 10.1364/OSAC.2.002362} {\bibfield  {journal}
  {\bibinfo  {journal} {OSA Continuum}\ }\textbf {\bibinfo {volume} {2}},\
  \bibinfo {pages} {2362} (\bibinfo {year} {2019})}\BibitemShut {NoStop}%
\bibitem [{\citenamefont {Lee}\ and\ \citenamefont
  {Peumans}(2010)}]{leeOE2010}%
  \BibitemOpen
  \bibfield  {author} {\bibinfo {author} {\bibfnamefont {J.-Y.}\ \bibnamefont
  {Lee}}\ and\ \bibinfo {author} {\bibfnamefont {P.}~\bibnamefont {Peumans}},\
  }\href {\doibase 10.1364/OE.18.010078} {\bibfield  {journal} {\bibinfo
  {journal} {Opt. Express}\ }\textbf {\bibinfo {volume} {18}},\ \bibinfo
  {pages} {10078} (\bibinfo {year} {2010})}\BibitemShut {NoStop}%
\bibitem [{\citenamefont {Nagel}\ and\ \citenamefont
  {Scarpulla}(2010)}]{nagelOE2010}%
  \BibitemOpen
  \bibfield  {author} {\bibinfo {author} {\bibfnamefont {J.~R.}\ \bibnamefont
  {Nagel}}\ and\ \bibinfo {author} {\bibfnamefont {M.~A.}\ \bibnamefont
  {Scarpulla}},\ }\href {\doibase 10.1364/OE.18.00A139} {\bibfield  {journal}
  {\bibinfo  {journal} {Opt. Express}\ }\textbf {\bibinfo {volume} {18}},\
  \bibinfo {pages} {A139} (\bibinfo {year} {2010})}\BibitemShut {NoStop}%
\bibitem [{\citenamefont {Chen}\ \emph {et~al.}(2019)\citenamefont {Chen},
  \citenamefont {Zhou}, \citenamefont {Li},\ and\ \citenamefont
  {Hong}}]{chenAPR2019}%
  \BibitemOpen
  \bibfield  {author} {\bibinfo {author} {\bibfnamefont {L.}~\bibnamefont
  {Chen}}, \bibinfo {author} {\bibfnamefont {Y.}~\bibnamefont {Zhou}}, \bibinfo
  {author} {\bibfnamefont {Y.}~\bibnamefont {Li}}, \ and\ \bibinfo {author}
  {\bibfnamefont {M.}~\bibnamefont {Hong}},\ }\href {\doibase
  10.1063/1.5082215} {\bibfield  {journal} {\bibinfo  {journal} {Applied
  Physics Reviews}\ }\textbf {\bibinfo {volume} {6}},\ \bibinfo {pages}
  {021304} (\bibinfo {year} {2019})},\ \Eprint
  {http://arxiv.org/abs/https://doi.org/10.1063/1.5082215}
  {https://doi.org/10.1063/1.5082215} \BibitemShut {NoStop}%
\bibitem [{\citenamefont {Mishchenko}(2008{\natexlab{a}})}]{mishchenkoOE2008}%
  \BibitemOpen
  \bibfield  {author} {\bibinfo {author} {\bibfnamefont {M.~I.}\ \bibnamefont
  {Mishchenko}},\ }\href {\doibase 10.1364/OE.16.002288} {\bibfield  {journal}
  {\bibinfo  {journal} {Opt. Express}\ }\textbf {\bibinfo {volume} {16}},\
  \bibinfo {pages} {2288} (\bibinfo {year} {2008}{\natexlab{a}})}\BibitemShut
  {NoStop}%
\bibitem [{\citenamefont
  {Mishchenko}(2008{\natexlab{b}})}]{mishchenkoJQSRT2008}%
  \BibitemOpen
  \bibfield  {author} {\bibinfo {author} {\bibfnamefont {M.~I.}\ \bibnamefont
  {Mishchenko}},\ }\href {\doibase https://doi.org/10.1016/j.jqsrt.2008.05.006}
  {\bibfield  {journal} {\bibinfo  {journal} {Journal of Quantitative
  Spectroscopy and Radiative Transfer}\ }\textbf {\bibinfo {volume} {109}},\
  \bibinfo {pages} {2386 } (\bibinfo {year} {2008}{\natexlab{b}})}\BibitemShut
  {NoStop}%
\bibitem [{\citenamefont {Durant}\ \emph
  {et~al.}(2007{\natexlab{a}})\citenamefont {Durant}, \citenamefont
  {Calvo-Perez}, \citenamefont {Vukadinovic},\ and\ \citenamefont
  {Greffet}}]{durantJOSAA2007a}%
  \BibitemOpen
  \bibfield  {author} {\bibinfo {author} {\bibfnamefont {S.}~\bibnamefont
  {Durant}}, \bibinfo {author} {\bibfnamefont {O.}~\bibnamefont {Calvo-Perez}},
  \bibinfo {author} {\bibfnamefont {N.}~\bibnamefont {Vukadinovic}}, \ and\
  \bibinfo {author} {\bibfnamefont {J.-J.}\ \bibnamefont {Greffet}},\ }\href
  {\doibase 10.1364/JOSAA.24.002943} {\bibfield  {journal} {\bibinfo  {journal}
  {J. Opt. Soc. Am. A}\ }\textbf {\bibinfo {volume} {24}},\ \bibinfo {pages}
  {2943} (\bibinfo {year} {2007}{\natexlab{a}})}\BibitemShut {NoStop}%
\bibitem [{\citenamefont {Durant}\ \emph
  {et~al.}(2007{\natexlab{b}})\citenamefont {Durant}, \citenamefont
  {Calvo-Perez}, \citenamefont {Vukadinovic},\ and\ \citenamefont
  {Greffet}}]{durantJOSAA2007}%
  \BibitemOpen
  \bibfield  {author} {\bibinfo {author} {\bibfnamefont {S.}~\bibnamefont
  {Durant}}, \bibinfo {author} {\bibfnamefont {O.}~\bibnamefont {Calvo-Perez}},
  \bibinfo {author} {\bibfnamefont {N.}~\bibnamefont {Vukadinovic}}, \ and\
  \bibinfo {author} {\bibfnamefont {J.-J.}\ \bibnamefont {Greffet}},\ }\href
  {\doibase 10.1364/JOSAA.24.002953} {\bibfield  {journal} {\bibinfo  {journal}
  {J. Opt. Soc. Am. A}\ }\textbf {\bibinfo {volume} {24}},\ \bibinfo {pages}
  {2953} (\bibinfo {year} {2007}{\natexlab{b}})}\BibitemShut {NoStop}%
\bibitem [{\citenamefont {Dick}\ and\ \citenamefont
  {Ivanov}(1999)}]{dickJOSAA1999}%
  \BibitemOpen
  \bibfield  {author} {\bibinfo {author} {\bibfnamefont {V.~P.}\ \bibnamefont
  {Dick}}\ and\ \bibinfo {author} {\bibfnamefont {A.~P.}\ \bibnamefont
  {Ivanov}},\ }\href {\doibase 10.1364/JOSAA.16.001034} {\bibfield  {journal}
  {\bibinfo  {journal} {J. Opt. Soc. Am. A}\ }\textbf {\bibinfo {volume}
  {16}},\ \bibinfo {pages} {1034} (\bibinfo {year} {1999})}\BibitemShut
  {NoStop}%
\bibitem [{\citenamefont {Xiao}\ \emph {et~al.}(2019)\citenamefont {Xiao},
  \citenamefont {Hu}, \citenamefont {Gartner}, \citenamefont {Yang},
  \citenamefont {Li}, \citenamefont {Jayaraman}, \citenamefont {Gianneschi},
  \citenamefont {Shawkey},\ and\ \citenamefont {Dhinojwala}}]{xiaoeSciAdv2019}%
  \BibitemOpen
  \bibfield  {author} {\bibinfo {author} {\bibfnamefont {M.}~\bibnamefont
  {Xiao}}, \bibinfo {author} {\bibfnamefont {Z.}~\bibnamefont {Hu}}, \bibinfo
  {author} {\bibfnamefont {T.~E.}\ \bibnamefont {Gartner}}, \bibinfo {author}
  {\bibfnamefont {X.}~\bibnamefont {Yang}}, \bibinfo {author} {\bibfnamefont
  {W.}~\bibnamefont {Li}}, \bibinfo {author} {\bibfnamefont {A.}~\bibnamefont
  {Jayaraman}}, \bibinfo {author} {\bibfnamefont {N.~C.}\ \bibnamefont
  {Gianneschi}}, \bibinfo {author} {\bibfnamefont {M.~D.}\ \bibnamefont
  {Shawkey}}, \ and\ \bibinfo {author} {\bibfnamefont {A.}~\bibnamefont
  {Dhinojwala}},\ }\href {\doibase 10.1126/sciadv.aax1254} {\bibfield
  {journal} {\bibinfo  {journal} {Science Advances}\ }\textbf {\bibinfo
  {volume} {5}} (\bibinfo {year} {2019}),\ 10.1126/sciadv.aax1254},\ \Eprint
  {http://arxiv.org/abs/https://advances.sciencemag.org/content/5/9/eaax1254.full.pdf}
  {https://advances.sciencemag.org/content/5/9/eaax1254.full.pdf} \BibitemShut
  {NoStop}%
\bibitem [{\citenamefont {Twersky}(1978)}]{twerskyJASA1978}%
  \BibitemOpen
  \bibfield  {author} {\bibinfo {author} {\bibfnamefont {V.}~\bibnamefont
  {Twersky}},\ }\href {\doibase 10.1121/1.382150} {\bibfield  {journal}
  {\bibinfo  {journal} {The Journal of the Acoustical Society of America}\
  }\textbf {\bibinfo {volume} {64}},\ \bibinfo {pages} {1710} (\bibinfo {year}
  {1978})},\ \Eprint {http://arxiv.org/abs/https://doi.org/10.1121/1.382150}
  {https://doi.org/10.1121/1.382150} \BibitemShut {NoStop}%
\bibitem [{\citenamefont {Holthoff}\ \emph {et~al.}(1997)\citenamefont
  {Holthoff}, \citenamefont {Borkovec},\ and\ \citenamefont
  {Schurtenberger}}]{holthoffPRE1997}%
  \BibitemOpen
  \bibfield  {author} {\bibinfo {author} {\bibfnamefont {H.}~\bibnamefont
  {Holthoff}}, \bibinfo {author} {\bibfnamefont {M.}~\bibnamefont {Borkovec}},
  \ and\ \bibinfo {author} {\bibfnamefont {P.}~\bibnamefont {Schurtenberger}},\
  }\href {\doibase 10.1103/PhysRevE.56.6945} {\bibfield  {journal} {\bibinfo
  {journal} {Phys. Rev. E}\ }\textbf {\bibinfo {volume} {56}},\ \bibinfo
  {pages} {6945} (\bibinfo {year} {1997})}\BibitemShut {NoStop}%
\bibitem [{\citenamefont {Conley}\ \emph {et~al.}(2014)\citenamefont {Conley},
  \citenamefont {Burresi}, \citenamefont {Pratesi}, \citenamefont {Vynck},\
  and\ \citenamefont {Wiersma}}]{conleyPRL2014}%
  \BibitemOpen
  \bibfield  {author} {\bibinfo {author} {\bibfnamefont {G.~M.}\ \bibnamefont
  {Conley}}, \bibinfo {author} {\bibfnamefont {M.}~\bibnamefont {Burresi}},
  \bibinfo {author} {\bibfnamefont {F.}~\bibnamefont {Pratesi}}, \bibinfo
  {author} {\bibfnamefont {K.}~\bibnamefont {Vynck}}, \ and\ \bibinfo {author}
  {\bibfnamefont {D.~S.}\ \bibnamefont {Wiersma}},\ }\href {\doibase
  10.1103/PhysRevLett.112.143901} {\bibfield  {journal} {\bibinfo  {journal}
  {Phys. Rev. Lett.}\ }\textbf {\bibinfo {volume} {112}},\ \bibinfo {pages}
  {143901} (\bibinfo {year} {2014})}\BibitemShut {NoStop}%
\bibitem [{\citenamefont {Lagendijk}\ \emph {et~al.}(1997)\citenamefont
  {Lagendijk}, \citenamefont {Nienhuis}, \citenamefont {van Tiggelen},\ and\
  \citenamefont {de~Vries}}]{Lagendijk1997PRL}%
  \BibitemOpen
  \bibfield  {author} {\bibinfo {author} {\bibfnamefont {A.}~\bibnamefont
  {Lagendijk}}, \bibinfo {author} {\bibfnamefont {B.}~\bibnamefont {Nienhuis}},
  \bibinfo {author} {\bibfnamefont {B.~A.}\ \bibnamefont {van Tiggelen}}, \
  and\ \bibinfo {author} {\bibfnamefont {P.}~\bibnamefont {de~Vries}},\ }\href
  {\doibase 10.1103/PhysRevLett.79.657} {\bibfield  {journal} {\bibinfo
  {journal} {Phys. Rev. Lett.}\ }\textbf {\bibinfo {volume} {79}},\ \bibinfo
  {pages} {657} (\bibinfo {year} {1997})}\BibitemShut {NoStop}%
\bibitem [{\citenamefont {Mallet}\ \emph {et~al.}(2005)\citenamefont {Mallet},
  \citenamefont {Gu\'erin},\ and\ \citenamefont {Sentenac}}]{malletPRB2005}%
  \BibitemOpen
  \bibfield  {author} {\bibinfo {author} {\bibfnamefont {P.}~\bibnamefont
  {Mallet}}, \bibinfo {author} {\bibfnamefont {C.~A.}\ \bibnamefont
  {Gu\'erin}}, \ and\ \bibinfo {author} {\bibfnamefont {A.}~\bibnamefont
  {Sentenac}},\ }\href {\doibase 10.1103/PhysRevB.72.014205} {\bibfield
  {journal} {\bibinfo  {journal} {Phys. Rev. B}\ }\textbf {\bibinfo {volume}
  {72}},\ \bibinfo {pages} {014205} (\bibinfo {year} {2005})}\BibitemShut
  {NoStop}%
\bibitem [{\citenamefont {Grimes}\ and\ \citenamefont
  {Grimes}(1991)}]{grimesPRB1991}%
  \BibitemOpen
  \bibfield  {author} {\bibinfo {author} {\bibfnamefont {C.~A.}\ \bibnamefont
  {Grimes}}\ and\ \bibinfo {author} {\bibfnamefont {D.~M.}\ \bibnamefont
  {Grimes}},\ }\href {\doibase 10.1103/PhysRevB.43.10780} {\bibfield  {journal}
  {\bibinfo  {journal} {Phys. Rev. B}\ }\textbf {\bibinfo {volume} {43}},\
  \bibinfo {pages} {10780} (\bibinfo {year} {1991})}\BibitemShut {NoStop}%
\bibitem [{\citenamefont {Chaumet}\ and\ \citenamefont
  {Rahmani}(2009)}]{chaumetJQSRT2009}%
  \BibitemOpen
  \bibfield  {author} {\bibinfo {author} {\bibfnamefont {P.~C.}\ \bibnamefont
  {Chaumet}}\ and\ \bibinfo {author} {\bibfnamefont {A.}~\bibnamefont
  {Rahmani}},\ }\href {\doibase https://doi.org/10.1016/j.jqsrt.2008.09.004}
  {\bibfield  {journal} {\bibinfo  {journal} {Journal of Quantitative
  Spectroscopy and Radiative Transfer}\ }\textbf {\bibinfo {volume} {110}},\
  \bibinfo {pages} {22 } (\bibinfo {year} {2009})}\BibitemShut {NoStop}%
\bibitem [{\citenamefont {Ruppin}(2000)}]{ruppinOC2000}%
  \BibitemOpen
  \bibfield  {author} {\bibinfo {author} {\bibfnamefont {R.}~\bibnamefont
  {Ruppin}},\ }\href {\doibase https://doi.org/10.1016/S0030-4018(00)00825-7}
  {\bibfield  {journal} {\bibinfo  {journal} {Optics Communications}\ }\textbf
  {\bibinfo {volume} {182}},\ \bibinfo {pages} {273 } (\bibinfo {year}
  {2000})}\BibitemShut {NoStop}%
\bibitem [{\citenamefont {Wheeler}\ \emph {et~al.}(2009)\citenamefont
  {Wheeler}, \citenamefont {Aitchison}, \citenamefont {Chen}, \citenamefont
  {Ozin},\ and\ \citenamefont {Mojahedi}}]{wheelerPRB2009}%
  \BibitemOpen
  \bibfield  {author} {\bibinfo {author} {\bibfnamefont {M.~S.}\ \bibnamefont
  {Wheeler}}, \bibinfo {author} {\bibfnamefont {J.~S.}\ \bibnamefont
  {Aitchison}}, \bibinfo {author} {\bibfnamefont {J.~I.~L.}\ \bibnamefont
  {Chen}}, \bibinfo {author} {\bibfnamefont {G.~A.}\ \bibnamefont {Ozin}}, \
  and\ \bibinfo {author} {\bibfnamefont {M.}~\bibnamefont {Mojahedi}},\ }\href
  {\doibase 10.1103/PhysRevB.79.073103} {\bibfield  {journal} {\bibinfo
  {journal} {Phys. Rev. B}\ }\textbf {\bibinfo {volume} {79}},\ \bibinfo
  {pages} {073103} (\bibinfo {year} {2009})}\BibitemShut {NoStop}%
\bibitem [{\citenamefont {Wang}\ and\ \citenamefont
  {Zhao}(2019)}]{wangPhotonAsia2019}%
  \BibitemOpen
  \bibfield  {author} {\bibinfo {author} {\bibfnamefont {B.~X.}\ \bibnamefont
  {Wang}}\ and\ \bibinfo {author} {\bibfnamefont {C.~Y.}\ \bibnamefont
  {Zhao}},\ }in\ \href {\doibase 10.1117/12.2537226} {\emph {\bibinfo
  {booktitle} {Quantum and Nonlinear Optics VI}}},\ Vol.\ \bibinfo {volume}
  {11195},\ \bibinfo {editor} {edited by\ \bibinfo {editor} {\bibfnamefont
  {Q.}~\bibnamefont {Gong}}, \bibinfo {editor} {\bibfnamefont {G.-C.}\
  \bibnamefont {Guo}}, \ and\ \bibinfo {editor} {\bibfnamefont {B.~S.}\
  \bibnamefont {Ham}}},\ \bibinfo {organization} {International Society for
  Optics and Photonics}\ (\bibinfo  {publisher} {SPIE},\ \bibinfo {year}
  {2019})\ pp.\ \bibinfo {pages} {27 -- 36}\BibitemShut {NoStop}%
\bibitem [{\citenamefont {Karal}\ and\ \citenamefont
  {Keller}(1964)}]{karalJMP1964}%
  \BibitemOpen
  \bibfield  {author} {\bibinfo {author} {\bibfnamefont {F.~C.}\ \bibnamefont
  {Karal}}\ and\ \bibinfo {author} {\bibfnamefont {J.~B.}\ \bibnamefont
  {Keller}},\ }\href {\doibase 10.1063/1.1704145} {\bibfield  {journal}
  {\bibinfo  {journal} {Journal of Mathematical Physics}\ }\textbf {\bibinfo
  {volume} {5}},\ \bibinfo {pages} {537} (\bibinfo {year} {1964})},\ \Eprint
  {http://arxiv.org/abs/https://doi.org/10.1063/1.1704145}
  {https://doi.org/10.1063/1.1704145} \BibitemShut {NoStop}%
\bibitem [{\citenamefont {Keller}(1964)}]{keller1964stochastic}%
  \BibitemOpen
  \bibfield  {author} {\bibinfo {author} {\bibfnamefont {J.~B.}\ \bibnamefont
  {Keller}},\ }\href@noop {} {\bibfield  {journal} {\bibinfo  {journal}
  {Stochastic processes in mathematical physics and engineering}\ }\textbf
  {\bibinfo {volume} {16}},\ \bibinfo {pages} {145} (\bibinfo {year}
  {1964})}\BibitemShut {NoStop}%
\bibitem [{\citenamefont {Keller}\ and\ \citenamefont
  {Karal}(1966)}]{kellerJMP1966}%
  \BibitemOpen
  \bibfield  {author} {\bibinfo {author} {\bibfnamefont {J.~B.}\ \bibnamefont
  {Keller}}\ and\ \bibinfo {author} {\bibfnamefont {F.~C.}\ \bibnamefont
  {Karal}},\ }\href {\doibase 10.1063/1.1704979} {\bibfield  {journal}
  {\bibinfo  {journal} {Journal of Mathematical Physics}\ }\textbf {\bibinfo
  {volume} {7}},\ \bibinfo {pages} {661} (\bibinfo {year} {1966})},\ \Eprint
  {http://arxiv.org/abs/https://doi.org/10.1063/1.1704979}
  {https://doi.org/10.1063/1.1704979} \BibitemShut {NoStop}%
\bibitem [{\citenamefont {Hespel}\ \emph {et~al.}(2001)\citenamefont {Hespel},
  \citenamefont {Mainguy},\ and\ \citenamefont {Greffet}}]{hespelJOSAA2001}%
  \BibitemOpen
  \bibfield  {author} {\bibinfo {author} {\bibfnamefont {L.}~\bibnamefont
  {Hespel}}, \bibinfo {author} {\bibfnamefont {S.}~\bibnamefont {Mainguy}}, \
  and\ \bibinfo {author} {\bibfnamefont {J.-J.}\ \bibnamefont {Greffet}},\
  }\href {\doibase 10.1364/JOSAA.18.003072} {\bibfield  {journal} {\bibinfo
  {journal} {J. Opt. Soc. Am. A}\ }\textbf {\bibinfo {volume} {18}},\ \bibinfo
  {pages} {3072} (\bibinfo {year} {2001})}\BibitemShut {NoStop}%
\bibitem [{\citenamefont {Ishimaru}\ and\ \citenamefont
  {Kuga}(1982)}]{ishimaruJOSA1982}%
  \BibitemOpen
  \bibfield  {author} {\bibinfo {author} {\bibfnamefont {A.}~\bibnamefont
  {Ishimaru}}\ and\ \bibinfo {author} {\bibfnamefont {Y.}~\bibnamefont
  {Kuga}},\ }\href {\doibase 10.1364/JOSA.72.001317} {\bibfield  {journal}
  {\bibinfo  {journal} {J. Opt. Soc. Am.}\ }\textbf {\bibinfo {volume} {72}},\
  \bibinfo {pages} {1317} (\bibinfo {year} {1982})}\BibitemShut {NoStop}%
\bibitem [{\citenamefont {Derode}\ \emph {et~al.}(2006)\citenamefont {Derode},
  \citenamefont {Mamou},\ and\ \citenamefont {Tourin}}]{derodePRE2006}%
  \BibitemOpen
  \bibfield  {author} {\bibinfo {author} {\bibfnamefont {A.}~\bibnamefont
  {Derode}}, \bibinfo {author} {\bibfnamefont {V.}~\bibnamefont {Mamou}}, \
  and\ \bibinfo {author} {\bibfnamefont {A.}~\bibnamefont {Tourin}},\ }\href
  {\doibase 10.1103/PhysRevE.74.036606} {\bibfield  {journal} {\bibinfo
  {journal} {Phys. Rev. E}\ }\textbf {\bibinfo {volume} {74}},\ \bibinfo
  {pages} {036606} (\bibinfo {year} {2006})}\BibitemShut {NoStop}%
\bibitem [{\citenamefont {Chanal}\ \emph {et~al.}(2006)\citenamefont {Chanal},
  \citenamefont {Segaud}, \citenamefont {Borderies},\ and\ \citenamefont
  {Saillard}}]{chanalJOSAA2006}%
  \BibitemOpen
  \bibfield  {author} {\bibinfo {author} {\bibfnamefont {H.}~\bibnamefont
  {Chanal}}, \bibinfo {author} {\bibfnamefont {J.~P.}\ \bibnamefont {Segaud}},
  \bibinfo {author} {\bibfnamefont {P.}~\bibnamefont {Borderies}}, \ and\
  \bibinfo {author} {\bibfnamefont {M.}~\bibnamefont {Saillard}},\ }\href
  {\doibase 10.1364/JOSAA.23.000370} {\bibfield  {journal} {\bibinfo  {journal}
  {J. Opt. Soc. Am. A}\ }\textbf {\bibinfo {volume} {23}},\ \bibinfo {pages}
  {370} (\bibinfo {year} {2006})}\BibitemShut {NoStop}%
\bibitem [{\citenamefont {Bringi}\ \emph {et~al.}(1982)\citenamefont {Bringi},
  \citenamefont {Varadan},\ and\ \citenamefont {Varadan}}]{bringi1982coherent}%
  \BibitemOpen
  \bibfield  {author} {\bibinfo {author} {\bibfnamefont {V.}~\bibnamefont
  {Bringi}}, \bibinfo {author} {\bibfnamefont {V.}~\bibnamefont {Varadan}}, \
  and\ \bibinfo {author} {\bibfnamefont {V.}~\bibnamefont {Varadan}},\ }\href
  {\doibase 10.1029/RS017i005p00946} {\bibfield  {journal} {\bibinfo  {journal}
  {Radio Science}\ }\textbf {\bibinfo {volume} {17}},\ \bibinfo {pages} {946}
  (\bibinfo {year} {1982})}\BibitemShut {NoStop}%
\bibitem [{\citenamefont {Ma}\ \emph {et~al.}(1988)\citenamefont {Ma},
  \citenamefont {Varadan},\ and\ \citenamefont {Varadan}}]{maAO1988}%
  \BibitemOpen
  \bibfield  {author} {\bibinfo {author} {\bibfnamefont {Y.}~\bibnamefont
  {Ma}}, \bibinfo {author} {\bibfnamefont {V.~V.}\ \bibnamefont {Varadan}}, \
  and\ \bibinfo {author} {\bibfnamefont {V.~K.}\ \bibnamefont {Varadan}},\
  }\href {\doibase 10.1364/AO.27.002469} {\bibfield  {journal} {\bibinfo
  {journal} {Applied Optics}\ }\textbf {\bibinfo {volume} {27}},\ \bibinfo
  {pages} {2469} (\bibinfo {year} {1988})}\BibitemShut {NoStop}%
\bibitem [{\citenamefont {West}\ \emph {et~al.}(1994)\citenamefont {West},
  \citenamefont {Gibbs}, \citenamefont {Tsang},\ and\ \citenamefont
  {Fung}}]{westJOSAA1994}%
  \BibitemOpen
  \bibfield  {author} {\bibinfo {author} {\bibfnamefont {R.}~\bibnamefont
  {West}}, \bibinfo {author} {\bibfnamefont {D.}~\bibnamefont {Gibbs}},
  \bibinfo {author} {\bibfnamefont {L.}~\bibnamefont {Tsang}}, \ and\ \bibinfo
  {author} {\bibfnamefont {A.~K.}\ \bibnamefont {Fung}},\ }\href {\doibase
  10.1364/JOSAA.11.001854} {\bibfield  {journal} {\bibinfo  {journal} {J. Opt.
  Soc. Am. A}\ }\textbf {\bibinfo {volume} {11}},\ \bibinfo {pages} {1854}
  (\bibinfo {year} {1994})}\BibitemShut {NoStop}%
\bibitem [{\citenamefont {Nashashibi}\ and\ \citenamefont
  {Sarabandi}(1999)}]{Nashashibi1999}%
  \BibitemOpen
  \bibfield  {author} {\bibinfo {author} {\bibfnamefont {A.}~\bibnamefont
  {Nashashibi}}\ and\ \bibinfo {author} {\bibfnamefont {K.}~\bibnamefont
  {Sarabandi}},\ }\href {\doibase 10.1109/8.793326} {\bibfield  {journal}
  {\bibinfo  {journal} {IEEE Transactions on Antennas and Propagation}\
  }\textbf {\bibinfo {volume} {47}},\ \bibinfo {pages} {1454} (\bibinfo {year}
  {1999})}\BibitemShut {NoStop}%
\bibitem [{\citenamefont {Prasher}(2007)}]{prasherJAP2007}%
  \BibitemOpen
  \bibfield  {author} {\bibinfo {author} {\bibfnamefont {R.}~\bibnamefont
  {Prasher}},\ }\href {\doibase 10.1063/1.2794703} {\bibfield  {journal}
  {\bibinfo  {journal} {Journal of Applied Physics}\ }\textbf {\bibinfo
  {volume} {102}},\ \bibinfo {pages} {074316} (\bibinfo {year}
  {2007})}\BibitemShut {NoStop}%
\bibitem [{\citenamefont {{Vander Meulen}}\ \emph {et~al.}(2001)\citenamefont
  {{Vander Meulen}}, \citenamefont {Feuillard}, \citenamefont {Matar},
  \citenamefont {Levassort},\ and\ \citenamefont {Lethiecq}}]{meulenJASA2001}%
  \BibitemOpen
  \bibfield  {author} {\bibinfo {author} {\bibfnamefont {F.}~\bibnamefont
  {{Vander Meulen}}}, \bibinfo {author} {\bibfnamefont {G.}~\bibnamefont
  {Feuillard}}, \bibinfo {author} {\bibfnamefont {O.~B.}\ \bibnamefont
  {Matar}}, \bibinfo {author} {\bibfnamefont {F.}~\bibnamefont {Levassort}}, \
  and\ \bibinfo {author} {\bibfnamefont {M.}~\bibnamefont {Lethiecq}},\ }\href
  {\doibase 10.1121/1.1404435} {\bibfield  {journal} {\bibinfo  {journal}
  {Journal of the Acoustical Society of America}\ }\textbf {\bibinfo {volume}
  {110}},\ \bibinfo {pages} {2301} (\bibinfo {year} {2001})}\BibitemShut
  {NoStop}%
\bibitem [{\citenamefont {Gyorffy}(1970)}]{gyorffyPRB1970}%
  \BibitemOpen
  \bibfield  {author} {\bibinfo {author} {\bibfnamefont {B.~L.}\ \bibnamefont
  {Gyorffy}},\ }\href {\doibase 10.1103/PhysRevB.1.3290} {\bibfield  {journal}
  {\bibinfo  {journal} {Phys. Rev. B}\ }\textbf {\bibinfo {volume} {1}},\
  \bibinfo {pages} {3290} (\bibinfo {year} {1970})}\BibitemShut {NoStop}%
\bibitem [{\citenamefont {Kwong}\ \emph {et~al.}(2019)\citenamefont {Kwong},
  \citenamefont {Wilkowski}, \citenamefont {Delande},\ and\ \citenamefont
  {Pierrat}}]{kwongPRA2019}%
  \BibitemOpen
  \bibfield  {author} {\bibinfo {author} {\bibfnamefont {C.~C.}\ \bibnamefont
  {Kwong}}, \bibinfo {author} {\bibfnamefont {D.}~\bibnamefont {Wilkowski}},
  \bibinfo {author} {\bibfnamefont {D.}~\bibnamefont {Delande}}, \ and\
  \bibinfo {author} {\bibfnamefont {R.}~\bibnamefont {Pierrat}},\ }\href
  {\doibase 10.1103/PhysRevA.99.043806} {\bibfield  {journal} {\bibinfo
  {journal} {Phys. Rev. A}\ }\textbf {\bibinfo {volume} {99}},\ \bibinfo
  {pages} {043806} (\bibinfo {year} {2019})}\BibitemShut {NoStop}%
\bibitem [{\citenamefont {Wang}\ and\ \citenamefont
  {Zhao}(2018{\natexlab{d}})}]{wang2018role}%
  \BibitemOpen
  \bibfield  {author} {\bibinfo {author} {\bibfnamefont {B.~X.}\ \bibnamefont
  {Wang}}\ and\ \bibinfo {author} {\bibfnamefont {C.~Y.}\ \bibnamefont
  {Zhao}},\ }\href@noop {} {\bibfield  {journal} {\bibinfo  {journal} {arXiv
  preprint arXiv:1807.09953}\ } (\bibinfo {year}
  {2018}{\natexlab{d}})}\BibitemShut {NoStop}%
\bibitem [{\citenamefont {Soven}(1967)}]{sovenPR1967}%
  \BibitemOpen
  \bibfield  {author} {\bibinfo {author} {\bibfnamefont {P.}~\bibnamefont
  {Soven}},\ }\href {\doibase 10.1103/PhysRev.156.809} {\bibfield  {journal}
  {\bibinfo  {journal} {Phys. Rev.}\ }\textbf {\bibinfo {volume} {156}},\
  \bibinfo {pages} {809} (\bibinfo {year} {1967})}\BibitemShut {NoStop}%
\bibitem [{\citenamefont {Roth}(1972)}]{rothPRL1972}%
  \BibitemOpen
  \bibfield  {author} {\bibinfo {author} {\bibfnamefont {L.~M.}\ \bibnamefont
  {Roth}},\ }\href {\doibase 10.1103/PhysRevLett.28.1570} {\bibfield  {journal}
  {\bibinfo  {journal} {Phys. Rev. Lett.}\ }\textbf {\bibinfo {volume} {28}},\
  \bibinfo {pages} {1570} (\bibinfo {year} {1972})}\BibitemShut {NoStop}%
\bibitem [{\citenamefont {Slovick}\ \emph {et~al.}(2014)\citenamefont
  {Slovick}, \citenamefont {Yu},\ and\ \citenamefont
  {Krishnamurthy}}]{slovickPRB2014}%
  \BibitemOpen
  \bibfield  {author} {\bibinfo {author} {\bibfnamefont {B.~A.}\ \bibnamefont
  {Slovick}}, \bibinfo {author} {\bibfnamefont {Z.~G.}\ \bibnamefont {Yu}}, \
  and\ \bibinfo {author} {\bibfnamefont {S.}~\bibnamefont {Krishnamurthy}},\
  }\href {\doibase 10.1103/PhysRevB.89.155118} {\bibfield  {journal} {\bibinfo
  {journal} {Phys. Rev. B}\ }\textbf {\bibinfo {volume} {89}},\ \bibinfo
  {pages} {155118} (\bibinfo {year} {2014})}\BibitemShut {NoStop}%
\bibitem [{\citenamefont {Slovick}(2017)}]{slovickPRB2017}%
  \BibitemOpen
  \bibfield  {author} {\bibinfo {author} {\bibfnamefont {B.~A.}\ \bibnamefont
  {Slovick}},\ }\href {\doibase 10.1103/PhysRevB.95.094202} {\bibfield
  {journal} {\bibinfo  {journal} {Phys. Rev. B}\ }\textbf {\bibinfo {volume}
  {95}},\ \bibinfo {pages} {094202} (\bibinfo {year} {2017})}\BibitemShut
  {NoStop}%
\bibitem [{\citenamefont {Huang}\ \emph {et~al.}(2018)\citenamefont {Huang},
  \citenamefont {Wang},\ and\ \citenamefont {Zhao}}]{huangJQSRT2018}%
  \BibitemOpen
  \bibfield  {author} {\bibinfo {author} {\bibfnamefont {T.}~\bibnamefont
  {Huang}}, \bibinfo {author} {\bibfnamefont {B.}~\bibnamefont {Wang}}, \ and\
  \bibinfo {author} {\bibfnamefont {C.}~\bibnamefont {Zhao}},\ }\href {\doibase
  https://doi.org/10.1016/j.jqsrt.2018.04.030} {\bibfield  {journal} {\bibinfo
  {journal} {Journal of Quantitative Spectroscopy and Radiative Transfer}\
  }\textbf {\bibinfo {volume} {214}},\ \bibinfo {pages} {82 } (\bibinfo {year}
  {2018})}\BibitemShut {NoStop}%
\bibitem [{\citenamefont {Soukoulis}\ \emph {et~al.}(1994)\citenamefont
  {Soukoulis}, \citenamefont {Datta},\ and\ \citenamefont
  {Economou}}]{soukoulisPRB1994}%
  \BibitemOpen
  \bibfield  {author} {\bibinfo {author} {\bibfnamefont {C.~M.}\ \bibnamefont
  {Soukoulis}}, \bibinfo {author} {\bibfnamefont {S.}~\bibnamefont {Datta}}, \
  and\ \bibinfo {author} {\bibfnamefont {E.~N.}\ \bibnamefont {Economou}},\
  }\href {\doibase 10.1103/PhysRevB.49.3800} {\bibfield  {journal} {\bibinfo
  {journal} {Phys. Rev. B}\ }\textbf {\bibinfo {volume} {49}},\ \bibinfo
  {pages} {3800} (\bibinfo {year} {1994})}\BibitemShut {NoStop}%
\bibitem [{\citenamefont {Busch}\ and\ \citenamefont
  {Soukoulis}(1995)}]{buschPRL1995}%
  \BibitemOpen
  \bibfield  {author} {\bibinfo {author} {\bibfnamefont {K.}~\bibnamefont
  {Busch}}\ and\ \bibinfo {author} {\bibfnamefont {C.~M.}\ \bibnamefont
  {Soukoulis}},\ }\href {\doibase 10.1103/PhysRevLett.75.3442} {\bibfield
  {journal} {\bibinfo  {journal} {Phys. Rev. Lett.}\ }\textbf {\bibinfo
  {volume} {75}},\ \bibinfo {pages} {3442} (\bibinfo {year}
  {1995})}\BibitemShut {NoStop}%
\bibitem [{\citenamefont {Busch}\ and\ \citenamefont
  {Soukoulis}(1996)}]{buschPRB1996}%
  \BibitemOpen
  \bibfield  {author} {\bibinfo {author} {\bibfnamefont {K.}~\bibnamefont
  {Busch}}\ and\ \bibinfo {author} {\bibfnamefont {C.~M.}\ \bibnamefont
  {Soukoulis}},\ }\href {\doibase 10.1103/PhysRevB.54.893} {\bibfield
  {journal} {\bibinfo  {journal} {Phys. Rev. B}\ }\textbf {\bibinfo {volume}
  {54}},\ \bibinfo {pages} {893} (\bibinfo {year} {1996})}\BibitemShut
  {NoStop}%
\bibitem [{\citenamefont {Schertel}\ \emph
  {et~al.}(2019{\natexlab{a}})\citenamefont {Schertel}, \citenamefont {Wimmer},
  \citenamefont {Besirske}, \citenamefont {Aegerter}, \citenamefont {Maret},
  \citenamefont {Polarz},\ and\ \citenamefont {Aubry}}]{schertelPRMat2019}%
  \BibitemOpen
  \bibfield  {author} {\bibinfo {author} {\bibfnamefont {L.}~\bibnamefont
  {Schertel}}, \bibinfo {author} {\bibfnamefont {I.}~\bibnamefont {Wimmer}},
  \bibinfo {author} {\bibfnamefont {P.}~\bibnamefont {Besirske}}, \bibinfo
  {author} {\bibfnamefont {C.~M.}\ \bibnamefont {Aegerter}}, \bibinfo {author}
  {\bibfnamefont {G.}~\bibnamefont {Maret}}, \bibinfo {author} {\bibfnamefont
  {S.}~\bibnamefont {Polarz}}, \ and\ \bibinfo {author} {\bibfnamefont {G.~J.}\
  \bibnamefont {Aubry}},\ }\href {\doibase 10.1103/PhysRevMaterials.3.015203}
  {\bibfield  {journal} {\bibinfo  {journal} {Phys. Rev. Materials}\ }\textbf
  {\bibinfo {volume} {3}},\ \bibinfo {pages} {015203} (\bibinfo {year}
  {2019}{\natexlab{a}})}\BibitemShut {NoStop}%
\bibitem [{\citenamefont {Schertel}\ \emph
  {et~al.}(2019{\natexlab{b}})\citenamefont {Schertel}, \citenamefont
  {Siedentop}, \citenamefont {Meijer}, \citenamefont {Keim}, \citenamefont
  {Aegerter}, \citenamefont {Aubry},\ and\ \citenamefont
  {Maret}}]{schertelADOM2019}%
  \BibitemOpen
  \bibfield  {author} {\bibinfo {author} {\bibfnamefont {L.}~\bibnamefont
  {Schertel}}, \bibinfo {author} {\bibfnamefont {L.}~\bibnamefont {Siedentop}},
  \bibinfo {author} {\bibfnamefont {J.-M.}\ \bibnamefont {Meijer}}, \bibinfo
  {author} {\bibfnamefont {P.}~\bibnamefont {Keim}}, \bibinfo {author}
  {\bibfnamefont {C.~M.}\ \bibnamefont {Aegerter}}, \bibinfo {author}
  {\bibfnamefont {G.~J.}\ \bibnamefont {Aubry}}, \ and\ \bibinfo {author}
  {\bibfnamefont {G.}~\bibnamefont {Maret}},\ }\href {\doibase
  10.1002/adom.201900442} {\bibfield  {journal} {\bibinfo  {journal} {Advanced
  Optical Materials}\ }\textbf {\bibinfo {volume} {7}},\ \bibinfo {pages}
  {1900442} (\bibinfo {year} {2019}{\natexlab{b}})},\ \Eprint
  {http://arxiv.org/abs/https://onlinelibrary.wiley.com/doi/pdf/10.1002/adom.201900442}
  {https://onlinelibrary.wiley.com/doi/pdf/10.1002/adom.201900442} \BibitemShut
  {NoStop}%
\bibitem [{\citenamefont {Peng}\ and\ \citenamefont
  {Dinsmore}(2007)}]{Peng2007}%
  \BibitemOpen
  \bibfield  {author} {\bibinfo {author} {\bibfnamefont {X.~T.}\ \bibnamefont
  {Peng}}\ and\ \bibinfo {author} {\bibfnamefont {A.~D.}\ \bibnamefont
  {Dinsmore}},\ }\href {\doibase 10.1103/PhysRevLett.99.143902} {\bibfield
  {journal} {\bibinfo  {journal} {Phys. Rev. Lett.}\ }\textbf {\bibinfo
  {volume} {99}},\ \bibinfo {pages} {143902} (\bibinfo {year}
  {2007})}\BibitemShut {NoStop}%
\bibitem [{\citenamefont {Bekshaev}\ \emph {et~al.}(2013)\citenamefont
  {Bekshaev}, \citenamefont {Bliokh},\ and\ \citenamefont
  {Nori}}]{bekshaevOE2013}%
  \BibitemOpen
  \bibfield  {author} {\bibinfo {author} {\bibfnamefont {A.~Y.}\ \bibnamefont
  {Bekshaev}}, \bibinfo {author} {\bibfnamefont {K.~Y.}\ \bibnamefont
  {Bliokh}}, \ and\ \bibinfo {author} {\bibfnamefont {F.}~\bibnamefont
  {Nori}},\ }\href {\doibase 10.1364/OE.21.007082} {\bibfield  {journal}
  {\bibinfo  {journal} {Opt. Express}\ }\textbf {\bibinfo {volume} {21}},\
  \bibinfo {pages} {7082} (\bibinfo {year} {2013})}\BibitemShut {NoStop}%
\bibitem [{\citenamefont {Auger}\ \emph {et~al.}(2009)\citenamefont {Auger},
  \citenamefont {Martinez},\ and\ \citenamefont {Stout}}]{augerJCTR2009}%
  \BibitemOpen
  \bibfield  {author} {\bibinfo {author} {\bibfnamefont {J.-C.}\ \bibnamefont
  {Auger}}, \bibinfo {author} {\bibfnamefont {V.~A.}\ \bibnamefont {Martinez}},
  \ and\ \bibinfo {author} {\bibfnamefont {B.}~\bibnamefont {Stout}},\ }\href
  {\doibase 10.1007/s11998-008-9116-6} {\bibfield  {journal} {\bibinfo
  {journal} {Journal of Coatings Technology and Research}\ }\textbf {\bibinfo
  {volume} {6}},\ \bibinfo {pages} {89} (\bibinfo {year} {2009})}\BibitemShut
  {NoStop}%
\bibitem [{\citenamefont {Lallich}\ \emph {et~al.}(2009)\citenamefont
  {Lallich}, \citenamefont {Enguehard},\ and\ \citenamefont
  {Baillis}}]{lallichJHT2009}%
  \BibitemOpen
  \bibfield  {author} {\bibinfo {author} {\bibfnamefont {S.}~\bibnamefont
  {Lallich}}, \bibinfo {author} {\bibfnamefont {F.}~\bibnamefont {Enguehard}},
  \ and\ \bibinfo {author} {\bibfnamefont {D.}~\bibnamefont {Baillis}},\ }\href
  {\doibase 10.1115/1.3109999} {\bibfield  {journal} {\bibinfo  {journal}
  {Journal of Heat Transfer}\ }\textbf {\bibinfo {volume} {131}},\ \bibinfo
  {pages} {082701} (\bibinfo {year} {2009})}\BibitemShut {NoStop}%
\bibitem [{\citenamefont {Qiu}\ \emph {et~al.}(2015)\citenamefont {Qiu},
  \citenamefont {Liu},\ and\ \citenamefont {f.~Hsu}}]{qiuJQSRT2015}%
  \BibitemOpen
  \bibfield  {author} {\bibinfo {author} {\bibfnamefont {J.}~\bibnamefont
  {Qiu}}, \bibinfo {author} {\bibfnamefont {L.}~\bibnamefont {Liu}}, \ and\
  \bibinfo {author} {\bibfnamefont {P.}~\bibnamefont {f.~Hsu}},\ }\href
  {\doibase https://doi.org/10.1016/j.jqsrt.2015.01.014} {\bibfield  {journal}
  {\bibinfo  {journal} {Journal of Quantitative Spectroscopy and Radiative
  Transfer}\ }\textbf {\bibinfo {volume} {158}},\ \bibinfo {pages} {101 }
  (\bibinfo {year} {2015})},\ \bibinfo {note} {special Issue on the Second
  International Workshop on Micro-Nano Thermal Radiation}\BibitemShut {NoStop}%
\bibitem [{\citenamefont {Dyachenko}\ \emph {et~al.}(2014)\citenamefont
  {Dyachenko}, \citenamefont {do~Rosário}, \citenamefont {Leib}, \citenamefont
  {Petrov}, \citenamefont {Kubrin}, \citenamefont {Schneider}, \citenamefont
  {Weller}, \citenamefont {Vossmeyer},\ and\ \citenamefont
  {Eich}}]{dyachenkoACSPhoton2014}%
  \BibitemOpen
  \bibfield  {author} {\bibinfo {author} {\bibfnamefont {P.~N.}\ \bibnamefont
  {Dyachenko}}, \bibinfo {author} {\bibfnamefont {J.~J.}\ \bibnamefont
  {do~Rosário}}, \bibinfo {author} {\bibfnamefont {E.~W.}\ \bibnamefont
  {Leib}}, \bibinfo {author} {\bibfnamefont {A.~Y.}\ \bibnamefont {Petrov}},
  \bibinfo {author} {\bibfnamefont {R.}~\bibnamefont {Kubrin}}, \bibinfo
  {author} {\bibfnamefont {G.~A.}\ \bibnamefont {Schneider}}, \bibinfo {author}
  {\bibfnamefont {H.}~\bibnamefont {Weller}}, \bibinfo {author} {\bibfnamefont
  {T.}~\bibnamefont {Vossmeyer}}, \ and\ \bibinfo {author} {\bibfnamefont
  {M.}~\bibnamefont {Eich}},\ }\href {\doibase 10.1021/ph500224r} {\bibfield
  {journal} {\bibinfo  {journal} {ACS Photonics}\ }\textbf {\bibinfo {volume}
  {1}},\ \bibinfo {pages} {1127} (\bibinfo {year} {2014})},\ \Eprint
  {http://arxiv.org/abs/https://doi.org/10.1021/ph500224r}
  {https://doi.org/10.1021/ph500224r} \BibitemShut {NoStop}%
\bibitem [{\citenamefont {Chen}\ \emph
  {et~al.}(2018{\natexlab{a}})\citenamefont {Chen}, \citenamefont {Zhao},\ and\
  \citenamefont {Wang}}]{chenJQSRT2018}%
  \BibitemOpen
  \bibfield  {author} {\bibinfo {author} {\bibfnamefont {X.~W.}\ \bibnamefont
  {Chen}}, \bibinfo {author} {\bibfnamefont {C.~Y.}\ \bibnamefont {Zhao}}, \
  and\ \bibinfo {author} {\bibfnamefont {B.~X.}\ \bibnamefont {Wang}},\ }\href
  {\doibase https://doi.org/10.1016/j.jqsrt.2018.02.009} {\bibfield  {journal}
  {\bibinfo  {journal} {Journal of Quantitative Spectroscopy and Radiative
  Transfer}\ }\textbf {\bibinfo {volume} {210}},\ \bibinfo {pages} {116 }
  (\bibinfo {year} {2018}{\natexlab{a}})}\BibitemShut {NoStop}%
\bibitem [{\citenamefont {Yu}\ \emph {et~al.}(2014)\citenamefont {Yu},
  \citenamefont {Liu}, \citenamefont {Duan},\ and\ \citenamefont
  {Wang}}]{yuOE2014}%
  \BibitemOpen
  \bibfield  {author} {\bibinfo {author} {\bibfnamefont {H.}~\bibnamefont
  {Yu}}, \bibinfo {author} {\bibfnamefont {D.}~\bibnamefont {Liu}}, \bibinfo
  {author} {\bibfnamefont {Y.}~\bibnamefont {Duan}}, \ and\ \bibinfo {author}
  {\bibfnamefont {X.}~\bibnamefont {Wang}},\ }\href {\doibase
  10.1364/OE.22.007925} {\bibfield  {journal} {\bibinfo  {journal} {Opt.
  Express}\ }\textbf {\bibinfo {volume} {22}},\ \bibinfo {pages} {7925}
  (\bibinfo {year} {2014})}\BibitemShut {NoStop}%
\bibitem [{\citenamefont {Auger}\ and\ \citenamefont
  {McLoughlin}(2015)}]{augerJCTR2015}%
  \BibitemOpen
  \bibfield  {author} {\bibinfo {author} {\bibfnamefont {J.-C.}\ \bibnamefont
  {Auger}}\ and\ \bibinfo {author} {\bibfnamefont {D.}~\bibnamefont
  {McLoughlin}},\ }\href {\doibase 10.1007/s11998-015-9677-0} {\bibfield
  {journal} {\bibinfo  {journal} {Journal of Coatings Technology and Research}\
  }\textbf {\bibinfo {volume} {12}},\ \bibinfo {pages} {693} (\bibinfo {year}
  {2015})}\BibitemShut {NoStop}%
\bibitem [{\citenamefont {Tsang}\ \emph {et~al.}(1992)\citenamefont {Tsang},
  \citenamefont {Mandt},\ and\ \citenamefont {Ding}}]{tsangOL1992}%
  \BibitemOpen
  \bibfield  {author} {\bibinfo {author} {\bibfnamefont {L.}~\bibnamefont
  {Tsang}}, \bibinfo {author} {\bibfnamefont {C.~E.}\ \bibnamefont {Mandt}}, \
  and\ \bibinfo {author} {\bibfnamefont {K.~H.}\ \bibnamefont {Ding}},\ }\href
  {\doibase 10.1364/OL.17.000314} {\bibfield  {journal} {\bibinfo  {journal}
  {Opt. Lett.}\ }\textbf {\bibinfo {volume} {17}},\ \bibinfo {pages} {314}
  (\bibinfo {year} {1992})}\BibitemShut {NoStop}%
\bibitem [{\citenamefont {Zurk}\ \emph {et~al.}(1995)\citenamefont {Zurk},
  \citenamefont {Tsang}, \citenamefont {Ding},\ and\ \citenamefont
  {Winebrenner}}]{zurkJOSAA1995}%
  \BibitemOpen
  \bibfield  {author} {\bibinfo {author} {\bibfnamefont {L.~M.}\ \bibnamefont
  {Zurk}}, \bibinfo {author} {\bibfnamefont {L.}~\bibnamefont {Tsang}},
  \bibinfo {author} {\bibfnamefont {K.~H.}\ \bibnamefont {Ding}}, \ and\
  \bibinfo {author} {\bibfnamefont {D.~P.}\ \bibnamefont {Winebrenner}},\
  }\href {\doibase 10.1364/JOSAA.12.001772} {\bibfield  {journal} {\bibinfo
  {journal} {J. Opt. Soc. Am. A}\ }\textbf {\bibinfo {volume} {12}},\ \bibinfo
  {pages} {1772} (\bibinfo {year} {1995})}\BibitemShut {NoStop}%
\bibitem [{\citenamefont {Wang}\ and\ \citenamefont
  {Zhao}(2018{\natexlab{e}})}]{wangJQSRT2018}%
  \BibitemOpen
  \bibfield  {author} {\bibinfo {author} {\bibfnamefont {B.~X.}\ \bibnamefont
  {Wang}}\ and\ \bibinfo {author} {\bibfnamefont {C.~Y.}\ \bibnamefont
  {Zhao}},\ }\href {\doibase https://doi.org/10.1016/j.jqsrt.2018.07.010}
  {\bibfield  {journal} {\bibinfo  {journal} {Journal of Quantitative
  Spectroscopy and Radiative Transfer}\ }\textbf {\bibinfo {volume} {218}},\
  \bibinfo {pages} {72 } (\bibinfo {year} {2018}{\natexlab{e}})}\BibitemShut
  {NoStop}%
\bibitem [{\citenamefont {Sapienza}\ \emph {et~al.}(2007)\citenamefont
  {Sapienza}, \citenamefont {Garc\'{\i}a}, \citenamefont {Bertolotti},
  \citenamefont {Mart\'{\i}n}, \citenamefont {Blanco}, \citenamefont {Vi\~na},
  \citenamefont {L\'opez},\ and\ \citenamefont {Wiersma}}]{sapienzaPRL2007}%
  \BibitemOpen
  \bibfield  {author} {\bibinfo {author} {\bibfnamefont {R.}~\bibnamefont
  {Sapienza}}, \bibinfo {author} {\bibfnamefont {P.~D.}\ \bibnamefont
  {Garc\'{\i}a}}, \bibinfo {author} {\bibfnamefont {J.}~\bibnamefont
  {Bertolotti}}, \bibinfo {author} {\bibfnamefont {M.~D.}\ \bibnamefont
  {Mart\'{\i}n}}, \bibinfo {author} {\bibfnamefont {A.}~\bibnamefont {Blanco}},
  \bibinfo {author} {\bibfnamefont {L.}~\bibnamefont {Vi\~na}}, \bibinfo
  {author} {\bibfnamefont {C.}~\bibnamefont {L\'opez}}, \ and\ \bibinfo
  {author} {\bibfnamefont {D.~S.}\ \bibnamefont {Wiersma}},\ }\href {\doibase
  10.1103/PhysRevLett.99.233902} {\bibfield  {journal} {\bibinfo  {journal}
  {Phys. Rev. Lett.}\ }\textbf {\bibinfo {volume} {99}},\ \bibinfo {pages}
  {233902} (\bibinfo {year} {2007})}\BibitemShut {NoStop}%
\bibitem [{\citenamefont {Pattelli}\ \emph {et~al.}(2018)\citenamefont
  {Pattelli}, \citenamefont {Egel}, \citenamefont {Lemmer},\ and\ \citenamefont
  {Wiersma}}]{pattelliOptica2018}%
  \BibitemOpen
  \bibfield  {author} {\bibinfo {author} {\bibfnamefont {L.}~\bibnamefont
  {Pattelli}}, \bibinfo {author} {\bibfnamefont {A.}~\bibnamefont {Egel}},
  \bibinfo {author} {\bibfnamefont {U.}~\bibnamefont {Lemmer}}, \ and\ \bibinfo
  {author} {\bibfnamefont {D.~S.}\ \bibnamefont {Wiersma}},\ }\href {\doibase
  10.1364/OPTICA.5.001037} {\bibfield  {journal} {\bibinfo  {journal} {Optica}\
  }\textbf {\bibinfo {volume} {5}},\ \bibinfo {pages} {1037} (\bibinfo {year}
  {2018})}\BibitemShut {NoStop}%
\bibitem [{\citenamefont {Varadan}\ \emph {et~al.}(1983)\citenamefont
  {Varadan}, \citenamefont {Bringi}, \citenamefont {Varadan},\ and\
  \citenamefont {Ishimaru}}]{varadanRS1983}%
  \BibitemOpen
  \bibfield  {author} {\bibinfo {author} {\bibfnamefont {V.~K.}\ \bibnamefont
  {Varadan}}, \bibinfo {author} {\bibfnamefont {V.~N.}\ \bibnamefont {Bringi}},
  \bibinfo {author} {\bibfnamefont {V.~V.}\ \bibnamefont {Varadan}}, \ and\
  \bibinfo {author} {\bibfnamefont {A.}~\bibnamefont {Ishimaru}},\ }\href
  {\doibase 10.1029/RS018i003p00321} {\bibfield  {journal} {\bibinfo  {journal}
  {Radio Science}\ }\textbf {\bibinfo {volume} {18}},\ \bibinfo {pages} {321}
  (\bibinfo {year} {1983})},\ \Eprint
  {http://arxiv.org/abs/https://agupubs.onlinelibrary.wiley.com/doi/pdf/10.1029/RS018i003p00321}
  {https://agupubs.onlinelibrary.wiley.com/doi/pdf/10.1029/RS018i003p00321}
  \BibitemShut {NoStop}%
\bibitem [{\citenamefont {Green}\ and\ \citenamefont
  {Lumme}(2001)}]{greenAO2001}%
  \BibitemOpen
  \bibfield  {author} {\bibinfo {author} {\bibfnamefont {K.}~\bibnamefont
  {Green}}\ and\ \bibinfo {author} {\bibfnamefont {K.}~\bibnamefont {Lumme}},\
  }\href {\doibase 10.1364/AO.40.003711} {\bibfield  {journal} {\bibinfo
  {journal} {Appl. Opt.}\ }\textbf {\bibinfo {volume} {40}},\ \bibinfo {pages}
  {3711} (\bibinfo {year} {2001})}\BibitemShut {NoStop}%
\bibitem [{\citenamefont {Roux}\ \emph {et~al.}(2001)\citenamefont {Roux},
  \citenamefont {Mareschal}, \citenamefont {Vukadinovic}, \citenamefont
  {Thibaud},\ and\ \citenamefont {Greffet}}]{rouxJOSAA2001}%
  \BibitemOpen
  \bibfield  {author} {\bibinfo {author} {\bibfnamefont {L.}~\bibnamefont
  {Roux}}, \bibinfo {author} {\bibfnamefont {P.}~\bibnamefont {Mareschal}},
  \bibinfo {author} {\bibfnamefont {N.}~\bibnamefont {Vukadinovic}}, \bibinfo
  {author} {\bibfnamefont {J.-B.}\ \bibnamefont {Thibaud}}, \ and\ \bibinfo
  {author} {\bibfnamefont {J.-J.}\ \bibnamefont {Greffet}},\ }\href {\doibase
  10.1364/JOSAA.18.000374} {\bibfield  {journal} {\bibinfo  {journal} {J. Opt.
  Soc. Am. A}\ }\textbf {\bibinfo {volume} {18}},\ \bibinfo {pages} {374}
  (\bibinfo {year} {2001})}\BibitemShut {NoStop}%
\bibitem [{\citenamefont {Sch\"{a}fer}\ and\ \citenamefont
  {Kienle}(2008)}]{schaferOL2008}%
  \BibitemOpen
  \bibfield  {author} {\bibinfo {author} {\bibfnamefont {J.}~\bibnamefont
  {Sch\"{a}fer}}\ and\ \bibinfo {author} {\bibfnamefont {A.}~\bibnamefont
  {Kienle}},\ }\href {\doibase 10.1364/OL.33.002413} {\bibfield  {journal}
  {\bibinfo  {journal} {Opt. Lett.}\ }\textbf {\bibinfo {volume} {33}},\
  \bibinfo {pages} {2413} (\bibinfo {year} {2008})}\BibitemShut {NoStop}%
\bibitem [{\citenamefont {Voit}\ \emph {et~al.}(2009)\citenamefont {Voit},
  \citenamefont {Sch\"{a}fer},\ and\ \citenamefont {Kienle}}]{voitOL2009}%
  \BibitemOpen
  \bibfield  {author} {\bibinfo {author} {\bibfnamefont {F.}~\bibnamefont
  {Voit}}, \bibinfo {author} {\bibfnamefont {J.}~\bibnamefont {Sch\"{a}fer}}, \
  and\ \bibinfo {author} {\bibfnamefont {A.}~\bibnamefont {Kienle}},\ }\href
  {\doibase 10.1364/OL.34.002593} {\bibfield  {journal} {\bibinfo  {journal}
  {Opt. Lett.}\ }\textbf {\bibinfo {volume} {34}},\ \bibinfo {pages} {2593}
  (\bibinfo {year} {2009})}\BibitemShut {NoStop}%
\bibitem [{\citenamefont {Nalitov}\ \emph {et~al.}(2015)\citenamefont
  {Nalitov}, \citenamefont {Solnyshkov},\ and\ \citenamefont
  {Malpuech}}]{nalitovPRL2015}%
  \BibitemOpen
  \bibfield  {author} {\bibinfo {author} {\bibfnamefont {A.~V.}\ \bibnamefont
  {Nalitov}}, \bibinfo {author} {\bibfnamefont {D.~D.}\ \bibnamefont
  {Solnyshkov}}, \ and\ \bibinfo {author} {\bibfnamefont {G.}~\bibnamefont
  {Malpuech}},\ }\href {\doibase 10.1103/PhysRevLett.114.116401} {\bibfield
  {journal} {\bibinfo  {journal} {Phys. Rev. Lett.}\ }\textbf {\bibinfo
  {volume} {114}},\ \bibinfo {pages} {116401} (\bibinfo {year}
  {2015})}\BibitemShut {NoStop}%
\bibitem [{\citenamefont {Ma}\ \emph {et~al.}(1990)\citenamefont {Ma},
  \citenamefont {Varadan},\ and\ \citenamefont {Varadan}}]{ma1990enhanced}%
  \BibitemOpen
  \bibfield  {author} {\bibinfo {author} {\bibfnamefont {Y.}~\bibnamefont
  {Ma}}, \bibinfo {author} {\bibfnamefont {V.}~\bibnamefont {Varadan}}, \ and\
  \bibinfo {author} {\bibfnamefont {V.}~\bibnamefont {Varadan}},\ }\href
  {\doibase 10.1115/1.2910391} {\bibfield  {journal} {\bibinfo  {journal}
  {Journal of heat transfer}\ }\textbf {\bibinfo {volume} {112}},\ \bibinfo
  {pages} {402} (\bibinfo {year} {1990})}\BibitemShut {NoStop}%
\bibitem [{\citenamefont {Wei}\ \emph {et~al.}(2012)\citenamefont {Wei},
  \citenamefont {Fedorov}, \citenamefont {Luo},\ and\ \citenamefont
  {Ni}}]{weiAO2012}%
  \BibitemOpen
  \bibfield  {author} {\bibinfo {author} {\bibfnamefont {W.}~\bibnamefont
  {Wei}}, \bibinfo {author} {\bibfnamefont {A.~G.}\ \bibnamefont {Fedorov}},
  \bibinfo {author} {\bibfnamefont {Z.}~\bibnamefont {Luo}}, \ and\ \bibinfo
  {author} {\bibfnamefont {M.}~\bibnamefont {Ni}},\ }\href {\doibase
  10.1364/AO.51.006159} {\bibfield  {journal} {\bibinfo  {journal} {Appl.
  Opt.}\ }\textbf {\bibinfo {volume} {51}},\ \bibinfo {pages} {6159} (\bibinfo
  {year} {2012})}\BibitemShut {NoStop}%
\bibitem [{\citenamefont {Taylor}\ \emph {et~al.}(2013)\citenamefont {Taylor},
  \citenamefont {Coulombe}, \citenamefont {Otanicar}, \citenamefont {Phelan},
  \citenamefont {Gunawan}, \citenamefont {Lv}, \citenamefont {Rosengarten},
  \citenamefont {Prasher},\ and\ \citenamefont {Tyagi}}]{taylorJAP2013}%
  \BibitemOpen
  \bibfield  {author} {\bibinfo {author} {\bibfnamefont {R.}~\bibnamefont
  {Taylor}}, \bibinfo {author} {\bibfnamefont {S.}~\bibnamefont {Coulombe}},
  \bibinfo {author} {\bibfnamefont {T.}~\bibnamefont {Otanicar}}, \bibinfo
  {author} {\bibfnamefont {P.}~\bibnamefont {Phelan}}, \bibinfo {author}
  {\bibfnamefont {A.}~\bibnamefont {Gunawan}}, \bibinfo {author} {\bibfnamefont
  {W.}~\bibnamefont {Lv}}, \bibinfo {author} {\bibfnamefont {G.}~\bibnamefont
  {Rosengarten}}, \bibinfo {author} {\bibfnamefont {R.}~\bibnamefont
  {Prasher}}, \ and\ \bibinfo {author} {\bibfnamefont {H.}~\bibnamefont
  {Tyagi}},\ }\href {\doibase 10.1063/1.4754271} {\bibfield  {journal}
  {\bibinfo  {journal} {Journal of Applied Physics}\ }\textbf {\bibinfo
  {volume} {113}},\ \bibinfo {pages} {011301} (\bibinfo {year}
  {2013})}\BibitemShut {NoStop}%
\bibitem [{\citenamefont {Said}\ \emph {et~al.}(2013)\citenamefont {Said},
  \citenamefont {Sajid}, \citenamefont {Saidur}, \citenamefont
  {Kamalisarvestani},\ and\ \citenamefont {Rahim}}]{saidICHMT2013}%
  \BibitemOpen
  \bibfield  {author} {\bibinfo {author} {\bibfnamefont {Z.}~\bibnamefont
  {Said}}, \bibinfo {author} {\bibfnamefont {M.}~\bibnamefont {Sajid}},
  \bibinfo {author} {\bibfnamefont {R.}~\bibnamefont {Saidur}}, \bibinfo
  {author} {\bibfnamefont {M.}~\bibnamefont {Kamalisarvestani}}, \ and\
  \bibinfo {author} {\bibfnamefont {N.}~\bibnamefont {Rahim}},\ }\href
  {\doibase https://doi.org/10.1016/j.icheatmasstransfer.2013.05.013}
  {\bibfield  {journal} {\bibinfo  {journal} {International Communications in
  Heat and Mass Transfer}\ }\textbf {\bibinfo {volume} {46}},\ \bibinfo {pages}
  {74 } (\bibinfo {year} {2013})}\BibitemShut {NoStop}%
\bibitem [{\citenamefont {Xuan}\ \emph {et~al.}(2014)\citenamefont {Xuan},
  \citenamefont {Duan},\ and\ \citenamefont {Li}}]{xuanRSCA2014}%
  \BibitemOpen
  \bibfield  {author} {\bibinfo {author} {\bibfnamefont {Y.}~\bibnamefont
  {Xuan}}, \bibinfo {author} {\bibfnamefont {H.}~\bibnamefont {Duan}}, \ and\
  \bibinfo {author} {\bibfnamefont {Q.}~\bibnamefont {Li}},\ }\href {\doibase
  10.1039/C4RA00630E} {\bibfield  {journal} {\bibinfo  {journal} {RSC Adv.}\
  }\textbf {\bibinfo {volume} {4}},\ \bibinfo {pages} {16206} (\bibinfo {year}
  {2014})}\BibitemShut {NoStop}%
\bibitem [{\citenamefont {Liu}\ and\ \citenamefont
  {Xuan}(2017)}]{liuNanoscale2017}%
  \BibitemOpen
  \bibfield  {author} {\bibinfo {author} {\bibfnamefont {X.}~\bibnamefont
  {Liu}}\ and\ \bibinfo {author} {\bibfnamefont {Y.}~\bibnamefont {Xuan}},\
  }\href {\doibase 10.1039/C7NR03912C} {\bibfield  {journal} {\bibinfo
  {journal} {Nanoscale}\ }\textbf {\bibinfo {volume} {9}},\ \bibinfo {pages}
  {14854} (\bibinfo {year} {2017})}\BibitemShut {NoStop}%
\bibitem [{\citenamefont {Gao}\ \emph {et~al.}(2017)\citenamefont {Gao},
  \citenamefont {Zhao},\ and\ \citenamefont {Wang}}]{gaoJAP2017}%
  \BibitemOpen
  \bibfield  {author} {\bibinfo {author} {\bibfnamefont {J.~D.}\ \bibnamefont
  {Gao}}, \bibinfo {author} {\bibfnamefont {C.~Y.}\ \bibnamefont {Zhao}}, \
  and\ \bibinfo {author} {\bibfnamefont {B.~X.}\ \bibnamefont {Wang}},\ }\href
  {\doibase 10.1063/1.4978418} {\bibfield  {journal} {\bibinfo  {journal}
  {Journal of Applied Physics}\ }\textbf {\bibinfo {volume} {121}},\ \bibinfo
  {pages} {113105} (\bibinfo {year} {2017})}\BibitemShut {NoStop}%
\bibitem [{\citenamefont {Psaltis}\ \emph {et~al.}(2006)\citenamefont
  {Psaltis}, \citenamefont {Quake},\ and\ \citenamefont
  {Yang}}]{psaltisNature2006}%
  \BibitemOpen
  \bibfield  {author} {\bibinfo {author} {\bibfnamefont {D.}~\bibnamefont
  {Psaltis}}, \bibinfo {author} {\bibfnamefont {S.~R.}\ \bibnamefont {Quake}},
  \ and\ \bibinfo {author} {\bibfnamefont {C.}~\bibnamefont {Yang}},\ }\href
  {\doibase 10.1038/nature05060} {\bibfield  {journal} {\bibinfo  {journal}
  {Nature}\ }\textbf {\bibinfo {volume} {442}},\ \bibinfo {pages} {381}
  (\bibinfo {year} {2006})}\BibitemShut {NoStop}%
\bibitem [{\citenamefont {Erickson}\ \emph {et~al.}(2011)\citenamefont
  {Erickson}, \citenamefont {Sinton},\ and\ \citenamefont
  {Psaltis}}]{ericksonNPhoton2011}%
  \BibitemOpen
  \bibfield  {author} {\bibinfo {author} {\bibfnamefont {D.}~\bibnamefont
  {Erickson}}, \bibinfo {author} {\bibfnamefont {D.}~\bibnamefont {Sinton}}, \
  and\ \bibinfo {author} {\bibfnamefont {D.}~\bibnamefont {Psaltis}},\ }\href
  {\doibase 10.1038/nphoton.2011.209} {\bibfield  {journal} {\bibinfo
  {journal} {Nature Photonics}\ }\textbf {\bibinfo {volume} {5}},\ \bibinfo
  {pages} {583} (\bibinfo {year} {2011})}\BibitemShut {NoStop}%
\bibitem [{\citenamefont {{Sheremet}}\ \emph {et~al.}(2020)\citenamefont
  {{Sheremet}}, \citenamefont {{Pierrat}},\ and\ \citenamefont
  {{Carminati}}}]{sheremet2020absorption}%
  \BibitemOpen
  \bibfield  {author} {\bibinfo {author} {\bibfnamefont {A.}~\bibnamefont
  {{Sheremet}}}, \bibinfo {author} {\bibfnamefont {R.}~\bibnamefont
  {{Pierrat}}}, \ and\ \bibinfo {author} {\bibfnamefont {R.}~\bibnamefont
  {{Carminati}}},\ }\href@noop {} {\bibfield  {journal} {\bibinfo  {journal}
  {arXiv e-prints}\ ,\ \bibinfo {eid} {arXiv:2001.06227}} (\bibinfo {year}
  {2020})},\ \Eprint {http://arxiv.org/abs/2001.06227} {arXiv:2001.06227
  [physics.optics]} \BibitemShut {NoStop}%
\bibitem [{\citenamefont {Bigourdan}\ \emph {et~al.}(2019)\citenamefont
  {Bigourdan}, \citenamefont {Pierrat},\ and\ \citenamefont
  {Carminati}}]{bigourdanOE2019}%
  \BibitemOpen
  \bibfield  {author} {\bibinfo {author} {\bibfnamefont {F.}~\bibnamefont
  {Bigourdan}}, \bibinfo {author} {\bibfnamefont {R.}~\bibnamefont {Pierrat}},
  \ and\ \bibinfo {author} {\bibfnamefont {R.}~\bibnamefont {Carminati}},\
  }\href {\doibase 10.1364/OE.27.008666} {\bibfield  {journal} {\bibinfo
  {journal} {Opt. Express}\ }\textbf {\bibinfo {volume} {27}},\ \bibinfo
  {pages} {8666} (\bibinfo {year} {2019})}\BibitemShut {NoStop}%
\bibitem [{\citenamefont {Sun}\ \emph {et~al.}(2014)\citenamefont {Sun},
  \citenamefont {He}, \citenamefont {Zeng}, \citenamefont {Du}, \citenamefont
  {Guo}, \citenamefont {Liu}, \citenamefont {Wu}, \citenamefont {He},\ and\
  \citenamefont {Ma}}]{sunBOE2014}%
  \BibitemOpen
  \bibfield  {author} {\bibinfo {author} {\bibfnamefont {M.}~\bibnamefont
  {Sun}}, \bibinfo {author} {\bibfnamefont {H.}~\bibnamefont {He}}, \bibinfo
  {author} {\bibfnamefont {N.}~\bibnamefont {Zeng}}, \bibinfo {author}
  {\bibfnamefont {E.}~\bibnamefont {Du}}, \bibinfo {author} {\bibfnamefont
  {Y.}~\bibnamefont {Guo}}, \bibinfo {author} {\bibfnamefont {S.}~\bibnamefont
  {Liu}}, \bibinfo {author} {\bibfnamefont {J.}~\bibnamefont {Wu}}, \bibinfo
  {author} {\bibfnamefont {Y.}~\bibnamefont {He}}, \ and\ \bibinfo {author}
  {\bibfnamefont {H.}~\bibnamefont {Ma}},\ }\href {\doibase
  10.1364/BOE.5.004223} {\bibfield  {journal} {\bibinfo  {journal} {Biomed.
  Opt. Express}\ }\textbf {\bibinfo {volume} {5}},\ \bibinfo {pages} {4223}
  (\bibinfo {year} {2014})}\BibitemShut {NoStop}%
\bibitem [{\citenamefont {Bao}\ \emph {et~al.}(2017)\citenamefont {Bao},
  \citenamefont {Yan}, \citenamefont {Wang}, \citenamefont {Fang},
  \citenamefont {Zhao},\ and\ \citenamefont {Ruan}}]{baoSEMSC2017}%
  \BibitemOpen
  \bibfield  {author} {\bibinfo {author} {\bibfnamefont {H.}~\bibnamefont
  {Bao}}, \bibinfo {author} {\bibfnamefont {C.}~\bibnamefont {Yan}}, \bibinfo
  {author} {\bibfnamefont {B.~X.}\ \bibnamefont {Wang}}, \bibinfo {author}
  {\bibfnamefont {X.}~\bibnamefont {Fang}}, \bibinfo {author} {\bibfnamefont
  {C.~Y.}\ \bibnamefont {Zhao}}, \ and\ \bibinfo {author} {\bibfnamefont
  {X.~L.}\ \bibnamefont {Ruan}},\ }\href {\doibase
  https://doi.org/10.1016/j.solmat.2017.04.020} {\bibfield  {journal} {\bibinfo
   {journal} {Solar Energy Materials and Solar Cells}\ }\textbf {\bibinfo
  {volume} {168}},\ \bibinfo {pages} {78 } (\bibinfo {year}
  {2017})}\BibitemShut {NoStop}%
\bibitem [{\citenamefont {Baillis}\ \emph {et~al.}(2004)\citenamefont
  {Baillis}, \citenamefont {Pilon}, \citenamefont {Randrianalisoa},
  \citenamefont {Gomez},\ and\ \citenamefont {Viskanta}}]{baillisJOSAA2004}%
  \BibitemOpen
  \bibfield  {author} {\bibinfo {author} {\bibfnamefont {D.}~\bibnamefont
  {Baillis}}, \bibinfo {author} {\bibfnamefont {L.}~\bibnamefont {Pilon}},
  \bibinfo {author} {\bibfnamefont {H.}~\bibnamefont {Randrianalisoa}},
  \bibinfo {author} {\bibfnamefont {R.}~\bibnamefont {Gomez}}, \ and\ \bibinfo
  {author} {\bibfnamefont {R.}~\bibnamefont {Viskanta}},\ }\href {\doibase
  10.1364/JOSAA.21.000149} {\bibfield  {journal} {\bibinfo  {journal} {J. Opt.
  Soc. Am. A}\ }\textbf {\bibinfo {volume} {21}},\ \bibinfo {pages} {149}
  (\bibinfo {year} {2004})}\BibitemShut {NoStop}%
\bibitem [{\citenamefont {Zaccanti}\ \emph {et~al.}(2003)\citenamefont
  {Zaccanti}, \citenamefont {Bianco},\ and\ \citenamefont
  {Martelli}}]{zaccantiAO2003}%
  \BibitemOpen
  \bibfield  {author} {\bibinfo {author} {\bibfnamefont {G.}~\bibnamefont
  {Zaccanti}}, \bibinfo {author} {\bibfnamefont {S.~D.}\ \bibnamefont
  {Bianco}}, \ and\ \bibinfo {author} {\bibfnamefont {F.}~\bibnamefont
  {Martelli}},\ }\href {\doibase 10.1364/AO.42.004023} {\bibfield  {journal}
  {\bibinfo  {journal} {Appl. Opt.}\ }\textbf {\bibinfo {volume} {42}},\
  \bibinfo {pages} {4023} (\bibinfo {year} {2003})}\BibitemShut {NoStop}%
\bibitem [{\citenamefont {Wang}\ \emph {et~al.}(1995)\citenamefont {Wang},
  \citenamefont {Liang}, \citenamefont {Galland}, \citenamefont {Ho},\ and\
  \citenamefont {Alfano}}]{wangOL1995}%
  \BibitemOpen
  \bibfield  {author} {\bibinfo {author} {\bibfnamefont {L.}~\bibnamefont
  {Wang}}, \bibinfo {author} {\bibfnamefont {X.}~\bibnamefont {Liang}},
  \bibinfo {author} {\bibfnamefont {P.}~\bibnamefont {Galland}}, \bibinfo
  {author} {\bibfnamefont {P.~P.}\ \bibnamefont {Ho}}, \ and\ \bibinfo {author}
  {\bibfnamefont {R.~R.}\ \bibnamefont {Alfano}},\ }\href {\doibase
  10.1364/OL.20.000913} {\bibfield  {journal} {\bibinfo  {journal} {Opt.
  Lett.}\ }\textbf {\bibinfo {volume} {20}},\ \bibinfo {pages} {913} (\bibinfo
  {year} {1995})}\BibitemShut {NoStop}%
\bibitem [{\citenamefont {Tong}\ \emph {et~al.}(2011)\citenamefont {Tong},
  \citenamefont {Yang}, \citenamefont {Si}, \citenamefont {Tan}, \citenamefont
  {Chen}, \citenamefont {Yi},\ and\ \citenamefont {Hou}}]{tongOpteng2011}%
  \BibitemOpen
  \bibfield  {author} {\bibinfo {author} {\bibfnamefont {J.}~\bibnamefont
  {Tong}}, \bibinfo {author} {\bibfnamefont {Y.}~\bibnamefont {Yang}}, \bibinfo
  {author} {\bibfnamefont {J.}~\bibnamefont {Si}}, \bibinfo {author}
  {\bibfnamefont {W.}~\bibnamefont {Tan}}, \bibinfo {author} {\bibfnamefont
  {F.}~\bibnamefont {Chen}}, \bibinfo {author} {\bibfnamefont {W.}~\bibnamefont
  {Yi}}, \ and\ \bibinfo {author} {\bibfnamefont {X.}~\bibnamefont {Hou}},\
  }\href {\doibase 10.1117/1.3567069} {\bibfield  {journal} {\bibinfo
  {journal} {Optical Engineering}\ }\textbf {\bibinfo {volume} {50}},\ \bibinfo
  {pages} {1 } (\bibinfo {year} {2011})}\BibitemShut {NoStop}%
\bibitem [{\citenamefont {d'Abzac}\ \emph {et~al.}(2012)\citenamefont
  {d'Abzac}, \citenamefont {Kervella}, \citenamefont {Hespel},\ and\
  \citenamefont {Dartigalongue}}]{dabzacOE2012}%
  \BibitemOpen
  \bibfield  {author} {\bibinfo {author} {\bibfnamefont {F.-X.}\ \bibnamefont
  {d'Abzac}}, \bibinfo {author} {\bibfnamefont {M.}~\bibnamefont {Kervella}},
  \bibinfo {author} {\bibfnamefont {L.}~\bibnamefont {Hespel}}, \ and\ \bibinfo
  {author} {\bibfnamefont {T.}~\bibnamefont {Dartigalongue}},\ }\href {\doibase
  10.1364/OE.20.009604} {\bibfield  {journal} {\bibinfo  {journal} {Opt.
  Express}\ }\textbf {\bibinfo {volume} {20}},\ \bibinfo {pages} {9604}
  (\bibinfo {year} {2012})}\BibitemShut {NoStop}%
\bibitem [{\citenamefont {WANG}\ \emph {et~al.}(1991)\citenamefont {WANG},
  \citenamefont {HO}, \citenamefont {LIU}, \citenamefont {ZHANG},\ and\
  \citenamefont {ALFANO}}]{wangScience1991}%
  \BibitemOpen
  \bibfield  {author} {\bibinfo {author} {\bibfnamefont {L.}~\bibnamefont
  {WANG}}, \bibinfo {author} {\bibfnamefont {P.~P.}\ \bibnamefont {HO}},
  \bibinfo {author} {\bibfnamefont {C.}~\bibnamefont {LIU}}, \bibinfo {author}
  {\bibfnamefont {G.}~\bibnamefont {ZHANG}}, \ and\ \bibinfo {author}
  {\bibfnamefont {R.~R.}\ \bibnamefont {ALFANO}},\ }\href {\doibase
  10.1126/science.253.5021.769} {\bibfield  {journal} {\bibinfo  {journal}
  {Science}\ }\textbf {\bibinfo {volume} {253}},\ \bibinfo {pages} {769}
  (\bibinfo {year} {1991})},\ \Eprint
  {http://arxiv.org/abs/https://science.sciencemag.org/content/253/5021/769.full.pdf}
  {https://science.sciencemag.org/content/253/5021/769.full.pdf} \BibitemShut
  {NoStop}%
\bibitem [{\citenamefont {Xu}\ \emph {et~al.}(2015)\citenamefont {Xu},
  \citenamefont {Tan}, \citenamefont {Si}, \citenamefont {Zhan}, \citenamefont
  {Tong},\ and\ \citenamefont {Hou}}]{xuOE2015}%
  \BibitemOpen
  \bibfield  {author} {\bibinfo {author} {\bibfnamefont {S.}~\bibnamefont
  {Xu}}, \bibinfo {author} {\bibfnamefont {W.}~\bibnamefont {Tan}}, \bibinfo
  {author} {\bibfnamefont {J.}~\bibnamefont {Si}}, \bibinfo {author}
  {\bibfnamefont {P.}~\bibnamefont {Zhan}}, \bibinfo {author} {\bibfnamefont
  {J.}~\bibnamefont {Tong}}, \ and\ \bibinfo {author} {\bibfnamefont
  {X.}~\bibnamefont {Hou}},\ }\href {\doibase 10.1364/OE.23.001800} {\bibfield
  {journal} {\bibinfo  {journal} {Opt. Express}\ }\textbf {\bibinfo {volume}
  {23}},\ \bibinfo {pages} {1800} (\bibinfo {year} {2015})}\BibitemShut
  {NoStop}%
\bibitem [{\citenamefont {Aernouts}\ \emph {et~al.}(2013)\citenamefont
  {Aernouts}, \citenamefont {Zamora-Rojas}, \citenamefont {Beers},
  \citenamefont {Watt\'{e}}, \citenamefont {Wang}, \citenamefont {Tsuta},
  \citenamefont {Lammertyn},\ and\ \citenamefont {Saeys}}]{aernoutsOE2013}%
  \BibitemOpen
  \bibfield  {author} {\bibinfo {author} {\bibfnamefont {B.}~\bibnamefont
  {Aernouts}}, \bibinfo {author} {\bibfnamefont {E.}~\bibnamefont
  {Zamora-Rojas}}, \bibinfo {author} {\bibfnamefont {R.~V.}\ \bibnamefont
  {Beers}}, \bibinfo {author} {\bibfnamefont {R.}~\bibnamefont {Watt\'{e}}},
  \bibinfo {author} {\bibfnamefont {L.}~\bibnamefont {Wang}}, \bibinfo {author}
  {\bibfnamefont {M.}~\bibnamefont {Tsuta}}, \bibinfo {author} {\bibfnamefont
  {J.}~\bibnamefont {Lammertyn}}, \ and\ \bibinfo {author} {\bibfnamefont
  {W.}~\bibnamefont {Saeys}},\ }\href {\doibase 10.1364/OE.21.032450}
  {\bibfield  {journal} {\bibinfo  {journal} {Opt. Express}\ }\textbf {\bibinfo
  {volume} {21}},\ \bibinfo {pages} {32450} (\bibinfo {year}
  {2013})}\BibitemShut {NoStop}%
\bibitem [{\citenamefont {Prahl}(2011)}]{prahl2011iad}%
  \BibitemOpen
  \bibfield  {author} {\bibinfo {author} {\bibfnamefont {S.}~\bibnamefont
  {Prahl}},\ }\href@noop {} {\bibfield  {journal} {\bibinfo  {journal} {Oregon
  Medical Laser Center, St. Vincent Hospital}\ ,\ \bibinfo {pages} {1}}
  (\bibinfo {year} {2011})}\BibitemShut {NoStop}%
\bibitem [{\citenamefont {Kuhn}\ \emph {et~al.}(1993)\citenamefont {Kuhn},
  \citenamefont {Korder}, \citenamefont {Arduini‐Schuster}, \citenamefont
  {Caps},\ and\ \citenamefont {Fricke}}]{kuhnRSI1993}%
  \BibitemOpen
  \bibfield  {author} {\bibinfo {author} {\bibfnamefont {J.}~\bibnamefont
  {Kuhn}}, \bibinfo {author} {\bibfnamefont {S.}~\bibnamefont {Korder}},
  \bibinfo {author} {\bibfnamefont {M.~C.}\ \bibnamefont {Arduini‐Schuster}},
  \bibinfo {author} {\bibfnamefont {R.}~\bibnamefont {Caps}}, \ and\ \bibinfo
  {author} {\bibfnamefont {J.}~\bibnamefont {Fricke}},\ }\href {\doibase
  10.1063/1.1143914} {\bibfield  {journal} {\bibinfo  {journal} {Review of
  Scientific Instruments}\ }\textbf {\bibinfo {volume} {64}},\ \bibinfo {pages}
  {2523} (\bibinfo {year} {1993})},\ \Eprint
  {http://arxiv.org/abs/https://doi.org/10.1063/1.1143914}
  {https://doi.org/10.1063/1.1143914} \BibitemShut {NoStop}%
\bibitem [{\citenamefont {Göbel}\ \emph {et~al.}(1995)\citenamefont {Göbel},
  \citenamefont {Kuhn},\ and\ \citenamefont {Fricke}}]{gobelWRM1995}%
  \BibitemOpen
  \bibfield  {author} {\bibinfo {author} {\bibfnamefont {G.}~\bibnamefont
  {Göbel}}, \bibinfo {author} {\bibfnamefont {J.}~\bibnamefont {Kuhn}}, \ and\
  \bibinfo {author} {\bibfnamefont {J.}~\bibnamefont {Fricke}},\ }\href
  {\doibase 10.1088/0959-7174/5/4/003} {\bibfield  {journal} {\bibinfo
  {journal} {Waves in Random Media}\ }\textbf {\bibinfo {volume} {5}},\
  \bibinfo {pages} {413} (\bibinfo {year} {1995})},\ \Eprint
  {http://arxiv.org/abs/https://doi.org/10.1088/0959-7174/5/4/003}
  {https://doi.org/10.1088/0959-7174/5/4/003} \BibitemShut {NoStop}%
\bibitem [{\citenamefont {Yang}\ and\ \citenamefont
  {Zhao}(2015)}]{yangJHT2015}%
  \BibitemOpen
  \bibfield  {author} {\bibinfo {author} {\bibfnamefont {G.}~\bibnamefont
  {Yang}}\ and\ \bibinfo {author} {\bibfnamefont {C.}~\bibnamefont {Zhao}},\
  }\href@noop {} {\bibfield  {journal} {\bibinfo  {journal} {Journal of Heat
  Transfer}\ }\textbf {\bibinfo {volume} {137}},\ \bibinfo {pages} {091024}
  (\bibinfo {year} {2015})}\BibitemShut {NoStop}%
\bibitem [{\citenamefont {Mazzamuto}\ \emph {et~al.}(2016)\citenamefont
  {Mazzamuto}, \citenamefont {Pattelli}, \citenamefont {Toninelli},\ and\
  \citenamefont {Wiersma}}]{mazzamutoNJP2016}%
  \BibitemOpen
  \bibfield  {author} {\bibinfo {author} {\bibfnamefont {G.}~\bibnamefont
  {Mazzamuto}}, \bibinfo {author} {\bibfnamefont {L.}~\bibnamefont {Pattelli}},
  \bibinfo {author} {\bibfnamefont {C.}~\bibnamefont {Toninelli}}, \ and\
  \bibinfo {author} {\bibfnamefont {D.~S.}\ \bibnamefont {Wiersma}},\ }\href
  {\doibase 10.1088/1367-2630/18/2/023036} {\bibfield  {journal} {\bibinfo
  {journal} {New Journal of Physics}\ }\textbf {\bibinfo {volume} {18}},\
  \bibinfo {pages} {023036} (\bibinfo {year} {2016})}\BibitemShut {NoStop}%
\bibitem [{\citenamefont {Anderson}(1985)}]{andersonPMB1985}%
  \BibitemOpen
  \bibfield  {author} {\bibinfo {author} {\bibfnamefont {P.~W.}\ \bibnamefont
  {Anderson}},\ }\href {\doibase 10.1080/13642818508240619} {\bibfield
  {journal} {\bibinfo  {journal} {Philosophical Magazine B}\ }\textbf {\bibinfo
  {volume} {52}},\ \bibinfo {pages} {505} (\bibinfo {year} {1985})},\ \Eprint
  {http://arxiv.org/abs/https://doi.org/10.1080/13642818508240619}
  {https://doi.org/10.1080/13642818508240619} \BibitemShut {NoStop}%
\bibitem [{\citenamefont {Scheffold}\ \emph {et~al.}(1999)\citenamefont
  {Scheffold}, \citenamefont {Lenke}, \citenamefont {Tweer},\ and\
  \citenamefont {Maret}}]{scheffoldNature1999}%
  \BibitemOpen
  \bibfield  {author} {\bibinfo {author} {\bibfnamefont {F.}~\bibnamefont
  {Scheffold}}, \bibinfo {author} {\bibfnamefont {R.}~\bibnamefont {Lenke}},
  \bibinfo {author} {\bibfnamefont {R.}~\bibnamefont {Tweer}}, \ and\ \bibinfo
  {author} {\bibfnamefont {G.}~\bibnamefont {Maret}},\ }\href {\doibase
  10.1038/18347} {\bibfield  {journal} {\bibinfo  {journal} {Nature}\ }\textbf
  {\bibinfo {volume} {398}},\ \bibinfo {pages} {206} (\bibinfo {year}
  {1999})}\BibitemShut {NoStop}%
\bibitem [{\citenamefont {Wiersma}\ \emph {et~al.}(1999)\citenamefont
  {Wiersma}, \citenamefont {Rivas}, \citenamefont {Bartolini}, \citenamefont
  {Lagendijk},\ and\ \citenamefont {Righini}}]{wiersmaNature1999reply}%
  \BibitemOpen
  \bibfield  {author} {\bibinfo {author} {\bibfnamefont {D.~S.}\ \bibnamefont
  {Wiersma}}, \bibinfo {author} {\bibfnamefont {J.~G.}\ \bibnamefont {Rivas}},
  \bibinfo {author} {\bibfnamefont {P.}~\bibnamefont {Bartolini}}, \bibinfo
  {author} {\bibfnamefont {A.}~\bibnamefont {Lagendijk}}, \ and\ \bibinfo
  {author} {\bibfnamefont {R.}~\bibnamefont {Righini}},\ }\href {\doibase
  10.1038/18350} {\bibfield  {journal} {\bibinfo  {journal} {Nature}\ }\textbf
  {\bibinfo {volume} {398}},\ \bibinfo {pages} {207} (\bibinfo {year}
  {1999})}\BibitemShut {NoStop}%
\bibitem [{\citenamefont {Sperling}\ \emph {et~al.}(2016)\citenamefont
  {Sperling}, \citenamefont {Schertel}, \citenamefont {Ackermann},
  \citenamefont {Aubry}, \citenamefont {Aegerter},\ and\ \citenamefont
  {Maret}}]{sperlingNJP2016}%
  \BibitemOpen
  \bibfield  {author} {\bibinfo {author} {\bibfnamefont {T.}~\bibnamefont
  {Sperling}}, \bibinfo {author} {\bibfnamefont {L.}~\bibnamefont {Schertel}},
  \bibinfo {author} {\bibfnamefont {M.}~\bibnamefont {Ackermann}}, \bibinfo
  {author} {\bibfnamefont {G.~J.}\ \bibnamefont {Aubry}}, \bibinfo {author}
  {\bibfnamefont {C.~M.}\ \bibnamefont {Aegerter}}, \ and\ \bibinfo {author}
  {\bibfnamefont {G.}~\bibnamefont {Maret}},\ }\href {\doibase
  10.1088/1367-2630/18/1/013039} {\bibfield  {journal} {\bibinfo  {journal}
  {New Journal of Physics}\ }\textbf {\bibinfo {volume} {18}},\ \bibinfo
  {pages} {013039} (\bibinfo {year} {2016})}\BibitemShut {NoStop}%
\bibitem [{\citenamefont {Germer}\ \emph {et~al.}(2014)\citenamefont {Germer},
  \citenamefont {Stover},\ and\ \citenamefont
  {Schröder}}]{germer2014angleresolved}%
  \BibitemOpen
  \bibfield  {author} {\bibinfo {author} {\bibfnamefont {T.~A.}\ \bibnamefont
  {Germer}}, \bibinfo {author} {\bibfnamefont {J.~C.}\ \bibnamefont {Stover}},
  \ and\ \bibinfo {author} {\bibfnamefont {S.}~\bibnamefont {Schröder}},\ }in\
  \href {\doibase https://doi.org/10.1016/B978-0-12-386022-4.00008-X} {\emph
  {\bibinfo {booktitle} {Spectrophotometry}}},\ \bibinfo {series} {Experimental
  Methods in the Physical Sciences}, Vol.~\bibinfo {volume} {46},\ \bibinfo
  {editor} {edited by\ \bibinfo {editor} {\bibfnamefont {T.~A.}\ \bibnamefont
  {Germer}}, \bibinfo {editor} {\bibfnamefont {J.~C.}\ \bibnamefont
  {Zwinkels}}, \ and\ \bibinfo {editor} {\bibfnamefont {B.~K.}\ \bibnamefont
  {Tsai}}}\ (\bibinfo  {publisher} {Academic Press},\ \bibinfo {year} {2014})\
  pp.\ \bibinfo {pages} {291 -- 331}\BibitemShut {NoStop}%
\bibitem [{\citenamefont {Milandri}\ \emph {et~al.}(2002)\citenamefont
  {Milandri}, \citenamefont {Asllanaj},\ and\ \citenamefont
  {Jeandel}}]{milandriJQSRT2002}%
  \BibitemOpen
  \bibfield  {author} {\bibinfo {author} {\bibfnamefont {A.}~\bibnamefont
  {Milandri}}, \bibinfo {author} {\bibfnamefont {F.}~\bibnamefont {Asllanaj}},
  \ and\ \bibinfo {author} {\bibfnamefont {G.}~\bibnamefont {Jeandel}},\ }\href
  {\doibase https://doi.org/10.1016/S0022-4073(01)00276-X} {\bibfield
  {journal} {\bibinfo  {journal} {Journal of Quantitative Spectroscopy and
  Radiative Transfer}\ }\textbf {\bibinfo {volume} {74}},\ \bibinfo {pages}
  {637 } (\bibinfo {year} {2002})}\BibitemShut {NoStop}%
\bibitem [{\citenamefont {Randrianalisoa}\ \emph {et~al.}(2006)\citenamefont
  {Randrianalisoa}, \citenamefont {Baillis},\ and\ \citenamefont
  {Pilon}}]{randrianalisoaJTHT2006}%
  \BibitemOpen
  \bibfield  {author} {\bibinfo {author} {\bibfnamefont {J.}~\bibnamefont
  {Randrianalisoa}}, \bibinfo {author} {\bibfnamefont {D.}~\bibnamefont
  {Baillis}}, \ and\ \bibinfo {author} {\bibfnamefont {L.}~\bibnamefont
  {Pilon}},\ }\href@noop {} {\bibfield  {journal} {\bibinfo  {journal} {Journal
  of thermophysics and heat transfer}\ }\textbf {\bibinfo {volume} {20}},\
  \bibinfo {pages} {871} (\bibinfo {year} {2006})}\BibitemShut {NoStop}%
\bibitem [{\citenamefont {Mishchenko}\ \emph {et~al.}(2011)\citenamefont
  {Mishchenko}, \citenamefont {Yatskiv}, \citenamefont {Rosenbush},\ and\
  \citenamefont {Videen}}]{mishchenko2011polarimetric}%
  \BibitemOpen
  \bibfield  {author} {\bibinfo {author} {\bibfnamefont {M.~I.}\ \bibnamefont
  {Mishchenko}}, \bibinfo {author} {\bibfnamefont {Y.~S.}\ \bibnamefont
  {Yatskiv}}, \bibinfo {author} {\bibfnamefont {V.~K.}\ \bibnamefont
  {Rosenbush}}, \ and\ \bibinfo {author} {\bibfnamefont {G.}~\bibnamefont
  {Videen}},\ }\href@noop {} {\emph {\bibinfo {title} {Polarimetric detection,
  characterization and remote sensing}}}\ (\bibinfo  {publisher} {Springer},\
  \bibinfo {year} {2011})\BibitemShut {NoStop}%
\bibitem [{\citenamefont {Riviere}\ \emph {et~al.}(2013)\citenamefont
  {Riviere}, \citenamefont {Ceolato},\ and\ \citenamefont
  {Hespel}}]{riviereJQSRT2013}%
  \BibitemOpen
  \bibfield  {author} {\bibinfo {author} {\bibfnamefont {N.}~\bibnamefont
  {Riviere}}, \bibinfo {author} {\bibfnamefont {R.}~\bibnamefont {Ceolato}}, \
  and\ \bibinfo {author} {\bibfnamefont {L.}~\bibnamefont {Hespel}},\ }\href
  {\doibase https://doi.org/10.1016/j.jqsrt.2013.04.019} {\bibfield  {journal}
  {\bibinfo  {journal} {Journal of Quantitative Spectroscopy and Radiative
  Transfer}\ }\textbf {\bibinfo {volume} {131}},\ \bibinfo {pages} {88 }
  (\bibinfo {year} {2013})},\ \bibinfo {note} {concepts in electromagnetic
  scattering for particulate-systems characterization}\BibitemShut {NoStop}%
\bibitem [{\citenamefont {Ceolato}\ and\ \citenamefont
  {Riviere}(2018)}]{ceolato2018spectropolarimetric}%
  \BibitemOpen
  \bibfield  {author} {\bibinfo {author} {\bibfnamefont {R.}~\bibnamefont
  {Ceolato}}\ and\ \bibinfo {author} {\bibfnamefont {N.}~\bibnamefont
  {Riviere}},\ }\enquote {\bibinfo {title} {Advances in spectro-polarimetric
  light-scattering by particulate media},}\ in\ \href {\doibase
  10.1007/978-3-319-70808-9_2} {\emph {\bibinfo {booktitle} {Springer Series in
  Light Scattering: Volume 2: Light Scattering, Radiative Transfer and Remote
  Sensing}}},\ \bibinfo {editor} {edited by\ \bibinfo {editor} {\bibfnamefont
  {A.}~\bibnamefont {Kokhanovsky}}}\ (\bibinfo  {publisher} {Springer
  International Publishing},\ \bibinfo {address} {Cham},\ \bibinfo {year}
  {2018})\ pp.\ \bibinfo {pages} {55--107}\BibitemShut {NoStop}%
\bibitem [{\citenamefont {Fiebig}(2010)}]{fiebig2010coherent}%
  \BibitemOpen
  \bibfield  {author} {\bibinfo {author} {\bibfnamefont {S.}~\bibnamefont
  {Fiebig}},\ }\emph {\bibinfo {title} {Coherent backscattering from multiple
  scattering systems}},\ \href@noop {} {Ph.D. thesis} (\bibinfo {year}
  {2010})\BibitemShut {NoStop}%
\bibitem [{\citenamefont {Aegerter}\ and\ \citenamefont
  {Maret}(2009)}]{Aegerter2008}%
  \BibitemOpen
  \bibfield  {author} {\bibinfo {author} {\bibfnamefont {C.~M.}\ \bibnamefont
  {Aegerter}}\ and\ \bibinfo {author} {\bibfnamefont {G.}~\bibnamefont
  {Maret}}\ }(\bibinfo  {publisher} {Elsevier},\ \bibinfo {year} {2009})\ pp.\
  \bibinfo {pages} {1 -- 62}\BibitemShut {NoStop}%
\bibitem [{\citenamefont {Gross}\ \emph {et~al.}(2007)\citenamefont {Gross},
  \citenamefont {Störzer}, \citenamefont {Fiebig}, \citenamefont {Clausen},
  \citenamefont {Maret},\ and\ \citenamefont {Aegerter}}]{grossRSI2007}%
  \BibitemOpen
  \bibfield  {author} {\bibinfo {author} {\bibfnamefont {P.}~\bibnamefont
  {Gross}}, \bibinfo {author} {\bibfnamefont {M.}~\bibnamefont {Störzer}},
  \bibinfo {author} {\bibfnamefont {S.}~\bibnamefont {Fiebig}}, \bibinfo
  {author} {\bibfnamefont {M.}~\bibnamefont {Clausen}}, \bibinfo {author}
  {\bibfnamefont {G.}~\bibnamefont {Maret}}, \ and\ \bibinfo {author}
  {\bibfnamefont {C.~M.}\ \bibnamefont {Aegerter}},\ }\href {\doibase
  10.1063/1.2712943} {\bibfield  {journal} {\bibinfo  {journal} {Review of
  Scientific Instruments}\ }\textbf {\bibinfo {volume} {78}},\ \bibinfo {pages}
  {033105} (\bibinfo {year} {2007})},\ \Eprint
  {http://arxiv.org/abs/https://doi.org/10.1063/1.2712943}
  {https://doi.org/10.1063/1.2712943} \BibitemShut {NoStop}%
\bibitem [{\citenamefont {Weiner}(2011)}]{weiner2011ultrafast}%
  \BibitemOpen
  \bibfield  {author} {\bibinfo {author} {\bibfnamefont {A.}~\bibnamefont
  {Weiner}},\ }\href@noop {} {\emph {\bibinfo {title} {Ultrafast optics}}},\
  Vol.~\bibinfo {volume} {72}\ (\bibinfo  {publisher} {John Wiley \& Sons},\
  \bibinfo {year} {2011})\BibitemShut {NoStop}%
\bibitem [{\citenamefont {Watson}\ \emph {et~al.}(1987)\citenamefont {Watson},
  \citenamefont {Fleury},\ and\ \citenamefont {McCall}}]{watsonPRL1987}%
  \BibitemOpen
  \bibfield  {author} {\bibinfo {author} {\bibfnamefont {G.~H.}\ \bibnamefont
  {Watson}}, \bibinfo {author} {\bibfnamefont {P.~A.}\ \bibnamefont {Fleury}},
  \ and\ \bibinfo {author} {\bibfnamefont {S.~L.}\ \bibnamefont {McCall}},\
  }\href {\doibase 10.1103/PhysRevLett.58.945} {\bibfield  {journal} {\bibinfo
  {journal} {Phys. Rev. Lett.}\ }\textbf {\bibinfo {volume} {58}},\ \bibinfo
  {pages} {945} (\bibinfo {year} {1987})}\BibitemShut {NoStop}%
\bibitem [{\citenamefont {Vreeker}\ \emph {et~al.}(1988)\citenamefont
  {Vreeker}, \citenamefont {Albada}, \citenamefont {Sprik},\ and\ \citenamefont
  {Lagendijk}}]{vreekerPLA1988}%
  \BibitemOpen
  \bibfield  {author} {\bibinfo {author} {\bibfnamefont {R.}~\bibnamefont
  {Vreeker}}, \bibinfo {author} {\bibfnamefont {M.~V.}\ \bibnamefont {Albada}},
  \bibinfo {author} {\bibfnamefont {R.}~\bibnamefont {Sprik}}, \ and\ \bibinfo
  {author} {\bibfnamefont {A.}~\bibnamefont {Lagendijk}},\ }\href {\doibase
  https://doi.org/10.1016/0375-9601(88)90438-0} {\bibfield  {journal} {\bibinfo
   {journal} {Physics Letters A}\ }\textbf {\bibinfo {volume} {132}},\ \bibinfo
  {pages} {51 } (\bibinfo {year} {1988})}\BibitemShut {NoStop}%
\bibitem [{\citenamefont {Drake}\ and\ \citenamefont
  {Genack}(1989)}]{drakePRL1989}%
  \BibitemOpen
  \bibfield  {author} {\bibinfo {author} {\bibfnamefont {J.~M.}\ \bibnamefont
  {Drake}}\ and\ \bibinfo {author} {\bibfnamefont {A.~Z.}\ \bibnamefont
  {Genack}},\ }\href {\doibase 10.1103/PhysRevLett.63.259} {\bibfield
  {journal} {\bibinfo  {journal} {Phys. Rev. Lett.}\ }\textbf {\bibinfo
  {volume} {63}},\ \bibinfo {pages} {259} (\bibinfo {year} {1989})}\BibitemShut
  {NoStop}%
\bibitem [{\citenamefont {Genack}\ and\ \citenamefont
  {Drake}(1990)}]{genackEPL1990}%
  \BibitemOpen
  \bibfield  {author} {\bibinfo {author} {\bibfnamefont {A.~Z.}\ \bibnamefont
  {Genack}}\ and\ \bibinfo {author} {\bibfnamefont {J.~M.}\ \bibnamefont
  {Drake}},\ }\href {\doibase 10.1209/0295-5075/11/4/007} {\bibfield  {journal}
  {\bibinfo  {journal} {Europhysics Letters ({EPL})}\ }\textbf {\bibinfo
  {volume} {11}},\ \bibinfo {pages} {331} (\bibinfo {year} {1990})}\BibitemShut
  {NoStop}%
\bibitem [{\citenamefont {Yoo}\ \emph {et~al.}(1990)\citenamefont {Yoo},
  \citenamefont {Liu},\ and\ \citenamefont {Alfano}}]{yooPRL1990}%
  \BibitemOpen
  \bibfield  {author} {\bibinfo {author} {\bibfnamefont {K.~M.}\ \bibnamefont
  {Yoo}}, \bibinfo {author} {\bibfnamefont {F.}~\bibnamefont {Liu}}, \ and\
  \bibinfo {author} {\bibfnamefont {R.~R.}\ \bibnamefont {Alfano}},\ }\href
  {\doibase 10.1103/PhysRevLett.64.2647} {\bibfield  {journal} {\bibinfo
  {journal} {Phys. Rev. Lett.}\ }\textbf {\bibinfo {volume} {64}},\ \bibinfo
  {pages} {2647} (\bibinfo {year} {1990})}\BibitemShut {NoStop}%
\bibitem [{\citenamefont {Kop}\ and\ \citenamefont {Sprik}(1995)}]{kopRSI1995}%
  \BibitemOpen
  \bibfield  {author} {\bibinfo {author} {\bibfnamefont {R.~H.~J.}\
  \bibnamefont {Kop}}\ and\ \bibinfo {author} {\bibfnamefont {R.}~\bibnamefont
  {Sprik}},\ }\href {\doibase 10.1063/1.1146069} {\bibfield  {journal}
  {\bibinfo  {journal} {Review of Scientific Instruments}\ }\textbf {\bibinfo
  {volume} {66}},\ \bibinfo {pages} {5459} (\bibinfo {year} {1995})},\ \Eprint
  {http://arxiv.org/abs/https://doi.org/10.1063/1.1146069}
  {https://doi.org/10.1063/1.1146069} \BibitemShut {NoStop}%
\bibitem [{\citenamefont {Kop}\ \emph {et~al.}(1997)\citenamefont {Kop},
  \citenamefont {de~Vries}, \citenamefont {Sprik},\ and\ \citenamefont
  {Lagendijk}}]{kopPRL1997}%
  \BibitemOpen
  \bibfield  {author} {\bibinfo {author} {\bibfnamefont {R.~H.~J.}\
  \bibnamefont {Kop}}, \bibinfo {author} {\bibfnamefont {P.}~\bibnamefont
  {de~Vries}}, \bibinfo {author} {\bibfnamefont {R.}~\bibnamefont {Sprik}}, \
  and\ \bibinfo {author} {\bibfnamefont {A.}~\bibnamefont {Lagendijk}},\ }\href
  {\doibase 10.1103/PhysRevLett.79.4369} {\bibfield  {journal} {\bibinfo
  {journal} {Phys. Rev. Lett.}\ }\textbf {\bibinfo {volume} {79}},\ \bibinfo
  {pages} {4369} (\bibinfo {year} {1997})}\BibitemShut {NoStop}%
\bibitem [{\citenamefont {Chabanov}\ \emph {et~al.}(2003)\citenamefont
  {Chabanov}, \citenamefont {Zhang},\ and\ \citenamefont
  {Genack}}]{chabanovPRL2003}%
  \BibitemOpen
  \bibfield  {author} {\bibinfo {author} {\bibfnamefont {A.~A.}\ \bibnamefont
  {Chabanov}}, \bibinfo {author} {\bibfnamefont {Z.~Q.}\ \bibnamefont {Zhang}},
  \ and\ \bibinfo {author} {\bibfnamefont {A.~Z.}\ \bibnamefont {Genack}},\
  }\href {\doibase 10.1103/PhysRevLett.90.203903} {\bibfield  {journal}
  {\bibinfo  {journal} {Phys. Rev. Lett.}\ }\textbf {\bibinfo {volume} {90}},\
  \bibinfo {pages} {203903} (\bibinfo {year} {2003})}\BibitemShut {NoStop}%
\bibitem [{\citenamefont {Johnson}\ \emph {et~al.}(2003)\citenamefont
  {Johnson}, \citenamefont {Imhof}, \citenamefont {Bret}, \citenamefont
  {Rivas},\ and\ \citenamefont {Lagendijk}}]{johnsonPRE2003}%
  \BibitemOpen
  \bibfield  {author} {\bibinfo {author} {\bibfnamefont {P.~M.}\ \bibnamefont
  {Johnson}}, \bibinfo {author} {\bibfnamefont {A.}~\bibnamefont {Imhof}},
  \bibinfo {author} {\bibfnamefont {B.~P.~J.}\ \bibnamefont {Bret}}, \bibinfo
  {author} {\bibfnamefont {J.~G.}\ \bibnamefont {Rivas}}, \ and\ \bibinfo
  {author} {\bibfnamefont {A.}~\bibnamefont {Lagendijk}},\ }\href {\doibase
  10.1103/PhysRevE.68.016604} {\bibfield  {journal} {\bibinfo  {journal} {Phys.
  Rev. E}\ }\textbf {\bibinfo {volume} {68}},\ \bibinfo {pages} {016604}
  (\bibinfo {year} {2003})}\BibitemShut {NoStop}%
\bibitem [{\citenamefont {Bestemyanov}\ \emph {et~al.}(2004)\citenamefont
  {Bestemyanov}, \citenamefont {Gordienko}, \citenamefont {Ivanov},
  \citenamefont {Konovalov},\ and\ \citenamefont
  {Podshivalov}}]{bestemyanovQE2004}%
  \BibitemOpen
  \bibfield  {author} {\bibinfo {author} {\bibfnamefont {K.~P.}\ \bibnamefont
  {Bestemyanov}}, \bibinfo {author} {\bibfnamefont {V.~M.}\ \bibnamefont
  {Gordienko}}, \bibinfo {author} {\bibfnamefont {A.~A.}\ \bibnamefont
  {Ivanov}}, \bibinfo {author} {\bibfnamefont {A.~N.}\ \bibnamefont
  {Konovalov}}, \ and\ \bibinfo {author} {\bibfnamefont {A.~A.}\ \bibnamefont
  {Podshivalov}},\ }\href {\doibase 10.1070/qe2004v034n07abeh002824} {\bibfield
   {journal} {\bibinfo  {journal} {Quantum Electronics}\ }\textbf {\bibinfo
  {volume} {34}},\ \bibinfo {pages} {666} (\bibinfo {year} {2004})}\BibitemShut
  {NoStop}%
\bibitem [{\citenamefont {Cheikh}\ \emph {et~al.}(2006)\citenamefont {Cheikh},
  \citenamefont {Nghi\^{e}m}, \citenamefont {Ettori}, \citenamefont {Tinet},
  \citenamefont {Avrillier},\ and\ \citenamefont {Tualle}}]{cheikhOL2006}%
  \BibitemOpen
  \bibfield  {author} {\bibinfo {author} {\bibfnamefont {M.}~\bibnamefont
  {Cheikh}}, \bibinfo {author} {\bibfnamefont {H.~L.}\ \bibnamefont
  {Nghi\^{e}m}}, \bibinfo {author} {\bibfnamefont {D.}~\bibnamefont {Ettori}},
  \bibinfo {author} {\bibfnamefont {E.}~\bibnamefont {Tinet}}, \bibinfo
  {author} {\bibfnamefont {S.}~\bibnamefont {Avrillier}}, \ and\ \bibinfo
  {author} {\bibfnamefont {J.-M.}\ \bibnamefont {Tualle}},\ }\href {\doibase
  10.1364/OL.31.002311} {\bibfield  {journal} {\bibinfo  {journal} {Opt.
  Lett.}\ }\textbf {\bibinfo {volume} {31}},\ \bibinfo {pages} {2311} (\bibinfo
  {year} {2006})}\BibitemShut {NoStop}%
\bibitem [{\citenamefont {Aegerter}\ \emph {et~al.}(2006)\citenamefont
  {Aegerter}, \citenamefont {Störzer},\ and\ \citenamefont
  {Maret}}]{aegerterEPL2006}%
  \BibitemOpen
  \bibfield  {author} {\bibinfo {author} {\bibfnamefont {C.~M.}\ \bibnamefont
  {Aegerter}}, \bibinfo {author} {\bibfnamefont {M.}~\bibnamefont {Störzer}},
  \ and\ \bibinfo {author} {\bibfnamefont {G.}~\bibnamefont {Maret}},\ }\href
  {\doibase 10.1209/epl/i2006-10144-3} {\bibfield  {journal} {\bibinfo
  {journal} {Europhysics Letters ({EPL})}\ }\textbf {\bibinfo {volume} {75}},\
  \bibinfo {pages} {562} (\bibinfo {year} {2006})}\BibitemShut {NoStop}%
\bibitem [{\citenamefont {Aegerter}\ \emph {et~al.}(2007)\citenamefont
  {Aegerter}, \citenamefont {St\"{o}rzer}, \citenamefont {Fiebig},
  \citenamefont {B\"{u}hrer},\ and\ \citenamefont {Maret}}]{aegerterJOSAA2007}%
  \BibitemOpen
  \bibfield  {author} {\bibinfo {author} {\bibfnamefont {C.~M.}\ \bibnamefont
  {Aegerter}}, \bibinfo {author} {\bibfnamefont {M.}~\bibnamefont
  {St\"{o}rzer}}, \bibinfo {author} {\bibfnamefont {S.}~\bibnamefont {Fiebig}},
  \bibinfo {author} {\bibfnamefont {W.}~\bibnamefont {B\"{u}hrer}}, \ and\
  \bibinfo {author} {\bibfnamefont {G.}~\bibnamefont {Maret}},\ }\href
  {\doibase 10.1364/JOSAA.24.000A23} {\bibfield  {journal} {\bibinfo  {journal}
  {J. Opt. Soc. Am. A}\ }\textbf {\bibinfo {volume} {24}},\ \bibinfo {pages}
  {A23} (\bibinfo {year} {2007})}\BibitemShut {NoStop}%
\bibitem [{\citenamefont {Conti}\ and\ \citenamefont
  {Fratalocchi}(2008)}]{contiNaturephys2008}%
  \BibitemOpen
  \bibfield  {author} {\bibinfo {author} {\bibfnamefont {C.}~\bibnamefont
  {Conti}}\ and\ \bibinfo {author} {\bibfnamefont {A.}~\bibnamefont
  {Fratalocchi}},\ }\href {\doibase 10.1038/nphys1035} {\bibfield  {journal}
  {\bibinfo  {journal} {Nature Physics}\ }\textbf {\bibinfo {volume} {4}},\
  \bibinfo {pages} {794} (\bibinfo {year} {2008})}\BibitemShut {NoStop}%
\bibitem [{\citenamefont {Borycki}\ \emph {et~al.}(2016)\citenamefont
  {Borycki}, \citenamefont {Kholiqov},\ and\ \citenamefont
  {Srinivasan}}]{boryckiOptica2016}%
  \BibitemOpen
  \bibfield  {author} {\bibinfo {author} {\bibfnamefont {D.}~\bibnamefont
  {Borycki}}, \bibinfo {author} {\bibfnamefont {O.}~\bibnamefont {Kholiqov}}, \
  and\ \bibinfo {author} {\bibfnamefont {V.~J.}\ \bibnamefont {Srinivasan}},\
  }\href {\doibase 10.1364/OPTICA.3.001471} {\bibfield  {journal} {\bibinfo
  {journal} {Optica}\ }\textbf {\bibinfo {volume} {3}},\ \bibinfo {pages}
  {1471} (\bibinfo {year} {2016})}\BibitemShut {NoStop}%
\bibitem [{\citenamefont {Cobus}\ \emph {et~al.}(2016)\citenamefont {Cobus},
  \citenamefont {Skipetrov}, \citenamefont {Aubry}, \citenamefont {van
  Tiggelen}, \citenamefont {Derode},\ and\ \citenamefont
  {Page}}]{cobusPRL2016}%
  \BibitemOpen
  \bibfield  {author} {\bibinfo {author} {\bibfnamefont {L.~A.}\ \bibnamefont
  {Cobus}}, \bibinfo {author} {\bibfnamefont {S.~E.}\ \bibnamefont
  {Skipetrov}}, \bibinfo {author} {\bibfnamefont {A.}~\bibnamefont {Aubry}},
  \bibinfo {author} {\bibfnamefont {B.~A.}\ \bibnamefont {van Tiggelen}},
  \bibinfo {author} {\bibfnamefont {A.}~\bibnamefont {Derode}}, \ and\ \bibinfo
  {author} {\bibfnamefont {J.~H.}\ \bibnamefont {Page}},\ }\href {\doibase
  10.1103/PhysRevLett.116.193901} {\bibfield  {journal} {\bibinfo  {journal}
  {Phys. Rev. Lett.}\ }\textbf {\bibinfo {volume} {116}},\ \bibinfo {pages}
  {193901} (\bibinfo {year} {2016})}\BibitemShut {NoStop}%
\bibitem [{\citenamefont {Lyons}\ \emph {et~al.}(2019)\citenamefont {Lyons},
  \citenamefont {Tonolini}, \citenamefont {Boccolini}, \citenamefont {Repetti},
  \citenamefont {Henderson}, \citenamefont {Wiaux},\ and\ \citenamefont
  {Faccio}}]{lyonsNaturephoton2019}%
  \BibitemOpen
  \bibfield  {author} {\bibinfo {author} {\bibfnamefont {A.}~\bibnamefont
  {Lyons}}, \bibinfo {author} {\bibfnamefont {F.}~\bibnamefont {Tonolini}},
  \bibinfo {author} {\bibfnamefont {A.}~\bibnamefont {Boccolini}}, \bibinfo
  {author} {\bibfnamefont {A.}~\bibnamefont {Repetti}}, \bibinfo {author}
  {\bibfnamefont {R.}~\bibnamefont {Henderson}}, \bibinfo {author}
  {\bibfnamefont {Y.}~\bibnamefont {Wiaux}}, \ and\ \bibinfo {author}
  {\bibfnamefont {D.}~\bibnamefont {Faccio}},\ }\href {\doibase
  10.1038/s41566-019-0439-x} {\bibfield  {journal} {\bibinfo  {journal} {Nature
  Photonics}\ }\textbf {\bibinfo {volume} {13}},\ \bibinfo {pages} {575}
  (\bibinfo {year} {2019})}\BibitemShut {NoStop}%
\bibitem [{\citenamefont {Duran-Ledezma}\ \emph {et~al.}(2018)\citenamefont
  {Duran-Ledezma}, \citenamefont {Jacinto-M\'{e}ndez},\ and\ \citenamefont
  {Rojas-Ochoa}}]{duran-ledezmaAO2018}%
  \BibitemOpen
  \bibfield  {author} {\bibinfo {author} {\bibfnamefont {A.~A.}\ \bibnamefont
  {Duran-Ledezma}}, \bibinfo {author} {\bibfnamefont {D.}~\bibnamefont
  {Jacinto-M\'{e}ndez}}, \ and\ \bibinfo {author} {\bibfnamefont {L.~F.}\
  \bibnamefont {Rojas-Ochoa}},\ }\href {\doibase 10.1364/AO.57.000208}
  {\bibfield  {journal} {\bibinfo  {journal} {Appl. Opt.}\ }\textbf {\bibinfo
  {volume} {57}},\ \bibinfo {pages} {208} (\bibinfo {year} {2018})}\BibitemShut
  {NoStop}%
\bibitem [{\citenamefont {Guo}\ \emph {et~al.}(2002)\citenamefont {Guo},
  \citenamefont {Aber}, \citenamefont {Garetz},\ and\ \citenamefont
  {Kumar}}]{guoJQSRT2002}%
  \BibitemOpen
  \bibfield  {author} {\bibinfo {author} {\bibfnamefont {Z.}~\bibnamefont
  {Guo}}, \bibinfo {author} {\bibfnamefont {J.}~\bibnamefont {Aber}}, \bibinfo
  {author} {\bibfnamefont {B.~A.}\ \bibnamefont {Garetz}}, \ and\ \bibinfo
  {author} {\bibfnamefont {S.}~\bibnamefont {Kumar}},\ }\href {\doibase
  https://doi.org/10.1016/S0022-4073(01)00203-5} {\bibfield  {journal}
  {\bibinfo  {journal} {Journal of Quantitative Spectroscopy and Radiative
  Transfer}\ }\textbf {\bibinfo {volume} {73}},\ \bibinfo {pages} {159 }
  (\bibinfo {year} {2002})},\ \bibinfo {note} {third International Symposium on
  Radiative Transfer}\BibitemShut {NoStop}%
\bibitem [{\citenamefont {Wan}\ \emph {et~al.}(2004)\citenamefont {Wan},
  \citenamefont {Guo}, \citenamefont {Kumar}, \citenamefont {Aber},\ and\
  \citenamefont {Garetz}}]{wanJQSRT2004}%
  \BibitemOpen
  \bibfield  {author} {\bibinfo {author} {\bibfnamefont {S.~K.}\ \bibnamefont
  {Wan}}, \bibinfo {author} {\bibfnamefont {Z.}~\bibnamefont {Guo}}, \bibinfo
  {author} {\bibfnamefont {S.}~\bibnamefont {Kumar}}, \bibinfo {author}
  {\bibfnamefont {J.}~\bibnamefont {Aber}}, \ and\ \bibinfo {author}
  {\bibfnamefont {B.~A.}\ \bibnamefont {Garetz}},\ }\href {\doibase
  https://doi.org/10.1016/S0022-4073(03)00266-8} {\bibfield  {journal}
  {\bibinfo  {journal} {Journal of Quantitative Spectroscopy and Radiative
  Transfer}\ }\textbf {\bibinfo {volume} {84}},\ \bibinfo {pages} {493 }
  (\bibinfo {year} {2004})},\ \bibinfo {note} {eurotherm Seminar 73 -
  Computational Thermal Radiation in Participating Media}\BibitemShut {NoStop}%
\bibitem [{\citenamefont {Calba}\ \emph {et~al.}(2008)\citenamefont {Calba},
  \citenamefont {M\'{e}\`{e}s}, \citenamefont {Roz\'{e}},\ and\ \citenamefont
  {Girasole}}]{calbaJOSAA2008}%
  \BibitemOpen
  \bibfield  {author} {\bibinfo {author} {\bibfnamefont {C.}~\bibnamefont
  {Calba}}, \bibinfo {author} {\bibfnamefont {L.}~\bibnamefont {M\'{e}\`{e}s}},
  \bibinfo {author} {\bibfnamefont {C.}~\bibnamefont {Roz\'{e}}}, \ and\
  \bibinfo {author} {\bibfnamefont {T.}~\bibnamefont {Girasole}},\ }\href
  {\doibase 10.1364/JOSAA.25.001541} {\bibfield  {journal} {\bibinfo  {journal}
  {J. Opt. Soc. Am. A}\ }\textbf {\bibinfo {volume} {25}},\ \bibinfo {pages}
  {1541} (\bibinfo {year} {2008})}\BibitemShut {NoStop}%
\bibitem [{\citenamefont {Guo}\ and\ \citenamefont
  {Hunter}(2013)}]{guoHTR2013}%
  \BibitemOpen
  \bibfield  {author} {\bibinfo {author} {\bibfnamefont {Z.}~\bibnamefont
  {Guo}}\ and\ \bibinfo {author} {\bibfnamefont {B.}~\bibnamefont {Hunter}},\
  }\href@noop {} {\bibfield  {journal} {\bibinfo  {journal} {Heat Transfer
  Research}\ }\textbf {\bibinfo {volume} {44}},\ \bibinfo {pages} {303}
  (\bibinfo {year} {2013})}\BibitemShut {NoStop}%
\bibitem [{\citenamefont {Elaloufi}\ \emph {et~al.}(2003)\citenamefont
  {Elaloufi}, \citenamefont {Carminati},\ and\ \citenamefont
  {Greffet}}]{elaloufiJOSAA2003}%
  \BibitemOpen
  \bibfield  {author} {\bibinfo {author} {\bibfnamefont {R.}~\bibnamefont
  {Elaloufi}}, \bibinfo {author} {\bibfnamefont {R.}~\bibnamefont {Carminati}},
  \ and\ \bibinfo {author} {\bibfnamefont {J.-J.}\ \bibnamefont {Greffet}},\
  }\href {\doibase 10.1364/JOSAA.20.000678} {\bibfield  {journal} {\bibinfo
  {journal} {J. Opt. Soc. Am. A}\ }\textbf {\bibinfo {volume} {20}},\ \bibinfo
  {pages} {678} (\bibinfo {year} {2003})}\BibitemShut {NoStop}%
\bibitem [{\citenamefont {Patterson}\ \emph {et~al.}(1989)\citenamefont
  {Patterson}, \citenamefont {Chance},\ and\ \citenamefont
  {Wilson}}]{pattersonAO1989}%
  \BibitemOpen
  \bibfield  {author} {\bibinfo {author} {\bibfnamefont {M.~S.}\ \bibnamefont
  {Patterson}}, \bibinfo {author} {\bibfnamefont {B.}~\bibnamefont {Chance}}, \
  and\ \bibinfo {author} {\bibfnamefont {B.~C.}\ \bibnamefont {Wilson}},\
  }\href {\doibase 10.1364/AO.28.002331} {\bibfield  {journal} {\bibinfo
  {journal} {Appl. Opt.}\ }\textbf {\bibinfo {volume} {28}},\ \bibinfo {pages}
  {2331} (\bibinfo {year} {1989})}\BibitemShut {NoStop}%
\bibitem [{\citenamefont {Cheung}\ \emph {et~al.}(2004)\citenamefont {Cheung},
  \citenamefont {Zhang}, \citenamefont {Zhang}, \citenamefont {Chabanov},\ and\
  \citenamefont {Genack}}]{cheungPRL2004}%
  \BibitemOpen
  \bibfield  {author} {\bibinfo {author} {\bibfnamefont {S.~K.}\ \bibnamefont
  {Cheung}}, \bibinfo {author} {\bibfnamefont {X.}~\bibnamefont {Zhang}},
  \bibinfo {author} {\bibfnamefont {Z.~Q.}\ \bibnamefont {Zhang}}, \bibinfo
  {author} {\bibfnamefont {A.~A.}\ \bibnamefont {Chabanov}}, \ and\ \bibinfo
  {author} {\bibfnamefont {A.~Z.}\ \bibnamefont {Genack}},\ }\href {\doibase
  10.1103/PhysRevLett.92.173902} {\bibfield  {journal} {\bibinfo  {journal}
  {Phys. Rev. Lett.}\ }\textbf {\bibinfo {volume} {92}},\ \bibinfo {pages}
  {173902} (\bibinfo {year} {2004})}\BibitemShut {NoStop}%
\bibitem [{\citenamefont {St\"orzer}\ \emph
  {et~al.}(2006{\natexlab{b}})\citenamefont {St\"orzer}, \citenamefont
  {Aegerter},\ and\ \citenamefont {Maret}}]{storzerPRE2006}%
  \BibitemOpen
  \bibfield  {author} {\bibinfo {author} {\bibfnamefont {M.}~\bibnamefont
  {St\"orzer}}, \bibinfo {author} {\bibfnamefont {C.~M.}\ \bibnamefont
  {Aegerter}}, \ and\ \bibinfo {author} {\bibfnamefont {G.}~\bibnamefont
  {Maret}},\ }\href {\doibase 10.1103/PhysRevE.73.065602} {\bibfield  {journal}
  {\bibinfo  {journal} {Phys. Rev. E}\ }\textbf {\bibinfo {volume} {73}},\
  \bibinfo {pages} {065602} (\bibinfo {year} {2006}{\natexlab{b}})}\BibitemShut
  {NoStop}%
\bibitem [{\citenamefont {Störzer}(2006)}]{storzer2006anderson}%
  \BibitemOpen
  \bibfield  {author} {\bibinfo {author} {\bibfnamefont {M.}~\bibnamefont
  {Störzer}},\ }\emph {\bibinfo {title} {Anderson Localization of Light}},\
  \href@noop {} {Ph.D. thesis},\ \bibinfo  {school} {Universität Konstanz},
  \bibinfo {address} {Konstanz} (\bibinfo {year} {2006})\BibitemShut {NoStop}%
\bibitem [{\citenamefont {Sperling}(2015)}]{sperling2015experimental}%
  \BibitemOpen
  \bibfield  {author} {\bibinfo {author} {\bibfnamefont {T.}~\bibnamefont
  {Sperling}},\ }\emph {\bibinfo {title} {The Experimental Search for Anderson
  Localisation of Light in Three Dimensions}},\ \href@noop {} {Ph.D. thesis},\
  \bibinfo  {school} {Universität Konstanz}, \bibinfo {address} {Konstanz}
  (\bibinfo {year} {2015})\BibitemShut {NoStop}%
\bibitem [{\citenamefont {Torricelli}\ \emph {et~al.}(2005)\citenamefont
  {Torricelli}, \citenamefont {Pifferi}, \citenamefont {Spinelli},
  \citenamefont {Cubeddu}, \citenamefont {Martelli}, \citenamefont
  {Del~Bianco},\ and\ \citenamefont {Zaccanti}}]{torricelliPRL2005}%
  \BibitemOpen
  \bibfield  {author} {\bibinfo {author} {\bibfnamefont {A.}~\bibnamefont
  {Torricelli}}, \bibinfo {author} {\bibfnamefont {A.}~\bibnamefont {Pifferi}},
  \bibinfo {author} {\bibfnamefont {L.}~\bibnamefont {Spinelli}}, \bibinfo
  {author} {\bibfnamefont {R.}~\bibnamefont {Cubeddu}}, \bibinfo {author}
  {\bibfnamefont {F.}~\bibnamefont {Martelli}}, \bibinfo {author}
  {\bibfnamefont {S.}~\bibnamefont {Del~Bianco}}, \ and\ \bibinfo {author}
  {\bibfnamefont {G.}~\bibnamefont {Zaccanti}},\ }\href {\doibase
  10.1103/PhysRevLett.95.078101} {\bibfield  {journal} {\bibinfo  {journal}
  {Phys. Rev. Lett.}\ }\textbf {\bibinfo {volume} {95}},\ \bibinfo {pages}
  {078101} (\bibinfo {year} {2005})}\BibitemShut {NoStop}%
\bibitem [{\citenamefont {Sperling}\ \emph {et~al.}(2013)\citenamefont
  {Sperling}, \citenamefont {B{\"u}hrer}, \citenamefont {Aegerter},\ and\
  \citenamefont {Maret}}]{sperlingNaturephoton2013}%
  \BibitemOpen
  \bibfield  {author} {\bibinfo {author} {\bibfnamefont {T.}~\bibnamefont
  {Sperling}}, \bibinfo {author} {\bibfnamefont {W.}~\bibnamefont
  {B{\"u}hrer}}, \bibinfo {author} {\bibfnamefont {C.~M.}\ \bibnamefont
  {Aegerter}}, \ and\ \bibinfo {author} {\bibfnamefont {G.}~\bibnamefont
  {Maret}},\ }\href {\doibase 10.1038/nphoton.2012.313} {\bibfield  {journal}
  {\bibinfo  {journal} {Nature Photonics}\ }\textbf {\bibinfo {volume} {7}},\
  \bibinfo {pages} {48} (\bibinfo {year} {2013})}\BibitemShut {NoStop}%
\bibitem [{\citenamefont {Selb}\ \emph {et~al.}(2006)\citenamefont {Selb},
  \citenamefont {Joseph},\ and\ \citenamefont {Boas}}]{selbJBO2006}%
  \BibitemOpen
  \bibfield  {author} {\bibinfo {author} {\bibfnamefont {J.~J.}\ \bibnamefont
  {Selb}}, \bibinfo {author} {\bibfnamefont {D.~K.}\ \bibnamefont {Joseph}}, \
  and\ \bibinfo {author} {\bibfnamefont {D.~A.}\ \bibnamefont {Boas}},\ }\href
  {\doibase 10.1117/1.2337320} {\bibfield  {journal} {\bibinfo  {journal}
  {Journal of Biomedical Optics}\ }\textbf {\bibinfo {volume} {11}},\ \bibinfo
  {pages} {1 } (\bibinfo {year} {2006})}\BibitemShut {NoStop}%
\bibitem [{\citenamefont {Contini}\ \emph {et~al.}(2006)\citenamefont
  {Contini}, \citenamefont {Torricelli}, \citenamefont {Pifferi}, \citenamefont
  {Spinelli}, \citenamefont {Paglia},\ and\ \citenamefont
  {Cubeddu}}]{continiOE2006}%
  \BibitemOpen
  \bibfield  {author} {\bibinfo {author} {\bibfnamefont {D.}~\bibnamefont
  {Contini}}, \bibinfo {author} {\bibfnamefont {A.}~\bibnamefont {Torricelli}},
  \bibinfo {author} {\bibfnamefont {A.}~\bibnamefont {Pifferi}}, \bibinfo
  {author} {\bibfnamefont {L.}~\bibnamefont {Spinelli}}, \bibinfo {author}
  {\bibfnamefont {F.}~\bibnamefont {Paglia}}, \ and\ \bibinfo {author}
  {\bibfnamefont {R.}~\bibnamefont {Cubeddu}},\ }\href {\doibase
  10.1364/OE.14.005418} {\bibfield  {journal} {\bibinfo  {journal} {Opt.
  Express}\ }\textbf {\bibinfo {volume} {14}},\ \bibinfo {pages} {5418}
  (\bibinfo {year} {2006})}\BibitemShut {NoStop}%
\bibitem [{\citenamefont {Pifferi}\ \emph {et~al.}(2016)\citenamefont
  {Pifferi}, \citenamefont {Contini}, \citenamefont {Mora}, \citenamefont
  {Farina}, \citenamefont {Spinelli},\ and\ \citenamefont
  {Torricelli}}]{pifferiJBO2016}%
  \BibitemOpen
  \bibfield  {author} {\bibinfo {author} {\bibfnamefont {A.}~\bibnamefont
  {Pifferi}}, \bibinfo {author} {\bibfnamefont {D.}~\bibnamefont {Contini}},
  \bibinfo {author} {\bibfnamefont {A.~D.}\ \bibnamefont {Mora}}, \bibinfo
  {author} {\bibfnamefont {A.}~\bibnamefont {Farina}}, \bibinfo {author}
  {\bibfnamefont {L.}~\bibnamefont {Spinelli}}, \ and\ \bibinfo {author}
  {\bibfnamefont {A.}~\bibnamefont {Torricelli}},\ }\href {\doibase
  10.1117/1.JBO.21.9.091310} {\bibfield  {journal} {\bibinfo  {journal}
  {Journal of Biomedical Optics}\ }\textbf {\bibinfo {volume} {21}},\ \bibinfo
  {pages} {1 } (\bibinfo {year} {2016})}\BibitemShut {NoStop}%
\bibitem [{\citenamefont {Alayed}\ and\ \citenamefont
  {Deen}(2017)}]{alayedSensors2017}%
  \BibitemOpen
  \bibfield  {author} {\bibinfo {author} {\bibfnamefont {M.}~\bibnamefont
  {Alayed}}\ and\ \bibinfo {author} {\bibfnamefont {M.~J.}\ \bibnamefont
  {Deen}},\ }\href {\doibase 10.3390/s17092115} {\bibfield  {journal} {\bibinfo
   {journal} {Sensors}\ }\textbf {\bibinfo {volume} {17}} (\bibinfo {year}
  {2017}),\ 10.3390/s17092115}\BibitemShut {NoStop}%
\bibitem [{\citenamefont {Tosi}\ \emph {et~al.}(2011)\citenamefont {Tosi},
  \citenamefont {Mora}, \citenamefont {Zappa}, \citenamefont {Gulinatti},
  \citenamefont {Contini}, \citenamefont {Pifferi}, \citenamefont {Spinelli},
  \citenamefont {Torricelli},\ and\ \citenamefont {Cubeddu}}]{tosiOE2011}%
  \BibitemOpen
  \bibfield  {author} {\bibinfo {author} {\bibfnamefont {A.}~\bibnamefont
  {Tosi}}, \bibinfo {author} {\bibfnamefont {A.~D.}\ \bibnamefont {Mora}},
  \bibinfo {author} {\bibfnamefont {F.}~\bibnamefont {Zappa}}, \bibinfo
  {author} {\bibfnamefont {A.}~\bibnamefont {Gulinatti}}, \bibinfo {author}
  {\bibfnamefont {D.}~\bibnamefont {Contini}}, \bibinfo {author} {\bibfnamefont
  {A.}~\bibnamefont {Pifferi}}, \bibinfo {author} {\bibfnamefont
  {L.}~\bibnamefont {Spinelli}}, \bibinfo {author} {\bibfnamefont
  {A.}~\bibnamefont {Torricelli}}, \ and\ \bibinfo {author} {\bibfnamefont
  {R.}~\bibnamefont {Cubeddu}},\ }\href {\doibase 10.1364/OE.19.010735}
  {\bibfield  {journal} {\bibinfo  {journal} {Opt. Express}\ }\textbf {\bibinfo
  {volume} {19}},\ \bibinfo {pages} {10735} (\bibinfo {year}
  {2011})}\BibitemShut {NoStop}%
\bibitem [{\citenamefont {Popescu}\ and\ \citenamefont
  {Dogariu}(1999)}]{popescuOL1999}%
  \BibitemOpen
  \bibfield  {author} {\bibinfo {author} {\bibfnamefont {G.}~\bibnamefont
  {Popescu}}\ and\ \bibinfo {author} {\bibfnamefont {A.}~\bibnamefont
  {Dogariu}},\ }\href {\doibase 10.1364/OL.24.000442} {\bibfield  {journal}
  {\bibinfo  {journal} {Opt. Lett.}\ }\textbf {\bibinfo {volume} {24}},\
  \bibinfo {pages} {442} (\bibinfo {year} {1999})}\BibitemShut {NoStop}%
\bibitem [{\citenamefont {Popescu}\ \emph {et~al.}(2000)\citenamefont
  {Popescu}, \citenamefont {Mujat},\ and\ \citenamefont
  {Dogariu}}]{popescuPRE2000}%
  \BibitemOpen
  \bibfield  {author} {\bibinfo {author} {\bibfnamefont {G.}~\bibnamefont
  {Popescu}}, \bibinfo {author} {\bibfnamefont {C.}~\bibnamefont {Mujat}}, \
  and\ \bibinfo {author} {\bibfnamefont {A.}~\bibnamefont {Dogariu}},\ }\href
  {\doibase 10.1103/PhysRevE.61.4523} {\bibfield  {journal} {\bibinfo
  {journal} {Phys. Rev. E}\ }\textbf {\bibinfo {volume} {61}},\ \bibinfo
  {pages} {4523} (\bibinfo {year} {2000})}\BibitemShut {NoStop}%
\bibitem [{\citenamefont {Tualle}\ \emph {et~al.}(2001)\citenamefont {Tualle},
  \citenamefont {Tinet},\ and\ \citenamefont {Avrillier}}]{tualleOC2001}%
  \BibitemOpen
  \bibfield  {author} {\bibinfo {author} {\bibfnamefont {J.-M.}\ \bibnamefont
  {Tualle}}, \bibinfo {author} {\bibfnamefont {E.}~\bibnamefont {Tinet}}, \
  and\ \bibinfo {author} {\bibfnamefont {S.}~\bibnamefont {Avrillier}},\ }\href
  {\doibase https://doi.org/10.1016/S0030-4018(01)01045-8} {\bibfield
  {journal} {\bibinfo  {journal} {Optics Communications}\ }\textbf {\bibinfo
  {volume} {189}},\ \bibinfo {pages} {211 } (\bibinfo {year}
  {2001})}\BibitemShut {NoStop}%
\bibitem [{\citenamefont {Weiss}\ \emph {et~al.}(2013)\citenamefont {Weiss},
  \citenamefont {van Leeuwen},\ and\ \citenamefont {Kalkman}}]{weissPRE2013}%
  \BibitemOpen
  \bibfield  {author} {\bibinfo {author} {\bibfnamefont {N.}~\bibnamefont
  {Weiss}}, \bibinfo {author} {\bibfnamefont {T.~G.}\ \bibnamefont {van
  Leeuwen}}, \ and\ \bibinfo {author} {\bibfnamefont {J.}~\bibnamefont
  {Kalkman}},\ }\href {\doibase 10.1103/PhysRevE.88.042312} {\bibfield
  {journal} {\bibinfo  {journal} {Phys. Rev. E}\ }\textbf {\bibinfo {volume}
  {88}},\ \bibinfo {pages} {042312} (\bibinfo {year} {2013})}\BibitemShut
  {NoStop}%
\bibitem [{\citenamefont {Badon}\ \emph {et~al.}(2015)\citenamefont {Badon},
  \citenamefont {Lerosey}, \citenamefont {Boccara}, \citenamefont {Fink},\ and\
  \citenamefont {Aubry}}]{badonPRL2015}%
  \BibitemOpen
  \bibfield  {author} {\bibinfo {author} {\bibfnamefont {A.}~\bibnamefont
  {Badon}}, \bibinfo {author} {\bibfnamefont {G.}~\bibnamefont {Lerosey}},
  \bibinfo {author} {\bibfnamefont {A.~C.}\ \bibnamefont {Boccara}}, \bibinfo
  {author} {\bibfnamefont {M.}~\bibnamefont {Fink}}, \ and\ \bibinfo {author}
  {\bibfnamefont {A.}~\bibnamefont {Aubry}},\ }\href {\doibase
  10.1103/PhysRevLett.114.023901} {\bibfield  {journal} {\bibinfo  {journal}
  {Phys. Rev. Lett.}\ }\textbf {\bibinfo {volume} {114}},\ \bibinfo {pages}
  {023901} (\bibinfo {year} {2015})}\BibitemShut {NoStop}%
\bibitem [{\citenamefont {Badon}\ \emph {et~al.}(2016)\citenamefont {Badon},
  \citenamefont {Li}, \citenamefont {Lerosey}, \citenamefont {Boccara},
  \citenamefont {Fink},\ and\ \citenamefont {Aubry}}]{badonOptica2016}%
  \BibitemOpen
  \bibfield  {author} {\bibinfo {author} {\bibfnamefont {A.}~\bibnamefont
  {Badon}}, \bibinfo {author} {\bibfnamefont {D.}~\bibnamefont {Li}}, \bibinfo
  {author} {\bibfnamefont {G.}~\bibnamefont {Lerosey}}, \bibinfo {author}
  {\bibfnamefont {A.~C.}\ \bibnamefont {Boccara}}, \bibinfo {author}
  {\bibfnamefont {M.}~\bibnamefont {Fink}}, \ and\ \bibinfo {author}
  {\bibfnamefont {A.}~\bibnamefont {Aubry}},\ }\href {\doibase
  10.1364/OPTICA.3.001160} {\bibfield  {journal} {\bibinfo  {journal} {Optica}\
  }\textbf {\bibinfo {volume} {3}},\ \bibinfo {pages} {1160} (\bibinfo {year}
  {2016})}\BibitemShut {NoStop}%
\bibitem [{\citenamefont {Sankaran}\ \emph {et~al.}(2000)\citenamefont
  {Sankaran}, \citenamefont {Walsh},\ and\ \citenamefont
  {Maitland}}]{sankaranOL2000}%
  \BibitemOpen
  \bibfield  {author} {\bibinfo {author} {\bibfnamefont {V.}~\bibnamefont
  {Sankaran}}, \bibinfo {author} {\bibfnamefont {J.~T.}\ \bibnamefont {Walsh}},
  \ and\ \bibinfo {author} {\bibfnamefont {D.~J.}\ \bibnamefont {Maitland}},\
  }\href {\doibase 10.1364/OL.25.000239} {\bibfield  {journal} {\bibinfo
  {journal} {Opt. Lett.}\ }\textbf {\bibinfo {volume} {25}},\ \bibinfo {pages}
  {239} (\bibinfo {year} {2000})}\BibitemShut {NoStop}%
\bibitem [{\citenamefont {Groenhuis}\ \emph
  {et~al.}(1983{\natexlab{a}})\citenamefont {Groenhuis}, \citenamefont
  {Ferwerda},\ and\ \citenamefont {Bosch}}]{groenhuisAO1983a}%
  \BibitemOpen
  \bibfield  {author} {\bibinfo {author} {\bibfnamefont {R.~A.~J.}\
  \bibnamefont {Groenhuis}}, \bibinfo {author} {\bibfnamefont {H.~A.}\
  \bibnamefont {Ferwerda}}, \ and\ \bibinfo {author} {\bibfnamefont {J.~J.~T.}\
  \bibnamefont {Bosch}},\ }\href {\doibase 10.1364/AO.22.002456} {\bibfield
  {journal} {\bibinfo  {journal} {Appl. Opt.}\ }\textbf {\bibinfo {volume}
  {22}},\ \bibinfo {pages} {2456} (\bibinfo {year}
  {1983}{\natexlab{a}})}\BibitemShut {NoStop}%
\bibitem [{\citenamefont {Groenhuis}\ \emph
  {et~al.}(1983{\natexlab{b}})\citenamefont {Groenhuis}, \citenamefont
  {Bosch},\ and\ \citenamefont {Ferwerda}}]{groenhuisAO1983b}%
  \BibitemOpen
  \bibfield  {author} {\bibinfo {author} {\bibfnamefont {R.~A.~J.}\
  \bibnamefont {Groenhuis}}, \bibinfo {author} {\bibfnamefont {J.~J.~T.}\
  \bibnamefont {Bosch}}, \ and\ \bibinfo {author} {\bibfnamefont {H.~A.}\
  \bibnamefont {Ferwerda}},\ }\href {\doibase 10.1364/AO.22.002463} {\bibfield
  {journal} {\bibinfo  {journal} {Appl. Opt.}\ }\textbf {\bibinfo {volume}
  {22}},\ \bibinfo {pages} {2463} (\bibinfo {year}
  {1983}{\natexlab{b}})}\BibitemShut {NoStop}%
\bibitem [{\citenamefont {Farrell}\ \emph {et~al.}(1992)\citenamefont
  {Farrell}, \citenamefont {Patterson},\ and\ \citenamefont
  {Wilson}}]{farrellMP1992}%
  \BibitemOpen
  \bibfield  {author} {\bibinfo {author} {\bibfnamefont {T.~J.}\ \bibnamefont
  {Farrell}}, \bibinfo {author} {\bibfnamefont {M.~S.}\ \bibnamefont
  {Patterson}}, \ and\ \bibinfo {author} {\bibfnamefont {B.}~\bibnamefont
  {Wilson}},\ }\href {\doibase 10.1118/1.596777} {\bibfield  {journal}
  {\bibinfo  {journal} {Medical Physics}\ }\textbf {\bibinfo {volume} {19}},\
  \bibinfo {pages} {879} (\bibinfo {year} {1992})},\ \Eprint
  {http://arxiv.org/abs/https://aapm.onlinelibrary.wiley.com/doi/pdf/10.1118/1.596777}
  {https://aapm.onlinelibrary.wiley.com/doi/pdf/10.1118/1.596777} \BibitemShut
  {NoStop}%
\bibitem [{\citenamefont {Durian}\ and\ \citenamefont
  {Rudnick}(1999)}]{durianJOSAA1999}%
  \BibitemOpen
  \bibfield  {author} {\bibinfo {author} {\bibfnamefont {D.~J.}\ \bibnamefont
  {Durian}}\ and\ \bibinfo {author} {\bibfnamefont {J.}~\bibnamefont
  {Rudnick}},\ }\href {\doibase 10.1364/JOSAA.16.000837} {\bibfield  {journal}
  {\bibinfo  {journal} {J. Opt. Soc. Am. A}\ }\textbf {\bibinfo {volume}
  {16}},\ \bibinfo {pages} {837} (\bibinfo {year} {1999})}\BibitemShut
  {NoStop}%
\bibitem [{\citenamefont {Kim}\ \emph {et~al.}(2013)\citenamefont {Kim},
  \citenamefont {Sridharan},\ and\ \citenamefont {Popescu}}]{kim2013fourier}%
  \BibitemOpen
  \bibfield  {author} {\bibinfo {author} {\bibfnamefont {T.}~\bibnamefont
  {Kim}}, \bibinfo {author} {\bibfnamefont {S.}~\bibnamefont {Sridharan}}, \
  and\ \bibinfo {author} {\bibfnamefont {G.}~\bibnamefont {Popescu}},\
  }\enquote {\bibinfo {title} {Fourier transform light scattering of
  tissues},}\ in\ \href {\doibase 10.1007/978-1-4614-5176-1_7} {\emph {\bibinfo
  {booktitle} {Handbook of Coherent-Domain Optical Methods: Biomedical
  Diagnostics, Environmental Monitoring, and Materials Science}}},\ \bibinfo
  {editor} {edited by\ \bibinfo {editor} {\bibfnamefont {V.~V.}\ \bibnamefont
  {Tuchin}}}\ (\bibinfo  {publisher} {Springer New York},\ \bibinfo {address}
  {New York, NY},\ \bibinfo {year} {2013})\ pp.\ \bibinfo {pages}
  {259--290}\BibitemShut {NoStop}%
\bibitem [{\citenamefont {Nguyen}\ \emph {et~al.}(2013)\citenamefont {Nguyen},
  \citenamefont {Faber}, \citenamefont {van~der Pol}, \citenamefont {van
  Leeuwen},\ and\ \citenamefont {Kalkman}}]{nguyenOE2013}%
  \BibitemOpen
  \bibfield  {author} {\bibinfo {author} {\bibfnamefont {V.~D.}\ \bibnamefont
  {Nguyen}}, \bibinfo {author} {\bibfnamefont {D.~J.}\ \bibnamefont {Faber}},
  \bibinfo {author} {\bibfnamefont {E.}~\bibnamefont {van~der Pol}}, \bibinfo
  {author} {\bibfnamefont {T.~G.}\ \bibnamefont {van Leeuwen}}, \ and\ \bibinfo
  {author} {\bibfnamefont {J.}~\bibnamefont {Kalkman}},\ }\href {\doibase
  10.1364/OE.21.029145} {\bibfield  {journal} {\bibinfo  {journal} {Opt.
  Express}\ }\textbf {\bibinfo {volume} {21}},\ \bibinfo {pages} {29145}
  (\bibinfo {year} {2013})}\BibitemShut {NoStop}%
\bibitem [{\citenamefont {Almasian}\ \emph {et~al.}(2015)\citenamefont
  {Almasian}, \citenamefont {Bosschaart}, \citenamefont {van Leeuwen},\ and\
  \citenamefont {Faber}}]{almasianJBO2015}%
  \BibitemOpen
  \bibfield  {author} {\bibinfo {author} {\bibfnamefont {M.}~\bibnamefont
  {Almasian}}, \bibinfo {author} {\bibfnamefont {N.}~\bibnamefont
  {Bosschaart}}, \bibinfo {author} {\bibfnamefont {T.~G.}\ \bibnamefont {van
  Leeuwen}}, \ and\ \bibinfo {author} {\bibfnamefont {D.~J.}\ \bibnamefont
  {Faber}},\ }\href {\doibase 10.1117/1.JBO.20.12.121314} {\bibfield  {journal}
  {\bibinfo  {journal} {Journal of Biomedical Optics}\ }\textbf {\bibinfo
  {volume} {20}},\ \bibinfo {pages} {1 } (\bibinfo {year} {2015})}\BibitemShut
  {NoStop}%
\bibitem [{\citenamefont {Fishkin}\ \emph {et~al.}(1996)\citenamefont
  {Fishkin}, \citenamefont {Fantini}, \citenamefont {vandeVen},\ and\
  \citenamefont {Gratton}}]{fishkinPRE1996}%
  \BibitemOpen
  \bibfield  {author} {\bibinfo {author} {\bibfnamefont {J.~B.}\ \bibnamefont
  {Fishkin}}, \bibinfo {author} {\bibfnamefont {S.}~\bibnamefont {Fantini}},
  \bibinfo {author} {\bibfnamefont {M.~J.}\ \bibnamefont {vandeVen}}, \ and\
  \bibinfo {author} {\bibfnamefont {E.}~\bibnamefont {Gratton}},\ }\href
  {\doibase 10.1103/PhysRevE.53.2307} {\bibfield  {journal} {\bibinfo
  {journal} {Phys. Rev. E}\ }\textbf {\bibinfo {volume} {53}},\ \bibinfo
  {pages} {2307} (\bibinfo {year} {1996})}\BibitemShut {NoStop}%
\bibitem [{\citenamefont {Sun}\ \emph {et~al.}(2002)\citenamefont {Sun},
  \citenamefont {Huang},\ and\ \citenamefont {Sevick-Muraca}}]{sunRSI2002}%
  \BibitemOpen
  \bibfield  {author} {\bibinfo {author} {\bibfnamefont {Z.}~\bibnamefont
  {Sun}}, \bibinfo {author} {\bibfnamefont {Y.}~\bibnamefont {Huang}}, \ and\
  \bibinfo {author} {\bibfnamefont {E.~M.}\ \bibnamefont {Sevick-Muraca}},\
  }\href {\doibase 10.1063/1.1427303} {\bibfield  {journal} {\bibinfo
  {journal} {Review of Scientific Instruments}\ }\textbf {\bibinfo {volume}
  {73}},\ \bibinfo {pages} {383} (\bibinfo {year} {2002})},\ \Eprint
  {http://arxiv.org/abs/https://doi.org/10.1063/1.1427303}
  {https://doi.org/10.1063/1.1427303} \BibitemShut {NoStop}%
\bibitem [{\citenamefont {O'Sullivan}\ \emph {et~al.}(2012)\citenamefont
  {O'Sullivan}, \citenamefont {Cerussi}, \citenamefont {Tromberg},\ and\
  \citenamefont {Cuccia}}]{osullivanJBO2012}%
  \BibitemOpen
  \bibfield  {author} {\bibinfo {author} {\bibfnamefont {T.~D.}\ \bibnamefont
  {O'Sullivan}}, \bibinfo {author} {\bibfnamefont {A.~E.}\ \bibnamefont
  {Cerussi}}, \bibinfo {author} {\bibfnamefont {B.~J.}\ \bibnamefont
  {Tromberg}}, \ and\ \bibinfo {author} {\bibfnamefont {D.~J.}\ \bibnamefont
  {Cuccia}},\ }\href {\doibase 10.1117/1.JBO.17.7.071311} {\bibfield  {journal}
  {\bibinfo  {journal} {Journal of Biomedical Optics}\ }\textbf {\bibinfo
  {volume} {17}},\ \bibinfo {pages} {1 } (\bibinfo {year} {2012})}\BibitemShut
  {NoStop}%
\bibitem [{\citenamefont {Cuccia}\ \emph {et~al.}(2005)\citenamefont {Cuccia},
  \citenamefont {Bevilacqua}, \citenamefont {Durkin},\ and\ \citenamefont
  {Tromberg}}]{cucciaOL2005}%
  \BibitemOpen
  \bibfield  {author} {\bibinfo {author} {\bibfnamefont {D.~J.}\ \bibnamefont
  {Cuccia}}, \bibinfo {author} {\bibfnamefont {F.}~\bibnamefont {Bevilacqua}},
  \bibinfo {author} {\bibfnamefont {A.~J.}\ \bibnamefont {Durkin}}, \ and\
  \bibinfo {author} {\bibfnamefont {B.~J.}\ \bibnamefont {Tromberg}},\ }\href
  {\doibase 10.1364/OL.30.001354} {\bibfield  {journal} {\bibinfo  {journal}
  {Opt. Lett.}\ }\textbf {\bibinfo {volume} {30}},\ \bibinfo {pages} {1354}
  (\bibinfo {year} {2005})}\BibitemShut {NoStop}%
\bibitem [{\citenamefont {Cuccia}\ \emph {et~al.}(2009)\citenamefont {Cuccia},
  \citenamefont {Bevilacqua}, \citenamefont {Durkin}, \citenamefont {Ayers},\
  and\ \citenamefont {Tromberg}}]{cucciaJBO2009}%
  \BibitemOpen
  \bibfield  {author} {\bibinfo {author} {\bibfnamefont {D.~J.}\ \bibnamefont
  {Cuccia}}, \bibinfo {author} {\bibfnamefont {F.~P.}\ \bibnamefont
  {Bevilacqua}}, \bibinfo {author} {\bibfnamefont {A.~J.}\ \bibnamefont
  {Durkin}}, \bibinfo {author} {\bibfnamefont {F.~R.}\ \bibnamefont {Ayers}}, \
  and\ \bibinfo {author} {\bibfnamefont {B.~J.}\ \bibnamefont {Tromberg}},\
  }\href {\doibase 10.1117/1.3088140} {\bibfield  {journal} {\bibinfo
  {journal} {Journal of Biomedical Optics}\ }\textbf {\bibinfo {volume} {14}},\
  \bibinfo {pages} {1 } (\bibinfo {year} {2009})}\BibitemShut {NoStop}%
\bibitem [{\citenamefont {Kanick}\ \emph {et~al.}(2014)\citenamefont {Kanick},
  \citenamefont {McClatchy}, \citenamefont {Krishnaswamy}, \citenamefont
  {Elliott}, \citenamefont {Paulsen},\ and\ \citenamefont
  {Pogue}}]{kanickBOE2014}%
  \BibitemOpen
  \bibfield  {author} {\bibinfo {author} {\bibfnamefont {S.~C.}\ \bibnamefont
  {Kanick}}, \bibinfo {author} {\bibfnamefont {D.~M.}\ \bibnamefont
  {McClatchy}}, \bibinfo {author} {\bibfnamefont {V.}~\bibnamefont
  {Krishnaswamy}}, \bibinfo {author} {\bibfnamefont {J.~T.}\ \bibnamefont
  {Elliott}}, \bibinfo {author} {\bibfnamefont {K.~D.}\ \bibnamefont
  {Paulsen}}, \ and\ \bibinfo {author} {\bibfnamefont {B.~W.}\ \bibnamefont
  {Pogue}},\ }\href {\doibase 10.1364/BOE.5.003376} {\bibfield  {journal}
  {\bibinfo  {journal} {Biomed. Opt. Express}\ }\textbf {\bibinfo {volume}
  {5}},\ \bibinfo {pages} {3376} (\bibinfo {year} {2014})}\BibitemShut
  {NoStop}%
\bibitem [{\citenamefont {Gioux}\ \emph {et~al.}(2019)\citenamefont {Gioux},
  \citenamefont {Mazhar},\ and\ \citenamefont {Cuccia}}]{giouxJBO2019}%
  \BibitemOpen
  \bibfield  {author} {\bibinfo {author} {\bibfnamefont {S.}~\bibnamefont
  {Gioux}}, \bibinfo {author} {\bibfnamefont {A.}~\bibnamefont {Mazhar}}, \
  and\ \bibinfo {author} {\bibfnamefont {D.~J.}\ \bibnamefont {Cuccia}},\
  }\href {\doibase 10.1117/1.JBO.24.7.071613} {\bibfield  {journal} {\bibinfo
  {journal} {Journal of Biomedical Optics}\ }\textbf {\bibinfo {volume} {24}},\
  \bibinfo {pages} {1 } (\bibinfo {year} {2019})}\BibitemShut {NoStop}%
\bibitem [{\citenamefont {Berne}\ and\ \citenamefont
  {Pecora}(2000)}]{berne2000dynamic}%
  \BibitemOpen
  \bibfield  {author} {\bibinfo {author} {\bibfnamefont {B.~J.}\ \bibnamefont
  {Berne}}\ and\ \bibinfo {author} {\bibfnamefont {R.}~\bibnamefont {Pecora}},\
  }\href@noop {} {\emph {\bibinfo {title} {{Dynamic light scattering: with
  applications to chemistry, biology, and physics}}}},\ Dover Books on Physics\
  (\bibinfo  {publisher} {Dover},\ \bibinfo {address} {New York, NY},\ \bibinfo
  {year} {2000})\BibitemShut {NoStop}%
\bibitem [{\citenamefont {Gulari}\ \emph {et~al.}(1979)\citenamefont {Gulari},
  \citenamefont {Gulari}, \citenamefont {Tsunashima},\ and\ \citenamefont
  {Chu}}]{gulariJCP1979}%
  \BibitemOpen
  \bibfield  {author} {\bibinfo {author} {\bibfnamefont {E.}~\bibnamefont
  {Gulari}}, \bibinfo {author} {\bibfnamefont {E.}~\bibnamefont {Gulari}},
  \bibinfo {author} {\bibfnamefont {Y.}~\bibnamefont {Tsunashima}}, \ and\
  \bibinfo {author} {\bibfnamefont {B.}~\bibnamefont {Chu}},\ }\href {\doibase
  10.1063/1.437950} {\bibfield  {journal} {\bibinfo  {journal} {The Journal of
  Chemical Physics}\ }\textbf {\bibinfo {volume} {70}},\ \bibinfo {pages}
  {3965} (\bibinfo {year} {1979})},\ \Eprint
  {http://arxiv.org/abs/https://doi.org/10.1063/1.437950}
  {https://doi.org/10.1063/1.437950} \BibitemShut {NoStop}%
\bibitem [{\citenamefont {Pusey}\ and\ \citenamefont {van
  Megen}(1984)}]{puseyJCP1984}%
  \BibitemOpen
  \bibfield  {author} {\bibinfo {author} {\bibfnamefont {P.~N.}\ \bibnamefont
  {Pusey}}\ and\ \bibinfo {author} {\bibfnamefont {W.}~\bibnamefont {van
  Megen}},\ }\href {\doibase 10.1063/1.447195} {\bibfield  {journal} {\bibinfo
  {journal} {The Journal of Chemical Physics}\ }\textbf {\bibinfo {volume}
  {80}},\ \bibinfo {pages} {3513} (\bibinfo {year} {1984})},\ \Eprint
  {http://arxiv.org/abs/https://doi.org/10.1063/1.447195}
  {https://doi.org/10.1063/1.447195} \BibitemShut {NoStop}%
\bibitem [{\citenamefont {Maret}\ and\ \citenamefont
  {Wolf}(1987)}]{maretZPB1987}%
  \BibitemOpen
  \bibfield  {author} {\bibinfo {author} {\bibfnamefont {G.}~\bibnamefont
  {Maret}}\ and\ \bibinfo {author} {\bibfnamefont {P.~E.}\ \bibnamefont
  {Wolf}},\ }\href {\doibase 10.1007/BF01303762} {\bibfield  {journal}
  {\bibinfo  {journal} {Zeitschrift f{\"u}r Physik B Condensed Matter}\
  }\textbf {\bibinfo {volume} {65}},\ \bibinfo {pages} {409} (\bibinfo {year}
  {1987})}\BibitemShut {NoStop}%
\bibitem [{\citenamefont {Viasnoff}\ \emph {et~al.}(2002)\citenamefont
  {Viasnoff}, \citenamefont {Lequeux},\ and\ \citenamefont
  {Pine}}]{viasnoffRSI2002}%
  \BibitemOpen
  \bibfield  {author} {\bibinfo {author} {\bibfnamefont {V.}~\bibnamefont
  {Viasnoff}}, \bibinfo {author} {\bibfnamefont {F.}~\bibnamefont {Lequeux}}, \
  and\ \bibinfo {author} {\bibfnamefont {D.~J.}\ \bibnamefont {Pine}},\ }\href
  {\doibase 10.1063/1.1476699} {\bibfield  {journal} {\bibinfo  {journal}
  {Review of Scientific Instruments}\ }\textbf {\bibinfo {volume} {73}},\
  \bibinfo {pages} {2336} (\bibinfo {year} {2002})},\ \Eprint
  {http://arxiv.org/abs/https://doi.org/10.1063/1.1476699}
  {https://doi.org/10.1063/1.1476699} \BibitemShut {NoStop}%
\bibitem [{\citenamefont {Morin}\ \emph {et~al.}(2002)\citenamefont {Morin},
  \citenamefont {Borrega}, \citenamefont {Cloitre},\ and\ \citenamefont
  {Durian}}]{morinAO2002}%
  \BibitemOpen
  \bibfield  {author} {\bibinfo {author} {\bibfnamefont {F.}~\bibnamefont
  {Morin}}, \bibinfo {author} {\bibfnamefont {R.}~\bibnamefont {Borrega}},
  \bibinfo {author} {\bibfnamefont {M.}~\bibnamefont {Cloitre}}, \ and\
  \bibinfo {author} {\bibfnamefont {D.}~\bibnamefont {Durian}},\ }\href
  {\doibase 10.1364/AO.41.007294} {\bibfield  {journal} {\bibinfo  {journal}
  {Appl. Opt.}\ }\textbf {\bibinfo {volume} {41}},\ \bibinfo {pages} {7294}
  (\bibinfo {year} {2002})}\BibitemShut {NoStop}%
\bibitem [{\citenamefont {Yeh}\ and\ \citenamefont
  {Cummins}(1964)}]{yehAPL1964}%
  \BibitemOpen
  \bibfield  {author} {\bibinfo {author} {\bibfnamefont {Y.}~\bibnamefont
  {Yeh}}\ and\ \bibinfo {author} {\bibfnamefont {H.~Z.}\ \bibnamefont
  {Cummins}},\ }\href {\doibase 10.1063/1.1753925} {\bibfield  {journal}
  {\bibinfo  {journal} {Applied Physics Letters}\ }\textbf {\bibinfo {volume}
  {4}},\ \bibinfo {pages} {176} (\bibinfo {year} {1964})},\ \Eprint
  {http://arxiv.org/abs/https://doi.org/10.1063/1.1753925}
  {https://doi.org/10.1063/1.1753925} \BibitemShut {NoStop}%
\bibitem [{\citenamefont {Kienle}\ \emph {et~al.}(1996)\citenamefont {Kienle},
  \citenamefont {Patterson}, \citenamefont {Ott},\ and\ \citenamefont
  {Steiner}}]{kienleAO1996}%
  \BibitemOpen
  \bibfield  {author} {\bibinfo {author} {\bibfnamefont {A.}~\bibnamefont
  {Kienle}}, \bibinfo {author} {\bibfnamefont {M.~S.}\ \bibnamefont
  {Patterson}}, \bibinfo {author} {\bibfnamefont {L.}~\bibnamefont {Ott}}, \
  and\ \bibinfo {author} {\bibfnamefont {R.}~\bibnamefont {Steiner}},\ }\href
  {\doibase 10.1364/AO.35.003404} {\bibfield  {journal} {\bibinfo  {journal}
  {Appl. Opt.}\ }\textbf {\bibinfo {volume} {35}},\ \bibinfo {pages} {3404}
  (\bibinfo {year} {1996})}\BibitemShut {NoStop}%
\bibitem [{\citenamefont {Savo}\ \emph {et~al.}(2017)\citenamefont {Savo},
  \citenamefont {Pierrat}, \citenamefont {Najar}, \citenamefont {Carminati},
  \citenamefont {Rotter},\ and\ \citenamefont {Gigan}}]{savoScience2017}%
  \BibitemOpen
  \bibfield  {author} {\bibinfo {author} {\bibfnamefont {R.}~\bibnamefont
  {Savo}}, \bibinfo {author} {\bibfnamefont {R.}~\bibnamefont {Pierrat}},
  \bibinfo {author} {\bibfnamefont {U.}~\bibnamefont {Najar}}, \bibinfo
  {author} {\bibfnamefont {R.}~\bibnamefont {Carminati}}, \bibinfo {author}
  {\bibfnamefont {S.}~\bibnamefont {Rotter}}, \ and\ \bibinfo {author}
  {\bibfnamefont {S.}~\bibnamefont {Gigan}},\ }\href {\doibase
  10.1126/science.aan4054} {\bibfield  {journal} {\bibinfo  {journal}
  {Science}\ }\textbf {\bibinfo {volume} {358}},\ \bibinfo {pages} {765}
  (\bibinfo {year} {2017})},\ \Eprint
  {http://arxiv.org/abs/https://science.sciencemag.org/content/358/6364/765.full.pdf}
  {https://science.sciencemag.org/content/358/6364/765.full.pdf} \BibitemShut
  {NoStop}%
\bibitem [{\citenamefont {Hass}\ \emph {et~al.}(2013)\citenamefont {Hass},
  \citenamefont {M\"{u}nzberg}, \citenamefont {Bressel},\ and\ \citenamefont
  {Reich}}]{hassAO2013}%
  \BibitemOpen
  \bibfield  {author} {\bibinfo {author} {\bibfnamefont {R.}~\bibnamefont
  {Hass}}, \bibinfo {author} {\bibfnamefont {M.}~\bibnamefont {M\"{u}nzberg}},
  \bibinfo {author} {\bibfnamefont {L.}~\bibnamefont {Bressel}}, \ and\
  \bibinfo {author} {\bibfnamefont {O.}~\bibnamefont {Reich}},\ }\href
  {\doibase 10.1364/AO.52.001423} {\bibfield  {journal} {\bibinfo  {journal}
  {Appl. Opt.}\ }\textbf {\bibinfo {volume} {52}},\ \bibinfo {pages} {1423}
  (\bibinfo {year} {2013})}\BibitemShut {NoStop}%
\bibitem [{\citenamefont {Kim}\ \emph {et~al.}(2019)\citenamefont {Kim},
  \citenamefont {{\c S}enbil}, \citenamefont {Zhang}, \citenamefont
  {Scheffold},\ and\ \citenamefont {Mason}}]{kimPNAS2019}%
  \BibitemOpen
  \bibfield  {author} {\bibinfo {author} {\bibfnamefont {H.~S.}\ \bibnamefont
  {Kim}}, \bibinfo {author} {\bibfnamefont {N.}~\bibnamefont {{\c S}enbil}},
  \bibinfo {author} {\bibfnamefont {C.}~\bibnamefont {Zhang}}, \bibinfo
  {author} {\bibfnamefont {F.}~\bibnamefont {Scheffold}}, \ and\ \bibinfo
  {author} {\bibfnamefont {T.~G.}\ \bibnamefont {Mason}},\ }\href {\doibase
  10.1073/pnas.1817029116} {\bibfield  {journal} {\bibinfo  {journal}
  {Proceedings of the National Academy of Sciences}\ }\textbf {\bibinfo
  {volume} {116}},\ \bibinfo {pages} {7766} (\bibinfo {year} {2019})},\ \Eprint
  {http://arxiv.org/abs/https://www.pnas.org/content/116/16/7766.full.pdf}
  {https://www.pnas.org/content/116/16/7766.full.pdf} \BibitemShut {NoStop}%
\bibitem [{\citenamefont {Baudouin}\ \emph {et~al.}(2014)\citenamefont
  {Baudouin}, \citenamefont {Guerin},\ and\ \citenamefont
  {Kaiser}}]{baudouin2014cold}%
  \BibitemOpen
  \bibfield  {author} {\bibinfo {author} {\bibfnamefont {Q.}~\bibnamefont
  {Baudouin}}, \bibinfo {author} {\bibfnamefont {W.}~\bibnamefont {Guerin}}, \
  and\ \bibinfo {author} {\bibfnamefont {R.}~\bibnamefont {Kaiser}},\ }in\
  \href@noop {} {\emph {\bibinfo {booktitle} {Annual Review of Cold Atoms and
  Molecules}}}\ (\bibinfo  {publisher} {World Scientific},\ \bibinfo {year}
  {2014})\ pp.\ \bibinfo {pages} {251--311}\BibitemShut {NoStop}%
\bibitem [{\citenamefont {Sokolov}\ and\ \citenamefont
  {Guerin}(2019)}]{sokolovJOSAB2019}%
  \BibitemOpen
  \bibfield  {author} {\bibinfo {author} {\bibfnamefont {I.~M.}\ \bibnamefont
  {Sokolov}}\ and\ \bibinfo {author} {\bibfnamefont {W.}~\bibnamefont
  {Guerin}},\ }\href {\doibase 10.1364/JOSAB.36.002030} {\bibfield  {journal}
  {\bibinfo  {journal} {J. Opt. Soc. Am. B}\ }\textbf {\bibinfo {volume}
  {36}},\ \bibinfo {pages} {2030} (\bibinfo {year} {2019})}\BibitemShut
  {NoStop}%
\bibitem [{\citenamefont {Guerin}\ \emph {et~al.}(2017)\citenamefont {Guerin},
  \citenamefont {Rouabah},\ and\ \citenamefont {Kaiser}}]{guerin2017light}%
  \BibitemOpen
  \bibfield  {author} {\bibinfo {author} {\bibfnamefont {W.}~\bibnamefont
  {Guerin}}, \bibinfo {author} {\bibfnamefont {M.}~\bibnamefont {Rouabah}}, \
  and\ \bibinfo {author} {\bibfnamefont {R.}~\bibnamefont {Kaiser}},\
  }\href@noop {} {\bibfield  {journal} {\bibinfo  {journal} {Journal of Modern
  Optics}\ }\textbf {\bibinfo {volume} {64}},\ \bibinfo {pages} {895} (\bibinfo
  {year} {2017})}\BibitemShut {NoStop}%
\bibitem [{\citenamefont {Lagendijk}\ \emph {et~al.}(2009)\citenamefont
  {Lagendijk}, \citenamefont {Tiggelen},\ and\ \citenamefont
  {Wiersma}}]{lagendijkPT2009}%
  \BibitemOpen
  \bibfield  {author} {\bibinfo {author} {\bibfnamefont {A.}~\bibnamefont
  {Lagendijk}}, \bibinfo {author} {\bibfnamefont {B.~v.}\ \bibnamefont
  {Tiggelen}}, \ and\ \bibinfo {author} {\bibfnamefont {D.~S.}\ \bibnamefont
  {Wiersma}},\ }\href {\doibase 10.1063/1.3206091} {\bibfield  {journal}
  {\bibinfo  {journal} {Physics Today}\ }\textbf {\bibinfo {volume} {62}},\
  \bibinfo {pages} {24} (\bibinfo {year} {2009})},\ \Eprint
  {http://arxiv.org/abs/https://doi.org/10.1063/1.3206091}
  {https://doi.org/10.1063/1.3206091} \BibitemShut {NoStop}%
\bibitem [{\citenamefont {Dogariu}\ and\ \citenamefont
  {Carminati}(2015)}]{Dogariu2015}%
  \BibitemOpen
  \bibfield  {author} {\bibinfo {author} {\bibfnamefont {A.}~\bibnamefont
  {Dogariu}}\ and\ \bibinfo {author} {\bibfnamefont {R.}~\bibnamefont
  {Carminati}},\ }\href {\doibase 10.1016/j.physrep.2014.11.003} {\bibfield
  {journal} {\bibinfo  {journal} {Phys. Rep.}\ }\textbf {\bibinfo {volume}
  {559}},\ \bibinfo {pages} {1} (\bibinfo {year} {2015})}\BibitemShut {NoStop}%
\bibitem [{\citenamefont {Berkovits}\ and\ \citenamefont
  {Feng}(1994)}]{berkovitsPhysRep1994}%
  \BibitemOpen
  \bibfield  {author} {\bibinfo {author} {\bibfnamefont {R.}~\bibnamefont
  {Berkovits}}\ and\ \bibinfo {author} {\bibfnamefont {S.}~\bibnamefont
  {Feng}},\ }\href {\doibase https://doi.org/10.1016/0370-1573(94)90079-5}
  {\bibfield  {journal} {\bibinfo  {journal} {Physics Reports}\ }\textbf
  {\bibinfo {volume} {238}},\ \bibinfo {pages} {135 } (\bibinfo {year}
  {1994})}\BibitemShut {NoStop}%
\bibitem [{\citenamefont {Mishchenko}(2006)}]{mishchenkoJQSRT2006}%
  \BibitemOpen
  \bibfield  {author} {\bibinfo {author} {\bibfnamefont {M.~I.}\ \bibnamefont
  {Mishchenko}},\ }\href {\doibase https://doi.org/10.1016/j.jqsrt.2006.02.065}
  {\bibfield  {journal} {\bibinfo  {journal} {Journal of Quantitative
  Spectroscopy and Radiative Transfer}\ }\textbf {\bibinfo {volume} {101}},\
  \bibinfo {pages} {540 } (\bibinfo {year} {2006})},\ \bibinfo {note} {light in
  Planetary Atmospheres and Other Particulate Media}\BibitemShut {NoStop}%
\bibitem [{\citenamefont {Fiebig}\ \emph {et~al.}(2008)\citenamefont {Fiebig},
  \citenamefont {Aegerter}, \citenamefont {Bührer}, \citenamefont {Störzer},
  \citenamefont {Akkermans}, \citenamefont {Montambaux},\ and\ \citenamefont
  {Maret}}]{fiebigEPL2008}%
  \BibitemOpen
  \bibfield  {author} {\bibinfo {author} {\bibfnamefont {S.}~\bibnamefont
  {Fiebig}}, \bibinfo {author} {\bibfnamefont {C.~M.}\ \bibnamefont
  {Aegerter}}, \bibinfo {author} {\bibfnamefont {W.}~\bibnamefont {Bührer}},
  \bibinfo {author} {\bibfnamefont {M.}~\bibnamefont {Störzer}}, \bibinfo
  {author} {\bibfnamefont {E.}~\bibnamefont {Akkermans}}, \bibinfo {author}
  {\bibfnamefont {G.}~\bibnamefont {Montambaux}}, \ and\ \bibinfo {author}
  {\bibfnamefont {G.}~\bibnamefont {Maret}},\ }\href {\doibase
  10.1209/0295-5075/81/64004} {\bibfield  {journal} {\bibinfo  {journal} {{EPL}
  (Europhysics Letters)}\ }\textbf {\bibinfo {volume} {81}},\ \bibinfo {pages}
  {64004} (\bibinfo {year} {2008})}\BibitemShut {NoStop}%
\bibitem [{\citenamefont {Wolf}\ and\ \citenamefont
  {Maret}(1985)}]{wolfPRL1985}%
  \BibitemOpen
  \bibfield  {author} {\bibinfo {author} {\bibfnamefont {P.-E.}\ \bibnamefont
  {Wolf}}\ and\ \bibinfo {author} {\bibfnamefont {G.}~\bibnamefont {Maret}},\
  }\href {\doibase 10.1103/PhysRevLett.55.2696} {\bibfield  {journal} {\bibinfo
   {journal} {Phys. Rev. Lett.}\ }\textbf {\bibinfo {volume} {55}},\ \bibinfo
  {pages} {2696} (\bibinfo {year} {1985})}\BibitemShut {NoStop}%
\bibitem [{\citenamefont {Etemad}\ \emph {et~al.}(1987)\citenamefont {Etemad},
  \citenamefont {Thompson}, \citenamefont {Andrejco}, \citenamefont {John},\
  and\ \citenamefont {MacKintosh}}]{etemadPRL1987}%
  \BibitemOpen
  \bibfield  {author} {\bibinfo {author} {\bibfnamefont {S.}~\bibnamefont
  {Etemad}}, \bibinfo {author} {\bibfnamefont {R.}~\bibnamefont {Thompson}},
  \bibinfo {author} {\bibfnamefont {M.~J.}\ \bibnamefont {Andrejco}}, \bibinfo
  {author} {\bibfnamefont {S.}~\bibnamefont {John}}, \ and\ \bibinfo {author}
  {\bibfnamefont {F.~C.}\ \bibnamefont {MacKintosh}},\ }\href {\doibase
  10.1103/PhysRevLett.59.1420} {\bibfield  {journal} {\bibinfo  {journal}
  {Phys. Rev. Lett.}\ }\textbf {\bibinfo {volume} {59}},\ \bibinfo {pages}
  {1420} (\bibinfo {year} {1987})}\BibitemShut {NoStop}%
\bibitem [{\citenamefont {Daozhong}\ \emph {et~al.}(1994)\citenamefont
  {Daozhong}, \citenamefont {Wei}, \citenamefont {Youlong}, \citenamefont
  {Zhaolin}, \citenamefont {Bingying},\ and\ \citenamefont
  {Guozhen}}]{zhangPRB1994}%
  \BibitemOpen
  \bibfield  {author} {\bibinfo {author} {\bibfnamefont {Z.}~\bibnamefont
  {Daozhong}}, \bibinfo {author} {\bibfnamefont {H.}~\bibnamefont {Wei}},
  \bibinfo {author} {\bibfnamefont {Z.}~\bibnamefont {Youlong}}, \bibinfo
  {author} {\bibfnamefont {L.}~\bibnamefont {Zhaolin}}, \bibinfo {author}
  {\bibfnamefont {C.}~\bibnamefont {Bingying}}, \ and\ \bibinfo {author}
  {\bibfnamefont {Y.}~\bibnamefont {Guozhen}},\ }\href {\doibase
  10.1103/PhysRevB.50.9810} {\bibfield  {journal} {\bibinfo  {journal} {Phys.
  Rev. B}\ }\textbf {\bibinfo {volume} {50}},\ \bibinfo {pages} {9810}
  (\bibinfo {year} {1994})}\BibitemShut {NoStop}%
\bibitem [{\citenamefont {Sheinfux}\ \emph {et~al.}(2017)\citenamefont
  {Sheinfux}, \citenamefont {Lumer}, \citenamefont {Ankonina}, \citenamefont
  {Genack}, \citenamefont {Bartal},\ and\ \citenamefont
  {Segev}}]{sheinfuxScience2017}%
  \BibitemOpen
  \bibfield  {author} {\bibinfo {author} {\bibfnamefont {H.~H.}\ \bibnamefont
  {Sheinfux}}, \bibinfo {author} {\bibfnamefont {Y.}~\bibnamefont {Lumer}},
  \bibinfo {author} {\bibfnamefont {G.}~\bibnamefont {Ankonina}}, \bibinfo
  {author} {\bibfnamefont {A.~Z.}\ \bibnamefont {Genack}}, \bibinfo {author}
  {\bibfnamefont {G.}~\bibnamefont {Bartal}}, \ and\ \bibinfo {author}
  {\bibfnamefont {M.}~\bibnamefont {Segev}},\ }\href {\doibase
  10.1126/science.aah6822} {\bibfield  {journal} {\bibinfo  {journal}
  {Science}\ }\textbf {\bibinfo {volume} {356}},\ \bibinfo {pages} {953}
  (\bibinfo {year} {2017})},\ \Eprint
  {http://arxiv.org/abs/https://science.sciencemag.org/content/356/6341/953.full.pdf}
  {https://science.sciencemag.org/content/356/6341/953.full.pdf} \BibitemShut
  {NoStop}%
\bibitem [{\citenamefont {Shi}\ \emph {et~al.}(2018{\natexlab{b}})\citenamefont
  {Shi}, \citenamefont {Liu}, \citenamefont {Peng}, \citenamefont {Xu},
  \citenamefont {Zhang}, \citenamefont {Jing}, \citenamefont {Fan},
  \citenamefont {Huang}, \citenamefont {Wang},\ and\ \citenamefont
  {Wang}}]{shiNL2018}%
  \BibitemOpen
  \bibfield  {author} {\bibinfo {author} {\bibfnamefont {W.-B.}\ \bibnamefont
  {Shi}}, \bibinfo {author} {\bibfnamefont {L.-Z.}\ \bibnamefont {Liu}},
  \bibinfo {author} {\bibfnamefont {R.}~\bibnamefont {Peng}}, \bibinfo {author}
  {\bibfnamefont {D.-H.}\ \bibnamefont {Xu}}, \bibinfo {author} {\bibfnamefont
  {K.}~\bibnamefont {Zhang}}, \bibinfo {author} {\bibfnamefont
  {H.}~\bibnamefont {Jing}}, \bibinfo {author} {\bibfnamefont {R.-H.}\
  \bibnamefont {Fan}}, \bibinfo {author} {\bibfnamefont {X.-R.}\ \bibnamefont
  {Huang}}, \bibinfo {author} {\bibfnamefont {Q.-J.}\ \bibnamefont {Wang}}, \
  and\ \bibinfo {author} {\bibfnamefont {M.}~\bibnamefont {Wang}},\ }\href
  {\doibase 10.1021/acs.nanolett.7b05191} {\bibfield  {journal} {\bibinfo
  {journal} {Nano Letters}\ }\textbf {\bibinfo {volume} {18}},\ \bibinfo
  {pages} {1896} (\bibinfo {year} {2018}{\natexlab{b}})},\ \bibinfo {note}
  {pMID: 29432022},\ \Eprint
  {http://arxiv.org/abs/https://doi.org/10.1021/acs.nanolett.7b05191}
  {https://doi.org/10.1021/acs.nanolett.7b05191} \BibitemShut {NoStop}%
\bibitem [{\citenamefont {Caselli}\ \emph {et~al.}(2017)\citenamefont
  {Caselli}, \citenamefont {Intonti}, \citenamefont {La~China}, \citenamefont
  {Biccari}, \citenamefont {Riboli}, \citenamefont {Gerardino}, \citenamefont
  {Li}, \citenamefont {Linfield}, \citenamefont {Pagliano}, \citenamefont
  {Fiore},\ and\ \citenamefont {Gurioli}}]{caselliAPL2017}%
  \BibitemOpen
  \bibfield  {author} {\bibinfo {author} {\bibfnamefont {N.}~\bibnamefont
  {Caselli}}, \bibinfo {author} {\bibfnamefont {F.}~\bibnamefont {Intonti}},
  \bibinfo {author} {\bibfnamefont {F.}~\bibnamefont {La~China}}, \bibinfo
  {author} {\bibfnamefont {F.}~\bibnamefont {Biccari}}, \bibinfo {author}
  {\bibfnamefont {F.}~\bibnamefont {Riboli}}, \bibinfo {author} {\bibfnamefont
  {A.}~\bibnamefont {Gerardino}}, \bibinfo {author} {\bibfnamefont
  {L.}~\bibnamefont {Li}}, \bibinfo {author} {\bibfnamefont {E.~H.}\
  \bibnamefont {Linfield}}, \bibinfo {author} {\bibfnamefont {F.}~\bibnamefont
  {Pagliano}}, \bibinfo {author} {\bibfnamefont {A.}~\bibnamefont {Fiore}}, \
  and\ \bibinfo {author} {\bibfnamefont {M.}~\bibnamefont {Gurioli}},\ }\href
  {\doibase 10.1063/1.4976747} {\bibfield  {journal} {\bibinfo  {journal}
  {Applied Physics Letters}\ }\textbf {\bibinfo {volume} {110}},\ \bibinfo
  {pages} {081102} (\bibinfo {year} {2017})},\ \Eprint
  {http://arxiv.org/abs/https://doi.org/10.1063/1.4976747}
  {https://doi.org/10.1063/1.4976747} \BibitemShut {NoStop}%
\bibitem [{\citenamefont {Schwartz}\ \emph {et~al.}(2007)\citenamefont
  {Schwartz}, \citenamefont {Bartal}, \citenamefont {Fishman},\ and\
  \citenamefont {Segev}}]{Schwartz2007}%
  \BibitemOpen
  \bibfield  {author} {\bibinfo {author} {\bibfnamefont {T.}~\bibnamefont
  {Schwartz}}, \bibinfo {author} {\bibfnamefont {G.}~\bibnamefont {Bartal}},
  \bibinfo {author} {\bibfnamefont {S.}~\bibnamefont {Fishman}}, \ and\
  \bibinfo {author} {\bibfnamefont {M.}~\bibnamefont {Segev}},\ }\href
  {\doibase 10.1038/nature05623} {\bibfield  {journal} {\bibinfo  {journal}
  {Nature (London)}\ }\textbf {\bibinfo {volume} {446}},\ \bibinfo {pages} {52}
  (\bibinfo {year} {2007})}\BibitemShut {NoStop}%
\bibitem [{\citenamefont {Skipetrov}\ and\ \citenamefont
  {Page}(2016)}]{skipetrovNJP2016}%
  \BibitemOpen
  \bibfield  {author} {\bibinfo {author} {\bibfnamefont {S.~E.}\ \bibnamefont
  {Skipetrov}}\ and\ \bibinfo {author} {\bibfnamefont {J.~H.}\ \bibnamefont
  {Page}},\ }\href {\doibase 10.1088/1367-2630/18/2/021001} {\bibfield
  {journal} {\bibinfo  {journal} {New Journal of Physics}\ }\textbf {\bibinfo
  {volume} {18}},\ \bibinfo {pages} {021001} (\bibinfo {year}
  {2016})}\BibitemShut {NoStop}%
\bibitem [{\citenamefont {Scheffold}\ and\ \citenamefont
  {Wiersma}(2013)}]{scheffoldNaturephoton2013}%
  \BibitemOpen
  \bibfield  {author} {\bibinfo {author} {\bibfnamefont {F.}~\bibnamefont
  {Scheffold}}\ and\ \bibinfo {author} {\bibfnamefont {D.}~\bibnamefont
  {Wiersma}},\ }\href {\doibase 10.1038/nphoton.2013.210} {\bibfield  {journal}
  {\bibinfo  {journal} {Nature Photonics}\ }\textbf {\bibinfo {volume} {7}},\
  \bibinfo {pages} {934} (\bibinfo {year} {2013})}\BibitemShut {NoStop}%
\bibitem [{\citenamefont {John}(1987)}]{johnPRL1987}%
  \BibitemOpen
  \bibfield  {author} {\bibinfo {author} {\bibfnamefont {S.}~\bibnamefont
  {John}},\ }\href {\doibase 10.1103/PhysRevLett.58.2486} {\bibfield  {journal}
  {\bibinfo  {journal} {Phys. Rev. Lett.}\ }\textbf {\bibinfo {volume} {58}},\
  \bibinfo {pages} {2486} (\bibinfo {year} {1987})}\BibitemShut {NoStop}%
\bibitem [{\citenamefont {Jeon}\ \emph {et~al.}(2017)\citenamefont {Jeon},
  \citenamefont {Kwon},\ and\ \citenamefont {Hur}}]{jeonNaturephys2017}%
  \BibitemOpen
  \bibfield  {author} {\bibinfo {author} {\bibfnamefont {S.-Y.}\ \bibnamefont
  {Jeon}}, \bibinfo {author} {\bibfnamefont {H.}~\bibnamefont {Kwon}}, \ and\
  \bibinfo {author} {\bibfnamefont {K.}~\bibnamefont {Hur}},\ }\href {\doibase
  10.1038/nphys4002} {\bibfield  {journal} {\bibinfo  {journal} {Nature
  Physics}\ }\textbf {\bibinfo {volume} {13}},\ \bibinfo {pages} {363}
  (\bibinfo {year} {2017})}\BibitemShut {NoStop}%
\bibitem [{\citenamefont {Bromberg}\ and\ \citenamefont
  {Cao}(2014)}]{brombergPRL2014}%
  \BibitemOpen
  \bibfield  {author} {\bibinfo {author} {\bibfnamefont {Y.}~\bibnamefont
  {Bromberg}}\ and\ \bibinfo {author} {\bibfnamefont {H.}~\bibnamefont {Cao}},\
  }\href {\doibase 10.1103/PhysRevLett.112.213904} {\bibfield  {journal}
  {\bibinfo  {journal} {Phys. Rev. Lett.}\ }\textbf {\bibinfo {volume} {112}},\
  \bibinfo {pages} {213904} (\bibinfo {year} {2014})}\BibitemShut {NoStop}%
\bibitem [{\citenamefont {Shapiro}(1986)}]{shapiroPRL1986}%
  \BibitemOpen
  \bibfield  {author} {\bibinfo {author} {\bibfnamefont {B.}~\bibnamefont
  {Shapiro}},\ }\href {\doibase 10.1103/PhysRevLett.57.2168} {\bibfield
  {journal} {\bibinfo  {journal} {Phys. Rev. Lett.}\ }\textbf {\bibinfo
  {volume} {57}},\ \bibinfo {pages} {2168} (\bibinfo {year}
  {1986})}\BibitemShut {NoStop}%
\bibitem [{\citenamefont {Genack}\ \emph {et~al.}(1990)\citenamefont {Genack},
  \citenamefont {Garcia},\ and\ \citenamefont {Polkosnik}}]{genackPRL1990}%
  \BibitemOpen
  \bibfield  {author} {\bibinfo {author} {\bibfnamefont {A.~Z.}\ \bibnamefont
  {Genack}}, \bibinfo {author} {\bibfnamefont {N.}~\bibnamefont {Garcia}}, \
  and\ \bibinfo {author} {\bibfnamefont {W.}~\bibnamefont {Polkosnik}},\ }\href
  {\doibase 10.1103/PhysRevLett.65.2129} {\bibfield  {journal} {\bibinfo
  {journal} {Phys. Rev. Lett.}\ }\textbf {\bibinfo {volume} {65}},\ \bibinfo
  {pages} {2129} (\bibinfo {year} {1990})}\BibitemShut {NoStop}%
\bibitem [{\citenamefont {Emiliani}\ \emph {et~al.}(2003)\citenamefont
  {Emiliani}, \citenamefont {Intonti}, \citenamefont {Cazayous}, \citenamefont
  {Wiersma}, \citenamefont {Colocci}, \citenamefont {Aliev},\ and\
  \citenamefont {Lagendijk}}]{emilianiPRL2003}%
  \BibitemOpen
  \bibfield  {author} {\bibinfo {author} {\bibfnamefont {V.}~\bibnamefont
  {Emiliani}}, \bibinfo {author} {\bibfnamefont {F.}~\bibnamefont {Intonti}},
  \bibinfo {author} {\bibfnamefont {M.}~\bibnamefont {Cazayous}}, \bibinfo
  {author} {\bibfnamefont {D.~S.}\ \bibnamefont {Wiersma}}, \bibinfo {author}
  {\bibfnamefont {M.}~\bibnamefont {Colocci}}, \bibinfo {author} {\bibfnamefont
  {F.}~\bibnamefont {Aliev}}, \ and\ \bibinfo {author} {\bibfnamefont
  {A.}~\bibnamefont {Lagendijk}},\ }\href {\doibase
  10.1103/PhysRevLett.90.250801} {\bibfield  {journal} {\bibinfo  {journal}
  {Phys. Rev. Lett.}\ }\textbf {\bibinfo {volume} {90}},\ \bibinfo {pages}
  {250801} (\bibinfo {year} {2003})}\BibitemShut {NoStop}%
\bibitem [{\citenamefont {Carminati}(2010)}]{carminatiPRA2010}%
  \BibitemOpen
  \bibfield  {author} {\bibinfo {author} {\bibfnamefont {R.}~\bibnamefont
  {Carminati}},\ }\href {\doibase 10.1103/PhysRevA.81.053804} {\bibfield
  {journal} {\bibinfo  {journal} {Phys. Rev. A}\ }\textbf {\bibinfo {volume}
  {81}},\ \bibinfo {pages} {053804} (\bibinfo {year} {2010})}\BibitemShut
  {NoStop}%
\bibitem [{\citenamefont {Boas}\ and\ \citenamefont
  {Dunn}(2010)}]{boasJBO2010}%
  \BibitemOpen
  \bibfield  {author} {\bibinfo {author} {\bibfnamefont {D.~A.}\ \bibnamefont
  {Boas}}\ and\ \bibinfo {author} {\bibfnamefont {A.~K.}\ \bibnamefont
  {Dunn}},\ }\href {\doibase 10.1117/1.3285504} {\bibfield  {journal} {\bibinfo
   {journal} {Journal of Biomedical Optics}\ }\textbf {\bibinfo {volume}
  {15}},\ \bibinfo {pages} {1 } (\bibinfo {year} {2010})}\BibitemShut {NoStop}%
\bibitem [{\citenamefont {Briers}\ \emph {et~al.}(2013)\citenamefont {Briers},
  \citenamefont {Duncan}, \citenamefont {Hirst}, \citenamefont {Kirkpatrick},
  \citenamefont {Larsson}, \citenamefont {Steenbergen}, \citenamefont
  {Stromberg},\ and\ \citenamefont {Thompson}}]{biersJBO2013}%
  \BibitemOpen
  \bibfield  {author} {\bibinfo {author} {\bibfnamefont {D.}~\bibnamefont
  {Briers}}, \bibinfo {author} {\bibfnamefont {D.~D.}\ \bibnamefont {Duncan}},
  \bibinfo {author} {\bibfnamefont {E.~R.}\ \bibnamefont {Hirst}}, \bibinfo
  {author} {\bibfnamefont {S.~J.}\ \bibnamefont {Kirkpatrick}}, \bibinfo
  {author} {\bibfnamefont {M.}~\bibnamefont {Larsson}}, \bibinfo {author}
  {\bibfnamefont {W.}~\bibnamefont {Steenbergen}}, \bibinfo {author}
  {\bibfnamefont {T.}~\bibnamefont {Stromberg}}, \ and\ \bibinfo {author}
  {\bibfnamefont {O.~B.}\ \bibnamefont {Thompson}},\ }\href {\doibase
  10.1117/1.JBO.18.6.066018} {\bibfield  {journal} {\bibinfo  {journal}
  {Journal of Biomedical Optics}\ }\textbf {\bibinfo {volume} {18}},\ \bibinfo
  {pages} {1 } (\bibinfo {year} {2013})}\BibitemShut {NoStop}%
\bibitem [{\citenamefont {Heeman}\ \emph {et~al.}(2019)\citenamefont {Heeman},
  \citenamefont {Steenbergen}, \citenamefont {van Dam},\ and\ \citenamefont
  {Boerma}}]{heemanJBO2019}%
  \BibitemOpen
  \bibfield  {author} {\bibinfo {author} {\bibfnamefont {W.}~\bibnamefont
  {Heeman}}, \bibinfo {author} {\bibfnamefont {W.}~\bibnamefont {Steenbergen}},
  \bibinfo {author} {\bibfnamefont {G.~M.}\ \bibnamefont {van Dam}}, \ and\
  \bibinfo {author} {\bibfnamefont {E.~C.}\ \bibnamefont {Boerma}},\ }\href
  {\doibase 10.1117/1.JBO.24.8.080901} {\bibfield  {journal} {\bibinfo
  {journal} {Journal of Biomedical Optics}\ }\textbf {\bibinfo {volume} {24}},\
  \bibinfo {pages} {1 } (\bibinfo {year} {2019})}\BibitemShut {NoStop}%
\bibitem [{\citenamefont {Cheng}\ \emph {et~al.}(2018)\citenamefont {Cheng},
  \citenamefont {Tamborini}, \citenamefont {Carp}, \citenamefont {Shatrovoy},
  \citenamefont {Zimmerman}, \citenamefont {Tyulmankov}, \citenamefont
  {Siegel}, \citenamefont {Blackwell}, \citenamefont {Franceschini},\ and\
  \citenamefont {Boas}}]{chengOL2018}%
  \BibitemOpen
  \bibfield  {author} {\bibinfo {author} {\bibfnamefont {X.}~\bibnamefont
  {Cheng}}, \bibinfo {author} {\bibfnamefont {D.}~\bibnamefont {Tamborini}},
  \bibinfo {author} {\bibfnamefont {S.~A.}\ \bibnamefont {Carp}}, \bibinfo
  {author} {\bibfnamefont {O.}~\bibnamefont {Shatrovoy}}, \bibinfo {author}
  {\bibfnamefont {B.}~\bibnamefont {Zimmerman}}, \bibinfo {author}
  {\bibfnamefont {D.}~\bibnamefont {Tyulmankov}}, \bibinfo {author}
  {\bibfnamefont {A.}~\bibnamefont {Siegel}}, \bibinfo {author} {\bibfnamefont
  {M.}~\bibnamefont {Blackwell}}, \bibinfo {author} {\bibfnamefont {M.~A.}\
  \bibnamefont {Franceschini}}, \ and\ \bibinfo {author} {\bibfnamefont
  {D.~A.}\ \bibnamefont {Boas}},\ }\href {\doibase 10.1364/OL.43.002756}
  {\bibfield  {journal} {\bibinfo  {journal} {Opt. Lett.}\ }\textbf {\bibinfo
  {volume} {43}},\ \bibinfo {pages} {2756} (\bibinfo {year}
  {2018})}\BibitemShut {NoStop}%
\bibitem [{\citenamefont {Katz}\ \emph {et~al.}(2014)\citenamefont {Katz},
  \citenamefont {Heidmann}, \citenamefont {Fink},\ and\ \citenamefont
  {Gigan}}]{katzNaturephoton2014}%
  \BibitemOpen
  \bibfield  {author} {\bibinfo {author} {\bibfnamefont {O.}~\bibnamefont
  {Katz}}, \bibinfo {author} {\bibfnamefont {P.}~\bibnamefont {Heidmann}},
  \bibinfo {author} {\bibfnamefont {M.}~\bibnamefont {Fink}}, \ and\ \bibinfo
  {author} {\bibfnamefont {S.}~\bibnamefont {Gigan}},\ }\href
  {https://doi.org/10.1038/nphoton.2014.189} {\bibfield  {journal} {\bibinfo
  {journal} {Nature Photonics}\ }\textbf {\bibinfo {volume} {8}},\ \bibinfo
  {pages} {784 EP } (\bibinfo {year} {2014})},\ \bibinfo {note}
  {article}\BibitemShut {NoStop}%
\bibitem [{\citenamefont {Salhov}\ \emph {et~al.}(2018)\citenamefont {Salhov},
  \citenamefont {Weinberg},\ and\ \citenamefont {Katz}}]{salhovOL2018}%
  \BibitemOpen
  \bibfield  {author} {\bibinfo {author} {\bibfnamefont {O.}~\bibnamefont
  {Salhov}}, \bibinfo {author} {\bibfnamefont {G.}~\bibnamefont {Weinberg}}, \
  and\ \bibinfo {author} {\bibfnamefont {O.}~\bibnamefont {Katz}},\ }\href
  {\doibase 10.1364/OL.43.005528} {\bibfield  {journal} {\bibinfo  {journal}
  {Opt. Lett.}\ }\textbf {\bibinfo {volume} {43}},\ \bibinfo {pages} {5528}
  (\bibinfo {year} {2018})}\BibitemShut {NoStop}%
\bibitem [{\citenamefont {Stern}\ and\ \citenamefont
  {Katz}(2019)}]{sternOL2019}%
  \BibitemOpen
  \bibfield  {author} {\bibinfo {author} {\bibfnamefont {G.}~\bibnamefont
  {Stern}}\ and\ \bibinfo {author} {\bibfnamefont {O.}~\bibnamefont {Katz}},\
  }\href {\doibase 10.1364/OL.44.000143} {\bibfield  {journal} {\bibinfo
  {journal} {Opt. Lett.}\ }\textbf {\bibinfo {volume} {44}},\ \bibinfo {pages}
  {143} (\bibinfo {year} {2019})}\BibitemShut {NoStop}%
\bibitem [{\citenamefont {McCabe}\ \emph {et~al.}(2011)\citenamefont {McCabe},
  \citenamefont {Tajalli}, \citenamefont {Austin}, \citenamefont {Bondareff},
  \citenamefont {Walmsley}, \citenamefont {Gigan},\ and\ \citenamefont
  {Chatel}}]{mccabeNaturecomms2011}%
  \BibitemOpen
  \bibfield  {author} {\bibinfo {author} {\bibfnamefont {D.~J.}\ \bibnamefont
  {McCabe}}, \bibinfo {author} {\bibfnamefont {A.}~\bibnamefont {Tajalli}},
  \bibinfo {author} {\bibfnamefont {D.~R.}\ \bibnamefont {Austin}}, \bibinfo
  {author} {\bibfnamefont {P.}~\bibnamefont {Bondareff}}, \bibinfo {author}
  {\bibfnamefont {I.~A.}\ \bibnamefont {Walmsley}}, \bibinfo {author}
  {\bibfnamefont {S.}~\bibnamefont {Gigan}}, \ and\ \bibinfo {author}
  {\bibfnamefont {B.}~\bibnamefont {Chatel}},\ }\href {\doibase
  10.1038/ncomms1434} {\bibfield  {journal} {\bibinfo  {journal} {Nature
  Communications}\ }\textbf {\bibinfo {volume} {2}},\ \bibinfo {pages} {447}
  (\bibinfo {year} {2011})}\BibitemShut {NoStop}%
\bibitem [{\citenamefont {Aulbach}\ \emph {et~al.}(2011)\citenamefont
  {Aulbach}, \citenamefont {Gjonaj}, \citenamefont {Johnson}, \citenamefont
  {Mosk},\ and\ \citenamefont {Lagendijk}}]{aulbachPRL2011}%
  \BibitemOpen
  \bibfield  {author} {\bibinfo {author} {\bibfnamefont {J.}~\bibnamefont
  {Aulbach}}, \bibinfo {author} {\bibfnamefont {B.}~\bibnamefont {Gjonaj}},
  \bibinfo {author} {\bibfnamefont {P.~M.}\ \bibnamefont {Johnson}}, \bibinfo
  {author} {\bibfnamefont {A.~P.}\ \bibnamefont {Mosk}}, \ and\ \bibinfo
  {author} {\bibfnamefont {A.}~\bibnamefont {Lagendijk}},\ }\href {\doibase
  10.1103/PhysRevLett.106.103901} {\bibfield  {journal} {\bibinfo  {journal}
  {Phys. Rev. Lett.}\ }\textbf {\bibinfo {volume} {106}},\ \bibinfo {pages}
  {103901} (\bibinfo {year} {2011})}\BibitemShut {NoStop}%
\bibitem [{\citenamefont {Riboli}\ \emph {et~al.}(2014)\citenamefont {Riboli},
  \citenamefont {Caselli}, \citenamefont {Vignolini}, \citenamefont {Intonti},
  \citenamefont {Vynck}, \citenamefont {Barthelemy}, \citenamefont {Gerardino},
  \citenamefont {Balet}, \citenamefont {Li}, \citenamefont {Fiore},
  \citenamefont {Gurioli},\ and\ \citenamefont
  {Wiersma}}]{riboliNaturemat2014}%
  \BibitemOpen
  \bibfield  {author} {\bibinfo {author} {\bibfnamefont {F.}~\bibnamefont
  {Riboli}}, \bibinfo {author} {\bibfnamefont {N.}~\bibnamefont {Caselli}},
  \bibinfo {author} {\bibfnamefont {S.}~\bibnamefont {Vignolini}}, \bibinfo
  {author} {\bibfnamefont {F.}~\bibnamefont {Intonti}}, \bibinfo {author}
  {\bibfnamefont {K.}~\bibnamefont {Vynck}}, \bibinfo {author} {\bibfnamefont
  {P.}~\bibnamefont {Barthelemy}}, \bibinfo {author} {\bibfnamefont
  {A.}~\bibnamefont {Gerardino}}, \bibinfo {author} {\bibfnamefont
  {L.}~\bibnamefont {Balet}}, \bibinfo {author} {\bibfnamefont {L.~H.}\
  \bibnamefont {Li}}, \bibinfo {author} {\bibfnamefont {A.}~\bibnamefont
  {Fiore}}, \bibinfo {author} {\bibfnamefont {M.}~\bibnamefont {Gurioli}}, \
  and\ \bibinfo {author} {\bibfnamefont {D.~S.}\ \bibnamefont {Wiersma}},\
  }\href {https://doi.org/10.1038/nmat3966} {\bibfield  {journal} {\bibinfo
  {journal} {Nature Materials}\ }\textbf {\bibinfo {volume} {13}},\ \bibinfo
  {pages} {720 EP } (\bibinfo {year} {2014})},\ \bibinfo {note}
  {article}\BibitemShut {NoStop}%
\bibitem [{\citenamefont {Bruck}\ \emph {et~al.}(2016)\citenamefont {Bruck},
  \citenamefont {Vynck}, \citenamefont {Lalanne}, \citenamefont {Mills},
  \citenamefont {Thomson}, \citenamefont {Mashanovich}, \citenamefont {Reed},\
  and\ \citenamefont {Muskens}}]{bruckOptica2016}%
  \BibitemOpen
  \bibfield  {author} {\bibinfo {author} {\bibfnamefont {R.}~\bibnamefont
  {Bruck}}, \bibinfo {author} {\bibfnamefont {K.}~\bibnamefont {Vynck}},
  \bibinfo {author} {\bibfnamefont {P.}~\bibnamefont {Lalanne}}, \bibinfo
  {author} {\bibfnamefont {B.}~\bibnamefont {Mills}}, \bibinfo {author}
  {\bibfnamefont {D.~J.}\ \bibnamefont {Thomson}}, \bibinfo {author}
  {\bibfnamefont {G.~Z.}\ \bibnamefont {Mashanovich}}, \bibinfo {author}
  {\bibfnamefont {G.~T.}\ \bibnamefont {Reed}}, \ and\ \bibinfo {author}
  {\bibfnamefont {O.~L.}\ \bibnamefont {Muskens}},\ }\href {\doibase
  10.1364/OPTICA.3.000396} {\bibfield  {journal} {\bibinfo  {journal} {Optica}\
  }\textbf {\bibinfo {volume} {3}},\ \bibinfo {pages} {396} (\bibinfo {year}
  {2016})}\BibitemShut {NoStop}%
\bibitem [{\citenamefont {Ramezanpour}\ and\ \citenamefont
  {Mackowski}(2019)}]{ramezanpourJQSRT2019}%
  \BibitemOpen
  \bibfield  {author} {\bibinfo {author} {\bibfnamefont {B.}~\bibnamefont
  {Ramezanpour}}\ and\ \bibinfo {author} {\bibfnamefont {D.~W.}\ \bibnamefont
  {Mackowski}},\ }\href {\doibase https://doi.org/10.1016/j.jqsrt.2018.12.012}
  {\bibfield  {journal} {\bibinfo  {journal} {Journal of Quantitative
  Spectroscopy and Radiative Transfer}\ }\textbf {\bibinfo {volume} {224}},\
  \bibinfo {pages} {537 } (\bibinfo {year} {2019})}\BibitemShut {NoStop}%
\bibitem [{\citenamefont {Tallon}\ \emph {et~al.}(2017)\citenamefont {Tallon},
  \citenamefont {Brunet},\ and\ \citenamefont {Page}}]{tallonPRL2017}%
  \BibitemOpen
  \bibfield  {author} {\bibinfo {author} {\bibfnamefont {B.}~\bibnamefont
  {Tallon}}, \bibinfo {author} {\bibfnamefont {T.}~\bibnamefont {Brunet}}, \
  and\ \bibinfo {author} {\bibfnamefont {J.~H.}\ \bibnamefont {Page}},\ }\href
  {\doibase 10.1103/PhysRevLett.119.164301} {\bibfield  {journal} {\bibinfo
  {journal} {Phys. Rev. Lett.}\ }\textbf {\bibinfo {volume} {119}},\ \bibinfo
  {pages} {164301} (\bibinfo {year} {2017})}\BibitemShut {NoStop}%
\bibitem [{\citenamefont {Cohen-Tannoudji}\ and\ \citenamefont
  {Gu{\'e}ry-Odelin}(2011)}]{cohentannoudji2011}%
  \BibitemOpen
  \bibfield  {author} {\bibinfo {author} {\bibfnamefont {C.}~\bibnamefont
  {Cohen-Tannoudji}}\ and\ \bibinfo {author} {\bibfnamefont {D.}~\bibnamefont
  {Gu{\'e}ry-Odelin}},\ }\href@noop {} {\emph {\bibinfo {title} {Advances in
  atomic physics: an overview}}}\ (\bibinfo  {publisher} {World Scientific},\
  \bibinfo {year} {2011})\BibitemShut {NoStop}%
\bibitem [{\citenamefont {Labeyrie}\ \emph {et~al.}(2000)\citenamefont
  {Labeyrie}, \citenamefont {Müller}, \citenamefont {Wiersma}, \citenamefont
  {Miniatura},\ and\ \citenamefont {Kaiser}}]{labeyrieJOB2000}%
  \BibitemOpen
  \bibfield  {author} {\bibinfo {author} {\bibfnamefont {G.}~\bibnamefont
  {Labeyrie}}, \bibinfo {author} {\bibfnamefont {C.~A.}\ \bibnamefont
  {Müller}}, \bibinfo {author} {\bibfnamefont {D.~S.}\ \bibnamefont
  {Wiersma}}, \bibinfo {author} {\bibfnamefont {C.}~\bibnamefont {Miniatura}},
  \ and\ \bibinfo {author} {\bibfnamefont {R.}~\bibnamefont {Kaiser}},\
  }\href@noop {} {\bibfield  {journal} {\bibinfo  {journal} {Journal of Optics
  B: Quantum and Semiclassical Optics}\ }\textbf {\bibinfo {volume} {2}},\
  \bibinfo {pages} {672} (\bibinfo {year} {2000})}\BibitemShut {NoStop}%
\bibitem [{\citenamefont {Skipetrov}\ and\ \citenamefont
  {Sokolov}(2015)}]{Skipetrov2015}%
  \BibitemOpen
  \bibfield  {author} {\bibinfo {author} {\bibfnamefont {S.~E.}\ \bibnamefont
  {Skipetrov}}\ and\ \bibinfo {author} {\bibfnamefont {I.~M.}\ \bibnamefont
  {Sokolov}},\ }\href {\doibase 10.1103/PhysRevLett.114.053902} {\bibfield
  {journal} {\bibinfo  {journal} {Phys. Rev. Lett.}\ }\textbf {\bibinfo
  {volume} {114}},\ \bibinfo {pages} {053902} (\bibinfo {year}
  {2015})}\BibitemShut {NoStop}%
\bibitem [{\citenamefont {Moreira}\ \emph {et~al.}(2019)\citenamefont
  {Moreira}, \citenamefont {Kaiser},\ and\ \citenamefont
  {Bachelard}}]{moreira2019localization}%
  \BibitemOpen
  \bibfield  {author} {\bibinfo {author} {\bibfnamefont {N.~A.}\ \bibnamefont
  {Moreira}}, \bibinfo {author} {\bibfnamefont {R.}~\bibnamefont {Kaiser}}, \
  and\ \bibinfo {author} {\bibfnamefont {R.}~\bibnamefont {Bachelard}},\ }\href
  {\doibase 10.1209/0295-5075/127/54003} {\bibfield  {journal} {\bibinfo
  {journal} {{EPL} (Europhysics Letters)}\ }\textbf {\bibinfo {volume} {127}},\
  \bibinfo {pages} {54003} (\bibinfo {year} {2019})}\BibitemShut {NoStop}%
\bibitem [{\citenamefont {Baudouin}\ \emph {et~al.}(2013)\citenamefont
  {Baudouin}, \citenamefont {Mercadier}, \citenamefont {Guarrera},
  \citenamefont {Guerin},\ and\ \citenamefont
  {Kaiser}}]{baudouinNaturephys2013}%
  \BibitemOpen
  \bibfield  {author} {\bibinfo {author} {\bibfnamefont {Q.}~\bibnamefont
  {Baudouin}}, \bibinfo {author} {\bibfnamefont {N.}~\bibnamefont {Mercadier}},
  \bibinfo {author} {\bibfnamefont {V.}~\bibnamefont {Guarrera}}, \bibinfo
  {author} {\bibfnamefont {W.}~\bibnamefont {Guerin}}, \ and\ \bibinfo {author}
  {\bibfnamefont {R.}~\bibnamefont {Kaiser}},\ }\href@noop {} {\bibfield
  {journal} {\bibinfo  {journal} {Nature Physics}\ }\textbf {\bibinfo {volume}
  {9}},\ \bibinfo {pages} {357} (\bibinfo {year} {2013})}\BibitemShut {NoStop}%
\bibitem [{\citenamefont {Pereira}\ \emph {et~al.}(2004)\citenamefont
  {Pereira}, \citenamefont {Martinho},\ and\ \citenamefont
  {Berberan-Santos}}]{pereiraPRL2004}%
  \BibitemOpen
  \bibfield  {author} {\bibinfo {author} {\bibfnamefont {E.}~\bibnamefont
  {Pereira}}, \bibinfo {author} {\bibfnamefont {J.~M.~G.}\ \bibnamefont
  {Martinho}}, \ and\ \bibinfo {author} {\bibfnamefont {M.~N.}\ \bibnamefont
  {Berberan-Santos}},\ }\href {\doibase 10.1103/PhysRevLett.93.120201}
  {\bibfield  {journal} {\bibinfo  {journal} {Phys. Rev. Lett.}\ }\textbf
  {\bibinfo {volume} {93}},\ \bibinfo {pages} {120201} (\bibinfo {year}
  {2004})}\BibitemShut {NoStop}%
\bibitem [{\citenamefont {Mercadier}\ \emph {et~al.}(2013)\citenamefont
  {Mercadier}, \citenamefont {Chevrollier}, \citenamefont {Guerin},\ and\
  \citenamefont {Kaiser}}]{mercadierPRA2013}%
  \BibitemOpen
  \bibfield  {author} {\bibinfo {author} {\bibfnamefont {N.}~\bibnamefont
  {Mercadier}}, \bibinfo {author} {\bibfnamefont {M.}~\bibnamefont
  {Chevrollier}}, \bibinfo {author} {\bibfnamefont {W.}~\bibnamefont {Guerin}},
  \ and\ \bibinfo {author} {\bibfnamefont {R.}~\bibnamefont {Kaiser}},\ }\href
  {\doibase 10.1103/PhysRevA.87.063837} {\bibfield  {journal} {\bibinfo
  {journal} {Phys. Rev. A}\ }\textbf {\bibinfo {volume} {87}},\ \bibinfo
  {pages} {063837} (\bibinfo {year} {2013})}\BibitemShut {NoStop}%
\bibitem [{\citenamefont {Labeyrie}\ \emph {et~al.}(2006)\citenamefont
  {Labeyrie}, \citenamefont {Delande}, \citenamefont {Kaiser},\ and\
  \citenamefont {Miniatura}}]{labeyriePRL2006}%
  \BibitemOpen
  \bibfield  {author} {\bibinfo {author} {\bibfnamefont {G.}~\bibnamefont
  {Labeyrie}}, \bibinfo {author} {\bibfnamefont {D.}~\bibnamefont {Delande}},
  \bibinfo {author} {\bibfnamefont {R.}~\bibnamefont {Kaiser}}, \ and\ \bibinfo
  {author} {\bibfnamefont {C.}~\bibnamefont {Miniatura}},\ }\href {\doibase
  10.1103/PhysRevLett.97.013004} {\bibfield  {journal} {\bibinfo  {journal}
  {Phys. Rev. Lett.}\ }\textbf {\bibinfo {volume} {97}},\ \bibinfo {pages}
  {013004} (\bibinfo {year} {2006})}\BibitemShut {NoStop}%
\bibitem [{\citenamefont {Zhu}\ \emph {et~al.}(2016)\citenamefont {Zhu},
  \citenamefont {Cooper}, \citenamefont {Ye},\ and\ \citenamefont
  {Rey}}]{zhuPRA2016}%
  \BibitemOpen
  \bibfield  {author} {\bibinfo {author} {\bibfnamefont {B.}~\bibnamefont
  {Zhu}}, \bibinfo {author} {\bibfnamefont {J.}~\bibnamefont {Cooper}},
  \bibinfo {author} {\bibfnamefont {J.}~\bibnamefont {Ye}}, \ and\ \bibinfo
  {author} {\bibfnamefont {A.~M.}\ \bibnamefont {Rey}},\ }\href {\doibase
  10.1103/PhysRevA.94.023612} {\bibfield  {journal} {\bibinfo  {journal} {Phys.
  Rev. A}\ }\textbf {\bibinfo {volume} {94}},\ \bibinfo {pages} {023612}
  (\bibinfo {year} {2016})}\BibitemShut {NoStop}%
\bibitem [{\citenamefont {Corman}\ \emph {et~al.}(2017)\citenamefont {Corman},
  \citenamefont {Ville}, \citenamefont {Saint-Jalm}, \citenamefont
  {Aidelsburger}, \citenamefont {Bienaim\'e}, \citenamefont {Nascimb\`ene},
  \citenamefont {Dalibard},\ and\ \citenamefont {Beugnon}}]{cormanPRA2017}%
  \BibitemOpen
  \bibfield  {author} {\bibinfo {author} {\bibfnamefont {L.}~\bibnamefont
  {Corman}}, \bibinfo {author} {\bibfnamefont {J.~L.}\ \bibnamefont {Ville}},
  \bibinfo {author} {\bibfnamefont {R.}~\bibnamefont {Saint-Jalm}}, \bibinfo
  {author} {\bibfnamefont {M.}~\bibnamefont {Aidelsburger}}, \bibinfo {author}
  {\bibfnamefont {T.}~\bibnamefont {Bienaim\'e}}, \bibinfo {author}
  {\bibfnamefont {S.}~\bibnamefont {Nascimb\`ene}}, \bibinfo {author}
  {\bibfnamefont {J.}~\bibnamefont {Dalibard}}, \ and\ \bibinfo {author}
  {\bibfnamefont {J.}~\bibnamefont {Beugnon}},\ }\href {\doibase
  10.1103/PhysRevA.96.053629} {\bibfield  {journal} {\bibinfo  {journal} {Phys.
  Rev. A}\ }\textbf {\bibinfo {volume} {96}},\ \bibinfo {pages} {053629}
  (\bibinfo {year} {2017})}\BibitemShut {NoStop}%
\bibitem [{\citenamefont {Schilder}\ \emph {et~al.}(2016)\citenamefont
  {Schilder}, \citenamefont {Sauvan}, \citenamefont {Hugonin}, \citenamefont
  {Jennewein}, \citenamefont {Sortais}, \citenamefont {Browaeys},\ and\
  \citenamefont {Greffet}}]{Schilder2016}%
  \BibitemOpen
  \bibfield  {author} {\bibinfo {author} {\bibfnamefont {N.~J.}\ \bibnamefont
  {Schilder}}, \bibinfo {author} {\bibfnamefont {C.}~\bibnamefont {Sauvan}},
  \bibinfo {author} {\bibfnamefont {J.-P.}\ \bibnamefont {Hugonin}}, \bibinfo
  {author} {\bibfnamefont {S.}~\bibnamefont {Jennewein}}, \bibinfo {author}
  {\bibfnamefont {Y.~R.~P.}\ \bibnamefont {Sortais}}, \bibinfo {author}
  {\bibfnamefont {A.}~\bibnamefont {Browaeys}}, \ and\ \bibinfo {author}
  {\bibfnamefont {J.-J.}\ \bibnamefont {Greffet}},\ }\href {\doibase
  10.1103/PhysRevA.93.063835} {\bibfield  {journal} {\bibinfo  {journal} {Phys.
  Rev. A}\ }\textbf {\bibinfo {volume} {93}},\ \bibinfo {pages} {063835}
  (\bibinfo {year} {2016})}\BibitemShut {NoStop}%
\bibitem [{\citenamefont {Keaveney}\ \emph {et~al.}(2012)\citenamefont
  {Keaveney}, \citenamefont {Sargsyan}, \citenamefont {Krohn}, \citenamefont
  {Hughes}, \citenamefont {Sarkisyan},\ and\ \citenamefont
  {Adams}}]{keaveneyPRL2012}%
  \BibitemOpen
  \bibfield  {author} {\bibinfo {author} {\bibfnamefont {J.}~\bibnamefont
  {Keaveney}}, \bibinfo {author} {\bibfnamefont {A.}~\bibnamefont {Sargsyan}},
  \bibinfo {author} {\bibfnamefont {U.}~\bibnamefont {Krohn}}, \bibinfo
  {author} {\bibfnamefont {I.~G.}\ \bibnamefont {Hughes}}, \bibinfo {author}
  {\bibfnamefont {D.}~\bibnamefont {Sarkisyan}}, \ and\ \bibinfo {author}
  {\bibfnamefont {C.~S.}\ \bibnamefont {Adams}},\ }\href {\doibase
  10.1103/PhysRevLett.108.173601} {\bibfield  {journal} {\bibinfo  {journal}
  {Phys. Rev. Lett.}\ }\textbf {\bibinfo {volume} {108}},\ \bibinfo {pages}
  {173601} (\bibinfo {year} {2012})}\BibitemShut {NoStop}%
\bibitem [{\citenamefont {Meir}\ \emph {et~al.}(2014)\citenamefont {Meir},
  \citenamefont {Schwartz}, \citenamefont {Shahmoon}, \citenamefont {Oron},\
  and\ \citenamefont {Ozeri}}]{meirPRL2014}%
  \BibitemOpen
  \bibfield  {author} {\bibinfo {author} {\bibfnamefont {Z.}~\bibnamefont
  {Meir}}, \bibinfo {author} {\bibfnamefont {O.}~\bibnamefont {Schwartz}},
  \bibinfo {author} {\bibfnamefont {E.}~\bibnamefont {Shahmoon}}, \bibinfo
  {author} {\bibfnamefont {D.}~\bibnamefont {Oron}}, \ and\ \bibinfo {author}
  {\bibfnamefont {R.}~\bibnamefont {Ozeri}},\ }\href {\doibase
  10.1103/PhysRevLett.113.193002} {\bibfield  {journal} {\bibinfo  {journal}
  {Phys. Rev. Lett.}\ }\textbf {\bibinfo {volume} {113}},\ \bibinfo {pages}
  {193002} (\bibinfo {year} {2014})}\BibitemShut {NoStop}%
\bibitem [{\citenamefont {Chang}\ \emph {et~al.}(2014)\citenamefont {Chang},
  \citenamefont {Vuleti{\'c}},\ and\ \citenamefont
  {Lukin}}]{changNaturephoton2014}%
  \BibitemOpen
  \bibfield  {author} {\bibinfo {author} {\bibfnamefont {D.~E.}\ \bibnamefont
  {Chang}}, \bibinfo {author} {\bibfnamefont {V.}~\bibnamefont {Vuleti{\'c}}},
  \ and\ \bibinfo {author} {\bibfnamefont {M.~D.}\ \bibnamefont {Lukin}},\
  }\href@noop {} {\bibfield  {journal} {\bibinfo  {journal} {Nature Photonics}\
  }\textbf {\bibinfo {volume} {8}},\ \bibinfo {pages} {685} (\bibinfo {year}
  {2014})}\BibitemShut {NoStop}%
\bibitem [{\citenamefont {Ruostekoski}\ and\ \citenamefont
  {Javanainen}(1999)}]{ruostekoskiPRL1999}%
  \BibitemOpen
  \bibfield  {author} {\bibinfo {author} {\bibfnamefont {J.}~\bibnamefont
  {Ruostekoski}}\ and\ \bibinfo {author} {\bibfnamefont {J.}~\bibnamefont
  {Javanainen}},\ }\href {\doibase 10.1103/PhysRevLett.82.4741} {\bibfield
  {journal} {\bibinfo  {journal} {Phys. Rev. Lett.}\ }\textbf {\bibinfo
  {volume} {82}},\ \bibinfo {pages} {4741} (\bibinfo {year}
  {1999})}\BibitemShut {NoStop}%
\bibitem [{\citenamefont {Ruostekoski}(2000)}]{ruostekoskiPRA2000}%
  \BibitemOpen
  \bibfield  {author} {\bibinfo {author} {\bibfnamefont {J.}~\bibnamefont
  {Ruostekoski}},\ }\href {\doibase 10.1103/PhysRevA.61.033605} {\bibfield
  {journal} {\bibinfo  {journal} {Phys. Rev. A}\ }\textbf {\bibinfo {volume}
  {61}},\ \bibinfo {pages} {033605} (\bibinfo {year} {2000})}\BibitemShut
  {NoStop}%
\bibitem [{\citenamefont {Javanainen}\ \emph {et~al.}(2014)\citenamefont
  {Javanainen}, \citenamefont {Ruostekoski}, \citenamefont {Li},\ and\
  \citenamefont {Yoo}}]{javanainenPRL2014}%
  \BibitemOpen
  \bibfield  {author} {\bibinfo {author} {\bibfnamefont {J.}~\bibnamefont
  {Javanainen}}, \bibinfo {author} {\bibfnamefont {J.}~\bibnamefont
  {Ruostekoski}}, \bibinfo {author} {\bibfnamefont {Y.}~\bibnamefont {Li}}, \
  and\ \bibinfo {author} {\bibfnamefont {S.-M.}\ \bibnamefont {Yoo}},\ }\href
  {\doibase 10.1103/PhysRevLett.112.113603} {\bibfield  {journal} {\bibinfo
  {journal} {Phys. Rev. Lett.}\ }\textbf {\bibinfo {volume} {112}},\ \bibinfo
  {pages} {113603} (\bibinfo {year} {2014})}\BibitemShut {NoStop}%
\bibitem [{\citenamefont {Jenkins}\ \emph {et~al.}(2016)\citenamefont
  {Jenkins}, \citenamefont {Ruostekoski}, \citenamefont {Javanainen},
  \citenamefont {Bourgain}, \citenamefont {Jennewein}, \citenamefont
  {Sortais},\ and\ \citenamefont {Browaeys}}]{jenkinsPRL2016}%
  \BibitemOpen
  \bibfield  {author} {\bibinfo {author} {\bibfnamefont {S.~D.}\ \bibnamefont
  {Jenkins}}, \bibinfo {author} {\bibfnamefont {J.}~\bibnamefont
  {Ruostekoski}}, \bibinfo {author} {\bibfnamefont {J.}~\bibnamefont
  {Javanainen}}, \bibinfo {author} {\bibfnamefont {R.}~\bibnamefont
  {Bourgain}}, \bibinfo {author} {\bibfnamefont {S.}~\bibnamefont {Jennewein}},
  \bibinfo {author} {\bibfnamefont {Y.~R.~P.}\ \bibnamefont {Sortais}}, \ and\
  \bibinfo {author} {\bibfnamefont {A.}~\bibnamefont {Browaeys}},\ }\href
  {\doibase 10.1103/PhysRevLett.116.183601} {\bibfield  {journal} {\bibinfo
  {journal} {Phys. Rev. Lett.}\ }\textbf {\bibinfo {volume} {116}},\ \bibinfo
  {pages} {183601} (\bibinfo {year} {2016})}\BibitemShut {NoStop}%
\bibitem [{\citenamefont {Javanainen}\ \emph {et~al.}(2017)\citenamefont
  {Javanainen}, \citenamefont {Ruostekoski}, \citenamefont {Li},\ and\
  \citenamefont {Yoo}}]{javanainenPRA2017}%
  \BibitemOpen
  \bibfield  {author} {\bibinfo {author} {\bibfnamefont {J.}~\bibnamefont
  {Javanainen}}, \bibinfo {author} {\bibfnamefont {J.}~\bibnamefont
  {Ruostekoski}}, \bibinfo {author} {\bibfnamefont {Y.}~\bibnamefont {Li}}, \
  and\ \bibinfo {author} {\bibfnamefont {S.-M.}\ \bibnamefont {Yoo}},\ }\href
  {\doibase 10.1103/PhysRevA.96.033835} {\bibfield  {journal} {\bibinfo
  {journal} {Phys. Rev. A}\ }\textbf {\bibinfo {volume} {96}},\ \bibinfo
  {pages} {033835} (\bibinfo {year} {2017})}\BibitemShut {NoStop}%
\bibitem [{\citenamefont {Morice}\ \emph {et~al.}(1995)\citenamefont {Morice},
  \citenamefont {Castin},\ and\ \citenamefont {Dalibard}}]{moricePRA1995}%
  \BibitemOpen
  \bibfield  {author} {\bibinfo {author} {\bibfnamefont {O.}~\bibnamefont
  {Morice}}, \bibinfo {author} {\bibfnamefont {Y.}~\bibnamefont {Castin}}, \
  and\ \bibinfo {author} {\bibfnamefont {J.}~\bibnamefont {Dalibard}},\ }\href
  {\doibase 10.1103/PhysRevA.51.3896} {\bibfield  {journal} {\bibinfo
  {journal} {Phys. Rev. A}\ }\textbf {\bibinfo {volume} {51}},\ \bibinfo
  {pages} {3896} (\bibinfo {year} {1995})}\BibitemShut {NoStop}%
\bibitem [{\citenamefont {Schilder}\ \emph {et~al.}(2017)\citenamefont
  {Schilder}, \citenamefont {Sauvan}, \citenamefont {Sortais}, \citenamefont
  {Browaeys},\ and\ \citenamefont {Greffet}}]{schilderPRA2017}%
  \BibitemOpen
  \bibfield  {author} {\bibinfo {author} {\bibfnamefont {N.~J.}\ \bibnamefont
  {Schilder}}, \bibinfo {author} {\bibfnamefont {C.}~\bibnamefont {Sauvan}},
  \bibinfo {author} {\bibfnamefont {Y.~R.~P.}\ \bibnamefont {Sortais}},
  \bibinfo {author} {\bibfnamefont {A.}~\bibnamefont {Browaeys}}, \ and\
  \bibinfo {author} {\bibfnamefont {J.-J.}\ \bibnamefont {Greffet}},\ }\href
  {\doibase 10.1103/PhysRevA.96.013825} {\bibfield  {journal} {\bibinfo
  {journal} {Phys. Rev. A}\ }\textbf {\bibinfo {volume} {96}},\ \bibinfo
  {pages} {013825} (\bibinfo {year} {2017})}\BibitemShut {NoStop}%
\bibitem [{\citenamefont {Jennewein}\ \emph {et~al.}(2016)\citenamefont
  {Jennewein}, \citenamefont {Besbes}, \citenamefont {Schilder}, \citenamefont
  {Jenkins}, \citenamefont {Sauvan}, \citenamefont {Ruostekoski}, \citenamefont
  {Greffet}, \citenamefont {Sortais},\ and\ \citenamefont
  {Browaeys}}]{jenneweinPRL2016}%
  \BibitemOpen
  \bibfield  {author} {\bibinfo {author} {\bibfnamefont {S.}~\bibnamefont
  {Jennewein}}, \bibinfo {author} {\bibfnamefont {M.}~\bibnamefont {Besbes}},
  \bibinfo {author} {\bibfnamefont {N.~J.}\ \bibnamefont {Schilder}}, \bibinfo
  {author} {\bibfnamefont {S.~D.}\ \bibnamefont {Jenkins}}, \bibinfo {author}
  {\bibfnamefont {C.}~\bibnamefont {Sauvan}}, \bibinfo {author} {\bibfnamefont
  {J.}~\bibnamefont {Ruostekoski}}, \bibinfo {author} {\bibfnamefont {J.-J.}\
  \bibnamefont {Greffet}}, \bibinfo {author} {\bibfnamefont {Y.~R.~P.}\
  \bibnamefont {Sortais}}, \ and\ \bibinfo {author} {\bibfnamefont
  {A.}~\bibnamefont {Browaeys}},\ }\href {\doibase
  10.1103/PhysRevLett.116.233601} {\bibfield  {journal} {\bibinfo  {journal}
  {Phys. Rev. Lett.}\ }\textbf {\bibinfo {volume} {116}},\ \bibinfo {pages}
  {233601} (\bibinfo {year} {2016})}\BibitemShut {NoStop}%
\bibitem [{\citenamefont {Javanainen}\ and\ \citenamefont
  {Ruostekoski}(2016)}]{ruostekoskiOE2016}%
  \BibitemOpen
  \bibfield  {author} {\bibinfo {author} {\bibfnamefont {J.}~\bibnamefont
  {Javanainen}}\ and\ \bibinfo {author} {\bibfnamefont {J.}~\bibnamefont
  {Ruostekoski}},\ }\href {\doibase 10.1364/OE.24.000993} {\bibfield  {journal}
  {\bibinfo  {journal} {Opt. Express}\ }\textbf {\bibinfo {volume} {24}},\
  \bibinfo {pages} {993} (\bibinfo {year} {2016})}\BibitemShut {NoStop}%
\bibitem [{\citenamefont {Parkins}\ and\ \citenamefont
  {Walls}(1998)}]{parkinsPhysrep1998}%
  \BibitemOpen
  \bibfield  {author} {\bibinfo {author} {\bibfnamefont {A.}~\bibnamefont
  {Parkins}}\ and\ \bibinfo {author} {\bibfnamefont {D.}~\bibnamefont
  {Walls}},\ }\href {\doibase https://doi.org/10.1016/S0370-1573(98)00014-3}
  {\bibfield  {journal} {\bibinfo  {journal} {Physics Reports}\ }\textbf
  {\bibinfo {volume} {303}},\ \bibinfo {pages} {1 } (\bibinfo {year}
  {1998})}\BibitemShut {NoStop}%
\bibitem [{\citenamefont {Peyrot}\ \emph {et~al.}(2018)\citenamefont {Peyrot},
  \citenamefont {Sortais}, \citenamefont {Browaeys}, \citenamefont {Sargsyan},
  \citenamefont {Sarkisyan}, \citenamefont {Keaveney}, \citenamefont {Hughes},\
  and\ \citenamefont {Adams}}]{peyrotPRL2018}%
  \BibitemOpen
  \bibfield  {author} {\bibinfo {author} {\bibfnamefont {T.}~\bibnamefont
  {Peyrot}}, \bibinfo {author} {\bibfnamefont {Y.~R.~P.}\ \bibnamefont
  {Sortais}}, \bibinfo {author} {\bibfnamefont {A.}~\bibnamefont {Browaeys}},
  \bibinfo {author} {\bibfnamefont {A.}~\bibnamefont {Sargsyan}}, \bibinfo
  {author} {\bibfnamefont {D.}~\bibnamefont {Sarkisyan}}, \bibinfo {author}
  {\bibfnamefont {J.}~\bibnamefont {Keaveney}}, \bibinfo {author}
  {\bibfnamefont {I.~G.}\ \bibnamefont {Hughes}}, \ and\ \bibinfo {author}
  {\bibfnamefont {C.~S.}\ \bibnamefont {Adams}},\ }\href {\doibase
  10.1103/PhysRevLett.120.243401} {\bibfield  {journal} {\bibinfo  {journal}
  {Phys. Rev. Lett.}\ }\textbf {\bibinfo {volume} {120}},\ \bibinfo {pages}
  {243401} (\bibinfo {year} {2018})}\BibitemShut {NoStop}%
\bibitem [{\citenamefont {Pellegrino}\ \emph {et~al.}(2014)\citenamefont
  {Pellegrino}, \citenamefont {Bourgain}, \citenamefont {Jennewein},
  \citenamefont {Sortais}, \citenamefont {Browaeys}, \citenamefont {Jenkins},\
  and\ \citenamefont {Ruostekoski}}]{pellegrinoPRL2014}%
  \BibitemOpen
  \bibfield  {author} {\bibinfo {author} {\bibfnamefont {J.}~\bibnamefont
  {Pellegrino}}, \bibinfo {author} {\bibfnamefont {R.}~\bibnamefont
  {Bourgain}}, \bibinfo {author} {\bibfnamefont {S.}~\bibnamefont {Jennewein}},
  \bibinfo {author} {\bibfnamefont {Y.~R.~P.}\ \bibnamefont {Sortais}},
  \bibinfo {author} {\bibfnamefont {A.}~\bibnamefont {Browaeys}}, \bibinfo
  {author} {\bibfnamefont {S.~D.}\ \bibnamefont {Jenkins}}, \ and\ \bibinfo
  {author} {\bibfnamefont {J.}~\bibnamefont {Ruostekoski}},\ }\href {\doibase
  10.1103/PhysRevLett.113.133602} {\bibfield  {journal} {\bibinfo  {journal}
  {Phys. Rev. Lett.}\ }\textbf {\bibinfo {volume} {113}},\ \bibinfo {pages}
  {133602} (\bibinfo {year} {2014})}\BibitemShut {NoStop}%
\bibitem [{\citenamefont {Endres}\ \emph {et~al.}(2016)\citenamefont {Endres},
  \citenamefont {Bernien}, \citenamefont {Keesling}, \citenamefont {Levine},
  \citenamefont {Anschuetz}, \citenamefont {Krajenbrink}, \citenamefont
  {Senko}, \citenamefont {Vuletic}, \citenamefont {Greiner},\ and\
  \citenamefont {Lukin}}]{endresScience2016}%
  \BibitemOpen
  \bibfield  {author} {\bibinfo {author} {\bibfnamefont {M.}~\bibnamefont
  {Endres}}, \bibinfo {author} {\bibfnamefont {H.}~\bibnamefont {Bernien}},
  \bibinfo {author} {\bibfnamefont {A.}~\bibnamefont {Keesling}}, \bibinfo
  {author} {\bibfnamefont {H.}~\bibnamefont {Levine}}, \bibinfo {author}
  {\bibfnamefont {E.~R.}\ \bibnamefont {Anschuetz}}, \bibinfo {author}
  {\bibfnamefont {A.}~\bibnamefont {Krajenbrink}}, \bibinfo {author}
  {\bibfnamefont {C.}~\bibnamefont {Senko}}, \bibinfo {author} {\bibfnamefont
  {V.}~\bibnamefont {Vuletic}}, \bibinfo {author} {\bibfnamefont
  {M.}~\bibnamefont {Greiner}}, \ and\ \bibinfo {author} {\bibfnamefont
  {M.~D.}\ \bibnamefont {Lukin}},\ }\href@noop {} {\bibfield  {journal}
  {\bibinfo  {journal} {Science}\ ,\ \bibinfo {pages} {aah3752}} (\bibinfo
  {year} {2016})}\BibitemShut {NoStop}%
\bibitem [{\citenamefont {Barredo}\ \emph {et~al.}(2016)\citenamefont
  {Barredo}, \citenamefont {de~L{\'e}s{\'e}leuc}, \citenamefont {Lienhard},
  \citenamefont {Lahaye},\ and\ \citenamefont {Browaeys}}]{lahayeScience2016}%
  \BibitemOpen
  \bibfield  {author} {\bibinfo {author} {\bibfnamefont {D.}~\bibnamefont
  {Barredo}}, \bibinfo {author} {\bibfnamefont {S.}~\bibnamefont
  {de~L{\'e}s{\'e}leuc}}, \bibinfo {author} {\bibfnamefont {V.}~\bibnamefont
  {Lienhard}}, \bibinfo {author} {\bibfnamefont {T.}~\bibnamefont {Lahaye}}, \
  and\ \bibinfo {author} {\bibfnamefont {A.}~\bibnamefont {Browaeys}},\
  }\href@noop {} {\bibfield  {journal} {\bibinfo  {journal} {Science}\ }\textbf
  {\bibinfo {volume} {354}},\ \bibinfo {pages} {1021} (\bibinfo {year}
  {2016})}\BibitemShut {NoStop}%
\bibitem [{\citenamefont {Gonz{\'a}lez-Tudela}\ \emph
  {et~al.}(2015)\citenamefont {Gonz{\'a}lez-Tudela}, \citenamefont {Hung},
  \citenamefont {Chang}, \citenamefont {Cirac},\ and\ \citenamefont
  {Kimble}}]{gonzalezNaturephoton2015}%
  \BibitemOpen
  \bibfield  {author} {\bibinfo {author} {\bibfnamefont {A.}~\bibnamefont
  {Gonz{\'a}lez-Tudela}}, \bibinfo {author} {\bibfnamefont {C.-L.}\
  \bibnamefont {Hung}}, \bibinfo {author} {\bibfnamefont {D.~E.}\ \bibnamefont
  {Chang}}, \bibinfo {author} {\bibfnamefont {J.~I.}\ \bibnamefont {Cirac}}, \
  and\ \bibinfo {author} {\bibfnamefont {H.}~\bibnamefont {Kimble}},\
  }\href@noop {} {\bibfield  {journal} {\bibinfo  {journal} {Nature Photonics}\
  }\textbf {\bibinfo {volume} {9}},\ \bibinfo {pages} {320} (\bibinfo {year}
  {2015})}\BibitemShut {NoStop}%
\bibitem [{\citenamefont {Gullans}\ \emph {et~al.}(2012)\citenamefont
  {Gullans}, \citenamefont {Tiecke}, \citenamefont {Chang}, \citenamefont
  {Feist}, \citenamefont {Thompson}, \citenamefont {Cirac}, \citenamefont
  {Zoller},\ and\ \citenamefont {Lukin}}]{gullansPRL2012}%
  \BibitemOpen
  \bibfield  {author} {\bibinfo {author} {\bibfnamefont {M.}~\bibnamefont
  {Gullans}}, \bibinfo {author} {\bibfnamefont {T.~G.}\ \bibnamefont {Tiecke}},
  \bibinfo {author} {\bibfnamefont {D.~E.}\ \bibnamefont {Chang}}, \bibinfo
  {author} {\bibfnamefont {J.}~\bibnamefont {Feist}}, \bibinfo {author}
  {\bibfnamefont {J.~D.}\ \bibnamefont {Thompson}}, \bibinfo {author}
  {\bibfnamefont {J.~I.}\ \bibnamefont {Cirac}}, \bibinfo {author}
  {\bibfnamefont {P.}~\bibnamefont {Zoller}}, \ and\ \bibinfo {author}
  {\bibfnamefont {M.~D.}\ \bibnamefont {Lukin}},\ }\href {\doibase
  10.1103/PhysRevLett.109.235309} {\bibfield  {journal} {\bibinfo  {journal}
  {Phys. Rev. Lett.}\ }\textbf {\bibinfo {volume} {109}},\ \bibinfo {pages}
  {235309} (\bibinfo {year} {2012})}\BibitemShut {NoStop}%
\bibitem [{\citenamefont {López}(2018)}]{lopezADOM2018}%
  \BibitemOpen
  \bibfield  {author} {\bibinfo {author} {\bibfnamefont {C.}~\bibnamefont
  {López}},\ }\href {\doibase 10.1002/adom.201800439} {\bibfield  {journal}
  {\bibinfo  {journal} {Advanced Optical Materials}\ }\textbf {\bibinfo
  {volume} {6}},\ \bibinfo {pages} {1800439} (\bibinfo {year} {2018})},\
  \Eprint
  {http://arxiv.org/abs/https://onlinelibrary.wiley.com/doi/pdf/10.1002/adom.201800439}
  {https://onlinelibrary.wiley.com/doi/pdf/10.1002/adom.201800439} \BibitemShut
  {NoStop}%
\bibitem [{\citenamefont {Anderson}(1972)}]{andersonScience1972}%
  \BibitemOpen
  \bibfield  {author} {\bibinfo {author} {\bibfnamefont {P.~W.}\ \bibnamefont
  {Anderson}},\ }\href {\doibase 10.1126/science.177.4047.393} {\bibfield
  {journal} {\bibinfo  {journal} {Science}\ }\textbf {\bibinfo {volume}
  {177}},\ \bibinfo {pages} {393} (\bibinfo {year} {1972})},\ \Eprint
  {http://arxiv.org/abs/https://science.sciencemag.org/content/177/4047/393.full.pdf}
  {https://science.sciencemag.org/content/177/4047/393.full.pdf} \BibitemShut
  {NoStop}%
\bibitem [{\citenamefont {Tian}(2013)}]{tianPhysicaE2013}%
  \BibitemOpen
  \bibfield  {author} {\bibinfo {author} {\bibfnamefont {C.}~\bibnamefont
  {Tian}},\ }\href {\doibase https://doi.org/10.1016/j.physe.2013.01.021}
  {\bibfield  {journal} {\bibinfo  {journal} {Physica E: Low-dimensional
  Systems and Nanostructures}\ }\textbf {\bibinfo {volume} {49}},\ \bibinfo
  {pages} {124 } (\bibinfo {year} {2013})}\BibitemShut {NoStop}%
\bibitem [{\citenamefont {Fang}\ \emph {et~al.}(2019)\citenamefont {Fang},
  \citenamefont {Tian}, \citenamefont {Zhao}, \citenamefont {Bliokh},
  \citenamefont {Freilikher},\ and\ \citenamefont {Nori}}]{fangPRB2019}%
  \BibitemOpen
  \bibfield  {author} {\bibinfo {author} {\bibfnamefont {P.}~\bibnamefont
  {Fang}}, \bibinfo {author} {\bibfnamefont {C.}~\bibnamefont {Tian}}, \bibinfo
  {author} {\bibfnamefont {L.}~\bibnamefont {Zhao}}, \bibinfo {author}
  {\bibfnamefont {Y.~P.}\ \bibnamefont {Bliokh}}, \bibinfo {author}
  {\bibfnamefont {V.}~\bibnamefont {Freilikher}}, \ and\ \bibinfo {author}
  {\bibfnamefont {F.}~\bibnamefont {Nori}},\ }\href {\doibase
  10.1103/PhysRevB.99.094202} {\bibfield  {journal} {\bibinfo  {journal} {Phys.
  Rev. B}\ }\textbf {\bibinfo {volume} {99}},\ \bibinfo {pages} {094202}
  (\bibinfo {year} {2019})}\BibitemShut {NoStop}%
\bibitem [{\citenamefont {Abrahams}\ \emph {et~al.}(1979)\citenamefont
  {Abrahams}, \citenamefont {Anderson}, \citenamefont {Licciardello},\ and\
  \citenamefont {Ramakrishnan}}]{abrahamsPRL1979}%
  \BibitemOpen
  \bibfield  {author} {\bibinfo {author} {\bibfnamefont {E.}~\bibnamefont
  {Abrahams}}, \bibinfo {author} {\bibfnamefont {P.~W.}\ \bibnamefont
  {Anderson}}, \bibinfo {author} {\bibfnamefont {D.~C.}\ \bibnamefont
  {Licciardello}}, \ and\ \bibinfo {author} {\bibfnamefont {T.~V.}\
  \bibnamefont {Ramakrishnan}},\ }\href {\doibase 10.1103/PhysRevLett.42.673}
  {\bibfield  {journal} {\bibinfo  {journal} {Phys. Rev. Lett.}\ }\textbf
  {\bibinfo {volume} {42}},\ \bibinfo {pages} {673} (\bibinfo {year}
  {1979})}\BibitemShut {NoStop}%
\bibitem [{\citenamefont {Jacucci}\ \emph {et~al.}(2019)\citenamefont
  {Jacucci}, \citenamefont {Bertolotti},\ and\ \citenamefont
  {Vignolini}}]{jacucciADOM2019}%
  \BibitemOpen
  \bibfield  {author} {\bibinfo {author} {\bibfnamefont {G.}~\bibnamefont
  {Jacucci}}, \bibinfo {author} {\bibfnamefont {J.}~\bibnamefont {Bertolotti}},
  \ and\ \bibinfo {author} {\bibfnamefont {S.}~\bibnamefont {Vignolini}},\
  }\href {\doibase 10.1002/adom.201900980} {\bibfield  {journal} {\bibinfo
  {journal} {Advanced Optical Materials}\ }\textbf {\bibinfo {volume} {7}},\
  \bibinfo {pages} {1900980} (\bibinfo {year} {2019})},\ \Eprint
  {http://arxiv.org/abs/https://onlinelibrary.wiley.com/doi/pdf/10.1002/adom.201900980}
  {https://onlinelibrary.wiley.com/doi/pdf/10.1002/adom.201900980} \BibitemShut
  {NoStop}%
\bibitem [{\citenamefont {Mackowski}\ and\ \citenamefont
  {Mishchenko}(2008)}]{mackowskiJHT2008}%
  \BibitemOpen
  \bibfield  {author} {\bibinfo {author} {\bibfnamefont {D.~W.}\ \bibnamefont
  {Mackowski}}\ and\ \bibinfo {author} {\bibfnamefont {M.~I.}\ \bibnamefont
  {Mishchenko}},\ }\href {\doibase 10.1115/1.2957596} {\bibfield  {journal}
  {\bibinfo  {journal} {Journal of Heat Transfer}\ }\textbf {\bibinfo {volume}
  {130}} (\bibinfo {year} {2008}),\ 10.1115/1.2957596},\ \bibinfo {note}
  {112702},\ \Eprint
  {http://arxiv.org/abs/https://asmedigitalcollection.asme.org/heattransfer/article-pdf/130/11/112702/5725891/112702\_1.pdf}
  {https://asmedigitalcollection.asme.org/heattransfer/article-pdf/130/11/112702/5725891/112702\_1.pdf}
  \BibitemShut {NoStop}%
\bibitem [{\citenamefont {Mishchenko}(2017)}]{mishchenkoJQSRT2017}%
  \BibitemOpen
  \bibfield  {author} {\bibinfo {author} {\bibfnamefont {M.~I.}\ \bibnamefont
  {Mishchenko}},\ }\href {\doibase https://doi.org/10.1016/j.jqsrt.2017.06.003}
  {\bibfield  {journal} {\bibinfo  {journal} {Journal of Quantitative
  Spectroscopy and Radiative Transfer}\ }\textbf {\bibinfo {volume} {200}},\
  \bibinfo {pages} {137 } (\bibinfo {year} {2017})}\BibitemShut {NoStop}%
\bibitem [{\citenamefont {{Rytov}}\ \emph {et~al.}(1989)\citenamefont
  {{Rytov}}, \citenamefont {{Kravtsov}},\ and\ \citenamefont
  {{Tatarskii}}}]{rytovvolume4}%
  \BibitemOpen
  \bibfield  {author} {\bibinfo {author} {\bibfnamefont {S.~M.}\ \bibnamefont
  {{Rytov}}}, \bibinfo {author} {\bibfnamefont {Y.~A.}\ \bibnamefont
  {{Kravtsov}}}, \ and\ \bibinfo {author} {\bibfnamefont {V.~I.}\ \bibnamefont
  {{Tatarskii}}},\ }\href@noop {} {\emph {\bibinfo {title} {{Principles of
  statistical radiophysics. 4. Wave propagation through random media.}}}}\
  (\bibinfo {year} {1989})\BibitemShut {NoStop}%
\bibitem [{\citenamefont {Latella}\ \emph {et~al.}(2018)\citenamefont
  {Latella}, \citenamefont {Biehs}, \citenamefont {Messina}, \citenamefont
  {Rodriguez},\ and\ \citenamefont {Ben-Abdallah}}]{latellaPRB2018}%
  \BibitemOpen
  \bibfield  {author} {\bibinfo {author} {\bibfnamefont {I.}~\bibnamefont
  {Latella}}, \bibinfo {author} {\bibfnamefont {S.-A.}\ \bibnamefont {Biehs}},
  \bibinfo {author} {\bibfnamefont {R.}~\bibnamefont {Messina}}, \bibinfo
  {author} {\bibfnamefont {A.~W.}\ \bibnamefont {Rodriguez}}, \ and\ \bibinfo
  {author} {\bibfnamefont {P.}~\bibnamefont {Ben-Abdallah}},\ }\href {\doibase
  10.1103/PhysRevB.97.035423} {\bibfield  {journal} {\bibinfo  {journal} {Phys.
  Rev. B}\ }\textbf {\bibinfo {volume} {97}},\ \bibinfo {pages} {035423}
  (\bibinfo {year} {2018})}\BibitemShut {NoStop}%
\bibitem [{\citenamefont {Dong}\ \emph {et~al.}(2017)\citenamefont {Dong},
  \citenamefont {Zhao},\ and\ \citenamefont {Liu}}]{dongPRB2017}%
  \BibitemOpen
  \bibfield  {author} {\bibinfo {author} {\bibfnamefont {J.}~\bibnamefont
  {Dong}}, \bibinfo {author} {\bibfnamefont {J.}~\bibnamefont {Zhao}}, \ and\
  \bibinfo {author} {\bibfnamefont {L.}~\bibnamefont {Liu}},\ }\href {\doibase
  10.1103/PhysRevB.95.125411} {\bibfield  {journal} {\bibinfo  {journal} {Phys.
  Rev. B}\ }\textbf {\bibinfo {volume} {95}},\ \bibinfo {pages} {125411}
  (\bibinfo {year} {2017})}\BibitemShut {NoStop}%
\bibitem [{\citenamefont {Zhu}\ \emph {et~al.}(2018)\citenamefont {Zhu},
  \citenamefont {Guo},\ and\ \citenamefont {Fan}}]{zhuPRB2018}%
  \BibitemOpen
  \bibfield  {author} {\bibinfo {author} {\bibfnamefont {L.}~\bibnamefont
  {Zhu}}, \bibinfo {author} {\bibfnamefont {Y.}~\bibnamefont {Guo}}, \ and\
  \bibinfo {author} {\bibfnamefont {S.}~\bibnamefont {Fan}},\ }\href {\doibase
  10.1103/PhysRevB.97.094302} {\bibfield  {journal} {\bibinfo  {journal} {Phys.
  Rev. B}\ }\textbf {\bibinfo {volume} {97}},\ \bibinfo {pages} {094302}
  (\bibinfo {year} {2018})}\BibitemShut {NoStop}%
\bibitem [{\citenamefont {Ben-Abdallah}\ \emph {et~al.}(2013)\citenamefont
  {Ben-Abdallah}, \citenamefont {Messina}, \citenamefont {Biehs}, \citenamefont
  {Tschikin}, \citenamefont {Joulain},\ and\ \citenamefont
  {Henkel}}]{benabdallahPRL2013}%
  \BibitemOpen
  \bibfield  {author} {\bibinfo {author} {\bibfnamefont {P.}~\bibnamefont
  {Ben-Abdallah}}, \bibinfo {author} {\bibfnamefont {R.}~\bibnamefont
  {Messina}}, \bibinfo {author} {\bibfnamefont {S.-A.}\ \bibnamefont {Biehs}},
  \bibinfo {author} {\bibfnamefont {M.}~\bibnamefont {Tschikin}}, \bibinfo
  {author} {\bibfnamefont {K.}~\bibnamefont {Joulain}}, \ and\ \bibinfo
  {author} {\bibfnamefont {C.}~\bibnamefont {Henkel}},\ }\href {\doibase
  10.1103/PhysRevLett.111.174301} {\bibfield  {journal} {\bibinfo  {journal}
  {Phys. Rev. Lett.}\ }\textbf {\bibinfo {volume} {111}},\ \bibinfo {pages}
  {174301} (\bibinfo {year} {2013})}\BibitemShut {NoStop}%
\bibitem [{\citenamefont {Chen}\ \emph
  {et~al.}(2018{\natexlab{b}})\citenamefont {Chen}, \citenamefont {Zhao},\ and\
  \citenamefont {Wang}}]{chenJQSRT2018b}%
  \BibitemOpen
  \bibfield  {author} {\bibinfo {author} {\bibfnamefont {J.}~\bibnamefont
  {Chen}}, \bibinfo {author} {\bibfnamefont {C.}~\bibnamefont {Zhao}}, \ and\
  \bibinfo {author} {\bibfnamefont {B.}~\bibnamefont {Wang}},\ }\href {\doibase
  https://doi.org/10.1016/j.jqsrt.2018.08.024} {\bibfield  {journal} {\bibinfo
  {journal} {Journal of Quantitative Spectroscopy and Radiative Transfer}\
  }\textbf {\bibinfo {volume} {219}},\ \bibinfo {pages} {304 } (\bibinfo {year}
  {2018}{\natexlab{b}})}\BibitemShut {NoStop}%
\bibitem [{\citenamefont {Shalaev}(2007)}]{shalaevNaturephoton2007}%
  \BibitemOpen
  \bibfield  {author} {\bibinfo {author} {\bibfnamefont {V.~M.}\ \bibnamefont
  {Shalaev}},\ }\href@noop {} {\bibfield  {journal} {\bibinfo  {journal}
  {Nature photonics}\ }\textbf {\bibinfo {volume} {1}},\ \bibinfo {pages} {41}
  (\bibinfo {year} {2007})}\BibitemShut {NoStop}%
\bibitem [{\citenamefont {Kildishev}\ \emph {et~al.}(2013)\citenamefont
  {Kildishev}, \citenamefont {Boltasseva},\ and\ \citenamefont
  {Shalaev}}]{shalaevScience2013}%
  \BibitemOpen
  \bibfield  {author} {\bibinfo {author} {\bibfnamefont {A.~V.}\ \bibnamefont
  {Kildishev}}, \bibinfo {author} {\bibfnamefont {A.}~\bibnamefont
  {Boltasseva}}, \ and\ \bibinfo {author} {\bibfnamefont {V.~M.}\ \bibnamefont
  {Shalaev}},\ }\href@noop {} {\bibfield  {journal} {\bibinfo  {journal}
  {Science}\ }\textbf {\bibinfo {volume} {339}},\ \bibinfo {pages} {1232009}
  (\bibinfo {year} {2013})}\BibitemShut {NoStop}%
\bibitem [{\citenamefont {Ozawa}\ \emph {et~al.}(2019)\citenamefont {Ozawa},
  \citenamefont {Price}, \citenamefont {Amo}, \citenamefont {Goldman},
  \citenamefont {Hafezi}, \citenamefont {Lu}, \citenamefont {Rechtsman},
  \citenamefont {Schuster}, \citenamefont {Simon}, \citenamefont {Zilberberg},\
  and\ \citenamefont {Carusotto}}]{ozawa2018topological}%
  \BibitemOpen
  \bibfield  {author} {\bibinfo {author} {\bibfnamefont {T.}~\bibnamefont
  {Ozawa}}, \bibinfo {author} {\bibfnamefont {H.~M.}\ \bibnamefont {Price}},
  \bibinfo {author} {\bibfnamefont {A.}~\bibnamefont {Amo}}, \bibinfo {author}
  {\bibfnamefont {N.}~\bibnamefont {Goldman}}, \bibinfo {author} {\bibfnamefont
  {M.}~\bibnamefont {Hafezi}}, \bibinfo {author} {\bibfnamefont
  {L.}~\bibnamefont {Lu}}, \bibinfo {author} {\bibfnamefont {M.~C.}\
  \bibnamefont {Rechtsman}}, \bibinfo {author} {\bibfnamefont {D.}~\bibnamefont
  {Schuster}}, \bibinfo {author} {\bibfnamefont {J.}~\bibnamefont {Simon}},
  \bibinfo {author} {\bibfnamefont {O.}~\bibnamefont {Zilberberg}}, \ and\
  \bibinfo {author} {\bibfnamefont {I.}~\bibnamefont {Carusotto}},\ }\href
  {\doibase 10.1103/RevModPhys.91.015006} {\bibfield  {journal} {\bibinfo
  {journal} {Rev. Mod. Phys.}\ }\textbf {\bibinfo {volume} {91}},\ \bibinfo
  {pages} {015006} (\bibinfo {year} {2019})}\BibitemShut {NoStop}%
\bibitem [{\citenamefont {Bloch}(2005)}]{blochNaturephys2005}%
  \BibitemOpen
  \bibfield  {author} {\bibinfo {author} {\bibfnamefont {I.}~\bibnamefont
  {Bloch}},\ }\href@noop {} {\bibfield  {journal} {\bibinfo  {journal} {Nature
  Physics}\ }\textbf {\bibinfo {volume} {1}},\ \bibinfo {pages} {23} (\bibinfo
  {year} {2005})}\BibitemShut {NoStop}%
\bibitem [{\citenamefont {Vellekoop}\ \emph {et~al.}(2010)\citenamefont
  {Vellekoop}, \citenamefont {Lagendijk},\ and\ \citenamefont
  {Mosk}}]{Vellekoop2010}%
  \BibitemOpen
  \bibfield  {author} {\bibinfo {author} {\bibfnamefont {I.~M.}\ \bibnamefont
  {Vellekoop}}, \bibinfo {author} {\bibfnamefont {A.}~\bibnamefont
  {Lagendijk}}, \ and\ \bibinfo {author} {\bibfnamefont {A.~P.}\ \bibnamefont
  {Mosk}},\ }\href {\doibase 10.1038/nphoton.2010.3} {\bibfield  {journal}
  {\bibinfo  {journal} {Nat. Photon.}\ }\textbf {\bibinfo {volume} {4}},\
  \bibinfo {pages} {320} (\bibinfo {year} {2010})}\BibitemShut {NoStop}%
\bibitem [{\citenamefont {Wang}\ \emph {et~al.}(2019)\citenamefont {Wang},
  \citenamefont {Liu}, \citenamefont {Huang},\ and\ \citenamefont
  {Zhao}}]{wangESEE2019}%
  \BibitemOpen
  \bibfield  {author} {\bibinfo {author} {\bibfnamefont {B.~X.}\ \bibnamefont
  {Wang}}, \bibinfo {author} {\bibfnamefont {M.~Q.}\ \bibnamefont {Liu}},
  \bibinfo {author} {\bibfnamefont {T.~C.}\ \bibnamefont {Huang}}, \ and\
  \bibinfo {author} {\bibfnamefont {C.~Y.}\ \bibnamefont {Zhao}},\ }\href
  {\doibase 10.30919/esee8c360} {\bibfield  {journal} {\bibinfo  {journal} {ES
  Energy \& Environment}\ }\textbf {\bibinfo {volume} {6}},\ \bibinfo {pages}
  {18} (\bibinfo {year} {2019})}\BibitemShut {NoStop}%
\bibitem [{\citenamefont {Castro-Lopez}\ \emph {et~al.}(2017)\citenamefont
  {Castro-Lopez}, \citenamefont {Gaio}, \citenamefont {Sellers}, \citenamefont
  {Gkantzounis}, \citenamefont {Florescu},\ and\ \citenamefont
  {Sapienza}}]{castrolopezAPLPhoton2017}%
  \BibitemOpen
  \bibfield  {author} {\bibinfo {author} {\bibfnamefont {M.}~\bibnamefont
  {Castro-Lopez}}, \bibinfo {author} {\bibfnamefont {M.}~\bibnamefont {Gaio}},
  \bibinfo {author} {\bibfnamefont {S.}~\bibnamefont {Sellers}}, \bibinfo
  {author} {\bibfnamefont {G.}~\bibnamefont {Gkantzounis}}, \bibinfo {author}
  {\bibfnamefont {M.}~\bibnamefont {Florescu}}, \ and\ \bibinfo {author}
  {\bibfnamefont {R.}~\bibnamefont {Sapienza}},\ }\href {\doibase
  10.1063/1.4983990} {\bibfield  {journal} {\bibinfo  {journal} {APL
  Photonics}\ }\textbf {\bibinfo {volume} {2}},\ \bibinfo {pages} {061302}
  (\bibinfo {year} {2017})},\ \Eprint
  {http://arxiv.org/abs/https://doi.org/10.1063/1.4983990}
  {https://doi.org/10.1063/1.4983990} \BibitemShut {NoStop}%
\bibitem [{\citenamefont {van Putten}\ \emph {et~al.}(2011)\citenamefont {van
  Putten}, \citenamefont {Akbulut}, \citenamefont {Bertolotti}, \citenamefont
  {Vos}, \citenamefont {Lagendijk},\ and\ \citenamefont
  {Mosk}}]{puttenPRL2011}%
  \BibitemOpen
  \bibfield  {author} {\bibinfo {author} {\bibfnamefont {E.~G.}\ \bibnamefont
  {van Putten}}, \bibinfo {author} {\bibfnamefont {D.}~\bibnamefont {Akbulut}},
  \bibinfo {author} {\bibfnamefont {J.}~\bibnamefont {Bertolotti}}, \bibinfo
  {author} {\bibfnamefont {W.~L.}\ \bibnamefont {Vos}}, \bibinfo {author}
  {\bibfnamefont {A.}~\bibnamefont {Lagendijk}}, \ and\ \bibinfo {author}
  {\bibfnamefont {A.~P.}\ \bibnamefont {Mosk}},\ }\href {\doibase
  10.1103/PhysRevLett.106.193905} {\bibfield  {journal} {\bibinfo  {journal}
  {Phys. Rev. Lett.}\ }\textbf {\bibinfo {volume} {106}},\ \bibinfo {pages}
  {193905} (\bibinfo {year} {2011})}\BibitemShut {NoStop}%
\bibitem [{\citenamefont {Park}\ \emph {et~al.}(2013)\citenamefont {Park},
  \citenamefont {Park}, \citenamefont {Yu}, \citenamefont {Park}, \citenamefont
  {Han}, \citenamefont {Shin}, \citenamefont {Ko}, \citenamefont {Nam},
  \citenamefont {Cho},\ and\ \citenamefont {Park}}]{parkNaturephoton2013}%
  \BibitemOpen
  \bibfield  {author} {\bibinfo {author} {\bibfnamefont {J.-H.}\ \bibnamefont
  {Park}}, \bibinfo {author} {\bibfnamefont {C.}~\bibnamefont {Park}}, \bibinfo
  {author} {\bibfnamefont {H.}~\bibnamefont {Yu}}, \bibinfo {author}
  {\bibfnamefont {J.}~\bibnamefont {Park}}, \bibinfo {author} {\bibfnamefont
  {S.}~\bibnamefont {Han}}, \bibinfo {author} {\bibfnamefont {J.}~\bibnamefont
  {Shin}}, \bibinfo {author} {\bibfnamefont {S.~H.}\ \bibnamefont {Ko}},
  \bibinfo {author} {\bibfnamefont {K.~T.}\ \bibnamefont {Nam}}, \bibinfo
  {author} {\bibfnamefont {Y.-H.}\ \bibnamefont {Cho}}, \ and\ \bibinfo
  {author} {\bibfnamefont {Y.}~\bibnamefont {Park}},\ }\href {\doibase
  10.1038/nphoton.2013.95} {\bibfield  {journal} {\bibinfo  {journal} {Nature
  Photonics}\ }\textbf {\bibinfo {volume} {7}},\ \bibinfo {pages} {454}
  (\bibinfo {year} {2013})}\BibitemShut {NoStop}%
\bibitem [{\citenamefont {Jang}\ \emph {et~al.}(2018)\citenamefont {Jang},
  \citenamefont {Horie}, \citenamefont {Shibukawa}, \citenamefont {Brake},
  \citenamefont {Liu}, \citenamefont {Kamali}, \citenamefont {Arbabi},
  \citenamefont {Ruan}, \citenamefont {Faraon},\ and\ \citenamefont
  {Yang}}]{jangNaturephoton2018}%
  \BibitemOpen
  \bibfield  {author} {\bibinfo {author} {\bibfnamefont {M.}~\bibnamefont
  {Jang}}, \bibinfo {author} {\bibfnamefont {Y.}~\bibnamefont {Horie}},
  \bibinfo {author} {\bibfnamefont {A.}~\bibnamefont {Shibukawa}}, \bibinfo
  {author} {\bibfnamefont {J.}~\bibnamefont {Brake}}, \bibinfo {author}
  {\bibfnamefont {Y.}~\bibnamefont {Liu}}, \bibinfo {author} {\bibfnamefont
  {S.~M.}\ \bibnamefont {Kamali}}, \bibinfo {author} {\bibfnamefont
  {A.}~\bibnamefont {Arbabi}}, \bibinfo {author} {\bibfnamefont
  {H.}~\bibnamefont {Ruan}}, \bibinfo {author} {\bibfnamefont {A.}~\bibnamefont
  {Faraon}}, \ and\ \bibinfo {author} {\bibfnamefont {C.}~\bibnamefont
  {Yang}},\ }\href {\doibase 10.1038/s41566-017-0078-z} {\bibfield  {journal}
  {\bibinfo  {journal} {Nature Photonics}\ }\textbf {\bibinfo {volume} {12}},\
  \bibinfo {pages} {84} (\bibinfo {year} {2018})}\BibitemShut {NoStop}%
\bibitem [{\citenamefont {Abramowitz}\ and\ \citenamefont
  {Stegun}(1964)}]{abramowitz1964handbook}%
  \BibitemOpen
  \bibfield  {author} {\bibinfo {author} {\bibfnamefont {M.}~\bibnamefont
  {Abramowitz}}\ and\ \bibinfo {author} {\bibfnamefont {I.~A.}\ \bibnamefont
  {Stegun}},\ }\href@noop {} {\emph {\bibinfo {title} {Handbook of Mathematical
  Functions: with Formulas, Graphs, and Mathematical Tables}}},\ Vol.~\bibinfo
  {volume} {55}\ (\bibinfo  {publisher} {Courier Corporation},\ \bibinfo {year}
  {1964})\BibitemShut {NoStop}%
\end{thebibliography}%

\end{document}